\pdfoutput=1
\documentclass[12pt,epsf]{report}

\renewcommand{\thefootnote}{\fnsymbol{footnote}}
\newlength{\pubnumber} 

\catcode`\@=11
\@addtoreset{equation}{section}

\def\section{\@startsection{section}{1}{\z@}{3.5ex plus 1ex minus .2ex}
 {2.3ex plus .2ex}{\large\bf}}
\def\subsection{\@startsection{subsection}{2}{\z@}{2.3ex plus .2ex}
 {2.3ex plus .2ex}{\bf}}

\usepackage{graphicx}% Include figure files
\usepackage{dcolumn}% Align table columns on decimal point
\usepackage{bm}% bold math

\usepackage[margin=1in]{geometry}

\usepackage{amsmath}
\usepackage{amsfonts}
\usepackage{multirow}
\usepackage{rotating}
\usepackage{palatino, url, multicol}
\usepackage{amssymb}
\usepackage{hyperref}
\usepackage[T1]{fontenc}
\usepackage{yfonts}
\usepackage{mathrsfs}

\def\beq{\begin{equation}}
\def\eeq{\end{equation}}
\def\beqn{\begin{eqnarray}}
\def\eeqn{\end{eqnarray}}

\def\nolabel{\nonumber }

\def\mod{{\rm mod\ }}

\def\nolabel{\nonumber }

\begin{document}

\renewcommand{\thefootnote}{\arabic{footnote}}

\begin{titlepage}
\setcounter{page}{1}
\rightline{}
\rightline{}
\rightline{\tt }

\vspace{.06in}
\begin{center}
{\Large \bf A Simple Introduction to Particle Physics \\
\rm \large Part II - Geometric Foundations and Relativity}\\
\vspace{.12in}

{\large
        Matthew B. Robinson \footnote{m\_robinson@baylor.edu},
        Tibra Ali \footnote{tibra\_ali@baylor.edu},
        Gerald B. Cleaver \footnote{gerald\_cleaver@baylor.edu}}
\\
\vspace{.12in}
{\it        Department of Physics, One Bear Place \# 97316\\
            Baylor University\\
            Waco, TX 76798-7316\\}
\vspace{.06in}
\end{center}

\begin{abstract}
This is the second in a series of papers intended to provide a basic overview of some of the major ideas in particle physics.  Part I \cite{Firstpaper} was primarily an algebraic exposition of gauge theories.  We developed the group theoretic tools needed to understand the basic construction of gauge theory, as well as the physical concepts and tools to understand the structure of the Standard Model of Particle Physics as a gauge theory.  

In this paper (and the paper to follow), we continue our emphasis on gauge theories, but we do so with a more geometrical approach.  We will conclude this paper with a brief discussion of general relativity, and save more advanced topics (including fibre bundles, characteristic classes, etc.) for the next paper in the series.  

We wish to reiterate that these notes are not intended to be a comprehensive introduction to any of the ideas contained in them.  Their purpose is to introduce the ``forest" rather than the ``trees".  The primary emphasis is on the algebraic/geometric/mathematical underpinnings rather than the calculational/phenomenological details.  The topics were chosen according to the authors' preferences and agenda.  

These notes are intended for a student who has completed the standard undergraduate physics and mathematics courses, as well as the material contained in the first paper in this series.  Having studied the material in the ``Further Reading" sections of \cite{Firstpaper} would be ideal, but the material in this series of papers is intended to be self-contained, and familiarity with the first paper will suffice.  
\end{abstract}

\end{titlepage}
\setcounter{footnote}{0}

%***********************************************************************************************

\tableofcontents

\chapter*{Preface}
\addcontentsline{toc}{chapter}{Preface}

Before diving in to the content of this paper we make a few comments about what we are doing with this series.  The ultimate origin of the first paper \cite{Firstpaper} was a series of lectures given to REU students at Baylor University between 2003 and 2009.  They began with short introductions to relativistic quantum mechanics and over time came to include particle physics, general relativity, and several other topics.  

I decided to type out the first part of these notes because at the time I was a graduate student and I knew I could save time in teaching the material if the students could just read it directly instead of waiting on me to write everything up on the board.  After doing so it was suggested that we "pretty them up" and post them online.  I found a few other people interested in coauthoring sections I wasn't as comfortable writing, and after posting, we received a tremendous amount of positive feedback about the results.   

I had no intentions of continuing the project, but since finishing graduate school (and taking a job unrelated to particle physics), I have found that working on notes like these provides a great hobby and gives me an excuse to keep thinking about these fascinating topics.  I originally planned on this second part of the "Simple Introduction" series to cover string theory, but after beginning to write found that I was spending too much time elaborating on mathematical ideas and the ideas were getting lost in the details.\footnote{This is the opposite problem most people have with the standard string theory texts - the authors assume the reader has a tremendous amount of math and therefore there are too few details.}  I therefore decided to make the second part cover some of the geometrical ideas necessary to talk about strings, thus pushing string theory to the third part.  However, in working on part II, I decided it would be fun to a do a full geometrical picture of what was done in \cite{Firstpaper}, but soon realized that if these notes were to be self-contained for a first year grad student this would be too long.  So, the "geometry of gauge theories" turned into a two part work, and string theory became part IV.  But because the first three parts would be a reasonably comprehensive introduction to non-perturbative quantum field theory and particle physics, why not include a volume on perturbative quantum field theory and particle physics (making string theory part V).  But if we've gone that far, it would be irresponsible not to include a discussion of supersymmetry and supergravity, and so on and so on.  

All that was to say that I've decided to make this a long term project as a post particle physicist hobby.  The positive side of this is that for those who seem to find the writing style, content, and approach helpful, there will be a plethora of topics covered by the time the project is finished.  The negative side is that it will take significantly longer for the series to be completed.  For those who don't like the writing style, content, or approach - it doesn't matter for them anyway.  

We'd like to again emphasize the point of these notes.  As the abstract indicates we aren't trying to provide a comprehensive (or even terribly in-depth) treatment of any of the topics we cover.  The origin of these notes is in the general plight of graduate students in theoretical particle physics who aren't able to understand everything they see immediately (I don't know anyone who can do this).  After completing the standard undergraduate courses in a physics B.S., one still has about two years of time consuming course work to get through the standard topics.  Then there's special relativity, relativistic quantum mechanics, quantum field theory, particle physics, gauge theory, general relativity, cosmology, astrophysics, gauge theories, conformal field theory, and so on.  And all that assumes a strong understanding (usually on a graduate level) of real analysis, complex analysis, linear algebra, ODE, PDE, topology, abstract algebra, algebraic topology, differential topology, differential geometry, algebraic geometry, Lie groups, complex geometry, fibre bundles, characteristic classes, and on and on and on,\footnote{As you likely realize, this list is woefully incomplete.}  And \it then \rm you're ready to start a text on string theory.  This is all extremely frustrating and overwhelming for the vast majority of us.  The level of formality in the standard texts on the more advanced topics makes learning them even more formidable.  Of course, graduate students can typically go quite a ways without understanding the nuances of each of these topics, but a fuller understanding would certainly be beneficial for most of us.  

While there are countless outstanding references on individual topics (and subtopics) on this list, gaining a sense of comfort in working with these ideas requires a holistic picture of all of them.  And unfortunately, to my knowledge there doesn't exist a single reference that puts all of this together in a single coherent picture.\footnote{We do make special mention of \cite{nakahara} and \cite{frankel}, which both come closer than most texts do, and are consequently standards on the bookshelf of most theorists.}  There are many excellent math texts on each topic, but physicists often have a hard time reading literature written by mathematicians, and the sheer number of different topics, which span countless authors with different styles, agendas, notations, etc., getting that holistic picture is rarely far removed from "frustrating and overwhelming", despite the excellence of individual texts.  Furthermore, there are many excellent physics texts on each topic, but rarely do physicists want to take the time to outline all of the details underlying the topic they're writing about.  They usually want to emphasize the physics (who can blame them), and they have a specific agenda in what they're writing - to teach the specific physical topic they've decided to write about.  Including volumes of extra math is unnecessary.  

But for those of us who can't make perfect holistic sense of these ideas the first time we come across them, getting that holistic picture is difficult.  Frequently grad students in their early year of grad school are left with a reading list of dozens, or even hundreds, of physics and math books and papers which they will have to read, understand, and assimilate into a single coherent picture in their minds.  Again - "frustrating and overwhelming" for most of us.  This is the root of this series.  

It is my hope that through the course of these notes we will cover, at least at a surface level, the majority of these topics.  As we mentioned above, these notes won't come close to a full treatment of these topics.  We warn anyone intimately familiar with any of these topics that these notes will likely annoy you due to a lack of rigor, precise language, etc.  Our goal, first and foremost, is to explain as clearly as possible where these ideas come from, what motivates them, and how they fit together.  When faced with a choice between rigor and clarity, between detail and clarity, between precision and clarity, etc., we chose clarity.  We hope that after going through these notes, someone who has never come across these ideas will have a general hook to hang each idea on to get their bearings and a context when the idea comes up later.  For example, we don't necessarily care if you understand every detail of what a harmonic differential form is, but we do want to you to be able to relate it to the star operator, Hodge decomposition, cohomology, Betti numbers, Euler characteristics, etc., and know what type of information can generally be gleaned from them.  

So, again, don't come to these notes expecting a detailed discussion of every nuance of these concepts. We wrote with someone who has never heard of a "co-exact form" or the "spin connection" before in mind.  We expect the reader to get a general picture, and then go get more details on a particular topic on their own, either through directed coursework or through relevant literature.  

Furthermore, a major purpose of these notes is explanation.  A frequent frustration students have with many of the standard texts is the lack of explanation of what certain ideas mean.  For example, most texts on general relativity will discuss the origin of the Riemann tensor (moving vectors around a rectangle, etc.).  However, after doing this they define the Ricci tensor and Ricci scalar as linear combinations of the Riemann tensor elements, completely ignoring the fact that both have their own unique geometrical meaning!  Another example is the meaning of a connection \it apart \rm from the metric connection.  Most students learn about connections from introductory general relativity or differential geometry texts, where almost the entire discussion is related to the metric connection.  Then, students come across connections on fibre bundles while trying to learn about gauge theories and have little to no context for how generalized connections work. Authors seem to assume some familiarity with them and cite the results with no explanation of meaning.  Yet another example is the modern definition of the energy-momentum tensor as the functional derivative of the Lagrangian with respect to the metric.  This list goes on an on.  We have tried to include discussions of each of these ideas rather than passing them by, followed by summaries of the main points so the details don't distract too much from the big picture.  

We do make the following warning - in \cite{Firstpaper} we were careful not to introduce a mathematical idea that wouldn't come up in the physics part.  This, we believed, worked just fine there.  However we can't make the same promise about this volume.  We will introduce several mathematical ideas\footnote{For example, Hodge decomposition, Lie groups, etc.} in the first part that won't necessarily come back into play heavily until later in the series.  

Also, we want to make the warning that the graphics for this volume were either done using graphical software or by hand with a paint software.  The the latter case we apologize in advance for the poor quality.  

The general content of this paper is as follows: We begin in chapter \ref{sec:preliminaryconcepts} with a brief overview of some terms that physicists often forget but will come in handy.  A cursory read is all that is necessary.  Chapter \ref{sec:manifolds} is then an overview of the main ideas surrounding differentiable manifolds.  The primary concepts are: manifolds, tangent spaces, cotangent spaces, vector and covector frames, tangent maps and pullbacks, integration, and Lie groups.  Chapter \ref{sec:topcons} is then a superficial overview of algebraic topology, where we begin with the most conceptually simple but computationally difficult idea (homotopy), and move to the most conceptually difficult but computationally simple idea (cohomology).\footnote{With homology in the middle.}  We discuss the relationships between these ideas throughout and relate them to the contents of chapter \ref{sec:manifolds}.  

Chapter \ref{sec:chapwithmet} then begins the real meat of these notes - differential geometry.  You can essentially view the contents of chapters \ref{sec:manifolds}-\ref{sec:topcons} as providing formalisms and tools for chapter \ref{sec:chapwithmet}.\footnote{This isn't to say that the ideas in chapters \ref{sec:manifolds}-\ref{sec:topcons} aren't important on their own.  They will be later in the series - it is only in \it this \rm paper that they play a secondary role.}  Chapter \ref{sec:chapwithmet} ties together several of the ideas in the previous chapters together to ultimately allow for a discussion of curvature through the four standard curvature tensors.\footnote{Our discussions are limited almost entirely to the Levi-Civita connection in this paper, but we do our best to prepare for how things may be generalized later.}  

Then, we begin the physics part of the paper in chapter \ref{sec:electrodynamics} with a very brief discussion of electrodynamics in various notations, how they relate, and the advantages of each.  We conclude this with an admittedly superficial discussion of the Aharonov-Bohm affect.  The primary point of discussion Aharonov-Bohm is to illustrate that electromagnetism takes a form where cohomology can be easily discussed, thus illustrating the rich topological structure to the theory.  

The final major chapter, chapter \ref{sec:gravity}, is a very surface level introduction to the main ideas of general relativity, including how the theory was formulated, what its core equations mean, its correspondence with Newtonian gravitation, and a few specific applications.  

We end chapter \ref{sec:gravity} with a discussion of general relativity as a gauge theory in section \ref{sec:generalrelativityasagaugetheory}.  This section, along with the final and very brief chapter \ref{sec:finalchap}, are the "whole point" of this volume as far as physics go.  As indicated, we discuss electrodynamics and general relativity largely to illustrate the geometry discussion in the first part of these notes.  The real agenda is to eventually get to general gauge theories.  In section \ref{sec:generalrelativityasagaugetheory} and especially chapter \ref{sec:finalchap} we introduce how the geometry of these notes generalizes to the tools needed to discuss gauge theories.  

Finally, we give some idea of where this series is going.  The first paper, \cite{Firstpaper}, introduced (mostly non-perturbative) quantum field theory and particle physics algebraically.  This paper acts as a primer for the geometry needed to reformulate non-perturbative quantum field theory geometrically.  Part III will then discuss non-perturbative quantum field theory and particle physics more fully.\footnote{We will include things like anomalies, instantons, monopoles, etc. there.}

Then, with non-perturbative quantum field theory discussed in as much detail as we feel necessary for this series, part IV will then be a detailed discussion of perturbative quantum field theory and particle physics, including discussions of evaluating Feynman diagrams, renormalization and the renormalization group, phenomenology, anomalies, etc.  

Both the third and fourth parts will likely be structured in the same way as this part - a section of mathematics followed by physics.  

Part V will then be an introduction to supersymmetry and supergravity, including both perturbative and non-perturbative topics.  After that we haven't decided entirely on the order - but the likely topics for individual parts will include: conformal field theory, introductory string theory, algebraic geometry, and advanced topics string theory.  We may also eventually include cosmology, astrophysics, and string cosmology.  

We welcome and encourage any and all questions, comments, or corrections.  While we have proofread both \cite{Firstpaper} and these notes, many mistakes were found in \cite{Firstpaper} since they were first posted, and there is little doubt that despite our best efforts this volume will be no different.  We encourage any suggestions regarding correcting mistakes of any kind.  We would also appreciate any comments regarding how we could improve these notes as far as pedagogy goes - the point is to be clear.  If you find a section particularly vague or unclear (or a section particularly clear), please let us know.  

\vspace{5mm}

Matt Robinson

\vspace{5mm}

m\_robinson@baylor.edu

\part{Mathematics}

\chapter{Preliminary Concepts}
\label{sec:preliminaryconcepts}

\section{Introductory Concepts}

In this paper an the next, our ultimate goal is to recast nearly everything we did in \cite{Firstpaper} in a more rigorous mathematical framework.  We spent the first paper learning about group theory, which was the necessary ``bare minimum" mathematical depth we needed to introduce the physical concepts we have looked at so far.  For what is to come, we will need considerably more mathematical detail, especially when we get to String Theory in the next paper.  We therefore begin this paper by looking at several foundational mathematical concepts.  You likely already have some familiarity with many of them, or have seen them in math courses before.  Our introduction here will be brief.  

\subsection{Sets}

The first mathematical object we need is called a \bf Set\rm.  A set is a collection of objects that do not necessarily have any additional structure or properties.  A collection of fruit, numbers, people, points in some space, or anything else can form a set.  While set theory is a very large and very deep branch of mathematics, we will not need to consider sets any further in what we are doing, and we therefore move on.  

\subsection{Groups}

The second mathematical objects we need are \bf Groups\rm.  We discussed groups in great length in \cite{Firstpaper}, and we therefore won't repeat that discussion.  We encourage you to reread the relevant sections there, or the references recommended in the Further Reading sections there.  

\subsection{Fields}

The third mathematical objects we need is the concept of a \bf Field\rm.  A field $F$ is a collection of objects $\{f_0,f_1,f_2,\ldots\}$ along with two operations, $+$, called addition, and $\star$, called multiplication, such that the following hold:\\
\indent $1)$ $(F,+)$ is an Abelian group with identity $f_0$.  \\
\indent $2)$ $f_i, f_j \in F \Rightarrow f_i\star f_j \in F$. \\
\indent $3)$ $f_i\star (f_j\star f_k) = (f_i\star f_j)\star f_k$. \\
\indent $4)$ $f_i\star 1 = 1 \star f_i = f_i$. \\
\indent $5)$ $f_i \star f_i^{-1} = f^{-1}_i \star f_i = 1 \; \forall i$. \\
\indent $6)$ $f_i \star (f_j+f_k) = f_i\star f_j + f_i \star f_k$ and \\
\indent $\qquad (f_i + f_j)\star f_k = f_i \star f_k + f_j \star f_k$.  \\
Then $F$ is a field.  Additionally, if\\
\indent $7)$ $f_i\star f_j = f_j \star f_i\; \forall i,j$, we say that $F$ is \bf Commutative\rm.  We do not say that $F$ is Abelian in this case because $F$ is a field, not a group.  

\subsection{Vector Spaces}
\label{sec:vectorspaces}

The fourth mathematical object we introduce is a \bf Vector Space\rm.  A vector space $V$ consists of a collection of objects $\{\bf v\it_0, \bf v\it_1, \bf v\it_2, \ldots\} \in V$ (we are using bold characters instead of the ``bar" vector notation), called vectors, and a field $F$ with $\{f_0,f_1,f_2,\ldots\}\in F$, as defined above such that the following hold: \\
\indent $1)$ $(V,+)$ is an Abelian group.\\
\indent $2)$ If $f_i \in F$ and $\bf v\it_j \in V$, then $f_i\bf v\it_j \in V$. \\
\indent $3)$ $f_i(f_j\bf v\it_k) = (f_i \star f_j) \bf v\it_k$. \\
\indent $4)$ $1\bf v\it_i = \bf v\it_i  1 = \bf v\it_i$. \\
\indent $5)$ $f_i(\bf v\it_j + \bf v\it_k) = f_i \bf v\it_j + f_i \bf v\it_k$ and \\
\indent $\qquad (\bf v\it_i \it + \bf v\it_j\it) f_k = f_k \bf v\it_i \it + f_k \bf v\it_j\rm$. 

The most familiar example of a vector space is the ``physics I" example of a collection of objects with ``magnitude and direction".  In the usual three dimensions, the vectors are $\bf \hat i \it , \bf \hat j\it, \bf \hat k \it,$ and the field $F$ is the real numbers $\mathbb{R}$.  

Despite this first example, it is important to realize that although the word ``vector" is frequently used to mean something with ``magnitude and direction", in reality the idea of a vector and a vector field is a much more general concept.  The following examples should make that clear.  

Consider the field of real numbers with vector $1$.  This (almost trivially) forms a vector space.  The field of real numbers with the vectors $1$ and $i$ also form a vector space (which is the same as the vector space with field $\mathbb{C}$ and vector $1$).  

As a third example, consider some arbitrary linear differential operator, $\mathcal{D}$, and two functions $\phi_1$ and $\phi_2$ which are solutions to $\mathcal{D}\phi_1 = \mathcal{D}\phi_2 = 0$.  Then, any linear combination of $\phi_1$ and $\phi_2$ will also be a solution.  So, $\phi_1$ and $\phi_2$ form a vector space with the field $\mathbb{R}$, $\mathbb{C}$, or even $\mathbb{Z}$.  

A final example is the set of all $N\times M$ matrices with matrix addition.  

Now, we introduce a few definitions to make some familiar ideas more precise.  We begin with a definition you are likely already familiar with, that of \bf Linear Independence\rm.  The vectors $\bf v\it_0\it, \bf v\it_1\it, \bf v\it_2\it, \ldots$ are said to be linearly independent iff 
\begin{eqnarray}
\sum_i \alpha_i\bf v\it_i\it = 0 \rightarrow \alpha_i = 0 \; \forall \; i
\end{eqnarray}
For example, there is no linear combination of the vectors 
\begin{eqnarray}
\bf \hat e\it_1\it = \begin{pmatrix} 1 \\ 0 \end{pmatrix} \qquad \qquad \bf \hat e\it_2\rm = \begin{pmatrix} 0 \\ 1 \end{pmatrix}
\end{eqnarray}
that is zero unless the coefficients of both vectors are zero.  

Next, we define the \bf Dimension \rm of a vector space.  A vector space is said to be $N$-dimensional if it is possible to find $N$ non-zero linearly independent vectors, but any set of $N+1$ vectors in the space are not linearly independent.  

Finally, we define a \bf Basis\rm.  A basis of an $N$-dimensional vector space is a set of $N$ linearly independent non-zero vectors.  It is then possible to write \it any \rm vector in the space as a linear combination of the vectors in the basis.  Because you should already be familiar with this from several undergraduate courses, we will not consider examples of these ideas.  

\subsection{Algebras}
\label{sec:algebras}

The fifth and final mathematical object we introduce is an \bf Algebra\rm.  An algebra consists of a set of vectors $V$, a field $F$, and three operations, $+$ (addition), $\star$ (scalar multiplication), and $\circ$ (vector multiplication), subject to the following constraints:\\
\indent $1)$ $V,F,+,$ and $\star$ form a vector space.\\
\indent $2)$ If $\bf v\it_i, \bf v\it_j \in V$, then $\bf v\it_i \circ \bf v\it_j \in V$.  \\
\indent $3)$ $\bf v\it_i \circ (\bf v\it_j + \bf v\it_k) = \bf v\it_i  \circ \bf v\it_j + \bf v\it_i \circ \bf v\it_k $.\\
\indent $\qquad (\bf v\it_i + \bf v\it_j) \circ \bf v\it_k = \bf v\it_i \circ \bf v\it_k + \bf v\it_j  \circ \bf v\it_k $. 

To make this more precise, we consider a few examples.  The first is the set of all $n\times n$ matrices, with matrix addition, the usual scalar multiplication, and usual matrix multiplication.  

As a second example, consider the set of all \it symmetric \rm $n\times n$ matrices.  In this case, the matrix product of two symmetric matrices is not in general a symmetric matrix, and therefore this is not an algebra.  However, if instead of usual matrix multiplication, we define the vector product of two symmetric matrices to be $\bf v\it_i  \circ \bf v\it_j \equiv \{ \bf v\it_i, \bf v\it_j\}$ (anticommutation), then we do once again have an algebra, because the anticommutator of two symmetric matrices is a symmetric matrix.  

As a third example, consider the set of all \it antisymmetric \rm $n\times n$ matrices.  This will not form an algebra with matrix multiplication because the matrix product of two antisymmetric matrices is not in general antisymmetric.  However, if we define the vector multiplication to be the commutator, $\bf v\it_i  \circ \bf v\it_j  \equiv [\bf v\it_i, \bf v\it_j]$, we once again have an algebra.  Also, notice that this algebra does not have an identity element, in that there is no $\bf v\it_e$ such that $\bf v\it_e \circ \bf v\it_i = \bf v\it_i  \circ \bf v\it_e = \pm [\bf v\it_e, \bf v\it_i] = \bf v\it_i \; \forall i$.  Furthermore, notice that this algebra is not even associative.  Clearly, it does have an identity with addition $+$ and scalar multiplication $\star$, and it is associative with both $+$ and $\star$.  

We also have the choice to impose what is typically called the \bf Derivative Rule \rm (synonymously called the \bf Leibniz Rule\rm, or the \bf Product Rule\rm)
\begin{eqnarray}
\bf v\it_i \circ (\bf v\it_j \circ \bf v\it_k) = (\bf v\it_i  \circ \bf v\it_j ) \circ \bf v\it_k + \bf v\it_j  \circ (\bf v\it_i \circ \bf v\it_k)
\end{eqnarray}

If we once again consider the set of all antisymmetric $n\times n$ matrices, imposing the Leibniz rule with the antisymmetric matrices $A$, $B$, and $C$ gives
\begin{eqnarray}
[A,[B,C]] = [[A,B],C] + [B,[A,C]]
\end{eqnarray}
which has the more familiar form
\begin{eqnarray}
[A,[B,C]]+[C,[A,B]] + [B,[C,A]] = 0
\end{eqnarray}
When written in this form, the Leibniz rule is called the \bf Jacobi Identity\rm, which is nothing more than the Leibnitz Rule.  

Our final definition is that of a \bf Lie Algebra\rm.  A Lie algebra is the algebra of antisymmetric $n\times n$ matrices which obey the Jacobi Identity.  It turns out that this definition is actually equivalent to the one we gave in the first paper, though we will not delve into the details of the equivalence now.  

You can see that in the progression from sets to groups to fields to vectors space to algebras, we are adding a new level of structure to form each object.  It should be clear from the notes proceeding this section that each of these objects is extremely useful in physics.  

It may seem that we are being extremely abstract.  In some sense we are.  But, it will be necessary to understand these concepts to at least the depths we have discussed them here.  There is tremendous depth to each of these objects, but the primary definitions are all we will need.  We will provide examples and outline properties as we move through the remainder of the notes.  

\section{References and Further Reading}

The contents of this chapter are outlined in any book on abstract algebra.  We recommend \cite{fraleigh} for an introductory discussion of each or \cite{hungerford} for a more advanced discussion of each.  

\chapter{Differential Topology}
\label{sec:manifolds}

\section{Dual Space}
\label{sec:dualspace}

Before we dive into the geometry, we will present some new algebraic concepts which will prove vital to geometry.  In fact, you will see soon that this section is actually geometry, though it won't be obvious until later.  We are introducing the following ideas with an intentional abstraction, which will prove useful when we begin to look at these ideas in various specific contexts.  

We begin with the idea of a \bf Dual Space\rm.  Let $V$ be a (finite) $N$-dimensional vector space with field $\mathbb{R}$.\footnote{Try to resist the temptation to merely picture the already familiar Euclidian vector space, and rather see it as a mathematical object as defined above.}  Clearly, we can choose some set of basis vectors (as defined in the previous section), which we denote $\bf e\it_1,\bf e\it_2,\ldots,\bf e\it_N\rm$.  We are using subscripts to label basis \it vectors \rm deliberately.  

Now, we define a new space, called the \it Dual Space \rm to $V$, which we denote $V^{\star}$.  $V^{\star}$ will also be an $N$-dimensional vector space according to the definition in the previous section.  However, while the elements of $V$ are called \it vectors \rm (or \it contravariant \rm vectors), the elements of $V^{\star}$ are called \it covectors \rm (or \it covariant \rm vectors, or \it 1-forms\rm).  Because $V^{\star}$ is an $N$-dimensional vector space, we can choose a set of basis vectors for $V^{\star}$, which we will denote $\bf e\it^1,\bf e\it^2,\ldots,\bf e\it^N\rm$.  Again, we are using superscripts to label basis \it covectors \rm deliberately.  Furthermore, a given covector will have a well defined action \it on \rm a vector (which we will discuss later), in which a covector maps a vector to a real number $\mathbb{R}$.  We can therefore define the basis vectors in $V^{\star}$ so as to maintain the desired relationship with $V$ via the constraint
\begin{eqnarray}
\bf e\it^i (\bf e\it_j) \equiv \delta^i_j \label{eq:dualbasis}
\end{eqnarray}
where the parentheses are merely to say that $\bf e\it^i$ is acting on $\bf e\it_j$.  This constraint is simply saying a basis covector in a dual space is defined as the covector which maps the corresponding basis vector to $1\in \mathbb{R}$, and all other basis vectors to $0\in \mathbb{R}$.  

We are only working with the basis vectors (for now) because we are assuming these spaces are linear.  Therefore, knowing the behavior of the basis vectors tells us the behavior of any element of the space.  

The best way think about vectors and covectors is this; a covector \it acts \rm on a vector, and the result is in $\mathbb{R}$.  To write this in mathematical language:
\begin{eqnarray}
V^{\star} \equiv \{\phi:V \rightarrow \mathbb{R}, \; linear\}
\end{eqnarray}
which reads ``$V^{\star}$ is the set of all linear mappings $\phi$ which take an element of $V$ to a real number".  

As an extremely simple illustration of this using familiar objects, let $V$ be $3$-dimensional Euclidian space, with basis \it column \rm vectors
\begin{eqnarray}
\bf e\it_1 = 
\begin{pmatrix}
1 \\ 0 \\ 0
\end{pmatrix}\qquad \bf e\it_2 = 
\begin{pmatrix}
0 \\ 1 \\ 0
\end{pmatrix} \qquad \bf e\it_3 = 
\begin{pmatrix}
0 \\ 0 \\ 1
\end{pmatrix}
\end{eqnarray}
and general element
\begin{eqnarray}
\begin{pmatrix} a \\ b \\ c \end{pmatrix} = a\bf e\it_1 + b \bf e\it_2 + c \bf e\it_3
\end{eqnarray}
where $a,b,c \in \mathbb{R}$.  The dual space to $V$ will then be the $3$-dimensional vector space with the basis of \it row \rm (co)vectors
\begin{eqnarray}
\bf e\it^1 = \begin{pmatrix} 1 & 0 & 0 \end{pmatrix}\qquad
\bf e\it^2 = \begin{pmatrix} 0 & 1 & 0 \end{pmatrix}\qquad
\bf e\it^3 = \begin{pmatrix} 0 & 0 & 1 \end{pmatrix}
\end{eqnarray}
and general element
\begin{eqnarray}
\begin{pmatrix} a & b & c \end{pmatrix} = a\bf e\it^1 + b \bf e\it^2 + c \bf e\it^3
\end{eqnarray}
where again, $a,b,c\in \mathbb{R}$.  Notice that (\ref{eq:dualbasis}) is satisfied by this choice of basis for the dual space $V^{\star}$.  

So, we have a vector space $V$, and given any vector in this space, we can ``act on it" with any covector in $V^{\star}$, where ``acting on it" in this case means the usual dot product:
\begin{eqnarray}
\begin{pmatrix} a & b & c \end{pmatrix} \begin{pmatrix} d \\ e \\ f \end{pmatrix} = 
ad + be + cf
\end{eqnarray}
(where $ad+be+cf$ is clearly an element of $\mathbb{R}$).  

It may seem strange to treat Euclidian space spanned by column vectors and Euclidian space spanned by row vectors as two different spaces, but they are in fact fundamentally different.\footnote{Admittedly, the difference is not transparent in Euclidian space.  It will be apparent in less trivial spaces.}  We will see this more clearly as we proceed.  For now, recall that we said that covectors \it act \rm on vectors, resulting in a real number.  Notice that, following usual vector multiplication, the row vector always appears to the left of the column vector (it is \it acting \rm on the vector), as
\begin{eqnarray}
\begin{pmatrix} a & b & c \end{pmatrix} \begin{pmatrix} d \\ e \\ f \end{pmatrix} \in \mathbb{R}
\end{eqnarray}
However, switching the order does not result in a real number:
\begin{eqnarray}
\begin{pmatrix} d \\ e \\ f \end{pmatrix}\begin{pmatrix} a & b & c \end{pmatrix} \notin \mathbb{R}
\end{eqnarray}
This is our first indication that $V$ and $V^{\star}$ are in fact different spaces.  

We will continue to use the convention we have established that basis vectors have lowered indices, whereas basis covectors have raised indices.\footnote{If you\label{footnotepage} are familiar with differential geometry and/or general relativity, you are likely more familiar with (contravariant) vectors having raised indices and (covariant) covectors having lowered indices.  Notice that here the indices label vectors/covectors, not the components of a vector/covector.  Eventually, we will be using the standard notation.}  It is important to realize that the column vectors and row covectors are only one example of vectors and covectors, not the \it defining \rm example.  As the examples in the last section showed there are many different types of objects that can be vectors and covectors.  

\label{saidnodotproductbetweenvectors}Incidentally, there is no such thing as a dot product between two vectors.\footnote{This is only mostly true.  We will find in chapter \ref{sec:chapwithmet} that a metric provides a way of taking a dot product of two vectors.  However, it does this by turning a vector into a covector, so our statement here is technically true.} Rather, dot products can only be defined between vectors and covectors (or synonymously, between vectors and $1$-forms).  This is what you actually have been doing whenever you take a dot product - you are mapping vectors to $\mathbb{R}$ using the dual space.  

As a second example, consider the vector space where both $V$ and the field are $\mathbb{R}$.\footnote{Recall the definition of vector space given previously. }  Clearly, the dual space will be $V^{\star} = \mathbb{R}$, because any real number multiplied by another real number will be a real number.  You can see this as a consequence of the first example because a $1$-dimensional row vector is the same as a $1$-dimensional column vector.  

A third example is Dirac's ``bra and ket" space.  The kets $|\Psi\rangle$ are the vectors, whereas the bras $\langle \Psi|$ are the covectors in the dual space.  And, as you are familiar with, $\langle \Psi_1|\Psi_2\rangle \in \mathbb{R}$.  

Now, let's say we want to use another basis for $V$ besides $\bf e_i\rm$.  We can switch to a new basis by simply doing a standard linear transformation with some non-singluar $N\times N$ matrix $T$.\footnote{This is not a change of \it coordinates\rm, but rather a \it linear \rm change of basis at a point.  Such a linear transformation may be the result of a nonlinear change of coordinates, but not necessarily.}  So, the new basis vectors will now be
\begin{eqnarray}
\bf e\it'_a = \sum_{k=1}^N(T)^k_a\bf e\it_k  \label{eq:changebasis}
\end{eqnarray}
Notice that the set of all possible basis vectors we can transform to is the set of all non-singular $N\times N$ matrices $T$, which is simply the group $GL(N)$.  

Clearly, if we have transformed into the new basis $\bf e\it'_a$, equation (\ref{eq:dualbasis}) will no longer hold if we use the untransformed basis for $V^{\star}$.  So, we need to transform the covector basis as well.  The obvious way to do this, while preserving (\ref{eq:dualbasis}), is to use the inverse of $T$, or
\begin{eqnarray}
\bf e\it'^a = \sum_{k=1}^N (T^{-1})^a_k \bf e\it^k \label{eq:otherchangebasis}
\end{eqnarray}

So now equation (\ref{eq:dualbasis}) becomes (using the summation convention where an upper and lower index being the same means they are to be summed)
\begin{eqnarray}
\bf e\it'^a(\bf e\it'_b) &=& (T^{-1})^a_k\bf e\it^k (T)^i_b\bf e\it_i \nolabel \\
&=& \bf e\it^k\bf e\it_i (T^{-1})^a_k(T)^i_b \nolabel \\
&=& \delta^k_i (T^{-1})^a_k(T)^i_b \nolabel \\ 
&=& (T^{-1})^a_k(T)^k_b \nolabel \\
&=& \delta^a_b
\end{eqnarray}
which is exactly what we would expect.  

So, we have defined vector spaces spanned by vectors which are more or less familiar objects.  Then, we defined dual spaces spanned by covectors (which are also vector spaces).  The idea is that covectors act on vectors, mapping them to real numbers.  Next, we will express this idea in a more algebraic way.  

As a final comment, because we have denoted the basis of a vector space as vectors with \it lower \rm indices, in order to preserve the summation convention (where one upper and one lower index is summed), we denote the \it components \rm of a vector with upper indices.  That way, a general vector $\bf v\rm$ with components $v^1,v^2,\ldots$ and basis $\bf e\it_1,\bf e\it_2,\ldots$ can be written as
\begin{eqnarray}
\bf v\it = v^i\bf e\it_i
\end{eqnarray}
We will discuss the transformation properties of upper and lower \it component \rm indices together shortly.  

\section{Forms}
\label{sec:forms}

Consider some vector space $V$.  We can form the product space of $p$ copies of this space as
\begin{eqnarray}
V^{\otimes p} \equiv V\otimes V \otimes V \otimes \cdots \otimes V \qquad \rm p\; times
\end{eqnarray}

In the previous section, we defined covectors as objects which take single vectors to real numbers.  These covectors form a vector space, as we saw above.  Now that we have defined the product space of several copies of a vector space, we generalize the concept of a covector as well.  We will do so in the obvious way, with one additional rule.  

First, define the vector space $\Lambda^pV$ as the $p$-linear antisymmetric product of 1-forms (covectors) which map $V^{\otimes p}$ to $\mathbb{R}$.  Or, in more mathematical language, 
\begin{eqnarray}
\Lambda^pV \equiv \{\phi: V^{\otimes p} \rightarrow \mathbb{R}, \; p \; \rm linear, \; antisymmetric\}
\end{eqnarray}
Saying that $\Lambda^pV$ is ``$p$-linear" means that it is linear in each of the $p$ variables:\footnote{This notation ($\phi (V^{\otimes p})$) means simply that the covectors of $\phi$ each act on the vectors of $V^{\otimes p}$.  For example if $p=2$ and $V = |\psi \rangle$, then an element of $\Lambda^2 V$ could be expressed as $\phi = \langle \psi'_1|\otimes \langle \psi'_2|$.  So, the action of $\phi$ on $V\otimes V$ would be 
\begin{eqnarray}
\phi(V\otimes V) = \big(\langle \psi'_1|\otimes \langle \psi'_2|\big)\big(|\psi_1\rangle \otimes \psi_2\rangle) = \big(\langle \psi'_1|\psi_1\rangle \otimes \langle \psi'_2|\psi_2\rangle\big) \nolabel 
\end{eqnarray}
},\footnote{The superscripts on the $\bf v \it$'s are merely labels, not indicators of contravariance or covariance.  Because many of our indices will be geometric (covariant/contravariant) indices, while others won't, we will differentiate them by putting indices which are merely labels (and therefore carry no geometric significance) in parentheses, and the geometric indices without parentheses.  Nonetheless, all non-geometric (purely label) indices on vectors will be superscripts, and on covectors will be subscripts.}
\begin{eqnarray}
& & \phi(V\otimes \cdots \otimes (a \bf v\it^{(i)}  + b \bf v\it^{(j)})\otimes \cdots \otimes V) \nolabel \\
&=& a \phi (V\otimes \cdots \otimes \bf v\it^{(i)} \otimes \cdots \otimes V) +b \phi (V\otimes \cdots \otimes \bf v\it^{(j)} \otimes \cdots \otimes V) 
\end{eqnarray}

The antisymmetry rule (which may seem strange to demand now, but will prove very important later) simply means that the sign changes when any two of the vectors are swapped:
\begin{eqnarray}
\phi (V\otimes \cdots \otimes \bf v\it^{(i)} \otimes \cdots \otimes \bf v\it^{(j)} \otimes \cdots \otimes V) = -\phi (V\otimes \cdots \otimes \bf v\it^{(j)} \otimes \cdots \otimes \bf v\it^{(i)} \otimes \cdots \otimes V) \nolabel
\end{eqnarray}
Clearly, a consequence of the antisymmetry is that if any vector in $V^{\otimes p}$ is a linear combination of any others, then $\phi(V^{\otimes p}) = 0$.  In other words, if the vectors in $V^{\otimes p}$ are not linearly independent, then $\phi(V^{\otimes p}) = 0$.  

We call any object that carries upper and lower indices like this a \bf Tensor\rm.  Simply put, a \label{pagewherewetalkabouttensorofvariousranks}tensor with $n$ upper indices and $m$ lower indices is called a ``tensor of rank $(n,m)$".  A tensor of rank $(n,m)$ is an object that maps a tensor of rank $(m,n)$ to $\mathbb{R}$.  For example, a tensor of rank $(0,0)$ (no indices) is simply a scalar and is already in $\mathbb{R}$.  A tensor of type $(1,0)$ is a vector, and a tensor of type $(0,1)$ is a covector.  A $p$-form is a special type of rank $(0,p)$ tensor, namely one that is totally antisymmetric.  

As another example, consider the tensor of rank $(2,1)$.  This will have two upper (contracovariant) indices and one lower (covariant) index.  In other words, it will consist of two vectors and one covector.  Therefore it will map any object with one vector and two covectors to $\mathbb{R}$.  

Obviously, $\Lambda^1V = V^{\star}$ always.  And, if $p>N$ (the dimension of $V$) then $\phi(V^{\otimes p}) = 0$ because $p>N$ vectors cannot be linearly independent.  By convention we denote $\Lambda^0V = \mathbb{R}$.  We will refer to an element of $\Lambda^pV$ as a ``p-form".  

Now, we want a way of forming tensor products of $p$-forms.  Obviously we can't do this by merely forming the straightforward product because of the antisymmetry requirement.  So, we define what is called the \bf wedge product\rm, denoted $\wedge$, defined as 
\begin{eqnarray}
\wedge : \Lambda^pV \otimes \Lambda^pV \rightarrow \Lambda^{p+q}V
\end{eqnarray}
So, $\wedge$ takes the set of all $p$-forms and the set of all $q$-forms to the set of all $(p+q)$-forms, by creating the totally antisymmetric sum of all $p$-forms tensored with all $q$-forms.  Or, more specifically, if $\phi$ is a $p$-form and $\psi$ is a $q$-form, then $\psi \wedge \psi$ is a $(p+q)$-form.  This definition preserves the antisymmetry.

To illustrate this consider a 3-dimensional vector space with basis $\bf e\it_1, \bf e\it_2,\bf e\it_3$ (try not to think of these as the standard basis in 3-dimensional Euclidian space, but rather as general vectors as defined above - they can be Euclidian vectors, matrices, or anything else which obeys the definition of a vector space).  We then naturally define a basis for the dual space as $\bf e\it^1, \bf e\it^2, \bf e\it^3$, obeying (\ref{eq:dualbasis}).  Each $\bf e\it^i$ is a 1-form, and therefore $\bf e\it^i  \in \Lambda^1V = V^{\star}$, $\forall i$, as we already know.  This just means that each 1-form $\bf e\it^i$ takes a vector $\bf e\it_j$ to $\mathbb{R}$.  

Then, we can wedge any two 1-forms together by taking the totally antisymmetric sum of each 1-form tensored together.  For example,
\begin{eqnarray}
\bf e\it^1 \wedge \bf e\it^2 &=& \bf e\it^1 \otimes \bf e\it^2 - \bf e\it^2 \otimes \bf e\it^1 \nolabel \\
\bf e\it^1 \wedge \bf e\it^3 &=& \bf e\it^1 \otimes \bf e\it^3 - \bf e\it^3 \otimes \bf e\it^1  \nolabel \\
\bf e\it^2 \wedge \bf e\it^3 &=& \bf e\it^2 \otimes \bf e\it^3 - \bf e\it^3 \otimes \bf e\it^2 \nolabel \\
\end{eqnarray}
(clearly, $\bf e\it^1 \wedge \bf e\it^1 \equiv 0$, etc. because $\bf e\it^i \wedge \bf e\it^j = -\bf e\it^j \wedge \bf e\it^i$).  So for example the element $\bf e\it_1 \otimes \bf e\it_2 \in V\otimes V$ will be operated on according to
\begin{eqnarray}
(\bf e\it^1 \wedge \bf e\it^2 )(\bf e\it_1  \otimes \bf e\it_2) &=& (\bf e\it^1 \otimes \bf e\it^2 - \bf e\it^2 \otimes \bf e\it^1 )(\bf e\it_1 \otimes \bf e\it_2) \nolabel \\
&=& \bf e\it^1 (\bf e\it_1) \otimes \bf e\it^2 (\bf e\it_2) - \bf e\it^2 (\bf e\it_1)\otimes \bf e\it^1 (\bf e\it_2) \nolabel \\
&=& 1 \otimes 1 - 0 \otimes 0 = 1-0 = 1
\end{eqnarray}
and $\bf e\it_2 \otimes \bf e\it_1$ according to
\begin{eqnarray}
(\bf e\it^1 \wedge \bf e\it^2 )(\bf e\it_2  \otimes \bf e\it_1) &=& (\bf e\it^1 \otimes \bf e\it^2 - \bf e\it^2 \otimes \bf e\it^1 )(\bf e\it_2 \otimes \bf e\it_1) \nolabel \\
&=& \bf e\it^1 (\bf e\it_2) \otimes \bf e\it^2 (\bf e\it_1) - \bf e\it^2 (\bf e\it_2)\otimes \bf e\it^1 (\bf e\it_1) \nolabel \\
&=& 0 \otimes 0 - 1 \otimes 1 =0-1= -1
\end{eqnarray}
We are only discussing the behavior of the basis vectors because of linearity - knowing the behavior of the basis vectors tells us how any point in the space acts.  If we denote the 1-forms as $\phi_1$, then an element $\phi_1\wedge \phi_1\in \Lambda^2V$ is denoted $\phi_2$.  

We can then form $\phi_1 \wedge \phi_1 \wedge \phi_1 = \phi_1 \wedge \phi_2 = \phi_3 \in \Lambda^3V$ in the same way:
\begin{eqnarray}
\bf e\it^1 \wedge \bf e\it^2  \wedge \bf e\it^3 &=& \bf e\it^1 \otimes \bf e\it^2 \otimes \bf e\it^3 + \bf  e\it^2 \otimes \bf e\it^3 \otimes \bf e\it^1 + \bf e\it^3 \otimes \bf e\it^1 \otimes \bf e\it^2 \nolabel \\
& & -\bf e\it^1 \otimes \bf e\it^3 \otimes \bf e\it^2 - \bf e\it^3 \otimes \bf e\it^2 \otimes \bf e\it^1 - \bf e\it^2 \otimes \bf e\it^1 \otimes \bf e\it^3 \label{eq:lambda3in3d}
\end{eqnarray}
(clearly, $\bf e\it^i \wedge \bf e\it^j  \wedge \bf e\it^j \equiv 0$ etc. because of antisymmetry).  So, (\ref{eq:lambda3in3d}) will be the only element of $\Lambda^3V$ in three dimensions (up to a permutation with a sign change).  

And arbitrary element of $V\otimes V\otimes V$ will be acted on according to
\begin{eqnarray}
& & (\bf e\it^1 \wedge \bf e\it^2 \wedge \bf e\it^3 )(\bf e\it_1 \otimes \bf e\it_2 \otimes \bf e\it_3) = 1 \nolabel \\
& & (\bf e\it^1 \wedge \bf e\it^2  \wedge \bf e\it^3 )(\bf e\it_1 \otimes \bf e\it_3  \otimes \bf e\it_2) = -1 \nolabel \\
& &(\bf e\it^1 \wedge \bf e\it^2  \wedge \bf e\it^3 )(\bf e\it_1 \otimes \bf e\it_1  \otimes \bf e\it_3) = 0 \nolabel \\
& & etc.
\end{eqnarray}

The idea here is that the 0-forms $\phi_0$ form a one dimensional vector space with arbitrary element (in three-dimensions)
\begin{eqnarray}
a \in \mathbb{R}
\end{eqnarray}
The 1-forms $\phi_1$ form a basis for a three dimensional vector space with arbitrary element (in three-dimensions)
\begin{eqnarray}
a(\bf e\it^1\it) + b (\bf e\it^2 \it) + c(\bf e\it^3\it)
\end{eqnarray}
(where $a,b,c\in \mathbb{R}$).  The 2-forms $\phi_2$ form a three dimensional vector space with arbitrary element (in three-dimensions)
\begin{eqnarray}
a(\bf e\it^1\it \wedge \bf e\it^2 \it) + b (\bf e\it^1\it \wedge \bf e\it^3\it) + c (\bf e\it^2\it \wedge \bf e\it^3\rm) \label{eq:asdfadsfasdfadsfasdf}
\end{eqnarray}
(where $a,b,c\in\mathbb{R}$).  Then, the 3-forms $\phi_3$ form a one dimensional vector space with arbitrary element (in three dimensions)
\begin{eqnarray}
a (\bf e\it^1\it \wedge \bf e\it^2\it \wedge \bf e\it^3\it)
\end{eqnarray}
(where $a\in \mathbb{R}$).  

So, we have four vector spaces summarized as follows:

\begin{center}
\begin{tabular}{|c||c|c|c|}
\hline
p & Dimension & Basis & Equivalent Total Space \\
\hline
\hline
0 & 1 & 1 & $\mathbb{R}$ \\
\hline
1 & 3 & $\bf e\it^i\it$ & $\mathbb{R}^3$ \\
\hline
2 & 3 & $\bf e\it^i\it \wedge \bf e\it^j\it$ & $\mathbb{R}^3$ \\
\hline
3 & 1 & $\bf e\it^i\it \wedge \bf e\it^j \it \wedge \bf e\it^k\it$ & $\mathbb{R}$ \\
\hline
\end{tabular}
\end{center}

The symmetry in the dimensionality of each space is not a coincidence.  It turns out that
\begin{eqnarray}
\dim(\Lambda^pV) = \begin{pmatrix} N \\ p \end{pmatrix} = {N! \over p!(N-p)!} \label{eq:combin}
\end{eqnarray}
(you can work out the combinatorics to derive this yourself).  Much of this may seem abstract, but all we are doing is using wedge products of covectors as basis vectors for other spaces.  For example we could define
\begin{eqnarray}
\bf i\it^3 \equiv \bf e\it^1 \wedge \bf e\it^2 \nolabel \\
\bf i\it^2 \equiv \bf e\it^1 \wedge \bf e\it^3 \nolabel \\
\bf i\it^1 \equiv \bf e\it^2 \wedge \bf e\it^2
\end{eqnarray}
and then rewrite (\ref{eq:asdfadsfasdfadsfasdf}) as
\begin{eqnarray}
c \bf i\it^1 + b \bf i\it^2 + a \bf i\it^3
\end{eqnarray}

By generalizing to a vector space of arbitrary dimension $N$, we can easily see that the wedge product is\\
\indent 1) linear - $\phi_p \wedge (a \phi_q + b \phi_r) = a \phi_p \wedge \phi_q + b \phi_p \wedge \phi_r$, where $a,b \in \mathbb{R}$ \\
\indent 2) associative - $\phi_p \wedge (\phi_q \wedge \phi_r) = (\phi_p \wedge \phi_q) \wedge \phi_r$ \\
\indent 3) graded commutative - $\phi_p \wedge \phi_q = (-1)^{pq}\phi_q \wedge \phi_p$ \\
A consequence of the third (graded commutative) is that $\phi_p \wedge \phi_p \equiv 0 $ if $p$ and $p$ are both odd.\footnote{If you have never heard of graded commutative, don't lose sleep about.  It is merely defined as stated, and knowing the definition is all that is necessary for now.}

Once we have a basis for $\Lambda^pV$, we can expand an arbitrary element in terms of this basis.  Admittedly, we have been sloppy with the normalization so far, but we were merely trying to illustrate the major points, not the details.  We will now include the proper normalization.  

In general, an arbitrary element of $\Lambda^pV$ is
\begin{eqnarray}
\phi_p = {1\over p!} \sum_{i_1,i_2,\ldots,i_p}^N \omega_{i_1,i_2,\ldots,i_p}\bf e\it^{i_1}\it \wedge \bf e\it^{i_2}\it \wedge \ldots \wedge \bf e\it^{i_p}\it \label{eq:coefficientsofeuclidianforms}
\end{eqnarray}
where the wedge products force $\omega_{i_1,i_2,\ldots,i_p}$ to be totally antisymmetric in all indices.  The reason for this normalization can be illustrated in three dimensions.  A general two-form would be expanded as
\begin{eqnarray}
\omega_{12}\bf e\it^1\wedge \bf e\it^2 + \omega_{21}\bf e\it^2\wedge\bf e\it^1 + \omega_{13}\bf e\it^1 \wedge \bf e\it^3 + \omega_{31}\bf e\it^3\wedge \bf e\it^1 + \cdots
\end{eqnarray}
The antisymmetry of the wedge product ($\bf e\it^i\wedge \bf e\it^j = - \bf e\it^j \wedge \bf e\it^i$) demands that $\omega_{ij} = -\omega_{ji}$, and therefore
\begin{eqnarray}
\omega_{12}\bf e\it^1\wedge \bf e\it^2 + \omega_{21}\bf e\it^2\wedge\bf e\it^1 + \cdots &=& \omega_{12}\bf e\it^1\wedge \bf e\it^2 - \omega_{12}(-\bf e\it^1\wedge\bf e\it^2) + \cdots \nolabel \\
&=& \omega_{12}\bf e\it^1\wedge \bf e\it^2 + \omega_{21}\bf e\it^1\wedge\bf e\it^2 + \cdots \nolabel \\
&=& 2\omega_{12} \bf e\it^1 \wedge \bf e\it^2 + \cdots
\end{eqnarray}
So we normalize this with a factor of $1/2$.  You can write out more examples to see that we will always have a factor of $1/p!$.  

Looking at our three dimensions example above, we know that an element of $\Lambda^0V = \mathbb{R}$ will simply be
\begin{eqnarray}
\phi_0 = \omega
\end{eqnarray}
where $\omega \in \mathbb{R}$.  Clearly, there is only one degree of freedom here - $\omega$ is one dimensional (corresponding to the one dimensionality of $\Lambda^0V$ in the table above).  

Then, we know that an arbitrary element of $\Lambda^1V = V^{\star}$ is
\begin{eqnarray}
\phi_1 = \sum_{i=1}^N \omega_i \bf e\it^i \label{eq:arboneform}
\end{eqnarray}
where $\omega_i$ is a real number.  The index $i$ runs from $1$ to $3$, so this has 3 degrees of freedom (corresponding to the three dimensionality of $\Lambda^1V$ in the table above).  

Then, an arbitrary element of $\Lambda^2V$ is
\begin{eqnarray}
\phi_2 = {1\over 2} \sum_{i,j=1}^N\omega_{ij} \bf e\it^i\it \wedge \bf e\it^j\it \label{eq:arbtwoform}
\end{eqnarray}
where $\omega_{ij}\in \mathbb{R}$, and the wedge product forces $\omega_{ij}$ to be antisymmetric in $i$, $j$ ($\omega_{ij} = -\omega_{ji}$).  Because $i$ and $j$ both run from $1$ to $3$, but $\omega_{ij}$ is antisymmetric, we once again have three degrees of freedom (corresponding to the three dimensionality of $\Lambda^2V$ in the table above).  

Finally, an arbitrary element of $\Lambda^3V$ is
\begin{eqnarray}
\phi_3 = {1\over 3!} \sum_{i,j,k =1}^N \omega_{ijk}\; \bf e\it^i\it \wedge \bf e\it^j\it \wedge \bf e\it^k\it \label{eq:arbthreeform}
\end{eqnarray}
where $\omega_{ijk} \in \mathbb{R}$, and the wedge product forces it to be antisymmetric in all three indices.  So, because $i$, $j$, and $k$ all run $1$ to $3$, the antisymmetry only allows a single degree of freedom (once you have chosen one non-zero value of $\omega_{ijk}$, the rest are determined by antisymmetry).  This corresponds to the one dimensionality of $\Lambda^3V$ in the table above.  

Before moving on we briefly consider the transformation properties of $\phi_i$.  It is clear that we can express a form in terms of a basis as in (\ref{eq:arboneform}), (\ref{eq:arbtwoform}), and (\ref{eq:arbthreeform}).  On the other hand, if the basis is understood, we can drop the explicit reference to the basis vectors and merely use indicial notation, referring to $\omega_i$, $\omega_{ij}$, $\omega_{ijk}$, etc.\footnote{This is exactly analogous to the choice to write a normal vector $\bf v\rm$ in terms of a basis ($\bf v\rm = a\hat i + b \hat j + c \hat k$) or in indicial notation ($\bf v\rm \dot = \{ v^i\}$, where $v^1=a$, $v^2=b$, and $v^3=c$).  However, if one were merely given the vector $(1,2,3)^T$ with no other information, we wouldn't know if this was in Cartesian, spherical, or any other coordinates.  That is why the basis must be known.}  We must merely be careful to note the range of each index and the fact that each element $\omega_{i_1,i_2,\ldots,i_p} \in \phi_p$ is totally antisymmetric in each index.  

We\label{pagewhereargumentbegins} saw above in (\ref{eq:changebasis}) and (\ref{eq:otherchangebasis}) how $\bf e\it^i\rm$ and $\bf e\it_i$ transform under some transformation $T$.  We now want to see how $\omega_{i_1,i_2,\ldots,i_p}$ transforms.  Consider $\omega_i \in \phi_1$.  Now imagine an arbitrary covector (in $\mathbb{R}^2$) graphed as
\begin{center}
\includegraphics[scale=.4]{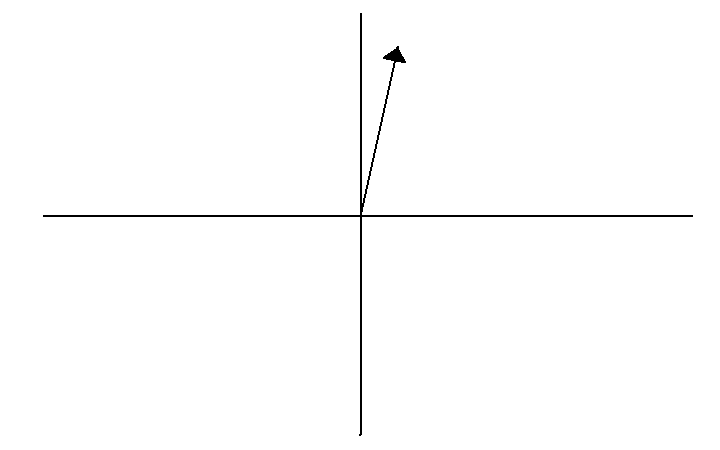}
\end{center}
If we transform the basis vectors through, say, a $45^o$ rotation without rotating the coordinates $\omega_i$, we will have
\begin{center}
\includegraphics[scale=.4]{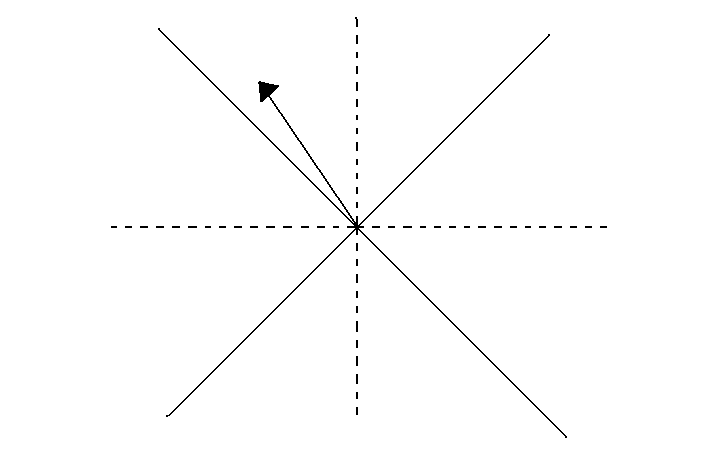}
\end{center} 
which is clearly a different location in the space.  So, in order to return the vector to its original location (described in terms of the transformed basis), we must rotate it back $45^o$:
\begin{center}
\includegraphics[scale=.4]{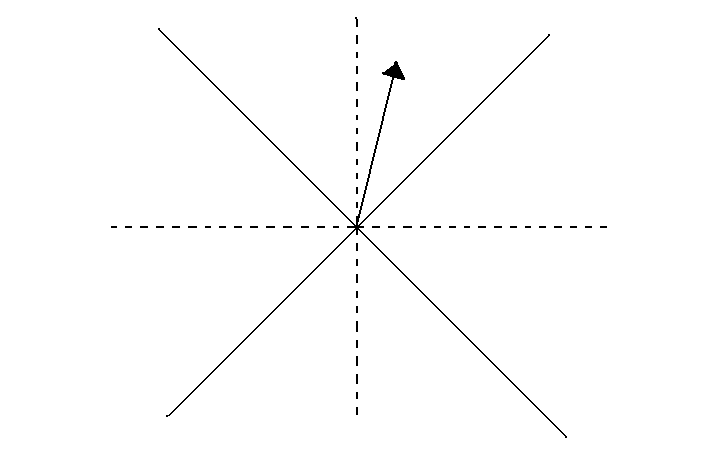}
\end{center}
So, we can see that if we rotate the basis covectors $\bf e\it^i$ according to (\ref{eq:otherchangebasis}), we must rotate the components $\omega_i$ in the ``opposite direction", or by (\ref{eq:changebasis}).  

This generalizes exactly to $\phi_p$, where all indices of $\omega_{i_1,i_2,\ldots,i_p}$ transform separately:
\begin{eqnarray}
\omega'_{i_1,i_2,\ldots,i_p} = (T)_{i_1}^{j_1}(T)_{i_2}^{j_2}\cdots(T)_{i_p}^{j_p} \omega_{j_1,j_2,\ldots,j_p}
\end{eqnarray}

Finally, by the same line of reasoning it can be seen that the components of a vector transform as
\begin{eqnarray}
v'^i = (T^{-1})^i_j v^j \label{eq:howvectorcompstransformfirsttime}
\end{eqnarray}

Before moving on, we want to once again emphasize that these are transformations of \it components\rm, not transformations of \it coordinates\rm.  For example, if we are talking about tensors in two dimensions, we may do a \it coordinate \rm transformation from Cartesian to polar coordinates.  This is a non-linear transformation that cannot be expressed using a linear transformation like $(T)^i_j$.  However, for any vector (or covector) living in this space (or its dual), we can rewrite the \it components \rm of that vector in terms of the new coordinates using the linear transformation $(T)^i_j$ at \it that \rm point.  The nonlinear \it coordinate \rm transformation induces a linear \it component \rm transformation \it at each point\rm.  We will discuss this more later.  

We can summarize what we have said so far as follows: vector spaces have bases with lowered indices which transform as in (\ref{eq:changebasis}) for some linear transformation $T$.  Dual spaces, or covector spaces, have bases with raised indices which transform as in (\ref{eq:otherchangebasis}).  An arbitrary element of a vector space (in terms of basis $\bf e\it_1$, $\bf e\it_2$, etc.) will have components with raised indices (for example $v^i$).  An arbitrary element of a covector/dual space (in terms of basis $\bf e\it^1$, $\bf e\it^2$, etc) will have components with lowered indices (for example $\omega_{i_1,i_2\ldots,i_p}$).  Objects with raised indices (vector space components and covector/dual space bases) transform the same way, while objects with lowered indices (covector/dual space components and vector space bases) transform the same way.  The transformation properties of each are related by inversion - the transformation of raised (lowered) index objets is the inverse transformation as the lowered (raised) index objects.  

Recall the footnote on page \pageref{footnotepage} where we commented on the raised/lowered index convention we are using.  The familiar notation where vectors have raised components and covectors have lowered vectors is now apparent in our construction.  The \it basis \rm vectors have ``switched" indices (basis vectors are lowered and basis covectors are raised), but the components, which are usually what is being discussed in physics, align with convention.  We are merely building these ideas in a more fundamental/formal way than is common in physics texts.  We will be very careful to maintain consistency with our notation throughout the rest of these notes - we will use raised indices when talking about vectors using indicial notation, but lowered indices when talking about the basis of a vector space.  We will use lowered indices when talking about covectors using indicial notation, but raised indices when talking about the basis of a covector/dual space (and indices in parentheses when when they are merely labels and neither is intended)

\label{whereItalkaboutcontraandcov}To further tie this into familiar concepts, we say anything with raised indices transforms in a \bf contravariant \rm way (i.e. (\ref{eq:changebasis})), and anything with lowered indices transforms in a \bf covariant \rm way (i.e. (\ref{eq:otherchangebasis})).  So, vectors and covector space bases transform contravariantly, while covectors and vector bases transform covariantly.  If you are familiar with general relativity or tensor calculus, you have likely seen much of this in the form of tensors with upper and lower indices.  Tensors with raised indices ($T^i$) are contravariant, and tensors with lowered indices ($T_i$) are covariant.  Everything that is typically done in the tensor formulation translates very naturally to the formulation we are using, with the only difference being that covariant tensors of rank $p>1$ are totally antisymmetry (which will be explained and justified later).  What is typically not discussed in general relativity is the underlying structure of the dual space and its relationship to the tangent space: $V^{\star}: V \longrightarrow \mathbb{R}$.  

Admittedly it is not obvious at this point why we want the basis vectors and the components to have opposite indices (raised/lowered).  Nor is it obvious how any of this relates to geometry.  Both of these ideas will be made clearer as we proceed.  

\subsection{Exterior Algebras}
\label{sec:exterioralgebras}

So, $\Lambda^pV$ is a vector space for each $p$.  We will find it useful to combine each $\Lambda^pV$ into one larger vector space.  Therefore, we can take a direct sum of all such spaces for a given $N$-dimensional vector space $V$.  We denote this space $\Lambda V$, and it is defined as
\begin{eqnarray}
\Lambda V \equiv \bigoplus_{p=0}^N \Lambda^p V
\end{eqnarray}
Clearly the dimension of $\Lambda V$ is the sum of the dimensions of $\Lambda^pV$ for each $p$ (for example above it was $1+3+3+1=8$).  We can also find the general dimension of an arbitrary $\Lambda V$ (using equation (\ref{eq:combin})):
\begin{eqnarray}
\dim(\Lambda V) = \sum_{p=1}^N \dim(\Lambda^p V) = \sum_{p=1}^N \begin{pmatrix} N \\ p \end{pmatrix} = 2^N
\end{eqnarray}

It turns out that the $2^N$-dimensional vector space $\Lambda V$, along with the wedge product $\wedge$, satisfies the definition of an algebra (cf section \ref{sec:algebras}).  Such algebras are called \bf Exterior Algebras \rm (they are also sometimes referred to as \bf Grassmann Algebras\rm).  

We have not spent a great deal of time building the concept of the exterior algebra, which (as the name implies) is a purely algebraic idea - not an obviously geometrical one.  However, as we will see, this exterior algebra will prove to be a powerful geometric tool.  

Before we begin the uphill climb towards a full geometric theory, we will look at a few examples of how these simple forms (which we will discuss in $\mathbb{R}^3$ for simplicity and familiarity in this section) can be looked at geometrically.  

Consider, in $\mathbb{R}^3$, the set of all 1-forms ($\phi_1$, c.f. equation (\ref{eq:arboneform})), where we are assuming the basis $\bf e\it^1$, $\bf e\it^2$, and $\bf e\it^3$ is given.  Consider two specific vectors in this vector space:\footnote{We are temporarily limiting ourselves to two dimensions here for simplicity - we will build to three dimensions shortly.}
\begin{eqnarray}
\bf \Omega\it_1 &\equiv& \omega_1\bf e\it^1 + \omega_2\bf e\it^2 \nolabel \\
\bf \Omega\it_2 &\equiv& \omega'_1\bf e\it^1 + \omega'_2\bf e\it^2 \label{eq:firstcovectors}
\end{eqnarray}
Then, the wedge product will be
\begin{eqnarray}
\bf \Omega\it_1 \wedge \bf \Omega\it_2 &=& (\omega_1\bf e\it^1 + \omega_2\bf e\it^2) \wedge (\omega'_1\bf e\it^1 + \omega'_2\bf e\it^2) \nolabel \\
&=& \omega_1\omega'_1 (\bf e\it^1 \wedge \bf e\it^1) + \omega_1\omega'_2 (\bf e\it^1 \wedge \bf e\it^2) + \omega_2 \omega'_1 (\bf e\it^2 \wedge \bf e\it^1) + \omega_2\omega'_2 (\bf e\it^2 \wedge \bf e\it^2) \nolabel \\
&=& (\omega_1\omega'_2 - \omega_2\omega'_1)(\bf e\it^1 \wedge \bf e\it^2)
\end{eqnarray}
where we made use of the antisymmetry of the wedge product to get the last equality.  

Notice that this has the form
\begin{eqnarray}
\bf \Omega\it_1 \wedge \bf \Omega\it_2 = \det\begin{pmatrix} \omega_1 & \omega_2 \\ \omega'_1 & \omega'_2 \end{pmatrix} (\bf e\it^1 \wedge \bf e\it^2) \label{eq:firstdet}
\end{eqnarray}
and for this reason we can see that the coefficient of $(\bf e\it^1 \wedge \bf e\it^2)$ is the area of the parallelogram formed by the covectors (\ref{eq:firstcovectors}).\footnote{Note that this area is \it directed\rm, meaning that it can be negative - this is an important property and we will say more on this later when we discuss orientation.}
\begin{center}
\includegraphics[scale=.4]{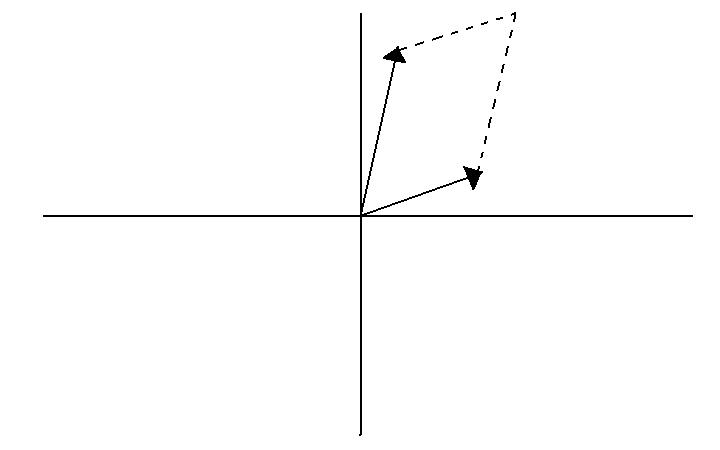}
\end{center}

Now, consider the three dimensional case with
\begin{eqnarray}
\bf \Omega\it_1 &\equiv& \omega_1 \bf e\it^1+ \omega_2 \bf e\it^2 + \omega_3 \bf e\it^3 \nolabel \\
\bf \Omega\it_2 &\equiv& \omega'_1 \bf e\it^1+ \omega'_2 \bf e\it^2 + \omega'_3 \bf e\it^3 \nolabel \\
\bf \Omega\it_3 &\equiv& \omega''_1 \bf e\it^1+ \omega''_2 \bf e\it^2 + \omega''_3 \bf e\it^3 \label{eq:threecovectors}
\end{eqnarray}
So, we can see
\begin{eqnarray}
\bf \Omega\it_1 \wedge \bf \Omega\it_2 &=& (\omega_1 \bf e\it^1+ \omega_2 \bf e\it^2 + \omega_3 \bf e\it^3) \wedge (\omega'_1 \bf e\it^1+ \omega'_2 \bf e\it^2 + \omega'_3 \bf e\it^3) \nolabel \\
&=& (\omega_1 \omega'_2 - \omega_2\omega'_1)(\bf e\it^1 \wedge \bf e\it^2) + (\omega_1 \omega'_3 - \omega_3\omega'_1)(\bf e\it^1 \wedge \bf e\it^3) \nolabel \\
& & + \;\;(\omega_2 \omega'_3 - \omega_3\omega'_2)(\bf e\it^2 \wedge \bf e\it^3)
\end{eqnarray}
Notice that, similarly to (\ref{eq:firstdet}), we can write this as\footnote{Note that this looks very similar to the form of the cross product $\bf A\rm \times \bf B$ learned in $E\&M$.  While this is not exactly a cross-product, it is related, and we will discuss the relationship later in these notes.}
\begin{eqnarray}
\bf \Omega\it_1 \wedge \bf \Omega\it_2  &=& \det \begin{pmatrix} \omega_1 & \omega_2 \\ \omega'_1 & \omega'_2 \end{pmatrix}(\bf e\it^1 \wedge \bf e\it^2) + \det \begin{pmatrix} \omega_1 & \omega_3 \\ \omega'_1 & \omega'_3 \end{pmatrix}(\bf e\it^1 \wedge \bf e\it^3) \nolabel \\
& & +\det \begin{pmatrix} \omega_2 & \omega_3 \\ \omega'_2 & \omega'_3 \end{pmatrix}(\bf e\it^2 \wedge \bf e\it^3) \nolabel \\
&=& \det 
\begin{pmatrix} 
\bf e\it^2 \wedge \bf e\it^3 & \bf e\it^1 \wedge \bf e\it^3& \bf e\it^1 \wedge \bf e\it^2 \\
\omega_1 & \omega_2 & \omega_3 \\
\omega'_1 & \omega'_2 & \omega'_3 \\
\end{pmatrix} \label{eq:earlierequationwheretwoomegaswedgedtogetherlookslikecrossproduct}
\end{eqnarray}
Then, we can write (sparing some tedium)
\begin{eqnarray}
& &\bf \Omega\it_1 \wedge \bf \Omega\it_2 \wedge \bf \Omega\it_3 \nolabel \\
& & = \big[ \omega_1(\omega'_2\omega''_3 - \omega'_3\omega''_2) - \omega_2(\omega'_1\omega''_3 - \omega'_3\omega''_1)+ \omega_3(\omega'_1\omega''_2 - \omega'_2\omega''_1) \big] (\bf e\it^1 \wedge \bf e\it^2 \wedge \bf e\it^3) \nolabel \\
\end{eqnarray}
We can rewrite this as
\begin{eqnarray}
& &\bf \Omega\it_1 \wedge \bf \Omega\it_2 \wedge \bf \Omega\it_3\rm \nolabel \\
& & = \det \begin{pmatrix}
\omega_1 & \omega_2 & \omega_3 \\
\omega'_1 & \omega'_2 & \omega'_3 \\
\omega''_1 & \omega''_2 & \omega''_3 \\
\end{pmatrix}(\bf e\it^1 \wedge \bf e\it^2 \wedge \bf e\it^3)
\end{eqnarray}
which we can associate with the volume of the (three-dimensional) parallelepiped spanned by the covectors (\ref{eq:threecovectors}):
\begin{center}
\includegraphics[scale=.25]{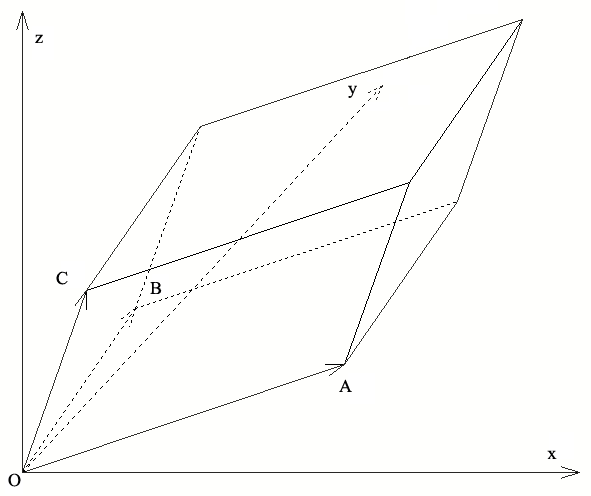}
\end{center}

We are finding that wedge products can be written in terms of determinants.  This is not a coincidence, but rather is part of much deeper geometric machinery.  We will explore this later in these notes.\label{wherewetalkaboutdeterminantsinwedgeproducts}  

In summary for this section, we see that even in these simple examples of forms in $\mathbb{R}^3$ there are clear geometric interpretations.  However, our reasoning here has depended on the fact that the space, or manifold,\footnote{If you aren't familiar with what a ``manifold" is, just replace that word with ``space".  We will talk about manifolds later in our exposition.} we are in is $\mathbb{R}^3$.  For less trivial spaces, we will have more work to do in later sections.  We are merely trying to give some idea of how these objects can have geometric meaning before branching into a fuller geometric exposition.  

The point for now is to understand the the set of all forms, made of antisymmetric tensor products (or wedge products) of the covectors in the dual space, form an algebra (as defined in section \ref{sec:algebras}), where each element contains both algebraic and geometric information about the types of objects that can exist in the space.  

\subsection{Vector Space Valued Forms}

Previously we discussed $p$-forms which act on vectors in some vector space $V$, mapping them to $\mathbb{R}$.  For example (in two dimensions) the tensor product 
\begin{eqnarray}
a^{11}(\bf e\it_1 \otimes \bf e\it_1) +\it a^{12} (\bf e\it_1 \otimes \bf e\it_2) +\it a^{21}(\bf e\it_2 \otimes \bf e\it_1)+ \it a^{22}(\bf e\it_2\otimes \bf e\it_2) \label{eq:firstvector}
\end{eqnarray}
(where $a^{ij}\in \mathbb{R} \; \forall \; i,j$) is acted on by the $2$-form 
\begin{eqnarray}
b_{12}(\bf e\it^1 \wedge \bf e\it^2) =\it  b_{12} (\bf e\it^1 \otimes \bf e\it^2 - \bf e\it^2 \otimes \bf e\it^1) \label{eq:firstform}
\end{eqnarray}
(where $b_{12} \in \mathbb{R}$) to give
\begin{eqnarray}
b_{12}(\bf e\it^1 &\wedge& \bf e\it^2\it) (a^{11}(\bf e\it_1 \otimes \bf e\it_1) +\it a^{12} (\bf e\it_1 \otimes \bf e\it_2) +\it a^{21}(\bf e\it_2 \otimes \bf e\it_1)+ \it a^{22}(\bf e\it_2\otimes \bf e\it_2)) \nolabel \\
&=& b_{12}a^{11}(1\otimes 0) + b_{12}a^{12}(1\otimes 1) - b_{12}a^{21}(1 \otimes 1) + b_{12}a^{22}(0 \otimes 1) \nolabel \\
&=& b_{12}a^{12} - b_{12}a^{21} \in \mathbb{R}
\end{eqnarray}
In other words, the form (\ref{eq:firstform}) maps the tensor (\ref{eq:firstvector}) to a real number.  We now want to consider the possibility of a form mapping a tensor to an element of a vector space.  Specifically, we will be working with the vector spaces $\mathbb{C}$, a Lie algebra, or a vector space which carries a representation of some Lie algebra (meaning that it transforms under that representation of that Lie group).  

Consider vector spaces $V$ and $W$.  $V$ will be the vector space whose dual space contains the forms we will be considering (in other words, $V$ means the same thing it did above).  $W$ will be the vector space the forms will map $V$ to.  

Our notation will be as follows:
\begin{eqnarray}
\Lambda^p(V,W) = \{ \phi : V^{\otimes p} \longrightarrow W,\; \it multilinear,\; antisymmetric\}
\end{eqnarray}
which reads ``$\Lambda^p(V,W)$ consists of the set of all multilinear and antisymmetric functions $\phi$ which map tensor products on the vector space $V$ to elements in the vector space $W$.  So $\Lambda^1(V,W)$ is simply the set of all linear maps from $V$ to $W$.  

The best way to think about this is that an element of $\Lambda^p(V,W)$ is a \it vector \rm in $W$ whose components are not real numbers, but rather real valued $p$-forms.  To see this assume the dimension of $W$ is $w$, and let $\bf a\it^i,\; i=1,\cdots w$, be a basis for $W$.  Then, we can express any $\phi \in \Lambda^p(V,W)$ as 
\begin{eqnarray}
\phi = \sum_{i=1}^w \bf a\it^i \phi_i \label{eq:expansionofphi}
\end{eqnarray}
where $\phi_i \in \Lambda^pV$.  

As a simple example, consider $V=\mathbb{R}^3$ and $W=\mathbb{R}^2$ and $p=2$.  Then (using (\ref{eq:expansionofphi}) as a guide, we want $\phi_i \in \Lambda^2V$.  Then, we choose the standard basis ($\bf\hat a\it^1,\; \bf \hat a\it^2$) for $W=\mathbb{R}^2$.  So, an example of an element of $\Lambda^2(\mathbb{R}^3,\mathbb{R}^2)$ would be\footnote{This is not the most general element, just an example.}
\begin{eqnarray}
\bf \hat a\it^1 (b_{12}\bf e\it^1 \wedge \bf e\it^2) + \bf \hat a\it^2 (c_{23}\bf e\it^2 \wedge \bf e\it^3) \label{eq:secondform}
\end{eqnarray}
(where $b_{12},c_{23}\in\mathbb{R}$).  If this acts on an arbitrary tensor \footnote{This is not the most general tensor, just an example.}
\begin{eqnarray}
d^{11}(\bf e\it_1 \otimes \bf e\it_1) + \it d^{12}(\bf e\it_1 \otimes \bf e\it_2) +\it  d^{32}(\bf e\it_3 \otimes \bf e\it_2) \label{eq:secondtensor}
\end{eqnarray}
(where $d^{12},d^{12},d^{32}\in \mathbb{R}$), we get
\begin{eqnarray}
\big( \bf \hat a\it^1 (b_{12}\bf e\it^1 \wedge \bf e\it^2) &+& \bf \hat a\it^2 (c_{32}\bf e\it^2 \wedge \bf e\it^3)\big) \big(\it d^{11}(\bf e\it_1 \otimes \bf e\it_1) + \it d^{12}(\bf e\it_1 \otimes \bf e\it_2) +\it  d^{32}(\bf e\it_3 \otimes \bf e\it_2) \big) \nolabel \\
&=& b_{12}d^{12}\bf \hat a\it^1 (1 \otimes 1) - c_{23}d^{21} \bf \hat a\it^2 (1 \otimes 1) \nolabel \\
&=& b_{12}d^{12} \bf \hat a\it^1 - c_{23}d^{21} \bf \hat a\it^2 \in \mathbb{R}^2
\end{eqnarray}
So, the form (\ref{eq:secondform}) maps tensor (\ref{eq:secondtensor}) to $\mathbb{R}^2$.  

Notice that the form (\ref{eq:secondform}) has a problem if we want to define a wedge product between these forms.  Consider, for example, trying to wedge (\ref{eq:secondform}) with, say,  
\begin{eqnarray}
\bf \hat a\it^1 (f_1 \bf e\it^1) + \bf \hat a\it^3 (f_3\bf e\it^3)
\end{eqnarray}
we get 
\begin{eqnarray}
\big[ \bf \hat a\it^1 (b_{12}\bf e\it^1 \wedge \bf e\it^2) + \bf \hat a\it^2 (c_{23}\bf e\it^2 \wedge \bf e\it^3) \big] \wedge \big[(\bf \hat a\it^1 (f_1 \bf e\it^1) + \bf \hat a\it^3 (f_3\bf e\it^3) \big] \label{eq:newwedge}
\end{eqnarray}
This cannot be defined unless we have some well-defined way of multiplying (unit) vectors in $W$.  If we do assume such a multiplication (denoted $\bf \hat a\it^i \circ \bf \hat a\it^j$), then (\ref{eq:newwedge}) becomes
\begin{eqnarray}
& &b_{12}f_3(\bf \hat a\it^1 \circ \bf \hat a\it^3) (\bf e\it^1 \wedge \bf e\it^2 \wedge \bf e\it^3) + c_{21}f_1 (\bf \hat a\it^2 \circ \bf \hat a\it^1) (\bf e\it^1 \wedge \bf e\it^2 \wedge \bf e\it^3) \nolabel \\
& =& \big[b_{12}f_3 (\bf \hat a\it^1 \circ \bf \hat a\it^3) + c_{21}f_1 (\bf \hat a\it^2 \circ \bf \hat a\it^1)\big](\bf e\it^1 \wedge \bf e\it^2 \wedge \bf e\it^3) \label{eq:circwedge}
\end{eqnarray}

As another example, let $W$ be the Lie algebra of, say, $SU(2)$.  This algebra consists of three generators $T^1,T^2$ and $T^3$.\footnote{We could specify a specific representation, for example the fundamental representation with the Paul matrices, but we will maintain generality for now.}  In this case, the multiplication of vectors in (\ref{eq:circwedge}) is the commutator.  So, (rewriting $\bf \hat a\it^i$ as $T^i$), we have 
\begin{eqnarray}
& &\big[b_{12}f_3[T^1,T^3] + c_{21}f_1[T^2,T^1]\big] (\bf e\it^1 \wedge \bf e\it^2 \wedge \bf e\it^3) \nolabel \\
&=& -i\big[b_{12}f_3 T^2 +c_{21}f_1 T^3\big](\bf e\it^1 \wedge \bf e\it^2 \wedge \bf e\it^3)
\end{eqnarray}

So, when we generalize to vector space valued forms, we must have a clear way of multiplying the vectors in $W$.  

\subsection{Pullbacks}
\label{sec:pullbacks}

Consider two vectors spaces $M$ and $N$, of dimension $m$ and $n$ respectively.  Now, consider a linear map $f$ from $M$ to $N$:
\begin{eqnarray}
f:M\longrightarrow N \qquad linear \nolabel 
\end{eqnarray}
In general, this map cannot be uniquely inverted.  Consider the very simple example where $M=\mathbb{R}^2,\; N=\mathbb{R}$, and $f(x,y)=(x+y)$, where $x,y,f(x,y) \in \mathbb{R}$.  Given, say, $f(x,y)=8$, there are an infinite number of choices for $x,y$ which produce this ($(8,0)$, $(10,-2)$, etc).    

While we cannot in general invert the mapping $f$, it turns out that $f$ does create a well-defined map from the \it dual space \rm of $N$ to the \it dual space \rm of $M$.  We denote this ``induced" map $f^{\star}$, and say
\begin{eqnarray}
f^{\star}: \Lambda^pN \longrightarrow \Lambda^pM
\end{eqnarray}
Note that $f$ mapped from $M$ to $N$, whereas $f^{\star}$ maps from $N$ to $M$.  

We define $f^{\star}$ as follows: given a tensor product of $p$ vectors in $M$, which we will denote $(\bf m\it^{(1)}\it, \bf m\it^{(2)}\it, \ldots, \bf m\it^{(p)})$, we know that it is mapped to a tensor product of $p$ vectors in $N$ by $f$ as
\begin{eqnarray}
(\bf m\it^{(1)}, \bf m\it^{(2)}\it, \ldots, \bf m\it^{(p)}) \longmapsto (f(\bf m\it^{(1)}), f(\bf m\it^{(2)}), \ldots, f(\bf m\it^{(p)})) \label{eq:righthandsidemmm}
\end{eqnarray}
(where $f(\bf m\it^{(i)})$ is understood to be a vector).  Acting on the right hand side of (\ref{eq:righthandsidemmm}) with a $p$-form in the dual space of $N$ (denoted $\phi_p$), it is mapped to $\mathbb{R}$:
\begin{eqnarray}
\phi_p(f(\bf m\it^{(1)}), f(\bf m\it^{(2)}), \ldots, f(\bf m\it^{(p)})) \longmapsto \mathbb{R}
\end{eqnarray}
The map $f^{\star}$ will, as stated above, map a $p$-form $\phi_p$ in the dual space of $N$ (denoted $N^{\star}$) to a $p$-form in the dual space of $M$ (denoted $M^{\star}$), which we denote $\psi_p$.  We define the exact action of $f^{\star}$ on $\phi_p \in N^{\star}$ as follows: 
\begin{eqnarray}
(f^{\star}\phi_p)(\bf m\it^{(1)}, \bf m\it^{(2)}, \ldots, \bf m\it^{(p)}) &=& \psi_p(\bf m\it^{(1)}, \bf m\it^{(2)}, \ldots, \bf m\it^{(p)}) \nolabel \\
&=& \phi_p(f(\bf m\it^{(1)}), f(\bf m\it^{(2)}), \ldots, f(\bf m\it^{(p)})) \label{eq:defpullback}
\end{eqnarray}

We call $f^{\star}$ the \bf pullback \rm induced by $f$.  Admittedly, this definition is cumbersome and difficult to follow.  Bear with us for a few more lines, and then we will consider an example.  

It will be helpful to write out $f^{\star}$ in terms of specific bases of $M$ and $N$.  Again taking the dimension of $M$ to be $m$ and the dimension of $N$ to be $n$, let $\bf a\it_i$ for $i=1,\ldots, m$ be a basis for $M$, and let $\bf b\it_i$ for $i=1,\ldots, n$ be a basis for $N$.  Then let $\bf A\it^i\it$ for $i=1,\ldots, m$ be a dual basis for $M^{\star}$, and let $\bf B\it^i\it$ for $i=1,\ldots,n$ be a dual basis for $N^{\star}$.  Knowing $f$, it is easy to write out the mappings from the basis for $M$ to the basis for $N$, generating an $n\times m$ matrix $f^i_j$ (where $i$ runs $1$ to $n$ and $j$ runs $1$ to $m$):
\begin{eqnarray}
f(\bf a\it_j\it) = \sum_{i=1}^n f^i_j \bf b\it_i\it \label{eq:fmatrix}
\end{eqnarray}
The components of $f^{\star}$ with respect to the dual bases is then given by
\begin{eqnarray}
f^{\star}(\bf B\it^i\it) = \sum_{j=1}^m f^i_j \bf A\it^j\it \label{eq:fstarmatrix}
\end{eqnarray}
Due to linearity, knowing how $f^{\star}$ acts on the basis covectors tells us how it acts on the entire space.  

We now consider an example.  

Let $M=\mathbb{R}^2$ and $N=\mathbb{R}^3$.  Then let $f$ be defined by 
\begin{eqnarray}
\it f:c^1\bf a\it_1\it + c^2 \bf a\it_2\it \longmapsto (c^1+3c^2)\bf b\it_1\it + (c^2)\bf b\it_2\it + (c^1+2c^2)\bf b\it_3\it
\end{eqnarray}
Clearly this cannot be in inverted in a well-defined way.  But, using (\ref{eq:fmatrix}) it is easy to write out the $3\times 2$ matrix $f^i_j$.  We have 
\begin{eqnarray}
\bf a\it_1\it &\longmapsto& \bf b\it_1\it + \bf b\it_3\it \nolabel \\
\bf a\it_2\it &\longmapsto& \it 3\bf b\it_1\it+\bf b\it_2\it+ 2 \bf b\it_3\it 
\end{eqnarray}
so
\begin{eqnarray}
f^i_j \dot = 
\begin{pmatrix}
1 & 3 \\
0 & 1 \\
1 & 2 \\
\end{pmatrix}
\end{eqnarray}
So, using (\ref{eq:fstarmatrix}) we can write out the action of $f^{\star}$ on the bases of the exterior algebras:
\begin{eqnarray}
f^{\star}(\bf B\it^1\it) &=& \bf A\it^1\it + 3\bf A\it^2\it \nolabel \\
f^{\star}(\bf B\it^2\it) &=& \bf A\it^2\it \nolabel \\
f^{\star}(\bf B\it^3\it) &=& \bf A\it^1\it + 2\bf A\it^2\it  \label{eq:actionoffstaronbasis}
\end{eqnarray}

To see what we have done, consider the vector
\begin{eqnarray}
2\bf a\it_1\it - \it 3 \bf a\it_2\it \in M \label{eq:thisvector}
\end{eqnarray}
This will be mapped by $f$ to
\begin{eqnarray}
\it -7\bf b\it_1\it - 3 \bf b\it_2\it - 4 \bf b\it_3\it \in N 
\end{eqnarray}
Now consider the $1$-form
\begin{eqnarray}
2\bf B\it^1\it -2\bf B\it^2\it + \bf B\it^3 \it \in \Lambda^1N \label{eq:thisform}
\end{eqnarray}
This will map (\ref{eq:thisvector}) to $\mathbb{R}$ as
\begin{eqnarray}
& &\big(2\bf B\it^1\it -2\bf B\it^2\it + \bf B\it^3 \it \big)\big(-7\bf b\it_1\it - 3 \bf b\it_2\it - 4 \bf b\it_3\it \big) \nolabel  \\
&&=-14+6-4 = -12\label{eq:finalanswerMN}
\end{eqnarray}

Now we can use (\ref{eq:actionoffstaronbasis}) to induce $f^{\star}$ on (\ref{eq:thisform}), giving
\begin{eqnarray}
f^{\star}(2\bf B\it^1\it -2\bf B\it^2\it + \bf B\it^3 \it ) &=& 2f^{\star}(\bf B\it^1\it) - 2f^{\star}(\bf B\it^2\it) + f^{\star}(\bf B\it^3\it) \nolabel \\
&=& 2(\bf A\it^1\it + 3\bf A\it^2\it) - 2(\bf A\it^2\it) + (\bf A\it^1\it + 2\bf A\it^2\it)\nolabel \\
&=& 3\bf A\it^1\it + 6\bf A\it^2\it \label{eq:finallyaforminM}
\end{eqnarray}
Finally, acting on (\ref{eq:thisvector}) with (\ref{eq:finallyaforminM}), we get
\begin{eqnarray}
\big(3\bf A\it^1\it + 6\bf A\it^2\it\big)\big(2\bf a\it_1\it - \it 3 \bf a\it_2\it\big) = 6-18 = -12
\end{eqnarray}
which agrees with (\ref{eq:finalanswerMN}).  

It is instructive to think about this example in terms of the more general definition of $f^{\star}$ given in (\ref{eq:defpullback}).  

Once we know how how $f^{\star}$ behaves on $\Lambda^1N$, we can extend it to arbitrary $\Lambda^pN$ using the following properties:\\
\indent 1) linearity: $f^{\star}(a \phi_p + b \phi'_q) = af^{\star}(\phi_p) + b f^{\star}(\phi'_q)$ \\
\indent 2) $f^{\star}(\phi_p \wedge \phi'_q) = (f^{\star} \phi_p) \wedge (f^{\star}\phi'_q)$ 

Admittedly, the \it point \rm of the pullback $f^{\star}$ likely isn't clear yet.  It will, however, become more apparent as we delve further into differential geometry.  The important thing to remember for now is that if we have two vector spaces $M$ and $N$ and a map between them, then any $p$-form on the target manifold can be ``pulled back" to a well-defined $p$-form on the original manifold.  We are introducing the idea now to have the building blocks established when we come back to them.  We ask for your patience until we can make their clearer.

\section{Differential Manifolds}
\label{sec:differentialmanifolds}

\subsection{Informal Definition}
\label{sec:informaldefinition}

In the next section, we will begin studying vectors, tensors, and forms that live on more general differential\footnote{We will drop the word ``differential" and merely refer to manifolds, though we will be working with differential manifolds throughout this paper.} manifolds than $\mathbb{R}^n$.  But before looking at what types of objects can live on a general manifold, we must understand what a manifold is.  

To get a feel for what a manifold is, we begin with an informal, more intuitive description/definition.  

In simple terms, an $n$-dimensional manifold is a space $\mathcal{M}$ where \it any \rm point $x\in \mathcal{M}$ has a neighborhood that looks like $\mathbb{R}^n$.  It is not necessary, however, that $\mathcal{M}$ look \it globally \rm like $\mathbb{R}^n$.  For example consider the circle (denoted $S^1$).  Clearly $S^1$ is a one-dimensional space, but globally it (obviously) looks nothing like $\mathbb{R}^1$, the real line.  However, if you zoom in on a small section of $S^1$ around any point $x\in S^1$, it does look \it locally \rm like a small section of $\mathbb{R}^1$.  And the more you ``zoom in", the more it looks like $\mathbb{R}^1$ (that is to say the more ``flat" it looks).  
\begin{center}
\includegraphics[scale=.5]{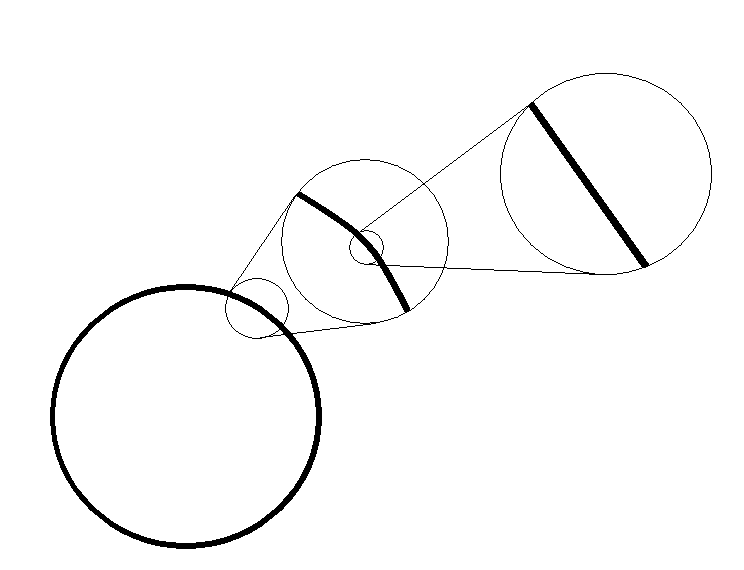}
\end{center}

Another example is the Earth.  From most locations on the Earth, it appears that the surface is flat.  So it is locally similar to $\mathbb{R}^2$.  As we know, however, the Earth does not have the global structure of $\mathbb{R}^2$, but rather of the two dimensional sphere, denoted $S^2$.  

Another example of a manifold is the two dimensional torus $T^2$.  
\begin{center}
\includegraphics[scale=.5]{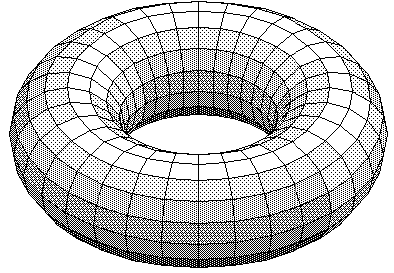} \label{pagewithtorusinchapter2}
\end{center}

Locally, it also looks like $\mathbb{R}^2$, but globally it has a very different structure.  

On the other hand, the ``figure-eight" space shown below 
\begin{center}
\includegraphics[scale=.5]{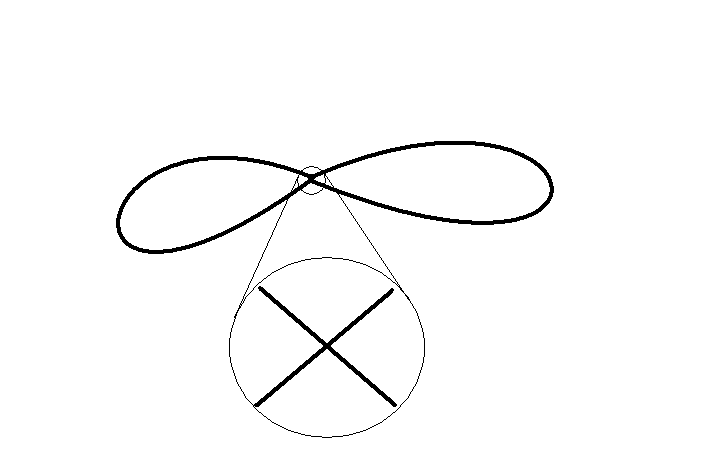}
\end{center}
is not a manifold.  Every point in this space is locally similar to $\mathbb{R}^1$ except the point in the middle.  No matter how small a neighborhood you choose, it maintains the general structure of a cross, which is not similar to $\mathbb{R}^n$ for any $n$.  

Notice that a space being a manifold depends on its \it topological \rm structure, not on its geometric structure.  This means that it depends on the \it qualitative \rm shape rather than the \it quantitative \rm shape.  To illustrate this, consider a square and a circle.  They are topologically, or qualitatively, the same thing.  They are both one dimensional loops.  In fact, circles, squares, rectangles, ovals, \it anything \rm that a circle can be stretched, twisted, etc. into without breaking it.  
\begin{center}
\includegraphics[scale=.45]{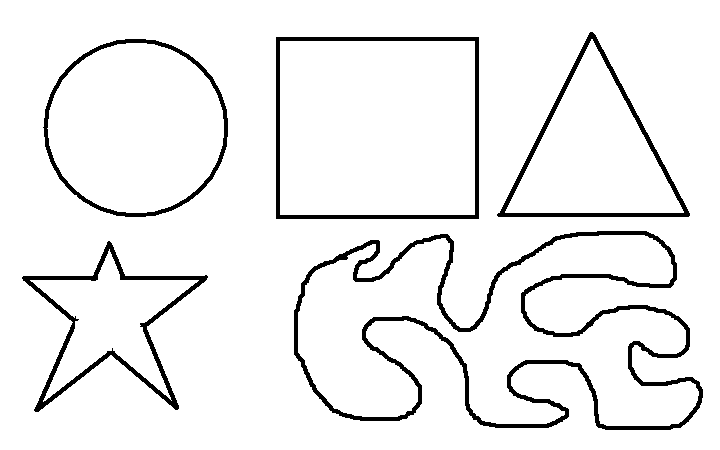}
\end{center}
Quantities like size, angle, distance, etc. don't matter.  Quantitatively, however, circles and squares are obviously very different.  In this sense squares and circles are geometrically different.  

\subsection{A Slightly More Formal Definition}
\label{sec:formaldefofmanifolds}

We now give a formal definition of manifolds to make the previous section more precise.  

An $n$-dimensional manifold is \\
\indent 1) A topological space $\mathcal{M}$.\footnote{A topological space is a formal mathematical idea, which you are encouraged to read about if you are not already familiar.  However, because we are doing physics, not mathematics, all of our spaces will trivially meet the requirements of a topological space (spacetime, Lie group space, etc).  You should eventually understand these things well, but doing so is not necessary for this paper.}\\
\indent 2) $\mathcal{M}$ has a family of pairs $\{(U_i, \boldsymbol\phi\it^{(i)})\}$ indexed by $i$.\footnote{We have set $\boldsymbol\phi\it^{(i)}$ in bold font because it is an $n$-dimensional vector. However, the index is merely a label, not an indication of covariance or contravariance, as indicated by the parentheses.}\\
\indent 3) $\{U_i\}$ is a family of open sets which covers $\mathcal{M}$.\footnote{Saying that the set of all $\{U_i\}$'s ``covers" $\mathcal{M}$ means that the union of all of the $U_i$'s is equal to $\mathcal{M}$: $$ \bigcup_i U_i = \mathcal{M}$$
In other words, we have split $\mathcal{M}$ up into a collection of (generally overlapping) sections $U_i$, and when you take the union of all of those sections, you have the entire space $\mathcal{M}$.}\\
\indent 4) $ \boldsymbol\phi\it^{(i)}$ is a homeomorphic\footnote{A map is ``homeomorphic" if it is continuous and if it has an inverse that is continuous.  The intuitive meaning is that two spaces are ``homeomorphic" to each other if they are topologically equivalent.  For example the circle and the oval (and the square and the star) are homeomorphic to each other, and therefore one can construct a homeomorphic map between them.} map from $U_i$ onto an open\footnote{An ``open" subset is a subset such that for any point $x\in V_i\subseteq \mathbb{R}^n$, there exists some $\epsilon>0$ such that the $n$-dimensional ball of radius $\epsilon$ around $x$ is entirely contained in $V_i$.  For example, consider the subset $[0,1]\subset \mathbb{R}^1$.  This contains the point $\{1\}$, and any $\epsilon$-neighborhood around $\{1\}$ will contain the point $1+\epsilon$, which is not in $[0,1]$.  On the other hand, the subset $(0,1)$ does not contain the point $\{1\}$, but rather any point of the form $1-\delta$ where $0<\delta< 1$.  Clearly, we can define $\epsilon \equiv {\delta \over 2}$, and the one dimensional ``ball" defined by all points $p$ satisfying $1-\delta-\epsilon < p < 1-\delta + \epsilon$ is contained in $(0,1)$.} subset $V_i$ of $\mathbb{R}^n$.  \\
\indent 5) Given two ``sections" of $\mathcal{M}$, $U_i$ and $U_j$, such that the intersection $U_i \cap U_j \neq 0$, the map\footnote{the map $\phi^{-1}$ is not bold because it is mapping a vector $\bf x\it$ in $\mathbb{R}^n$ to a \it point \rm in $\mathcal{M}$, which we do not take as a vector.} $\boldsymbol\psi\it^{(ij)}(\bf x\it) \equiv \boldsymbol\phi\it^{(i)} (\phi\it^{-1(j)}(\bf x\it))$ (where $\bf x\it \in V_j\subseteq \mathbb{R}^n$ and $\bf x \it$ is an $n$-dimensional vector in $V_j$) is smooth.\footnote{Smooth means that it is infinitely differentiable.  To a physicist, this simply means that it is a well-behaved function.}  

That definition is quite a bit to take in, so we will spend some time explaining what it means.  We did our best to relegate secondary mathematical ideas to the footnotes.  If you aren't familiar with them, we encourage you to read them in the recommended further reading, or at least the relevant Wikipedia pages.  

The picture we will be working with is the following:
\begin{center}
\includegraphics[scale=.7]{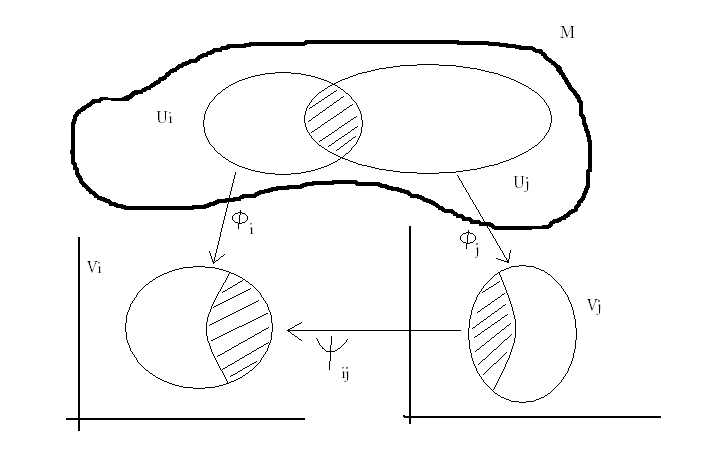}
\end{center}
We have some space $\mathcal{M}$ that is our $n$-dimensional manifold.  We divide it into several sections $U_i$ which will generally each cover a different part of $\mathcal{M}$ with some overlap.  All together, the $U_i$'s cover the entire space $\mathcal{M}$.  

For each of these sections $U_i$, there is a homeomorphism\footnote{We should make a somewhat formal comment at this point.  We referred to the $\boldsymbol\phi\it^{(i)}$'s as \it vectors \rm above.  This isn't entirely accurate.  There is a formal mathematical distinction made between ``points" in $\mathbb{R}^n$ and ``vectors" in the vector space $\mathbb{R}^n$.  A vector is an object satisfying the properties in section \ref{sec:vectorspaces}.  A point is merely an $n$-tuple which defines a point in the space.  In physics we typically blur this distinction.  For example we treat the point $\bf x\it = (x,y,z)^T$ as a ``position vector", when in reality it is the vector extending from the origin to the point that is truly the vector.  This distinction is, for us, not important.  We make this comment simply to acknowledge the distinction.  While we actually mean that $\boldsymbol\phi\it^{(i)}$ is merely an $n$-tuple in $\mathbb{R}^n$, we will think of it as a vector in everything that follows.  This lack of mathematical formality will have no affect on the contents of this paper.} $\boldsymbol\phi\it^{(i)}$ that maps points in $U_i \subset \mathcal{M}$ to points in an open subset of a copy of $\mathbb{R}^n$.  For example, if the manifold is $3$-dimensional, then each point $p\in \mathcal{M}$, will have an $x,y$, and $z$ component (in Cartesian coordinates).  So, there will be $\phi^{(i),x}$, $\phi^{(i),y}$ and $\phi^{(i),z}$ for every $i$.  

The fact that this map is a homeomorphism simply means that the section $U_i$ has the same topological structure as $\mathbb{R}^n$.  This in turn means that when we choose how to break $\mathcal{M}$ up into sections ($U_i$'s), we must only choose sections that are homeomorphic (topologically the same) as a section of $\mathbb{R}^n$.  

The pair $(U_i,\boldsymbol\phi\it^{(i)})$ is called a \bf chart\rm, while the collection of all charts $\{(U_i,\boldsymbol\phi\it^{(i)})\}$ is called an \bf atlas\rm.  The section $U_i$ is called a \bf coordinate neighborhood\rm, while $\boldsymbol\phi\it^{(i)}$ is called the \bf coordinate function \rm, or simply the \bf coordinates\rm, of the section $U_i$.  

So, we have a section of $U_i \subset \mathcal{M}$  that is homeomorphic to some section of $\mathbb{R}^n$, and then some map $\boldsymbol\phi\it^{(i)}$ that assigns each point $p\in U_i$ to a point in that section of $\mathbb{R}^n$.  In other words, $\boldsymbol\phi\it^{(i)}$ allows us us to label every point in $U_i$ with a unique $n$-dimensional point in $\mathbb{R}^n$ (which is why we call $\boldsymbol\phi\it^{(i)}$ the ``coordinates" of $U_i$).  

This is the content of the first 4 parts of the definition of a manifold.  

The meaning of the fifth requirement is a bit less obvious.  If two coordinate neighborhoods $U_i$ and $U_j$ overlap, there is some collection of points $p\in \mathcal{M}$ such that $p\in U_i \cap U_j$.  Therefore, the coordinate functions $\boldsymbol\phi\it^{(i)}$ and $\boldsymbol\phi\it^{(j)}$ will each map $p$ to a point in an open subset of $\mathbb{R}^n$ (cf the picture above), and there is nothing to guarantee that they are the same point in $\mathbb{R}^n$ (in general they will not be the same point in $\mathbb{R}^n$).  So, start with the point $p$ mapped to by either coordinate function, say $\boldsymbol\phi\it^{(j)}$, in $V_j\subset \mathbb{R}^n$.  Then, we can take this point in $V_j$ and use $\boldsymbol\phi^{-1(j)}$ to map it back to $p \in U_j \cap U_i$.  So, we have gone from $V_j$ to $\mathcal{M}$.  Now, we can take the point $p$ and map it \it back \rm to $\mathbb{R}^n$, but this time using $\boldsymbol\phi\it^{(i)}$ (instead of $\boldsymbol\phi^{(j)}$), taking $p$ to $V_i \subset \mathbb{R}^n$.  

So what we have done is mapped points in $V_j \subset \mathbb{R}^n$ to points in $V_i \subset \mathbb{R}^n$.  We call this composite mapping (from $V_j$ to $V_i$) $\boldsymbol\psi^{(ji)}$.  It is merely a way of moving points around in open subsets of $\mathbb{R}^n$.  In order for $\mathcal{M}$ to be a differentiable manifold, we demand that it be possible to choose our atlas so that $\boldsymbol\psi\it^{(ij)}$ is infinitely differentiable for all $i$ and $j$ (if $U_i$ and $U_j$ don't overlap then there this condition is met for $\boldsymbol\psi\it^{(ij)}$ trivially).  

To make this more clear, we consider a few examples.  The simplest example is obviously $\mathbb{R}^n$.  We can choose our atlas to consist of the single chart with $U_1 = \mathbb{R}^n$ and $\boldsymbol\phi\it^{(1)}$ the identity map.  

As a second example, consider the circle $S^1$:
\begin{center}
\includegraphics[scale=.2]{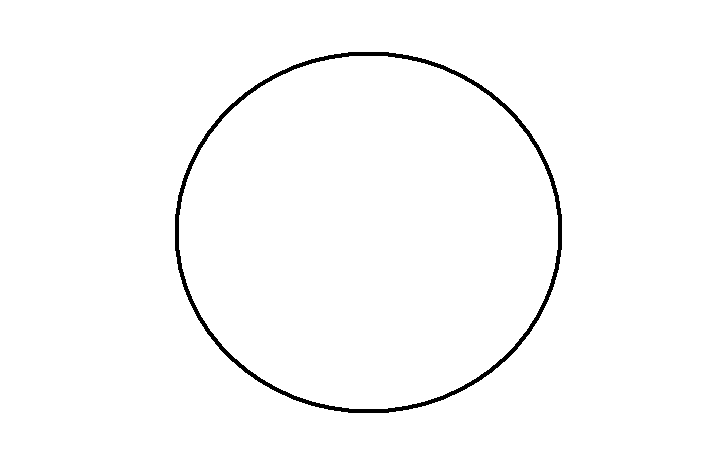}
\end{center}
Clearly it it a one-dimensional space, so we want to find an atlas that maps coordinate neighborhoods to $\mathbb{R}^1$.  An initial (naive) guess might be to use only one coordinate neighborhood going from a point on the circle all the way around, as follows:
\begin{center}
\includegraphics[scale=.3]{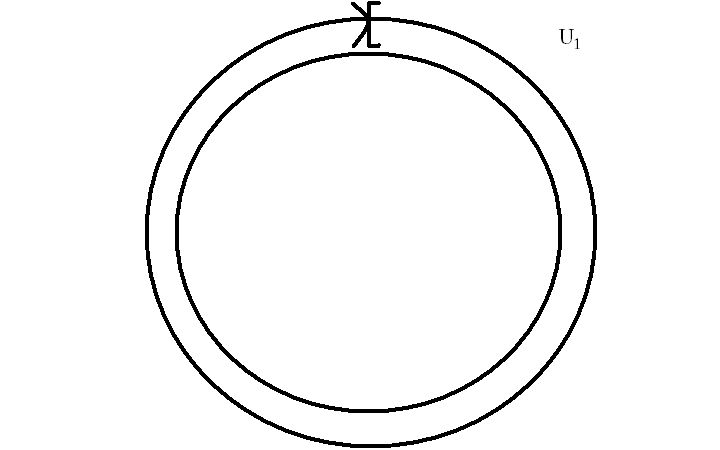}
\end{center}
We then have a lot of freedom with how we choose $\phi$ (a scalar because this is one dimensional).  The natural choice would be to define $\phi$  in terms of the angle around the circle:
\begin{center}
\includegraphics[scale=.3]{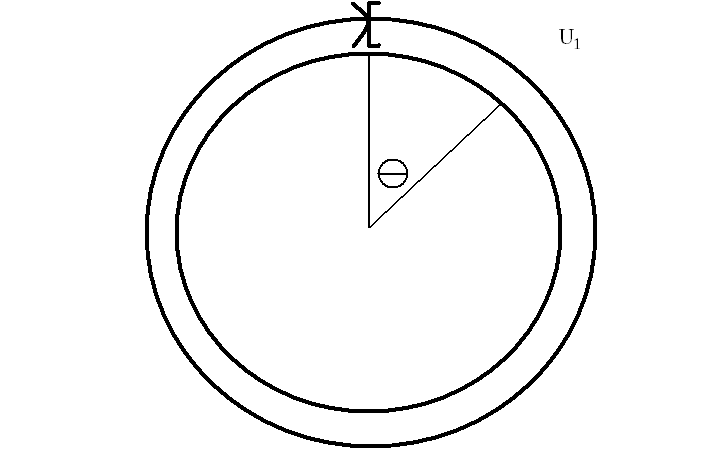}
\end{center}
so that $\phi$ maps to the interval $[0,2\pi)$.  But this is the problem.  The definition above states that the coordinate functions must map points in $\mathcal{M}$ to \it open \rm subsets of $\mathbb{R}^n$.  But $[0,2\pi)$ is not an open interval. So, this first guess doesn't work.  

It turns out that, because of the topology of $S^1$, we cannot define the atlas using only one coordinate neighborhood.  We must use at least two.  One possibility is to define two coordinate neighborhoods, each like the one in the naive approach above, with the one difference that they are open on both ends.  We could allow their ``origin" to be at any two places on the circle, but without loss of generality we take them to be on opposite sides of the circle:
\begin{center}
\includegraphics[scale=.5]{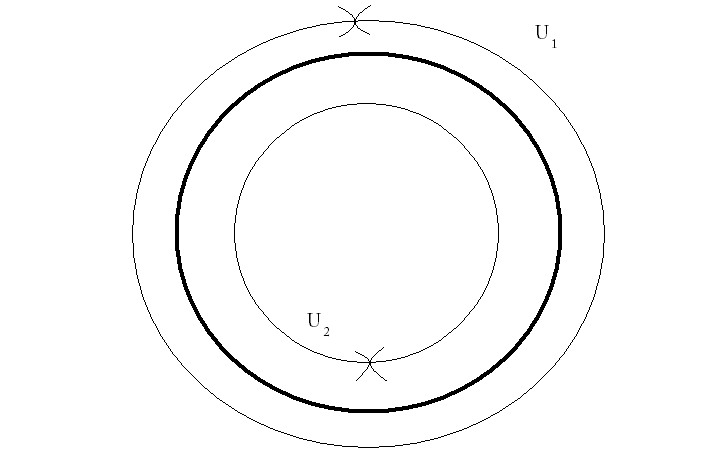}
\end{center}
Obviously, neither neighborhood alone covers the circle (they are both missing the point at their respective origins).  But, equally obvious is the fact that their union covers the entire circle.  For either neighborhood's missing point, the other neighborhood covers it.  

So how do we define $\phi^{(1)}$ and $\phi^{(2)}$?  We can define them in the same way as in the naive example above (except there will be no $0$ for either of them).  Now every point except the two origin points will be in $U_1 \cap U_2$.  So, consider an arbitrary point $p$ in the overlap:
\begin{center}
\includegraphics[scale=.5]{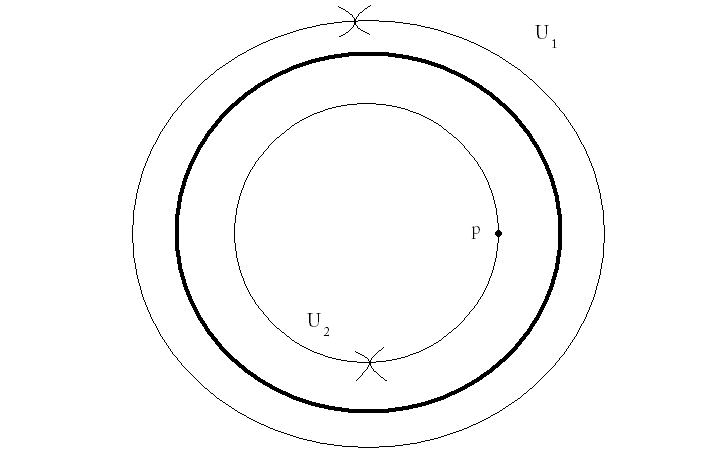}
\end{center}
This will be mapped to ${\pi \over 4}$ by $\phi^{(1)}$, and ${3\pi \over 4}$ by $\phi^{(2)}$.  More generally, any point on the right side of the picture will be mapped to $\theta$ by $\phi^{(1)}$ and ${\pi \over 2} + \theta$ by $\phi^{(2)}$.  Points on the left side are mapped to $\theta$ by $\phi^{(2)}$ and ${\pi \over 2} + \theta$ by $\phi^{(1)}$.  Obviously these maps are linear, and therefore smooth (infinitely differentiable).  

Another example is the sphere $S^2$:
\begin{center}
\includegraphics[scale=.4]{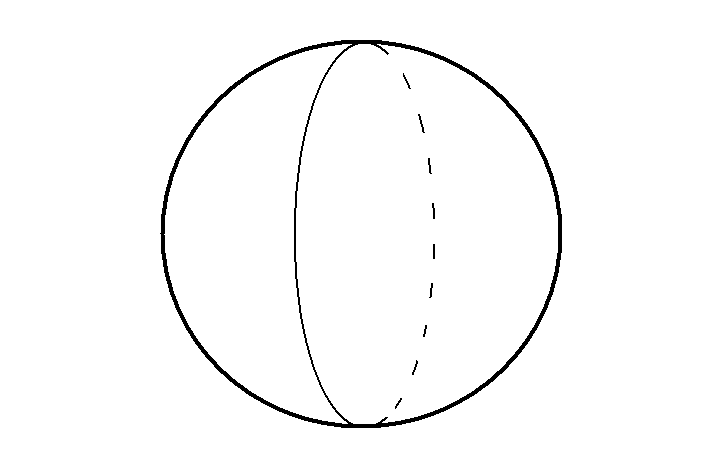}
\end{center}
It is a two-dimensional space, so we expect the coordinate functions map to $\mathbb{R}^2$.  Once again, the topology prevents us from using only one coordinate neighborhood.  Once again, however, we can use two: one including every point except the ``north pole", and another including every point except the ``south pole".  Each of these neighborhoods can be mapped to an open neighborhood of $\mathbb{R}^2$.  You can write out your own maps $\boldsymbol\phi\it^{(1)}$ and $\boldsymbol\phi^{(2)}$ and check to see that they are smooth.  

It would be instructive to figure out on your own how many coordinate neighborhoods are necessary for the torus $T^2$ (pictured above on page \pageref{pagewithtorusinchapter2}).  See if you can write out the coordinate functions for each patch and show that the $\boldsymbol\psi\it^{(ij)}$'s are smooth.  

We should comment that it is also possible to have a manifold with a \bf boundary\rm.  This allows that some or all of the coordinate neighborhoods $U_i$ are homeomorphic to an open subset of $\mathbb{R}^n$ where one of the coordinates is greater than or equal to $0$.  We won't be dealing with manifolds of this type in this paper, so we won't spend any more time discussing them.  

The reason we define manifolds in this way, with a set of abstract points and collections of charts homeomorphic to subsets of $\mathbb{R}^n$, is that it allows us to discuss manifolds as \it coordinate system free objects\rm.  In other words, we can talk about points in $p$ without having to define a coordinate system.  We know that all physics should be coordinate system independent, so this formulation is especially useful.  The manifold exists apart from any coordinate system designation.  

On the other hand, because we have coordinate neighborhoods that map homeomorphically to $\mathbb{R}^n$, and because we know how to deal with $\mathbb{R}^n$, we haven't lost calculational proficiency.  We can map the manifold to sections of $\mathbb{R}^n$, and do any calculations we may want there, knowing that both the homeomorphism structure and the smooth transitions between the target spaces of coordinate functions will preserve any and all structure.  

A few moments reflection should make it clear that this more formal definition is exactly equivalent to the informal one given in section \ref{sec:informaldefinition}.  Requiring that each $U_i$ be homeomorphic to an open subset of $\mathbb{R}^n$ is analogous to saying that every point has a neighborhood that looks like $\mathbb{R}^n$.  The motivation for demanding that the maps $\boldsymbol\psi\it^{(ij)}$ be smooth will become apparent when we begin doing calculus on manifolds.  

Also, recall that at this point we have the freedom to define the coordinate functions however we want.  We can map a particular $U_i$ to a small (open) $n$-sphere in $\mathbb{R}^n$ or to a large oddly shaped subset.  Because the map need only be a homeomorphism, you can literally choose \it any \rm open subset of $\mathbb{R}^n$ that is homeomorphic to $U_i$ without doing any damage to the manifold structure.  This leads to obvious shortcomings - how can you define distance on the manifold?  For example, mapping a coordinate neighborhood of $S^2$ to an open disk in $\mathbb{R}^2$ with radius 1 is just as good as mapping it to an open disk in $\mathbb{R}^2$ with radius 1,000,000,000.  Clearly the distance between the points in the Euclidian space will be different in the two cases, but that doesn't change the "distance" between them on $\mathcal{M}$.  \label{talkingaboutlackofgeometryonmanifolds}

Also, because we are merely demanding that the coordinate functions are homeomorphisms, they don't really tell us anything about the geometry of the manifold.  We only have information about its topology.  For example, there is no way to differentiate between a ``geometrically perfect" sphere $S^2$ and an ``egg" manifold (a sphere stretched out a bit).  

We will eventually deal with these and other problems when we impose the appropriate structures on $\mathcal{M}$.  For now, we will merely work with the structure we have, where $\mathcal{M}$ is simply a topological idea.  

\subsection{Tangent Spaces and Frames}
\label{sec:tangentspacesandframes}

In section \ref{sec:forms} we talked about tensor fields and forms that exist in $\mathbb{R}^n$.  This was interesting, but we want to begin generalizing to less trivial manifolds.  This section will begin that process.  

Consider some $n$-dimensional manifold $\mathcal{M}$.  For every point $p$ in some coordinate neighborhood $U_i$ with coordinate functions $\boldsymbol\phi_i$ mapping to $\mathbb{R}^n$, we can define the \bf Tangent Space \rm at point $p$ as follows:\footnote{In a slight violation of our notation, we are merely labeling the vectors in $\mathbb{R}^n$ in this definition using bold characters.  They should be considered to be \it vectors\rm, not covectors.  We will ``clean up" our notation later.  For now, just take them to be vectors in the vector space $\mathbb{R}^n$.}
\begin{eqnarray}
T_p\mathcal{M} = \{p\} \otimes \mathbb{R}^n = \{(p,\bf v\it)\; \big| \; \bf v \it \in \mathbb{R}^n\} \label{eq:tangentspacedef}
\end{eqnarray}
This means that at the point $p$ we have attached a copy of the vector space $\mathbb{R}^n$.\footnote{We are here referring to the point $p \in \mathcal{M}$, not to the target point $\boldsymbol\phi\it^{(i)}(p) \in \mathbb{R}^n$.}  Tangent spaces are real linear spaces with addition given by 
$$ a(p,\bf v\it) + b(p,\bf u\it) = (p,a\bf v\it + b\bf u\it)$$
You can effectually think of the first element in this notation (the $p$) as merely a label specifying what point in $\mathcal{M}$ the vector $\bf v\it$ is associated with.  Note that the $p$ in the definition of $T_p\mathcal{M}$ is in $\mathcal{M}$, not a point in an open subset of $\mathbb{R}^n$ that a coordinate function maps to.  Defining $T_p\mathcal{M}$ this way, in terms of point in $\mathcal{M}$ rather than in terms of where some $\boldsymbol\phi\it^{(i)}$ maps $p$, prevents redundancy issues regarding which $\boldsymbol\phi$ was used in case $p\in U_i \cap U_j$.  

For example, consider $S^1$.  Coordinate neighborhoods on $S^1$ are homeomorphic to open subsets of $\mathbb{R}^1$, so for a given point $p$, we attach every vector in the vector space $\mathbb{R}^1$.\footnote{Note that the tangent space is \it all \rm of $\mathbb{R}^n$, whereas the neighborhoods of $\mathcal{M}$ are homeomorphic to \it open subsets \rm of $\mathbb{R}^n$.}  
\begin{center}
\includegraphics[scale=.4]{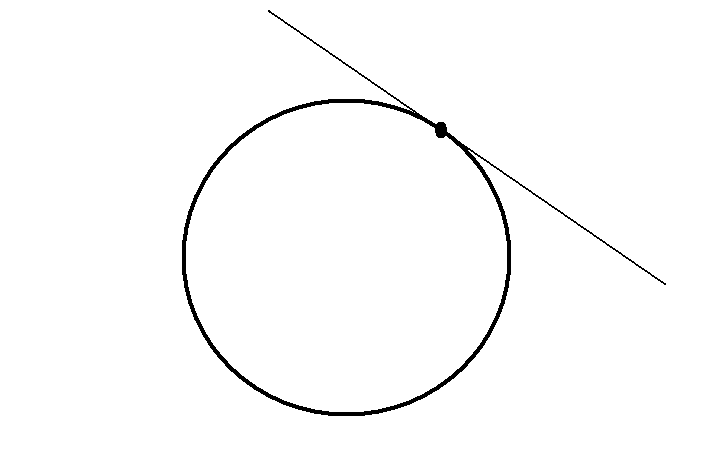}
\end{center}
In other words, we are attaching a line ($\mathbb{R}^1$) to every point.  So, each point has its own copy of $\mathbb{R}^1$,
\begin{center}
\includegraphics[scale=.4]{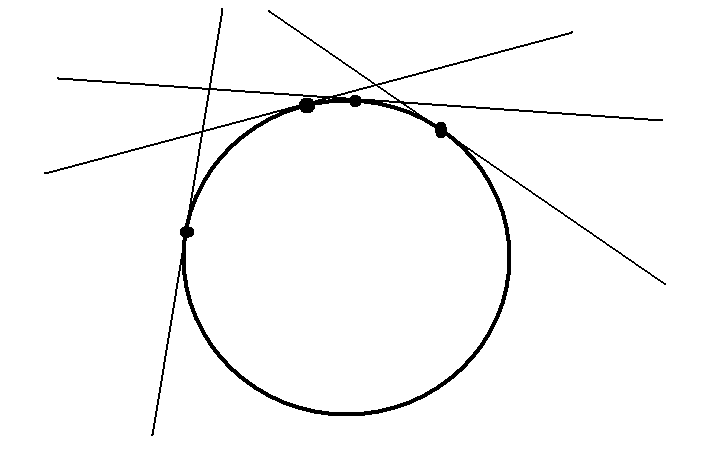}
\end{center}

Another example is $S^2$.  Coordinate neighborhoods are homeomorphic to open subsets of $\mathbb{R}^2$, so each point has a copy of $\mathbb{R}^2$ attached:
\begin{center}
\includegraphics[scale=1.5]{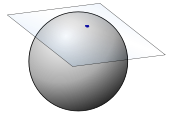}
\end{center}

As one final example, consider the ``open ball" in three dimensions.  To see this, imagine the sphere $S^2$, fill in the interior, and then remove the boundary of the sphere.  In other words this is the set of all points $(x,y,z)$ in three dimensions such that $x^2+y^2+z^2 < r$ for some radius $r>0$.\footnote{On the other hand, the ``closed ball" in three dimensions is the set of all point $(x,y,z)$ in three dimensions such that $x^2+y^2+z^2 \leq r>0$.  One includes the boundary, the other doesn't.  Clearly, the open ball is an example of a manifold without a boundary, while the closed ball is a manifold with a boundary.  The coordinate neighborhood of the closed ball that covers the boundary will be mapped to a subset of $\mathbb{R}^3$ where one coordinate is greater than or equal to zero.}  We only need one coordinate neighborhood $U_1$ for this manifold, and $\boldsymbol\phi\it^{(1)}$ can merely be the identity map to an open subset $V_1\subset \mathbb{R}^3$.  So, because the coordinate neighborhood is homeomorphic to $\mathbb{R}^3$, the tangent space will be $\mathbb{R}^3$.  So, each point will have a copy of $\mathbb{R}^3$ ``attached" to it.  

For reasons made clear by the above examples, the elements of $T_p\mathcal{M}$ are called the \bf tangent vectors \rm to the point $p$.  The tangent space associated with each point is the space of all directions you can move \it in \rm in the manifold at the point $p$.  

However, we make the important note that, mathematically, it is best not to think of the tangent spaces as ``attached" to $\mathcal{M}$ in a direct geometrical way, despite the indication of the pictures above.  The tangent space is a \it different \rm space than the manifold.  

So, we are working with multiples spaces.  First is the manifold $\mathcal{M}$ which is merely a collection of points with certain topological properties and has no specific coordinate system. $\mathcal{M}$ is then divided into coordinate neighborhoods, each of which are a part of $\mathcal{M}$, but are required to be homeomorphic to an open subset of $\mathbb{R}^n$.  Then, for every coordinate neighborhood, we have \it another \rm space, $\mathbb{R}^n$.  There is a separate copy of $\mathbb{R}^n$ for each coordinate neighborhood, and they are all distinct spaces both from each other as well as from $\mathcal{M}$.  Finally, we have the tangent spaces, which consists of an infinite number of copies of $\mathbb{R}^n$, one for each point in $\mathcal{M}$.  Again, each of these copies of $\mathbb{R}^n$ should be considered distinct from each other, $\mathcal{M}$, and from the open subsets of $\mathbb{R}^n$ which the coordinate functions map to.  

Let's consider a simple physical illustration of a tangent space.  Imagine a fly constrained to move on the surface of $S^2$.  For some path he moves through, at each instant we can assign a position vector, which will be a point $p \in S^2$ with coordinates $\boldsymbol\phi_i \in \mathbb{R}^2$.  Also, at each point he will have a velocity, which will be a vector ``tangent" to the sphere, and is therefore in the tangent space of $S^2$.  Notice that the fly cannot have a velocity vector that is not tangent to the point because it is constrained to fly in the space only, and therefore the space of all velocity vectors at each point is two dimensional.  So, the phase space of the motion of the fly will include, for each instant, a point in the space along with a vector in the tangent space of \it that \rm point.  

In the same way, if the fly is constrained to the open ball in $\mathbb{R}^3$, it can be at any point in the open ball, but it is free to have a velocity vector anywhere in the \it three \rm dimensional space.  

Phase space as in the two previous examples represents one of the simplest physical examples of a tangent space.  However, we should note an important mathematical point.  We said above that  we should think of the tangent space as a distinct space from the actual manifold, not as being geometrically attached to it.  The fly examples illustrate this in that the vector representing the velocity of the fly (which is a vector in the tangent space) does not depend at all on the location of the fly.  If we speak in terms of the coordinate functions $\boldsymbol\phi_i$ and assign coordinates to the location of the fly, there will be a vector in $\mathbb{R}^n$ (the target space of a $\boldsymbol\phi\it^{(i)}$) which represents the location, and an independent vector in a \it different \rm copy of $\mathbb{R}^n$ (the tangent space at that point).  The two vectors are not related, and for \it this \rm reason we think of the tangent space at a point $p$ being ``attached" to the manifold at $p$, but not geometrically dependent on it.  

We write tangent vectors in pairs (as in (\ref{eq:tangentspacedef})) because a particular tangent vector refers specifically to a point in $\mathcal{M}$.  For example it would make no sense to add tangent vectors at two different points:
$$ a(p,\bf v\it) + b(q,\bf u\it) = \;?$$
There is no natural connection between the tangent vectors at $p$ and $q$ (unless, of course, $p=q$).  For this reason, we take $T_p\mathcal{M}$ and $T_q\mathcal{M}$ to be completely different vector spaces.  This is what we meant above when we said that ``each point $p\in \mathcal{M}$ gets its own copy of the tangent space $\mathbb{R}^n$".  

However\label{firstmentiontangentbundlesthisisforthesectionongravasgaugetheory}, we can consider the union of all these vector spaces, which we refer to as the \bf tangent bundle \rm $T\mathcal{M}$ of $\mathcal{M}$:
\begin{eqnarray}
T\mathcal{M} = \bigcup_{p\in \mathcal{M}} T_p \mathcal{M} = \mathcal{M}\otimes \mathbb{R}^n = \{(p,\bf v\it)\big| p\in \mathcal{M},\; \bf v\it \in \mathbb{R}^n\} \label{eq:firsttimewementionthetangentbundlethisisforGRasgaugetheoyrsection}
\end{eqnarray}

At this point, we are using the term ``bundle" is a loose sense.  A tangent bundle is a very simple example of a \bf fibre bundle\rm, which we will discuss in more detail later in these notes.  For now, just think of it as attaching a space ($\mathbb{R}^n$) to every point in another space ($\mathcal{M}$), creating a space of dimension $2n$.

Before moving on we mention one idea that will seem unhelpful and even a little silly, but will be extraordinarily useful later.  We want to introduce it now so that it isn't entirely new when we come back to it.  As we said, each point $p\in \mathcal{M}$ has a copy of the tangent space $\mathbb{R}^n$.  So reversing this, a given vector in the tangent bundle, $(p,\bf v\it)$, can be associated with a specific point $p$.  We therefore define the \bf projection map \rm $\pi$, which takes a tangent vector and maps it to the point in $\mathcal{M}$ it is ``attached" to.  So,
\begin{eqnarray}
\pi:T_p\mathcal{M} &\longrightarrow& \mathcal{M} \nolabel \\
(p,\bf v\it) &\longmapsto & p
\end{eqnarray}
Obviously for any two tangent vectors $\bf v\it^{(i)}$ and $\bf v\it^{(j)}$ (with $\bf v\it^{(i)} \neq \bf v\it^{(j)}$), $$ \pi((p,\bf v\it^{(i)})) = \pi((p,\bf v\it^{(j)}))$$
because they are both attached to the same point $p$.  

Also, for any point $p$, we can use the inverse of $\pi$ to recover the entire tangent space:
\begin{eqnarray}
\pi^{-1}(p) = T_p\mathcal{M}
\end{eqnarray}
As we said above, taking time to specifically define a map that does something so apparently trivial may seem unprofitable at this point.  When we come back to it, however, it will be a very powerful tool.  

Moving on, we have now discussed manifolds, coordinate neighborhoods and coordinate functions, and tangent spaces.  A tangent space at a point is, as we have said, a copy of $\mathbb{R}^n$.  And, as you should be quite familiar with, we can represent any point in $\mathbb{R}^n$ with a set of basis vectors.  It is then natural to seek a way of finding a basis set for a tangent space at a point.  The way this is done will prove to be one of the most important and foundational results in differential geometry.  

Using our standard notation, considering a manifold $\mathcal{M}$ with coordinate neighborhoods $U_i$ and coordinate functions $\boldsymbol\phi\it^{(i)}$, we can take the $\boldsymbol\phi\it^{(i)}$'s to be whatever coordinate system we want.  We could take $\phi^{(i),1} = x$, $\phi^{(i),2}=y$, $\phi^{(i),3}=z$, etc. for Cartesian coordinates, or we could set $\phi^{(i),1}=r$, $\phi^{(i),2}=\theta$, etc for spherical coordinates, and so on.  We can choose any coordinate system we want in how we map elements $p \in \mathcal{M}$ to open subsets of $\mathbb{R}^n$.  

So, given a point $p$ with chosen coordinates $\boldsymbol\phi\it^{(i)}$ (in an $n$-dimensional manifold there will be $n$ components, so $\boldsymbol\phi\it^{(i)}$ is an $n$-component vector), we can define the set of vectors
\begin{eqnarray}
{\partial \over \partial \phi^{(i),j}} \bigg|_p  \equiv \bigg(p,{\partial \boldsymbol\phi\it^{(i)} \over \partial \phi^{(i),j}}\bigg|_p\bigg) \in T_p\mathcal{M} \label{eq:defframevectors}
\end{eqnarray}

To illustrate this, let $\mathcal{M}=S^2$, and let the $\boldsymbol\phi\it^{(i)}$ map to Cartesian coordinates.  Then, $\boldsymbol\phi\it^{(i)} = (x,y)^T$ (because the coordinate neighborhoods of $S^2$ are homeomorphic to $\mathbb{R}^2$).  So, the vectors are
\begin{eqnarray}
{\partial \over \partial \phi^{(i),1}}\bigg|_p &=& {\partial \over \partial x}\bigg|_p =\bigg( p,\bigg({\partial x \over \partial x},{\partial y \over \partial x}\bigg)\bigg) =  \big(p,(1,0)^T\big) \nolabel \\
{\partial \over \partial \phi^{(i),2}}\bigg|_p &=& {\partial \over \partial y}\bigg|_p = \bigg( p,\bigg({\partial x \over \partial y},{\partial y \over \partial y}\bigg)\bigg) =\big(p,(0,1)^T\big) \label{eq:spanr2}
\end{eqnarray}
for any $p \in U_i$.  Notice that the two left most expressions in (\ref{eq:spanr2}) form a basis for the $\mathbb{R}^2$ tangent space, so any vector in the tangent space at a point $p$ can be written as 
\begin{eqnarray}
\bf v \it \in T_p\mathcal{M} = v^1\big(p,(1,0)^T\big) + v^1\big(p,(0,1)^T\big) = v^j {\partial \over \partial \phi^{(i),j}} \label{eq:generalvectorintermsofdifferentialoperatorframes}
\end{eqnarray}
where the summation convention is in effect and $v^1,v^2\in \mathbb{R}$.  We have dropped the $\big|_p$ in the notation for now, but this should be understood as referring to a a specific point.  

The form of (\ref{eq:generalvectorintermsofdifferentialoperatorframes}) seems to beg for something on the right for these differential operators to act on.  This intuition is correct, but we aren't ready to discuss what they act on.  We will get to that later in these notes.  

Notice that the index on the left hand side of (\ref{eq:spanr2}) is lowered,\footnote{When an object with an index appears in the denominator, the index is ``switched".  For example the upper index in $x^i$ is a lower index in ${\partial \over \partial x^i}$, and the lower index in $x_i$ is an upper index in ${\partial \over \partial x_i}$.} which is consistent with our convention that vector components have raised indices ($\phi^{(i),j}$), whereas the basis vectors of a vector space have lowered indices.  

The general form (\ref{eq:defframevectors}) will always produce a set of basis vectors which, for any point $p\in\mathcal{M}$, will form a basis for the tangent space at $p$.  Furthermore, because of the properties of the coordinate functions, this basis will vary smoothly from point to point in $\mathcal{M}$.  Such a smoothly varying basis over the tangent bundle is called a \bf frame \rm on the tangent space of the manifold.  

Of course, the basis ${\partial \over \partial \phi^{(j),i}}$ is entirely dependent on the coordinate functions you choose.  We used the standard $x$ and $y$ above.  But what if we wanted to use a different basis instead?  We know from the definition of a manifold that the transformation functions $\boldsymbol\psi\it^{(ij)}$ are smooth.  Therefore for any coordinates $x^i$ there will exist some smooth mapping to some new coordinates $x'^j = x'^j (x^i)$ (meaning that $x'^j$ is a function of all of the $x^i$).  This map, from $x$ to $x'$, is exactly the $\boldsymbol\psi^{(ij)}$ discussed above, and therefore we know that it will be smooth (infinitely differentiable).  So, whereas the original frame is
\begin{eqnarray}
{\partial \over \partial x^i}
\end{eqnarray}
using the chain rule and exploiting the knowledge that the $\boldsymbol\psi\it^{(ij)}$'s are smooth, we can write the new frame as
\begin{eqnarray}
{\partial \over \partial x'^i} = {\partial x^j \over \partial x'^i} {\partial \over \partial x^j} \label{eq:changeofbasisvectorusingchainrulethingy}
\end{eqnarray}
Note that this is a linear transformation (${\partial x^j \over \partial x'^i}$ is exactly analogous to $(T)^j_i$ from (\ref{eq:changebasis})), whereas $x'^j = x'^j(x^i)$ is not necessarily linear.  For example, consider the (nonlinear) change of coordinates from Cartesian to polar (in two dimensions):
\begin{eqnarray}
x = r \cos \theta \qquad \qquad y = r \sin \theta \label{eq:polartocartesiantrans}
\end{eqnarray}
This is a nonlinear mapping of \it coordinates\rm, not vectors.  In polar coordinates, the frame is
\begin{eqnarray}
{\partial \over \partial x} \qquad \rm and \it \qquad {\partial \over \partial y}
\end{eqnarray}
Using (\ref{eq:changeofbasisvectorusingchainrulethingy}) we can easily write the new frame:
\begin{eqnarray}
{\partial \over \partial r} &=& \bigg({\partial x\over \partial r}\bigg){\partial \over \partial x} + \bigg({\partial y\over \partial r} \bigg){\partial \over \partial y} \nolabel \\
&=& \bigg({\partial \over \partial r}(r\cos \theta)\bigg){\partial \over \partial x} + \bigg({\partial \over \partial r}(r\sin\theta)\bigg){\partial \over \partial y} \nolabel \\
&=& (\cos \theta){\partial \over \partial x} + (\sin\theta){\partial \over \partial y} \nolabel \\
{\partial \over \partial \theta} &=& \bigg({\partial x \over \partial \theta}\bigg) {\partial \over \partial x} + \bigg({\partial y \over \partial \theta}\bigg) {\partial \over \partial y} \nolabel \\
&=& -(r\sin\theta){\partial \over \partial x} + (r\cos\theta){\partial \over \partial y}
\end{eqnarray}
\begin{center}
\includegraphics[scale=.7]{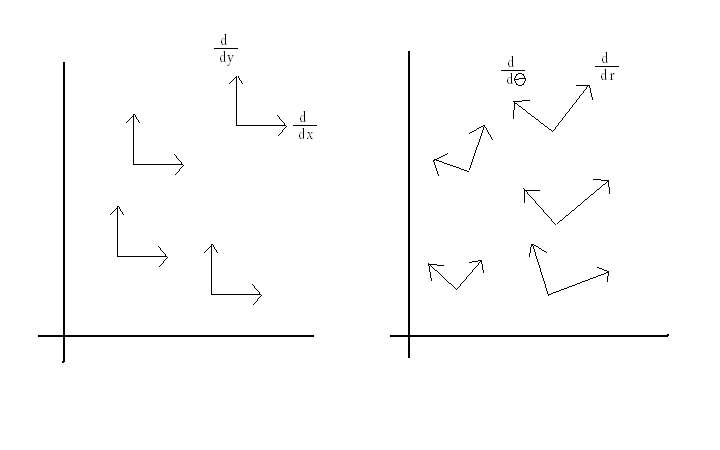}
\end{center}
So, the \underline{nonlinear} transformation (\ref{eq:polartocartesiantrans}) on the \it coordinates \rm provides the \underline{linear} transformation on the \it frame \rm, or \it basis vectors\rm at each point.  

Incidentally, notice that we can literally choose any non-singluar matrix ${\partial x^j \over \partial x'^i}$ to transform the basis.  Therefore, the set of all possible such matrices forms a general linear group.  And, because we have this freedom to choose \it any \rm basis we want without changing the physics of the system we are describing, we have found a gauge group of the system.  Simply by the definition we are using for manifolds, a gauge invariance has come about in our freedom to choose the coordinate functions however we want.  This will be used in much more detail later, in section \ref{sec:noncoordinatebases}.  

Moving on, just as the basis vectors transform according to (\ref{eq:changeofbasisvectorusingchainrulethingy}), the components of a vector must also transform (cf the argument beginning on page \pageref{pagewhereargumentbegins}).  We naturally expect the components to transform under the inverse transformation as the basis vectors (again, see argument beginning on page \pageref{pagewhereargumentbegins} or equations (\ref{eq:changebasis}) and (\ref{eq:otherchangebasis})).  So, if $x'^j = x'^j(x^i)$, then the inverse will be $x^i = x^i(x'^j)$.  So, the inverse of the matrix ${\partial x^j \over \partial x'^i}$ will be the matrix ${\partial x'^i \over \partial x^j}$.  So, the components of a vector, $v^i$, will transform according to
\begin{eqnarray}
v'^i = {\partial x'^i \over \partial x^j}\bigg|_p v^j \label{eq:transformationofcomponentsonmanifold}
\end{eqnarray}
(we included the $p$ because the components must be evaluated point-wise).  

To illustrate this, consider once again the transformation from polar to Cartesian, (\ref{eq:polartocartesiantrans}).  Then consider some arbitrary vector with components (in polar coordinates) $(1, {\pi \over 2})$.  So, the vector is
\begin{eqnarray}
{\partial \over \partial r} + {\pi \over 2} {\partial \over \partial \theta}
\end{eqnarray}
This will appear as on the graph below:
\begin{center}
\includegraphics[scale=.7]{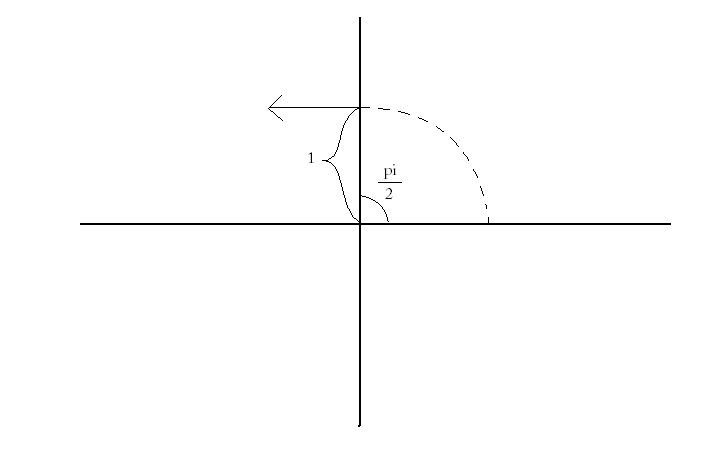}
\end{center}
Using (\ref{eq:transformationofcomponentsonmanifold}), we have
\begin{eqnarray}
v^x &=& {\partial x \over \partial r}\bigg|_p v^r + {\partial x \over \partial \theta}\bigg|_p v^{\theta} \nolabel \\
&=& \bigg(\cos\theta -{\pi \over 2}r\sin\theta\bigg)\bigg|_{(1,{\pi \over 2})}  = -{\pi \over 2} \nolabel \\
v^y &=& {\partial y \over \partial r} v^r + {\partial y \over \partial \theta} v^{\theta} \nolabel \\
&=&\bigg( \sin\theta + {\pi \over 2} r \cos \theta \bigg)\bigg|_{(1,{\pi \over 2})}= 1
\end{eqnarray}
So the vector in the transformed coordinates will be
\begin{eqnarray}
-{\pi \over 2} {\partial \over \partial x} + {\partial \over \partial y}
\end{eqnarray}
which is the same thing we had before.  

So, we have basis vectors which transform with lowered indices, and components which transform according to the inverse transformation with upper indices.  This matches exactly what we said at the end of section \ref{sec:forms}, justifying the convention declared there.  

Before moving on to the next section, we make a brief comment about how all of this may generalize.  A tangent space is built by attaching a copy of $\mathbb{R}^n$ to each point on a manifold.  We discussed the physical example of phase space - every spatial location (which is represented by a vector in $\mathbb{R}^n$) a fly (or any object) is in throughout its trajectory has a velocity (which is a vector in another copy of $\mathbb{R}^n$), which is the tangent space.  

\label{talkaboutfibresforthefirsttime} We also mentioned that tangent bundles are very simple examples of fibre bundles.  Another way of saying this is that a tangent space is a very simple example of a fibre.  While there is a more rigorous definition that we will discuss later in this series, a fibre is basically a space you attach at every point of another space.  In physics, the base manifold is usually spacetime.  We then attach a fibre at each point which corresponds to some physical property of the object that we are interested in.  We already discussed the tangent space fibre as a useful tool for describing the objects velocity.  The idea is that for each point on a spatial manifold, there is a point in the fibre which corresponds to a specific physical property.  

Another example may be a "rotation fibre".  Consider a baseball flying through the air.  At each point in space, there is a specific location (on the manifold), a velocity (a point in the tangent space fibre), \it and \rm typically the ball has some rotational ``spin" around an axis.  We can therefore attach another copy of $\mathbb{R}^3$ in addition to the tangent space.  A point in this fibre will then correspond to an angular velocity vector.  So, we have attached two fibres to the space manifold - one for its speed and another for its spin.  

As a more exotic example, consider a helicopter.  The base manifold will again be three dimensional space, and there will naturally be a tangent space for its speed.  We could then attach an additional fibre, not of $\mathbb{R}^3$, but an $S^1$ fibre.  So, at a given spatial point, we have a point in the tangent space telling us how fast the helicopter is moving, but we also have a point on $S^1$ which may tell us which direction the helicopter is facing at that location.  

While these are extremely, extremely simple examples of fibres and fibre bundles, you can see the basic idea.  It turns out that the theory of fibre bundles provides an extraordinarily powerful tool through which to make sense of nearly everything in particle physics.  We only mention these ideas to give you an idea of why we are doing things the way we are now.  We are building towards something much more general.  

\subsection{The Tangent Mapping}
\label{sec:tangentmapping}

Before moving on, there is an important property of mappings between manifolds we must consider.  Let $\mathcal{M}_1$ be a manifold of dimension $n_1$ and $\mathcal{M}_2$ be a manifold of dimension $n_2$.  We denote the coordinate functions $\boldsymbol\phi\it^{(i)}$ on $\mathcal{M}_1$ as $x^i$, $i=1,\ldots,n_1$ (dropping the index in parenthesis labeling a particular coordinate neighborhood for notational simplicity), and the coordinate functions in $\mathcal{M}_2$ as $y^i$, $i=1,\ldots, n_2$.  

Now consider a smooth map $f$ between these two manifolds: 
\begin{eqnarray}
f:\mathcal{M}_1 \longrightarrow \mathcal{M}_2
\end{eqnarray}

It turns out that we can use $f:\mathcal{M}_1 \longrightarrow \mathcal{M}_2$ to induce a well-defined map $T_pf$ which maps from the \it tangent \rm space at $p\in\mathcal{M}_1$ to the \it tangent \rm space at $f(p) \in \mathcal{M}_2$.  In other words, 
\begin{eqnarray}
f:\mathcal{M}_1 \longrightarrow \mathcal{M}_2 \quad \Longrightarrow\quad  T_pf:T_p\mathcal{M}_1 \longrightarrow T_{f(p)}\mathcal{M}_2
\end{eqnarray}
We call $T_pf$ the \bf tangent map \rm at $p$.  

We find the exact form of this map as follows: Let $\tilde Q$ be a homeomorphism from $[a,b]\subset \mathbb{R}$ to $\mathcal{M}_1$, such that $\tilde Q(\tau) \in \mathcal{M}_1 \; \forall \; \tau \in [a,b]$.  Obviously this map is a one dimensional path in $\mathcal{M}_1$ (in other words it is locally homeomorphic to $\mathbb{R}$).

Then, use the coordinate functions $\bf x\it$ on $\mathcal{M}_1$ to map the point $\tilde Q(\tau) \in \mathcal{M}_1\; \forall \tau$ to $\mathbb{R}^n$ (for simplicity, we assume for now that the entire curve is contained in a single coordinate neighborhood).  This induces a homeomorphic map\footnote{The reason we are using \it both \rm $\bf Q\it$ and $\tilde Q$ is merely that we are being careful.  Formally, $\tilde Q$ maps from $[a,b]$ to $\mathcal{M}_1$, which is coordinate free.  We are going to want to take derivatives this map and we therefore want to put it into a space where we know how to take derivatives, namely $\mathbb{R}^n$.  So, $\tilde Q$ maps from $[a,b]$ to $\mathcal{M}_1$, and $\bf x\it$ maps from $\mathcal{M}_1$ to $\mathbb{R}^n$.  We merely define $\bf Q\it$ as the composite map which takes $[a,b]$ directly to $\mathbb{R}^n$, which we can easily take derivatives of.}
\begin{eqnarray}
\bf Q\it(\tau) \equiv \bf x \it (\tilde Q(\tau)) : [a,b]\subset \mathbb{R} \longrightarrow \mathbb{R}^n
\end{eqnarray}
So, $\bf Q\it(\tau)$ is a parametric expression for a curve in $\mathbb{R}^n$.  We write it in bold because it is a point in the vector space $\mathbb{R}^n$.  

We can use the curve $\bf Q\it(\tau) \in \mathbb{R}^n$ to specify a tangent vector in $\mathcal{M}_1$ by taking the derivative of $\bf Q\it$ at a point.  If we take the curve in $\mathcal{M}_1$ defined by $\tilde Q(\tau)$ to pass through $p\in \mathcal{M}_1$ at, say, $\tau_0$ (so $\tilde Q(\tau_0) = p$), then the curve in $\mathbb{R}^n$ defined by $\bf Q\it(\tau)$ passes through $\bf x \it (p) \in \mathbb{R}^n$ at $\bf Q\it(\tau_0)$.  
\begin{center}
\includegraphics[scale=.7]{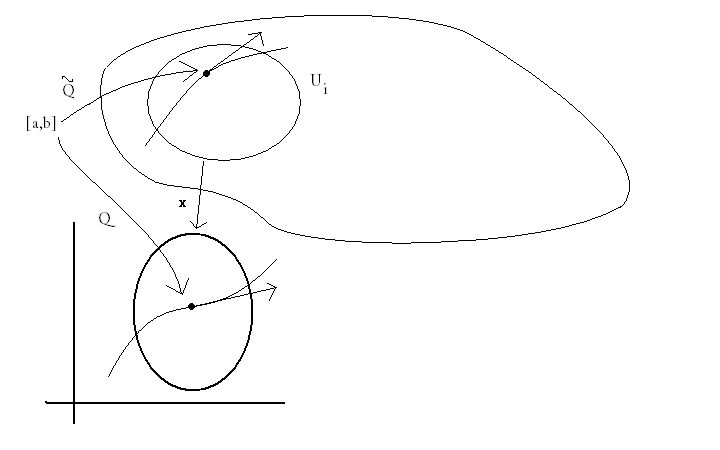}
\end{center}
Now, by taking a derivative of $\bf Q\it$ at $\tau_0$, we have a vector in $T_p\mathcal{M}_1$ in the direction of the curve:
\begin{eqnarray}
\bigg(p,{d \bf Q\it(\tau) \over d \tau }\bigg|_{\tau = \tau_0}\bigg) \in T_p\mathcal{M}_1 \label{eq:tangentvectorasaresultofQ}
\end{eqnarray}

We can express this vector in terms of the frame induced by the coordinate functions $\bf x \it $, so that the $i^{th}$ component is given by ${d Q^i(\tau)\over d \tau}\big|_{\tau=\tau_0}$.  So the vector in $T_p\mathcal{M}_1$ defined by the curve $\tilde Q$ is given by 
\begin{eqnarray}
{d Q^i \over d\tau} {\partial \over \partial x^i}\bigg|_p
\end{eqnarray}
(where the summation convention is being used), so the components are ${dQ^i \over d \tau}$ with basis vectors ${\partial \over \partial x^i}$ at point $p \in \mathcal{M}_1$.  

So a curve $\tilde Q:[a,b]\longrightarrow \mathcal{M}_1$ determines a well defined tangent vector in $T_p\mathcal{M}_1$, and any tangent vector in $T_p\mathcal{M}_1$ can be specified by some curve.  Now, using the map $f$ we can map this curve in $\mathcal{M}_1$ to a curve in $\mathcal{M}_2$.  Because $f$ is smooth this will be well-defined.  In other words, we have the composite map
\begin{eqnarray}
\tilde Q'(\tau) \equiv f(\tilde Q(\tau))
\end{eqnarray}
which maps $[a,b] \longrightarrow \mathcal{M}_2$.  Then, using the same argument as above, using the coordinates functions $\bf y \it$ on $\mathcal{M}_2$ we can map this curve from $\mathcal{M}_2$ to a curve in $\mathbb{R}^n$.  We call this map
\begin{eqnarray}
\bf Q\it'(\tau) \equiv \bf y\it(\tilde Q'(\tau)) = \bf y\it(f(\tilde Q(\tau))) \label{eq:defofboldQprimed}
\end{eqnarray}
\begin{center}
\includegraphics[scale=.5]{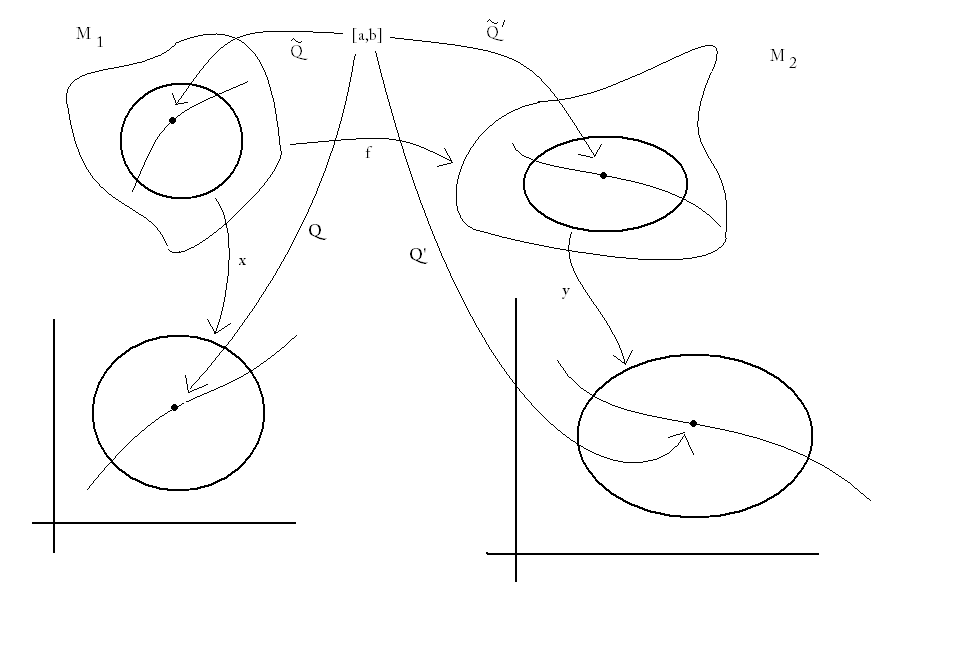}
\end{center}
If the point $p \in \mathcal{M}_1$ maps to the point $f(p) \in \mathcal{M}_1$, we can use $\bf Q\it'$ to define a vector in $T_{f(p)}\mathcal{M}_2$.  Namely, this will be the vector
\begin{eqnarray}
\bigg(f(p), {d\bf Q\it'(\tau) \over d\tau}\bigg|_{\tau=\tau_0}\bigg) \in T_{f(p)}\mathcal{M}_2
\end{eqnarray}
We can write this vector in terms of the frame induced by $\bf y\it$ 
\begin{eqnarray}
{dQ'^i\over d\tau} {\partial \over \partial y^i} \bigg|_{f(p)}
\end{eqnarray}

So, with a smooth map $f$ from $\mathcal{M}_1$ to $\mathcal{M}_2$, any tangent vector in $\mathcal{M}_1$ can be defined by some curve $\tilde Q$ in $\mathcal{M}_1$.  This curve will naturally induce a well defined map of the tangent vector induced by $\tilde Q$ at any point $p$ to a tangent vector in $\mathcal{M}_2$ at the point $f(p)$.  This map is called the \bf Tangent Mapping \rm induced by $f$.  It is also called the \bf pushforward \rm because it is dual to the pullback in the sense that the pushforward maps a vector to a vector while the pullback maps a form to a form.  

We consider an example of this before moving on.  Consider the manifolds $S^2$ and $\mathbb{R}^2$.  Working with only one hemisphere of $S^2$, we will choose spherical coordinates.  So a point in $S^2$ (minus one point at a pole, see section \ref{sec:formaldefofmanifolds}) will be mapped to some $(\theta, \phi)$.  We then choose the standard Cartesian coordinates on $\mathbb{R}^2$.  

Now we want to define the map $f$.  Obviously it will be easier to work with a coordinate representation of $f$.  This is done in the natural way: a point $p\in \mathcal{M}_1$ is mapped to $f(p)$ by $f$ and to $\bf x\it(p)$ by the coordinate functions.  Define the smooth map $\bf F\it:\bf x\it(S^2 - \{pole\}) \longrightarrow \bf y\it(\mathbb{R}^2)$ ($\bf F\it$ is a two component vector in this case because $\mathbb{R}^2$ is two dimensional) so that
\begin{eqnarray}
f:p\longmapsto f(p) \iff \bf F\it: \bf x\it(p) \longmapsto \bf y\it (f(p))
\end{eqnarray}
The form of $\bf F\it$ will be
\begin{eqnarray}
(x^1,x^2,\ldots,x^{n_1}) \mapsto \big(F^1(x^1,x^2,\ldots,x^{n_1}),\ldots,F^{n_2}(x^1,x^2,\ldots,x^{n_1})\big)
\end{eqnarray}

So, working with this ``coordinate representation" of $f$, we can define a map from $S^2$ to $\mathbb{R}^2$.  Let $\bf F\it$ take\footnote{This bijective map takes points on the sphere to points on the plane.  It is typically called a \bf stereographic projection\rm.}
\begin{eqnarray}
F^1 &=& x = {\cos\theta\sin\phi \over 1-\cos\phi} \nolabel \\
F^2 &=& y = {\sin\theta\sin\phi \over 1-\cos\phi} \nolabel \\
\end{eqnarray}
This map is smooth and can be easily inverted to
\begin{eqnarray}
F^{-1,1} &=& \theta = \tan^{-1}\bigg({y\over x}\bigg) \nolabel \\
F^{-1,2} &=& \phi = 2\tan^{-1}\bigg({1\over \sqrt{x^2+y^2}}\bigg)
\end{eqnarray}
which is also smooth.  

Now define a curve $\bf Q\it$ in $S^2$ as
\begin{eqnarray}
Q^1(\tau) &=& \theta(\tau) = \tau \nolabel \\
Q^2(\tau) &=& \phi(\tau) = 2\tau
\end{eqnarray}
where $\tau \in [0,\pi]$

Now let's consider the point $(\theta,\phi) = (\pi/4,\pi/2)$ (corresponding to $\tau=\pi/2$).  In $S^2$ this will define the tangent vector
\begin{eqnarray}
\bf v\it &=& {dQ^1 \over d\tau} {\partial \over \partial \theta}\bigg|_{\tau={\pi \over 2}} + {dQ^2 \over d\tau}{\partial \over \partial \phi}\bigg|_{\tau = {\pi \over 2}} \nolabel \\
&=& {\partial \over \partial \theta} + 2{\partial \over \partial \phi}
\end{eqnarray}

Now we can map the curve $\bf Q\it$ to $\mathbb{R}^2$ using $\bf F\it$, giving the curve $\bf Q\it'$ in $\mathbb{R}^2$
\begin{eqnarray}
Q'^1(\tau) &=& x(\tau) = {\cos(\tau)\sin(2\tau) \over 1-\cos(2\tau)} \nolabel \\
Q'^2(\tau) &=& y(\tau) = {\sin(\tau)\sin(2\tau) \over 1-\cos(2\tau)}
\end{eqnarray}

This will produce the vector
\begin{eqnarray}
\bf v\it' &=& {d \over d\tau}\bigg({\cos(\tau)\sin(2\tau) \over 1-\cos(2\tau)}\bigg){\partial \over \partial x}\bigg|_{\tau={\pi \over 2}} + {d \over d\tau}\bigg({\sin(\tau)\sin(2\tau) \over 1-\cos(2\tau)}\bigg){\partial \over \partial y}\bigg|_{\tau = {\pi \over 2}} \nolabel \\
&=& -{3\over \sqrt{2}}{\partial \over \partial x} - {1 \over \sqrt{2}} {\partial \over \partial y}
\end{eqnarray}
It would be instructive to graph these vectors in the respective spaces to see that they do in fact line up.  

So to summarize, for any smooth map $f$ between two manifolds $\mathcal{M}_1$ and $\mathcal{M}_2$, any vector in $T_p\mathcal{M}_1$ can be mapped to a vector in $T_{f(p)}\mathcal{M}_2$ in a well defined way by using the tangent mapping $T_pf$ at $p$.  

As one final (very important) comment for this section, it may seem that having to define a curve for every vector you want to pushforward is a bit tedious.  It turns out that it is possible to define the pushforward without having to refer to a curve.  This is done as follows.  In $\mathcal{M}_1$ we have the tangent vector
\begin{eqnarray}
\bigg(\tilde Q(\tau), {d \bf Q\it \over d\tau}\bigg|_{\tau = \tau_0}\bigg) = {d Q^i \over d\tau} {\partial \over \partial x^i}\bigg|_{Q(\tau_0)}
\end{eqnarray}
in $\mathcal{M}_1$ being mapped to the tangent vector
\begin{eqnarray}
\bigg(\tilde Q'(\tau),{d\bf Q\it' \over d\tau}\bigg|_{\tau = \tau_0}\bigg) = {d Q'^i \over d\tau} {\partial \over \partial y^i}\bigg|_{Q'(\tau_0)}
\end{eqnarray}
in $\mathcal{M}_2$.
But, using (\ref{eq:defofboldQprimed}) (but using the coordinate representation $\bf F\it$ instead of $f$), we can rewrite this as
\begin{eqnarray}
\bf Q\it'(\tau) = \bf F\it(\bf Q\it(\tau)) \quad \Longrightarrow \quad \bigg(\tilde Q'(\tau),{d \over d\tau}\bf F\it(\bf Q\it(\tau))\bigg|_{\tau = \tau_0}\bigg) = {d \over d\tau}(F^i(\bf Q\it(\tau))) {\partial \over \partial y^i}\bigg|_{Q'(\tau_0)}
\end{eqnarray}
And then using the chain rule,
\begin{eqnarray}
{d Q'^i \over d\tau} &=& {d \over d \tau} (F^i(\bf Q\it(\tau))) \nolabel \\
&=& {\partial F^i \over \partial x^j} {dQ^j \over d\tau}
\end{eqnarray}
Note that we inserted the identity operator
\begin{eqnarray}
{d Q^j \over d x^j} = 1
\end{eqnarray}
because the $Q^j$ are simply spatial coordinates.

Summarizing this, the tangent map takes the tangent vector
\begin{eqnarray}
\bigg(\tilde Q(\tau), {d \bf Q\it \over d\tau}\bigg|_{\tau = \tau_0}\bigg) = {d Q^i \over d\tau} {\partial \over \partial x^i}\bigg|_{Q(\tau_0)} \label{eq:firsttangentvectortangentmapping}
\end{eqnarray}
to the tangent vector
\begin{eqnarray}
\bigg(\tilde Q'(\tau),{\partial \bf F\it \over \partial x^j}{dQ^j \over d\tau}\bigg|_{\tau=\tau_0}\bigg) = {dQ^j \over d\tau} {\partial F^i \over \partial x^j}{\partial \over \partial y^i}\bigg|_{Q'(\tau_0)} \label{eq:secondtangentvectortangentmapping}
\end{eqnarray}

Now consider an arbitrary vector $\bf v\it$ with components $v^i \equiv {d Q^i \over d\tau}\big|_{\tau=\tau_0}$.  We can rephrase the previous paragraph about (\ref{eq:firsttangentvectortangentmapping}) and (\ref{eq:secondtangentvectortangentmapping}) by saying that the tangent map takes the tangent vector 
\begin{eqnarray}
(\tilde Q(\tau),\bf v \it) = v^i{\partial \over \partial x^i}
\end{eqnarray}
to the tangent vector
\begin{eqnarray}
\bigg(\tilde Q'(\tau),v^j{\partial \bf F\it \over \partial x^j}\bigg) = v^j {\partial F^i \over \partial x^j} {\partial \over \partial y^i}
\end{eqnarray}
Notice that the right hand sides make no reference to $\bf Q\it $ or $\bf Q'\it$, but does hold for any arbitrary vector $\bf v\it \in \mathcal{M}_1$.  

So, in general, we can say that for an arbitrary vector $v^i{\partial \over \partial x^i}$ at $p\in \mathcal{M}_1$, the pushforward induced by $f$ is given by
\begin{eqnarray}
(T_pf)\bigg(v^i{\partial \over \partial x^i}\bigg) = v^j{\partial F^i \over \partial x^j}{\partial \over \partial y^i} \label{eq:curvefreedefinitionofpushforward}
\end{eqnarray}

You can approach the $f:S^2\longrightarrow \mathbb{R}^2$ example above yourself using (\ref{eq:curvefreedefinitionofpushforward}) to see that you do indeed get the same answer.

\subsection{Cotangent Space}

We have now defined manifolds and tangent spaces.  Following what we did in section \ref{sec:dualspace}, we now seek to define the dual space to a tangent space, which is called the \bf Cotangent Space\rm.  The primary difference between what we did in section \ref{sec:dualspace} and what we will do here for arbitrary manifolds is that everything is done point-wise.  Just as each point $p \in \mathcal{M}$ has its own copy of the tangent space $\mathbb{R}^n$, each point also has its own copy of the cotangent space $\mathbb{R}^{n\star}$.  

To be clear, the cotangent space of a point has the same ``vector space" structure as the tangent space.  You can picture covectors in cotangent spaces just as much as you can picture vectors in tangent spaces.  The difference is that one is the dual of the other (cf section \ref{sec:dualspace})

In other words, for each point $p \in \mathcal{M}$, define the vector space $\Lambda^q T_p\mathcal{M}$ as the $q$-linear antisymmetric product of $1$-forms (covectors) in the cotangent space at $p$ which map $(T_p\mathcal{M})^{\otimes q}$ to $\mathbb{R}$.  Or, in more mathematical language,
\begin{eqnarray}
\Lambda^qT_p\mathcal{M} \equiv \{\phi:(T_p\mathcal{M})^{\otimes q} \rightarrow \mathbb{R},\; q\; \rm linear,\; antisymmetric \}
\end{eqnarray}
All of the properties we discussed in section \ref{sec:dualspace} will hold point-wise in this case, including the formation of the exterior algebra (at each point $p$), wedge products (at each point), vector space valued forms (at each point), transformation laws between covector bases, etc.  

However, we need a frame to provide a basis for $\Lambda^qT_p\mathcal{M}$ \it locally \rm just as 
${\partial \over \partial \phi^{(i),j} }(p)$ formed a local frame for $T_p\mathcal{M}$ locally.  For a given set of coordinate functions $\boldsymbol\phi\it^{(i)}$ (or components ($\phi^{(i),1}$, $\phi^{(i),2}$, $\ldots$)$^T$), we have a frame for the tangent space $T_p\mathcal{M}$ given by 
$$ {\partial \over \partial \phi^{(i),j} } $$
The basis for the cotangent space will then be given by the differential 
\begin{eqnarray}
d\phi^{(i),j} \label{eq:cotangentspacebasis}
\end{eqnarray}
such that (cf (\ref{eq:dualbasis}))
\begin{eqnarray}
d\phi^{(i),j} \bigg({\partial \over \partial \phi^{(i),k}}\bigg) = \delta^j_k \label{eq:ruleforbasisanddualbasisthatthedotprodmustbekronecker}
\end{eqnarray}

The expression $d\phi^{(i),j}$ should be taken to be a differential (like the things you integrate over in Calculus I) just as much as ${\partial \over \partial \phi^{(i),j}}$ is a derivative operator.  We will talk about the connection between (\ref{eq:cotangentspacebasis}) and the differential from in calculus soon.  This relationship is, in many ways, at the heart of \it differential \rm geometry.  

All of the properties of forms discussed above in section \ref{sec:forms} will hold true here.  The biggest difference is that differential forms are defined only at a single point.  

As a quick example, consider a manifold with coordinate neighborhoods homeomorphic to $\mathbb{R}^3$.  If we set the coordinate functions to be Cartesian coordinates ($\boldsymbol\phi\it^{(i)} \dot = (x,y,z)^T$), then the dual space basis covectors are $dx$, $dy$, and $dz$.  So an arbitrary element of $\Lambda^0T_p\mathcal{M}$ will simply be a real number.  
An arbitrary element of $\Lambda^1T_p\mathcal{M}$ will be
\begin{eqnarray}
a (dx) + b (dy) + c (dz)
\end{eqnarray}
An arbitrary element of $\Lambda^2T_p\mathcal{M}$ will be
\begin{eqnarray}
a (dx \wedge dy) + b (dx \wedge dz) + c( dy \wedge dz)
\end{eqnarray}
An arbitrary element of $\Lambda^3T_p\mathcal{M}$ will be
\begin{eqnarray}
a( dx \wedge dy \wedge dz)
\end{eqnarray}
The action of, say, an arbitrary $2$-form on some arbitrary element of $T_p\mathcal{M} \otimes T_p\mathcal{M}$ is then \footnote{This is not the most general vector - just a random example.}
\begin{eqnarray}
& & \big(a (dx \wedge dy) + b (dx \wedge dz) + c( dy \wedge dz)\big) \bigg(A \bigg({\partial \over \partial x} \otimes {\partial \over \partial z}\bigg) + B\bigg( {\partial \over \partial z} \otimes {\partial \over \partial y}\bigg)\bigg) \nolabel \\
& &= bA - cB \in \mathbb{R}
\end{eqnarray}

And just as in section \ref{sec:forms}, we can express an arbitrary element of $\Lambda^qT_p \mathcal{M}$ as
\begin{eqnarray}
{1 \over q!} \omega_{i_1,i_2,\ldots,i_q} d\phi^{i_1} \wedge d\phi^{i_2} \wedge \cdots \wedge d\phi^{i_q} \label{eq:arbitrarydifferentialqform}
\end{eqnarray}
(compare this to (\ref{eq:coefficientsofeuclidianforms})).  Of course the summation convention is in effect.  

So, for some change of coordinates, how will the covector basis transform?  Again, using the chain rule, the transformation law will be
\begin{eqnarray}
d\phi'^{j} = {\partial \phi'^{j} \over \partial \phi^i} d\phi^i \label{eq:changeofcovectorbasis}
\end{eqnarray}
which is the same transformation law for the \it components \rm of vector as in equation (\ref{eq:transformationofcomponentsonmanifold}).

For example, the change of coordinates (\ref{eq:polartocartesiantrans}) will result in the cotangent space frame
$$ dr \qquad \rm and \it \qquad d\theta$$
being transformed (linearly) to
\begin{eqnarray}
dx &=& {\partial x \over \partial r} dr + {\partial x \over \partial \theta} d\theta \nolabel \\
&=& (\cos\theta) dr -(r\sin\theta)d\theta \nolabel \\
dy &=& {\partial y \over \partial r} dr + {\partial y \over \partial \theta} d\theta \nolabel \\
&=& (\sin \theta)dr + (r\cos\theta)d\theta 
\end{eqnarray}

We can also write out the wedge product in terms of the new basis:
\begin{eqnarray}
dx \wedge dy &=& \big((\cos \theta)dr - (r\sin\theta)d\theta\big) \wedge \big((\sin \theta)dr + (r\cos\theta)d\theta \big) \nolabel \\
&=& (r\cos^2\theta + r\sin^2\theta)\; dr \wedge d\theta \nolabel \\
&=& r\; dr \wedge d\theta \label{eq:areaformfortwodimensionsusingcartesianandpolar}
\end{eqnarray}
We recognize both sides of this expression as looking a lot like area terms in their respective coordinate systems.  

\subsection{The Pullback of Differential Forms}
\label{sec:pullbackdiff}

Now we can generalize what we did in section \ref{sec:pullbacks} to differential forms.  We will follow what we did there fairly closely.  You are encouraged to go back and re-read that section in parallel with this one.  

Consider some map $f$ from manifold $\mathcal{M}_1$ (of dimension $n_1$) to manifold $\mathcal{M}_2$ (of dimension $n_2$):
$$f : \mathcal{M}_1 \longrightarrow \mathcal{M}_2$$
In general this map cannot be uniquely inverted.  

But, while we cannot in general invert the mapping $f$, it turns out that $f$ does create a well defined map from the \it cotangent \rm space of $\mathcal{M}_2$ to the \it cotangent \rm space of $\mathcal{M}_1$.  We denote this induced map $f^{\star}$, and say
\begin{eqnarray}
f^{\star} : \Lambda^qT_{f(p)}\mathcal{M}_2 \longrightarrow \Lambda^qT_p\mathcal{M}_1
\end{eqnarray}
Note that $f$ mapped from $1\mapsto 2$, whereas $f^{\star}$ maps from $2 \mapsto 1$.  

We define $f^{\star}$ as follows: given some tensor product of $q$ vectors in $T_p\mathcal{M}_1$, which we will denote $(\bf m\it^{(1)},\bf m\it^{(2)},\ldots,\bf m\it^{(q)})$, we know that it is mapped to a tensor product of $q$ vectors in $T_{f(p)}\mathcal{M}_2$ by the tangent mapping $T_pf$ as
\begin{eqnarray}
(\bf m\it^{(1)},\bf m\it^{(2)},\ldots,\bf m\it^{(q)}) \longmapsto \big(T_pf(\bf m\it^{(1)}),T_pf(\bf m\it^{(2)}),\ldots ,T_pf(\bf m\it^{(q)})\big) \label{eq:examplegeneraltensorproductinmanifoldforpullbackdef}
\end{eqnarray}
(where $T_pf(\bf m\it^{(i)})$ is understood to be a vector in $T_{f(p)}\mathcal{M}_2$).  Acting on the right side of (\ref{eq:examplegeneraltensorproductinmanifoldforpullbackdef}) with a $q$-form in $\Lambda^qT_{f(p)}\mathcal{M}_2$ (denoted $\phi_q$) maps it to $\mathbb{R}$:
\begin{eqnarray}
\phi_q(T_pf(\bf m\it^{(1)}),T_pf(\bf m\it^{(2)}),\ldots ,T_pf(\bf m\it^{(q)})) \longmapsto \mathbb{R}
\end{eqnarray}
The map $f^{\star}$ will, as state above, map a $q$-form $\phi_q$ in $\Lambda^qT_{f(p)}\mathcal{M}_2$ to a $q$ form in $\Lambda^qT_p\mathcal{M}_1$, which we denote $\psi_q$.  We define the exact action of $f^{\star}$ on $\phi_q\in \Lambda^qT_{f(p)}\mathcal{M}_2$ as follows:
\begin{eqnarray}
(f^{\star}\phi_q)(\bf m\it^{(1)},\bf m\it^{(2)},\ldots,\bf m\it^{(q)}) &=& \psi_q(\bf m\it^{(1)},\bf m\it^{(2)},\ldots,\bf m\it^{(q)}) \nolabel \\
&=& \phi_q (T_pf(\bf m\it^{(1)}),T_pf(\bf m\it^{(2)}),\ldots ,T_pf(\bf m\it^{(q)})) \label{eq:definitionofpullbackofdifferentialforms}
\end{eqnarray}

The remainder of the analogous discussion in section \ref{sec:pullbacks} will hold for the meaning and application of (\ref{eq:definitionofpullbackofdifferentialforms}).  

One very important property is that, in the case where $q=0$ (zero forms, or real functions), the pullback ``commutes" with the action of a one form:
\begin{eqnarray}
(f^{\star}\phi_0)(x) = \phi_0(f(x))
\end{eqnarray}

As an illustration we calculate an example.  Once again take $f:\mathcal{M}_1\longrightarrow \mathcal{M}_2$, where $\bf x\it$ are the coordinate functions on $\mathcal{M}_1$ and $\bf y\it$ are the coordinate functions on $\mathcal{M}_2$.  The coordinate representation of $f$, again denoted $\bf F\it:\bf x\it(\mathcal{M}_1) \longrightarrow \bf y \it(\mathcal{M}_2)$ will be used as in section \ref{sec:tangentmapping}.  Using (\ref{eq:curvefreedefinitionofpushforward}), a vector  $v^i{\partial \over \partial x^i}$ in $\mathcal{M}_1$ is mapped to a vector $v^j{\partial F^i \over \partial x^j}{\partial \over \partial y^i}$ in $\mathcal{M}_2$.  We want to find the pullback of the one-form $dy^j \in \Lambda^1T_{f(p)}\mathcal{M}_2$, which will be a one-form $(f^{\star}dy^j) \in \Lambda^1T_p\mathcal{M}_1$.  Using the definition (\ref{eq:definitionofpullbackofdifferentialforms}), we have
\begin{eqnarray}
(f^{\star}dy^j)\bigg(v^i{\partial \over \partial x^i}\bigg) &=& dy^j\bigg((T_pf)\bigg(v^i{\partial \over \partial x^i}\bigg)\bigg) \nolabel \\
&=& dy^j v^i{\partial F^k \over \partial x^i}{\partial \over \partial y^k} \nolabel \\
&=& v^i {\partial F^k \over \partial x^i} \delta^j_k \nolabel \\
&=& v^i{\partial F^j \over \partial x^i}
\end{eqnarray}
Or in terms of the basis vectors
\begin{eqnarray}
(f^{\star} dy^j)\bigg({\partial \over \partial x^i}\bigg) = {\partial F^j \over \partial x^i} \label{eq:intermsofthebasisvectorsdifferentialpullback}
\end{eqnarray}
The meaning of this is as follows: $(f^{\star}dy^j)$ is a form in $\mathcal{M}_1$ (the $\bf x\it$ space), and will therefore be some linear combination of the $dx^i$'s.  We will call it
\begin{eqnarray}
(f^{\star}dy^j) = A_i^jdx^i \label{eq:jasdhflkjahsdflkjh}
\end{eqnarray}
for some $A_i^j$ (proportionality terms).  If we act on a unit vector ${\partial \over \partial x^k}$ with this (using (\ref{eq:intermsofthebasisvectorsdifferentialpullback})), we have
\begin{eqnarray}
(f^{\star} dy^j)\bigg({\partial \over \partial x^k}\bigg) = A^j_idx^i\bigg({\partial \over \partial x^k}\bigg)= A^j_i\delta^i_k = A^j_k \equiv {\partial F^j \over \partial x^k}
\end{eqnarray}
So, using (\ref{eq:jasdhflkjahsdflkjh}),
\begin{eqnarray}
(f^{\star}dy^j) = {\partial F^j \over \partial x^i} dx^i \label{eq:howtoexpressyintermsofxdiffpullback}
\end{eqnarray}
In other words, (\ref{eq:howtoexpressyintermsofxdiffpullback}) is telling us how to express a form in $\bf y\it$ space in terms of forms in $\bf x\it$ space.  

You are already quite familiar with all of this, even though that may not be apparent.  What we essentially have is some space with coordinates $\bf x\it$, and then a map which assigns a point the space with coordinates $\bf y\it$ for every point in $\bf x\it$.  We call this map $\bf F\it$, and it has the form
\begin{eqnarray}
y^i=y^i(x^1,x^2,\ldots) = F^i(x^1,x^2,\ldots) 
\end{eqnarray}
(note that this is a map \it from \rm $\bf x\it$ space \it to \rm $\bf y\it$ space).  We can then take the differential of this expression, getting
\begin{eqnarray}
dy^i = {\partial F^i \over \partial x^j} dx^j
\end{eqnarray}
which is simply a slightly less formal statement of (\ref{eq:howtoexpressyintermsofxdiffpullback}) - relating the expression of forms in $\bf x\it$ space ($\mathcal{M}_1$) to $\bf y\it$ space ($\mathcal{M}_2$).  We have done nothing more than formalize the partial derivative.  While this may seem like a lot of work for such a simple concept, remember that we are going to be doing calculus on much more general spaces than $\mathbb{R}^n$, and therefore we will need all the formalism we are currently developing.  

\subsection{Exterior Derivatives}
\label{sec:exteriorderivatives}

We are now ready to discuss an extremely important idea that will provide a major part of the the framework for physics.  The \bf Exterior Derivative\rm, which we denote $d$, is a function which maps $q$-forms to $(q+1)$-forms:
\begin{eqnarray}
d: \Lambda^qT_p\mathcal{M} \longrightarrow \Lambda^{q+1}T_p\mathcal{M}
\end{eqnarray}

We will begin with the simplest example and build the general definition from there.  We will work on a manifold $\mathcal{M}$ with coordinate functions $\bf x\it$.  Starting with an element $\phi_0 \in \Lambda^0T_p\mathcal{M}$ (a zero-form, or merely a real function), we define the exterior derivative as 
\begin{eqnarray}
d\phi_0 \equiv {\partial \phi_0 \over \partial x^i} dx^i \label{eq:firstexampleexteriorderivativeofafunction}
\end{eqnarray}
Note that if we transform the coordinate covector basis basis (using (\ref{eq:changeofbasisvectorusingchainrulethingy}) and (\ref{eq:changeofcovectorbasis})), we have
\begin{eqnarray}
{\partial \over \partial x^i} \rightarrow {\partial y^j \over \partial x^i} {\partial \over \partial y^j} \qquad \rm and \it \qquad dx^i \rightarrow {\partial x^i \over \partial y^j} dy^j
\end{eqnarray}
So
\begin{eqnarray}
d\phi_0 = {\partial \phi_0 \over \partial x^i}dx^i &\longrightarrow& {\partial \phi_0 \over \partial y^j}\bigg({\partial y^j \over \partial x^i} {\partial x^i \over \partial y^k}\bigg) dy^k \nolabel \\
&=& {\partial \phi_0 \over \partial y^j} \delta ^j_k dy^k \nolabel \\
&=& {\partial \phi_0 \over \partial y^j} dy^j
\end{eqnarray}
which is the exact same form as in the $\bf x\rm$ coordinates.  So, the definition of the exterior derivative is \it coordinate system independent\rm!  That is an extremely important property which will prove extraordinarily useful later.  

Because $d\phi_0$ is a $(0+1)$ form on $\mathcal{M}$, we can act on a vector $\bf v\it = v^j{\partial \over \partial x^j}$ to get
\begin{eqnarray}
d\phi_0(\bf v\it) &=& {\partial \phi_0 \over \partial x^i}dx^i \bigg(v^j{\partial \over \partial x^j}\bigg) = v^j {\partial \phi_0 \over \partial x^i} dx^i\bigg({\partial \over \partial x^j}\bigg) \nolabel \\
&=& v^j{\partial \phi_0 \over \partial x^i}\delta^i_j = v^i {\partial \phi_0 \over \partial x^i}
\end{eqnarray}
This expression is interpreted as being the derivative of $\phi_0$ \it in the direction \rm of $\bf v\it$, or the directional derivative of $\phi_0$.  We will discuss the geometrical meaning of this (and other results) shortly, though you should already be somewhat familiar.  

Now we generalize the definition of $d$.  We will take $d$ to have the following properties:\footnote{Each of these can be proven very easily and straightforwardly, and we therefore won't write out the definitions.  You are encouraged to convince yourself that each of these properties are true.}\\
\indent 1) Leibniz - $d(fg) = fdg+gdf$, where $f,g$ are functions.  \\
\indent 2) Linear - $d(af+bg) = adf+bdg$, where $f,g$ are functions. \\
\indent 3) $d(\Lambda^qT_p\mathcal{M})\subseteq \Lambda^{q+1}T_p\mathcal{M}$ \\
\indent 4) Generalized Leibniz - $d(\phi_p \wedge \psi_q) = (d\phi_p)\wedge \psi_q + (-1)^p\phi_p\wedge (d\psi_q)$ \label{pagewithrulesforexteriorderivative}\\
\indent 5) Nilpotent - $d^2\phi_p = d(d \phi_p) = 0$ for any $p$-form $\phi_p$.  

With a little tedium, it can be shown that the exterior derivative of an arbitrary $p$-form $\phi_p$ (cf (\ref{eq:arbitrarydifferentialqform})), is
\begin{eqnarray}
d\phi_p =  {\partial \phi_{i_1,i_2,\ldots,i_p} \over \partial x^j} \; dx^j\wedge dx^{i_1}\wedge dx^{i_2} \wedge \cdots \wedge dx^{i_p} \label{eq:generalexpressionfortheexternalderivativeofapform}
\end{eqnarray}
(Check the indices very carefully!), where the summation convention is, as always, in effect.  The wedge product will ensure that this new $(p+1)$-form will still be antisymmetric.  

Once again we can see that this (\ref{eq:generalexpressionfortheexternalderivativeofapform}) is coordinate system independent.  

The nilpotency of $d$:
\begin{eqnarray}
d^2 = 0 \label{eq:nilpotentencyofd}
\end{eqnarray}
is an extremely important property that will be discussed in much, much greater depth later in these notes as well as throughout the rest of this series.  

Also, consider a map $f:\mathcal{M}_1\longrightarrow \mathcal{M}_2$.  Then let $(T_p\mathcal{M}_1)^{\otimes q}$ be an arbitrary tensor product of $q$ vectors in $T_p\mathcal{M}_1$.  We can then take the exterior derivative of the induced pullback:
\begin{eqnarray}
d(f^{\star}\phi_q)\big((T_p\mathcal{M}_1)^{\otimes q}\big) &=& (d\phi_q)\big(T_pf(T_p\mathcal{M}_1)^{\otimes q}\big) \nolabel \\
&=& f^{\star}(d\phi_q)\big((T_p\mathcal{M}_1)^{\otimes q}\big)
\end{eqnarray}
So the exterior derivative and the pullback commute with each other.\footnote{This isn't a proof, but it illustrates the point well enough.  A more formal proof is a straightforward though tedius application of the chain rule.}

Before moving on we will consider a few examples of exterior derivatives in the usual three-dimensional Euclidian space we are familiar with.  

Consider a real function $f$ in $\mathbb{R}^3$.  This is just a zero-form, so the exterior derivative will be as in (\ref{eq:firstexampleexteriorderivativeofafunction}):
\begin{eqnarray}
df = {\partial f \over \partial x^i}dx^i = {\partial f \over \partial x}dx + {\partial f \over \partial y} dy + {\partial f \over \partial z}dz \label{eq:gradintermsofexterior}
\end{eqnarray}
This is the exact expression for the \bf gradient \rm of the function $f$, usually denoted $\boldsymbol\nabla\it f$ (remember that the $dx^i$ should be thought of as unit vectors in the cotangent space, which is simply a vector space - so $dx^i$ is analogous to the row standard basis unit vectors.  So (\ref{eq:gradintermsofexterior}) would be written as $$
{\partial f \over \partial x} \hat i + {\partial f \over \partial y}\hat j + {\partial f \over \partial z}\hat k $$ in the more familiar vector calculus notation, where $\hat i,\hat j$, and $\hat k$ are understood as row vectors).  

Next consider the two form $\phi_2$,
\begin{eqnarray}
\phi_2 = \phi_x\;dy\wedge dz + \phi_y\;dz \wedge dx + \phi_z \;dx \wedge dy
\end{eqnarray}
Taking the exterior derivative,
\begin{eqnarray}
d\phi_2 = \bigg({\partial \phi_x \over \partial x} + {\partial \phi_y \over \partial y} + {\partial \phi_z \over \partial z}\bigg)\; dx \wedge dy \wedge dz \label{eq:firstdivergence}
\end{eqnarray}
If we recognize $\phi_2$ as a vector in a three dimensional space, this expression is its divergence.  

Now consider the one-form $\phi_1$
\begin{eqnarray}
\phi_1 = \phi_xdx+\phi_ydy+\phi_zdz
\end{eqnarray}
Taking the exterior derivative,
\begin{eqnarray}
d\phi_1 = \bigg({\partial \phi_z \over \partial y} - {\partial \phi_y \over \partial z}\bigg) dy \wedge dz + \bigg({\partial \phi_x \over \partial z} - {\partial \phi_z \over \partial x}\bigg) dz \wedge dx + \bigg({\partial \phi_y \over \partial x} - {\partial \phi_x \over \partial y}\bigg) dx \wedge dy  \label{eq:firstcurl}
\end{eqnarray}
Which we recognize as looking a lot like the curl of the vector field $\phi_1$.  

These examples allow us to see that, in three dimensions, the nilpotency of $d$ ($d^2\phi_0=0$) is equivalent to $\boldsymbol\nabla\it \times (\boldsymbol\nabla\it \phi_p) = 0$.  And, $d^2\phi_1 = 0$ is equivalent to $\boldsymbol\nabla\it \cdot (\boldsymbol\nabla\it \times \phi_1)=0$.  

As a final comment, it probably isn't clear how ``unit vectors" like $dy\wedge dz$ in (\ref{eq:firstcurl}) relate to the unit vectors you are used to seeing from vector calculus.  We will discuss this relationship later.

\subsection{Integration of Differential Forms}
\label{sec:intdiffforms}

As we mentioned above, a form $d\phi$ is very much a differential, just like the things you integrate over in introductory calculus.  We have now come to a point where we can discuss how forms are integrated.  

However we must first discuss the notion of \bf orientation\rm.  As an example to motivate this discussion, consider the M\"obius Strip $\mathcal{M}$
\begin{center}
\includegraphics[scale=.7]{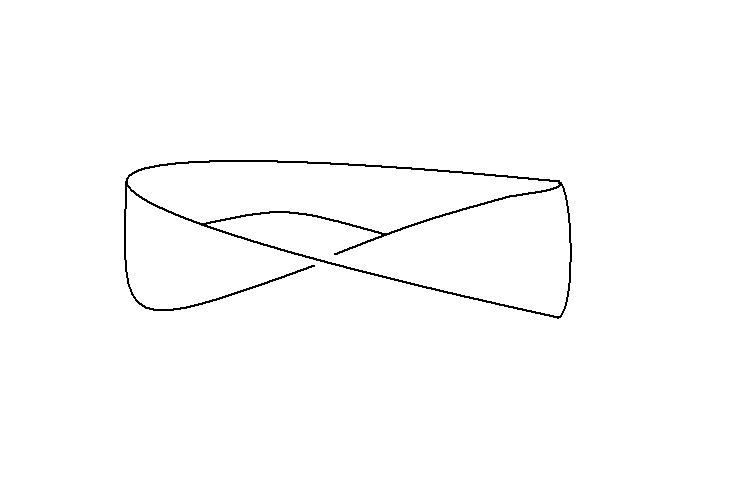}
\end{center}
We can treat this as a manifold (with a boundary - don't worry about this detail for now) and put the usual $x$ and $y$ Cartesian coordinates on it.  
\begin{center}
\includegraphics[scale=.7]{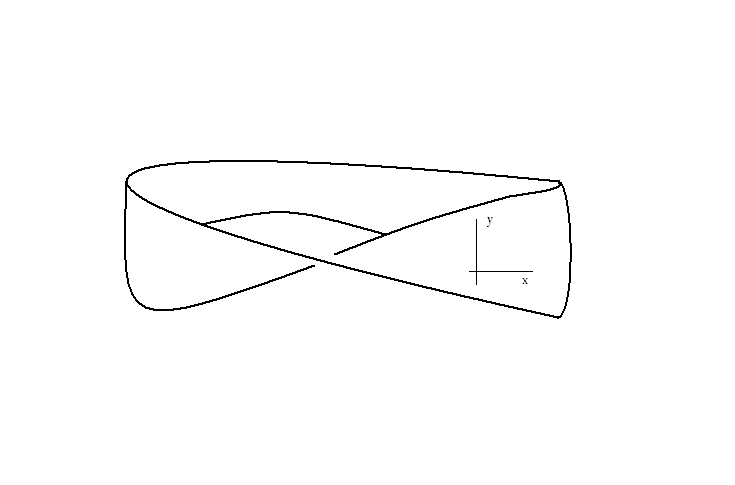}
\end{center}
Now let's say we want to integrate some real function $f$ (a $0$-form) over the entire surface of M\"obius Strip (we are taking $f$ to be a function of an abstract point on the manifold so it is $f(p)$ rather than $f(\boldsymbol\phi\it^{(i)}(p))$).  
$$ \int_{\mathcal{M}} f \;dxdy $$
But consider moving the coordinate system (the $x$ and $y$ axes) around the strip:
\begin{center}
\includegraphics[scale=.7]{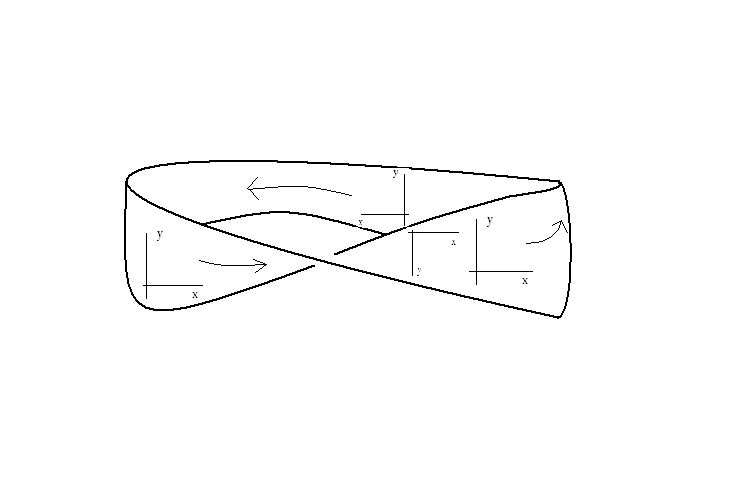}
\end{center}
After returning to where we started the coordinate system has reversed orientation - it started as a right handed system and has ended up a left handed system.  There is clearly no way to rotate the axes back to their original configuration without rotation out of the space or going all the way back around the strip.  

So, because moving around $\mathcal{M}$ has changed the orientation of the strip, and because there is no way to transform the new coordinates back to the old coordinates, the M\"obius Strip is an \bf unoriented \rm manifold.  Obviously this ambiguity will cause problems for integration - at any point there are two inequivalent coordinate systems to choose from, and we don't have any well defined way of choosing one.  Of course, we could choose a single coordinate neighborhood on $\mathcal{M}$, and the coordinate system would be well defined, and the integral would make sense.  

The standard cylinder $\mathcal{C}$
\begin{center}
\includegraphics[scale=.7]{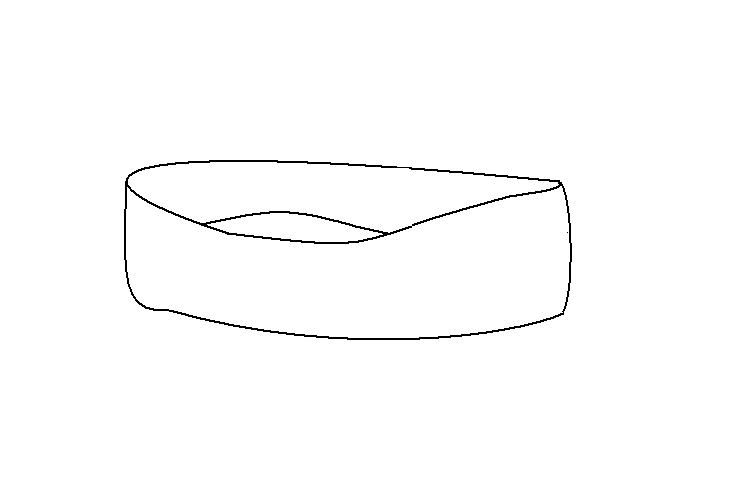}
\end{center}
will not have any problem with orientation.  Moving the coordinate axes around this brings them back to where they started.  Therefore $\mathcal{C}$ is an \bf oriented \rm manifold.  

Other examples of oriented manifolds are the $n$-sphere, the $n$-torus, $\mathbb{R}^n$.  Examples of non-oriented manifolds are M\"obius Strips, Klein Bottles, real projective planes.\footnote{If you aren't familiar with these manifolds that's not a problem.  We won't be discussing them further, though you are encouraged to reference them on your own.}  

Now we want to state this in a more mathematically well-defined way.  In order to do this, we must finally discuss the brief comment we made on page \pageref{wherewetalkaboutdeterminantsinwedgeproducts} about the observation that wedge products are expressible in terms of determinants.  Given some $n$-dimensional manifold $\mathcal{M}$ with some particular coordinate neighborhood $U_i$ and coordinate functions $\boldsymbol\phi\it^{(i)} = \bf x\it$, we know that $\Lambda^nT_p\mathcal{M}$ is one dimensional (cf equation (\ref{eq:combin})).  So, an arbitrary element of $\Lambda^nT_p\mathcal{M}$ can be expressed as
\begin{eqnarray}
f\; dx^1\wedge dx^2\wedge \cdots \wedge dx^n \in \Lambda^nT_p\mathcal{M}
\end{eqnarray}
where $f$ is some real function.  

Now let's assume $p$ is also in the coordinate neighborhood $U_j$ with coordinate functions $\bf y\it$.  We can easily form the linear transformation from $\bf x\it$ to $\bf y\it$ (\ref{eq:changeofcovectorbasis}).  However, because $\Lambda^nT_p\mathcal{M}$ is one dimensional, and therefore the transformed $n$-form will be proportional to the untransformed $n$-form.  We denote the constant of proportionality the \bf Jacobian \rm of the transformation, and it will be equal to the determinant of the transformation:
\begin{eqnarray}
f\; dx^1\wedge dx^2 \wedge \cdots \wedge dx^n &=& f\; {\partial x^1 \over \partial y^{i_1}} dy^{i_1} \wedge {\partial x^2 \over \partial y^{i_2}}dy^{i_2} \wedge \cdots \wedge {\partial x^n \over \partial x^{i_n}}dy^{i_n} \nolabel \\
&=& f \det \bigg({\partial x^i \over \partial y^j}\bigg) dy^1 \wedge dy^2 \wedge \cdots \wedge dy^n \label{eq:definitionofdeterminant}
\end{eqnarray}

For example, consider the three-form in $\mathbb{R}^3$ (with $f=1$)
\begin{eqnarray}
dx \wedge dy \wedge dz
\end{eqnarray}
We can transform this to spherical coordinates according to
\begin{eqnarray}
x &=& r\cos\theta\sin\phi \nolabel \\
y &=& r\sin\theta\sin\phi \nolabel \\
z &=& r\cos\phi
\end{eqnarray}
The transformation matrix will be
\begin{eqnarray}
{\partial x^i \over \partial y^j} = 
\begin{pmatrix}
{\partial x \over \partial r} & {\partial x \over \partial \theta} & {\partial x \over \partial \phi} \\
{\partial y \over \partial r} & {\partial y \over \partial \theta} & {\partial y \over \partial \phi} \\
{\partial z \over \partial r} & {\partial z \over \partial \theta} & {\partial z \over \partial \phi} \\
\end{pmatrix} = 
\begin{pmatrix}
\cos\theta\sin\phi & -r\sin\theta\sin\phi & r\cos\theta\cos\phi \\
\sin\theta\sin\phi & r\cos\theta\sin\phi & r\sin\theta\cos\phi \\
\cos\phi & 0 & -r\sin\phi
\end{pmatrix}
\end{eqnarray}
And the Jacobian (determinant) of this matrix is
\begin{eqnarray}
r^2\sin\phi
\end{eqnarray}
So
\begin{eqnarray}
dx\wedge dy\wedge dz = r^2\sin\phi \; dr \wedge d\theta \wedge d\phi \label{eq:volumeformsincartesianandspherical}
\end{eqnarray}
This makes perfect sense from standard vector calculus - we can recognize both sides as the volume element in their respective coordinate systems.  

Now, recall that with the M\"obius Strip above, the problem was that when you go around the strip the coordinate system changes orientation - in other words $dx\wedge dy \longrightarrow -dx \wedge dy$.  So, if we choose a point on the strip, it will be in the intersection of two different coordinate neighborhoods.  And while the two-forms in each coordinate neighborhood will be proportional to each other, they will differ in sign.  According to our above result (\ref{eq:definitionofdeterminant}), this is equivalent to saying that the Jacobian (the determinant of the transformation) is negative.  

Therefore we have the following definition - A manifold $\mathcal{M}$ is \bf orientable \rm if, for any overlapping coordinate neighborhoods $U_i$ and $U_j$, the Jacobian of the transformation between them, denoted $J$, satisfies $J>0$.\footnote{It obviously makes no sense to say that it must satisfy $J\geq 0$ because if $J=0$ then the transformation is singular and it cannot be inverted.}  

As a final comment about orientability, notice that if $\mathcal{M}$ is orientable, then according to what we have said above there must exist an $n$-form $\phi_n$ which never vanishes.  This form is called the \bf Volume Form\rm, and we denote it $\phi_V$.  Both sides of (\ref{eq:volumeformsincartesianandspherical}) are volume forms in their respective coordinates.  A volume form will play the role of the integration measure when we integrate over $\mathcal{M}$.  Any oriented manifold will admit two different equivalent classes of volume forms - one with a relative positive sign and another with a relative minus sign.  We call one of them \bf right handed \rm and the other \bf left handed\rm.  From now on, unless otherwise specified, you can assume that all manifolds we work with are oriented.  

We now turn to the idea of integrating a form over a manifold.  Let's say we have some function $F$ such that
\begin{eqnarray}
F:\mathcal{M} \longrightarrow \mathbb{R}
\end{eqnarray}
($F$ is just a real function on the manifold).  We will, for now, only worry about integrating over a single coordinate neighborhood $U_i$ of $\mathcal{M}$.  To integrate $F$ over $U_i$ we first multiply it by the volume form $\phi_V$ which acts as the integration \bf measure \rm for the integral.  The volume form will usually be some function $h$ (In three dimensions it was $h=1$ in Cartesian coordinates and $h=r^2\sin\phi$ in spherical coordinates, cf (\ref{eq:volumeformsincartesianandspherical})) times the wedge product of all basis covectors in some coordinates frame:
\begin{eqnarray}
\phi_V = h \; dx^1\wedge dx^2\wedge \cdots \wedge dx^n \label{eq:volumeforminintegrationsection}
\end{eqnarray}

So, the integral is written as
\begin{eqnarray}
\int_{U_i} F(p) \phi_V
\end{eqnarray}
(where $p\in U_i$).  While the integral will only be well-defined on the entire manifold if it is oriented, we can take a single coordinate neighborhood to be oriented because it is homeomorphic to $\mathbb{R}^n$ - if $\mathcal{M}$ is not oriented, then the orientation, or handedness, of $\phi_V$ will not be well defined in some region, and the integral cannot be globally defined.  

Once we have a function $F$ and a volume form (\ref{eq:volumeforminintegrationsection}), with the frame covectors $dx^i$ in a particular order, we define the integral as
\begin{eqnarray}
\int_{U_i} F(p)\phi_V = \int_{\boldsymbol\phi\it^{(i)}(U_i)} F(\phi^{-1,(i)}(x^1,x^2,\cdots,x^n))h dx^1dx^2\cdots dx^n \label{eq:definitionofintegralintermsofvolumeform}
\end{eqnarray}
Where the integral on the right hand side is understood as a normal integral as you learned about in an introductory Calculus course.  Notice that the integral on the left hand side ``lives" on $U_i$.  $F$ is a function directly from $\mathcal{M}$ to $\mathbb{R}$, and the volume form $\phi_V$ lives on the manifold.  However the right hand side ``lives" on the open subset of $\mathbb{R}^n$ that the coordinate functions map to.  Because $\boldsymbol\phi\it^{(i)}$ are homeomorphic they are invertible, and we can take any point in $\mathbb{R}^n$ and map it back to the manifold using $\phi\it^{-1,(i)}$.  

So the point is that an integral on the manifold is actually done on the open subset of $\mathbb{R}^n$ that the coordinate functions map to.  This isn't particularly profound - the manifold is an abstract, coordinate free space.  An integral needs more structure to make sense.  Therefore it only makes sense that we take the integral down to a space where we are used to doing integrals. 

Also notice that the integration measure $dx^1dx^2\cdots dx^n$ is ordered the same as the volume form (\ref{eq:volumeforminintegrationsection}).  To understand why this is important, recall that a particular wedge product is a unit vector in a particular vector space of the exterior algebra (see section \ref{sec:exterioralgebras}).  And just as $\hat i$ and $-\hat i$ represent two different (opposite) directions in $\mathbb{R}^3$, so also $dx\wedge dy$ and $dy\wedge dx$ ($=-dx\wedge dy$) represent two different (opposite) directions in the vector space $\Lambda^2T_p\mathbb{R}^3$.  The discussion around equations (\ref{eq:gradintermsofexterior}) through (\ref{eq:firstcurl}) may help clarify this.  So, the volume form on the left side of (\ref{eq:definitionofintegralintermsofvolumeform}) \it defines \rm the orientation of the coordinate system being integrated over.  Therefore if we were to switch two of the frame covectors, the volume form would be
\begin{eqnarray}
\phi'_V = -h\; dx^2\wedge dx^1 \wedge \cdots \wedge dx^n
\end{eqnarray}
And the integral would be
\begin{eqnarray}
\int_{U_i} F(p)\phi'_V = -\int_{\phi\it^{(i)}(U_i)} F(\phi^{-1,(i)}(x^1,x^2,\cdots,x^n)h dx^1dx^2\cdots dx^n
\end{eqnarray}
where the integral on the right hand side is understood as being over a coordinate system with the opposite orientation.  This is a generalization of the fact that
\begin{eqnarray}
\int_a^b f(x)dx = -\int_b^af(x)dx
\end{eqnarray}
from introductory calculus. 

Moving on, it may appear that (\ref{eq:definitionofintegralintermsofvolumeform}) depends on the coordinates chosen.  However it does not.  Let's say we have a smooth transformation from coordinate neighborhood $U_i$ to $U_j$ with coordinates $\bf y\it$, and for simplicity $U_i$ and $U_j$ cover the same portion of the manifold.  The volume form will transform according to (\ref{eq:definitionofdeterminant}), and therefore in the new coordinates the integral will be
\begin{eqnarray}
\int_{U_i}F(p)\phi_V &=& \int_{\phi^{(i)}(U_i)} F(\phi\it^{-1,(i)}(x^1,x^2,\ldots,x^n)h \; dx^1dx^2\cdots dx^n \nolabel \\
&=& \int_{\boldsymbol\phi\it^{(j)}(U_j)} F(\phi^{-1,(j)}(y^1,y^2,\ldots,y^n)h\det \bigg({\partial x^a \over \partial y^b}\bigg) dy^1dy^2\cdots dy^n \label{eq:generalchangeofvariableswhendoinganintegral}
\end{eqnarray}
where, because of the smoothness of the transition functions between coordinate's (the $\boldsymbol\psi\it^{(ij)}$), we could also have expressed $F$ in the new coordinates as
\begin{eqnarray}
F(x^1(y^1,y^2,\ldots,y^n),x^2(y^1,y^2,\ldots,y^n),\ldots,x^n(y^1,y^2,\ldots,y^n))
\end{eqnarray}

So, in (\ref{eq:generalchangeofvariableswhendoinganintegral}) we find the general equation for a change of variables in an integral.  You can see this clearly from (\ref{eq:volumeformsincartesianandspherical}) and (\ref{eq:areaformfortwodimensionsusingcartesianandpolar}).  In (\ref{eq:areaformfortwodimensionsusingcartesianandpolar}) we had the volume form\footnote{Volume in two dimensions is the same as area - area is a two-dimensional volume.} $dx \wedge dy$.  If we wanted to integrate a function $f$ over a coordinate neighborhood, it would have the form 
\begin{eqnarray}
\int_{U_i} f \; dx \wedge dy = \int_{\boldsymbol\phi\it^{(i)}(U_i)} f(x,y)dxdy
\end{eqnarray}
The right side of (\ref{eq:areaformfortwodimensionsusingcartesianandpolar}) then corresponds to the integral 
\begin{eqnarray}
\int_{U_j} f\; r\; dr \wedge d\theta = \int_{\boldsymbol\phi\it^{(j)}(U_j)} f(r,\theta)\; rdrd\theta
\end{eqnarray}
This corresponds exactly to what is done in an introductory vector calculus course.  You can do the same thing for (\ref{eq:volumeformsincartesianandspherical}).  

So far all we have discussed in this section is the special case in which we are only integrating over a single coordinate neighborhood $U_i$.  Now we can discuss how to perform an integral over multiple coordinate neighborhoods.  

Take some atlas $\{(U_i,\boldsymbol\phi\it^{(i)})\}$ such that each $p\in\mathcal{M}$ is covered by a finite number of coordinate neighborhoods (we assume this is always the case).  Then define a set of differentiable functions $\epsilon_i(p)$ satisfying the following:\\
\indent 1) $\epsilon_i(p) = 0$ if $p \not \in U_i$.\\
\indent 2) $0\leq \epsilon_i(p) \leq 1\; \forall \; p$ and $i$.  \\
\indent 3) $\sum_i \epsilon_i(p) = 1\; \forall \; p \in \mathcal{M}$.\\
Such a set of functions is called a \bf Partition of Unity\rm.  

The integral of a function $F$ across the entire manifold can then be written as
\begin{eqnarray}
\int_{\mathcal{M}} F \phi_V = \sum_i \int_{U_i} \epsilon_i(p)f(p) \phi_V \label{eq:particianofunityintegral}
\end{eqnarray}
It can be shown that the integral defined by (\ref{eq:particianofunityintegral}) is independent of the choice of coordinate functions on any of the neighborhoods, and is independent of the partition of unity.  

For example, consider the manifold $S^1$ and the function $F(\theta)=\cos^2\theta$.  Of course we can do the integral in the normal way:
\begin{eqnarray}
\int_{0}^{2\pi} d\theta \cos^2\theta = \pi
\end{eqnarray}
But we can reproduce this using a partition of unity.  Let
\begin{eqnarray}
\epsilon_1(\theta) &=& \sin^2(\theta/2) \nolabel \\
\epsilon_2(\theta) &=& \cos^2(\theta/2)
\end{eqnarray}
Then the integral is
\begin{eqnarray}
\int_{S^1} d\theta \cos^2\theta &=& \int_0^{2\pi} d\theta \sin^2(\theta/2)\cos^2\theta + \int_{-\pi}^{\pi} d\theta \cos^2(\theta/2)\cos^2\theta \nolabel \\
&=& {1\over 2} \pi + {1\over 2} \pi = \pi
\end{eqnarray}

So, on any orientable manifold, a function can be integrated across the manifold by mapping the entire integral into a coordinate system.  The volume form on the manifold determines the orientation of the integral, and the integral is independent of the coordinate functions chosen - the Jacobian allows us to transform the integral from one coordinate system to another.  If we must integrate across multiple coordinate neighborhoods, a partition of unity allows this to be done in a well defined way, and the integral does not depend on the value of the partition of unity.  

\subsection{Vector Fields and Lie Derivatives}
\label{sec:flowandlie}

We have discussed tangent \it spaces \rm at every point $p\in \mathcal{M}$.  However it will also prove useful to discuss \bf vector fields \rm on manifolds.  A vector field is a \it smooth \rm map from $\mathcal{M}$ to the tangent bundle $T\mathcal{M}$.  We define a smooth map $\bf \xi\it$ from $\mathcal{M}$ (or a subset of $\mathcal{M}$) into $T\mathcal{M}$, called a \bf vector field\rm, as
\begin{eqnarray}
p \longmapsto (p,\bf v\it(p)) = \bf \xi\it(p) \in T_p\mathcal{M}
\end{eqnarray}

In other words, at each point $p$, $\bf v\it$ selects one single vector out of the $\mathbb{R}^n$ which comprises the tangent space.  We then demand that this map be smooth in the above sense.  

An example of this is a fluid moving through some volume.  The manifold will be the volume the fluid is moving through.  Any single drop of fluid \it could \rm be moving in any direction in the tangent space (there are $n$ degrees of freedom in an $\mathbb{R}^n$ tangent space).  However, a particular vector field, or in other words a particular ``flow" of the fluid, will assign a unique velocity at each point in the manifold.  So, for each point $p$, we can assign a velocity $\bf v\it(p)$.  

Another example familiar in physics is an electric field $\bf E\it$.  The manifold in this case is usually just space, and the vector field assigns a specific ``magnitude and direction" to the electric field at each point.  

In section \ref{sec:tangentmapping}, we had a fairly in-depth discussion of the relationship between a curve $\tilde Q(\tau)\in \mathcal{M} \; \ni \tilde Q:[a,b]\longrightarrow \mathcal{M}$ and a tangent vector at some point on that curve.  You are encouraged to reread that section now and familiarize yourself with those ideas.  

There, we had a single curve defining tangent vectors at any point $\tau_0$ along the curve.  We also mentioned there that any vector can be defined by a curve passing through that point.  We will now exploit this one to one correspondence between curves and tangent vectors.  

Let us now define, instead of a single curve $\tilde Q \in \mathcal{M}$, a \it family \rm of curves, which we call $\tilde q(\tau,p) \in \mathcal{M}$, where $\tau\in[a,b]$ and $p\in \mathcal{M}$, and $\tau$ plays the exact same role as before.  We will define this so that $0\in[a,b]$, and so that $\tilde q(0,p)=p$.  In other words, this is a family of curves defined for every point $p\in \mathcal{M}$, parameterized such that the $\tau=0$ point is the identity map.  Then the point $\tilde q(\tau',p)$ will be another point on $\mathcal{M}$ further along the curve which passes through $p$.  Using the same type of argument as in section \ref{sec:tangentmapping}, this construction will define a tangent vector at every point of $\mathcal{M}$ (by taking the $\tau$ derivative of $\tilde q(\tau,p)$ at the point $p$), and consequently we have a vector field on $\mathcal{M}$ defined by our family of curves.  

One obvious necessary condition is that for every point $p\in \mathcal{M}$, there can be one and only one curve passing through $p$ - no two curves can intersect.  If there is an intersection then there will be two different tangent vectors associated with that point, which we don't want.  

So, given a family of curves with no intersections, we have a well-defined vector field - a single tangent vector at each point in $\mathcal{M}$.  We can, however, approach this in the opposite ``direction".  Let's say we have a vector field already defined on $\mathcal{M}$.  This implies that \it there exists \rm a family of curves $\tilde q(\tau,p)$ which define this vector field.  In other words, the $\tau$ derivative of $\tilde q(\tau,p)$ at a given point $p$ will define the tangent vector at that point.  We will use the coordinate functions $\bf x\it$ as in \ref{sec:tangentmapping}, resulting in the mapping 
\begin{eqnarray}
\bf q\it(\tau,\bf x\it(p))=\bf x\it(\tilde q(\tau,p)):[a,b]\longrightarrow \mathbb{R}^n
\end{eqnarray}
If we define a tangent vector at $p$ to have components $v^i(\bf x\it(p))$ (so the vector in the vector field at each point is $\bf v\it(\bf x \it(p)) = v^i(\bf x\it(p)){\partial \over \partial x^i}$),\footnote{We are letting the components $v^i$ depend on the \it coordinates \rm of $p$ rather than on $p$ for simplicity - it will be easier to keep everything in the same space, namely $\mathbb{R}^n$.} then we have
\begin{eqnarray}
{dq^i(\tau,\bf x\it(p)) \over d\tau}= v^i(\bf q\it(\tau,\bf x\it(p))) = v^i(\bf x\it(\tilde q(\tau,p))) 
\end{eqnarray}
or
\begin{eqnarray}
{d\bf q\it(\tau,\bf x\it(p)) \over d\tau}= \bf v\it(\bf q\it(\tau,\bf x\it(p))) = \bf v\it(\bf x\it(\tilde q(\tau,p))) \label{eq:relationshipbetweencurveandtangentvectorflow}
\end{eqnarray}
(compare this with (\ref{eq:tangentvectorasaresultofQ})).  Notice that this is a first order differential equation for $\bf q\it$ subject to the boundary condition 
\begin{eqnarray}
\bf q\it (0,\bf x\it(p)) = \bf x\it(p)
\end{eqnarray}
In other words, the vector field $v^i(\bf x\it(p))$ on $\mathcal{M}$ has produced a differential equation, the solutions of which will be the curves which ``flow" along the manifold and define the vector field.  

As an example, consider the vector field in $\mathcal{M}=\mathbb{R}^2$ given by
\begin{eqnarray}
\bf v\it = -y {\partial \over \partial x} + x{\partial \over \partial y}
\end{eqnarray}
So, we have $v^1=-y$ and $v^2=x$.  The differential equations for this will be
\begin{eqnarray}
{dq^1\over d\tau} &=& -y \; = -q^2\nolabel \\
{dq^2\over d\tau} &=& x \;= q^1 \label{eq:boundaryconditionsexampleliederiv}
\end{eqnarray}
Taking the second derivative of the first with $\tau$ and plugging in the second, we get
\begin{eqnarray}
{d^2q^1 \over d\tau^2} = -{dq^2 \over d\tau} = -q^1
\end{eqnarray}
which has solutions of $\sin$'s and $\cos$'s:
\begin{eqnarray}
q^1&=&A\sin\tau+B\cos\tau \nolabel \\
q^2 &=& C\sin\tau+ D\cos\tau
\end{eqnarray}

We will make our boundary condition at $\tau=0$ the point $(x_0,y_0)$.  We therefore have the conditions (cf (\ref{eq:boundaryconditionsexampleliederiv})):
\begin{eqnarray}
{dq^1\over d\tau}\bigg|_{\tau=0}&=&-y_0 \nolabel \\
{dq^2\over d\tau}\bigg|_{\tau=0}&=&x_0 \nolabel \\
q^1\big|_{\tau=0}&=&x_0 \nolabel \\
q^2\big|_{\tau=0}&=&y_0
\end{eqnarray}
This leads to the general solution:
\begin{eqnarray}
q^1(\tau) &=& -y_0\sin\tau + x_0\cos\tau \nolabel \\
q^2(\tau) &=& x_0 \sin\tau + y_0\cos\tau \label{eq:finalformofqinexampleforflow}
\end{eqnarray}
This is simply the parametric equation of a circle in $\mathbb{R}^2$, starting at $(x_0,y_0)$ and centered at the origin (in other words, a circle around the origin with radius $r^2=x_0^2+y_0^2$).  The total family of curves will be the set of such circles with all radii.  

Notice in (\ref{eq:finalformofqinexampleforflow}) that we can rewrite it as
\begin{eqnarray}
\begin{pmatrix}
q^1 \\ q^2
\end{pmatrix} = 
\begin{pmatrix}
\cos\tau & -\sin\tau \\
\sin\tau & \cos\tau
\end{pmatrix}
\begin{pmatrix}
x_0 \\ y_0
\end{pmatrix} \label{eq:elementofso2inliederivsection}
\end{eqnarray}
The matrix on the right hand side is an element of the two dimensional representation of the Lie Group $SO(2)$ (cf \cite{Firstpaper}).  So the vector field $\bf v\it(p)$ induces curves which take a point in $\mathcal{M}$ and act on it like $SO(2)$ - rotating it around the origin while preserving the radius.  

So, a family of curves over the entire manifold $\mathcal{M}$ is equivalent to a smooth vector field over the entire manifold $\mathcal{M}$.  We refer to the family of curves as the \bf flow \rm associated with the vector field $\bf v\it(\bf x\it(p))$.  

Notice that, in general, a flow $\tilde q(\tau,p)$ induces a Lie Group.  For any point $p\in\mathcal{M}$, we have\\
\indent 1) Closure - $\tilde q(\tau,\tilde q(\tau',p)) = \tilde q(\tau+\tau',p)$.\\
\indent 2) Associativity - obvious.\\
\indent 3) Identity - $\tilde q(0,p) = p$ \\
\indent 4) Inverse - $\tilde q(-\tau,\tilde q(\tau,p))=\tilde q(\tau-\tau,p)=\tilde q(0,p)=p$\\
Furthermore, this group (for a single point) is an Abelian Lie Group.  It will look locally like an additive group on $\mathbb{R}$, but not necessarily globally.  In the above example it was $SO(2)$, which looks locally like $\mathbb{R}$, but the identification of the $\tau$ and the $\tau+2\pi$ element shows that it is not globally like $\mathbb{R}$.  

Now consider an infinitesimal perturbation away from $p$ using the path $\tilde q$ (with induced vector field $\bf v\it(\bf x\it(p))$, where $p\rightarrow \tilde q(\epsilon,p)$ (where $\epsilon$ is infinitesimally small).  We want to do a Taylor expansion around $\tau=0$, so we use the coordinate functions $\bf x\it$ to map to $\mathbb{R}$, so now $\bf x\it(p) \longrightarrow \bf q\it(\epsilon,\bf x\it(p))$.  Now, doing the Taylor expansion,
\begin{eqnarray}
\bf q\it(\epsilon,\bf x\it(p)) &=& \bf q\it(\tau,\bf x\it(p))\big|_{\tau=0} + \epsilon{d \bf q\it(\tau,\bf x\it(p)) \over d\tau}\bigg|_{\tau=0}  + \cdots \nolabel \\
&=& \bf x\it(p) + \epsilon \bf v\it(\bf x\it(p)) + \cdots \label{eq:expandtofirstorderliederiv}
\end{eqnarray}
(where we have only kept terms to second order in $\epsilon$).  So we see that, geometrically, the vector $\bf v\it$ acts as a sort of ``generator" at $\bf x\it(p)$, pointing in the direction of the flow.  

Furthermore, consider now a finite (rather than infinitesimal) transformation from $\bf x\it(p)$:
\begin{eqnarray}
\bf q\it(\sigma,\bf x\it(p)) &=& \bf q\it(\tau,\bf x\it(p))\big|_{\tau=0} + \sigma {d\bf q\it(\tau,\bf x\it(p)) \over d\tau}\bigg|_{\tau=0} + {1\over 2!} \sigma^2 {d\bf q\it(\tau,\bf x\it(p)) \over d\tau}\bigg|_{\tau=0} + \cdots \nolabel \\
&=& \bigg(\sum_{n=0}^{\infty} {1\over n!}\sigma^n\bigg({d \over d\tau}\bigg)^n\bigg) \bf q\it(\tau,\bf x\it(p))\big|_{\tau=0} \nolabel \\
&=& e^{\sigma {d \over d\tau}}\bf q\it(\tau,\bf x\it(p))\big|_{\tau=0} \nolabel \\
&=& e^{\sigma \bf v\it}\bf x\it(p) \label{eq:vectorsaregeneratorsofcurvesthroughapoint}
\end{eqnarray}
To get the last equality, we have used the general equation that, for a vector $\bf v\it$  and operator $\hat{\mathcal{P}}$ with eigenvalue $\mathcal{P}$ (so $\hat{\mathcal{P}} \bf v\it = \mathcal{P}\bf v\it$), we have
\begin{eqnarray}
e^{\hat{\mathcal{P}}}\bf v\it = e^{\mathcal{P}}\bf v\it
\end{eqnarray}
Operating on $\bf q\it (\tau, \bf x\it(p))$ with ${d \over d\tau}$ gives the tangent vector $\bf v\it$, with only the starting point $\bf x\it(p)$ left.  

So the content of (\ref{eq:vectorsaregeneratorsofcurvesthroughapoint}) is that if we start at an arbitrary point $\bf x\it(p)$ on the manifold (mapped to its coordinates), the action of the class of curves $\tilde q(\tau,p)$ will move that point according to the exponentiation of the vector $\bf v\it(p)$ at the point $p$.  In other words, the vector $\bf v\it(p)$ is the \bf infinitesimal generator \rm of the flow at the point $p$.  

This should remind you of what we did in \cite{Firstpaper} with the generators of Lie groups.  It turns out that this is in fact the exact same thing.  The only difference is that we are approaching this geometrically rather than our purely algebraic approach in the first paper, and that our discussion right now is about general manifolds.  

Moving on, it is obvious that we have the following properties, which show the correspondence between using $\bf q\it$ and using the exponentiated vectors:\\
\indent 1) $\bf q\it(0,\bf x\it(p)) = \bf x\it(p) \qquad \iff \qquad e^{0\bf v\it}\bf x\it(p) = \bf x\it(p)$\\
\indent 2) ${d \bf q\it(\tau,\bf x\it(p)) \over d\tau}\big|_{\tau=0} = \bf v\it(\bf x\it(p)) \qquad \iff \qquad {d \over d\tau}e^{\tau\bf v\it}\big|_{\tau=0} \bf x\it(p) = \bf v\it e^{0\bf v\it} \bf x\it(p) = \bf v\it(\bf x\it(p)) $ \\
\indent 3) $\bf q\it(\tau,\bf q\it(\tau',p)) = \bf q\it(\tau+\tau',p) \qquad \iff \qquad e^{\tau\bf v\it}e^{\tau'\bf v\it}\bf x\it(p)=e^{(\tau+\tau')\bf v\it} \bf x\it(p) \\$
So we can see that these are indeed equivalent descriptions of the relationship between vector fields and flows.  

There is something interesting going on here - if we are at a point $p\in \mathcal{M}$ and we want to move along the ``flow" generated by a vector field, we can follow the flow line through $p$ as far as we want knowing \it only \rm the vector at $p$.  We don't need to know the vector at any other point except where we start.  As we mentioned above, this was the same situation with the generators of a Lie group in \cite{Firstpaper}.  The generators described the behavior of the \it group \rm near the starting point (the identity), and exponentiating the generators (along with some parameter) moved you anywhere in the group.  Here, the tangent vector at $p$ describes the behavior of the \it vector field \rm near the starting point (the point $p$), and exponentiating the tangent vector at $p$ (along with some parameter) moves you anywhere in the manifold along the flow line through $p$.  We have arrived at extremely similar ideas, first through purely algebraic means and now through purely geometric means.  The difference is that before the generators satisfied a specific algebra (defined by the structure constants), and there were multiple generating tangent vectors at each point.  

We will continue to explore the similarities between these two situations in this section, and in the next section we will see how we make the jump to a true Lie group structure, thus tying what we did in the group theory section of \cite{Firstpaper} with what we have done so far here.  

Moving on, let's say we have some manifold $\mathcal{M}$ with \it two \rm different vector fields, $\bf v\it^{(1)}=v^{(1),i}{\partial \over \partial x^i}$ and $\bf v\it^{(2)} = v^{(2),i}{\partial \over \partial x^i}$, which are not necessarily related to each other in any way.  

These two vector fields will naturally have associated families of curves, which we denote $\tilde q^{(1)}(\tau,p)$ and $\tilde q^{(2)}(\tau,p)$, respectively.  These will naturally satisfy (\ref{eq:relationshipbetweencurveandtangentvectorflow}) separately (mapping everything once again with the coordinate functions):
\begin{eqnarray}
{d\bf q\it^{(1)}(\tau,\bf x\it(p)) \over d\tau} &=& \bf v\it^{(1)}(\bf q\it^{(1)}(\tau,\bf x\it(p))) \nolabel \\
{d\bf q\it^{(2)}(\tau,\bf x\it(p)) \over d\tau} &=& \bf v\it^{(2)}(\bf q\it^{(2)}(\tau,\bf x\it(p))) \nolabel \\
\end{eqnarray}
So at the point $p$, there will be two different tangent vectors and two different flow lines.  Let's take the $\bf v\it^{(2)}$ tangent vector and the $\bf q\it^{(1)}$ flow line.  Our goal here is to calculate the derivative of $\bf v\it^{(2)}(\bf q\it^{(1)}(\tau,\bf x\it(p)))$ along the curve $\bf q\it^{(1)}$ (so we are, for this problem, not immediately interested in the vector field $\bf v\it^{(1)}$ or the curves $\bf q\it^{(2)}$).  

As in calculus I, we take any derivative of $\bf v\it^{(2)}$ at point $p$ by comparing it with $\bf v\it^{(2)}$ at a nearby point $\tilde q^{(1)}(\epsilon,p)$ and then taking the limit as $\epsilon \rightarrow 0$:
\begin{eqnarray}
\lim_{\epsilon \rightarrow 0} {\bf v\it^{(2)}(\bf q\it^{(1)}\big(\epsilon,\bf x\it(p))) - \bf v\it^{(2)}(\bf x\it(p)\big)  \over \epsilon} \label{eq:firstattemptatliederivative}
\end{eqnarray}
Equation\label{wherewediscussdifferenttangentspacesderiv} (\ref{eq:firstattemptatliederivative}) makes sense in principle, but it is not well defined.  There is an ambiguity in how to compare these two vectors.  The problem is that $\bf v\it^{(2)}(\bf x\it(p))$ is in the tangent space $T_p\mathcal{M}$, while $\bf v\it^{(2)}(\bf q\it^{(1)}(\epsilon,\bf x\it(p)))$ is in the tangent space $T_{\tilde q^{(1)}(\epsilon,p)}\mathcal{M}$, and there is no well defined way of comparing these two vectors.  

This problem is more obvious on curved manifolds, where there is no natural way at all compare vectors in two tangent places.  

The solution is to use the tangent map (cf section \ref{sec:tangentmapping}).  The problem is that the vectors aren't in the same tangent space, and so we want to move one of them to the tangent space of the other.  We have a smooth function $\tilde q^{(1)}$ which maps $p$ to $\tilde q^{(1)}(\epsilon,p)$ (or the map $\tilde q^{(1)}(-\epsilon,p')$ which will map $\tilde q^{(1)}(\epsilon, p)$ back to $p$), and therefore we can use what we learned in section \ref{sec:tangentmapping} to make (\ref{eq:firstattemptatliederivative}) make sense - by comparing the two vectors in the same tangent space, $T_p\mathcal{M}$.  

We make a warning here: the notation we will use in the next few pages will be hideous.  The meaning of the notation should be very clear at each point, but because of how many terms we will be writing in each expression it will appear very muddled.  Bear with us and trust that by the end everything will simplify very nicely.  

As we said above, the vector $\bf v\it^{(2)}(\bf q\it^{(1)}(\epsilon,\bf x\it(p)))$ is at the point $\tilde q^{(1)}(\epsilon,p) \in \mathcal{M}$.  We want to take the vector at this point back to the point $p$.  We do this by using the tangent map of $\tilde q^{(1)}(-\epsilon,p)$, which we denote $T_{\tilde q^{(1)}(\epsilon,p)}\tilde q^{(1)}(-\epsilon,\tilde q^{(1)}(\epsilon,p))$.  So, acting on the vector $\bf v\it^{(2)}(\bf q\it^{(1)}(\epsilon,\bf x\it(p)))$ with the tangent map $T_{\tilde q^{(1)}(\epsilon,p)}\tilde q^{(1)}(-\epsilon,\tilde q^{(1)}(\epsilon,p))$ will map $\bf v\it^{(2)}(\bf q\it^{(1)}(\epsilon,\bf x\it(p)))$ back to a vector in the tangent space $T_p\mathcal{M}$.  And therefore we can write the definition of the derivative of the $\bf v\it^{(2)}$ vector field along the $\bf q\it^{(1)}$ curve as
\begin{eqnarray}
\mathcal{L}_{\bf v\it^{(1)}}\bf v\it^{(2)}(\bf x\it(p)) = \lim_{\epsilon\rightarrow 0}   \bigg(   {T_{\tilde q^{(1)}(\epsilon,p)}\tilde q^{(1)}(-\epsilon,\tilde q^{(1)}(\epsilon,p))(\bf v\it^{(2)}(\bf q\it^{(1)}(\epsilon,\bf x\it(p)))) - \bf v\it^{(2)}(\bf x\it(p)) \over \epsilon }      \bigg)\nolabel \\ \label{eq:firstdefinitionofliederivative}
\end{eqnarray}
Admittedly this is a bit of a monstrosity, but it can greatly simplified as we will see.  All it means is that we are comparing elements of a vector field at $p$ and at $\tilde q^{(1)}(\epsilon,p)$, taking the difference and dividing by the ``distance" between them ($\epsilon$).  We are using the tangent map in order to make the comparison well defined.  

The derivative $\mathcal{L}_{\bf v\it^{(1)}}\bf v\it^{(2)}(\bf x\it(p))$, defined by (\ref{eq:firstdefinitionofliederivative}) is called the \bf Lie Derivative \rm of $\bf v\it^{(2)}(\bf x\it(p))$ at the point $p$ in the direction defined by the vector field $\bf v\it^{(1)}$.  Note that the subscript is the \it vector field \rm which defines the curves along which we take the derivative.  

We now begin to simplify (\ref{eq:firstdefinitionofliederivative}) by explicitly writing out the tangent map.  Because we are working with only an infinitesimal displacement $\epsilon$, we only need to keep terms to first order, as in (\ref{eq:expandtofirstorderliederiv}).  First, using (\ref{eq:expandtofirstorderliederiv}) we write out the vector 
\begin{eqnarray}
\bf v\it^{(2)}(\bf q\it^{(1)}(\epsilon,\bf x\it(p))) &=& \bf v\it^{(2)} \big(   \bf x\it(p) + \epsilon \bf v\it^{(1)}(\bf x\it(p))   \big)+ \cdots  \nolabel \\
&=& \bf v\it^{(2)}(\bf x\it(p)) + \epsilon  v^{(1),i}(\bf x\it(p)){\partial \bf v\it^{(2)} (\bf x\it(p))\over \partial x^i}+ \cdots \label{eq:expressionofcomponentsliederiv}
\end{eqnarray}
Keep in mind that this is a vector in the tangent space $T_{\tilde q^{(1)}(\epsilon,p)}\mathcal{M}$.  So we want to map this vector back to $T_p\mathcal{M}$ using the tangent mapping $T_{\tilde q^{(1)}(\epsilon,p)} \tilde q^{(1)}(-\epsilon,\tilde q^{(1)}(\epsilon,p))$.  

So, using (\ref{eq:curvefreedefinitionofpushforward}), where (\ref{eq:expressionofcomponentsliederiv}) plays the role of the original components (which were denoted $v^i$ in (\ref{eq:curvefreedefinitionofpushforward})) and $\bf q\it^{(1)}(\epsilon,\bf x\it(p))$ plays the role of the transformation function (what was denoted $F^i$ in (\ref{eq:curvefreedefinitionofpushforward})), we have (using the same coordinate functions and therefore the same basis vectors) that the tangent map gives the the vector with $k^{th}$ component (compare this closely to (\ref{eq:curvefreedefinitionofpushforward})):
\begin{eqnarray}
& &\bigg(v^{(2),j}(\bf x\it(p)) + \epsilon v^{(1),i}(\bf x\it(p)){\partial v^{(2),j}(\bf x\it(p)) \over \partial x^i}\bigg)\bigg({\partial q^{(1),k}(-\epsilon, \bf x\it(p)) \over \partial x^j}\bigg) \nolabel \\
&=&\bigg(v^{(2),j}(\bf x\it(p)) + \epsilon v^{(1),i}(\bf x\it(p)){\partial v^{(2),j}(\bf x\it(p)) \over \partial x^i}\bigg)\bigg({\partial \over \partial x^j} \big(x^k(p) - \epsilon v^{(1),k}(\bf x\it(p))\big)\bigg) \nolabel \\
&=&\bigg(v^{(2),j}(\bf x\it(p)) + \epsilon v^{(1),i}(\bf x\it(p)){\partial v^{(2),j}(\bf x\it(p)) \over \partial x^i}\bigg)\bigg(\delta^k_j - \epsilon {\partial v^{(1),k}(\bf x\it(p)) \over \partial x^j}\bigg) \nolabel \\
&=& v^{(2),j}(\bf x\it(p))\delta^k_j - \epsilon v^{(2),j} (\bf x\it(p)){\partial v^{(1),k}(\bf x\it(p)) \over \partial x^j} + \epsilon \delta^k_j v^{(1),i}(\bf x\it(p)) {\partial v^{(2),j}(\bf x\it(p)) \over \partial x^i} \nolabel \\
&=& v^{(2),k} + \epsilon \bigg[-v^{(\it 2),j}{\partial v^{(\it 1),k}\over \partial x^j} + v^{(\it1),i}{\partial v^{(\it 2),k} \over \partial x^i}\bigg] \label{eq:calculationtosimplifyliederivative}
\end{eqnarray}
where we have suppressed the arguments for notational clarity and used (\ref{eq:expandtofirstorderliederiv}) to get the second line.  Rewriting this as a vector equation (instead of merely components), we have
\begin{eqnarray}
\bf v\it^{(2)} + \epsilon\bigg[- v^{(2),j}{\partial \bf v\it^{(1)} \over \partial x^j} + v^{(1),j}{\partial \bf v\it^{(2)} \over \partial x^j} \bigg]
\end{eqnarray}

Now, plugging this into (\ref{eq:firstdefinitionofliederivative}), we have
\begin{eqnarray}
\mathcal{L}_{\bf v\it^{(1)}} \bf v\it^{(2)}(\bf x\it(p)) &=& \lim_{\epsilon\rightarrow 0}{1\over \epsilon}\bigg( \bf v\it^{(2)}+\epsilon\bigg[-v^{(2),j}{\partial \bf v\it^{(1)} \over \partial x^j} + v^{(1),j}{\partial \bf v\it^{(2)} \over \partial x^j} \bigg] - \bf v\it^{(2)}\bigg) \nolabel \\
&=& \lim_{\epsilon\rightarrow 0} {1\over \epsilon} \epsilon\bigg(-v^{(2),j}{\partial \bf v\it^{(1)} \over \partial x^j} + v^{(1),j}{\partial \bf v\it^{(2)} \over \partial x^j}\bigg) \nolabel \\
&=& -v^{(2),j}{\partial \bf v\it^{(1)} \over \partial x^j} + v^{(1),j} {\partial \bf v\it^{(2)} \over \partial x^j}
\end{eqnarray}

Now we can use the expansion of the vectors $\bf v\it^{(1)}$ and $\bf v\it^{(2)}$ in terms of the frame ${\partial \over \partial x^i}$ on $\mathcal{M}$, where $\bf v\it^{(1)} = v^{(1),i}{\partial \over \partial x^i}$ and $\bf v\it^{(2)} = v^{(2),i}{\partial \over \partial x^i}$, to write this as (rearranging the indices slightly)
\begin{eqnarray}
\mathcal{L}_{\bf v\it^{(1)}} \bf v\it^{(2)}(\bf x\it(p)) &=& -v^{(2),i}{\partial \over \partial x^i} v^{(1),j} {\partial \over \partial x^j} + v^{(1),j}{\partial \over \partial x^j} v^{(2),i} {\partial \over \partial x^i} \nolabel \\
&=&- \bigg( v^{(2),i}{\partial \over \partial x^i}\bigg)\bigg( v^{(1),j}{\partial \over \partial x^j}\bigg) + \bigg(v^{(1),j}{\partial \over \partial x^j}\bigg)\bigg( v^{(2),i}{\partial \over \partial x^i} \bigg)  \nolabel \\
&=& \bf v\it^{(1)} \bf v\it^{(2)} - \bf v\it^{(2)} \bf v\it^{(1)} \nolabel \\
&=& [\bf v\it^{(1)}, \bf v\it^{(2)}] \label{eq:finalformofliederivative}
\end{eqnarray}
where the brackets in the last line indicate a commutator.  

So, equation (\ref{eq:finalformofliederivative}) has given us a drastically simpler form than (\ref{eq:firstdefinitionofliederivative}) for the Lie derivative.  

So once again, the meaning of $\mathcal{L}_{\bf v\it^{(1)}} \bf v\it^{(2)}$ is the derivative of $\bf v\it^{(2)}$ in the direction of $\bf v\it^{(1)}$ at a given point $p\in \mathcal{M}$.  

To get a feel for the geometrical \it meaning \rm of the Lie derivative, consider the following situation.  Starting at some point $p\in \mathcal{M}$ (as usual) consider once again two vector fields $\bf v^{(1)}$ and $\bf v\it^{(2)}$ with respective families of curves $\tilde q^{(1)}$ and $\tilde q^{(2)}$.  Working with the coordinate maps once again, imagine starting from $\bf x\it(p)$ and going a small distance $\epsilon$ along $\tilde q^{(1)}$:
\begin{eqnarray}
\bf x\it(p) \longrightarrow \bf q\it^{(1)}(\epsilon, \bf x\it(p))
\end{eqnarray}
and then from $ \bf q\it^{(1)}(\epsilon,\bf x\it(p))$ going a small distance $\delta$ along $\bf q\it^{(2)}$:
\begin{eqnarray}
\bf q\it^{(1)}(\epsilon,\bf x\it(p)) \longrightarrow \bf q\it^{(2)}(\delta,\bf q\it^{(1)}(\epsilon,\bf x\it(p))) \label{eq:compositionofpaths}
\end{eqnarray}
Then, using (\ref{eq:expandtofirstorderliederiv}), 
\begin{eqnarray}
\bf q\it^{(2)}(\delta, \bf q\it^{(1)}(\epsilon, \bf x\it(p))) &=& \bf q\it^{(2)}(\delta, \bf x\it(p) + \epsilon \bf v\it^{(1)}(\bf x\it(p))) \nolabel \\
&=& \bf x\it(p) + \epsilon \bf v\it^{(1)}(\bf x\it(p)) +  \delta \big(\bf v\it^{(2)}(\bf x\it(p)+\epsilon \bf v\it^{(1)}(\bf x\it(p)))\big) \nolabel \\
&=& \bf x\it(p) + \epsilon \bf v\it^{(1)} (\bf x\it(p)) + \delta \bf v\it^{(2)}(\bf x\it(p)) + \delta \epsilon  v^{(1),i}(\bf x\it(p)){\partial \bf v\it^{(2)} (\bf x\it(p))\over \partial x^i}   \nolabel \\
&=& \bf x\it + \epsilon \bf v\it^{(1)} + \delta \bf v\it^{(2)} + \delta \epsilon v^{(1),i} {\partial \bf v\it^{(2)} \over \partial x^i} \label{eq:twoinfinitesimalpaths1}
\end{eqnarray}
Next, consider taking the paths in the opposite order - start with a small distance $\delta$ along $\bf q\it^{(2)}$ and and then a small distance $\epsilon$ along $\bf q\it^{(1)}$:
\begin{eqnarray}
\bf x\it (p) \longrightarrow \bf q\it^{(2)}(\delta, \bf x\it (p)) \longrightarrow \bf q\it^{(1)}(\epsilon,  \bf q\it(\delta,\bf x\it (p)))
\end{eqnarray}
This will result in
\begin{eqnarray}
\bf q\it^{(1)} (\epsilon, \bf q\it^{(2)}(\delta, \bf x\it(p))) = \cdots = \bf x\it + \epsilon \bf v\it^{(1)} + \delta \bf v\it^{(2)} + \delta \epsilon v^{(2),i} {\partial \bf v\it^{(1)} \over \partial x^i} \label{eq:twoinfinitesimalpaths2}
\end{eqnarray}

Now we want to know if these two vectors are different.  In other words if we move them along these two paths ($\tilde q^{(1)}(\epsilon)$, $\tilde q^{(2)}(\delta)$ and $\tilde q^{(2)}(\delta )$, $\tilde q^{(1)}(\epsilon)$), how will they compare to each other?  This will be given by simply taking the difference:
\begin{eqnarray}
\bf q\it^{(2)}(\delta, \bf q\it^{(1)}(\epsilon, \bf x\it(p))) &-& \bf q\it^{(1)}(\epsilon, \bf q\it^{(2)}(\delta, \bf x\it(p)))  \nolabel \\
&=& \bf x\it + \epsilon \bf v\it^{(1)} + \delta \bf v\it^{(2)} + \delta \epsilon v^{(1),i} {\partial \bf v\it^{(2)} \over \partial x^i} \nolabel \\
& & -\bf x\it -\epsilon \bf v\it^{(1)} - \delta \bf v\it^{(2)} - \delta \epsilon v^{(2),i} {\partial \bf v\it^{(1)} \over \partial x^i} \nolabel \\
&=& \delta \epsilon [\bf v\it^{(1)}, \bf v\it^{(2)}]  \nolabel \\
&=& \delta \epsilon \mathcal{L}_{\bf v\it^{(1)}} \bf v\it^{(2)} \label{eq:differenceexampleforliederiv}
\end{eqnarray}
\begin{center}
\includegraphics[scale=.75]{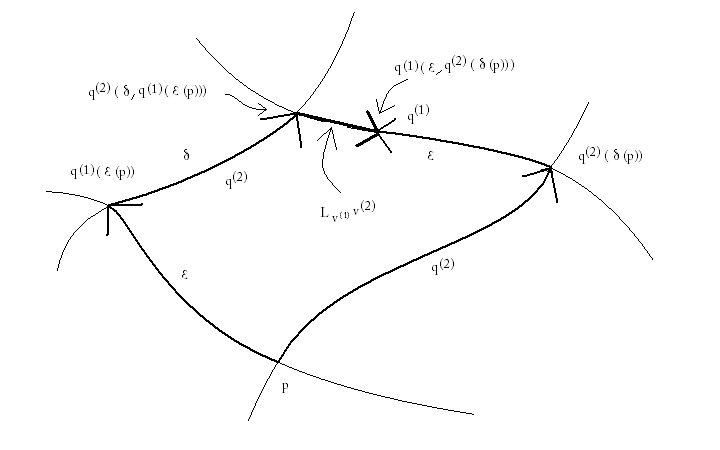}\label{pagewithpictureofliederivativenonclosure}
\end{center}
So the Lie derivative is a measure of how much two paths fail to commute.  This information will tell us very important information about the nature of the vector fields on $\mathcal{M}$.  We will explore this meaning later in this paper.  

Before moving on, notice that in considering vector fields on manifolds, we found in (\ref{eq:vectorsaregeneratorsofcurvesthroughapoint}) that the global behavior of the flow generated by a vector field is \it generated \rm by the element of the tangent vector at the starting point through exponentiation.  And now we have seen that, via the Lie derivative, the commutator of the tangent vectors (which act as generators) provides information about the structure of the vector fields on $\mathcal{M}$.  Once again, this should remind you of what we did with Lie groups in \cite{Firstpaper}.  We had a ``parameter space" where the behavior near the identity element is described by the generators, and we can see the global behavior through exponentiation of the tangent vectors at the starting point.  We have also seen that the general structure of the relationship between the vector fields is given by the commutation relation between them.  Once again, this is identical to the case with the elements of a Lie algebra.  

From a single point $p$ on a manifold $\mathcal{M}$, some set of vector fields will each define a specific tangent vector at $p$.  If there are $n$ such tangent vectors, we can move along $n$ different curves in $\mathcal{M}$.  What we will see is that through composition of the curves (like (\ref{eq:compositionofpaths})) we can move to any point on an $n$-dimensional subspace of $\mathcal{M}$.  If $n$ is the dimension of $\mathcal{M}$, then we can move to any point in $\mathcal{M}$ through compositions of curves.  

The exact nature of how these tangent vectors, which act as generators, relate to each other is then given by the commutation structure of the vectors.  For example consider starting at the point $p$ and first moving $\epsilon$ along $\bf v\it^{(1)}$ and then moving $\delta$ along $\bf v\it^{(2)}$ using equation (\ref{eq:vectorsaregeneratorsofcurvesthroughapoint}):
\begin{eqnarray}
e^{\delta \bf v\it^{(2)}} e^{\epsilon \bf v\it^{(1)}} \bf x\it(p)
\end{eqnarray}
This can be depicted similarly as in the picture above.  

Of course this operation should be equivalent to moving directly towards the new point, rather than moving along two different curves separately.  
\begin{center}
\includegraphics[scale=.75]{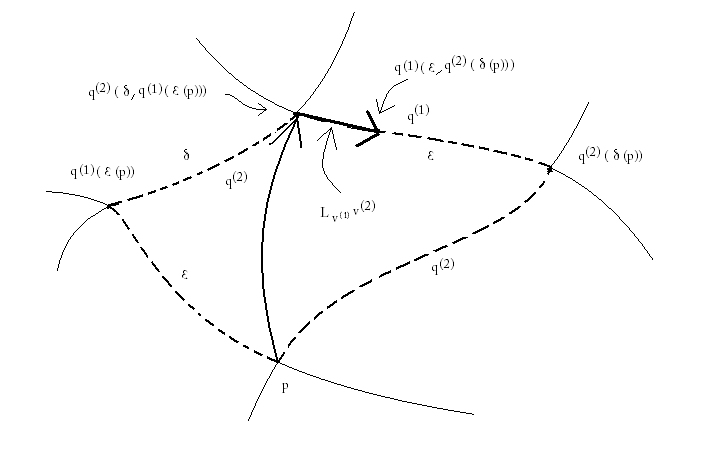}
\end{center}
So what is the relationship between the single tangent vector at $p$ and the distance in terms of these two tangent vector ($\bf v\it^{(1)}$ and $\bf v\it^{(2)}$) and distances ($\epsilon$ and $\delta$)?  By following the same calculation we did in \cite{Firstpaper} to derive the Baker-Hausdorff-Campbell formula we arrive at 
\begin{eqnarray}
e^{\delta \bf v\it^{(2)}} e^{\epsilon \bf v\it^{(1)}} \bf x\it(p) &=& e^{\delta \bf v\it^{(2)} + \epsilon \bf v\it^{(1)} - {1\over 2} [\delta \bf v\it^{(2)}, \epsilon \bf v\it^{(1)}]} \bf x\it(p) \nolabel \\
&=& e^{\delta \bf v\it^{(2)} + \epsilon \bf v\it^{(1)} - {1\over 2} \delta \epsilon \mathcal{L}_{\bf v\it^{(2)}} \bf v\it^{(1)}}\bf x\it(p)
\end{eqnarray}
We can see the geometric meaning of this easily.  Consider again the two paths pictured above.  One ambiguity is that the expression $e^{\delta \bf v\it^{(2)}} e^{\epsilon \bf v\it^{(1)}} \bf x\it(p))$ involves taking the tangent vector $\bf v\it^{(1)}$ first at $p$, and then the tangent vector $\bf v\it^{(2)}$ at the point $\tilde q^{(1)}(\epsilon,p)$.  However the expression $e^{\delta \bf v\it^{(2)} + \epsilon \bf v\it^{(1)} - {1\over 2} \delta \epsilon \mathcal{L}_{\bf v\it^{(2)}} \bf v\it^{(1)}}\bf x\it(p)$  involves both vectors being evaluated at $p$.  Obviously, because the curves may fail to commute (as measured by the Lie derivative), simply adding them together at $p$ may be wrong - and the amount it is off would be expected to be proportional to the Lie derivative.  So, $\mathcal{L}_{\bf v\it^{(2)}} \bf v\it^{(1)}$ provides the vector correction factor to compensate for the changes in the vector field evaluated at different points.  

Comparing what we have seen so far in this section to the definition of a Lie algebra in section \ref{sec:algebras}, you can see that what we have here is indeed the same.  

So, through purely geometrical reasoning, we can once again see the same type of structure as we saw through purely algebraic reasoning.  As we proceed this structure will become more and more apparent.  

Before moving on, however, we make one final comment.  We just considered how to use the tangent mapping to evaluate the Lie derivative of a vector along a curve.  We now consider briefly how to find the Lie derivative of a form.  

Given some covector field $\boldsymbol\omega\it(p) = \omega_i(p)dx^i$ and a vector field $\bf v\it(p) = v^i(p){\partial \over \partial x^i}$ generated by the family of curves $\tilde q(\tau,p)$, the derivation of the Lie derivative of $\boldsymbol\omega\it(p)$ along the path $\tilde q(\tau,p)$ is derived in almost the exact same way as the Lie derivative of a vector field.  The biggest difference is that we obviously can't use the tangent mapping (which maps vectors) to map the form $\boldsymbol\omega\it(p)$ at $\tilde q(\epsilon,p)$ back to the cotangent space at $p$.  Instead the obvious choice is to use the pullback $\tilde q^{\star}(\epsilon,p)$.  

Doing this and repeating what we did above, we find that the Lie derivative of $\boldsymbol\omega\it(p)$ is
\begin{eqnarray}
\mathcal{L}_{\bf v\it} \boldsymbol\omega\it(p) = v^i{\partial \omega_j \over \partial x^i} dx^j + \omega_i{\partial v^i \over \partial x^j} dx^j
\end{eqnarray}
Notice there is no minus sign in this expresion.  The reason for this is that the tangent mapping maps in the same direction as the map which induces it, whereas the pullback operates in the opposite direction.  So to map a vector from $\epsilon$ back to $p$, we must go in the $-\epsilon$ direction, hence the minus sign.  The pullback on the other hand goes in the \it opposite \rm direction, so to move from $\epsilon$ back to $p$ we start with the map which goes from $p$ to $\epsilon$ and therefore uses the $+\epsilon$ direction, hence the lack of a minus sign.  

Finally, given an arbitrary tensor of rank $(q,p)$, 
\begin{eqnarray}
\bf T\it(p) = T_{i_1,\cdots,i_p}^{j_1,\cdots,j_q}(p)  {\partial \over \partial x^{j_1}}\otimes \cdots \otimes {\partial \over \partial x^{j_q}}\otimes dx^{i_1}\wedge \cdots \wedge dx^{i_p}
\end{eqnarray}
we can take the Lie derivative along the curve defined by the vector field $\bf v\it(p) = v^i(p){\partial \over \partial x^i}$ as
\begin{eqnarray}
\mathcal{L}_{\bf v\it} \bf T\it(p) &=& v^i {\partial T_{i_1,\cdots,i_p}^{j_1,\cdots,j_q} \over \partial x^i} {\partial \over \partial x^{j_1}}\otimes \cdots \otimes {\partial \over \partial x^{j_q}}\otimes dx^{i_1}\wedge \cdots \wedge dx^{i_p} \nolabel \\
& & +  T_{i_1,\cdots,i_p}^{j_1,\cdots,j_q} \bigg(\mathcal{L}_{\bf v\it} {\partial \over \partial x^{j_1}}\bigg)\otimes \cdots \otimes {\partial \over \partial x^{j_p}} \otimes dx^{i_1}\wedge \cdots \wedge dx^{i_p} \nolabel \\
& & + \cdots  \nolabel \\
& & + T_{i_1,\cdots,i_p}^{j_1,\cdots,j_q} {\partial \over \partial x^{j_1}} \otimes \cdots \otimes \bigg(\mathcal{L}_{\bf v\it} {\partial \over \partial x^{j_p}}\bigg) \otimes dx^{i_1}\wedge \cdots \wedge dx^{i_p} \nolabel \\
& & + T_{i_1,\cdots,i_p}^{j_1,\cdots,j_q} {\partial \over \partial x^{j_1}} \otimes \cdots \otimes  {\partial \over \partial x^{j_p}} \otimes\bigg(\mathcal{L}_{\bf v\it} dx^{i_1}\bigg) \wedge \cdots \wedge dx^{i_p} \nolabel \\ 
& & + \cdots \nolabel \\
& & + T_{i_1,\cdots,i_p}^{j_1,\cdots,j_q} {\partial \over \partial x^{j_1}} \otimes \cdots \otimes  {\partial \over \partial x^{j_p}} \otimes dx^{i_1} \wedge \cdots \wedge \bigg(\mathcal{L}_{\bf v\it}dx^{i_p}\bigg) \nolabel \\ 
\end{eqnarray}

We now have the tools to consider Lie groups in fuller detail.  But first, in order to (hopefully) provide greater insight into the \it meaning \rm of the Lie derivative, we will consider another perspective on it.  

\subsection{Another Perspective on the Lie Derivative}
\label{sec:anotherperspectiveliederivatives}

Consider the standard basis for $\mathbb{R}^n$, $\bf e\it_i$ for $i=1,\ldots,n$.  By definition, if you move in the $\bf e\it_i$ direction, you \it aren't \rm moving at all in the $\bf e\it_j$ (for $i\neq j$) direction.  This is the point of the standard basis.  For example, in $\mathbb{R}^2$, moving in the $\bf e\it_x$ direction involves no motion at all in the $\bf e\it_y$ direction, or vice-versa.  Or in polar coordinates, moving in the $\bf e\it_r$ direction no motion at all in the $\bf e\it_{\phi}$ direction, or vice-versa.  

We can reformulate this in the following way.  Consider two vector fields in $\mathbb{R}^2$
\begin{eqnarray}
\bf v\it^{(1)}(x,y) &=& v^{(1),x}(x,y) \bf e\it_x + v^{(1),y}(x,y)\bf e\it_y \nolabel \\
\bf v\it^{(2)}(x,y) &=& v^{(2),x}(x,y) \bf e\it_x + v^{(2),y}(x,y)\bf e\it_y  \label{eq:generavectorfieldsforliederivativeintexamplesection}
\end{eqnarray}
If we take the special case $v^{(1),x} = v^{(2),y} = 1$ and $v^{(1),y} = v^{(2),x} = 0$, we have
\begin{eqnarray}
\bf v\it^{(1)}(x,y) &=& \bf e\it_x \nolabel \\
\bf v\it^{(2)}(x,y) &=& \bf e\it_y
\end{eqnarray}
Now consider some arbitrary point $p \rightarrow \bf x\it_0 = (x_0,y_0) \in \mathbb{R}^2$.  Repeating what we did in equations (\ref{eq:twoinfinitesimalpaths1}) and (\ref{eq:twoinfinitesimalpaths2}).  First we move from $\bf x\it_0$ a small displacement $\epsilon$ in the $\bf v\it^{(1)} = \bf e\it_x$ direction, then a small displacement $\delta$ in the $\bf v\it^{(2)} = \bf e\it_y$ direction:
\begin{eqnarray}
\bf x\it_0 &\rightarrow&\bf x\it'_0 =  \bf x\it_0 + \epsilon \bf v\it^{(1)} + \delta \bf v\it^{(2)} + \epsilon \delta v^{(1),i} {\partial \bf v\it^{(2)} \over \partial x^i} \nolabel \\
&=& \bf x\it_0 + \epsilon \bf v\it^{(1)} + \delta \bf v\it^{(2)} + \epsilon \delta v^{(1),i} {\partial \over \partial x^i} \bf e\it_y \nolabel \\
&=& \bf x\it_0 + \epsilon \bf v\it^{(1)} + \delta \bf v\it^{(2)} \label{eq:draggingxforthisexample}
\end{eqnarray}
Then, if we do this in the opposite order, we get
\begin{eqnarray}
\bf x \it_0 \rightarrow \bf x\it''_0 = \bf x\it_0 + \delta \bf v\it^{(2)} + \epsilon \bf v\it^{(1)} \label{eq:draggingxforthisexample2}
\end{eqnarray}
Consequently we can easily see
\begin{eqnarray}
\bf x\it'_0 - \bf x\it''_0 &=& \big(\bf x\it_0 + \epsilon \bf v\it^{(1)} + \delta \bf v\it^{(2)}\big) - \big(\bf x\it_0 + \epsilon \bf v\it^{(1)} + \delta \bf v\it^{(2)} \big) = 0 
\end{eqnarray}
and therefore (cf equation (\ref{eq:differenceexampleforliederiv}))
\begin{eqnarray}
\mathcal{L}_{\bf v\it^{(1)}} \bf v\it^{(2)} = \mathcal{L}_{\bf v\it^{(2)}} \bf v\it^{(1)} = 0 \label{eq:vanliederivexsec}
\end{eqnarray}
Notice that the origin of this Lie derivative vanishing\footnote{Of course we could have arrived at (\ref{eq:vanliederivexsec}) directly from the definition of a Lie derivative.  Using $\bf v\it^{(1)} = \bf e\it_x = {\partial \over \partial x}$ and $\bf v\it^{(2)} = \bf e\it_y = {\partial \over \partial y}$,
\begin{eqnarray}
\mathcal{L}_{\bf v\it^{(1)}} \bf v\it^{(2)} = \bigg[{\partial \over \partial x}, {\partial \over \partial y}\bigg] = 0 \label{eq:originofliederivvanishing}
\end{eqnarray}
} is in equation (\ref{eq:draggingxforthisexample}) - namely the vanishing of
\begin{eqnarray}
{\partial \bf v\it^{(2)} \over \partial x^i} = {\partial \over \partial x^i} \bf e\it_y = 0
\end{eqnarray}
(and the analogous term in equation (\ref{eq:draggingxforthisexample2})).  In other words, the vanishing of the Lie derivative is a consequence of the fact that moving $\bf v\it^{(1)} = \bf e\it_x$ in the $\bf v\it^{(2)} = \bf e\it_y$ direction doesn't change it, and vice versa.  Neither field changes when dragged along the other.  This is consistent with what we know of the Lie derivative - when you move $\bf v\it^{(1)}$ in the $\bf v\it^{(2)}$ direction it doesn't change, and therefore the derivative is zero.  

Now consider letting (\ref{eq:generavectorfieldsforliederivativeintexamplesection}) be defined by\footnote{Notice that these are simply the $\bf e\it_r$ and $\bf e\it_{\phi}$ basis vectors (polar coordinates) written in terms of Cartesian coordinates.}
\begin{eqnarray}
\bf v\it^{(1)}(x,y) &=& {x \over \sqrt{x^2+y^2}} \bf e\it_x + {y \over \sqrt{x^2+y^2}} \bf e\it_y \nolabel \\
&=& {x \over \sqrt{x^2+y^2}}{\partial \over \partial x} + {y \over \sqrt{x^2+y^2}} {\partial \over \partial y} \nolabel \\
\bf v\it^{(2)}(x,y) &=& -{y \over \sqrt{x^2+y^2}}\bf e\it_x + {x \over \sqrt{x^2+y^2}} \bf e\it_y \nolabel \\
&=& -{y \over \sqrt{x^2+y^2}}{\partial \over \partial x} + {x \over \sqrt{x^2+y^2}}{\partial \over \partial y}
\end{eqnarray}
We will calculate the Lie derivative directly from (\ref{eq:finalformofliederivative}) first:
\begin{eqnarray}
\mathcal{L}_{\bf v\it^{(1)}} \bf v\it^{(2)} &=& \bigg[{x \over \sqrt{x^2+y^2}}{\partial \over \partial x} + {y \over \sqrt{x^2+y^2}} {\partial \over \partial y} ,-{y \over \sqrt{x^2+y^2}}{\partial \over \partial x} + {x \over \sqrt{x^2+y^2}}{\partial \over \partial y}\bigg] \nolabel \\
&=& \cdots \nolabel \\
&=& {y \over x^2+y^2} {\partial \over \partial x} - {x \over x^2+ y^2} {\partial \over \partial y} \label{eq:firstcalcofliederivforpolarcoordsexample}
\end{eqnarray}
Then, again using (\ref{eq:twoinfinitesimalpaths1}) and (\ref{eq:twoinfinitesimalpaths2}),
\begin{eqnarray}
\bf x\it_0 &\rightarrow& \bf x\it'_0 = \bf x\it_0 + \epsilon \bf v\it^{(1)} + \delta \bf v\it^{(2)} + \epsilon \delta v^{(1),i} {\partial \bf v\it^{(2)} \over \partial x^i} \nolabel \\
&=& \bf x\it_0 + \epsilon \bf v\it^{(1)} + \delta \bf v\it^{(2)} + \epsilon \delta (0) \nolabel \\
&=& \bf x\it_0 + \epsilon \bf v\it^{(1)} + \delta \bf v\it^{(2)} 
\end{eqnarray}
and
\begin{eqnarray}
\bf x\it_0 &\rightarrow& \bf x\it''_0  = \bf x\it_0 + \delta \bf v\it^{(2)} + \epsilon \bf v\it^{(1)} + \epsilon \delta v^{(2),i} {\partial \bf v\it^{(1)} \over \partial x^i} \nolabel \\
&=& \bf x\it_0 + \delta \bf v\it^{(2)} + \epsilon \bf v\it^{(1)} + \epsilon \delta \bigg(-{y \over x^2+y^2}\bf e\it_x + {x \over x^2+y^2}\bf e\it_y\bigg)
\end{eqnarray}
So 
\begin{eqnarray}
\bf x\it'_0 - \bf x\it''_0 &=& \big( \bf x\it_0 + \epsilon \bf v\it^{(1)} + \delta \bf v\it^{(2)}\big) - \big(\bf x\it_0 + \delta \bf v\it^{(2)} + \epsilon \bf v\it^{(1)}\big) \nolabel \\
& & + \epsilon \delta \bigg({y \over x^2+y^2} \bf e\it_x - {x \over x^2+y^2}\bf e\it_y\bigg) \nolabel \\
&=& \epsilon \delta \bigg({y \over x^2+y^2} \bf e\it_x - {x \over x^2+y^2}\bf e\it_y\bigg)
\end{eqnarray}
which comparison with (\ref{eq:differenceexampleforliederiv}) gives the same Lie derivative as in (\ref{eq:firstcalcofliederivforpolarcoordsexample}). 

So, the meaning of the Lie derivative of one field in the direction of another is that it tells us how much one field changes as you move along the other.

\subsection{Lie Groups and Lie Algebras on Manifolds}

We begin this section by simply redefining a Lie group.  We gave significant detail as to what Lie groups are in \cite{Firstpaper} from a purely algebraic perspective.  Now we reformulate them using the geometry we have built up so far.  

Simply put, a Lie group is a special type of differentiable manifold which we denote $\mathcal{G}$.  What makes it special is that it it has a group structure (see \cite{Firstpaper} for more details about groups).  This means that we can think of every point $g\in\mathcal{G}$ as an element of a group.  Clearly there must exist some well-defined group multiplication law such that:\footnote{We aren't using any specific notation for the multiplication, such as $g_i\star g_j$ or $g_i\cdot g_j$, and instead are simply denote the product of two element $g_i$ and $g_j$ as $g_ig_j$.  This won't result in any ambiguity at any point in these notes.} \\
\indent 1) $g_i,g_j \in \mathcal{G} \; \Longrightarrow (g_ig_j) \in \mathcal{G}$. \\
\indent 2) $(g_ig_j)g_k = g_i(g_jg_k)$.\\
\indent 3) $\exists e \in \mathcal{G} \ni ge = eg = g\; \forall g\in \mathcal{G}$.\\
\indent 4) $\forall g\in \mathcal{G} \exists g^{-1}\in \mathcal{G} \ni gg^{-1}=g^{-1}g = e$.

As a simple example consider the manifold $S^1$.  It is not by itself a Lie group.  However we can give its coordinate $\theta$ an additive structure such that 
\begin{eqnarray}
\theta, \theta' \in \mathcal{G} \Longrightarrow \theta + \theta' \; (\mod 2\pi) \in \mathcal{G}
\end{eqnarray}
With this additional structure (the relationship between the coordinates, namely addition), we have turned $S^1$ into a Lie group.  Specifically, it is a representation of $SO(2)$.  

Notice that this is very similar to the definition we gave in \cite{Firstpaper}.  There we defined a Lie group as a group where the elements depend on a set of \it smooth \rm parameters.  This definition is not really different from the previous paper - it just now carries with it the additional ``baggage" of what we know about manifolds.  And just as in \cite{Firstpaper} we proceeded from that definition to study Lie groups algebraically, we now proceed to study them geometrically.  

Let's begin with an arbitrary point in $g\in\mathcal{G}$.  Keep in mind that this is both an element of the group \it and \rm a point on the manifold.  We can act on $g$ with any other point in $h\in \mathcal{G}$, resulting in the point $hg\in \mathcal{G}$ (we are taking multiplication in this case to be from the left - we could just as easily do all of this by acting from the right, and we would get the same results; we choose left multiplication for concreteness).  

Another way of thinking about this is that, for any element $h\in \mathcal{G}$, we can define a map $$f_h: \mathcal{G} \longrightarrow \mathcal{G}$$
such that $$f_h(g) = hg$$
Clearly there will be an infinite number of such maps - one for each $h \in \mathcal{G}$.  

Naturally this will induce a tangent map $T_gf_h$ (cf section \ref{sec:tangentmapping}) and a pullback $f_h^{\star}$ (cf section \ref{sec:pullbackdiff}).  

One of the primary differences between a Lie group, or a Lie manifold, and a normal manifold is that on a normal manifold there is no way of singling out particular vector fields.  On a Lie group/manifold there are certain vector fields which are ``special".  

Specifically, let $\bf v\it(g)$ be a vector field on $\mathcal{G}$.  Just as in section \ref{sec:tangentmapping}, the map $f_h$ will induce a tangent map $T_gf_h$ which takes a tangent vector at $g$ to a tangent vector at $hg$.  This allow us to define an \bf invariant vector field \rm as one which satisfies
\begin{eqnarray}
T_gf_h\big( \bf v\it(g)\big) = \bf v\it(hg) \label{eq:equationforinvariantvectorfieldliegroup}
\end{eqnarray}

As an illustration of this, consider the (simple) example of $S^1$ with an additive $SO(2)$ structure as described above.  We define the map 
\begin{eqnarray}
f_{\psi}: \theta \longmapsto \psi+\theta
\end{eqnarray}
So, given an arbitrary vector\footnote{a vector $v^{\theta}{\partial \over \partial \theta}$ is a vector with magnitude $v^{\theta}$ in the $\theta$ direction.} $v^{\theta}{\partial \over \partial \theta}$, the tangent map will be (using (\ref{eq:curvefreedefinitionofpushforward}))
\begin{eqnarray}
T_{\theta}f_{\psi}\bigg(v^{\theta}{\partial \over \partial \theta}\bigg) = v^{\theta}{\partial f_{\psi} \over \partial \theta} {\partial \over \partial \theta} = v^{\theta}{\partial (\psi + \theta) \over \partial \theta} {\partial \over \partial \theta} = v^{\theta}{\partial \over \partial \theta} \label{eq:tangentmapfromliegroupsectionexample}
\end{eqnarray}
\begin{center}
\includegraphics[scale=.5]{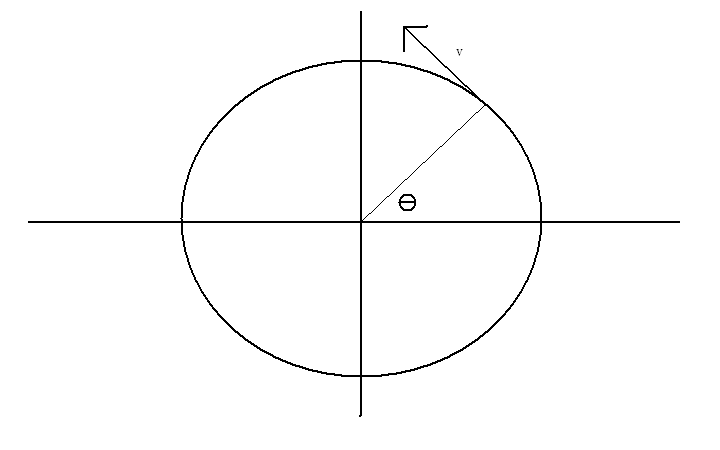}
\end{center}
This makes sense.  The tangent mapping simply rotates this vector around the circle without changing the magnitude.  So, consider the vector field $\bf v\it(\theta) = v^{\theta}(\theta){\partial \over \partial \theta} = v{\partial \over \partial \theta}$, where $v$ is a constant.  The tangent map (\ref{eq:tangentmapfromliegroupsectionexample}) will map this to $v{\partial \over \partial \theta}$ at the point $\psi+\theta$.  Also, we know that $\bf v\it(\psi + \theta) = \bf v\it(\theta)$ by the definition of the vector field we just gave.  
So, we have
\begin{eqnarray}
T_\theta f_{\psi}(\bf v\it(\theta)) &=& \bf v\it(\theta) = v{\partial \over \partial \theta} \nolabel \\
\bf v\it(\psi+\theta) &=& \bf v\it(\theta) = v{\partial \over \partial \theta} \nolabel \\
&\Longrightarrow& T_{\theta}f_{\psi}(\bf v\it(\theta)) = \bf v\it(\psi+\theta)
\end{eqnarray}
Which is simply (\ref{eq:equationforinvariantvectorfieldliegroup}) for this example.  Therefore, $\bf v\it(\theta)$ is in this case an invariant vector field.  

On the other hand consider the vector\footnote{The prime here is not a derivative - it is just a label.} field $\bf v\it'(\theta) = \cos(\theta){\partial \over \partial \theta}$.  Now, using the same $f_{\psi}$,
\begin{eqnarray}
T_{\theta}f_{\psi}(\bf v\it'(\theta)) &=& \bf v\it'(\theta) = \cos(\theta){\partial \over \partial \theta} \nolabel \\
\bf v\it'(\psi + \theta) &=& \cos(\psi+\theta){\partial \over \partial \theta} \nolabel \\
&\Longrightarrow& T_{\theta}f_{\psi}(\bf v\it'(\theta)) \neq \bf v\it'(\psi+\theta)
\end{eqnarray}
and so $\bf v\it'(\theta)$ is not an invariant vector field.  

So, for any point $g\in \mathcal{G}$, we can choose \it any \rm tangent vector $\bf \tilde v\it \in T_g\mathcal{G}$, and $\bf \tilde v\it$ will define a unique invariant vector field $\bf v\it$ (we suppress the argument $(g)$ because we are thinking of the \it entire \rm field here, not the field at a certain point) on $\mathcal{G}$.  If $\bf \tilde v\it$ is defined at $g$, then the vector field $\bf v\it$ is defined by
\begin{eqnarray}
\bf v\it \equiv \{T_gf_h(\bf \tilde v\it) | \forall h \in \mathcal{G}\}
\end{eqnarray}
In other words, the vector field at the point $a = hg \in \mathcal{G}$ is
\begin{eqnarray}
\bf v\it(a) = T_gf_h(\bf \tilde v\it)
\end{eqnarray}
Because $\mathcal{G}$ is a group, for \it any \rm starting point $g$ and desired point $a$, there exists an $h$ which will take you from $g$ to $a$.  Specifically, $h = ag^{-1}$, which is guaranteed to be in $\mathcal{G}$ by the definition of a group, will accomplish this.  

Thinking of this in the ``opposite direction", any invariant vector field $\bf v\it$ will define a unique vector at any arbitrary point $g \in \mathcal{G}$.  This vector at $g$ can be found by simply taking the tangent map from any point $a$ \it back \rm to $g$.  Specifically, we will be looking at the point in $\mathcal{G}$ which represents the identity element $e$ (this point isn't automatically chosen for us - it depends on how we choose our coordinate functions and is ultimately arbitrary).  So, any invariant vector field $\bf v\it$ on $\mathcal{G}$ will define a unique vector at $e$.  

Now consider the set of \it all \rm invariant vector fields on $\mathcal{G}$.  Each of these will define a unique vector at $e\in \mathcal{G}$.  Or on the other hand, every vector at $e$ will define an invariant vector field on $\mathcal{G}$.  We denote the set of all invariant vector fields \textgoth{g}.  Because \textgoth{g} and the set of all vectors at $e$ are one to one, we can think of \textgoth{g} as a vector space with the same dimension as $T_e\mathcal{G}$, which is the same dimension as $\mathcal{G}$.  Every vector in $T_e\mathcal{G}$ has a corresponding vector \it field \rm in \textgoth{g}.  

In the previous section, we defined the Lie derivative $\mathcal{L}_{\bf v\it^{(i)}} \bf v\it^{(j)} = [\bf v\it^{(i)}, \bf v\it^{(j)}] $, which is the derivative of $\bf v\it^{(j)}$ in the direction of $\bf v\it^{(i)}$, and is itself a vector field on the manifold.  Let's assume that $\bf v\it^{(i)}$ and $\bf v\it^{(j)}$ are both in \textgoth{g}.  Will $[\bf v\it^{(i)}, \bf v\it^{(j)}]$ be in \textgoth{g}?  We can find out by acting on $[\bf v\it^{(i)}, \bf v\it^{(j)}](g)$ with the tangent map $T_gf_h$:
\begin{eqnarray}
T_gf_h\big([\bf v\it^{(i)}, \bf v\it^{(j)}](g)\big) &=& \big[ T_gf_h(\bf v\it^{(i)}(g)), T_gf_h(\bf v\it^{(j)}(g))\big] \nolabel \\
&=& [\bf v\it^{(i)}(hg), \bf v\it^{(j)}(hg)] \nolabel \\
&=& [\bf v\it^{(i)}, \bf v\it^{(j)}](hg) \label{eq:closerofliederivativeundergothicg}
\end{eqnarray}
We used the general relationship $T_gf_h[\bf v\it^{(i)}, \bf v\it^{(j)}] = [T_gf_h (\bf v\it^{(i)}), T_gf_h(\bf v\it^{(j)})]$ (which you can easily convince yourself of by writing out the definition of the commutator and the tangent map), and to get the second equality we used the fact that $\bf v\it^{(i)}$ and $\bf v\it^{(j)}$ are each individually in \textgoth{g}.  

So, equation (\ref{eq:closerofliederivativeundergothicg}) tells us that \textgoth{g} is closed under the Lie derivative, or commutation relations between the tangent vectors at each point.  So, for any two elements of \textgoth{g}, we have the map
\begin{eqnarray}
[\;,\;] : \textgoth{g} \otimes \textgoth{g} \longrightarrow \textgoth{g}
\end{eqnarray}
We call the set of all invariant vector field \textgoth{g} along with the vector multiplication $[\;,\;]$ (which is just the commutator) the \bf Lie Algebra \rm of the Lie group/manifold $\mathcal{G}$.  It may be helpful to reread section \ref{sec:algebras} to see that this does in fact fit the proper description of an algebra.

For a given Lie group, we will denote its algebra by the same letters but in lower case.  For example the algebra of $SU(5)$ will be denoted $su(5)$.  Once again, we emphasize that every vector in $T_e\mathcal{G}$ corresponds to a specific invariant vector field on $\mathcal{G}$.  So it is natural to choose a set of vectors in $T_e\mathcal{G}$ which form a linearly independent basis set, or a frame, for $T_e\mathcal{G}$.  Once this is done, we can not only span $T_e\mathcal{G}$ with them, but the corresponding invariant vector fields also form a basis for \textgoth{g}.  Also, by exponentiating the elements of the frame at $e$, we can move from $e$ to any arbitrary point on $\mathcal{G}$.  This is why we refer to the specific frame vectors at $e$ as the \bf generators \rm of $\mathcal{G}$ - by starting at the identity they \it generate \rm all the other points in the group of the manifold through exponentiation.  

We can follow the same arguments we made in \cite{Firstpaper} or on the previous page to see that the the Lie derivative $[\bf v\it^{(i)},\bf v\it^{(j)}]$ at a point results in another vector at the point, and therefore if we consider the Lie derivative at $e$, then the commutator of two frame vectors must also be in $T_e\mathcal{G}$ and therefore must be equal to some linear combination of the frame vectors:
\begin{eqnarray}
[\bf v\it^{(i)},\bf v\it^{(j)}] = f_{ijk}\bf v\it^{(k)}
\end{eqnarray}
We call the values $f_{ijk}$ the \bf structure constants \rm of the algebra.  In that they define the value of the Lie derivative at $e$, which via exponentiation contains the information about the geometry of the flow throughout $\mathcal{G}$, the structure constants completely determine the structure of the manifold $\mathcal{G}$.  

To once again connect this to what we did in \cite{Firstpaper}, we had a parameter space of some dimension, and every point in the parameter space corresponded to a specific element of the group.  For example with $SO(3)$, the parameter space was three dimensional, corresponding to the three Euler angles: $\phi$, $\psi$, and $\theta$.  Then we had three ``vectors" (we called them generators), $J^1$, $J^2$, and $J^3$, with specific structure constants $ [J^i,J^j] = \epsilon_{ijk}J^k$.  The three dimensional parameter space was then spanned by the parameters with the generators as basis vectors.  So an arbitrary point in the parameter space was  
$$ \phi J^1 + \psi J^2  + \theta J^3 $$
This point in the three dimensional parameter space corresponded to the group element
$$ e^{i(\phi J^1 + \psi J^2  + \theta J^3)}$$
The factor of $i$ is simply from the fact that we were trying to make the generators Hermitian.  Recall that the generators were defined with an additional factor of $i$, so you can effectually ignore the $i$ when comparing this expression to what we have done so far in this paper.  

Restating the previous paragraph, we have a three dimensional manifold, denoted $\mathcal{G} = SO(3)$, with a group structure defined on it.  We have chosen three vectors in the tangent space $T_eSO(3)$ denoted $J^1$, $J^2$, and $J^3$, each of which correspond to a specific invariant vector field in \textgoth{g}.  So, to move from the point $e \in SO(3)$ to an arbitrary point on the manifold, we simply exponentiate the tangent vectors at $e$ (which are part of a vector field covering the entire $SO(3)$ manifold), and follow the curves defined by those vectors.  This moves us along the manifold to any arbitrary point.  The non-trivial geometry of the curves defined by the vector field which produced these generators is codified in the commutation relations, which are actually the Lie derivatives at each point - the way the field is changing.  

We chose three specific elements of \textgoth{g} to get these generators - we could have just as easily chosen any three invariant vector fields, resulting in an equivalent set of vectors at $e$.  In the language of \cite{Firstpaper} this would be doing a similarity transformation on the generators $J^i$.  The resulting generators would still be generators, but would correspond to different elements of \textgoth{g}.  

Everything we discussed in the previous sections about vector fields, flows and families of curves, the Abelian groups resulting from a single curve at a point, exponentiation, Lie derivatives, etc. will still hold true for Lie manifolds.  We can still have vector fields, form fields, tensor fields, etc.  The only difference between a normal manifold and a Lie group/manifold is that the Lie manifold allows us to single out specific vector fields as invariant, and the resulting details will be constrained by the invariance of the vector field under the action of the group.  This will (as you can recall from \cite{Firstpaper}) result in tremendously convenient and powerful properties when we begin to once again discuss physics using Lie groups.  

\subsection{Lie Group Manifolds Acting on Manifolds}

Finally, we want to extend these ideas from the geometry of a Lie manifold to the \it action \rm of a Lie manifold $\mathcal{G}$ on another manifold $\mathcal{M}$.  The \rm action \rm of $\mathcal{G}$ on $\mathcal{M}$ is a map $\sigma: \mathcal{G} \otimes \mathcal{M} \longrightarrow \mathcal{M}$.  In other words, a specific element $g\in \mathcal{G}$ takes a specific element of $\mathcal{M}$ to another point in $\mathcal{M}$.  To see this more clearly, recall near the beginning of section \ref{sec:flowandlie} we defined the familiar of curves $\tilde q(\tau,p)$ which took the point $p\in \mathcal{M}$ to another point $\tilde q(\tau,p)$ along the curve defined by $\tilde q$ passing through the point $p$ a ``distance" $\tau$.  

What we are doing here is very similar.  The difference is that, instead of an arbitrarily defined curve $\tilde q$, the curves in $\mathcal{M}$ are defined by the invariant vector fields of $\mathcal{G}$.  As a simple example, consider $\mathcal{M} = \mathbb{R}^2$ and $\mathcal{G} = SO(2)$.  For any point $p=(x,y)^T \in \mathcal{M} = \mathbb{R}^2$ we can can choose a point $g =\theta \in \mathcal{G} = SO(2)$ and act on $p$, moving it to another point in $\mathbb{R}^2$.  For example if we choose $(x,y)^T = (1,0)^T \in \mathbb{R}^2$, and the element ${\pi \over 2} \in SO(2)$, the action of ${\pi \over 2}$ on $(1,0)^T$ will be to take it to $(0,1)^T \in \mathbb{R}^2$.  Or in the language of the above paragraph, the action of $SO(2)$ on $\mathbb{R}^2$ is the map $\sigma: SO(2)\otimes \mathbb{R}^2 \longrightarrow \mathbb{R}^2$.  Then, for example, 
\begin{eqnarray}
\sigma\big(\pi /2, (1,0)^T\big) = (0,1)^T
\end{eqnarray}

More generally, $\sigma$ has the following properties: \\
\indent 1) $\sigma(e,p) = p$.\\
\indent 2) $\sigma(g_i, \sigma(g_j,p)) = \sigma(g_ig_j,p)$.

If the Lie manifold $\mathcal{G}$ acts on $\mathcal{M}$ in a well-defined way, we say that $\mathcal{M}$ ``carries", or ``sits in" the group $\mathcal{G}$.  

Now we can go back to everything we have done in these notes so far that depended on a smooth map, the tangent mapping, the pullback, flows, etc., and do them all over again where the smooth maps are the actions of $\mathcal{G}$ on $\mathcal{M}$.  

As an example, consider once again the action of $SO(2)$ on $\mathbb{R}^2$.  We will work with the $2\times 2$ matrix representation of $SO(2)$.  We want to use an invariant vector field on $SO(2)$, which is simply a constant tangent vector on the circle.  So obviously the element of this vector field at the identity $e$ of $SO(2)$ will be this constant vector.  $SO(2)$ has general element
\begin{eqnarray}
\begin{pmatrix}
\cos \theta & -\sin\theta \\ \sin\theta & \cos\theta
\end{pmatrix}
\end{eqnarray}
We know that the tangent vector to this at the identity $e$ will be (cf \cite{Firstpaper}) 
\begin{eqnarray}
v = \begin{pmatrix}
0 & -1 \\ 1 & 0
\end{pmatrix} \in T_eSO(2)
\end{eqnarray}
Exponentiation of this with the parameter $\theta$ will then properly give
\begin{eqnarray}
e^{\theta v}\bf x\it = 
\begin{pmatrix}
\cos \theta & -\sin\theta \\ \sin\theta & \cos\theta
\end{pmatrix}\bf x\it
\end{eqnarray}
This is the exact situation we had in the example leading to (\ref{eq:elementofso2inliederivsection}).  Working backwards, we can show that this action of $SO(2)$ on $\mathbb{R}^2$ has induced the flow of circles around the origin through some starting point.  And this flow will induce the vector field
\begin{eqnarray}
\bf v\it(x,y) = -y {\partial \over \partial x} + x {\partial \over \partial y}
\end{eqnarray}

We can do the same thing with the action of $SO(3)$ on $\mathbb{R}^3$ (see \cite{Firstpaper} for details), and find that there are \it three \rm induced vector fields given by
\begin{eqnarray}
\bf v\it^{(1)}(x,y,z) &=& -z{\partial \over \partial y} + y {\partial \over \partial z} \nolabel \\
\bf v\it^{(2)}(x,y,z) &=& -x {\partial \over \partial x} + z {\partial \over \partial x} \nolabel \\
\bf v\it^{(3)}(x,y,z) &=& -y {\partial \over \partial x} + x {\partial \over \partial y}  
\end{eqnarray}
We leave these details to you.  

It is these vector fields induced on a manifold $\mathcal{M}$ by the action of a Lie group $\mathcal{G}$ which will prove to be physically useful, as we will see later.  The \it point \rm of all of this is that the Lie group specifies certain vector fields on the manifold.

\subsection{Concluding Thoughts on Differential Topology}
\label{sec:concthoudiffman}

We have defined differentiable manifolds and discussed the types of objects that can live on them, the types of mappings between them, and how we can add structure to them.  However our discussion has left much to be wondered about.  

For example, while our complete dedication to a ``coordinate free" approach has been (and will be) useful, there is much lacking.  As we discussed briefly on page \pageref{talkingaboutlackofgeometryonmanifolds}, there is no way in our current approach of telling the difference between a perfect sphere and an egg.  In fact, we could stretch, twist, bend, and reshape $S^2$ all we want without any need to change the coordinate functions.  In other words, we only have descriptions of manifolds ``up to homeomorphism".  

What if we wanted to discuss the distance between two points on a manifold?  There is no well defined way of answering that question in what we have done so far.  Another ambiguity is angles.  If we have two lines intersecting, how can we determine the angle between them?  We could always stretch or twist the manifold however we want, thus changing the angle.  

Another difficulty is the relationship between vectors and covectors.  While we have discussed the differences between them, how are they related?  Certainly there must be \it some \rm relationship between the column vector $(a,b,c)^T$ and the row vector $(a,b,c)$.  

And finally, consider the manifold $\mathbb{R}^4$.  At first glance this is simply four dimensional Euclidian space.  But consider Minkowski space.  This is also a flat four dimensional manifold.  Topologically there is no real difference between $\mathbb{R}^4$ and Minkowski space, yet we know from physics that they are extremely different.  What is it that makes this difference?  

The answers to these questions will be dealt with when we introduce the idea of a \bf metric \rm on a manifold.  However before we do this, there are a few more things we can say about manifolds defined only ``up to homeomorphism".  For example, while we can't tell the difference between a sphere and an egg, or between $\mathbb{R}^4$ and Minkowski space (yet), we can certainly tell the difference between $S^2$ and the torus $T^2$.  They are \it topologically \rm different, and we don't need geometry to see the differences - there is no way to stretch, twist, or bend one to look like the other.  So we will take a short hiatus from our geometrical considerations to focus on a few key issues in topology.  These topics will allow us to categorize spaces based on \it qualitative \rm properties indicative of their topological structure, rather than on \it quantitative \rm properties indicative of their geometric structure.  

\section{References and Further Reading}

The primary source for this chapter was \cite{gockeler}, though we also made use of \cite{nakahara}.  For further reading we recommend \cite{appel}, \cite{bachman}, \cite{gilmore}, \cite{lee}, \cite{lovelock}, \cite{schutzDG}, or \cite{spivak}.  

\chapter{Algebraic Topology}
\label{sec:topcons}

Before diving in, we briefly discuss the \it point \rm of this chapter.  As we said previously, topology is concerned with the qualitative aspects of spaces, rather than the quantitative.  In other words, it isn't concerned with differences of length or angle, but of the general properties, or categories, of shapes.  

Another way of putting this is to ask the question ``when I stretch, compress, twist, bend, and generally reshape a space, but I do so without breaking the space or adding points to it, what doesn't change?"  For example, consider the annulus (a disk with a hole cut in it):
\begin{center}
\includegraphics[scale=.3]{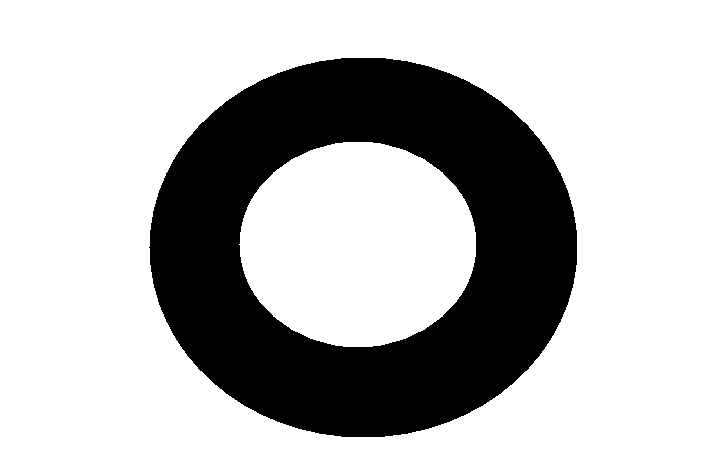}
\end{center}
One could categorize this as a two dimensional shape with a single hole.  Furthermore, no matter how I stretch it, twist it, bend it, etc., it will \it still \rm be a two dimensional shape with a single hole in it:
\begin{center}
\includegraphics[scale=.5]{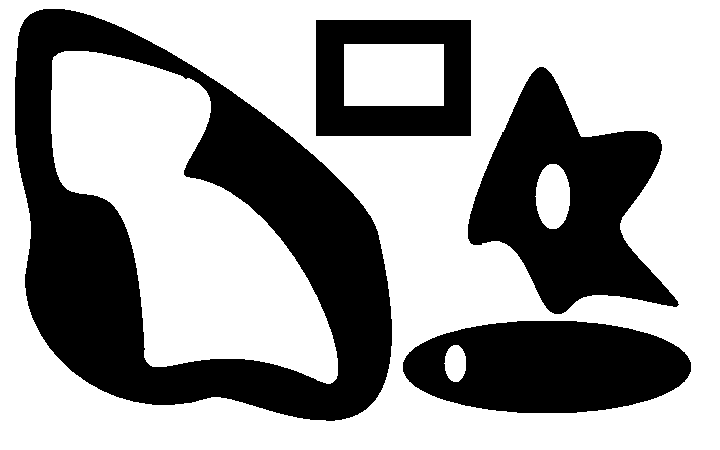}
\end{center}
We say that "a single hole in in it" is a \bf topological invariant \rm of this two dimensional shape.  

We could also consider the two dimensional disk with \it two \rm holes in it:
\begin{center}
\includegraphics[scale=.4]{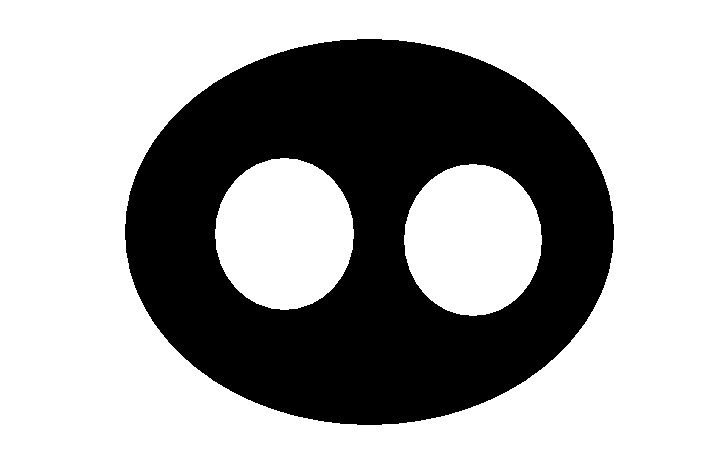}
\end{center}
Once again, we can stretch this all we want, but as long as we don't break it or add points to it, it will always be a two dimensional shape with two holes in it, and we can't make it look like the annulus by just reshaping it.  

So two dimensional holes are a topological invariant.  Now, consider the sphere
\begin{center}
\includegraphics[scale=.4]{sphere.PNG}
\end{center}
Clearly this has a hole in it, but it isn't a ``two dimensional" hole - the hole in this case is three dimensional.  Again we can deform this:
\begin{center}
\includegraphics[scale=.4]{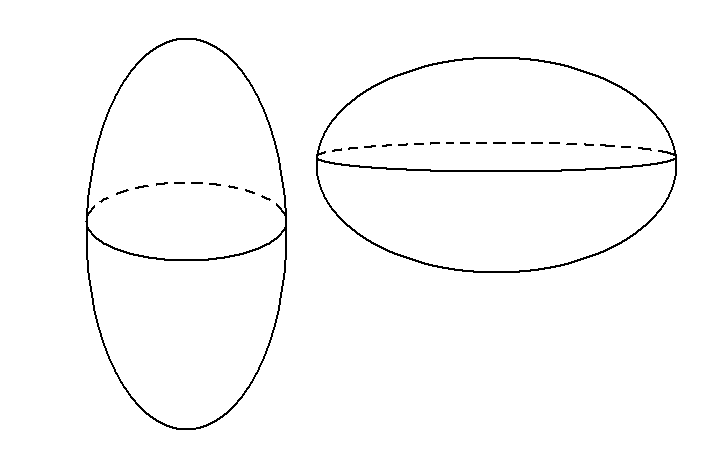}
\end{center}
but it is always a space with a three dimensional hole.  

But then consider the torus:
\begin{center}
\includegraphics[scale=.4]{torus.png}
\end{center}
This also has a three dimensional space missing from its volume, but it is obviously a very different type of three dimensional ``hole" than was in the sphere.  The ``double torus",
\begin{center}
\includegraphics[scale=.6]{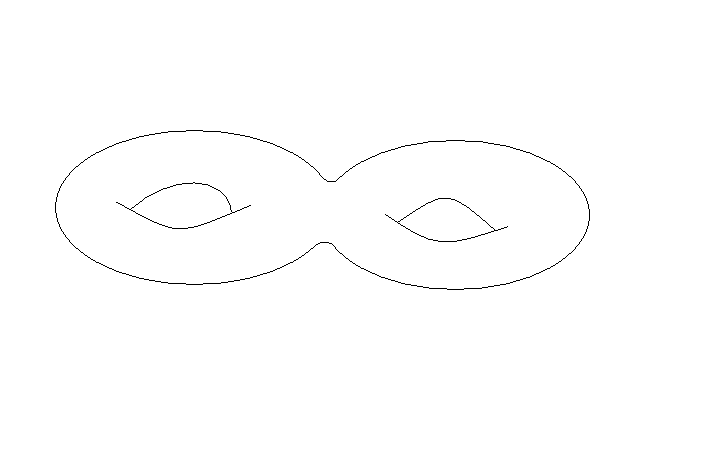}
\end{center}
is also different from both the torus and the sphere.

So, in that topology is concerned with \it categories \rm of spaces and shapes, topology is usually discussed in terms of ``what doesn't change when we stretch, twist, etc. a shape?"  This section is devoted to three of the most commonly discussed and powerful types of such topological invariants: homotopy, homology, and cohomology.  The ultimate fulfillment of this chapter will be when we finally get to more advanced topics in string theory later in this series (though we will also use them quite a bit when doing QFT in the next paper in this series).  The purpose of this chapter is not to give a comprehensive treatment, or even a particularly deep treatment.  We merely intend to give the basic idea of each and move on.  We will discuss all three in greater depth both in this paper and later in this series.  Many of the more interesting features of all three live in the merging of topology and geometry, especially in the theory of fibre bundles, which we will discuss later in these notes and in later papers.  For now we are merely interested in planting seeds.  It will take a while for them to grow into something useful, but we will see that topology provides some of the most powerful tools for advanced physics.  

When we do discuss these ideas again later in this paper, we will consider a few of the non-string theoretic applications of topology in physics.  

The general area of topology these topics are usually studied in is called ``algebraic topology".  It is an approach to topology in which we can take questions which are inherently topological but extremely difficult for topology to answer, and rephrase them in the language of algebra (usually making them easier to answer).  Once again, we will discuss this later in this paper and subsequent papers.  We mention this now to give the larger context of what we are doing.  

We will be talking about ``spaces" $X$ throughout this chapter.  We are doing this to echo the mathematics literature.  We will eventually be interested in specific types of ``spaces", namely manifolds.  For now bear with us as we talk about spaces, keeping in mind that we will eventually be talking about manifolds.  

As one final comment, we admit up front that this section will be lacking a tremendous amount of rigor.  A glaring example is that we aren't going to go through the usual introduction to topology.  We won't discuss things like topologies on sets, covers, compactness, and so on.  While we are losing quite a bit by skipping this material, but we are gaining brevity.  As we said before, the purpose of this series is to provide the ``forest" rather than the "trees".  We will depend a great deal on the intuitive ideas of "space" and other related concepts.  We will provide greater depth when absolutely necessary.  We do provide references for suitable introductory topology and algebraic topology texts, and we encourage the reader to study those (or take a course) after or during reading these notes.  

With all of that said, we can begin.  

\section{Homotopy}

\subsection{Qualitative Explanation of Homotopy}

Consider the spaces $\mathbb{R}$ and $\mathbb{R}^2$.  We can plainly see that these spaces are \it not \rm homeomorphic\footnote{Recall that a homeomorphism is a mapping where a space merely changes shape without changing topology.  In this chapter you can read ``homeomorphic" as ``having the same topological structure"} to each other.  But how could one prove this mathematically?  

One way would be to consider removing a single point from $\mathbb{R}$.  If this is done, the real line is broken into two disjoint pieces.  For example if we remove the point $0$, the space $\mathbb{R} - \{0\}$ is broken into two spaces, and if you are on one side, say at $-1$, it is impossible to move to the other side, say $+1$, without leaving the space.  

But with the plane $\mathbb{R}^2$, the removal of a point doesn't have the same affect.  Removing a single point leaves the plane very much intact.  This difference between $\mathbb{R}$ and $\mathbb{R}^2$ is a fundamental topological distinction, and it illustrates one very important topological idea - \bf connectedness\rm.  There are several different types of connectedness, and we will discuss several of them in this section.  For our purposes now, however, the intuitive idea of what you'd expect ``connected" to mean will work.  When $\mathbb{R}$ is split into two parts, it obviously isn't a connected space - there are two \it dis\rm connected parts.  $\mathbb{R}^2$ minus a point, on the other hand, is still one single space.  This idea of connectedness is our first example of a topological invariant for this chapter.  Of course, saying that space $X_1$ and space $X_2$ are both connected doesn't mean that they are homeomorphic, but if one is connected and the other isn't we can obviously say that they \it aren't \rm homeomorphic.  

Now consider the case of $\mathbb{R}^2$ and $\mathbb{R}^3$.  Obviously the removal of a single point from either leaves each of them connected, and therefore our method above of removing a single point doesn't help us.  However, consider removing a point from each and then drawing a \it loop \rm around the missing point in each.  Then ask the question ``can this loop be smoothly retracted to a point (other than the missing point)?"  With $\mathbb{R}^2$ the answer is ``no", because this would require the loop to pass through the missing point, which it cannot do:
\begin{center}
\includegraphics[scale=.5]{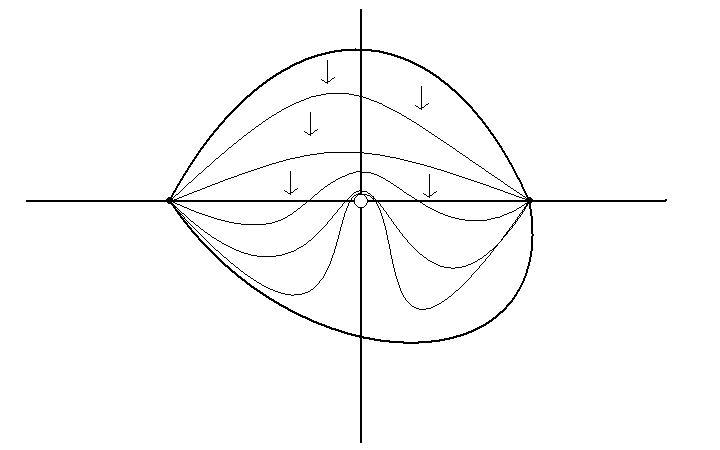}
\end{center}

On the other hand, with $\mathbb{R}^3$ this isn't a problem - we can simply use the extra dimension to move the loop around the missing point and then continue bringing it to a single point.  So this ``loop" approach has given us a way of distinguishing $\mathbb{R}^2$ and $\mathbb{R}^3$.\footnote{Of course there are many other ways - we are taking a route that serves our purposes for this section.}

If we move on to compare $\mathbb{R}^3$ to $\mathbb{R}^4$, drawing loops around a missing point once again fails us, because in either space we have the ``room" to simply slide the loop around the missing point.  But consider a \it circle \rm $S^2$ being drawn around the missing point.  In $\mathbb{R}^4$ there is an extra dimension to slide the circle around the missing point, but in $\mathbb{R}^3$ the circle is ``stuck" around the point.  

We can generalize this to say that to compare $\mathbb{R}^n$ to $\mathbb{R}^{n+1}$, we must remove a point from each and then use $S^{n-1}$'s around the missing point to see if they can be retracted down to a single point (other than the missing point).  In each case, the $S^{n-1}$ can be retracted in $\mathbb{R}^{n+1}$ but not in $\mathbb{R}^n$.  

As an example that doesn't use Euclidian space, consider the circle $S^1$.  If we choose a point on the circle and draw a loop around the circle (coming back to the same point), then we obviously can't retract this circle to a point without leaving the circle.  However, any loop on $S^2$ can be retracted in this way.  The torus $T^2$ is another example of space where some loops can be retracted and some cannot:
\begin{center}
\includegraphics[scale=.5]{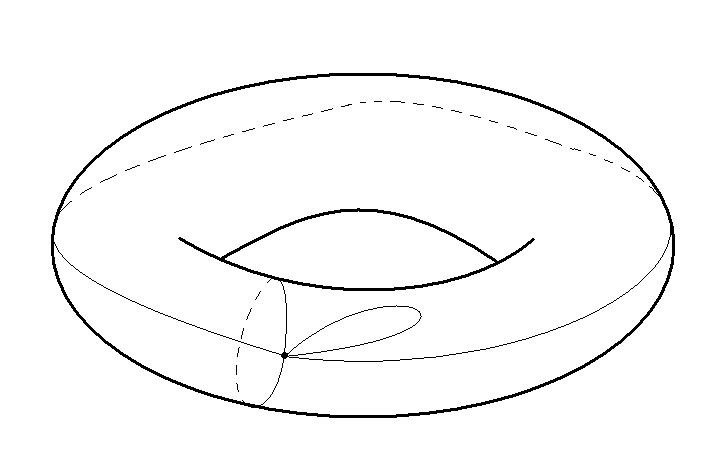}
\end{center}
The torus, however, provides two options for loops that cannot be retracted.  

All of these ideas are made precise with the introduction of \bf homotopy groups\rm.  For a given space $X$, we can define an infinite number of groups, which we denote $\pi_n(X)$ for $n=1,2,\ldots$, where each $\pi_n(X)$ is called the $n^{th}$ homotopy group of $X$.  The meaning of the $n^{th}$ homotopy group is that it keeps track of the number of distinct ways an $n$-sphere ($S^n$) can be mapped into $X$.  For example with $\mathbb{R}^2 - \{0\}$ there were two different ways to map $S^1$ into it - where it didn't go around the missing point and could therefore be retracted to a point, and where it did go around the missing point and therefore couldn't be retracted to a point.  

\subsection{Path Homotopy and the Fundamental Group}

We will restrict our discussion in this chapter to the first homotopy group $\pi_1(X)$ for two reasons: 1) it has a unique importance in physics, and 2) higher homotopy groups are extremely difficult to calculate and we don't have the machinery yet.  With that said, even to study $\pi_1(X)$ we need a few definitions first.  

The first definition is a specific idea of connectedness.  We say a space is \bf path connected \rm if for \underline{any} two points $x_i,x_j\in X$ there is a smooth path that can be drawn connecting them.  This lines up with the intuitive idea of connectedness, and the only examples of spaces that can be considered ``connected" but not path connected are mathematically interesting but not relevant for our purposes.  Any space we consider that is not made up of disjoint parts is path connected.  

Next we define a specific type of path.\footnote{It may seem strange that we defined path connectedness in terms of paths, and are now defining paths.  The notion of path above in the definition of path connectedness was intended to be more intuitive - it simply meant that if you are standing on $x_i$ then you can get to $x_j$ without leaving the space.  Here we are giving a more formal definition of a different (but admittedly related) concept.}  A \bf path \rm from $x_0$ to $x_1$ in a path-connected topological space $X$ (where $x_0,x_1\in X$) is a continuous map $\alpha$ such that
\begin{eqnarray}
\alpha:[0,1]\longrightarrow X
\end{eqnarray}
such that 
\begin{eqnarray}
\alpha(0) &=& x_0 \nolabel \\
\alpha(1) &=& x_1 \label{eq:constraintsonalphahomotopy}
\end{eqnarray}
The points $x_0$ is called the \bf initial \rm point, and $x_1$ is called the \bf terminal \rm point.  

In the previous section we talked loosely about ``retracting" a loop to a point.  We want to make this more clear.  Consider two different paths from $x_0$ to $x_1$, which we call $\alpha$ and $\beta$.  Obviously we will have $\alpha(0) = \beta(0) = x_0$ and $\alpha(1) = \beta(1) = x_1$, but $\alpha$ and $\beta$ can differ in between.  We say that $\alpha$ and $\beta$ are \bf path homotopic \rm relative to the set $\{0,1\}$ if there exists a continuous map $F$ such that
\begin{eqnarray}
F:[0,1]\times [0,1] \longrightarrow X
\end{eqnarray}
which satisfies 
\begin{eqnarray}
F(0,t) = x_0 \qquad and \qquad F(1,t) = x_1 \qquad \forall t \in [0,1] \nolabel \\
F(s,0) = \alpha(s) \qquad and \qquad F(s,1) = \beta(s) \quad \forall s \in [0,1] \label{eq:constraintsonFhomotopy}
\end{eqnarray}
To understand this more clearly, consider the two paths:
\begin{center}
\includegraphics[scale=.5]{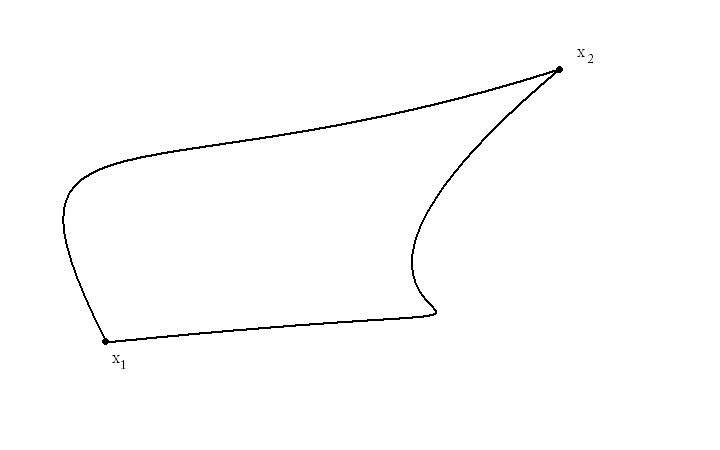}
\end{center}
The idea is that the first variable in $F(s,t)$ represents a point along the path.  The second variable represents a parameterization of an infinite number of paths ``in between" $\alpha$ and $\beta$.  
\begin{center}
\includegraphics[scale=.5]{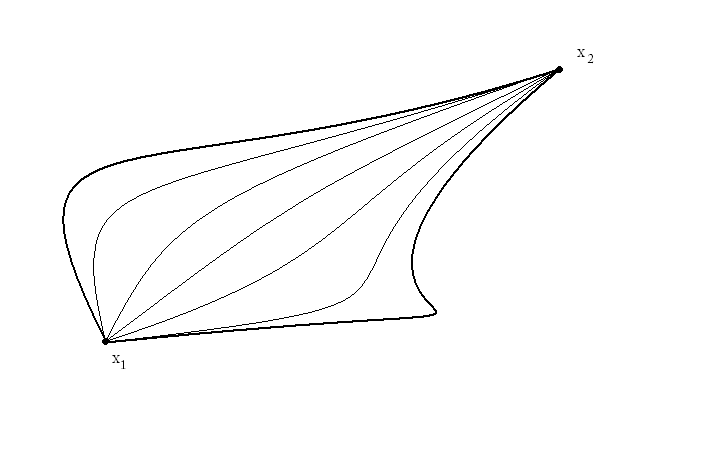}
\end{center}
For example the path at $t=1/2$ would be somewhere in between $\alpha$ and $\beta$
\begin{center}
\includegraphics[scale=.5]{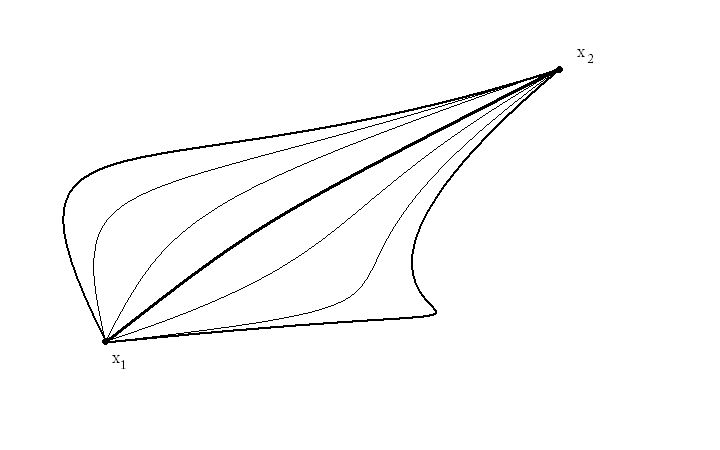}
\end{center}
There are an infinite number of paths from $x_0$ to $x_1$, one for each real number $t\in[0,1]$.  
We call the map $F(s,t)$ a \bf path homotopy\rm.  If such a map exists so that $\alpha$ and $\beta$ are path homotopic relative to $\{0,1\}$, we denote it
\begin{eqnarray}
\alpha \sim \beta :  \{0,1\}
\end{eqnarray}
The meaning of this is that the tilde $\sim$ indicates that $\alpha$ and $\beta$ are homotopic, and what is to the right of the colon is where they are identical - in this case they are only the same at $0$ and $1$.  

As a simple example, if $X=\mathbb{R}^n$, where $x_0$ and $x_1$ are any two points in $\mathbb{R}^n$, define $F(s,t)$ as
\begin{eqnarray}
F(s,t) = (1-t)\alpha(s) + t\beta(s)
\end{eqnarray}
This obviously satisfies (\ref{eq:constraintsonFhomotopy}) as long as $\alpha$ and $\beta$ each satisfy (\ref{eq:constraintsonalphahomotopy}).  

On the other hand consider once again $\mathbb{R}^2 - \{0\}$.  Let $x_0$ be on the positive $x$ axis and $x_1$ be on the negative $x$ axis.  Then let $\alpha$ be a path from $x_0$ to $x_1$ that goes \it above \rm the missing point at the origin, and then let $\beta$ be a path that goes \it under \rm the missing point.  
\begin{center}
\includegraphics[scale=.5]{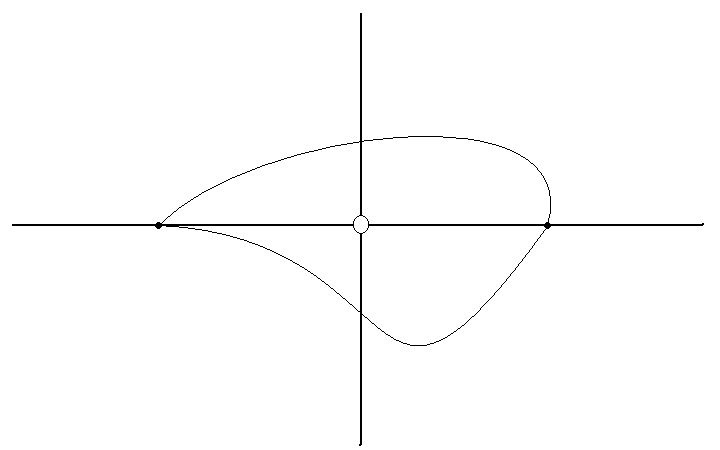}
\end{center}
Clearly no such $F(s,t)$ can be written because the homotopy would, at some point, have to pass over the missing point.  Therefore $F$ wouldn't be continuous, and the definition does not hold - in this case $\alpha$ and $\beta$ are not path homotopic.  

And in general, whether two paths from $x_0$ to $x_1$ are path homotopic is a question about the topology of the space $X$.  

The foundational idea of homotopy is that, for any points $x_0,x_1\in X$, the set of all paths from $x_0$ to $x_1$ which are path homotopic to each other forms an equivalence relation.\footnote{If you aren't familiar with equivalence relations or equivalence classes, you are encouraged to review them in one of the items in the references and further readings section.  The basic idea is to define a set of items which can be related to each other in some way.  In this case we say that they are ``equivalent", and all of the items which are equivalent form an ``equivalence class".  For example consider the disk - we could say that two points on the disk are ``equivalent" if they are at the same radius.  Then the equivalence class would be the collection of all points at a certain radius.  This would result in an infinite number of equivalence classes, all parameterized by the radius $r$.  On the other hand, we could impose the equivalence relation that all points on the same line from the center of the disk are equivalent.  Then there would be an infinite number of equivalence classes parameterized by the polar angle.  The standard notation for two things being ``equivalent" under some equivalence relation (which has been defined in the text) is that $a$ and $b$ are equivalent is $a\sim b$.  This footnote should suffice for what you need to know about equivalence classes for these notes.}  

To be more precise, if $X$ is a topological space and $\alpha,\beta$, and $\gamma$ are paths from $x_0$ to $x_1$, then we have \\
\indent 1) $\alpha \sim \alpha : \{0,1\}$ \\
\indent 2) $\alpha \sim \beta : \{0,1\} \iff \beta \sim \alpha: \{0,1\}$ \\
\indent 3) $\alpha \sim \beta:\{0,1\} \; and\; \beta \sim \gamma:\{0,1\} \Longrightarrow \alpha \sim \gamma:\{0,1\}$\\
All three of these should be intuitively clear, and we therefore omit a rigorous proof.  We will denote the set of all paths that are path homotopic (equivalent) to $\alpha$ as $[\alpha]$.  

As an example, consider again $\mathbb{R}^2 - \{0\}$ with $x_0$ and $x_1$ on the positive and negative $x$ axis, respectively.  It is clear that for any two paths that both go above the missing point at the origin, a path homotopy $F(s,t)$ can be written.  Therefore we can say that all paths which go above the origin are equivalent.  Similarly all paths that go below the origin are path homotopic and therefore equivalent.  Therefore we have two equivalence classes, or ``types of paths", from $x_0$ to $x_1$.  We could choose any (single) arbitrary path $\alpha$ above the origin and any (single) arbitrary path $\beta$ below it, and then the \it set \rm of all paths above the origin would be denoted $[\alpha]$, and the \it set \rm of all paths below the origin would be $[\beta]$.  This is not to say that \it there exist \rm only two elements.  One could also have a path which looped around the origin:
\begin{center}
\includegraphics[scale=.5]{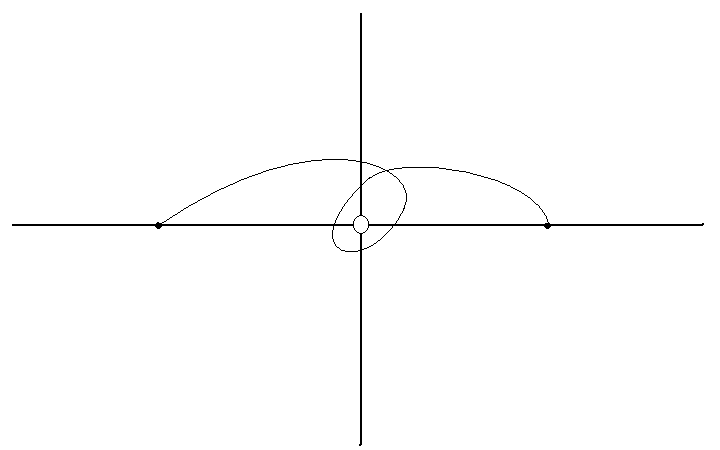}
\end{center}
This would define yet another equivalence class.  Notice that we could define clockwise as positive and counterclockwise as negative, and we could have paths which looped around the origin any integer number of times.  We will explore this relationship between loops and integers more soon.  

Now define the \bf backwards path \rm of $\alpha$, denoted $\alpha^{\leftarrow}$, as
\begin{eqnarray}
\alpha^{\leftarrow} \equiv \alpha(1-s)
\end{eqnarray}
The backwards path of a path $\alpha$ is again a map $\alpha^{\leftarrow}:[0,1]\longrightarrow X$, but it goes in the opposite direction as $\alpha$.  So if $\alpha$ is a path from $x_0$ to $x_1$, then $\alpha^{\leftarrow}$ is a path from $x_1$ to $x_0$.  We can see (though we do not prove because it is a fairly intuitive result) that if two paths $\alpha$ and $\beta$ are path homotopic, $\alpha \sim \beta: \{0,1\}$, then $\alpha^{\leftarrow} \sim \beta^{\leftarrow}:\{0,1\}$.  

We consider \it three \rm points in $X$, $x_0,x_1$ and $x_2$.  Let $\alpha$ be a path from $x_0$ to $x_1$, and let $\beta$ be a path from $x_1$ to $x_2$.  
Define the \bf composite map \rm $\alpha\beta:[0,1]\longrightarrow X$ as
\begin{eqnarray}
\alpha\beta(s) &\equiv& \alpha(2s) \; , \qquad \quad \; \;\; 0 \leq s \leq 1/2 \nolabel \\
& & \beta(2s-1)\;, \qquad 1/2 \leq s \leq 1
\end{eqnarray}
\begin{center}
\includegraphics[scale=.7]{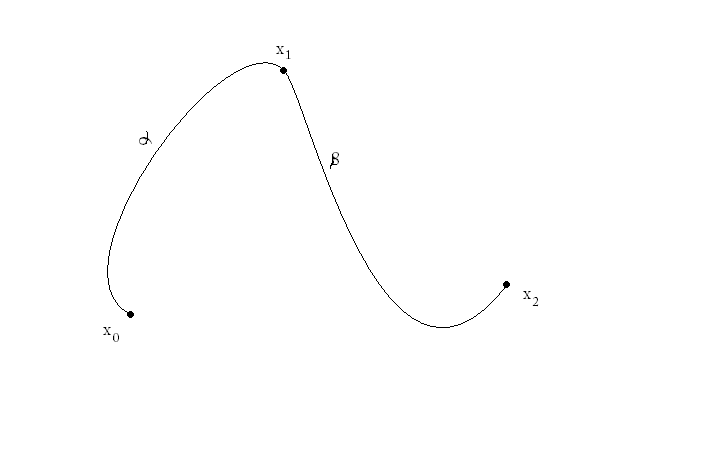}
\end{center}

It should also be clear (though again we do not prove) that if $\alpha_1$ and $\alpha_2$ are both paths from $x_0$ to $x_1$, and $\beta_1$ and $\beta_2$ are both paths from $x_1$ to $x_2$, then
\begin{eqnarray}
\alpha_1\sim\alpha_2:\{0,1\} \; and \; \beta_2\sim \beta_2:\{0,1\} \quad \iff \quad \alpha_1\beta_1 \sim \alpha_2 \beta_2 :\{0,1\}
\end{eqnarray}

We can define equivalence classes of backwards paths as well.  Naturally we will have the relation
\begin{eqnarray}
[\alpha]^{\leftarrow} = [\alpha^{\leftarrow}]
\end{eqnarray}
This definition does not depend at all on which element of $[\alpha]$ we choose.  

It is also easy to see that we can define the product of equivalence classes as
\begin{eqnarray}
[\alpha][\beta] = [\alpha\beta]
\end{eqnarray}
as long as $\beta$ begins where $\alpha$ ends.  

Now we get to the crux of homotopy.  If a path $\alpha$ satisfies 
\begin{eqnarray}
\alpha(0) = \alpha(1)
\end{eqnarray}
then we say that $\alpha$ is a \bf loop \rm at $\alpha(0)=\alpha(1)= x_0$.  Notice now that if $\alpha$ and $\beta$ are both loops at $x_0$, there will never be a problem with defining $\alpha\beta$.  

One loop we will work with a great deal is the \bf trivial loop\rm, denoted $c$, which is defined as
\begin{eqnarray}
c(s) = x_0 \; \forall s \in [0,1]
\end{eqnarray}
It is a ``loop" that never leaves the point $x_0$.  We will denote the set of all loops equivalent to $c$ as $[c]$.  

More generally, the set of all homotopy classes of loops at $x_0$ is denoted $\pi_1(X,x_0)$.  The most important result of this section is that this set has a group structure.  Once we recognize this structure, we call $\pi_1(X,x_0)$ the \bf Fundamental Group\rm, or the first homotopy group, of $X$, with base point $x_0$.  We define the group structure as follows:\\
\indent 1) A given element of the group is $[\alpha]$, where $\alpha$ is a loop at some point $x_0$.  It turns out that the point $x_0$ chosen doesn't matter as long as the space is path connected.  However, it is important to keep track of where the base point $x_0$ is.  While the choice of base point doesn't affect the fundamental group of the space, loops from two different base points are always two different points.  In other words, if a loop can only be continuously deformed to another by moving its base point, those two loops are not homotopic.  However, because the fundamental group doesn't depend on the choice of base point, we will omit it from our notation for most of this section.    \\
\indent 2) Group multiplication between two elements $[\alpha]$ and $[\beta]$ is $[\alpha][\beta] = [\alpha\beta]$.  \\
\indent 3) The identity is $[c]$.  \\
\indent 4) The inverse of an element is $[\alpha]^{-1} = [\alpha^{\leftarrow}]$.  

For example, consider once again the space $X=\mathbb{R}^2$.  Choose an arbitrary point, say the origin, and consider loops from the origin.  Clearly, because there are no holes in $\mathbb{R}^2$, any loop will be path homotopic with the trivial loop.
\begin{center}
\includegraphics[scale=.5]{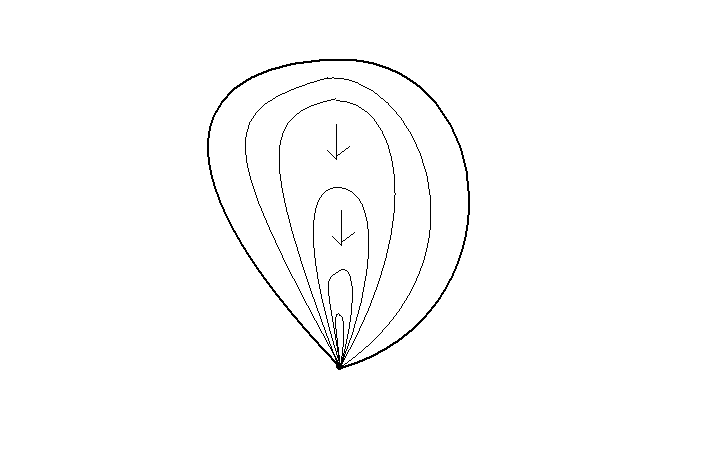}
\end{center}
Therefore every loop in $\pi_1(\mathbb{R}^2)$ will be in $[c]$.  Thus $\pi_1(\mathbb{R}^2)$ only has the single element ($[c]$), and is equal to the trivial group with one element (cf \cite{Firstpaper} if this is not familiar).  

Now consider $\mathbb{R}^2-\{0\}$.  Choose an arbitrary point, say at $(x,y)=(1,0)$ to be the point.  Clearly there will an infinite number of loops that are path homotopic to $c$.  We could also make a loop which goes around the missing point a single time counterclockwise.  We denote an arbitrary loop of this type $\alpha_{+1}$, and the equivalence class of all such loops $[\alpha_{+1}]$.  We could also go around counterclockwise twice - denote this equivalence class $[\alpha_{+2}]$.  Generalizing, we could go around counterclockwise any number $n$ times - we denote this equivalence class $[\alpha_{+n}]$.  It should be clear that $[\alpha_{+n}]$ and $[\alpha_{+m}]$ are not the same for $n\neq m$: it is not possible to continuously map a loop that goes around $n$ times to a loop that goes around $m$ times.  

We could also have a loop that goes around \it clockwise \rm some number $n$ times.  We denote this equivalence class $[\alpha_{-n}]$.  So the set of all equivalence classes will be
\begin{eqnarray}
\pi_1\big(\mathbb{R}^2-\{0\}\big) = \{\alpha_{0}, \alpha_{\pm 1}, \alpha_{\pm 2}, \cdots\}
\end{eqnarray}
Furthermore, if we have an arbitrary element of, say $[\alpha_{1}]$ and an arbitrary element of, say, $[\alpha_{-1}]$, then their composition will be
\begin{center}
\includegraphics[scale=.5]{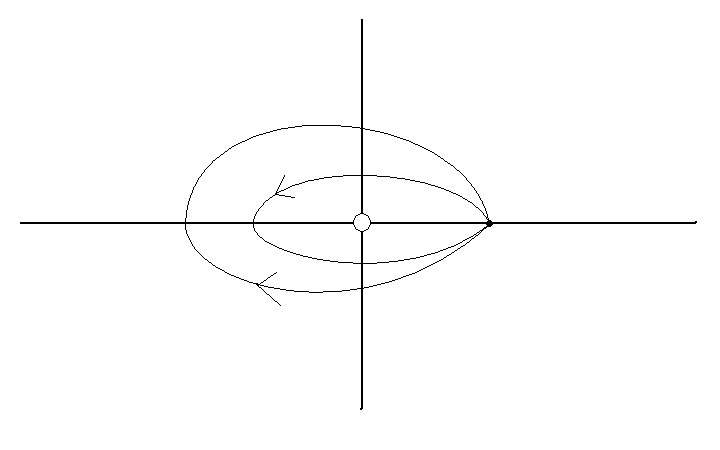}
\includegraphics[scale=.5]{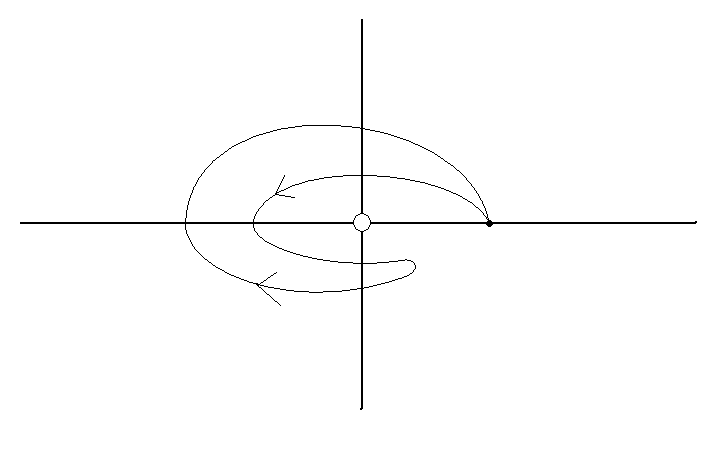}
\end{center}
which is expressed as
\begin{eqnarray}
[\alpha_{1}][\alpha_{-1}] = [\alpha_{1}\alpha_{-1}] = [\alpha_{1-1}] = [\alpha_{0}]
\end{eqnarray}
Or more generally,
\begin{eqnarray}
[\alpha_{n}][\alpha_m] = [\alpha_{n+m}]
\end{eqnarray}

So finally we can see that the fundamental group of $\mathbb{R}^2-\{0\}$ is the group $\mathbb{Z}$ (with addition).  

As one final definition for this section, we say that a space $X$ is \bf simply connected \rm if $\pi_1(X)$ is equal to the trivial group with one element.  With a little thought you can see that therefore $\mathbb{R}^n$ is simply connected for all $n$, and that $\mathbb{R}^n-\{0\}$ is simply connected for $n>2$.  The circle $S^1$ is not simply connected, but $S^n$ is for $n>1$.  The sphere $S^2$ is therefore simply connected, but so is $S^2 - \{p\}$ where $p$ is an arbitrary point on the sphere (take a moment to convince yourself that $S^2-\{p\}$ is homeomorphic to a disk).  However $S^2 - \{p\} - \{q\}$ (the sphere with two points missing, which is homeomorphic to a cylinder) also has fundamental group equal to $\mathbb{Z}$ (convince yourself of this) and is therefore not simply connected.  Another way of saying this is that "not all loops are contractible".

\subsection{More Examples of Fundamental Groups}
\label{sec:examplesfundamentalgroups}

Because detailed calculations of homotopy groups (even fundamental groups) can be extremely difficult, and we don't have the mathematical ``machinery" to do so, we will take a section to merely quote the fundamental groups of various spaces.  These examples should provide a good intuitive understanding of how homotopy groups, especially fundamental groups, behave.  

As our first (uninteresting) example, we mention once again that the fundamental group of the circle $S^1$ is 
\begin{eqnarray}
\pi_1(S^1) = \mathbb{Z}
\end{eqnarray}

As another example, consider two spaces $X_1$ and $X_2$.  We can form the product space $X = X_1 \otimes X_2$.  What will the relationship between $\pi_1(X_1)$, $\pi_1(X_2)$, and $\pi_1(X)$ be?  To see this, consider maps $p_i$ which ``project" from the product space $X$ onto the $X_i$:
\begin{eqnarray}
p_i: X_1\otimes X_2 &\longrightarrow& X_i \nolabel \\
p_i(x_1,x_2) &=& x_i \qquad for \qquad i=1,2
\end{eqnarray}
Any loop in $\alpha \in X$ can then be projected to either $X_i$:
\begin{eqnarray}
p_i(\alpha(s)) = \alpha_i(s) \in X_i
\end{eqnarray}
On the other hand, any two loops $\alpha_i \in X_i$ give a well defined loop in $X$.  So define a map
\begin{eqnarray}
\phi: \pi_1 (X) &\longrightarrow& \pi_1(X_1) \oplus \pi_1(X_2) \nolabel \\
\phi([\alpha]) &=& \big([\alpha_1],[\alpha_2]\big)
\end{eqnarray}
clearly $\phi$ preserves group structure:
\begin{eqnarray}
\phi([\alpha\beta]) = \big([\alpha_1\beta_1],[\alpha_2\beta_2]\big)
\end{eqnarray}
and can be inverted:
\begin{eqnarray}
\phi^{-1}\big([\alpha_1],[\alpha_2]\big) = [\alpha]
\end{eqnarray}
and therefore $\phi$ is an isomorphism\footnote{We will explain what an isomorphism is in the next section, but the basic idea is that it is a way of saying that two groups are the same.} between the fundamental group of $X$ and the direct sum of fundamental groups of $X_1$ and $X_2$.  In other words, we have the extremely useful result
\begin{eqnarray}
\pi_1(X) = \pi_1(X_1\otimes X_2) = \pi_1(X_1) \oplus \pi_2(X_2) \label{eq:relationshipbetweenfundamentalgroupsinproductspacehomotopy}
\end{eqnarray}

As a simple illustration of this consider the infinite cylinder.  You can likely guess (with a little thought) that the cylinder will have the same fundamental group as $S^1$.  But we can prove this using (\ref{eq:relationshipbetweenfundamentalgroupsinproductspacehomotopy}) as follows.  The cylinder can be written as $\mathbb{R} \otimes S^1$.  We know $\pi_1(\mathbb{R}) = 1$ and $\pi_1(S^1) = \mathbb{Z}$.  Therefore the fundamental group of the cylinder is $\pi_1(\mathbb{R} \otimes S^1) = \pi_1(\mathbb{R}) \oplus \pi_1(S^1) = 1 \oplus \mathbb{Z} = \mathbb{Z}$, which is what we expected.  

A less obvious example is the torus $T^2$.  This can be written as $T^2 = S^1\otimes S^1$.  And we know that $\pi_1(S^1) = \mathbb{Z}$, and therefore $\pi_1(T^2) = \mathbb{Z}\oplus \mathbb{Z}$.  So an arbitrary element of the fundamental group of $T^2$ will be $\big([\alpha_{n}],[\alpha_{m}]\big)$, where $\alpha_n$ is a loop around one $S^1$ ($n$ times), and $\alpha_m$ is a loop around the other $S^1$ ($m$ times).  

Another example is the $n$-dimensional sphere $S^n$.  The fundamental group is
\begin{eqnarray}
\pi_1(S^n) = 1
\end{eqnarray}
(the identity) for $n>1$.  

For the M\"obius strip $\mathcal{M}$, 
\begin{eqnarray}
\pi_1(\mathcal{M}) = \mathbb{Z}
\end{eqnarray}

As one final extremely important example, consider the ``figure-eight" \label{wherewediscussthefigureeightinhomotopy}
\begin{center}
\includegraphics[scale=.4]{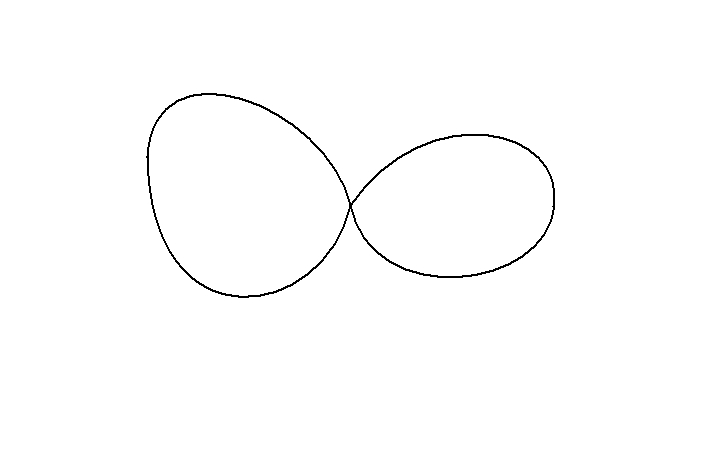}
\end{center}
This has a \it non-Abelian \rm fundamental group group.  Namely, given two elements $g_1$ and $g_2$,\footnote{The elements $g_1$ and $g_2$ correspond to wrapping around the two sides of the figure-eight some number of times - for example $g_1^4g_2^7$ would be wrapping around one side four times and the other side seven times.  } an element of the fundamental group of the figure-eight is the set all all elements made from products of these elements.  For example the identity would be $g_1^0 = g_2^0 = 1$, which would be the identity.  Another arbitrary element would be $g_1^2g_2^7g_1^{-5}$.  Yet another would be $g_1^5g_2^{-3}$.  The product of these two would be
\begin{eqnarray}
g_1^2g_2^7g_1^{-5}g_1^5g_2^{-3} = g_1^2g_2^4
\end{eqnarray}
which is also an element.  Obviously the order of this group is infinite, and it is non-Abelian:
\begin{eqnarray}
g_1g_2 \neq g_2g_1
\end{eqnarray}
You can see the non-Abelian nature of this group as follows: consider a loop $\alpha$ around the left side of the figure-eight, and another loop $\beta$ around the right side.  Consider the composition $\alpha\beta$:
\begin{center}
\includegraphics[scale=.7]{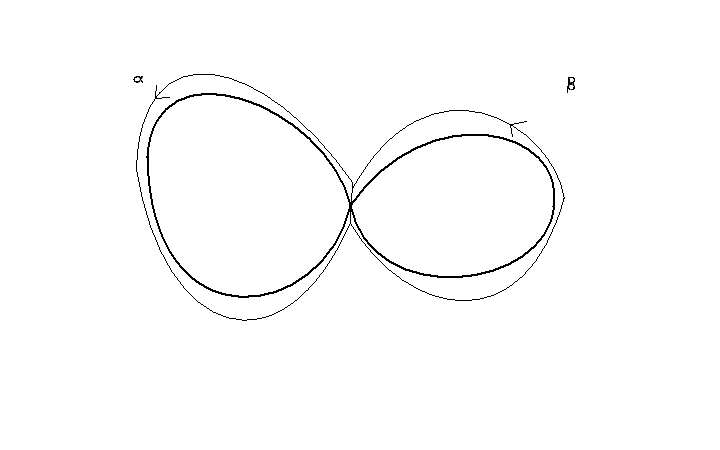}
\end{center}
This is the path where you go around the left side first and then the right side.  The opposite path, $\beta\alpha$, consists of going around the right side first and then the left side.  If you were to try to continuously deform $\alpha\beta$ into $\beta\alpha$, you would have to move the base point of both loops in order to do so.  As we mentioned in the previous section, if you must move the base point of a loop to move it to another loop, those loops are not homotopic.  Therefore $\alpha\beta$ is not homotopic to $\beta\alpha$.

\subsection{Homotopy and Homomorphisms}
\label{sec:homos}

The point of homotopy is the intuitively clear fact that the fundamental group is a topological invariant of $X$.  If $X$ has fundamental group $\pi_1(X)$, then now matter how you stretch, twist, etc. $X$ it will have the same fundamental group.  

This fact can be made more formal by considering maps from one space to another.  But first we need another definition.  A \bf homomorphism \rm is a map from one group to another that  preserves the group structure.  For example if $h$ is a homomorphism from group $G_1$ (with group multiplication $\star_1$) to group $G_2$ (with group multiplication $\star_2$), then for $g_i,g_j\in G_1$, it must be true that $h(g_i\star_1g_j) = h(g_i)\star_2 h(g_j)$.   This simply means that if you map every element of the group through $h$, you still have the same group structure.  

But notice that if $G_2$ is the trivial group, so $h(g_i) = 1 \; \forall i$, we still have a homomorphism.  In other words, $h$ doesn't have to be invertible to be a homomorphism.  This leads to the stricter idea of an \bf isomorpism\rm.  An isomorphism is an \it invertible \rm homomorphism from one group to another.  This type of map preserves the \it entire \rm group structure - nothing is lost.  So a map which takes every element of $G_1$ to the identity is a homomorphism but not an isomorphism.  You can generally think of an isomorphism as being a group theoretic way of saying ``equal".  If $G_1$ and $G_2$ are isomorphic, they describe the same general structure.  Of course, what they act on may be different - the $j=1/2$ and the $j=1$ representations of $SU(2)$ are isomorphic, but they act on very different things.  The groups $SO(2)$ and $U(1)$ are isomorphic, but they act on different things.  

So, to understand homotopy more clearly, we introduce a ``homomorphism induced by a continuous map".  Suppose now that $f:X_1 \longrightarrow X_2$ is a continuous map that carries $x_1 \in X_1$ to $x_2 \in X_2$.  We can introduce the notation
\begin{eqnarray}
f:(X_1,x_1) \longrightarrow (X_2,x_2)
\end{eqnarray}
where the first item in parentheses represents the spaces $f$ maps from and to, and the second item represents specific points $f$ maps to.  

If $\alpha(s)$ is a loop in $X_1$ with base point $x_1$, then the map $f(\alpha(s))$ is a loop in $X_2$ with base point $f(x_1) = x_2$.  Therefore the map $f$ defines a homomorphism, which we denote $f_{\star}$, from $\pi_1(X_1) \longrightarrow \pi_1(X_2)$.  

This leads to the natural result (and the point of this section) that if $f:(X_1,x_1) \longrightarrow (X_2,x_2) $is a homeomorphism, then $f_{\star}$ is an isomorphism between $\pi_1(X_1)$ and $\pi_1(X_2)$.  In other words, if $X_1$ and $X_2$ are homeomorphic, then they have the same fundamental group.  

Of course the converse is not necessarily true - two spaces having the same fundamental group does not necessarily mean that they are isomorphic.  For example $S_1$ and $\mathbb{R}^2-\{0\}$ both have $\pi_1(S^1) = \pi_1(\mathbb{R}^2-\{0\}) = \mathbb{Z}$, but they are not homeomorphic.  The invariance of the fundamental group allows us to say that, if two spaces do \it not \rm have the same fundamental group, they are not homeomorphic. 

\subsection{Homotopic Maps}

As one final idea regarding homotopy, we can not only apply homotopy to spaces, but also to maps between spaces.  Recall in the previous section that we mentioned that the the fundamental group is invariant under homeomorphisms, but two spaces having the same fundamental group doesn't guarantee that they are homeomorphic.  This section will illustrate why.  

Consider two different maps:
\begin{eqnarray}
f,g:X_1 \longrightarrow X_2
\end{eqnarray}
We say that these two maps are \bf homotopic maps \rm if there exists a continuous map 
\begin{eqnarray}
F:X_1\otimes [0,1] \longrightarrow X_2
\end{eqnarray}
such that 
\begin{eqnarray}
F(x,0) = f(x) \qquad and \qquad F(x,1) = g(x)
\end{eqnarray}
For example if $X_1=X_2=\mathbb{R}$, and $f(x) = \sin(x)$ and $g(x) = x^2$, then we could define
\begin{eqnarray}
F(x,t) = (1-t)\sin(x) + tx^2
\end{eqnarray}
which is well defined at every value of $x$ and $t$.  Therefore $\sin(x)$ and $x^2$ are homotopic maps from $\mathbb{R}$ to $\mathbb{R}$.  

now consider again two spaces $X_1$ and $X_2$.  We say that $X_1$ and $X_2$ are of the same \bf homotopy type\rm, which we denote $X_1 \simeq X_2$, if there exist continuous maps
\begin{eqnarray}
f:X_1 \longrightarrow X_2 \nolabel \\
g:X_2 \longrightarrow X_1
\end{eqnarray}
such that $f(g(x_2))$ and $g(f(x_1))$ are each homotopic maps with the identity.  

For example consider $\mathbb{R}$ and the point $p$ (we are treating $p$ as a space by itself, not a point in $\mathbb{R}$).  We can define $f:\mathbb{R} \longrightarrow p$ by $f(x) = p \; \forall x\in \mathbb{R}$, and then $g:p \longrightarrow \mathbb{R}$ by $g(p) = 0 \in \mathbb{R}$.  So, the map $f(g(p))$ takes $p$ to itself, and is trivially homotopic to the identity.  The map $g(f(x))$ will take every point $x \in \mathbb{R}$ to $0 \in \mathbb{R}$.  The identity map on $\mathbb{R}$ is $id_{\mathbb{R}}(x) = x$, so we can define
\begin{eqnarray}
F(x,t) = (1-t)x + t\; 0 =  (1-t)x
\end{eqnarray}
which is continuous and therefore $p$ and $\mathbb{R}$ are of the same homotopy type.  

Now consider $\mathbb{R}$ and $S^1$.  We can define $f:\mathbb{R} \longrightarrow S^1$ as $f(x) = (x\ \; \mod \; 2\pi) \; \in S^1$, and $g:S^1 \longrightarrow \mathbb{R}$ as $g(\theta) = \theta \in \mathbb{R}$.  Then $f(g(\theta)) = \theta$, which is obviously homotopic to the identity.  But $g(f(x)) = x\; \mod\; 2\pi$ is not homotopic to the identity.  Therefore $S^1$ and $\mathbb{R}$ are not of the same homotopy type.  

The amazing (and extremely useful) point of all of this is that if two spaces are of the same homotopy type, then they have the same fundamental group even if they are not homeomorphic.  

Clearly homotopy type forms a less strict classification of spaces than homeomorphism.  There is much, much more we could say about homotopy, but we will save those ideas for later.  We trust that this section has provided a basic understanding of what homotopy is, and how it can be used to classify spaces.  The two main ideas to take away from this section are that 1) the fundamental group $\pi_1(X)$ is invariant under homeomorphism, and 2) spaces of the same homotopy type form an equivalence relation among spaces - if two spaces are of the same homotopy type, they have the same fundamental group.  

\section{Homology}
\label{sec:homology}

\subsection{Qualitative Explanation of Homology}
\label{sec:qualexplhomology}

Consider the two-dimensional disk $D^2$ and $S^1$:
\begin{center}
\includegraphics[scale=.5]{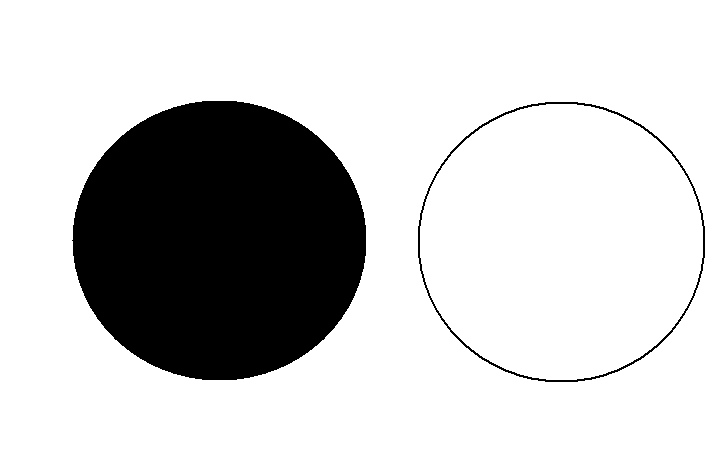}
\end{center}
Obviously they are topologically different, but how can we characterize this difference?  Homotopy is one obvious option (since $\pi_1(D^2) = 1$ and $\pi_1(S^1) = \mathbb{Z}$).  But as we saw in the previous section, spaces can be homotopic but not homeomorphic.  Therefore, if we want to classify spaces as much as possible, having more topological invariants would be helpful.  Also, as we mentioned above, homotopy groups can be extremely difficult to calculate.  Something easier would also be nice.  

With that in mind, one obvious thing we can notice is that $D^2$ and $S^1$ are the same, except $D^2$ contains its interior, whereas $S^1$ does not - $S^1$ is the boundary of $D^2$.  We can also think of this in the following way: $D^2$ is equivalent to $S^1$ with its interior filled in.  $S^1$ is $S^1$ without the interior filled in.  So $D^2$ has an $S^1$ that \it is \rm the boundary of something, whereas $S^1$ has an $S^1$ that \it is not \rm the boundary of something.  And the fact that in $S^1$, the $S^1$ is \it not \rm a boundary is what implies that there is a hole in $S^1$.\footnote{This may seem strange or even silly, but bear with us.}

On the other hand consider the annulus:
\begin{center}
\includegraphics[scale=.3]{annulus.PNG}
\end{center}
We can once again think of the \it outer \rm boundary as an $S^1$ that \it is \rm the boundary of something.  However we can also think of the inner boundary as an $S^1$, but \it this \rm $S^1$ is \it not \rm a boundary (there is nothing inside it).  Once again this implies a hole.  

More generally, we can take a loop (something homeomorphic to $S^1$) in a space such that the loop is \it not \rm a boundary of something to be an indicator of a hole in the space (if this is not obvious, think about it for a second).  

Generalizing once again, we can consider the three dimensional ball $B^3$ and the sphere $S^2$.  Here, $S^2$ is the boundary of $B^3$, but $S^2$ does not have a boundary itself.  So, the presence of an $S^2$ that is not the boundary of anything indicates a hole, but a different type of hole than the presence of a loop that is not the boundary of anything.  

This is the basic idea of homology - an object that does not itself have a boundary is an object that can be the boundary of something else.  Therefore an object that does not have a boundary, and is not the boundary of anything, will indicate a certain type of hole (a two dimensional hole as in $S^1$ or the annulus, or a three dimensional hole as in $S^2$, etc.).  

Intuitively it is likely clear that there must be some similarity between homology and homotopy somewhere under the surface.  Indeed there is, and we will discuss this similarity later.  

Once again, we are doing \it algebraic \rm topology, and therefore our goal will be to assign a group structure to the topological ideas we are dealing with.  We will therefore need a few additional algebraic ideas, but we will introduce them as we go instead of all at once up front.  We feel this will ease learning in this particular section.  

\subsection{Simplicial Complexes}
\label{sec:simpcomp}

We begin by considering a nice geometric way of representing almost any space.  A \bf simplex \rm is a building block for larger spaces.  We often preface the word ``simplex" with the dimensionality of the simplex.  So an $n$-dimensional simplex is an ``$n$-simplex".  

To get a feel for the structure of an $n$-simplex, we give an oversimplified definition that let's us get the basic idea.  We will define a $n$-simplex in $\mathbb{R}^{n+1}$ as the set of all points in $\mathbb{R}^{n+1}$ with non-negative coordinates which all add up to $1$.  

For example, the $0$-simplex will be the set of all points in $\mathbb{R}^{0+1} = \mathbb{R}$ with positive coordinates adding up to $1$.  This will obviously be the \it single point \rm at $1 \in \mathbb{R}$.  So, the $0$-simplex is a single point.  

The $1$-simplex in $\mathbb{R}^2$ will be the set of all points $(x,y)\in\mathbb{R}$ such that $x+y = 1$ (with $x,y \geq 0$).  In other words, this will be the points on the (all positive part of the) line $y=1-x$, or:
\begin{center}
\includegraphics[scale=.5]{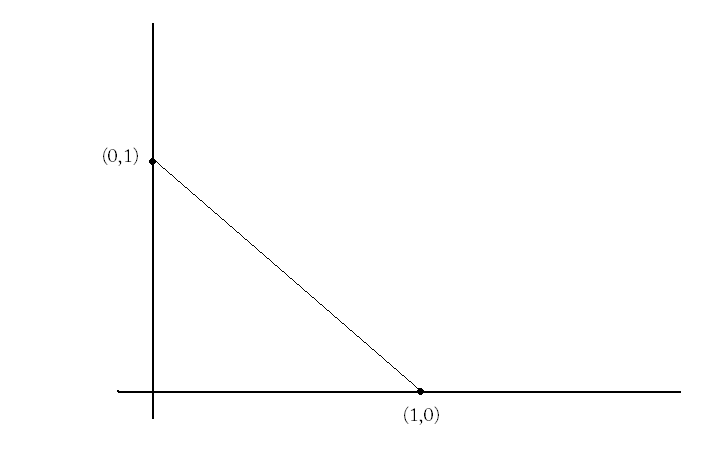}
\end{center}
So the $1$-simplex is a solid line.  

The $2$-simplex in $\mathbb{R}^3$ will be the set of all points $(x,y,z) \in \mathbb{R}^3$ such that $x+y+z = 1$ (with $x,y,z \geq 0$), or (the positive part of) the plane defined by $z = 1-x-y$, or:
\begin{center}
\includegraphics[scale=.6]{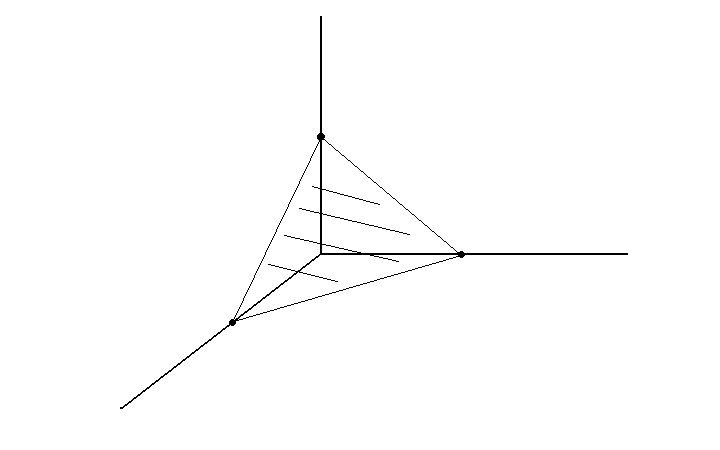}
\end{center}
So the $2$-simplex is a solid triangle.  

Continuing, the $3$-simplex in $\mathbb{R}^4$ will be a solid tetrahedron:
\begin{center}
\includegraphics[scale=.6]{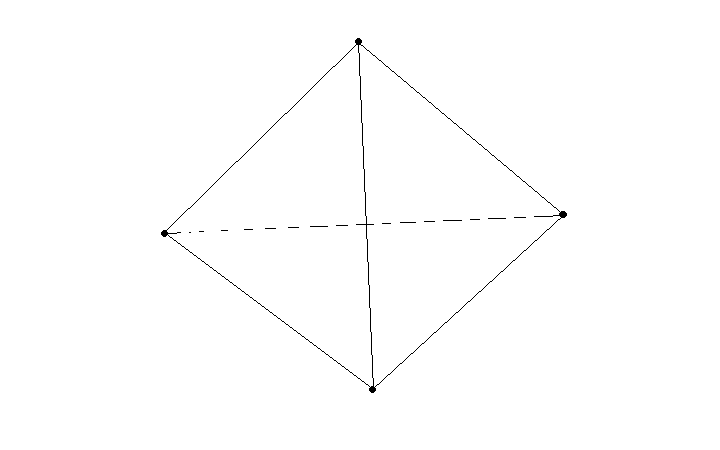}
\end{center}

These examples should give you a good idea of the basic structure of simplexes.  As a comment, we are not really interested in the \it specific \rm definitions given above, settled so neatly in $\mathbb{R}^n$.  We are interested in the general structure of simplexes (i.e. a $0$-simplex is a point, a $1$-simplex is a line, a $2$-simplex is a triangle, etc.), which this definition illustrates.  

Notice that in each case, the $n$-simplex is a collection of $(n-1)$-simplexes with the space between them filled in.  For example the $1$-simplex is two points ($0$-simplexes) with the one-dimensional space between them filled in:
\begin{center}
\includegraphics[scale=.5]{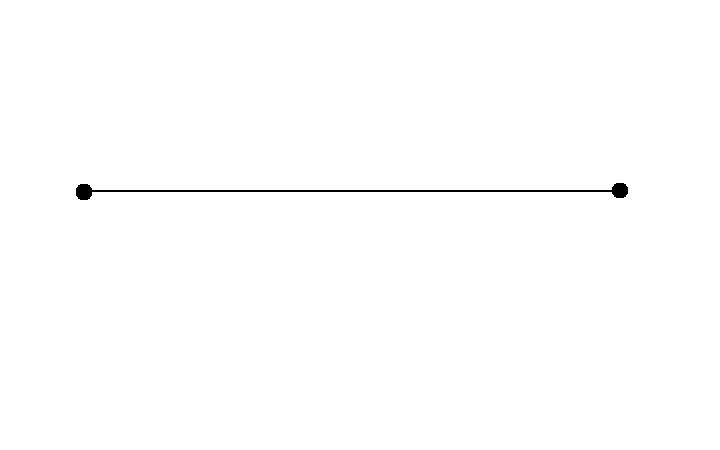}
\end{center}
The $2$-simplex is \it three \rm $1$-simplexes (lines) with the two-dimensional space between them filled in:
\begin{center}
\includegraphics[scale=.5]{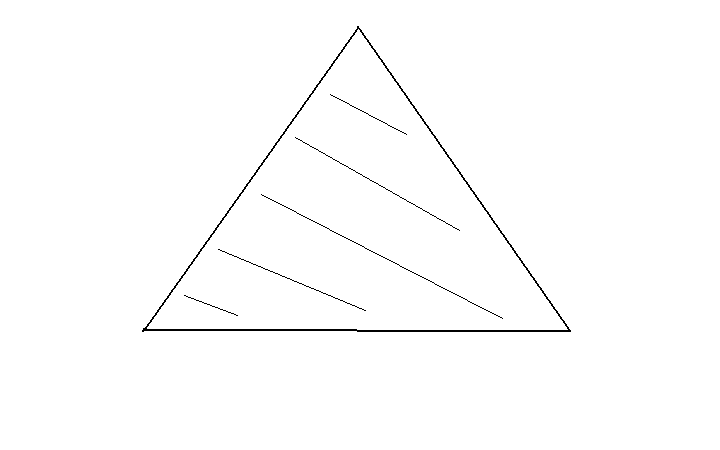}
\end{center}
The $3$-simplex is \it four \rm $2$-simplexes (triangles) with the three-dimensional space between them filled in, etc.  So in general, an $n$-simplex will be a collection $(n+1)$ $(n-1)$-simplexes arranged together with the $n$-dimensional space between them filled in.  Therefore, it is clear that an $n$-simplex will always be an $n$-dimensional space.  And, from this observation, notice that each $n$-simplex is an $n$-dimensional space with an $(n-1)$ dimensional boundary.  

We can generalize this overly simplified definition by considering, instead of nicely arranged points on the axes of $\mathbb{R}^{n+1}$, $n+1$ arbitrary geometrically independent\footnote{The term ``geometrically independent" means that no $n-1$ hyperplane contains all $n+1$ points.  For example a $2$-simplex in $\mathbb{R}^3$ will contain three points.  We demand that no single $2-1 = 1$ dimensional hyperplane, or line, contain all $2+1 = 3$ points.}  points in $\mathbb{R}^{m}$ for any $m\geq n$, which we label $p_0$, $p_1$, $p_2$, $\ldots$, $p_n$ (each with their own set of $m$-tuple coordinates denoted $\bf x\it_{p_i}$ for $i=0,\cdots n$).  Then, we define the simplex over these $n+1$ points similarly to the ``overly simplified" definition above.  Specifically, for the $n+1$ geometrically independent points $p_0,p_1,\ldots,p_n$, the $n$-simplex over these points, denoted $\sigma_n$, is 
\begin{eqnarray}
\sigma_n \equiv \bigg\{ \bf x\it \in \mathbb{R}^m\bigg| \bf x\it = \sum_{i=0}^{n} c_i\bf x\it_{p_i}, \; c_i \geq 0,\; \sum_{i=0}^n c_i =\rm 1\bigg\} \label{eq:firstdefofsimplex}
\end{eqnarray}
We also denote such an $n$-simplex over these $n+1$ points by 
\begin{eqnarray}
\sigma_n = \langle p_0p_1\cdots p_n\rangle
\end{eqnarray}

So, for example, consider two arbitrary points in $\mathbb{R}^2$ (we could also use $\mathbb{R}^1$ if we wanted), denoted $p_0$ and $p_1$:
\begin{center}
\includegraphics[scale=.5]{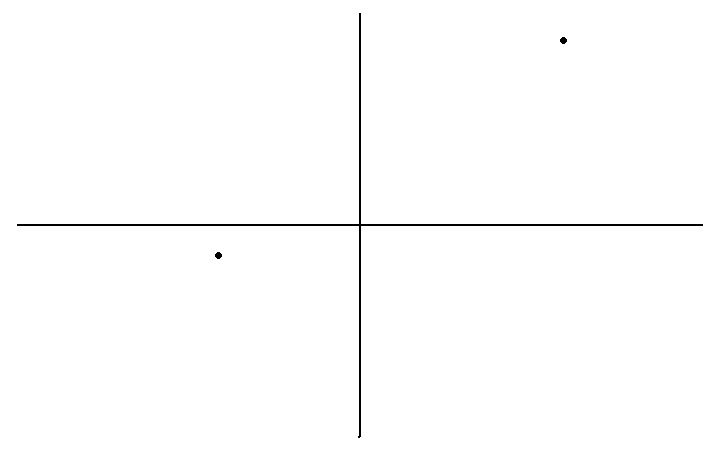}
\end{center}
This will create the $1$-simplex $\langle p_0p_1\rangle$ shown below:
\begin{center}
\includegraphics[scale=.5]{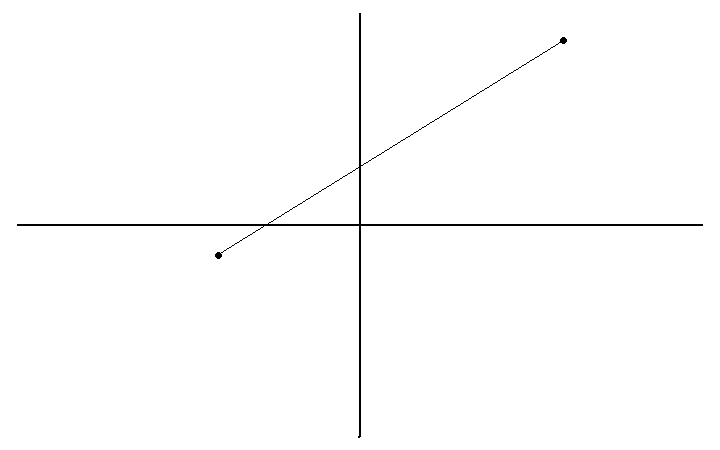}
\end{center}

We could also choose any three arbitrary points in $\mathbb{R}^2$ (or $\mathbb{R}^3$ or $\mathbb{R}^4$ or higher):
\begin{center}
\includegraphics[scale=.5]{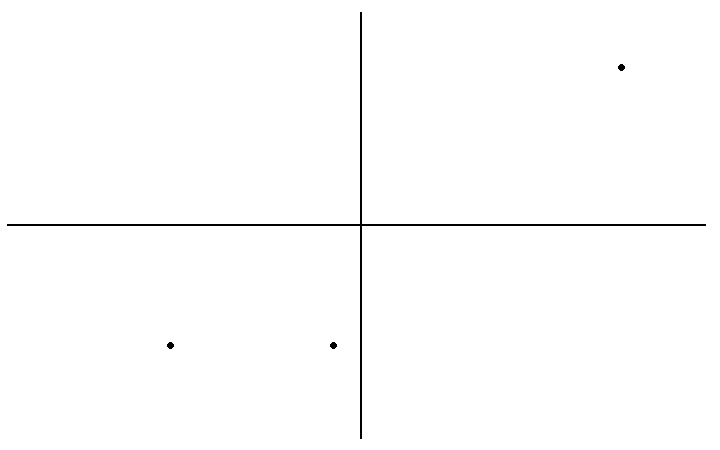}
\end{center}
These will create the $2$-simplex $\langle p_0p_1p_2\rangle$ shown below:
\begin{center}
\includegraphics[scale=.5]{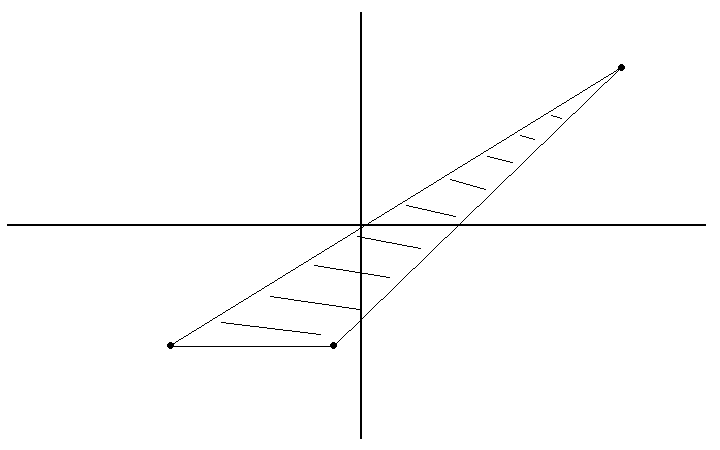}
\end{center}

Next we define the notion of a \bf simplicial complex\rm.  A simplicial complex is a set of complexes fitted together in such a way as to form a larger space.  For consistency we demand that, for any $n$-simplex that is in the complex, all the boundaries of that $n$-simplex are in the complex.  This means that you can't have, say, a $2$-complex without a boundary:
\begin{center}
\includegraphics[scale=.6]{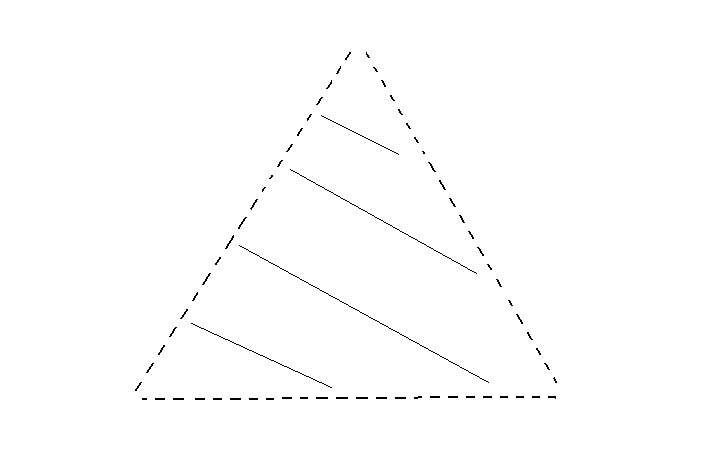}
\end{center}
is not permitted.  Nor can you have a $2$-simplex without the $0$-simplexes that makes the vertices:
\begin{center}
\includegraphics[scale=.6]{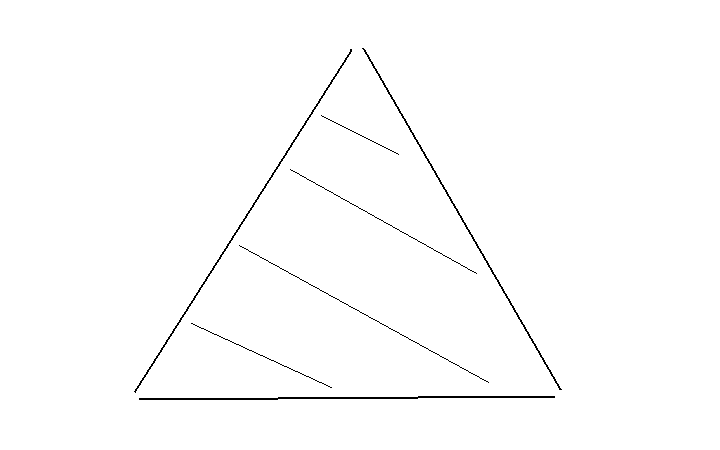}
\end{center}
is not permitted.  However, 
\begin{center}
\includegraphics[scale=.6]{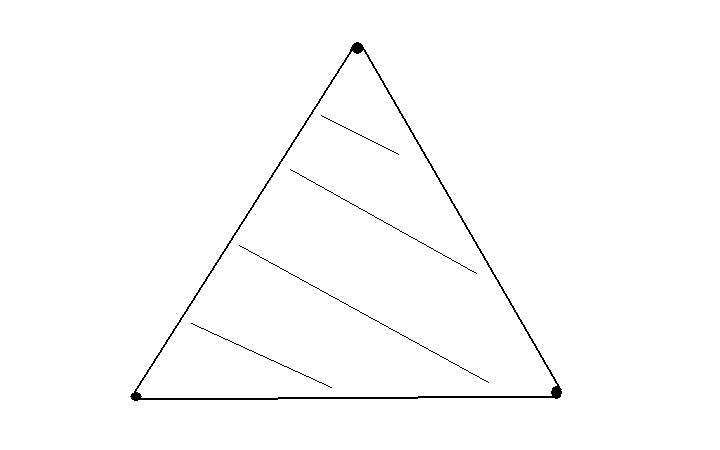}
\end{center}
is permitted.  

Another possible ``poorly" constructed complex would be something like
\begin{center}
\includegraphics[scale=.6]{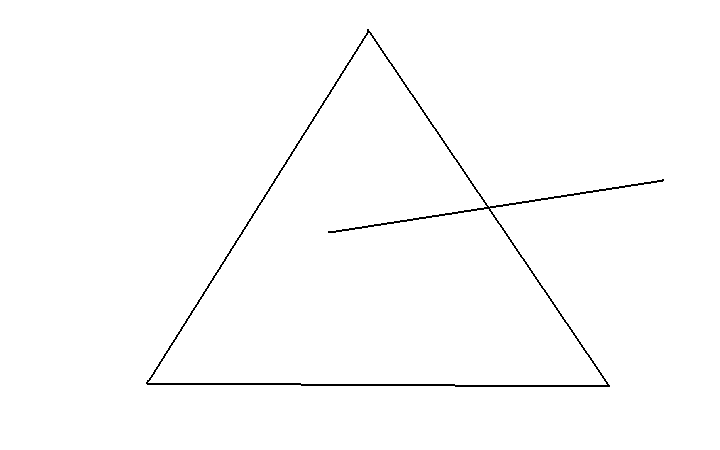}
\end{center}
To prevent things like this, we say that if $\sigma_1$ and $\sigma_2$ are both in the complex, then either $\sigma_1 \cap \sigma_2 = \emptyset$, or $\sigma_1 \cap \sigma_2$ is a face of (or equal to) $\sigma_1$, or $\sigma_1\cap\sigma_2$ is a face of (or equal to) $\sigma_2$.  Obviously the complex in the above diagram doesn't fit this requirement.  

It turns out that, for most spaces $X$ (especially the spaces we will be working with), there exists a simplicial complex $K_X$ that is exactly homeomorphic to $X$.  Such spaces are called \bf triangulable\rm, and the pair $(X,K_X)$ is call the \bf triangulation \rm of $X$.  Of course the triangulation of a space is not unique - for example we could denote $S^1$ in any of the following ways:
\begin{center}
\includegraphics[scale=.6]{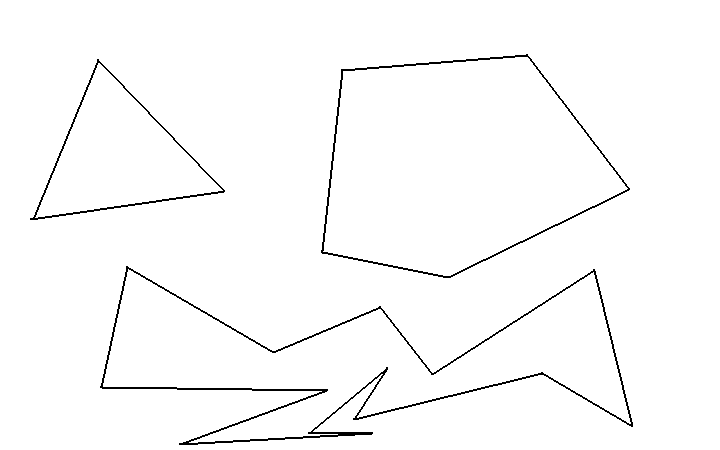}
\end{center}
Or the sphere in any of the following ways:
\begin{center}
\includegraphics[scale=.6]{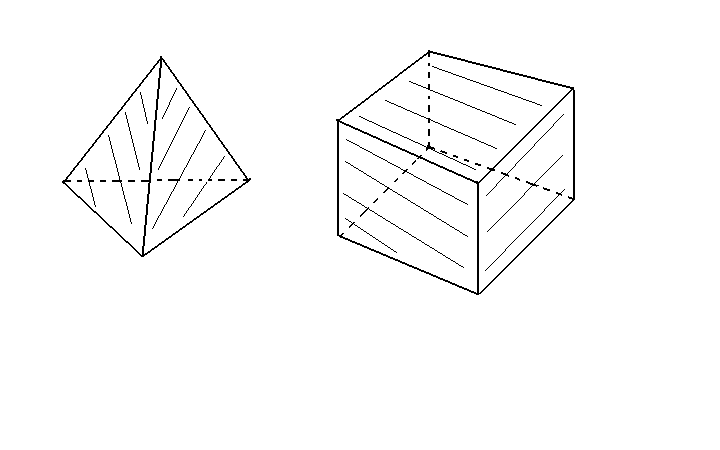}
\end{center}

We also introduce the convention that the simplexes $\langle p_0p_1\cdots p_n\rangle$ are \it directed \rm according to the order of the vertices.  In other words the $1$-simplex $\langle p_0p_1\rangle$ is directed as going \it from \rm $p_0$ \it to \rm $p_1$:
\begin{center}
\includegraphics[scale=.7]{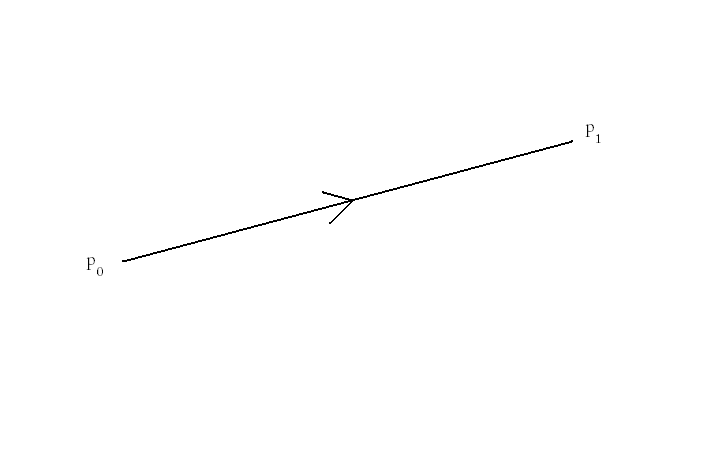}
\end{center}
It will be the ``opposite" of the $1$-simplex $\langle p_1p_0\rangle$
\begin{center}
\includegraphics[scale=.7]{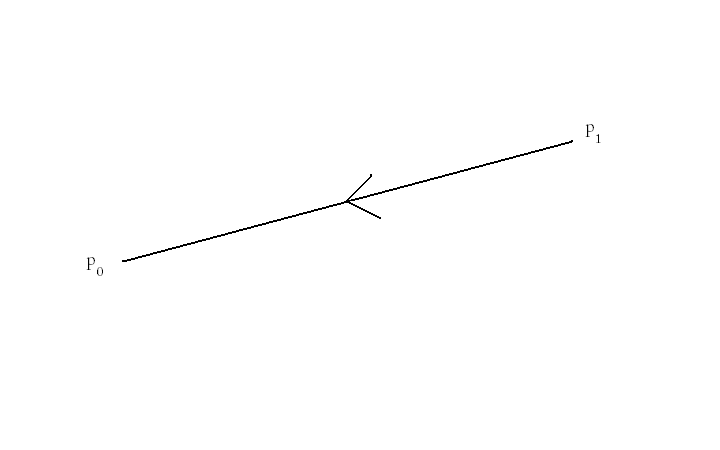}
\end{center}
We therefore write
\begin{eqnarray}
\langle p_0p_1\rangle = - \langle p_1p_0\rangle
\end{eqnarray}

The $2$-simplex $\langle p_0p_1p_2\rangle$ is also directed as
\begin{center}
\includegraphics[scale=.7]{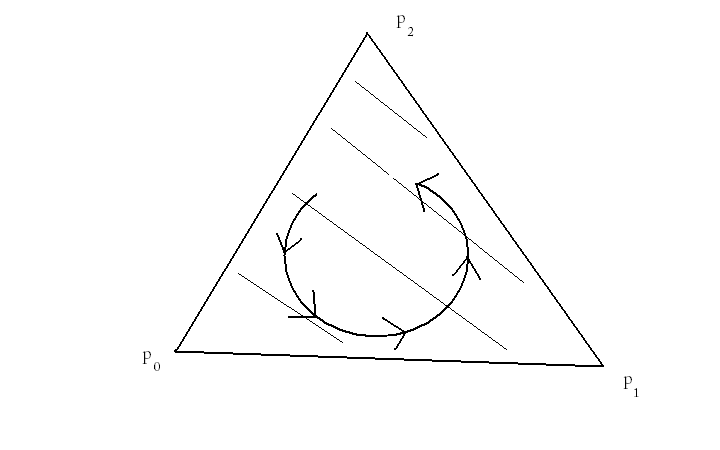}
\end{center}
So, here we find
\begin{eqnarray}
\langle p_0p_1p_2\rangle = \langle p_1p_2p_0\rangle &=& \langle p_2p_0p_1\rangle \nolabel \\
&=& -\langle p_1p_0p_2\rangle =- \langle p_0p_2p_1\rangle =- \langle p_2p_1p_0\rangle
\end{eqnarray}

This generalizes in the obvious way.  

\subsection{The Group Structure of Simplexes}
\label{sec:groupstructuresimplexes}

The construction of simplicial complexes in the last section allows us to define a group structure.  For a complex $K$, we define the \bf $n$-Chain Group\rm, which we denote $C_n(K)$, as the additive group over $\mathbb{Z}$ with elements 
\begin{eqnarray}
c = \sum_{i} c_i \sigma_{n,i}
\end{eqnarray}
where $c_i \in \mathbb{Z}$ and $\sigma_{n,i}$ is the $i^{th}$ $n$-simplex in $K$.  A single element $c \in C_n(K)$ is called an \bf $n$-chain\rm, and addition of $n$-chains $c = \sum_i c_i\sigma_{n,i}$ and $c' = \sum_ic'_i\sigma_{n,i}$ is
\begin{eqnarray}
(c+c') = \sum_i (c_i+c'_i) \sigma_{n,i}
\end{eqnarray}

For example we can consider the complex
\begin{center}
\includegraphics[scale=.7]{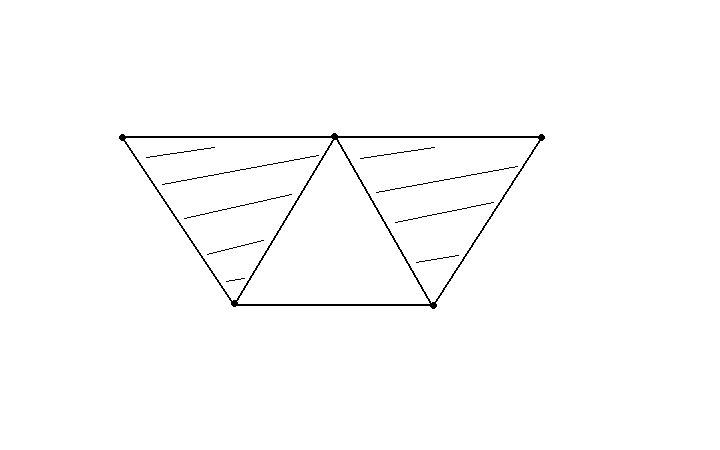}
\end{center}
There are five zero simplexes (the five vertices), so we will have $\sigma_{0,i}$ for $i=1,\ldots,5$, seven $1$-simplexes, so we have $\sigma_{1,i}$ with $i=1,\ldots,7$, and two $2$-simplexes, so we have $\sigma_{2,i}$ with $i=1,2$.  So, the elements of $C_0(K)$ will be of the form
\begin{eqnarray}
c \in C_0(K) = c_1\sigma_{0,1}+c_2\sigma_{0,2}+c_3\sigma_{0,3}+c_4\sigma_{0,4}+c_5\sigma_{0,5} = \sum_{i=1}^5c_i\sigma_{0,i}
\end{eqnarray}
So, consider the two elements
\begin{eqnarray}
c &=& 2\sigma_{0,1} - 12\sigma_{0,4} + \sigma_{0,5} \nolabel \\
c' &=& -\sigma_{0,2} + 4\sigma_{0,4} -\sigma_{0,5} \label{eq:firstexamplecinhomologydefsec}
\end{eqnarray}
These will give
\begin{eqnarray}
c+c' = 2\sigma_{0,1} - \sigma_{0,2} - 8\sigma_{0,4} \in C_0(K)
\end{eqnarray}

Don't loose sleep trying to picture what a given element ``looks like" geometrically on the graph.  For example the first term ($c$) in (\ref{eq:firstexamplecinhomologydefsec}) doesn't correspond to two copies of $\sigma_{0,1}$, negative twelve copies of $\sigma_{0,4}$ and one copy of $\sigma_{0,5}$ or anything like that.  We are effectually created a vector space with vectors $\sigma_{n,i}$ over the field of integers.  Each $n$-simplex is analogous to a ``unit vector" in the $\sigma_{n,i}$ direction.  There is no deeper geometric meaning you should concern yourself with.  

Our group structure is as follows:\\
\indent 1) For any $c,c'\in C_n(K)$, we will have $(c+c') \in C_n(K)$.  \\
\indent 2) Associativity trivially holds.  \\
\indent 3) The element $c=0$ will be the identity element of $C_n(K)\; \forall n$.  \\
\indent 4) For any element $c \in C_n(K)$, the element $-c$ will be the inverse of $c$.  So $c-c = 0$.    

Notice that because the group operation is additive, this will always be an Abelian group.  

\subsection{The Boundary Operator}
\label{sec:boundaryoperator}

While we do have a nice group structure, we admit this group isn't particularly interesting.  In order to proceed we need the \bf boundary operator\rm.  For a directed $n$-simplex $\sigma_n = \langle p_0,p_1\cdots p_n\rangle$, the boundary operator $\partial_n$, which results in the boundary of $\sigma_n$, acts as follow:
\begin{eqnarray}
\partial_n\sigma_n \equiv \sum_{i=0}^n (-1)^i\langle p_0p_1\cdots \hat p_i \cdots p_n\rangle \label{eq:definitionofboundaryoperator}
\end{eqnarray}
where the hat symbol $\;\hat{}\;$ indicates that the element under it should be omitted.  

For example consider the $0$-simplex $\langle p_0\rangle$.  This is obviously
\begin{eqnarray}
\partial_0\sigma_0 = 0 \label{eq:pointshavenoboundary}
\end{eqnarray}
In other words, $\sigma_0$ has no boundary.  This is intuitively clear - a single $0$-dimensional point has no boundary.  

Next consider $\sigma_1 = \langle p_0p_1\rangle$.  This will give
\begin{eqnarray}
\partial_1\sigma_1 = \langle p_1\rangle - \langle p_0\rangle
\end{eqnarray}
Looking at $\sigma_1$ again:
\begin{center}
\includegraphics[scale=.7]{p0top1.PNG}
\end{center}
it is the directed-ness that results in the minus sign.  This is simply the ``final minus initial" points of the directed line element.  The point is that the boundary of $\sigma_1$ is a linear combination of the points which make the boundary of the line - its endpoints.  

For $\sigma_2 = \langle p_0p_1p_2\rangle$, we have
\begin{eqnarray}
\partial_2\sigma_2 &=& \langle p_1p_2\rangle - \langle p_0p_2\rangle + \langle p_0p_1\rangle \nolabel \\
&=& \langle p_1p_2 \rangle + \langle p_2p_0\rangle + \langle p_0 p_1 \rangle
\end{eqnarray}
Once again this makes sense - the boundary of 
\begin{center}
\includegraphics[scale=.6]{directed2simplex.PNG}
\end{center}
will be 
\begin{center}
\includegraphics[scale=.6]{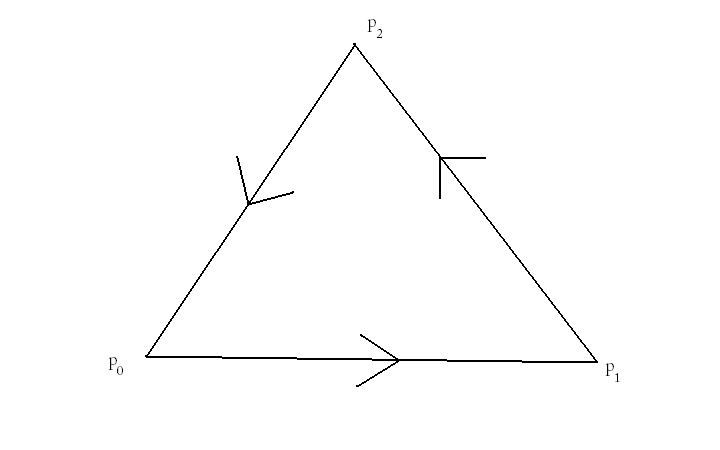}
\end{center}
where once again the directed-ness produces the minus sign.  

One consequence of the definition of the boundary operator is the following:
\begin{eqnarray}
\partial_n(\partial_{n+1}\sigma_{n+1}) &=& \partial_n\bigg(\sum_{i=0}^{n+1} (-1)^i\langle p_0\cdots \hat p_i \cdots p_{n+1}\rangle \bigg) \nolabel \\
&=& \sum_{i=0}^{n+1}\sum_{j=0}^{i-1} (-1)^i(-1)^j\langle p_0\cdots \hat p_j \cdots \hat p_i \cdots p_{n+1}\rangle \nolabel \\
& & + \sum_{i=0}^{n+1}\sum_{j=i+1}^{n+1}(-1)^i(-1)^{j-1}\langle p_0\cdots \hat p_i \cdots \hat p_j \cdots p_{n+1}\rangle \nolabel \\
&=& \sum_{j<i}(-1)^{i+j} \langle p_0 \cdots \hat p_j \cdots \hat p_i \cdots p_{n+1}\rangle \nolabel \\
& & - \sum_{j>i}(-1)^{i+j}\langle p_9\cdots \hat p_i \cdots \hat p_j \cdots p_{n+1}\rangle \nolabel \\
&\equiv & 0 \label{eq:nilpotencyofboundaryoperator}
\end{eqnarray}
In other words $\partial$ is nilpotent - $\partial^2 = 0 $.  Briefly look back at equation (\ref{eq:nilpotentencyofd}), where we pointed out that the exterior derivative operator $d$ is also nilpotent.  Even though $d$ and $\partial$ are very different, there is a very deep and rich correlation between them that we will be exploiting throughout the rest of this series.  

But moving on for now, we point out that the boundary operator acts linearly on elements of $C_n(K)$:
\begin{eqnarray}
\partial_n c = \sum_i c_i \partial_n \sigma_{n,i}
\end{eqnarray}

And more importantly, notice that in each example above, when $\partial_n$ acts on an $n$-simplex $\sigma_n$, the result is a linear combination of $\sigma_{n-1}$'s, which is an element of $C_{n-1}(K)$.  So,
\begin{eqnarray}
\partial_n : C_n(K) \longrightarrow C_{n-1}(K)
\end{eqnarray}
For example, consider again the $2$-dimensional complex K shown here with orientation indicated by the arrows:
\begin{center}
\includegraphics[scale=.5]{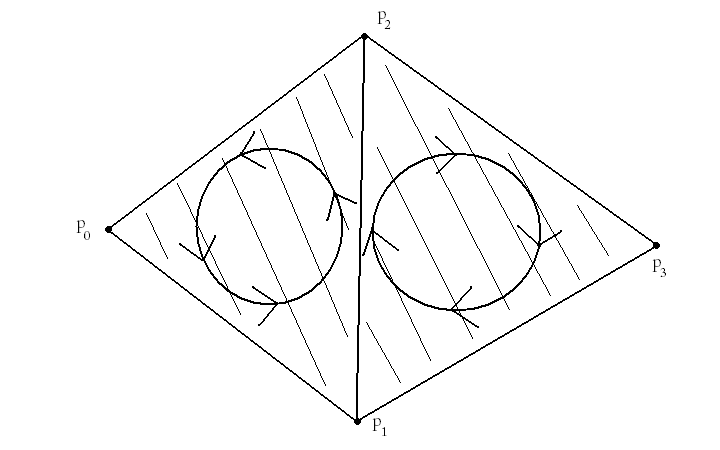}
\end{center}
will have two $\sigma_{2,i}$'s, so an arbitrary element $c\in C_2(K)$ will be $c = A\sigma_{2,1}+B\sigma_{2,2}$, and its boundary will be
\begin{eqnarray}
\partial_2c &=& A\partial_2\sigma_{2,1} + B\partial_2\sigma_{2,2} \nolabel \\
&=& A\;\partial_2 \; \langle p_0p_1p_2\rangle + B \;\partial_2\; \langle p_1p_2p_3\rangle \nolabel \\
&=& A\big(\langle p_1p_2\rangle - \langle p_0p_2\rangle + \langle p_0p_1\rangle \big) + B \big(\langle p_2p_3\rangle - \langle p_1p_3\rangle + \langle p_1p_2\rangle\big) \nolabel \\
& &\in \;  C_1(K)
\end{eqnarray}

Also, from the linearity of $\partial_n$,
\begin{eqnarray}
\partial_n(c+c') = \sum_i (c_i+c'_i)\partial_n\sigma_{n,i}
\end{eqnarray}
So the group structure of $C_n(K)$ is preserved when mapping down to $C_{n-1}(K)$ using $\partial_n$.  This means that $\partial_n$ is a homomorphism (cf section \ref{sec:homos}).  

All of this means that, for an $N$-dimensional simplicial complex $K$, we can form the \bf chain complex\rm, which is the following series of Abelian groups $C_n(K)$ and homomorphism:
\begin{eqnarray}
C_N(K) \xrightarrow{\partial_N} C_{N-1}(K) \xrightarrow{\partial_{N-1}} \cdots \xrightarrow{\partial_2} C_1(K) \xrightarrow{\partial_1} C_0(K) \xrightarrow{\partial_0} 0 \label{eq:chaincomplex}
\end{eqnarray}
We denote the chain complex $C(K)$.  

\subsection{Homology Groups}
\label{sec:homologygroups}

Above we defined the Chain Groups, which weren't all that interesting - just multidimensional additive groups over integers.  However the chain complex resulting from the boundary operator allows us to create a much more interesting group.  

We can choose a single $n$-chain $c \in C_n(K)$ at any $n$ along the chain complex.  It is clear from (\ref{eq:chaincomplex}) that many (but not all) of the $c \in C_n$ can be written as the image of something in $C_{n+1}(K)$ through $\partial_{n+1}$, and it is clear that any $c\in C_n(K)$ will be mapped to $C_{n-1}(K)$ by $\partial_n$.  

Consider specifically the set of all chains in $c \in C_n(K)$ that are mapped to $0$ in $C_{n-1} (K)$:\footnote{Mathematically, given a map $f$ from group $G_1$ to group $G_2$, we call the set of all elements of $G_1$ which are mapped by $f$ to the identity element of $G_2$ the \bf kernal \rm of $f$.  We will employ this notation later in these notes (and explain it in greater depth), but we are mentioning it now so it's not new later. \label{wherekernalsareintroduced}}
\begin{eqnarray}
\partial_nc = 0
\end{eqnarray}
Such chains are called \bf $n$-cycles\rm.  We denote the set of all $n$-cycles in $C_n(K)$ as $Z_n(K)$.  $Z_n(K)$ is a subgroup of $C_n(K)$:\\
\indent 1) If $c$ and $c'$ are in $Z_n(K)$ ($\partial_nc = \partial_nc' = 0$), then $c+c'$ will be also ($c+c' = \partial_n(c+c') = 0$).  \\
\indent 2) Associativity holds trivially.  \\
\indent 3) The $n$-cycle $0$ is a trivial identity element in $Z_n(K)$: $\partial_n \; 0 = 0$.  \\
\indent 4) If $c \in Z_n(K)$ (so $\partial_n c = 0$), then trivially the inverse $-c$ satisfies $\partial_n(-c) = 0$.  

Note that $Z_0(K)$ contains only $0$-chains, all of which trivially map to $0$ (cf (\ref{eq:pointshavenoboundary})), and therefore $Z_0(K) = C_0(K)$.  

As we said above, many $c\in C_n(K)$ can be written as the image of something in $C_{n+1}(K)$ mapped by $\partial_{n+1}$.  This is equivalent to saying that many of the $n$-simplexes in $C_n(K)$ are boundaries of $(n+1)$-simplexes in $C_{n+1}(K)$.  We want to focus on the $c\in C_n(K)$ that \it can \rm be written as the image of something in $C_{n+1}(K)$ through $\partial_{n+1}$.  Therefore we define an \bf $n$-boundary \rm as the set of all elements $c\in C_n(K)$ that can be written as the image of something in $C_{n+1}(K)$.  In other words, if $c\in C_n(K)$ can be written as
\begin{eqnarray}
c = \partial_{n+1} d
\end{eqnarray}
for some $d\in C_{n+1}(K)$, then $c$ is an $n$-boundary.  We denote the set of all $n$-boundaries $B_n(K)$.  This name is natural - if something in $C_{n+1}(K)$ is mapped by the \it boundary \rm operator to $c\in C_n(K)$, then it must be that $c$ is the boundary of that something.  Hence, it is an $n$-boundary.  

Now, from the nilpotency of $\partial$ as seen in equation (\ref{eq:nilpotencyofboundaryoperator}), we make an important observation - \it any $n$-boundary is an $n$-cycle\rm.  If it is possible to write $c$ as $\partial_{n+1}d$ (with $d\in C_{n+1}(K)$), then $\partial_nc = \partial_n(\partial_{n+1}d) = 0$.  We have therefore proven the intuitive fact that boundaries of things don't have boundaries.  We mentioned this fact (without proof) in section \ref{sec:qualexplhomology}.  

However the converse is not necessarily true - if an $n$-chain $c \in C_n(K)$ is an $n$-cycle, that doesn't necessarily meant hat it is an $n$-boundary.  Recall that in section \ref{sec:qualexplhomology} we also mentioned that our interest is in things that \it do not \rm have boundaries, but are not themselves boundaries of anything.  In other words, we are interested in chains that are elements of $Z_n(K)$ but are not in $B_n(K)$.  We therefore need some way of looking at $Z_n(K)$ without $B_n(K)$ ``getting in the way".  It is now apparent why we have introduced this (somewhat boring) chain group structure on $K$ - group theory has a tool which gives us this exactly - the \bf factor group \rm (see \cite{Firstpaper} For a review of this idea).  Recall that the idea behind a factor group $G/H$ is that, when $H$ is a normal subgroup of $G$, then $G/H$ is the set of all elements in $G$ with all of $H$ ``shrunk" to the identity element.  In other words, $G/H$ is $G$ with $H$ ``divided out".  

Here, we have the group $Z_n(K)$ and a subgroup $B_n(K)$:
\begin{eqnarray}
B_n(K) \subseteq Z_n(K) \label{eq:bnisasubsetofzn}
\end{eqnarray}
(which is automatically a normal subgroup because these groups are all Abelian), because all boundaries are cycles.  Therefore, if we want to look at things that are cycles but not boundaries, we can take the factor group $Z_n(K) / B_n(K)$, and we are done!  This will give us the answer we wanted in section \ref{sec:qualexplhomology}.  

We therefore define the $n^{th}$ \bf Homology Group\rm, denoted $H_n(K)$, as 
\begin{eqnarray}
H_n(K) \equiv Z_n(K) / B_n(K) \label{eq:homologydef}
\end{eqnarray}
Once again, the idea behind ``what this means" is that $Z_n(K)$ is the set of all things that are mapped to $0$ by the boundary operator - in other words the set of all things that do not have boundaries, and therefore \it could \rm be the boundary of something.\footnote{Recalling that something that \it could \rm be the boundary of something is anything that doesn't have a boundary itself.}  We know from (\ref{eq:nilpotencyofboundaryoperator}) that boundaries of things do not have boundaries.  Therefore we can think of $Z_n(K)$ as effectually finding boundaries for us.  However, among things that \it could be \rm boundaries, we want to differentiate between things that are the boundaries of something and things that are not.  For example 
\begin{center}
\includegraphics[scale=.6]{directedboundaryof2simplex.PNG}
\end{center}
is a collection of $1$-simplexes that \it could \rm be a boundary if the center was filled in originally, or could not be a boundary if the triangle wasn't filled in.  Topologically, the more interesting situation is when the triangle is \it not \rm filled in (because then we have a non trivial loop), and therefore we are more interested in the situation in which the thing that \it could \rm be a boundary is not actually a boundary.  In other words, we want something that will ``alert" us when 
\begin{center}
\includegraphics[scale=.6]{directedboundaryof2simplex.PNG}
\end{center}
is by itself (and is therefore not a boundary of anything) rather than ``merely" the boundary of 
\begin{center}
\includegraphics[scale=.6]{directed2simplex.PNG}
\end{center}
Therefore we use the factor group $Z_n(K)/ B_n(K)$, which takes the set of all things that \it could \rm be boundaries ($Z_n(K)$), and ``collapses" the set of all things that \it are \rm boundaries ($B_n(K)$) to the identity, because they are topologically trivial anyway (like the solid triangle above).  

\subsection{Fundamental Theorem of Homomorphisms}
\label{sec:fundthrmhomo}

Before moving on, we mention an extremely important relationship that will help simplify calculations of Homology groups.  Consider the general situation of a homomorphism $f:G_1\longrightarrow G_2$.  If we take $G_2$ to be a normal subgroup of $G_1$, then to study homology groups, the group we are interested in is the factor group $G_1/ker(f)$, where $ker(f)$ is the kernal of $f$ as introduced in the footnote on page \pageref{wherekernalsareintroduced}, which is $B_n(K)$.  

The \bf Fundamental Theorem of Homomorphisms\rm, which we will not prove, says that 
\begin{eqnarray}
G_1/ ker(f) = im(f) \label{eq:fundamentaltheoremofhomomorphisms}
\end{eqnarray}
where $im(f)$ is the image of $f$ in $G_2$.  In other words, $im(f) = f(G_1)$.  The equals sign in this case means ``isomorphic".  

As an example, consider $f:\mathbb{Z} \longrightarrow \mathbb{Z}_2$, where 
\begin{eqnarray}
f(a) &=& 0,\;\; a\; even \nolabel \\
f(a) &=& 1,\;\; a\; odd
\end{eqnarray}
This is clearly a homomorphism (but not an isomorphism).  The kernal of $f$ will be everything that maps to the identity in $\mathbb{Z}_2$, which in this case is $0$.  So, $ker(f) = 2\mathbb{Z}$.  The image of $f$, or $f(\mathbb{Z})$, is $\{0,1\}$.  So, the fundamental theorem of homomorphisms tells us
\begin{eqnarray}
\mathbb{Z}/ker(f) = \mathbb{Z}/2\mathbb{Z} = im(f) = \{0,1\} = \mathbb{Z}_2
\end{eqnarray}
which is what we found in \cite{Firstpaper}.  This makes sense - the set of all integers mod the even integers ($\mathbb{Z}/2\mathbb{Z}$) leaves only two possibilities - even or odd.  This is a group with two elements, $0$ or $1$ - hence $\mathbb{Z}_2$.  

We will make great use of (\ref{eq:fundamentaltheoremofhomomorphisms}) in calculating homology groups.  For example, from (\ref{eq:homologydef}) ($H_n(K) \equiv Z_n(K)/B_n(K)$), if we can express $B_n(K)$ as the kernal of some map $f$, then
\begin{eqnarray}
H_n(K) = Z_n(K)/B_n(K) = Z_n(K)/ker(f) = im(f)
\end{eqnarray}
which is generally much easier to calculate.  

But what type of function must $f$ be?  First of all, it must be a homomorphism, or else (\ref{eq:fundamentaltheoremofhomomorphisms}) doesn't apply.  Second, for (\ref{eq:fundamentaltheoremofhomomorphisms}) to apply, $f$ must map \it from \rm $Z_n(K)$ to a normal subgroup of $Z_n(K)$ (and because $Z_n(K)$ is Abelian, \it any \rm subgroup is normal).  The trick is to make sure that if $c\in B_n(K) \subseteq Z_n(K)$, then $f(c) = 0$.  We will discuss how this is done in the examples below.  

\subsection{Examples of Homology Groups}
\label{sec:examplesofhomologygroups}

Consider the $0$-simplex $K=\langle p_0\rangle$.  For this we have $C_0(K)$ consisting of all chains $c$ of the form $c_0 \langle p_0\rangle$ with $c_0 \in \mathbb{Z}$.  This is therefore isomorphic to the group $\mathbb{Z}$ with addition.  Obviously $\partial_0c = 0$ for any $c\in C_0(K)$, so $Z_0(K) = C_0(K)$.  Then, because there is no $C_1(K)$ in this case, we take $B_0(K)$ to be $0$ - the identity element.  So, we have
\begin{eqnarray}
H_0(K) = Z_0(K)/B_0(K) = C_0(K)/\{0\} = \mathbb{Z}/\{0\} = \mathbb{Z}
\end{eqnarray}
It is easy to show that $H_n(K) = 0$ for $n>0$ (we let you work this trivial result out yourself). 

Now consider the complex $K = \{\langle p_0\rangle, \langle p_1\rangle, \langle p_2\rangle, \langle p_0p_1\rangle, \langle p_1p_2\rangle, \langle p_2p_0\rangle, \langle p_0p_1p_2\rangle\}$, a solid filled-in triangle 
\begin{center}
\includegraphics[scale=.6]{directed2simplex.PNG}
\end{center}
$C_0(K)$ will consist of elements $c$ such that
\begin{eqnarray}
c = A\langle p_0\rangle + B \langle p_1\rangle + C \langle p_2\rangle
\end{eqnarray}
where $A,B,C \in \mathbb{Z}$, and therefore $C_0(K) = \mathbb{Z}\oplus \mathbb{Z}\oplus \mathbb{Z}$.  To find $Z_0(K)$ we find all of the $0$-cycles:
\begin{eqnarray}
\partial_0\big(A\langle p_0\rangle + B \langle p_1\rangle + C \langle p_2\rangle\big) \equiv 0 \label{eq:oqiuweyroiquyewr}
\end{eqnarray}
and therefore $Z_0(K) = C_0(K) =  \mathbb{Z}\oplus \mathbb{Z}\oplus \mathbb{Z}$.  Then, $B_0(K)$ will be the set of all chains that satisfy
\begin{eqnarray}
& &\partial_2 \big(D\langle p_0p_1\rangle + E\langle p_1p_2\rangle + F\langle p_2p_0\rangle\big) \nolabel \\
&& = D(\langle p_1\rangle - \langle p_0\rangle) + E(\langle p_2\rangle - \langle p_1\rangle) + F(\langle p_0\rangle - \langle p_2\rangle) \nolabel \\
&&= (F-D)\langle p_0\rangle + (D-E)\langle p_1\rangle + (E-F)\langle p_2\rangle \nolabel \\
&& \equiv D'\langle p_0\rangle +E'\langle p_1\rangle + F'\langle p_2\rangle \label{eq:generalb0k}
\end{eqnarray}
where $D,E,F \in \mathbb{Z}$.  The first line represents the boundary operator $\partial_2$ acting on an arbitrary 2-chain, and the third line is the image of $\partial_2$.  So, any $c\in C_0(K)$ of the form in the third line here will be a $0$-boundary, and therefore in $B_0(K)$.  Notice that, as we have redefined the coefficients of the $0$-chains, they satisfy
\begin{eqnarray}
-D'-E' = F' \qquad \Rightarrow \qquad F'+E'+D' = 0 \label{eq:hintfor0homologyoftriangle}
\end{eqnarray}
So, solving for $F'$, an arbitrary element of $B_0(K)$ will be of the form
\begin{eqnarray}
D'\langle p_0\rangle + E'\langle p_1\rangle -(D'+E')\langle p_2\rangle \label{eq:zxcvpoizucv}
\end{eqnarray}

Taking $H_0(K) = Z_0(K)/B_0(K)$ directly as we did above is not an obvious calculation.  Therefore we use the fundamental theorem of homomorphisms (\ref{eq:fundamentaltheoremofhomomorphisms}).  As we discussed at the end of section \ref{sec:fundthrmhomo} we want a homomorphism $f$ which maps $Z_0(K)$ to a normal subgroup of $Z_0(K)$ such that the kernal of $f$ is $B_n(K)$.  

The way we should construct $f$ is revealed by (\ref{eq:hintfor0homologyoftriangle}).  We know that $Z_0(K) = C_0(K)$ is isomorphic to $\mathbb{Z}\oplus \mathbb{Z}\oplus \mathbb{Z}$ by equation (\ref{eq:oqiuweyroiquyewr}) ff.  $B_0(K)$ also appears this way in the last line of (\ref{eq:generalb0k}).  However $B_0(K)$ actually has the addition constraint of (\ref{eq:hintfor0homologyoftriangle}).  And, because (\ref{eq:hintfor0homologyoftriangle}) is a constraint over $\mathbb{Z}$ (each term in (\ref{eq:hintfor0homologyoftriangle}) is an element of $\mathbb{Z}$), we can define $f:Z_0(K) \longrightarrow \mathbb{Z}$ as
\begin{eqnarray}
f(A\langle p_0\rangle + B\langle p_1\rangle + C\langle p_2\rangle) = A+B+C \label{eq:imageoffhomologyexamples}
\end{eqnarray}
Now any element of $Z_0(K)$ that is mapped to zero (in other words, is in the kernal of $f$) will satisfy
\begin{eqnarray}
A+B+C = 0 \qquad \Longrightarrow \qquad C=-(A+B)
\end{eqnarray}
which defines exactly an element of $B_0(K)$ (cf equation (\ref{eq:zxcvpoizucv})).  

So, 
\begin{eqnarray}
H_0(K) = Z_0(K)/B_0(K) = Z_n(K)/ker(f) = im(f) = \mathbb{Z}
\end{eqnarray}
(the image of $f$ is clearly seen to be $\mathbb{Z}$ from (\ref{eq:imageoffhomologyexamples})).  

Next we want to find $H_1(K)$.  First we find $Z_1(K)$ - for $c\in C_1(K)$,
\begin{eqnarray}
\partial_1c &=& \partial_1 \big(A\langle p_0p_1\rangle + B \langle p_1p_2\rangle + C \langle p_2p_0\rangle\big) \nolabel \\
&=& \cdots \nolabel \\
&=& (C-A)\langle p_0\rangle + (A-B)\langle p_1\rangle + (B-C)\langle p_2\rangle \label{eq:previousthingweareborrowingtheresultfromagain}
\end{eqnarray}
(we borrowed the result from (\ref{eq:generalb0k})) where, as usual, $A,B,C \in \mathbb{Z}$.  This will be $0$ only in the case
\begin{eqnarray}
A=B=C \label{eq:aequalsbequalsc}
\end{eqnarray}
Therefore, a general element of $Z_1(K)$ is
\begin{eqnarray}
A\big(\langle p_0p_1\rangle +\langle p_1p_2\rangle + \langle p_2p_0\rangle\big) \in Z_1(K) \label{eq:secondhomologysecondexamplehomology}
\end{eqnarray}

Now to find $B_1(K)$.  This will be an element that is the boundary of a two simplex:
\begin{eqnarray}
\partial_2D\langle p_0p_1p_2\rangle &=& D\big(\langle p_1p_2\rangle - \langle p_0p_2\rangle + \langle p_0p_1\rangle\big) \nolabel \\
&=& D\big(\langle p_0p_1\rangle + \langle p_1p_2\rangle + \langle p_2p_0\rangle\big) \label{eq:secondhomologysecondexamplehomology2}
\end{eqnarray}
where $D\in\mathbb{Z}$.  Comparing (\ref{eq:secondhomologysecondexamplehomology}) and (\ref{eq:secondhomologysecondexamplehomology2}) we see that they are the same.  Therefore 
\begin{eqnarray}
B_1(K) = Z_1(K)
\end{eqnarray}
which means
\begin{eqnarray}
H_1(K) = Z_1(K) /B_1(K) = 0
\end{eqnarray}
where $0$ is the identity element in the group.  

We leave it to you to show that $H_n(K) = 0$ for $n>0$ in this case.  

As a third example, consider the complex $K=\{\langle p_0\rangle, \langle p_1\rangle, \langle p_2\rangle, \langle p_0p_1\rangle, \langle p_1p_2\rangle, \langle p_2p_0\rangle\}$, a triangle without the inside filled in.  
\begin{center}
\includegraphics[scale=.6]{directedboundaryof2simplex.PNG}
\end{center}

An identical argument to the one above gives 
\begin{eqnarray}
H_0(K) = \mathbb{Z}
\end{eqnarray}
for this complex.  

$H_1(K)$, however, will be different.  First, $Z_1(K)$ will be the same as in (\ref{eq:previousthingweareborrowingtheresultfromagain}), (\ref{eq:aequalsbequalsc}), and (\ref{eq:secondhomologysecondexamplehomology}), and therefore isomorphic to $\mathbb{Z}$.  But unlike before, $B_1(K)=0$ here because there is no $2$-simplex to take the boundary of as in (\ref{eq:secondhomologysecondexamplehomology2}).  So,
\begin{eqnarray}
H_1(K) = Z_0(K)/\{0\} = Z_0(K) = \mathbb{Z}
\end{eqnarray}

As a few final examples which we leave to you to work out, the triangulation of $S^2$ (an empty tetrahedron), or $K=\{\langle p_0\rangle, \langle p_1\rangle, \langle p_2\rangle, \langle p_3\rangle, \langle p_0p_1\rangle, \langle p_0p_2\rangle, \langle p_0p_3\rangle, \langle p_1p_2\rangle$,
$\langle p_1p_3\rangle, \langle p_2p_3\rangle, \langle p_0p_1p_2\rangle, \langle p_0p_1p_3\rangle, \langle p_1p_2p_3\rangle\}$, will give
\begin{eqnarray}
H_0(K) = \mathbb{Z} \qquad \qquad H_1(K) = 0 \qquad \qquad H_2(K) = \mathbb{Z} \label{eq:qwemnrb}
\end{eqnarray}
and all higher groups $H_n(K)=0$.  

A triangulation of the M\"obius strip (you can write out $K$ yourself) produces
\begin{eqnarray}
H_0(K) = \mathbb{Z} \qquad \qquad H_1(K) = \mathbb{Z} \qquad \qquad H_2(K) = 0\qquad \qquad etc.
\end{eqnarray}

The torus $T^2$ produces
\begin{eqnarray}
H_0(T^2) = \mathbb{Z} \qquad \qquad H_1(K) = \mathbb{Z} \oplus \mathbb{Z} \qquad \qquad H_2(T^2) = \mathbb{Z}
\end{eqnarray}

The cylinder $\mathcal{C}$ produces
\begin{eqnarray}
H_0(\mathcal{C}) = \mathbb{Z} \qquad \qquad H_1(\mathcal{C}) = \mathbb{Z} \qquad \qquad H_n(\mathcal{C}) = 0, \; \forall n>1
\end{eqnarray}

A triangulation of the figure eight:
\begin{center}
\includegraphics[scale=.6]{fig82.PNG}
\end{center}
will be 
\begin{eqnarray}
H_0(K) = \mathbb{Z} \qquad \qquad H_1(K) = \mathbb{Z} \oplus \mathbb{Z} \qquad \qquad H_n(K)=0, \; \forall n>1 \label{eq:ytsdjfkh}
\end{eqnarray}

\subsection{Meaning of the $n^{th}$ Homology Group}

As we explained in the introduction to this section, the basic idea of homology is to find something that \it could be \rm a boundary, but isn't one - this indicates a hole.  The simplest example is two distinct points - they \it could \rm be the boundary of a line.  So for a line, the two points on the ends both \it could \rm be and \it are \rm boundary points.  However, two disjoint points with no line between them \it could \rm be and \it are not \rm boundary points.  

This is the idea behind homology groups.  The group $H_n(K)$ looks for ``$n$-dimensional holes" in a space by taking all of the things that could be boundaries ($Z_n(K)$) and mod-ing out all the things that actually are ($B_n(K)$).  What is left, $H_n(K)$, is a record of the $n$-dimensional holes.  

For example, a one-dimensional hole is a circle $S^1$.  Therefore any space with a one-dimensional hole will have a non-trivial $H_1(K)$.  A two-dimensional hole is $S^2$, and therefore any space with a two-dimensional hole will have a non-trivial $H_2(K)$ (cf equations (\ref{eq:qwemnrb})-(\ref{eq:ytsdjfkh})).  

Another way of thinking about this is as follows: for a one-dimensional hole, which is like $S^1$, imagine a circle living in the space, but it will move itself around to try to collapse itself down to a point.  If the space is such that there is some way a loop can live in it without being able to find a way of contracting itself down, then the $H_1$ homology group is non-trivial.  Obviously $S^1$, the M\"obius strip, and the figure eight all satisfy this (the figure eight had two ways of ``supporting" an $S^1$ in this way), whereas the other examples considered above did not.  

Of the spaces we considered as examples above, only the tetrahedron (which was homeomorphic to $S^2$ itself) had a non-trivial $H_2$ group.  Once again this makes intuitive sense.  

The $H_0$ groups are less obvious.  In essence $H_0(K)$ will be non-trivial whenever the space can separate two individual points ($S^0$).  All that is required for this is for the space $K$ to contain a single point.  Therefore $H_0(\emptyset) = 0$, but for any $K\neq \emptyset$, $H_0(K) \neq 0$.  In fact the following result can be proven:
\begin{eqnarray}
H_0(K) = \mathbb{Z}
\end{eqnarray}
for any \it connected \rm complex $K$.  If $K$ is not connected, then 
\begin{eqnarray}
H_0(K) = \mathbb{Z}\oplus \mathbb{Z} \oplus \cdots \oplus \mathbb{Z} \; p\; times
\end{eqnarray}
where $p$ is the number of disconnected components of $K$.  For example, if $K$ is three disjoint points,
\begin{eqnarray}
H_0(K) = \mathbb{Z} \oplus \mathbb{Z} \oplus \mathbb{Z}
\end{eqnarray}

For the general $H_n(K)$ group, the basic form will be
\begin{eqnarray}
H_n(K) = \mathbb{Z}\oplus \mathbb{Z} \oplus \cdots \oplus \mathbb{Z} \; p\; times \label{eq:generalhomologygroup}
\end{eqnarray}
where now $p$ is the number of $n$-dimensional holes in $K$.\footnote{As a note to more mathematically inclined readers, we are omitting discussion of the torsion subgroups for now.}

We recognize that this discussion has left a tremendous amount to be desired, and anyone already familiar with the topic is likely pulling their hair out due to the lack of rigor and often somewhat careless treatment of concepts.  We apologize for this, but want to reiterate that our only goal is to give the reader a small picture of the basic ideas behind these topics in algebraic topology.  Not a comprehensive (or even close to comprehensive) introduction.  We have done our best to maintain (pedagogical) clarity even if that has resulted in a lack of completeness.  We will address these ideas again later, after we have developed more sophisticated concepts, tools, techniques, and vocabulary.  

\subsection{Connection Between Homology and Homotopy}
\label{sec:connectionbetweenhomologyandhomotopoy}

Before moving on we take a moment to consider how homotopy groups and homology groups relate.  You no doubt noticed that there is some similarity between them.  To emphasize this we summarize several of our results so far in the following table.  

\begin{center}
\begin{tabular}{|c|c|c|}\hline  & $\pi_1(K)$ & $H_1(K)$ \\\hline Disk & 1 & 1 \\\hline Circle & $\mathbb{Z}$ & $\mathbb{Z}$ \\\hline Torus & $\mathbb{Z}\oplus \mathbb{Z}$ & $\mathbb{Z}\oplus \mathbb{Z}$ \\\hline Cylinder & $\mathbb{Z}$ & $\mathbb{Z}$ \\\hline 2-Sphere & 1 & 1 \\\hline M\"obius Strip & $\mathbb{Z}$ & $\mathbb{Z}$ \\\hline Figure Eight & Generated by $g_1,g_2$ (non-Abelian) & $\mathbb{Z}\oplus \mathbb{Z}$ \\\hline \end{tabular}\label{pagewithtablesummarizingfundamentalgroupsandhomology}
\end{center}

Notice that, in all cases, the fundamental group ($\pi_1$) and the first homology group ($H_1$) is the same except for the figure eight.  We will address the case of the figure eight shortly.  

For the other six spaces, the reason for the similarity is that $\pi_n$ and $H_n$ are essentially looking for the same thing - $n$-dimensional ``holes" in the spaces (for clarity, we will focus on $\pi_n$ and $H_n$ for $n>0$ only.  In the cases above, we are specifically considering $1$-dimensional holes, which are equivalent to circles.  It is clear by inspection that a disk and a $2$-sphere do not have any holes such that a circle cannot be retracted to a point.  This is why both $\pi_1$ and $H_1$ are trivial for each of them.  

On the other hand, the circle, the cylinder, and the M\"obius strip all have a single loop in them which would prevent a circle from being contracted, and therefore $\pi_1 = H_1$ for all of them.  

The torus, as we pointed out above, has \it two \rm holes in it, and therefore $\pi_1$ and $H_1$ both have \it two \rm copies of $\mathbb{Z}$.  

You can effectively think of the $\mathbb{Z}$'s in each corresponding to a hole in the sense of the number of times you can wrap around that hole.  

Higher homotopy groups ($\pi_n$) and homology groups ($H_n$) are similar.  They are both looking for the same things.  The difference is the group structure they assign based on what they find.  
	
So what about the figure-eight?  Why are $\pi_1$ and $H_1$ different in this case?  Furthermore, notice that $H_1$ for the figure-eight is equal to $H_1$ for the torus.  Yet, while $\pi_1$ for the torus is equal to $H_1$ for the torus, $\pi_1$ for the figure-eight is unique.  

The difference is in the nature of the groups generated.  As we mentioned above, the general form of $H_n(K)$ is (\ref{eq:generalhomologygroup}).  We can classify such groups more generally.  

Consider some group $G$ with elements $g_i$.  We will take the group operation to be $\star$.  Let us assume that $G$ is a group such that there exists a finite set of $n$ elements $h_1,\ldots,h_n \in G$ which act as \bf generators \rm of the group.  This means that any element $g_i$ can be written as some product of these $n$ elements:
\begin{eqnarray}
g_i = h_1^{i_1} h_2^{i_2} \cdots h_n^{i_n} \label{eq:whatexponentsmeaninrelationshipbetweenhomologyandhomotopysection}
\end{eqnarray}
where $h_i^j = h_i \star h_i \star \cdots \star h_i$, $j$ times.  We briefly discussed this when we talked about the homotopy group of the figure eight on page \pageref{wherewediscussthefigureeightinhomotopy}.  There there exist a finite set of such generators, we say that $G$ is \bf finitely generated\rm.  If the generators are linearly independent, we say they are \bf free \rm generators.  If the generators are linearly independent and free, then $G$ is necessarily Abelian, and we call $G$ a \bf finitely generated free Abelian group\rm.  The number of generators $n$ is called the \bf rank \rm of $G$.  In simpler terms, if $G$ is finitely generated by $n$ linearly independent generators, we call $G$ a \bf free Abelian group \rm or rank $n$.  

For example, consider the group $\mathbb{Z}\oplus \mathbb{Z}$ with addition.  An arbitrary element can be written as $(a,b)$, where $a,b\in \mathbb{Z}$.  This group is obviously infinite, but we can write any element in terms of the generators
\begin{eqnarray}
h_1 = (1,0) \qquad \qquad h_2(0,1)
\end{eqnarray}
Then, the group element $(a,b)$ is written as
\begin{eqnarray}
(a,b) = h_1^a h_2^b
\end{eqnarray}
(recall what the exponents mean here, cf (\ref{eq:whatexponentsmeaninrelationshipbetweenhomologyandhomotopysection}) ff, and that the group operation $\star$ here is merely addition, $+$).  

So, as the chart above indicates, $H_1(\rm figure\; eight\it) = \mathbb{Z} \oplus \mathbb{Z}$.  This is this group exactly - the free Abelian group of rank $2$.  We pointed out above than any arbitrary homology group can be written as in (\ref{eq:generalhomologygroup}), which are all free Abelian groups, where the rank (the number of generators) is equal to the number of holes.  So, all homology groups are free Abelian groups.  

The reason for this is that the definition of the $n^{th}$ homology group $H_n(K)$ is given in terms of $Z_n(K)$ and $B_n(K)$ (see (\ref{eq:homologydef})).  Notice that, by definition, $Z_n(K)$ and $B_n(K)$ are both free Abelian groups.  Therefore, as the factor group of a free Abelian group, it is obvious why $H_n(K)$ is a free Abelian group.  

Furthermore, notice from the table above (and from section \ref{sec:examplesfundamentalgroups}) that $\pi_1(\rm figure\; eight\it)$ is the \it non-Abelian \rm group with two generators.  The reason for this is again because of the nature of the definition.  Homology groups are \it defined \rm by free Abelian groups ($Z_n(K)$ and $B_n(K)$), whereas homotopy groups are \it defined \rm by loops, which (as we saw in section \ref{sec:examplesfundamentalgroups}), aren't always commutative.  Therefore it is possible to have non-Abelian fundamental groups.  The figure-eight is only one example of a space with a non-Abelian fundamental group.  Of course, the torus is an example of a space with a fundamental group that is ``bigger" than $\mathbb{Z}$, but is still Abelian.  

As a final comment, there is a general relationship between $\pi_1(K)$ and $H_1(K)$ where, if you have $\pi_1$, it is straightforward to calculate $H_1$.  First define the \bf commutator subgroup\rm, denoted $F$ of a group as the set of all elements of the form
\begin{eqnarray}
aba^{-1}b^{-1}
\end{eqnarray}
If we take the (generally non-Abelian) group $\pi_1(K)$ and then find the factor group 
\begin{eqnarray}
\pi_1(K) / F \label{eq:pi1modcomsubgroup}
\end{eqnarray}
then we have \bf Abelianized\label{pagewherewefirsttalkaboutabelianized} \rm $\pi_1(K)$.  Recall that the meaning of the factor group $G/H$ is to contract everything in $H$ to the identity.  So, (\ref{eq:pi1modcomsubgroup}) is setting everything in the commutator subgroup equal to the identity, or $1$.  In other words, we are making every element commute.\footnote{Don't be confused by the difference in notation between $1$ for the identity and $0$ for the identity.  We are speaking in the abstract here.  Contracting $F$ ``to the identity" simply means that we are making everything commute.}  And if every element commutes, we are ``Abelianizing" the non-Abelian group $\pi_1$.  

For example the fundamental group of the figure-eight was the non-Abelian group with two generators.  If we call this group $G$ and the commutator subgroup $F$, then we are effectually taking everything that doesn't commute in $\pi_1$ and making it commute. What will be left is an Abelian group with two generators, which is naturally written as $\mathbb{Z}\oplus \mathbb{Z}$.  

More generally, for any fundamental group $\pi_1(K)$ and its commutator subgroup $F_{\pi_1(K)}$, we have the relation
\begin{eqnarray}
\pi_1(K) / F_{\pi_1(K)} = H_1(K)
\end{eqnarray}

We have been extremely cursory in this section.  If you don't feel confident about the ideas we have discussed here, don't worry.  We will readdress everything again in greater detail when we need it.  We are only introducing these ideas here for completion.  

\subsection{The Euler Characteristic and Betti Numbers}
\label{sec:theeulercharacteristicandbettinumbers}

There are two final ideas we need to discuss in this section, and we will do so very briefly.  Before bringing homology into focus here, we first introduce a related idea, the Euler characteristic of a space.  

For any simplicial complex $K$, with simplexes $\sigma_n$, we can define the number function of simplexes $N$ such that $N(\sigma_n)$ equals the number of $n$-complexes $K$.  We can then define the \bf Euler Characteristic\rm, also called the \bf Euler Number\rm, of the space as
\begin{eqnarray}
\chi(K) = \sum_{i=0}^{\infty} (-1)^{i} N(\sigma_i) 
\end{eqnarray}

For example consider the filled in triangle
\begin{center}
\includegraphics[scale=.6]{directed2simplex.PNG}
\end{center}
there are $N(\sigma_0)=3$ zero-simplexes, $N(\sigma_1)=3$ one-simplexes, and $N(\sigma_2) = 1$ two-simplexes.  So, the Euler characteristic is
\begin{eqnarray}
\chi(\rm triangle\it) &=& \sum_{i=0}^{\infty} (-1)^iN(\sigma_i) \nolabel \\
 &=& N(\sigma_0)-N(\sigma_1)+N(\sigma_2) - N(\sigma_3) + N(\sigma_4)- N(\sigma_5)+\cdots \nolabel \\
&=& 3-3+1 -0+0-0 +\cdots \nolabel \\
&=& 1
\end{eqnarray}

How you triangulate a given space doesn't affect its Euler characteristic.  The triangle above is a triangulation of a disk.  We could also triangulate the disk as
\begin{center}
\includegraphics[scale=.6]{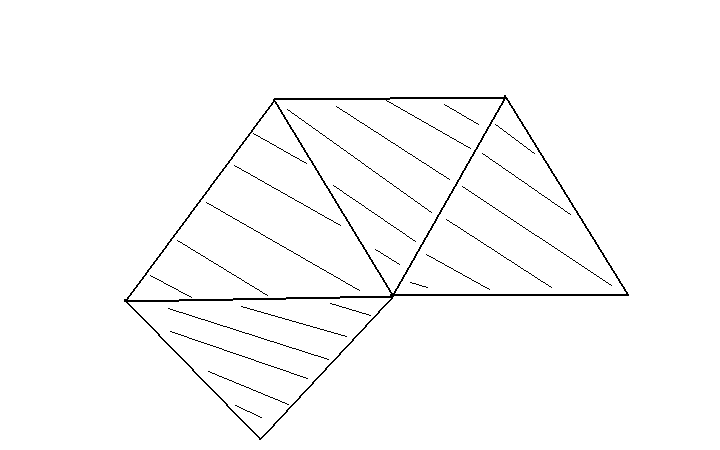}
\end{center}
Now,
\begin{eqnarray}
\chi(\rm triangle\it) &=& N(\sigma_0)-N(\sigma_1)+N(\sigma_2) - \cdots \nolabel \\
&=& 6- 9 + 4 \nolabel \\
&=& 1
\end{eqnarray}

Or, a triangulation of a circle:
\begin{center}
\includegraphics[scale=.4]{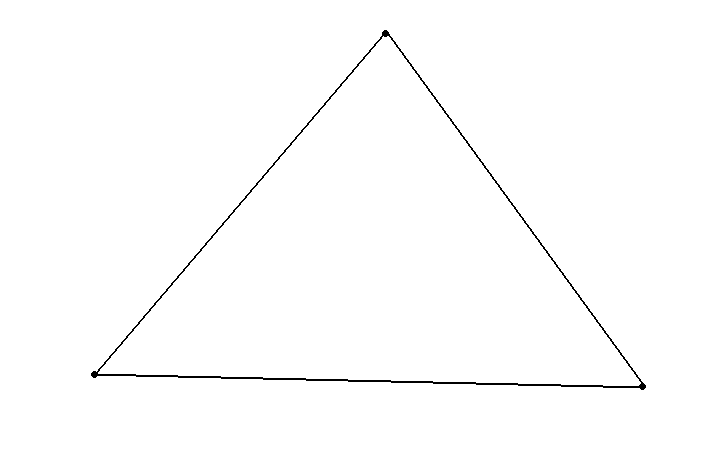}
\end{center}
will be
\begin{eqnarray}
\chi(S^1) = 3-3 =0
\end{eqnarray}
Or, another triangulation of a circle
\begin{center}
\includegraphics[scale=.5]{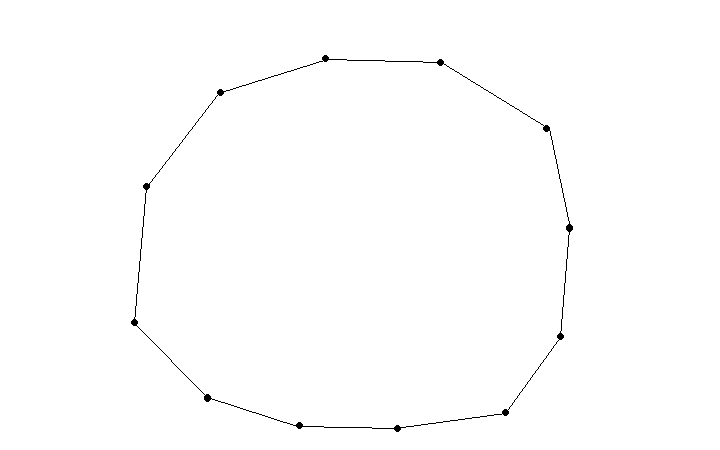}
\end{center}
will give
\begin{eqnarray}
\chi(S^1) = 12 - 12 = 0
\end{eqnarray}

It turns out that the Euler characteristic of a space, which is inherent to the space itself and not to how you triangulate it (in other words, the Euler characteristic of a disk is \it always \rm $1$, and the Euler characteristic of a circle is \it always \rm $0$), is a topological invariant.  

And, it turns out that the Euler characteristic is tied to the homology groups.  To see this relationship, we define the \bf Betti Number \rm $b_n(K)$ of a space $K$ as the \it dimension \rm of the $n^{th}$ homology group.  In other words, $b_n(K)$ is the rank of the free Abelian group $H_n(K)$.  In other, other words, the Betti number $b_n(K)$ is the number of topologically non-trivial $n$-dimensional spaces in $K$ ($b_1$ is the number of non-trivial loops, $b_2$ is the number of non-trivial $S^2$'s, etc.).  

For example, in the examples in section \ref{sec:examplesofhomologygroups}, we can make the following table
\begin{center}
\begin{tabular}{|c||c|c|c||c|c|c|}\hline  & $H_0$ & $H_1$ & $H_2$ & $b_0$ & $b_1$ & $b_2$ \\\hline \hline Point & $\mathbb{Z}$ & 0 & 0 & 1 & 0 & 0 \\\hline Disk & $\mathbb{Z}$ & 0 & 0 & 1 & 0 & 0 \\\hline Circle & $\mathbb{Z}$ & $\mathbb{Z}$ & 0 & 1 & 1 & 0 \\\hline Sphere & $\mathbb{Z}$ & 0 & $\mathbb{Z}$ & 1 & 0 & 1 \\\hline M\"obius Strip & $\mathbb{Z}$ & $\mathbb{Z}$ & 0 & 1 & 1 & 0 \\\hline Torus & $\mathbb{Z}$ & $\mathbb{Z}\oplus \mathbb{Z}$ & $\mathbb{Z}$ & 1 & 2 & 1 \\\hline Cylinder & $\mathbb{Z}$ & $\mathbb{Z}$ & 0 & 1 & 1 & 0 \\\hline Figure Eight & $\mathbb{Z}$ & $\mathbb{Z}\oplus \mathbb{Z}$ & 0 & 1 & 2 & 0 \\\hline \end{tabular}
\end{center}

The relationship between the homology groups and the Euler characteristic is given by the \bf Euler-Poincar\'e Theorem\rm, which states
\begin{eqnarray}
\chi(K) = \sum_{i=0}^{\infty} (-1)^ib_i(K) \label{eq:firstequationofeulernumber}
\end{eqnarray}
In other words, the Euler characteristic is equal to both the alternating sum of the number of simplexes, and the alternating sum of the Betti numbers.

While there is drastically more we could say about homology, we will stop here for now.  We will address these ideas again later in this series when we have more mathematical ``machinery" to work with.  For now we trust you have the basic ideas of what homology \it is\rm, and we move on to our final topological idea, cohomology.  

\section{Cohomology}
\label{sec:cohomology}

\subsection{Introduction to Cohomology}

So far in this chapter we have discussed homotopy and homology.  We saw that they effectively ``looked for" the same things - $n$-dimensional holes.  The difference between them is the particular group structure they assign to the holes they find.  Homotopy assigned (generally) non-Abelian groups, while homology assigned free Abelian groups.  

We discussed several examples of both homotopy groups (especially fundamental groups) and homology groups for a variety of simple spaces.  As we mentioned several times above, calculating general homotopy and homology groups can be extraordinary difficult.  Imagine trying to triangulate or picture loops in an $11$-dimensional space!  There are a number of tools and techniques which allow for the calculation of homology and homotopy for more complicated (and interesting) spaces, but we don't have the math we need to consider them at this point.  Our only goal here has been to introduce the basic ideas.  

But, in introducing these basic ideas, it has no doubt crossed your mind more than once that what we have been doing in this chapter seems to have little (if any) correlation to what we did previously with differentiable manifolds and all of the various operations on them.  This observation is true on the surface - there is no \it obvious \rm correlation between the topological considerations of this chapter and the geometric/analytic considerations of the previous.  

However, this is not ultimately true.  The ideas contained in this chapter and in the previous are actually related in extremely profound, rich, and deep ways.  We are now in a position to see this correlation.  Admittedly, nothing we have been doing with the (global) topological properties of the spaces we've considered in this chapter have looked like the geometric and analytic tools from above - we haven't been looking at forms or pullbacks or vector fields when considering loops and free Abelian groups.  This has been part of the reason that no connection is apparent.  

Cohomology provides the bridge between chapters \ref{sec:manifolds} and \ref{sec:topcons}.  We will see that, just as the name implies, \bf co\rm homology is indeed dual to homology, just as a \bf co\rm vector is dual to a vector.  This section will not only begin to look familiar, but also begin to tie together everything we have done so far. 

We will approach cohomology from a slightly different perspective than is standard in texts.  We will build it through direct analogy with homology, hopefully generating the ideas in a pedagogically logical way.  Once we have built the necessary motivation and (hopefully) intuition regarding what cohomology is, we will introduce the more rigorous ideas behind it, as well as discuss how exactly is the ``dual" of homology.  

\subsection{The Nilpotency of $d$ and $\partial$}

In the previous section, we saw that the boundary operator, which we defined by (\ref{eq:definitionofboundaryoperator}), is nilpotent (cf. (\ref{eq:nilpotencyofboundaryoperator})):
\begin{eqnarray}
\partial^2 = 0
\end{eqnarray}

Also, in section \ref{sec:exteriorderivatives} we also saw that the exterior derivative operator, which we defined in (\ref{eq:generalexpressionfortheexternalderivativeofapform}), is nilpotent as well (cf. (\ref{eq:nilpotentencyofd})):
\begin{eqnarray}
d^2 = 0
\end{eqnarray}

Recall that $\partial$ is a map 
\begin{eqnarray}
\partial : C_n(K) \longrightarrow C_{n-1}(K)
\end{eqnarray}
And that $d$ is a map
\begin{eqnarray}
d : \Lambda^nT_p\mathcal{M} \longrightarrow \Lambda^{n+1}T_p\mathcal{M}
\end{eqnarray}
In both cases we have a nilpotent operator that moves us from one dimensionality to another.  

With the boundary operator $\partial$, we considered two types of $n$-chains $c$: those that were ``cycles", meaning that they had no boundaries: $\partial c = 0$ (denoted $Z_n(K)$), and those that were ``boundaries", meaning that they could be written as the boundary of an $(n+1)$-object: $\partial d = c$ (denoted $B_n(K)$).  We then looked for all of the cycles that were not boundaries, $Z_n(K)/B_n(K)$.  

What made this work was specifically the nilpotency of $\partial$, which ensured that $B_n(K) \subseteq Z_n(K)$.  In other words, if a chain $c$ is a boundary, then $c=\partial d$, and therefore $\partial c = \partial (\partial d) = \partial^2d \equiv 0$.  We were looking at the things in $Z_n(K)$ ($\partial c = 0$) that are not in $B_n(K)$ ($\partial c = 0$, but not because $c=\partial d$).  

It is natural to do the exact same thing with $d$.  

\subsection{Cohomology Groups}
\label{sec:cohomologygroups}

We will build up the definition of a cohomology group by ``mere analogy" with homology groups.  We will then go back and explain what exactly they are/mean.  

Consider the set of all $n$-forms $\omega_n$ on some manifold $\mathcal{M}$.  We can define the \bf $n$-Chain Group\rm, which we denote $C^n(\mathcal{M})$, as the additive group over $\mathbb{R}$ with elements
\begin{eqnarray}
c = \sum_i c^i \omega_{n,i}
\end{eqnarray}
where $c^i \in \mathbb{R}$ and the $i$ subscript on $\omega_n$ is merely an index labeling a particular $n$-form.  The linearity and group construction follows exactly from section \ref{sec:groupstructuresimplexes}.  

Notice that the coefficients $c^i$ are now in $\mathbb{R}$, not in $\mathbb{Z}$ as with homology.  This is simply because it is more natural to use integers for the coefficients of simplexes and real numbers for the coefficients of forms (multi-dimensional functions).  We are using superscripts on the coefficients to emphasize that this will be the \it dual \rm construct to homology.  

Next we use the exterior derivative $d$ in a similar way to how we used $\partial$ with homology.  We say that an $n$-form $\omega_n$ on $\mathcal{M}$ is an \bf $n$-cycle \rm (also called an \bf closed \rm $n$-form) if it satisfies
\begin{eqnarray}
d\omega_n = 0
\end{eqnarray}
We denote the set of \it all \rm closed forms $Z^n(\mathcal{M})$.  We say an $n$-form $\omega_n$ on $\mathcal{M}$ is an \bf $n$-boundary \rm (also called an \bf exact \rm $n$-form) if it can be written \it globally\rm\footnote{The requirement that it be written \it globally \rm is vital.  This means that we can write $\omega_n = d\omega_{n-1}$ over the entire manifold - there is no point where it fails to hold.} as the exterior derivative of an $(n-1)$-form:
\begin{eqnarray}
\omega_n = d \omega_{n-1}
\end{eqnarray}
We denote the set of \it all \rm exact forms $B^n(\mathcal{M})$.  

Obviously, because of the nilpotency of $d$, any exact $n$-form is closed (or, any $n$-boundary is an $n$-cycle).  In other words, if $\omega_n = d\omega_{n-1}$, then
\begin{eqnarray}
d\omega_n = d(d\omega_{n-1}) = d^2 \omega_{n-1} \equiv 0
\end{eqnarray}
and therefore
\begin{eqnarray}
B^n(\mathcal{M}) \subseteq Z^n(\mathcal{M})
\end{eqnarray}
(compare this to (\ref{eq:bnisasubsetofzn})).  Again, we are using superscripts on $Z^n$ and $B^n$ intentionally to emphasize that these will form the \it dual \rm of the homology groups.  

Again, because $B^n$ and $Z^n$ are Abelian, $B^n$ is automatically a \it normal \rm subgroup of $Z^n$, and we have the a natural form of the $n^{th}$ \bf Cohomology Group\rm:
\begin{eqnarray}
H^n(\mathcal{M}) \equiv Z^n(\mathcal{M}) /B^n(\mathcal{M}) \label{eq:defofcohomology}
\end{eqnarray}

The definition of the homology groups was a bit easier, because the group structure was easier to visualize.  Making sense of this will require a bit of review.  

\subsection{Examples of Cohomology Groups}

As a quick preliminary note, we should point out that $H^n(\mathcal{M})$ is actually a \it vector space\rm, not really a group.  The factor group $Z^n/B^n$ is a set of equivalence classes on $Z^n$, which form a vector space.  We have assigned an additive group structure, but $H^n$ is still formally a vector space.  However the term ``Cohomology Group" is so embedded in the literature that any attempt to change it would only be confusing (and pointless).  We will therefore stick with convention and refer to ``cohomology groups".  We make this point just for clarity.  

Even though we don't have at this point a good intuitive understanding of what a cohomology group is, we will still calculate a few simple examples of them.  The next section will discuss in more detail what they mean.  

First consider $\mathcal{M} = \mathbb{R}$.  First we find $H^0(\mathbb{R})$.  The set $B^0(\mathbb{R})$ has no meaning because there are no $-1$ forms.  So $H^0(\mathbb{R}) = Z^0(\mathbb{R})$.  The set $Z^0(\mathbb{R})$ is the set of all $0$-forms $f$ that are closed, or $df = 0$.  The only way to have $df=0$ is if $f$ is constant.  And the set of all constant functions, or $0$-forms $f$ will be isomorphic to $\mathbb{R}$.  So, $H^0(\mathbb{R}) = Z^0(\mathbb{R})/B^0(\mathbb{R}) = \mathbb{R}/\{0\} = \mathbb{R}$. 

It turns out that, as with $H_0(K)$, as long as $\mathcal{M}$ is connected, $H^0(\mathcal{M})$ will be isomorphic to $\mathbb{R}$.  If $\mathcal{M}$ has $n$ connected components, then $H^0(\mathcal{M})$ will be isomorphic to $n$ copies of $\mathbb{R}$, or $\mathbb{R} \oplus \mathbb{R} \oplus \cdots \oplus \mathbb{R}$ ($n$ times).   

Next we find $H^1(\mathbb{R})$.  First, we know that because $\mathbb{R}$ is $1$-dimensional, any $1$-form on $\mathbb{R}$ is closed.  Furthermore, we can take any $1$-form $fdx$ and integrate it:
\begin{eqnarray}
F = \int f dx
\end{eqnarray}
so that
\begin{eqnarray}
f = dF
\end{eqnarray}
And therefore all $1$-forms are exact.  So, all one forms are closed and all one forms are exact.  Therefore
\begin{eqnarray}
H^1(\mathbb{R}) = Z^1(\mathbb{R})/B^1(\mathbb{R}) = \{0\}
\end{eqnarray}
the identity.  It is straightforward to show that $H^n(\mathbb{R}) = 0$ for $n\geq 2$.  

As another example, consider $\mathcal{M} = S^1$.  We can repeat a similar analysis as before and find that
\begin{eqnarray}
H^0(S^1) &=& \mathbb{R} \nolabel \\
H^1(S^1) &=& \mathbb{R} \nolabel \\
H^n(S^1) &=& \{0\},\qquad n\geq 2
\end{eqnarray}

And for $\mathcal{M} = S^2$:
\begin{eqnarray}
H^0(S^2) &=& \mathbb{R} \nolabel \\
H^1(S^2) &=& \{0\} \nolabel \\
H^2(S^2) &=& \mathbb{R} \nolabel \\
H^n(S^2) &=& \{0\},\qquad n\geq 3
\end{eqnarray}

And for $\mathcal{M} = S^n$:
\begin{eqnarray}
H^0(S^n) &=& \mathbb{R} \nolabel \\
H^n(S^n) &=& \mathbb{R} \nolabel \\
H^j(S^n) &=& \{0\},\qquad j \neq 0,n
\end{eqnarray}

And so on.  So once again, we see that the $n^{th}$ cohomology group ``detects" $n$-dimensional holes in exactly the same way as the $n^{th}$ homotopy group or the $n^{th}$ homology group.  Cohomology is quite remarkable - a purely (local) analytic statement - the types of forms that can exist on $\mathcal{M}$ - giving us (global) topological information about $\mathcal{M}$.

\subsection{Meaning of the $n^{th}$ Cohomology Group}
\label{sec:meaningofnthcohomologygroup}

The $n^{th}$ homology group $H_n(K)$ was defined as $H_n(K) \equiv Z_n(K) /B_n(K)$.  We could ``picture" this as the set of all closed cycles \it without \rm the set of all boundaries.  We thought of the factor group $Z_n(K)/B_n(K)$ as the set of everything in $Z_n(K)$ with everything in $B_n(K)$ ``contracted to the identity".  This picture doesn't work as well for cohomology, where $Z_n(\mathcal{M})$ is the space of all $n$-forms on $\mathcal{M}$.  So to understand this more clearly, we briefly review factor groups.  

Consider the Abelian group $\mathbb{Z} = \{0,\pm 1,\pm 2, \cdots\}$ with addition.  We can take the normal subgroup of all elements of $\mathbb{Z}$ of the form $4n$, where $n\in \mathbb{Z}$.  We also write this as $4\mathbb{Z} = \{0,\pm 4,\pm 8,\pm 12,\cdots\}$.  Obviously $4\mathbb{Z}$ obeys closure:
\begin{eqnarray}
4n + 4m = 4(m+n) = 4p\qquad p\in \mathbb{Z}
\end{eqnarray}
it is associative, there is an identity ($0$), and an inverse ($(4n) + (-4n) = 0$).  

We can find the factor group $\mathbb{Z}/4\mathbb{Z}$ by using the definition in \cite{Firstpaper}.  An arbitrary element of $\mathbb{Z}/4\mathbb{Z}$ is
\begin{eqnarray}
n(4\mathbb{Z}) = n\{0,\pm4,\pm 8,\pm12,\cdots\} = \{n+0,n\pm 4, n\pm 8, n\pm 12, \cdots\} \qquad n \in \mathbb{Z}
\end{eqnarray}
So we can write out:
\begin{eqnarray}
0(4\mathbb{Z}) &=& \{0+0,0\pm 4,0\pm 8,0\pm 12,\cdots\} = \{0,\pm 4,\pm 8,\pm 12,\cdots\} \nolabel \\
1(4\mathbb{Z}) &=& \{1+0,1\pm 4,1\pm 8,1\pm 12,\cdots\} = \{1,5,-3,9,-7,13,-11,\cdots\} \nolabel \\
2(4\mathbb{Z}) &=& \{2+0,2\pm 4,2\pm 8,2\pm 12,\cdots\} = \{2,-2,6,-6,10,-10,14,\cdots\}\nolabel \\
3(4\mathbb{Z}) &=& \{3+0,3\pm 4,3\pm 8,3\pm 12,\cdots\} = \{3,7,-1,11,-5,15,-9,\cdots\} \nolabel \\
4(4\mathbb{Z}) &=& \{4+0,4\pm 4,4\pm 8,4\pm 12,\cdots\} = \{4,0,8,-4,12,-8,16,\cdots\} \nolabel \\
etc.
\end{eqnarray}
Comparing the far left and right sides of these, we see that $0(4\mathbb{Z})$ and $4(4\mathbb{Z})$ are the same.  You can continue writing these out and you will find that any element of the form $(4n)(4\mathbb{Z})$ will be the same, where $n\in \mathbb{Z}$.  Furthermore, you will find that $1(4\mathbb{Z})$ and $5(4\mathbb{Z})$ will be the same, as will any element of the form $(4n+1)(4\mathbb{Z})$.  All elements of the form $(4n+2)(4\mathbb{Z})$ will be the same, and all elements of the form $(4n+3)(4\mathbb{Z})$ will be the same.  In other words, the factor group $\mathbb{Z}/4\mathbb{Z}$ breaks $\mathbb{Z}$ up into four equivalence classes.  There is the equivalence class represented by $0$, which we can denote $[0]$, by $1$ (denoted $[1]$), by $2$ (denoted $[2]$), and $3$ (denoted $[3]$).  This group is isomorphic to $\mathbb{Z}_4$, the integers $\mod\; 4$.  

You can write this out for any subgroup $n\mathbb{Z} \subseteq \mathbb{Z}$ and find that it breaks $\mathbb{Z}$ into $n$ equivalence classes.  

You could also take the group $\mathbb{R}$ under addition with the normal subgroup $\mathbb{Z}$, and the factor group $\mathbb{R}/\mathbb{Z}$ will break every element of $\mathbb{Z}$ into equivalence classes.  It would be instructive to write this out and see that there are an infinite number of equivalence classes in this case, all parameterized by the real numbers from the interval $[0,1)$.  An element $a \in [0,1)$ will be equivalent to any element $a+n$ where $n\in \mathbb{Z}$.  We can reword this to say that two elements $a,b\in \mathbb{Z}$ are equivalent in $\mathbb{R}/\mathbb{Z}$ if $a-b \in \mathbb{Z}$.  

Notice that the same can be said of $\mathbb{Z}/4\mathbb{Z}$ - two elements $a,b \in \mathbb{Z}$ are equivalent in $\mathbb{Z}/4\mathbb{Z}$ if $a-b \in 4\mathbb{Z}$.  

We can generalize this to arbitrary (additive) groups.  For any group $G$ with normal subgroup $H$, two elements $g_i,g_j \in G$ are equivalent in $G/H$ if $a-b \in H$.  This definition of $G/H$ is equivalent to the ones given before.  

This way of thinking about factor groups, where the subgroup $H$ defines an equivalence class of elements of $G$, and these equivalence classes are the elements of $G/H$, will be the most useful way of thinking about cohomology groups.  

So, looking at the definition of the $n^{th}$ cohomology group (\ref{eq:defofcohomology}), we have the group of all closed forms $Z^n(\mathcal{M})$ and the normal subgroup $B^n(\mathcal{M})$ of all exact forms.  And while an exact form is necessarily closed, a closed form is not necessarily exact.  This is what $H^n(\mathcal{M})$ is measuring - it is finding all forms that are closed and ``collapsing" the ones that are exact to the identity.  

Or in other words, $B^n(\mathcal{M})$ creates an equivalence class of forms in $Z^n(\mathcal{M})$ where two chains $c,c' \in Z^n(\mathcal{M})$ are equivalent in $Z^n(\mathcal{M})/B^n(\mathcal{M})$ if $(c-c') \in B^n(\mathcal{M})$.  We can speak in terms of forms (rather than linear combinations of forms, or chains), and say that two forms are equivalent in $Z^n(\mathcal{M})/B^n(\mathcal{M})$ if they differ by an exact form.  

So, $\omega_n$ and $\omega_n+d\omega'_{n-1}$ are equivalent\footnote{The prime does not indicate any sort of derivative - it is only notational.} in $Z^n(\mathcal{M})$.  An important consequence of this (as we will see) is that the exterior derivatives of two equivalent forms are equal:
\begin{eqnarray}
\omega''_n = \omega_n + d \omega'_{n-1} \qquad \Longrightarrow \qquad d\omega''_n &=& d(\omega_n + d\omega'_{n-1}) \nolabel \\
&=& d\omega_n + d^2\omega'_{n-1} \nolabel \\
&=& d\omega_n 
\end{eqnarray}
If two forms differ by an exact form in this way, they are called \bf cohomologous\rm.  

Incidentally, we could have thought of homology in this way.  The factor group
\begin{eqnarray}
H_n(K) = Z_n(K)/B_n(K)
\end{eqnarray}
divides $Z_n(K)$ up into similar equivalence classes.  Two $n$-simplexes $\sigma_n$ and $\sigma''_n$ are related such that their difference is an $(n+1)$-boundary
\begin{eqnarray}
\sigma''_n = \sigma_n + \partial_{n+1} \sigma'_{n+1}
\end{eqnarray}
then $\sigma''_n$ and $\sigma_n$ are \bf homologous\rm.  Obviously the boundary of $\sigma''_n$ and $\sigma_n$ are the same:
\begin{eqnarray}
\partial_n\sigma''_n &=& \partial_n(\sigma_n + \partial_n\partial_{n+1}\sigma'_{n+1}) \nolabel \\
&=& \partial_n\sigma_n \label{eq:boundariesarethesameforhomologycasedifferbyboundary}
\end{eqnarray}
However, unlike with the cohomology case, we can draw a picture of what is happening in (\ref{eq:boundariesarethesameforhomologycasedifferbyboundary}).  Let $\sigma''_n$ and $\sigma_n$ each be standard $2$-simplexes (triangles),
\begin{center}
\includegraphics[scale=.6]{2simplexnoboundary2.PNG}
\end{center}
and let $\sigma'_{n+1}$ be a standard $3$-simplex (a filled in tetrahedron).  Then, $\sigma''_n$ will again just be the triangle, while $\sigma_n+\partial_{n+1}\sigma'_{n+1}$ will be a triangle with the boundary of a tetrahedron attached:
\begin{center}
\includegraphics[scale=.5]{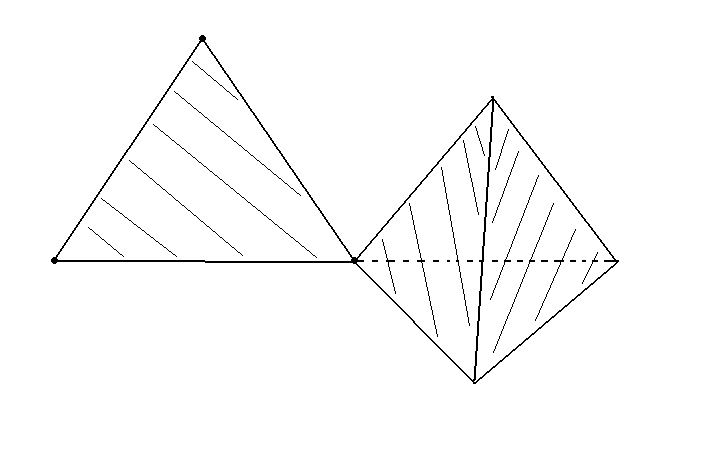}
\end{center}
Clearly the ``empty tetrahedron" attached to $\sigma_n+\partial_{n+1}\sigma'_{n+1}$ contributes nothing to its boundary, and the only boundary of $\sigma_n+\partial_{n+1}\sigma'_{n+1}$ is that of the triangle.  Therefore $\partial_n\sigma''_n = \partial_n\sigma_n$.  

We don't mean to get sidetracked with this rehashed discussion of homology.  We are merely trying to give some intuition regarding what we are doing with the exterior derivatives and cohomologous forms.  We know from the previous section that $H^n$ detects ``holes" in $\mathcal{M}$ - the question is `how?'  

To see \label{pagewherewediscusshowcohomologyworks} this more explicitly, consider the manifold $\mathcal{M} = \mathbb{R}^2$, and the form 
\begin{eqnarray}
\omega = {-y \over x^2+y^2} dx + {x \over x^2+y^2} dy
\end{eqnarray}
First of all
\begin{eqnarray}
d\omega &=& {\partial \over \partial y} \bigg({-y \over x^2+y^2}\bigg) dy dx + {\partial \over \partial x}\bigg({x \over x^2+y^2}\bigg) dxdy \nolabel \\
&=& -\bigg( {x^2+y^2 - 2y^2 \over (x^2+y^2)^2}\bigg) dydx + \bigg({x^2+y^2 - 2x^2 \over (x^2+y^2)^2} \bigg) dxdy \nolabel \\
&=& \bigg({x^2-y^2 \over (x^2+y^2)^2}\bigg) dxdy - \bigg({x^2-y^2 \over (x^2+y^2)^2}\bigg) dxdy \nolabel \\
&=& 0
\end{eqnarray}
(notice the order of the differentials in each line)  So $\omega$ is closed.  Also, consider the function
\begin{eqnarray}
F(x,y) = \tan^{-1}\bigg({y \over x}\bigg)
\end{eqnarray}
It is straightforward to show
\begin{eqnarray}
\omega = dF
\end{eqnarray}
So we are tempted to say that $\omega$ is also exact.  However $F$ is not defined on all of $\mathcal{M} = \mathbb{R}^2$, and we mentioned above that "exact" means that it must be \it globally \rm expressible as the exterior derivative of another form.  Here, it breaks down at $x=0$.  

But consider $\mathcal{M} = \mathbb{R}^2 - \{0\}$.  We can now repeat the same thing as before and we find that $F(x,y)$ is well-defined on all of $\mathcal{M}$.  So, we have a form that is closed on $\mathbb{R}^2$ and closed on $\mathbb{R}^2-\{0\}$, but is only exact on $\mathbb{R}^2-\{0\}$.  \it This \rm is a simple example of how $H^n(\mathcal{M})$ is able to detect holes in $\mathcal{M}$.  When the origin is included $\omega$ is not exact because it can't be globally written in terms of $F$.  When the origin is not included it can be written globally as $dF$.  So we see that the differential structure has given us topological information.  

Once again, there is \it much \rm more we could say about cohomology and how cohomology groups can be calculated, what they mean, etc.  However we postpone such discussions until later in this series.  For now, understanding the basic ideas - that the $n^{th}$ cohomology group $H^n(\mathcal{M})$ detects $n$-dimensional holes in $\mathcal{M}$ - is all that is necessary.  We will dive back into this once we have the necessary mathematical machinery.  

\subsection{Cohomology, Integration, and Stokes Theorem}

To introduce this section we point out an important difference between homology and cohomology.  Namely, notice that homology deals with \it spaces\rm, while cohomology relates to \it forms defined on spaces\rm.  Rewording this, cohomology relates to what you \it integrate\rm, while homology relates to what you integrate \it over\rm.  We will see that this fact is what makes homology and cohomology dual ideas.  

To understand this relationship we need to do some preliminary work.  We define the \bf standard $n$-simplex \rm as a simplex in $\mathbb{R}^n$ (\it not \rm $\mathbb{R}^{n+1}$ as before) with with the points 
\begin{eqnarray}
p_0 = \{0,0,\ldots,0\} \nolabel \\
p_1 = \{1,0,\ldots,0\} \nolabel \\
\vdots \qquad \qquad \nolabel \\
p_n = \{0,0,\ldots,1\} \label{eq:standardnsimplexvertices}
\end{eqnarray}
In other words, the standard $n$-simplex (denoted $\bar \sigma_n$) is
\begin{eqnarray}
\bar \sigma_n = \bigg\{ (x^1,x^2,\ldots, x^n) \in \mathbb{R}^n \big| x^i \geq 0 , \; \sum_i^n x^i \leq 1\bigg\} \label{eq:yetansasdfasd}
\end{eqnarray}
(We will generally drop the word ``standard" when talking about standard simplexes, and use the term ``simplex" and ``standard simplex" interchangeably) Compare this expression carefully to (\ref{eq:firstdefofsimplex}) especially the last ``summation" constraint at the end.  There the definition put a $0$-simplex at the number $1$ on the real line, the $1$-simplex ``diagonal" in $\mathbb{R}^2$, the $2$-simplex in $\mathbb{R}^3$, etc.  

This definition is slightly different - the $0$ simplex is defined in $\mathbb{R}^0$ - the point.  The $1$-simplex is defined in $\mathbb{R}^1$, the $2$-simplex is defined in $\mathbb{R}^2$, the $3$-simplex is defined in $\mathbb{R}^3$:
\begin{center}
\includegraphics[scale=.5]{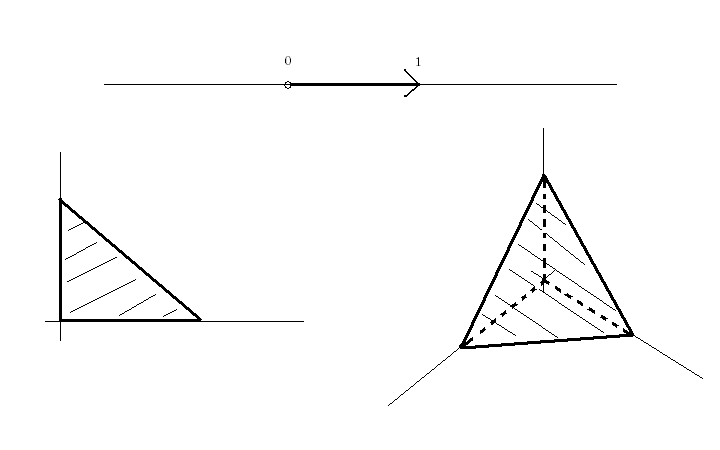}
\end{center}
Notice that in each case the general shape is the same - they are just situated in the Euclidian space differently.  

It is natural to think of these simplexes as differentiable manifolds.  They have a boundary, which we haven't discussed yet, but that won't be a problem for us.  The point of thinking of them as manifolds is that we spent the last chapter talking about how to define things on manifolds - vectors, forms, integration, etc.  We are only thinking of these are manifolds for continuity with that chapter - so that we can define integrals on them.  Obviously the coordinate functions for these simplexes are trivial because we have defined them in $\mathbb{R}^n$ - the coordinates are simply the coordinates we defined them in.  

It should be clear from the definition that $\bar \sigma_n$ is an \it orientable \rm manifold.  So it is natural to define the volume form (cf section \ref{sec:intdiffforms}) 
\begin{eqnarray}
\phi_V = h\; dx^1\wedge dx^2 \wedge \cdots \wedge dx^n
\end{eqnarray}
on them.  The function $h$ depends on the coordinates chosen.  We can then integrate the volume form over the simplex:
\begin{eqnarray}
\int_{\bar \sigma_n} \phi_V = \int_{\bar \sigma_n} h\; dx^1\wedge dx^2 \wedge \cdots \wedge dx^n
\end{eqnarray}
in the same way as as in section \ref{sec:intdiffforms}.  We used $\bar \sigma_n$ to describe both the manifold on the left hand side and the ``open subset of $\mathbb{R}^n$" on the right hand side simply because we \it defined \rm the manifold in $\mathbb{R}^n$ to begin with.  

Note that if we use Cartesian coordinates, the form of this integral will be (from equation (\ref{eq:yetansasdfasd}))
\begin{eqnarray}
\int_{\bar \sigma_n} h dx^1\wedge dx^2 \wedge \cdots \wedge dx^n &=& \int_0^1 dx^1 \int_0^{1-x^1} dx^2\int_0^{1-x^1-x^2} dx^3 \cdots \nolabel \\
& & \cdots \int_0^{1-\sum_{i=1}^{n-1} x^i}dx^n\;\; h \label{eq:generalformofintegraloverximplextoones}
\end{eqnarray}

So, we have $n$-simplexes and volume forms defined on them.  Now we want to ``put" these on some manifold $\mathcal{M}$.  We therefore define a map
\begin{eqnarray}
f:\bar \sigma_n \longrightarrow \mathcal{M}
\end{eqnarray}
Just as $\bar \sigma_n$ is called an $n$-simplex in $\mathbb{R}^n$, the image $f(\bar \sigma_n)\in \mathcal{M}$ is called the \bf singular $n$-simplex \rm in $\mathcal{M}$.  We denote this image 
\begin{eqnarray}
f(\bar \sigma_n) = s_n
\end{eqnarray}
So $s_n$ is a singular simplex in $\mathcal{M}$.  

Now, the standard operations on a simplex $\bar \sigma_n$ are defined on $\mathcal{M}$.  For example we can take the boundary of $s_n$ as
\begin{eqnarray}
\partial_n s_n = f(\partial_n \bar \sigma_n)
\end{eqnarray}
Obviously we again have $\partial_n(\partial_{n+1}s_{n+1}) = 0$.  

In this way, we can define chains, cycles, and boundaries in the exact same way as in the previous section, allowing us to define the \bf singular homology groups\rm, which are exactly isomorphic to the simplicial homology groups, and we therefore won't pursue them further (remember we are not really interested in homology in this section - we are doing all of this merely to prepare us for more discussion of cohomology).  

What this construction allows us to do is define the integration of some form $\omega_n$ on $\mathcal{M}$ over a chain $s_n$ on $\mathcal{M}$:
\begin{eqnarray}
\int_{s_n} \omega_n = \int_{\bar \sigma_n} f^{\star}( \omega_n)
\end{eqnarray}
where $f(\bar \sigma_n) = s_n$ and $f^{\star}$ is the pullback of $f$ (cf section \ref{sec:pullbackdiff}).  For a general $n$-chain $c = \sum_i c_i s_{n,i}$, we write
\begin{eqnarray}
\int_c \omega_n = \sum_i c_i \int_{s_{n,i}} \omega_n = \sum_ic_i \int_{\bar \sigma_{n,i}} f^{\star}(\omega_n)
\end{eqnarray}

Then, using (\ref{eq:generalformofintegraloverximplextoones}) to write out the general form of the integral, we can write this as
\begin{eqnarray}
\int_{\bar \sigma_n}\omega_n= \int_{\bar \sigma_n} f^{\star}(\omega_n) &=&  \int_0^1 dx^1 \int_0^{1-x^1} dx^2\int_0^{1-x^1-x^2} dx^3 \cdots \nolabel \\
& & \cdots \int_0^{1-\sum_{i=1}^{n-1}x^i}dx^n\;\; f^{\star}(\omega_n) \nolabel \\
&=& \int_0^1 dx^1 \int_0^{1-x^1} dx^2\int_0^{1-x^1-x^2} dx^3 \cdots \nolabel \\
& & \cdots \int_0^{1-\sum_{i=1}^{n-1}x^i}dx^n\;\; \psi_n
\end{eqnarray}
where $\psi_n = f^{\star}(\omega_n)$ is merely the $n$-form $\omega_n$ pulled back to the subset of $\mathbb{R}^n$ where the standard simplex $\bar \sigma_n$ sits.  

Now let's consider the situation in which $\psi_n$ is exact - it can be written as the exterior derivative of an $(n-1)$-form $\alpha_{n-1}$:
\begin{eqnarray}
\psi_n = d \alpha_{n-1}
\end{eqnarray}
If we write $\alpha_{n-1}$ out in components, we have
\begin{eqnarray}
\alpha_{n-1} = \alpha_{i_1,i_2,\ldots,i_{n-1}} \; dx^{i_1}\wedge dx^{i_2} \wedge \cdots \wedge dx^{i_{n-1}}
\end{eqnarray}
However, due to antisymmetry $\alpha_{i_1,i_2,\ldots,i_{n-1}}$ will only have $n-1$ independent components.\footnote{Remember that $\psi_n$ is a form in $\mathbb{R}^n$ and is therefore a volume element.}  For notational simplicity, we therefore give it a single index $i$ that runs from $1$ to $n-1$, and write it as
\begin{eqnarray}
\alpha_{n-1} = \alpha_i dx^1\wedge \cdots \wedge \widehat{dx^i} \wedge \cdots \wedge dx^n \label{eq:cohomologysectiondefinitionofalphasubnminusone}
\end{eqnarray}
where the hat indicates that the form below it is omitted.  The exterior derivative of $\alpha_{n-1}$ is then
\begin{eqnarray}
d\alpha_{n-1} = {\partial \alpha_i \over \partial x^j} dx^j \wedge dx^1 \wedge \cdots \wedge \widehat{dx^i} \wedge \cdots \wedge dx^n \label{eq:omittingthexitermincohomologysection}
\end{eqnarray}
We are summing over $j$, but obviously only a single term will survive (when $i=j$) because of antisymmetry.  Nonetheless we leave the summation in for now.

So
\begin{eqnarray}
\int_{\bar \sigma_n} \omega_n &=& \int_{\bar \sigma_n} f^{\star}(\omega_n) = \int_{\bar \sigma_n} \psi_n = \int_{\bar \sigma_n} d \alpha_{n-1} \nolabel \\
&=& \int_{\bar \sigma_n} {\partial \alpha_i \over \partial x^j} dx^j \wedge dx^1 \wedge \cdots \wedge \widehat{dx^i} \wedge \cdots \wedge dx^n \nolabel \\
&=& (-1)^{n-1} \int_{\bar \sigma_n} {\partial \alpha_i \over \partial x^j} dx^1 \wedge \cdots \wedge \widehat{dx^i} \wedge \cdots \wedge dx^n \wedge dx^j \nolabel \\
&=& (-1)^{n-1} \int_{\bar \sigma_n} dx^1 \wedge \cdots \wedge \widehat{dx^i} \wedge \cdots \wedge dx^n \wedge dx^j \; {\partial \alpha_i \over \partial x^j} \nolabel \\
&=& (-1)^{n-1}\int_0^1 dx^1 \int_0^{1-x^1}dx^2 \cdots  \int_0^{1-\sum_{q=1}^{i-1}x^q} \widehat{dx^i} \cdots \nolabel \\
& & \cdots \int_0^{1-\sum_{k=1}^{n-1} x^k} dx^n \int_0^{1-\sum_{p=1}^nx^p}dx^j\; {\partial \alpha_i \over \partial x^j} \nolabel \\
&=& (-1)^{n-1} \int_0^1 dx^1 \int_0^{1-x^1} dx^2\cdots  \int_0^{1-\sum_{q=1}^{i-1}x^q} \widehat{dx^i} \cdots \int_0^{1-\sum_{k=1}^{n-1} x^k} dx^n\alpha_i\big|_{x^j=0}^{x^j=1-\sum_{p=1}^nx^p} \nolabel \\
&=& (-1)^{n-1} \int_0^1 dx^1 \int_0^{1-x^1} dx^2 \cdots \int_0^{1-\sum_{q=1}^{i-2}x^q} dx^{i-1} \int_0^{1-\sum_{s=1}^{i} x^s} x^{i+1} \cdots \nolabel \\
& & \cdots  \int_0^{1-\sum_{k=1}^{n-1} x^k} dx^n\alpha_i\big|_{x^i=0}^{x^i=1-\sum_{p=1}^nx^p} \nolabel \\
& & \label{eq:uglyequationforstokestheorem}
\end{eqnarray}
Notice that, after the discussion immediately following equation (\ref{eq:omittingthexitermincohomologysection}), we have written the bounds in the last line of (\ref{eq:uglyequationforstokestheorem}) in terms of $x^i$ instead of $x^j$ as in the previous lines.  

This equation is quite ugly, but we can make some sense of it.  First of all, notice that in each of the \it standard \rm simplexes drawn above, an $n$-dimensional simplex ($n>0$) consists of several $(n-1)$-dimensional simplexes touching the origin, a single $(n-1)$-dimensional simplex not touching the origin, and all of the $n$-dimensional space ``between" each of these filled in.  The $n=0$ case is special because there are no $-1$-dimensional simplexes, but the idea still holds.  For example, the one-simplex is a $0$-dimensional simplex touching the origin (located at the origin), another $0$-dimensional simplex at the point $1 \in \mathbb{R}$, and then the one-dimensional space between them filled in.  The two-simplex has two $1$-simplexes at the origin, one along the $x$-axis and one along the $y$-axis, a third $1$-simplex not touching the origin stretching from $(0,1)$ to $(1,0)$, and then the $2$-dimensional space between these three $1$-simplexes filled in.  

Also, notice that in each of the standard simplexes, if we set one of the vertex vectors (\ref{eq:standardnsimplexvertices}) equal to $0$, the result is an $n-1$-simplex.  For example, with the standard $2$-simplex, if we set $y=0$ for every point, we ``project" the $2$-simplex down to the $x$-axis, resulting in a $1$-simplex.  We could also have set $x=0$ resulting in a projection of the $2$-simplex to the $y$-axis.  We have three ways of projecting the $3$-simplex to a $2$-simplex: set $x=0$ projecting it to the $y,z$ plane, set $y=0$ projecting it to the $x,z$ plane, or set $z=0$ projecting it to the $x,y$ plane.  

So in (\ref{eq:uglyequationforstokestheorem}), notice that each term in the sum over $i$ is an integral over the entire $n$-dimensional simplex, but with the $x^i$ coordinate either set to $0$ or to one minus the sum of all of the other $x$'s.  If we look at the term where $x^i$ is set to $0$, notice that we have exactly the ``projection" from the $n$-simplex onto an $(n-1)$-simplex as in the previous paragraph.  Therefore (\ref{eq:uglyequationforstokestheorem}) is a sum over every $(n-1)$-simplex forming the boundary of the $n$-simplex $\bar \sigma_n$ the integral was originally over.  

The term with $x^i$ set to one minus the sum of all of the other $x$'s on the other hand will be the $(n-1)$-simplex that is not touching the origin.  

So, each term in this sum over $i$ will result in each $(n-1)$-dimensional simplex forming the boundary of $\bar \sigma_n$ being integrated over.  The $(n-1)$-simplexes that touch the origin correspond to one of the $x$'s being set to $0$.  This means that The $\alpha_i$ component of $\alpha_{n-1}$ (which has no $x^i$ component, cf (\ref{eq:cohomologysectiondefinitionofalphasubnminusone})) will be zero on the $(n-1)$-simplex projected onto by setting $x^i=0$.  Then, we integrate \it each \rm component $\alpha_i$ over the $(n-1)$-dimensional simplex that is not attached to the origin, each of which may have a non-zero component on \it that \rm face.  

In other words, the integral $\int_{\bar \sigma_n}d\alpha_{n-1}$ is equal to integrating $\alpha_{n-1}$ over the \it boundary \rm of $\bar \sigma_n$.  This gives us the remarkable result
\begin{eqnarray}
\int_{\bar \sigma_n} d \alpha_{n-1} = \int_{\partial \bar \sigma_n} \alpha_{n-1} \label{eq:stokestheorem}
\end{eqnarray}
Notice that the integral on the left is an $n$-form integrated over an $n$-dimensional space, whereas the right is an $(n-1)$-form integrated over an $(n-1)$-dimensional space, so the integral makes sense on both sides.  Equation (\ref{eq:stokestheorem}) is called \bf Stokes Theorem\rm.  It is an extraordinarily powerful result, and we will see soon that it is actually very familiar to a physics student.  

To illustrate (\ref{eq:stokestheorem}) we consider a two-dimensional example.  We'll work in Cartesian coordiantes where $x^1=x$ and $x^2=y$.  We will integrate over the $2$-simplex
\begin{eqnarray}
\bar \sigma_2 = \{ x,y \in \mathbb{R}^2 \big| x,y \geq 0,\; x+y \leq 1\}
\end{eqnarray}
We then take $\alpha_1$ to be\footnote{Note the bad notation - $\alpha_1$ is both the name of the \it one \rm form as well as the $1$ component of the one form.  The context will make it obvious which we are talking about so this won't cause a problem.  We just wanted to bring it to your attention.}
\begin{eqnarray}
\alpha_1 = \alpha_1 dx^2  + \alpha_2 dx^1 \label{eq:definitionofalphaforstokestheoremexample}
\end{eqnarray}
The exterior derivative will be
\begin{eqnarray}
d\alpha_1 &=& {\partial \alpha_i \over \partial x^j} dx^j \wedge dx^i \nolabel \\
&=& {\partial \alpha_1 \over \partial x^2} dx^2 \wedge dx^1 + {\partial \alpha_2 \over \partial x^1} dx^1\wedge dx^2 \nolabel \\
&=& \bigg( {\partial \alpha_2 \over \partial x^1} - {\partial \alpha_1 \over \partial x^2}\bigg) dx^1 \wedge dx^2 \label{eq:oneformexampleforstokestheorem}
\end{eqnarray}
Our integral will then be (using the first line of (\ref{eq:oneformexampleforstokestheorem}))
\begin{eqnarray}
\int_{\bar \sigma_2} d \alpha_1 &=&  \int_{\bar \sigma_n} {\partial \alpha_i \over \partial x^j} dx^j \wedge dx^i \nolabel \\
&=& -\int_{\bar \sigma_n} {\partial \alpha_i \over \partial x^j} dx^i \wedge dx^j \nolabel \\
&=& -\int_{\bar \sigma_n} dx^i \wedge dx^j {\partial \alpha_i \over \partial x^j} \nolabel \\
&=& -\int_0^1 dx^2\int_0^{1-x^2} dx^1{\partial \alpha_1 \over \partial x^1} - \int_0^1 dx^1 \int_0^{1-x^1} dx^2{\partial \alpha_2 \over \partial x^2} \nolabel \\
&=& -\int_0^1 dx^2 \alpha_1\big|_{x^1=0}^{x^1=1-x^2} - \int_0^1dx^1 \alpha_2\big|_{x^2=0}^{x^2=1-x^1} \nolabel \\
&=& \int_0^1 dx^2 \alpha_1(x^1=0) + \int_0^1 dx^1 \alpha_2(x^2=0) \nolabel \\
& & -\int_0^1 dx^2\alpha_1(x^1 = 1-x^2) - \int_0^1 dx^1 \alpha_2(x^2 = 1-x^1)
\end{eqnarray}
The first term is integrating $\alpha_1$ up the $y$-axis at $x=0$ and the second is integrating $\alpha_2$ across the $x$-axis with $y=0$.  Notice in (\ref{eq:definitionofalphaforstokestheoremexample}) that $\alpha_1$ has no $x$-component and $\alpha_2$ has no $y$-component, and therefore there is no need to integrate $\alpha_2$ along the $y$-axis or $\alpha_1$ across the $x$-axis.  However, both $\alpha_1$ and $\alpha_2$ may be non-zero on the $1$-simplex not touching the origin (from $(1,0)$ to $(0,1)$), and therefore we integrate both $\alpha_1$ and $\alpha_2$ across this line in the third and fourth term.  The minus signs are merely to factor in the directed-ness of the simplex. 

So, Stokes theorem has give us a relationship between the integral of an exact form $\omega_n = d\alpha_{n-1}$ over some space $\bar \sigma_n$ and the integral of $\alpha_{n-1}$ over the boundary $\partial \bar \sigma_n$.  

Furthermore, as we said previously, we can map the standard simplexes into a manifold $\mathcal{M}$ to integrate over a neighborhood of $\mathcal{M}$, and so by using everything we know about integration, we can generalize this to integrals of forms on arbitrary manifolds.  

To see how Stokes theorem relates to standard physics, consider the covector $\boldsymbol \omega = \phi_x dx + \phi_y dy + \phi_z dz \; \in \mathbb{R}^3$.  We found the exterior derivative of $\boldsymbol \omega$ in equation (\ref{eq:firstcurl}), and recognized it as the curl of $\boldsymbol \omega$.  If we want to integrate $d\boldsymbol \omega = \boldsymbol \nabla \times \boldsymbol \omega$ over some two dimensional area $A$ with boundary $\partial A = C$, Stokes theorem gives us
\begin{eqnarray}
\int_A \boldsymbol \nabla \times \boldsymbol \omega \cdot d \boldsymbol S = \oint_C \boldsymbol \omega \cdot d \boldsymbol S
\end{eqnarray}
This is a standard result in $E\&M$ and is what physicists typically mean when they refer to ``Stokes theorem".  

Furthermore, if we imitate what we did in (\ref{eq:firstdivergence}), where 
\begin{eqnarray}
\boldsymbol \phi_2 = \phi_x\;dy\wedge dz + \phi_y\;dz \wedge dx + \phi_z \;dx \wedge dy
\end{eqnarray}
and
\begin{eqnarray}
d\boldsymbol \phi_2 = \bigg({\partial \phi_x \over \partial x} + {\partial \phi_y \over \partial y} + {\partial \phi_z \over \partial z}\bigg)\; dx \wedge dy \wedge dz
\end{eqnarray}
which we recognized as the divergence of $\boldsymbol \phi_2$, or $\boldsymbol \nabla \cdot \boldsymbol \phi_2$, then we can integrate this over some three dimensional volume $V$ with boundary $\partial V = S$.  Stokes theorem gives us
\begin{eqnarray}
\int_V \boldsymbol \nabla \cdot \boldsymbol \phi_2 d V = \oint_S \boldsymbol \phi_2 \cdot d\boldsymbol S \label{eq:generalgausslaw}
\end{eqnarray}
Equation (\ref{eq:generalgausslaw}) is another standard result in $E\&M$.  For example, consider the Maxwell equation
\begin{eqnarray}
\boldsymbol \nabla \cdot \boldsymbol E = {\rho \over \epsilon_0} \label{eq:maxwell1andff}
\end{eqnarray}
where $\rho$ is the charge density and the constant $\epsilon_0$ is the permittivity of free space.  We can integrate both sides over some volume $V$:
\begin{eqnarray}
\int_V \boldsymbol \nabla \cdot \boldsymbol E  = {1 \over \epsilon_0}\int_V \rho
\end{eqnarray}
The right side is the integral of the volume charge density over a volume and is therefore simple the total charge inside the volume, or $Q$.  We can apply Stokes theorem to the left side as in (\ref{eq:generalgausslaw}), getting
\begin{eqnarray}
\oint_S \boldsymbol E \cdot d \boldsymbol S = {Q \over \epsilon_0}
\end{eqnarray}
So the total electric charge \it inside \rm of some space is equal to the integral of the electric field across the \it surface \rm of the space.  This extraordinarily powerful result is referred to in physics literature as \bf Gauss' Law\rm.  

So, both Gauss' Law (which is typically used to do calculations with electric $\boldsymbol E$ fields) and (the physical version of) Stokes Law (which is typically used to do calculations with magnetic $\boldsymbol B$ fields) are both consequences of the same general law - (the mathematical version of) Stokes theorem, (\ref{eq:stokestheorem}).  

\subsection{The Duality of Homology and Cohomology}
\label{sec:dualityofhomologyandcohomology}

As we have stressed previously, homology relates to a space while cohomology relates to forms.  Put another way, homology relates to what you integrate \it over\rm, while cohomology relates to what you \it integrate\rm.  

As we discussed in section \ref{sec:dualspace}, given some space of objects $V$, the \it dual space \rm is the set of objects $V^{\star}$ which map elements of $V$ to $\mathbb{R}$.  We can naturally define this exactly within the context of homology and cohomology.  Namely, consider a form $\omega$.  Obviously $\omega$ is not generally an element of $\mathbb{R}$.  But forms are things we integrate, and when we integrate $\omega$ over some space $\mathcal{M}$, the result is in $\mathbb{R}$.  In other words, \it spaces \rm are dual to \it forms \rm through integration.  

To make this more precise, consider the manifold $\mathcal{M}$ with chain group $C_n(\mathcal{M})$ and set of $n$-forms $\Lambda^n\mathcal{M}$.  If $c \in C_n(\mathcal{M})$ and $\omega \in \Lambda^n\mathcal{M}$, then we define the inner product between $c$ and $\omega$ as
\begin{eqnarray}
\langle c | \omega \rangle = \int_c \omega \in \mathbb{R}
\end{eqnarray}

This definition is linear in both $c$ and $\omega$:
\begin{eqnarray}
\langle c|\omega + \omega'\rangle &=& \int_c (\omega + \omega') = \int_c \omega + \int_c \omega' \nolabel \\
\langle c+ c'|\omega\rangle &=& \int_{c+c'} \omega = \int_c \omega + \int_{c'} \omega
\end{eqnarray}

We can write Stokes theorem in the compact form
\begin{eqnarray}
\langle \partial c|\omega\rangle = \langle c | d\omega\rangle \label{eq:stokestheoremcompactform}
\end{eqnarray}

Then consider the operator $U_{d}$ which acts on a vector $|\omega\rangle$ as 
\begin{eqnarray}
U_{d} |\omega\rangle = |d \omega\rangle
\end{eqnarray}
Obviously it will be the conjugate of this that acts on the dual space.  So,
\begin{eqnarray}
\langle c|d\omega\rangle = \langle c|U_d|\omega\rangle = \langle \partial c|\omega \rangle 
\end{eqnarray}
So 
\begin{eqnarray}
\langle c|U_d = \langle \partial c| \qquad \Longrightarrow\qquad  (U_d)^{\dagger} |c\rangle = |\partial c\rangle
\end{eqnarray}
So
\begin{eqnarray}
(U_d)^{\dagger} = U_{\partial}
\end{eqnarray}
We will see this relationship in greater detail when we discuss Hodge Theory in the next chapter.  

Another obvious consequence of this duality is that
\begin{eqnarray}
\langle c | \omega\rangle = 0
\end{eqnarray}
if \it either \rm \\
\indent 1) $c \in B_n(\mathcal{M})$ and $\omega \in Z^n(\mathcal{M})$, or \\
\indent 2) $c \in Z_n(\mathcal{M})$ and $\omega \in B^n(\mathcal{M})$.

Also, recall that $H_n(\mathcal{M}) = Z_n(\mathcal{M})/B_n(\mathcal{M})$ and $H^n(\mathcal{M}) = Z^n(\mathcal{M})/B^n(\mathcal{M})$ both consist of equivalence classes of $Z_n(\mathcal{M})$ and $Z^n(\mathcal{M})$, respectively.  We can denote the set of all chains equivalent to $c$ by $[c]$, and the set of all closed forms equivalent to $\omega$ by $[\omega]$.  So, $[c]\in H_n(\mathcal{M})$ and $[\omega] \in H^n(\mathcal{M})$, and we have the natural inner product
\begin{eqnarray}
\langle [c]|[\omega]\rangle = \int_c\omega
\end{eqnarray}
And because of Stokes theorem, this inner product is well defined regardless of the choice of element in $[c]$ or $[\omega]$.  For example, if some other cycle $c''$ is equivalent to $c$ and hence in $[c]$, then $c$ and $c''$ differ by an exact form $\partial c'$:
\begin{eqnarray}
c'' = c + \partial c'
\end{eqnarray}
So the new dot product will be
\begin{eqnarray}
\int_{c+\partial c'} \omega &=& \int_c \omega + \int_{\partial c'}\omega \nolabel \\
&=& \int_c \omega + \int_{c'} d\omega \nolabel \\
&=& \int_c \omega
\end{eqnarray}
where we used Stokes theorem to get the second line and the fact that $\omega$ is closed to get the third.  Therefore, \it any \rm element of $[c]$ will give the same integral as $c$.  

Also, if $\omega''$ is equivalent to $\omega$, then 
\begin{eqnarray}
\omega'' = \omega + d \omega'
\end{eqnarray}
and
\begin{eqnarray}
\int_c \omega'' &=& \int_c (\omega + d\omega') \nolabel \\
&=& \int_c \omega + \int_{\partial c} \omega' \nolabel \\
&=& \int_c \omega
\end{eqnarray}
where we used the fact that $\partial c = 0$ (because $c \in Z_n(\mathcal{M})$ by definition).  

The above considerations show us that $H_n(\mathcal{M})$ is indeed the dual space to $H^n(\mathcal{M})$.  

\subsection{The Euler Characteristic and Betti Numbers}

As we have seen, the cohomology group ($H^n(\mathcal{M})$) is isomorphic to the homology group ($H_n(\mathcal{M})$), except $H_n$ is over $\mathbb{Z}$ and $H^n$ is over $\mathbb{R}$.  This similarity allows us to define the Betti numbers equivalently as
\begin{eqnarray}
b^n(\mathcal{M}) = dim(H^n(\mathcal{M})) = dim(H_n(\mathcal{M})) = b_n(\mathcal{M})
\end{eqnarray}
And therefore (following (\ref{eq:firstequationofeulernumber})) the Euler number is
\begin{eqnarray}
\chi(\mathcal{M}) = \sum_{i=0}^{\infty}(-1)^ib^i(\mathcal{M}) 
\end{eqnarray}
This equation is one of the most remarkable in all of mathematics.  The left hand side is a \it purely \rm topological statement about $\mathcal{M}$.  It relies entirely on global information about the qualitative shape of $\mathcal{M}$, and is completely independent of any geometrical information.  The right hand side, on the other hand, comes from purely \it analytic \rm and \it geometrical \rm statements about $\mathcal{M}$.  The basis of the cohomology groups (from which $b^n$ is defined) is the set of equivalence classes of $Z^n(\mathcal{M})$, which is the set of solutions to the \it differential equations \rm 
\begin{eqnarray}
d\omega = 0
\end{eqnarray}

This interplay between topology and analytic geometry provides one of the deepest, richest, and most powerful avenues for mathematics.  We will be diving much, much deeper into these types of relationships as we proceed though this series.  We will see that such considerations actually form the backbone of much of string theory and fundamental particle physics.  

\section{Concluding Thoughts on Algebraic Topology}

We have considered homotopy, homology, and cohomology.\footnote{The order we discussed these in was very deliberate.  We started with the conceptually easiest and went to the most conceptually difficult.  And, we started with the ``most topological" (read ``least geometry involved) and went to the most geometrical.}  We saw that each of these provide a way of calculating the various types of ``holes" that may appear in some space $\mathcal{M}$.  Homotopy simply told us what types of maps of $S^n$ can and cannot be contracted to a point, and in general formed non-Abelian groups.  This notion of ``what types of $S^n$ can be contracted" was the whole point of homotopy.  While this approach to classifying spaces is extremely easy to visualize, actually calculating homotopy groups is extremely difficult.\footnote{As we mentioned in the section on homotopy, there are a few very powerful methods of calculating homotopy groups, and we will discuss some of them later.  However, even with such ``machinery", such calculations remain notoriously difficult.}

Homology, like homotopy, also gives us information about the $n$-dimensional holes (places where $S^n$ cannot be contracted to a point), but there is a tradeoff.  Homology groups are much easier to calculate, but they provide less structure in that they are merely free Abelian groups, unlike the generally non-Abelian homotopy groups.  Also, the concept of what a homology group is is slightly more difficult to understand (the set of all things that could be boundaries not including the the things that actually are boundaries).  Fortunately the group structure we lost is not vital - often all we need to know is the number of holes - the added information in homotopy groups is not necessarily necessary.  Also, the use of simplexes in homology brought us a little closer to geometry, and in that sense a little familiarity.  

Cohomology in many ways provides exactly the same information as homology.  The only different in the information provided is that homology uses integers $\mathbb{Z}$ while cohomology uses the reals $\mathbb{R}$.  This difference isn't really substantial (especially for our purposes).  What makes cohomology interesting is \it how \rm it is calculated.  Homology was based on the space itself - this made it easier to visualize (though admittedly more difficult than homotopy), but less familiar to physicists.  It wasn't difficult to see directly from the definition how the topological information about $\mathcal{M}$ provided produced topological information about $\mathcal{M}$.  Knowledge of the simplicial complex (which is essentially topological information about $\mathcal{M}$) gave us a way of calculating the number and types of holes.  Topological information led to topological information.  

Cohomology, on the other hand, provided similar information as homology, but didn't require any topological information.  While \it how \rm this works wasn't obvious, (though we did provide an illustration starting on page \pageref{pagewherewediscusshowcohomologyworks}), the calculations involve knowing \it nothing \rm about the topology, but instead only analytic information about what types of forms make sense on $\mathcal{M}$.  This provides greater familiarity to a physicist (who is accustomed to the vector space structure of forms), and provides insights into the deep interplay between geometry and physics.  

\section{References and Further Reading}

The primary source for the section on homotopy was \cite{naberTop} and \cite{nash}, and the primary source for the sections on homology and cohomology were \cite{nakahara} and \cite{nash}.  For further reading in algebraic topology we recommend \cite{bredon}, \cite{croom}, \cite{frankel}, \cite{hatcher}, \cite{massey}, \cite{naberTop}, \cite{rotman}, and \cite{spanier}.  For introductions to general topology we recommend \cite{mendelson}, \cite{munkres}, and \cite{schwarz}.  

\chapter{Differential Geometry}
\label{sec:chapwithmet}

Now that we have given a brief introduction to some of the topological aspects of manifolds, we return to our exposition of the geometrical considerations necessary for mathematical particle physics.  

The basic idea of this chapter is the introduction to a differentiable manifold of a structure called a \bf metric\rm.  In section \ref{sec:concthoudiffman} we discussed several of the questions that still remained after that initial chapter on manifolds.  We currently have no way of telling the difference between a perfect sphere and an egg (and the topological considerations of the last chapter obviously don't help at all either).  Things like distance between points, angles, and the relationship between vectors and covectors are also out of our reach thus far.  

The introduction of a metric on $\mathcal{M}$ will provide explanations/answers to all of these things.  Whereas before we can stretch, twist, bend, etc. $\mathcal{M}$ into whatever (homeomorphic) shape we want, adding a metric essentially makes $\mathcal{M}$ rigid.  However it doesn't make it \it completely \rm rigid.  The metric limits us from being able to stretch $\mathcal{M}$ however we want to being able to only stretch $\mathcal{M}$ in directions perpendicular to itself.  A good analogy would be a piece of paper.  If you lay the paper down on a table, you can't stretch it \it along \rm the direction of the paper (it would tear).  However, you can bend the paper into a cylinder, etc.  This involves only bending that is \it not \rm along the direction of the manifold.  The paper is rigid, but not completely rigid.  

Another way of thinking about this is to take the same piece of paper and draw two intersecting lines.  You can measure the angle between those lines easily.  Then you can bend the paper however you want, and (assuming you don't tear it), the angle between those lines will be the same.  

These two equivalent ideas are the point of a metric.  We are adding a structure that prevents us from deforming $\mathcal{M}$ along its own dimensions, and/or we are only allowing deformations of $\mathcal{M}$ that do \it not \rm change the angle between two arbitrary lines.  The following sections will make this precise.  

\section{Metrics}
\label{sec:metrics}

A \bf Riemannian Metric \rm $g_{ij}$ on some differentiable manifold $\mathcal{M}$ is a type $(0,2)$ real symmetric tensor field (no upper indices, two lower indices, cf page \pageref{pagewherewetalkabouttensorofvariousranks}) defined at each point $p \in \mathcal{M}$.  For $p \in U_i \subseteq \mathcal{M}$ with coordinate functions $\boldsymbol\phi\it^{(i)} = \bf x\it$, we write the metric as
\begin{eqnarray}
g(p) = g_{ij}(p) dx^i \otimes dx^j \label{eq:firstmetric}
\end{eqnarray}
where the $dx^i$ are a coframe (basis covectors for the cotangent space) for $\Lambda^1 T_p\mathcal{M}$.  The metric (\ref{eq:firstmetric}) will act on tensor products of the basis for $T_p\mathcal{M}$ in the natural way:
\begin{eqnarray}
\bigg[ g_{ij}(p) dx^i \otimes dx^j\bigg] \bigg[ \bigg({\partial \over \partial x^a}\bigg)\otimes \bigg({\partial \over \partial x^b}\bigg)\bigg] &=& g_{ij}(p) \big[ \delta^i_a \otimes \delta^j_b\big] \nolabel \\
&=& g_{ab}(p)
\end{eqnarray}
More generally, for vectors 
\begin{eqnarray}
\bf v\it^{(1)} &=& v^{(1),i} {\partial \over \partial x^i} \nolabel \\
\bf v\it^{(2)} &=& v^{(2),i} {\partial \over \partial x^i} 
\end{eqnarray}
at $p \in U_i \subseteq \mathcal{M}$, we have
\begin{eqnarray}
g(p)(\bf v\it^{(1)},\bf v\it^{(2)}) &=& \bigg[g_{ij}(p) dx^i \otimes dx^j\bigg] \bigg[v^{(1),a} {\partial \over \partial x^a} \otimes v^{(2),b}{\partial \over \partial x^b}\bigg]  \nolabel \\
&=& g_{ij}(p) v^{(1),a} v^{(2),b} \big[ \delta^i_a \otimes \delta^j_b\big] \nolabel \\
&=& g_{ab}(p) v^{(1),a}v^{(2),b} \label{eq:generalformofmetricinnerproduct}
\end{eqnarray}

We will drop the $p$ from the notation for the metric unless doing so causes confusion.  

As stated above, we demand that $g_{ij}$ be symmetric:
\begin{eqnarray}
g_{ij} = g_{ji}
\end{eqnarray}

Mathematicians make the distinction between a true Riemannian metric and a \bf pseudo-Riemannian metric \rm by requiring that 
\begin{eqnarray}
g(\bf v\it^{(i)}, \bf v\it^{(j)}) \geq 0
\end{eqnarray}
(where the equality only holds when either $\bf v \it = 0$)  for all vectors at all points for a metric to be ``Riemannian".  This condition is relaxed for a pseudo-Riemannian metric - we only require
\begin{eqnarray}
g(\bf v\it^{(i)},\bf v\it^{(j)}) = 0 \quad \forall \quad  \bf v\it^{(i)} \in T_p\mathcal{M} \qquad \Longrightarrow \qquad \bf v\it^{(j)} = 0
\end{eqnarray}
Most physics is done on pseudo-Riemannian manifolds and we will therefore focus on them, though we will talk about both.  

\subsection{Metrics, Vectors, and Covectors}

We said on page \pageref{saidnodotproductbetweenvectors} that there is no such thing as a dot product between two vectors.  This was only somewhat true.  If we have a manifold $\mathcal{M}$ with a metric $g$, then we can take any two vectors and map them to $\mathbb{R}$.  In this sense, we have an inner product between two vectors:
\begin{eqnarray}
g(\bf v\it^{(i)},\bf v\it^{(j)}) \in \mathbb{R}
\end{eqnarray}

However, an equivalent (and better) way of thinking about this is that the metric provides an isomorphism between $T_p\mathcal{M}$ and $\Lambda^1T_p\mathcal{M}$.  To see this consider a vector $\bf v\it$ with components $v^{j}$ and a covectors $\bf u\it$ with components $u_{j}$.  The inner product between them (as defined in chapter $2$), is given by
\begin{eqnarray}
\big( u_{i} dx^i\big)\bigg(v^j {\partial \over \partial x^j}\bigg) &=& u_iv^j dx^i {\partial \over \partial x^j} \nolabel \\
&=& u_i v^j \delta^i_j \nolabel \\
&=& u_iv^i
\end{eqnarray}
So the inner product between them is the sum of the products of the components.  

Now, for two vectors $\bf v\it^{(1)}$ and $\bf v\it^{(2)}$ (with components $v^{(1),i}$ and $v^{(2),i}$ respectively), the inner product using the metric will be (using (\ref{eq:generalformofmetricinnerproduct}))
\begin{eqnarray}
g(\bf v\it^{(i)},\bf v\it^{(j)})  = g_{ij}v^{(1),i}v^{(2),j}
\end{eqnarray}
Now take the first part of the right hand side of this, $g_{ij}v^{(1),i}$.  We can treat this as a single object with only a single lower index (the $j$ index), because the $i$ index is a summed dummy index:
\begin{eqnarray}
g_{ij}v^{(1),i} \equiv a_{j}
\end{eqnarray}
But, we can recognize $a_j$ as the components of some covector in $\Lambda^1T_p\mathcal{M}$.  Therefore we can express the isomorphism the metric produces between $T_p\mathcal{M}$ and $\Lambda^1T_p\mathcal{M}$:
\begin{eqnarray}
\omega_i = g_{ij} v^j
\end{eqnarray}
where $\omega_i$ is a $1$-form in $\Lambda^1T_p\mathcal{M}$ and $v^j$ is a vector in $T_p\mathcal{M}$.  

Also, we require that the metric be non-singular, and we denote its inverse as
\begin{eqnarray}
g^{ij} = (g_{ij})^{-1}
\end{eqnarray}
and
\begin{eqnarray}
g^{jk}g_{ij} &=& g_{ij}g^{jk} = \delta_i^k \nolabel \\
g^{ij}g_{ij} &=& \delta^i_i = n \label{eq:thetraceofthemetricisequaltothedimensionofthemanifold}
\end{eqnarray}
where $n$ is the dimension of the manifold.  We can use this to see that we can also write
\begin{eqnarray}
\omega_i = g_{ij}v^j \Longrightarrow g^{ki}\omega_i = g^{ki}g_{ij}v^j \Longrightarrow g^{ki}\omega_i = v^k
\end{eqnarray}

Summarizing, we have
\begin{eqnarray}
\omega_i &=& g_{ij}v^j \nolabel \\
v^i &=& g^{ij}\omega_j
\end{eqnarray}
In other words, the metric is used to ``raise" or ``lower" the indices of a vector or a covector, turning one into the other, etc.  

Generalizing, for a general tensor with multiple indices, we can apply the metric to each index separately:
\begin{eqnarray}
\omega_{ijk} &=& g_{ia}g_{jb}g_{kc} v^{abc} \nolabel \\
v^{ijk} &=& g^{ia}g^{jb}g^{kc} \omega_{abc}
\end{eqnarray}
and so on.  

\subsection{Simple Examples of Metrics}

So what does a metric \it do\rm?  We will first consider this question in terms of a few simple examples.  

The simplest example of a metric is the \bf Euclidian Metric\rm, where
\begin{eqnarray}
g_{ij} = g^{ij} = \delta_{ij} = 
\begin{pmatrix}
1 & 0 & \cdots & 0 \\
0 & 1 & \cdots & 0 \\
\vdots & \vdots & \ddots & \cdots \\
0 & 0 & \vdots & 1
\end{pmatrix} \label{eq:matrixrepwrittenoutforeuclidianmetric}
\end{eqnarray}
Then using (\ref{eq:generalformofmetricinnerproduct}) the inner product between two vectors $\bf v\it^{(1)}$ and $\bf v\it^{(2)}$ is
\begin{eqnarray}
g(\bf v\it^{(1)},\bf v\it^{(2)}) = \sum_i v^{(1),i} v^{(2),i} \label{eq:firstdotproductoftwovectorseuclidianmetric}
\end{eqnarray}
Or, we can find the covector corresponding to $\bf v\it^{(1)}$:
\begin{eqnarray}
\omega_i = g_{ij} v^{(1),j} =\delta_{ij} v^{(1),j} =  v^{(1),i}
\end{eqnarray}
So the components are simply equal, and we therefore write the form
\begin{eqnarray}
v_{(1),i} = g_{ij}v^{(1),j} = v^{(1),i}
\end{eqnarray}
instead of $\omega_i$.  
So the inner product between $\bf v\it^{(1)}$ and $\bf v\it^{(2)}$ in (\ref{eq:firstdotproductoftwovectorseuclidianmetric}) is actually
\begin{eqnarray}
g(\bf v\it^{(1)},\bf v\it^{(2)}) = v_{(1),i}v^{(2),i} = v^{(1),i}v_{(2),i}
\end{eqnarray}
Once again this is actually an inner product between a vector and a covector (the metric turns vectors into covectors).  In this way our ``upper and lower indices are summed" convention holds.  

Another example is called the \bf Minkowski Metric\rm, 
\begin{eqnarray}
g_{ij} = g^{ij} = \eta_{ij} = 
\begin{pmatrix}
-1 & 0 & \cdots & 0 \\
0 & 1 & \cdots & 0 \\
\vdots & \vdots & \ddots & \cdots \\
0 & 0 & \vdots & 1
\end{pmatrix}
\end{eqnarray}
Now we have\footnote{We are taking the first component to have index $0$ instead of $1$ for later convenience.}  
\begin{eqnarray}
g(\bf v\it^{(1)},\bf v\it^{(2)}) = -v^{(1),0} v^{(2),0} + \sum_{i=1} v^{(1),i} v^{(2),i}
\end{eqnarray}
Furthermore, when we raise or lower indices, we find that 
\begin{eqnarray}
v_{(1),i} &=& v^{(1),i}\qquad i\neq 0 \nolabel \\
v_{(1),0} &=& -v^{(1),0} \qquad i =0
\end{eqnarray}
(and the same for $\bf v\it^{(2)}$).  Therefore the ``upper and lower index" form of the inner product,
\begin{eqnarray}
g(\bf v\it^{(1)},\bf v\it^{(2)}) = v_{(1),i}v^{(2),i} = -v^{(1),0} v^{(2),0} + \sum_{i=1} v^{(1),i} v^{(2),i}
\end{eqnarray}
still holds.  

We can also write down a more general metric with more than one negative component.  Consider a metric with $n$ $-1$'s and $m$ $+1$'s:
\begin{eqnarray}
g_{ij} = g^{ij} = 
\left(\begin{array}{ccccc}-1 & 0 & \cdots & 0 & 0 \\0 & -1 & \cdots & 0 & 0 \\\vdots & \vdots & \ddots & \cdots & \cdots \\0 & 0 & \vdots & 1 & 0 \\0 & 0 & \vdots & 0 & 1\end{array}\right)
\end{eqnarray}
Now the dot product will be 
\begin{eqnarray}
g(\bf v\it^{(1)},\bf v\it^{(2)}) =- \sum_{i=0}^{n-1} v^{(1),i}v^{(2),i} + \sum_{i=n}^{m+n-1} v^{(1),i}v^{(2),i}
\end{eqnarray}
Such a metric is said to have \bf index \rm $(n,m)$.  For the special case of $n=1$ we call $g$ a \bf Lorentz Metric\rm, or a \bf Minkowski Metric\rm.  

Finally, let's look one more time at the Euclidian metric (\ref{eq:matrixrepwrittenoutforeuclidianmetric}).  Because $g_{ij} = \delta_{ij}$ is simply a tensor, we can do a normal transformation to write it in any other coordinate system.  For example in $\mathbb{R}^2$ we have (in Cartesian coordinates)
\begin{eqnarray}
g_{ij} = \delta_{ij} = 
\begin{pmatrix}
1 & 0 \\ 0 & 1
\end{pmatrix} \label{eq:euclidianmetricintwodimensionsforexmaple}
\end{eqnarray}
If we wanted to transform this to, say, polar coordinates, we can use the standard transformation law for a rank 2 covariant tensor (\ref{eq:changeofcovectorbasis}) (using a tilde to represent the polar coordinates)
\begin{eqnarray}
\tilde g_{ij} = {\partial x^k \over \partial \tilde x^i} {\partial x^l \over \partial \tilde x^j} g_{kl} \label{eq:transofmetricblatheringblatherskyte}
\end{eqnarray}
So, using the transformations
\begin{eqnarray}
x &=& r\cos\theta \nolabel \\
y &=& r\sin\theta
\end{eqnarray}
we have
\begin{eqnarray}
\tilde g_{11} &=& {\partial x^k \over \partial \tilde x^1} {\partial x^l \over \partial \tilde x^1} g_{kl} \nolabel \\
&=& {\partial x^k \over \partial \tilde x^1}{\partial x^l \over \partial \tilde x^1} \delta_{kl} \nolabel \\
&=& {\partial x \over \partial r} {\partial x \over \partial r} + {\partial y \over \partial r} {\partial y \over \partial r} \nolabel \\
&=& \cos^2\theta + \sin^2\theta \nolabel \\
&=& 1
\end{eqnarray}
And similarly
\begin{eqnarray}
\tilde g_{22} &=& {\partial x^k \over \partial \theta} {\partial x^l \over \partial \theta} \delta_{kl} \nolabel \\
&=& {\partial x \over \partial \theta}{\partial x \over \partial \theta} + {\partial y \over \partial \theta} {\partial y \over \partial \theta} \nolabel \\
&=& r^2\sin^2\theta + r^2\cos^2\theta \nolabel \\
&=& r^2
\end{eqnarray}
So, in polar coordinates we have
\begin{eqnarray}
\tilde g_{ij}  = 
\begin{pmatrix}
1 & 0 \\ 0 & r^2
\end{pmatrix} \label{eq:metrictransfromcarttopolar}
\end{eqnarray}

Generalizing to $\mathbb{R}^3$, you can start with the three dimensional Euclidian metric on $\mathbb{R}^3$ in Cartesian coordinates, 
\begin{eqnarray}
g_{ij} = 
\begin{pmatrix}
1 & 0 & 0 \\ 0 & 1 & 0 \\ 0 & 0 & 1
\end{pmatrix}
\end{eqnarray}
and transform to show that in spherical coordinates this metric is
\begin{eqnarray}
g_{ij} = \begin{pmatrix}
1 & 0 & 0 \\ 0 & r^2 \sin^2 \phi & 0 \\ 0 & 0 & r^2
\end{pmatrix} \label{eq:sphericalcoordinatemetricinr3}
\end{eqnarray}

\subsection{The Metric and the Pythagorean Theorem}
\label{sec:TheMetricandthePythagoreanTheorem}

Consider a Euclidian manifold with metric $g_{ij} = \delta_{ij}$.  Then consider some vector
\begin{eqnarray}
\bf v\it = x^1 \bf e\it_1 + x^2 \bf e\it_2 + x^3 \bf e\it_3
\end{eqnarray}
We can use the metric to lower the index to find the corresponding covector $\bf w\it$:
\begin{eqnarray}
x_i = g_{ij}x^j &=& \delta_{ij}x^j \nolabel \\
&\Rightarrow& x_1 = x^1 \nolabel \\
&\Rightarrow& x_2 = x^2 \nolabel \\
&\Rightarrow& x_3 = x^3 \nolabel \\
\end{eqnarray}
So, the inner product between the vector $\bf v\it$ and its corresponding covector is
\begin{eqnarray}
g(\bf v\it,\bf v\it) = \bf w\it( \bf v\it) = x_ix^i = x^1x^1+x^2x^2+x^3x^3
\end{eqnarray}
We recognize the right hand side of this expression as the length squared of the vector $\bf v\it$.  If we had chosen a different metric, say
\begin{eqnarray}
g_{ij} \dot = 
\begin{pmatrix}
2 & 0 & 0 \\
0 & 1 & 0 \\
0 & 0 & 1 
\end{pmatrix}
\end{eqnarray}
then we would have a different length for the vector:
\begin{eqnarray}
g(\bf v\it,\bf v\it) = 2x^1x^1 + x^2x^2+x^3x^3
\end{eqnarray}
So these two metrics correspond to two different ``types" of $\mathbb{R}^3$.  The first is ``normal" three-dimensional Euclidian space, while the second has been contracted along one axis.  However, this can actually be thought of as merely a coordinate transformation where we relabel the $x$-coordinates.  

For this reason we can also think of a metric as relating to distance.  Specifically, consider moving an infinitesimal distance (generalizing to an arbitrary three-dimensional manifold)
\begin{eqnarray}
dx {\partial \over \partial x} + dy {\partial \over \partial y} + dz{\partial \over \partial z}
\end{eqnarray}
Obviously the infinitesimal displacement will be
\begin{eqnarray}
(dx)^2 + (dy)^2 + (dz)^2
\end{eqnarray}
If we want to know the distance between two points $a$ and $b$ along some path on $\mathcal{M}$, we simply do the integral\footnote{Generally we rewrite this as 
$$\int_a^b \sqrt{(dx)^2 + (dy)^2 + (dz)^2} = \int_a^b dx\sqrt{1 + y'^2 + z'^2 }$$
where the prime denotes a derivative with respect to $x$.  This expression is the one generally learned in an introductory Calculus course for distance.  }
\begin{eqnarray}
\int_a^b \sqrt{(dx)^2 + (dy)^2 + (dz)^2}
\end{eqnarray}
along the path from $a$ to $b$.  

More generally, for some infinitesimal displacement
\begin{eqnarray}
dx^i{\partial \over \partial x^i}
\end{eqnarray}
we can find the infinitesimal distance, denoted $ds^2$, as
\begin{eqnarray}
ds^2 &=& g\bigg( dx^i{\partial \over \partial x^i}, dx^j {\partial \over \partial x^j}\bigg) \nolabel \\
&=& dx^i dx^j g\bigg({\partial \over \partial x^i},{\partial \over \partial x^j}\bigg) \nolabel \\
&=& g_{ij}dx^idx^j
\end{eqnarray}
which is essentially the definition of the metric (\ref{eq:firstmetric}).  For this reason we will speak of the ``metric" and the ``infinitesimal displacement squared" interchangeably.  

\subsection{Induced Metrics}
\label{sec:inducedmetrics}

Finding a metric for a space is extremely important for most physical applications.  However doing so may not always be easy.  One very powerful tool for finding the metric of a given space is to find an \bf induced metric\rm.  

Writing a metric for, say $\mathbb{R}^2$ is easy - $g_{ij} = \delta_{ij}$.  But what is the metric for a circle?  This may not be so obvious.  However, it \it is \rm possible to write the circle as a subspace of $\mathbb{R}^2$:
\begin{eqnarray}
x =r \cos \theta \nolabel \\
y = r\sin \theta \label{eq:circleintor2forinducedmetricssection}
\end{eqnarray}
(for constant $r$).  We can think of (\ref{eq:circleintor2forinducedmetricssection}) as a map\footnote{It would be helpful to reread section \ref{sec:pullbackdiff} at this point.} $\bf F\it$ from $S^1$ into $\mathbb{R}^2$:
\begin{eqnarray}
& &\bf F\it: S^1 \longrightarrow \mathbb{R}^2 \nolabel \\
& &F^1(\theta) = r\cos\theta\nolabel \\
& & F^2 (\theta) =r \sin\theta
\end{eqnarray}
Therefore, any tensor product of forms on $\mathbb{R}^2$ can be mapped back to $S^1$ by using the pullback $f^{\star}$.  And because a metric (on $\mathbb{R}^2$) is a tensor product of forms, we can find its pullback on $S^1$:
\begin{eqnarray}
g_{S^1} = f^{\star} (g_{\mathbb{R}^2})
\end{eqnarray}
So applying (\ref{eq:howtoexpressyintermsofxdiffpullback}) to $g = \delta_{ij}dx^i\otimes dx^j$, 
\begin{eqnarray}
f^{\star}(\delta_{ij} dx^i\otimes dx^j) &=& \delta_{ij} f^{\star}(dx^i\otimes dx^j) \nolabel \\
&=& \delta_{ij} {\partial F^i \over \partial \theta}{\partial F^j \over \partial \theta} d\theta \otimes d\theta \nolabel \\
&=& \bigg( \bigg({\partial F^1 \over \partial \theta}\bigg)^2 + \bigg({\partial F^2 \over \partial \theta}\bigg)^2\bigg) d\theta \otimes d\theta \nolabel \\
&=& \big((-r\sin\theta)^2 + (r\cos\theta)^2\big) d\theta \otimes d\theta \nolabel \\
&=& r^2 \big(\sin^2\theta + \cos^2\theta\big)d\theta \otimes d\theta \nolabel \\
&=& r^2 d\theta \otimes d\theta \nolabel \\
&=& r^2 (d\theta)^2
\end{eqnarray}
This is exactly what we would have expected - this is saying that the distance you've moved if you walk some displacement around a circle is proportional to the radius and the angle.  In other words
\begin{eqnarray}
ds^2 = r^2(d\theta)^2 \quad \Longrightarrow \quad ds = rd\theta \label{eq:metricforcircle}
\end{eqnarray}
We can integrate this to get
\begin{eqnarray}
s = r\theta \label{eq:basicexpressionforcirc}
\end{eqnarray}
which is a standard result in introductory math and physics.  The most familiar case of (\ref{eq:basicexpressionforcirc}) is that if you walk around a full circumference $C$, or $\theta = 2\pi$ radians, then (\ref{eq:basicexpressionforcirc})) gives
\begin{eqnarray}
C = 2\pi r
\end{eqnarray}
The standard expression for the circumference of a circle.  

On the other hand, we could map $S^1$ into $\mathbb{R}^2$ in a different way.  Consider
\begin{eqnarray}
x &=& 2r\cos\theta \nolabel \\
y &=& r\sin\theta
\end{eqnarray}
(again for constant $r$).  This creates an ellipse with radius $2$ along the $x$ axis and radius $1$ along the $y$ axis.  
\begin{center}
\includegraphics[scale=.5]{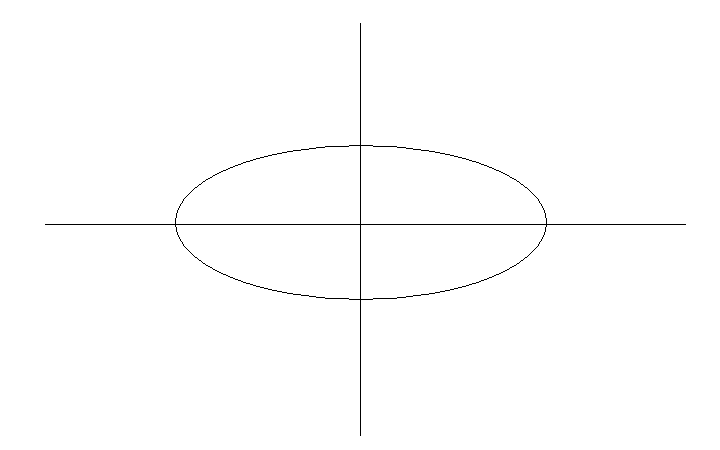}
\end{center}
Now we will have
\begin{eqnarray}
f^{\star}(\delta_{ij} dx^i \otimes dx^j) &=& \delta_{ij} {\partial F^i \over \partial \theta} {\partial F^j \over \partial \theta} (d\theta \otimes d\theta) \nolabel \\
&=& r^2\big(4\cos^2\theta + \sin^2\theta\big) d\theta^2
\end{eqnarray}

So for the circle we have 
\begin{eqnarray}
ds^2_{circle} = r^2 d\theta^2
\end{eqnarray}
and for the ellipse we have
\begin{eqnarray}
ds^2_{ellipse} = r^2(4\cos^2\theta + \sin^2\theta)d\theta^2 \label{eq:metricforoval}
\end{eqnarray}
We can graph each of these as a polar plot.  For the circle, we graph $R(r,\theta) = r^2$, which gives
\begin{center}
\includegraphics[scale=.5]{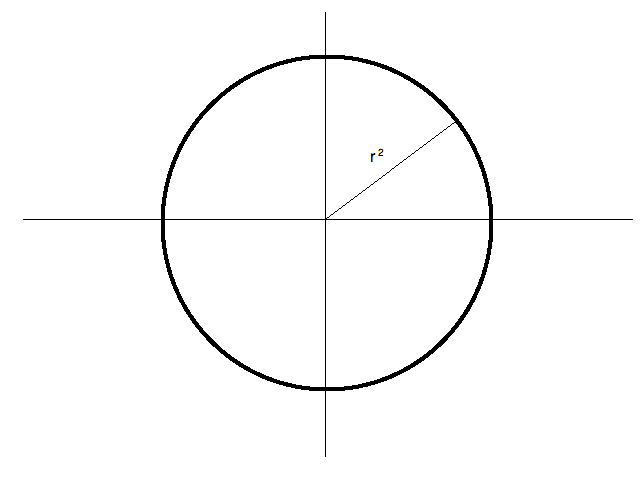}
\end{center}
And for the ellipse we can graph $R(r,\theta) = r^2(4\cos^2\theta + \sin^2\theta)$, which gives
\begin{center}
\includegraphics[scale=.5]{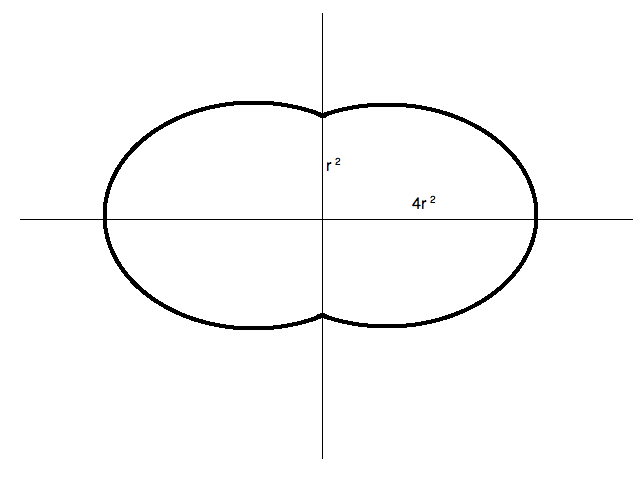}
\end{center}
Recall from the previous section that you can think of the metric as defining an infinitesimal distance.  In other words if you start from a point $\theta$ on $\mathcal{M}$ and move a small displacement $d\theta$, the distance (squared) you travel is given by $ds^2 = g_{ij}dx^idx^j$, or in our case $ds^2 = g_{\theta \theta} d\theta^2$.  For the circle this is simply constant - no matter where on the circle you are, a small displacement in $\theta$ takes you the same distance.  On the ellipse, however, a small displacement does not have the same result at any two points.  For example, if you are on the point on the positive $x$ axis ($\theta = 0$), you are very far from the origin and therefore a small change in $\theta$ will move you farther than if you were close, for example on the positive $y$ axis ($\theta = \pi / 2$).  Therefore the value of the metric is the greatest at the points farther from the origin, and the least at the points closest.  

Notice that these two maps are \it both \rm from $S^1$.  They are completely homeomorphic to each other.  The only difference is that one is stretched out more than the other \it when drawn in \rm $\mathbb{R}^2$.  One difficulty with our approach is that we are only talking about metrics that are \it induced \rm on a submanifold by the manifold it is embedded in.  With that said, should we really expect a difference between $ds^2_{circle}$ and $ds^2_{ellipse}$?  They are both one-dimensional manifolds, and moving along that one dimension a certain distance is simply that - moving that certain distance.  The reason we found different metrics is that we embedded $S^1$ in $\mathbb{R}^2$ in different ways, and as a result the value $d\theta$ had a different meaning in the two cases.  However, some given distance on the circle isn't really different than the same given distance on the ellipse.  We therefore suspect that there is no substantive difference between the two, and the different metrics are merely due to different coordinates (in other words, different $\theta$'s).  Sure enough, it is straightforward to show that if we start with the circle metric
\begin{eqnarray}
g_{\theta\theta} = r^2
\end{eqnarray}
We then take the coordinates on the ellipse to be $\theta'$.  We can do a coordinate transformation to get the metric on the ellipse from the metric on the circle:
\begin{eqnarray}
g_{\theta'\theta'} &=& {\partial \theta \over \partial \theta'}{\partial \theta \over \partial \theta'} g_{\theta\theta} \nolabel \\
&=& \bigg({\partial \theta \over \partial \theta'}\bigg) r^2 \nolabel \\
&=& r^2 (4\cos^2\theta'+\sin^2\theta')
\end{eqnarray}
which would come from the coordinate transformation
\begin{eqnarray}
\theta = \int d\theta' \sqrt{4\cos^2\theta' + \sin^2\theta'}
\end{eqnarray}
Evaluating this in closed form is difficult because it is an elliptic integral, but it can be done and the result would simply transform $g_{\theta\theta}$ into $g_{\theta'\theta'}$.  So we have shown that the metric $ds^2_{ellipse}$ and $ds^2_{circle}$ are actually the same metric, differing only in the coordinates being used.  

In general, if two manifolds (appear to) have different metrics, but there exists some coordinate transformation that takes one metric to the other, then the metrics are actually equivalent.  On the other hand, if no such transformation exists then the metrics are truly different.  In general such an approach is a highly nontrivial (and essentially impossible) problem and other techniques must be used.  Such ideas will be discussed later.

As another example, consider a map from $S^2$ (with unit radius) into $\mathbb{R}^3$ given by
\begin{eqnarray}
x &=& \sin\theta \cos \phi \nolabel \\
y &=& \sin\theta \sin \phi \nolabel \\
z &=& \cos\theta
\end{eqnarray}
We leave the details to you, but the pull-back of the Euclidian metric on $\mathbb{R}^3$ will give
\begin{eqnarray}
ds^2 = d\theta^2 + \sin^2\theta d\phi^2  \label{eq:metricfors2inmetricsection}
\end{eqnarray}

On the other hand, we could choose another map homeomorphic to $S^2$ but with a different metric.  Consider
\begin{eqnarray}
x &=& \sin\theta \cos \phi \nolabel \\
y &=& \sin\theta \sin \phi \nolabel \\
z &=& \lambda\cos\theta
\end{eqnarray}
for some arbitrary constant $\lambda$.  We again leave the details to you and merely quote the result.  
\begin{eqnarray}
ds^2 = {1 \over 2} \big((\lambda^2 + 1) - (\lambda^2 - 1)\cos(2\theta)\big)d\theta^2 + \sin^2\theta d\phi^2 \label{eq:metricfordeformeds2}
\end{eqnarray}
If we choose $\lambda$ to be greater than $1$ we get an egg:
\begin{center}
\includegraphics[scale=.5]{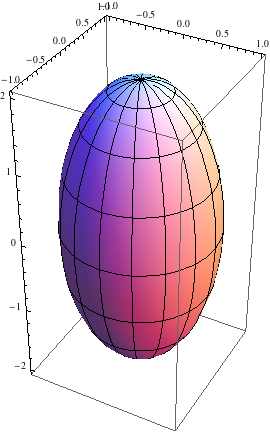} \label{pictureofeggelongatedalongzaxis}
\end{center}
On the other hand if we choose $\lambda$ to be smaller than $1$ we get a pancake:
\begin{center}
\includegraphics[scale=.6]{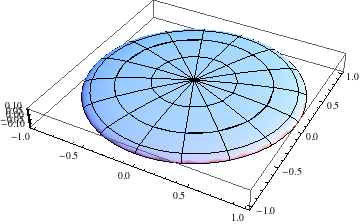} \label{eqpictureofpancakeshortatzaxiswithalmbda}
\end{center}

As a illustrative exercise, consider the $\lambda <<1$ regime, which gives a very flat pancake.  Here
\begin{eqnarray}
ds^2 = {1 \over 2}(1+\cos(2\theta)) d\theta^2 + \sin^2\theta d\phi^2
\end{eqnarray}
In the $\theta << 1 $ region of the manifold (in the center of the flat part), we have
\begin{eqnarray}
ds^2 = d\theta^2
\end{eqnarray}
or simply
\begin{eqnarray}
ds = d\theta
\end{eqnarray}
Here a change $d\theta$ corresponds to moving radially outward from the center of the flat part of the pancake, and this metric tells us that indeed the lines are flat.  Furthermore, in this limit, there is no real meaning to a change in $\phi$, as indicated by the metric.  Notice that $ds = d\theta$ is the same metric we would have for a circle of radius $1$.  In fact, in the $\theta = \pi/2$ regime we have 
\begin{eqnarray}
ds^2 = d\phi^2
\end{eqnarray}
The difference in the two relates to the global topology of the space we are considering.  The metric for $\mathbb{R}^1$ is $ds^2 = dx^2$, and the metric for a unit radius $S^1$ is $ds^2 = d\theta^2$.  But while they have the same metric, they have different topologies.  This illustrates an important point that we will discuss later - the metric is a \it local \rm structure.  It gives us information, point by point, about infinitesimal distances.  The metric is not, however, a global structure, and it therefore doesn't tell us anything about the topology of the manifold.  We will consider physical ramifications of this later.  

For now, however, we are merely trying to show what kind of information the metric can express by considering various limits of things homeomorphic to $S^2$.  

As a final example consider the map from the torus $T^2$ into $\mathbb{R}^3$ given by
\begin{eqnarray}
x &=& (R+r\cos\theta)\cos\phi \nolabel \\
y &=& (R+r\cos\theta)\sin\phi \nolabel \\
z &=& r\sin\theta
\end{eqnarray}
where $R$ and $r<R$ are the two radii of the torus.  We again leave the details to you to show that the Euclidian metric on $\mathbb{R}^3$ will pullback to induce
\begin{eqnarray}
ds^2 = r^2d\theta^2 + (R+r\cos\theta)^2d\phi^2 \label{eq:metricfortorusinmetricsection}
\end{eqnarray}

For any manifold $\mathcal{N}$ that you can write as a submanifold of another $\mathcal{M}$, you can easily find the metric on $\mathcal{N}$ that is induced by the metric on $\mathcal{M}$ by simply using the pullback.  

\section{Connections and Covariant Derivatives}
\label{sec:connectionsandcovariantderivatives}

\subsection{Parallel Transport}
\label{sec:paralleltransport}

In this section we will continue with our discussion of metrics, though it will not be immediately obvious that this is what we are doing.  Our goal will be to build up a more general calculus on manifolds.  But, recall from section \ref{sec:flowandlie} that we had a problem when we wanted to find the derivative of some vector field (cf the discussion on page \pageref{wherewediscussdifferenttangentspacesderiv}) over a manifold.  The problem was that derivatives involve comparing vectors at two different points, say $p$ and $q$.  However the value of a vector field at $p$ and the value at $q$ are (obviously) in two different tangent spaces.  And because there is no natural way to relate tangent spaces to each other, this comparison (between the vector at $p$ and the vector at $q$) is not well defined.  

One consequence of this is the transformation properties of a derivative.  In general, a vector is described by a set of numbers in some basis (i.e. $(x,y,z)$, $(r, \theta, \phi)$, $(\rho, \phi, z)$, etc.).  However this vector is independent of the coordinate system used.  This is why vectors have specific transformation rules under a coordinate transformation - the description retains its identity even though its description changes.  Such a transformation is called a \bf covariant transformation\rm.  We have considered such transformations extensively in previous sections.  

Recall from the discussion following (\ref{eq:howvectorcompstransformfirsttime}) that any (generally nonlinear) change of coordinates induces a linear change of components \it at each point\rm.  The basic idea is that we want things that transform in a tensorial way under such transformations (cf page \pageref{whereItalkaboutcontraandcov}).  A derivative, however, is by definition evaluated at two different points in the manifold (taking the difference between the tensor at two points and dividing by their distance).  For this reason the standard definition of a derivative will not transform in the nice linear tensorial way under a coordinate transformation.  To see this, consider a vector field $\bf v\it = v^i {\partial \over \partial x^i}$ defined over a manifold $\mathcal{M}$.  We can take the derivative of $v^i$, 
\begin{eqnarray}
{\partial v^i \over \partial x^j} = \bigg( {\partial \over \partial x^j}\bigg) (v^i) \label{eq:derivativevectortotransform}
\end{eqnarray}
If we want to apply a transformation to this, the derivative part will transform according to (\ref{eq:changeofbasisvectorusingchainrulethingy}), and the vector part will transform according to (\ref{eq:transformationofcomponentsonmanifold}).  So (\ref{eq:derivativevectortotransform}) will transform according as
\begin{eqnarray}
\bigg({\partial \over \partial x^j}\bigg) (v^i) &\longrightarrow& \bigg({\partial \over \partial x'^j}\bigg) (v'^i) \nolabel \\
&=& \bigg({\partial x^k \over \partial x'^j} {\partial \over \partial x^k}\bigg) \bigg({\partial x'^i \over \partial x^l} v^l\bigg) \nolabel \\
&=& {\partial x^k \over \partial x'^j} {\partial x'^i \over \partial x^l} {\partial v^l \over \partial x^k} +  {\partial x^k \over \partial x'^j} {\partial^2 x'^i \over \partial x^k \partial x^l} v^l \label{eq:transformationofderivativeofvector}
\end{eqnarray}
Notice that the first term in the last line would be the expected transformation law for an object with a single covariant and a single contravariant tensor (i.e. ${\partial v^i \over \partial x^j}$).  However the second term means that this is not a tensorial object.  

The reason, as we have said, is that a derivative involves evaluating two different points, and because of the general non-linearity of a coordinate transformation, the component transformation for a derivative that is induced is not linear.  

To illustrate this imagine a vector field $\bf w\it (\bf q\it (\epsilon , \bf x\it (p)))$ defined along the curve $\bf q\it (\epsilon, \bf x\it(p))$.  We can assume polar coordinates and take the partial derivative of, say $w^r$ with respect to $\phi$
\begin{eqnarray}
{\partial w^r \over \partial \phi} = \bigg( {\partial \over \partial \phi}\bigg) (w^r)
\end{eqnarray}
If we take 
\begin{eqnarray}
x^1 &=& x \qquad x^2 = y \nolabel \\
y^1 &=& r \qquad y^2 = \phi 
\end{eqnarray}
and the coordinates of $\bf w\it$ in Cartesian coordinates to be denoted $v^1 = v^x$ and $v^2 = v^y$, then the transformation of this from polar back into Cartesian is (following (\ref{eq:transformationofderivativeofvector}))  
\begin{eqnarray}
\bigg( {\partial \over \partial \phi}\bigg) (w^r) &=& \bigg({\partial \over \partial y^2}\bigg) (w^1)  \nolabel \\
&=& \bigg( {\partial x^k \over \partial y^2 }{\partial \over \partial x^k} \bigg) \bigg({\partial y^1 \over \partial x^l} v^l\bigg) \nolabel \\
&=& {\partial x^k \over \partial y^2} {\partial y^1 \over \partial x^l} {\partial v^l \over \partial x^k} + {\partial x^k \over \partial y^2} {\partial^2 y^1\over \partial x^k \partial x^l} v^l \label{eq:transformationofwfrompolartocart}
\end{eqnarray}
Again, the first part in the last line is what we would expect for a tensorial transformation.  Using the standard
\begin{eqnarray}
x^1 &=& x = r\cos \phi = y^1 \cos y^2 \nolabel \\
x^2 &=& y = r\sin \phi = y^1 \sin y^2 \nolabel \\
y^1 &=& r = \sqrt{x^2 + y^2} = \sqrt{(x^1)^2 + (x^2)^2} \nolabel \\
y^2 &=& \phi = \tan^{-1}\bigg({y \over x}\bigg) = \tan^{-1}\bigg({y^2 \over y^1}\bigg)
\end{eqnarray}
we can write the second non-tensorial term in (\ref{eq:transformationofwfrompolartocart}) as
\begin{eqnarray}
 {\partial x^k \over \partial y^2} {\partial^2 y^1\over \partial x^k \partial x^l} v^l &=& \bigg( {\partial x \over \partial \phi} {\partial^2 r \over \partial x \partial x} + {\partial y \over \partial \phi} {\partial^2 r \over \partial x \partial y}\bigg) v^1 \nolabel \\
& & + \bigg({\partial x \over \partial \phi} {\partial^2 r \over \partial x \partial y} + {\partial y \over \partial \phi} {\partial^2 r \over \partial y \partial y}\bigg) v^2 \nolabel \\
&=& \cdots \nolabel \\
&=& -\sin \phi v^1 + \cos \phi v^2 \label{eq:inhomotermforwsuperr}
\end{eqnarray}
So, as expected, the partial derivative ${\partial w^r \over \partial \phi}$ doesn't transform as a tensor, due to the inhomogeneous term (\ref{eq:inhomotermforwsuperr}).  in other words
\begin{eqnarray}
{\partial w^r \over \partial \phi} ={\partial w^1 \over \partial y^2} =  {\partial x^k \over \partial y^2} {\partial y^1 \over \partial x^l} {\partial v^l \over \partial x^k} - \sin\phi v^1 + \cos \phi v^2 \label{eq:partialwrwrtphiwithadditionalterms}
\end{eqnarray}
The additional two terms resulted from the fact that the derivative involved comparing points in two different tangent spaces.  It is for this reason that to take a derivative we must find a way of evaluating both parts in the same tangent space (in other words, at the same point), so as to avoid this problem.  In section \ref{sec:flowandlie} we solved this for the Lie derivative by taking the derivative of one vector field ($\bf v\it^{(1)}$) with respect to another vector field ($\bf v\it^{(2)}$).  Specifically, we used the curve induced by $\bf v\it^{(2)}$ to build the tangent mapping to map $\bf v\it^{(1)}$ at $p$ to its value at $q$ (where the curve induced by $\bf v\it^{(2)}$ goes from $p$ to $q$).  Doing this allowed us to consider both tensors at the same point, and therefore we had a derivative that would transform in ``nice" tensorial way.  

And while this derivative is well-defined and extremely useful for a variety of things, it is not the most general derivative we can define.  Therefore let us rethink how we can find derivatives on a manifold to solve this problem.  The fundamental problem is that we must be able to compare a vector in $T_p\mathcal{M}$ to a vector in $T_q\mathcal{M}$ where $p\neq q$.  This requires some way of ``transporting" a vector from $T_p\mathcal{M}$ to $T_q\mathcal{M}$.  While the Lie derivative provided one way of doing this, it is not the most general.  

Consider a manifold $\mathcal{M}$ and a curve through the point $p$, denoted $\bf q\it (\tau, \bf x\it( p))$.  The vector $\bf v\it(\bf q\it (\tau, \bf x\it (p)))$ corresponding to this curve at $p$ is 
\begin{eqnarray}
{d \bf q\it (\tau, \bf x\it (p)) \over d \tau} = \bf v\it(\bf q\it (\tau, \bf x\it (p)))
\end{eqnarray}
(cf equation (\ref{eq:relationshipbetweencurveandtangentvectorflow})).  Now consider some other (arbitrary) vector $\bf u\it (\bf q\it (\tau, \bf x\it (p)))$ at $p$, that is not necessarily part of a vector field - it is just a vector in $T_p\mathcal{M}$.  We want a way of transporting $\bf u\it$ along $\bf q\it$ (in the direction of $\bf v\it$) without changing it.  
\begin{center}
\includegraphics[scale=.4]{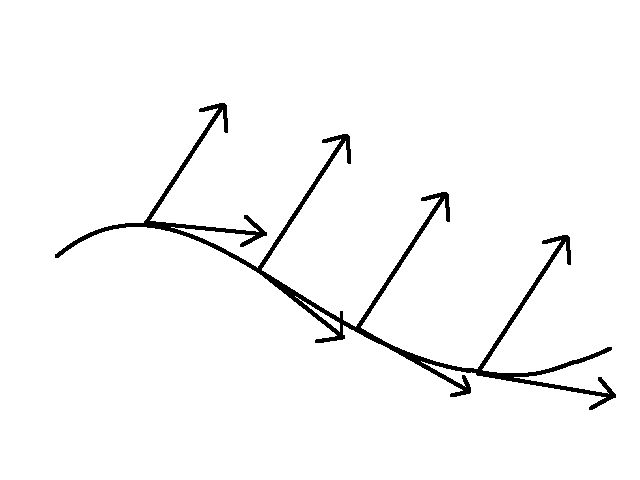}
\end{center}

But what do we mean by ``not changing it"?  This notion is not entirely precise and we must therefore specify some set of rules by which we move $\bf u \it$ around.  As a simple example, consider an arbitrary vector with components $\bf u \it = u^x \bf e\it_x + u^y \bf e\it_y$ in $\mathbb{R}^2$:
\begin{center}
\includegraphics[scale=.6]{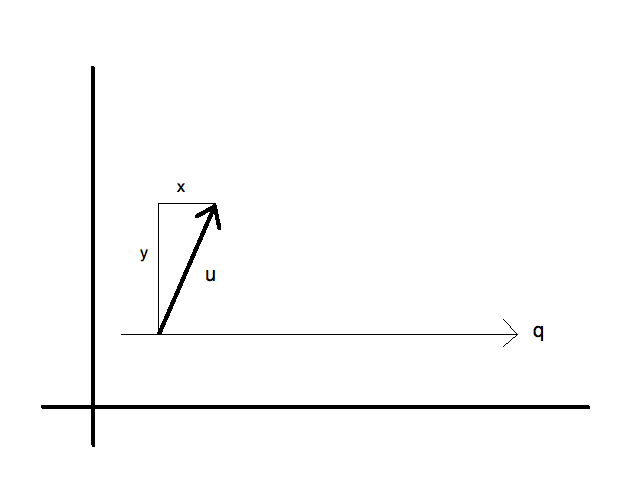}
\end{center}
Let's take the curve $\bf q\it$ to be the line parallel to the $x$ axis through the point $(x,y)$ so that $\bf v\it = \bf e\it_x$.  If we drag $\bf u \it$ along $\bf q\it$ to some other point:
\begin{center}
\includegraphics[scale=.6]{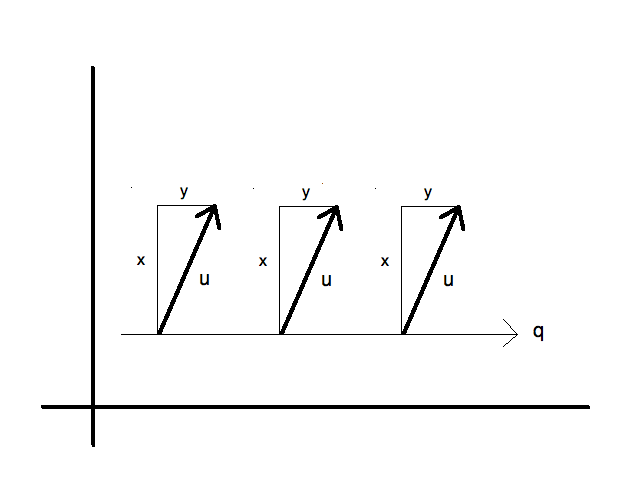}
\end{center}
notice that the components haven't changed: $\bf u \it = u^x \bf e\it_x + u^y \bf e\it_y$.  Similarly, had we taken $\bf q\it$ to be parallel to the $y$ axis we would have gotten the same result.  So, moving a vector $\bf u \it$ around $\mathbb{R}^2$ with the standard Cartesian basis is trivial - no matter where you move $\bf u\it$ the coordinates are the same:
\begin{eqnarray}
u^x \longrightarrow u^x \nolabel \\
u^y \longrightarrow u^y \label{eq:transformationofuincartcords}
\end{eqnarray}

Now consider the exact same vector, but let's use polar coordinates:
\begin{center}
\includegraphics[scale=.6]{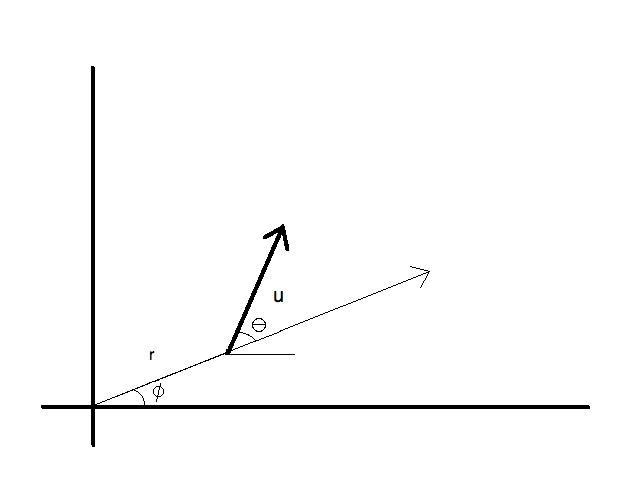}
\end{center}
Notice that we can break $\bf u \it = u^r \bf e\it_r +  u^{\phi} \bf e\it_{\phi}$ up into components according to
\begin{eqnarray}
u^r &=& |\bf u \it| \cos \theta \nolabel \\
ru^{\phi} &=& |\bf u\it| \sin \theta
\end{eqnarray}
Now let's take $\bf q\it$ to be along the $r$ direction:
\begin{center}
\includegraphics[scale=.6]{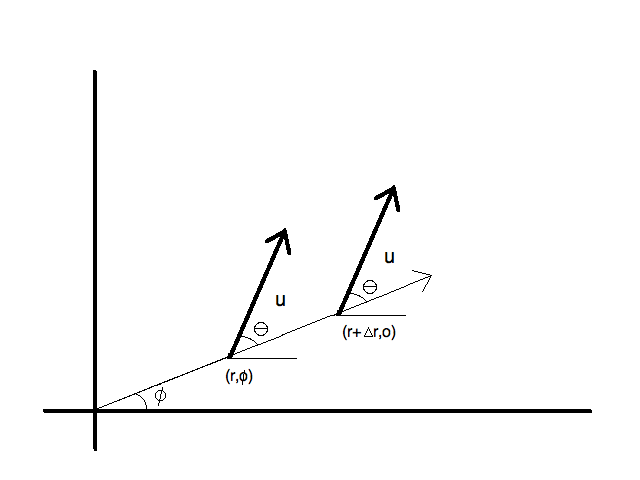}
\end{center}
Taking $r \rightarrow r+ \delta r$ has no affect on $u^r$, and therefore under this transformation 
\begin{eqnarray}
u^r \rightarrow u^r \label{eq:usubrdoesntchange}
\end{eqnarray}
However, 
\begin{eqnarray}
u^{\phi} = {|\bf u \it| \sin \theta \over r} &\longrightarrow& {|\bf u \it| \sin \theta \over r+ \delta r} \nolabel \\
&=& { 1 \over r} {|\bf u \it| \sin \theta \over 1 + {\delta r \over r}} \nolabel \\
&\approx& {1 \over r} \bigg( 1 - {\delta r \over r}\bigg) |\bf u \it| \sin \theta \nolabel \\
&=& {|\bf u \it| \sin \theta \over r}  - {\delta r \over r}{ |\bf u \it| \sin \theta \over r} \nolabel \\
&=& u^{\phi} - {\delta r \over r} u^{\phi} \label{eq:howusubthetachangeswithr}
\end{eqnarray}
So, when an arbitrary vector is moved along the $r$ direction the components don't stay the same.  An additional term is needed to ``correct" for the changing coordinate system.  

Similarly, if we take $\bf q\it$ to be along the $\phi$ direction:
\begin{center}
\includegraphics[scale=.6]{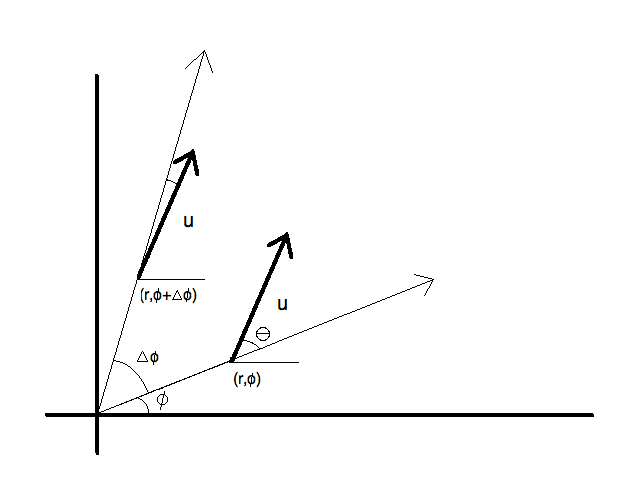}
\end{center}
we have
\begin{eqnarray}
u^r &=& |\bf u\it| \cos \theta \longrightarrow |\bf u\it| \cos (\theta - \delta \phi ) \approx u^r + u^{\phi} r \delta \phi \nolabel \\
u^{\phi} &=& {|\bf u \it| \sin \theta \over r} \longrightarrow {|\bf u \it| \sin (\theta - \delta \phi) \over r} \approx u^{\phi} - u^r {\delta \phi \over r} \label{eq:uranduphiwhenqisalongtheta}
\end{eqnarray}

So when Cartesian coordinates are used, transporting a vector around $\mathbb{R}^2$ is trivial - the coordinates don't change.  But when polar coordinates are used, we must include a term to correct for the changing coordinate system.  

So what is the use of what we have done?  We have written out a way to move an arbitrary vector around $\mathbb{R}^2$ without changing it.  We denote this way of dragging a vector around \bf parallel transporting \label{pagewherewetalkaboutparalleltransporting}\rm the vector.  Because parallel transportation is a way of moving a vector $\bf u \it$ along a path in the direction $\bf v\it$ without changing it, there must exist some type of derivative, denoted $\nabla_{\bf v\it}$, such that
\begin{eqnarray}
\nabla_{\bf v\it} \bf u \it = 0 \label{eq:firstcovderivativeexplanation}
\end{eqnarray}
We call $\nabla_{\bf v\it}$ the \bf covariant derivative \rm in the direction of $\bf v\it$.  
\begin{center}
\includegraphics[scale=.6]{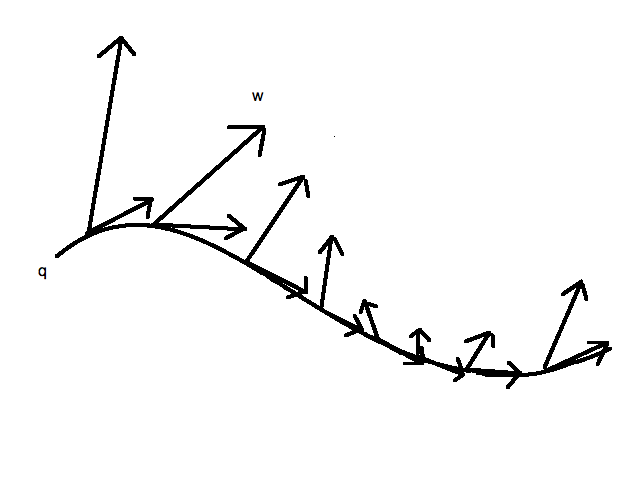}
\end{center}
Now again consider some vector field $\bf w \it (\bf q\it(\tau, \bf x\it(p)))$ defined on $\bf q\it \subset \mathbb{R}^2$:

Focusing on the non-trivial polar coordinate case, equations (\ref{eq:usubrdoesntchange}), (\ref{eq:howusubthetachangeswithr}), and (\ref{eq:uranduphiwhenqisalongtheta}) tell us how an arbitrary vector can be dragged along $\bf q\it$ without changing it.  This allows us to construct a derivative in a well-defined way.  We first Taylor expand to the new point, then we parallel transport - the difference defines the derivative.  

For example, if we want to find the derivative of $\bf w\it$ at point $\bf q\it (0,\bf x\it (p)) \in \mathbb{R}^2$, first move some small distance $\epsilon$ away from $\bf q\it(0,\bf x\it(p))$ and evaluate 
\begin{eqnarray}
\bf w\it (\bf q\it (\epsilon, \bf x\it (p))) &=& \bf w\it (\bf x\it (p) + \delta \bf x\it) \nolabel \\
&=& \bf w \it (\bf x\it (p)) + \delta x^j {\partial \bf w\it (\bf x\it (p)) \over \partial x^i} \label{eq:introducethedeltabfxit}
\end{eqnarray}
Or
\begin{eqnarray}
w^r (\bf x\it(p)+\delta \bf x\it) &=& w^r (\bf x\it (p)) + \delta r {\partial w^r (\bf x\it (p)) \over \partial r} + \delta \phi {\partial w^r (\bf x\it(p)) \over \partial \phi} \nolabel \\
w^{\phi} (\bf x\it(p) + \delta \bf x\it) &=& w^{\phi} (\bf x\it(p)) + \delta r {\partial w^{\phi} (\bf x\it (p)) \over \partial r} + \delta \phi {\partial w^{\phi} (\bf x\it(p)) \over \partial \phi} \label{eq:introcovderivfirstpart}
\end{eqnarray}
Then, using (\ref{eq:usubrdoesntchange}), (\ref{eq:howusubthetachangeswithr}), and (\ref{eq:uranduphiwhenqisalongtheta}), parallel transport $\bf w\it (\bf q\it (0, \bf x\it(p))$ from $\bf x\it (p)$ to $\bf x\it(p) + \delta \bf x\it$.  For example, if $\bf q\it$ is along the $\theta$ direction, 
\begin{eqnarray}
w^r(\bf x\it(p) + \delta \bf x\it) &=& w^r (\bf x\it(p))+ w^{\phi}(\bf x\it(p)) r \delta \phi \nolabel \\
w^{\phi}(\bf x\it(p) + \delta \bf x\it) &=& w^{\phi}(\bf x\it(p)) - w^r (\bf x\it(p)){\delta \phi \over r} \label{eq:introcovderivsecondpart}
\end{eqnarray}
Now we can build the covariant derivative of $\bf w\it$ by taking the difference between (\ref{eq:introcovderivfirstpart}) and (\ref{eq:introcovderivsecondpart}).  

So let's say we want to take the covariant derivative in the $\phi$ direction.  This will be\footnote{Suppressing the $\bf x\it(p)$ argument for notational simplicity.}
\begin{eqnarray}
\nabla_{\phi} w^r &=& \lim_{\delta \bf x\it \rightarrow 0} {1 \over \delta \phi}\bigg( w^r + \delta r {\partial w^r \over \partial r} + \delta \phi {\partial w^r \over \partial \phi} - w^r - w^{\phi} r \delta \phi\bigg) \nolabel \\
&=& \lim_{\delta \bf x\it \rightarrow 0} {1 \over \delta \phi}\bigg( \delta r {\partial w^r \over \partial r} + \delta \phi {\partial w^r \over \partial \phi} - w^{\phi} r \delta \phi\bigg) \nolabel \\
&=& {\partial w^r \over \partial \phi} - w^{\phi} r \nolabel \\
\nabla_{\phi}w^{\phi} &=& \lim_{\delta \bf x\it \rightarrow 0} {1 \over \delta \phi}\bigg( w^{\phi} + \delta r {\partial w^{\phi} \over \partial r} + \delta \phi {\partial w^{\phi} \over \partial \phi} - w^{\phi} + w^r {\delta \phi \over r}\bigg) \nolabel \\
&=& {\partial w^{\phi} \over \partial \phi} + {w^r \over r} \label{eq:firsttwocovderivsoffirstexample}
\end{eqnarray}
We can calculate $\nabla_r \bf w\it$ similarly:
\begin{eqnarray}
\nabla_r w^r &=& {\partial w^r \over \partial r} \nolabel \\
\nabla_r w^{\phi} &=& {\partial w^{\phi} \over \partial r} + {w^{\phi} \over r}
\end{eqnarray}  

Summarizing, 
\begin{eqnarray}
\nabla_{\phi} \bf w\it &=& \bigg({\partial w^r \over \partial \phi} - w^{\phi} r\bigg) \bf e\it_r + \bigg({\partial w^{\phi} \over \partial \phi} + {w^r \over r}\bigg)\bf e\it_{\phi} \nolabel \\
\nabla_{r} \bf w\it &=& \bigg({\partial w^r \over \partial r}\bigg) \bf e\it_r + \bigg({\partial w^{\phi} \over \partial r} + {w^{\phi} \over r} \bigg) \bf e\it_{\phi} \label{eq:summaryofcovderivsinfirstexample}
\end{eqnarray}

Now consider $\nabla_{\phi} w^r$ only (just as we considered ${\partial w^r \over \partial \phi}$ only in (\ref{eq:partialwrwrtphiwithadditionalterms})).  We know from (\ref{eq:partialwrwrtphiwithadditionalterms}) how ${\partial w^r \over \partial \phi}$ transforms.  We can also calculate how the second part will transform:
\begin{eqnarray}
r w^{\phi} &=& r {\partial \phi \over \partial x^l} v^l \nolabel \\
&=& r{\partial \phi \over \partial x} v^1 + r{\partial \phi \over \partial y} v^2 \nolabel \\
&=& -{ry \over x^2 + y^2} v^1 + {rx \over x^2+y^2} v^2 \nolabel \\
&=& -\sin \phi v^1 + \cos \phi v^2 \label{eq:wayadditionaltermtransformsfirstexample}
\end{eqnarray}
Notice that this is exactly what we found to be the ``problematic" term in (\ref{eq:partialwrwrtphiwithadditionalterms}).  This means that the transformation of $\nabla_{\phi} w^r$ will be
\begin{eqnarray}
\nabla_{\phi} w^r &=& {\partial w^r \over \partial \phi} - w^{\phi} r \nolabel \\
&=& {\partial x^k \over \partial y^2} {\partial y^1 \over \partial x^l} {\partial v^l \over \partial x^k} - \sin \phi v^1 + \cos \phi v^2 + \sin\phi v^1 - \cos \phi v^2 \nolabel \\
&=& {\partial x^k \over \partial y^2} {\partial y^1 \over \partial x^l} {\partial v^l \over \partial x^k} \label{eq:covderivtransformedwithconnectionsubtractingoffbadterm}
\end{eqnarray}
And from (\ref{eq:transformationofuincartcords}) we can see that
\begin{eqnarray}
{\partial v^l \over \partial x^k} = \nabla_k v^l
\end{eqnarray}
and therefore
\begin{eqnarray}
\nabla_{\phi}w^r = {\partial x^k \over \partial y^2} {\partial y^1 \over \partial x^l} \nabla_k v^l
\end{eqnarray}
So by constructing the covariant derivative we have created a derivative that transforms in a nice tensorial covariant way.  That is the meaning of the additional terms added to the derivative - they preserve the covariance of the derivative under arbitrary coordinate transformations by canceling the non-linear terms created when the partial derivative and the connection transform.  

To summarize, what we have done in this section is a very specific and very simple exposition of creating a derivative that transforms covariantly.  The fundamental problem with defining a derivative is that the definition of a derivative involves comparing tensors in different tangent spaces, and there is no way natural way of doing this.  

A fundamental symptom of this problem is the additional non-tensorial term in (\ref{eq:transformationofderivativeofvector}), which tells us that in general the derivative of a tensor doesn't transform in a ``nice" tensorial way under general coordinate transformations.  We then took a very simple vector field $\bf w\it$ in polar coordinates and looked at how the frame changes under translations (cf. equations (\ref{eq:usubrdoesntchange}), (\ref{eq:howusubthetachangeswithr}), and (\ref{eq:uranduphiwhenqisalongtheta})).  By using these terms we were able to construct a well-defined parallel transported vector from one tangent space to another, and using this we constructed a derivative that took into consideration the changes in the coordinates (cf equation (\ref{eq:summaryofcovderivsinfirstexample})).  We called this derivative the covariant derivative because we claimed that it would transform covariantly.  

We then took one of the components of the covariant derivative and, knowing the transformation laws of each part (${\partial w^r \over \partial \phi}$ and $w^{\phi}r$), wrote out the transformation law for $\nabla_{\phi} w^r$, and found that the additional term ($-w^{\phi}r$) transformed in a way that exactly cancelled the additional non-tensorial term in (\ref{eq:transformationofderivativeofvector}), proving that the covariant derivative does in fact transform covariantly as the name suggests.  

We want to generalize this approach to constructing covariant derivatives, but before doing so we make a few comments.  As we said, (\ref{eq:transformationofderivativeofvector}) suggests that the problem with a normal derivative is that you get an additional non-linear term when you transform a derivative.  The form of each term in (\ref{eq:summaryofcovderivsinfirstexample}) reveals what we will find to be a standard pattern - the covariant derivative is the normal partial derivative plus some other term.  And as equations (\ref{eq:wayadditionaltermtransformsfirstexample}) and (\ref{eq:covderivtransformedwithconnectionsubtractingoffbadterm}) indicate, the additional term will transform in a way that exactly cancels the additional non-linear term from the partial derivative.  

We will discuss all this in much greater detail and generality in later sections, but for now we comment that because the additional terms that are added onto the partial derivative to form the covariant derivative provide a way of comparing tensors in different tangent spaces in a well-defined way, or in other words of ``connecting" tangent spaces, we call the additional terms the \bf connection\label{pagewithfirstuseofthewordconnection} \rm terms.  

\subsection{Connections and The Covariant Derivative}
\label{sec:connections}

We now continue with a more quantitative explanation of what we did in the previous section.  We will, in essence, repeat much of what we did previously, but more abstractly and with more generality.  

As we said above, our goal is to construct a derivative via some way of comparing vectors in different tangent spaces in a well-defined way.  This will entail finding a way of taking a vector at $p$ in $T_p\mathcal{M}$ and mapping it to a vector at $p'$ in $T_{p'}\mathcal{M}$.  We can specify this new location as we did before via some curve $p = \tilde q(\tau,p)$.  Therefore, instead of (\ref{eq:curvefreedefinitionofpushforward}):
\begin{eqnarray}
v^i {\partial \over \partial x^i} \longmapsto v^j {\partial q^i \over \partial x^j} {\partial \over \partial x^i}
\end{eqnarray}
we will take a more general approach.  The basic content of (\ref{eq:curvefreedefinitionofpushforward}) is 
\begin{eqnarray}
v^i {\partial \over \partial x^i} \longmapsto v^i {\partial \over \partial y^i}
\end{eqnarray}
where 
\begin{eqnarray}
{\partial \over \partial y^i} \equiv {\partial q^j \over \partial x^i} {\partial \over \partial x^j}
\end{eqnarray}
In other words, this was a prescription for changing the frame when moving from one point to another (and therefore from one tangent space to another).  The use of the tangent mapping induced by $\tilde q(\tau,p)$ provided a way of ``connecting" two tangent spaces.  

To generalize this, consider a vector field $\bf v\it(p)$.  We want to compare the value at point $p$ and at point $\tilde q(\epsilon,p)$.  Or in coordinates, at point $\bf q\it(\bf x\it(p))$ and $\bf q\it(\epsilon,\bf x\it(p))
$.  But we can naturally rewrite $\bf q\it(\epsilon,\bf x\it(p))$ as we did in (\ref{eq:introducethedeltabfxit})
\begin{eqnarray}
\bf q\it(\epsilon,\bf x\it(p)) = \bf x\it(p) + \delta \bf x\it
\end{eqnarray}
for some appropriate $\delta \bf x\it$ (cf (\ref{eq:expandtofirstorderliederiv})).  In components this will be
\begin{eqnarray}
q^i(\epsilon, \bf x\it(p)) = x^i(p) + \delta x^i
\end{eqnarray}
Now we can write the two values of our vector field $\bf v\it(p)$ as
\begin{eqnarray}
\bf v\it(p) &=& \bf v\it(\bf x\it(p)) \nolabel \\
\bf v\it(\tilde q(\epsilon,p)) &=& \bf v\it (\bf q\it(\epsilon,\bf x\it(p))) = \bf v\it(\bf x\it(p) +  \delta \bf x\it)
\end{eqnarray}
The derivative in the $x^k$ direction will now be
\begin{eqnarray}
\lim_{\delta  \bf x\it \rightarrow 0} {1\over \delta  x^k} \big[\bf v\it(\bf x\it(\tilde q(\epsilon,p)))- \bf v\it(\bf x\it(p) + \delta \bf x\it)\big] \label{eq:generalderivativeforcovder}
\end{eqnarray}
Just as in section \ref{sec:flowandlie}, we have the problem that this is a valid form for a derivative in principle but there is no natural way of comparing the tangent space at $p$ to the tangent space at $\tilde q(\tau,p)$.  To fix the problem in section \ref{sec:flowandlie} we used the tangent mapping induced by the curve $\tilde q(\tau,p)$ to write $\bf v\it(\bf x\it(p) + \delta \bf x\it)$ as a vector in $T_p\mathcal{M}$ (as indicated by (\ref{eq:curvefreedefinitionofpushforward})).  And as we pointed out above, (\ref{eq:curvefreedefinitionofpushforward}) is simply a way of ``connecting" vectors in two tangent spaces.  

Now we can generalize.  Instead of using the tangent mapping induced by $\tilde q(\tau,p)$, let's say that
\begin{eqnarray}
v^i{\partial \over \partial x^i} \longmapsto v^i{\partial \over \partial y^i}
\end{eqnarray}
where the new frame is still some linear combination of the old frame.  Specifically, let's say
\begin{eqnarray}
v^i(\bf x\it(p) +  \delta \bf x\it) = v^i(\bf x\it(p)) -  \Gamma_{jk}^i \delta x^j v^k(\bf x\it(p)) \label{eq:introtogamma}
\end{eqnarray}
where the $\Gamma^i_{jk}$ are some particular set of functions (which may depend on $p$).  The minus sign is for later convenience.  The content of this is that if we take the vector $\bf v\it(\bf x\it(p) +  \delta \bf x\it)$, which is in the tangent space $T_{\tilde q(\epsilon,p)}\mathcal{M}$, and map it to the tangent space $T_p\mathcal{M}$, the new components are the old components plus a term proportional to the old components and the displacement.  Because $\Gamma^i_{jk}$ can be \it any \rm set of constants, this is a general expression.  

On the other hand, we can use the standard Taylor expansion for the second term in (\ref{eq:generalderivativeforcovder}):
\begin{eqnarray}
v^i(\bf x\it(\tilde q(\epsilon,p))) &=& v^i(\bf q\it(\epsilon,\bf x\it(p))) \nolabel \\
&=& v^i(\bf x\it(p) + \delta \bf x\it) \nolabel \\
&=& v^i(\bf x\it(p)) + \delta x^j{\partial v^i(\bf x\it(p)) \over \partial x^j} \label{eq:taylorofv}
\end{eqnarray}

Plugging (\ref{eq:introtogamma}) and (\ref{eq:taylorofv}) into (\ref{eq:generalderivativeforcovder}) (in component form), 
\begin{eqnarray}
& &\lim_{\delta \bf x\it  \rightarrow 0} {1\over \delta x^k} \big[\bf v\it(\bf x\it(\tilde q(\epsilon,p)))- \bf v\it(\bf x\it(p) + \delta \bf x\it) \big] \nolabel \\
&& =\lim_{\delta \bf x\it \rightarrow 0} {1\over \delta x^k} \bigg[ v^i(\bf x\it(p)) + \delta x^j {\partial v^i(\bf x\it(p)) \over \partial x^j}  - v^i(\bf x\it(p)) + \Gamma^i_{jn} \delta x^j v^n (\bf x\it(p))\bigg] \nolabel \\
&& =\lim_{\delta \bf x\it \rightarrow 0} {1\over \delta x^k} \delta x^j \bigg[  {\partial v^i \over \partial x^j} + \Gamma^i_{jn} v^n\bigg] \label{eq:takingcovderivwithtaylorexpansionandgamma}
\end{eqnarray}
The only term that will survive the $\delta \bf x\it \rightarrow 0$ limit is the $j=k$ term, so finally the $k^{th}$ derivative is
\begin{eqnarray}
{\partial v^i \over \partial x^k} + \Gamma^i_{kn}v^n \label{eq:firstexampleofacovariantderivativeofavectorfield}
\end{eqnarray}
Notice that this has the form we expect from the previous section - the partial derivative plus an extra term.  This is the general form of the covariant derivative, and the terms $\Gamma^i_{kn}$ are the connection terms because they provides a particular way of connecting two tangent spaces.  

We denote the covariant derivative of $v^i{\partial \over \partial x^i}$ in the $x^j$ direction as
\begin{eqnarray}
\nabla_j \bf v\it = \nabla_j \bigg(v^i{\partial \over \partial x^i}\bigg) \equiv \bigg({\partial v^i \over \partial x^j} + \Gamma^i_{jk}v^k\bigg){\partial \over \partial x^i} \label{eq:simpleexpressionforcovariantderivativeofsinglevector}
\end{eqnarray}
More generally we can find the derivative of a vector $\bf u\it = u^i{\partial \over \partial x^i}$ in the direction of a vector $\bf v\it = v^i {\partial \over \partial x^i}$:
\begin{eqnarray}
\nabla_{\bf v\it} \bf u\it = \nabla_{\bf v\it} \bigg(u^i{\partial \over \partial x^i}\bigg) =v^j \nabla_j \bigg( u^i {\partial \over \partial x^i}\bigg) = v^j\bigg({\partial u^i \over \partial x^j} + \Gamma^i_{jk}  u^k\bigg) {\partial \over \partial x^i} \label{eq:covariantderivativeintrosectionderivofvecalongvec}
\end{eqnarray}

Next we find the covariant derivative of a form $\boldsymbol\omega = \omega_idx^i$.  We know that $\nabla_{\bf v\it}$ is a derivative and we therefore expect that it obeys the Leibnitz rule.  So, taking the inner product of $\boldsymbol\omega$ with a vector $\bf v\it = v^i{\partial \over \partial x^i}$, we have
\begin{eqnarray}
& &\nabla_{\bf u\it} (\omega_i v^i) = \omega_i(\nabla_{\bf u\it} v^i) + (\nabla_{\bf u\it} \omega_i) v^i \nolabel \\
\Longrightarrow & & u^j{\partial \over \partial x^j} (\omega_i v^i) = \omega_i u^j\bigg({\partial v^i \over \partial x^j} + \Gamma^i_{jk}v^k\bigg) + (\nabla_{\bf u\it} \omega_i)v^i \nolabel \\
\Longrightarrow & & u^j\omega_i {\partial v^i \over \partial x^j} + u^jv^i {\partial \omega_i \over \partial x^j} = \omega_iu^j{\partial v^i \over \partial x^j} + \omega_i u^j \Gamma^i_{jk} v^k + (\nabla_{\bf u\it} \omega_i)v^i \nolabel \\
\Longrightarrow & & (\nabla_{\bf u\it} \omega_i)v^i = u^jv^i{\partial \omega_i \over \partial x^j} - \omega_i u^j \Gamma^i_{jk} v^k \nolabel
\end{eqnarray}
The left hand side of the second line comes from the fact that the covariant derivative of a scalar ($\omega_iv^i$ is a scalar) is simply the derivative - there are no indices to contract with the connection and therefore the connection does not have an effect on a scalar).  Then, we can take $v^i = 1$ for one value of $i=n$ and $0$ for the rest, leaving
\begin{eqnarray}
(\nabla_{\bf u\it} \omega_n) = u^j\bigg({\partial \omega_n \over \partial x^j} - \omega_i \Gamma^i_{jn}\bigg)
\end{eqnarray}
Or more formally
\begin{eqnarray}
\nabla_{\bf u\it} \omega_i dx^i = u^j\bigg({\partial \omega_i \over \partial x^j} - \omega_k \Gamma^k_{ji}\bigg) dx^i
\end{eqnarray}

We can easily extrapolate this definition to an arbitrary tenor of rank $(m,n)$ as follows:
\begin{eqnarray}
\nabla_i T^{j_1,j_2,\ldots,j_m}_{k_1,k_2,\ldots,k_n} &=& {\partial \over \partial x^i} T^{j_1,j_2,\ldots,j_m}_{k_1,k_2,\ldots,k_n} \nolabel \\
& & + \Gamma^{j_1}_{ip}T^{p,j_2,\ldots,j_m}_{k_1,k_2,\ldots,k_n} + \Gamma^{j_2}_{ip}T^{j_1,p,j_3,\ldots,j_m}_{k_1,k_2,\ldots,k_n} + \cdots + \Gamma^{j_m}_{ip}T^{j_1,j_2,\ldots,p}_{k_1,k_2,\ldots,k_n} \nolabel \\
& & - \Gamma^p_{ik_1}T^{j_1,j_2,\ldots,j_m}_{p,k_2,\ldots,k_n} - \Gamma^p_{ik_2}T^{j_1,j_2,\ldots,j_m}_{k_1,p,k_3,\ldots,k_n} - \cdots - \Gamma^p_{ik_n}T^{j_1,j_2,\ldots,j_m}_{k_1,k_2,\ldots,p} \nolabel \\ \label{eq:mostgeneralcovderoftensorpossible}
\end{eqnarray}

Finally, notice from (\ref{eq:simpleexpressionforcovariantderivativeofsinglevector}) that we can find the covariant derivative of an individual basis vector $\bf e\it_i = \delta^j_i {\partial \over \partial x^j}$:
\begin{eqnarray}
\nabla_k \bf e\it_i = \bigg( {\partial \delta ^j_i \over \partial x^k} + \Gamma^j_{kl} \delta^l_i\bigg){\partial \over \partial x^j} = \Gamma^j_{ki} {\partial \over \partial x^j} = \Gamma^j_{ki} \delta^l_j{\partial \over \partial x^l} = \Gamma^j_{ki} \bf e\it_j \label{eq:covderivofbasisvecs}
\end{eqnarray}
So as we move around in $\mathcal{M}$ the basis vectors change.  The \it way \rm a given basis vector $\bf e\it_i\in T_p\mathcal{M}$ is changing at $p \in \mathcal{M}$ is again a vector (the derivative).  And because the derivative is a vector in $T_p\mathcal{M}$, we can write it in terms of the basis of $T_p\mathcal{M}$, which is simply a linear combination of the basis vectors of $T_p\mathcal{M}$.  So we can think of $\Gamma^i_{jk}$ as the $i^{th}$ component of the derivative of $\bf e\it_k$ with respect to $x^j$.  We can actually think of (\ref{eq:covderivofbasisvecs}) as the \it definition \rm of the connection coefficients.  

For clarity, the example we did in the previous section with the vector field $\bf w\it$ in polar coordinates gave us (cf. equation (\ref{eq:introtogamma}))
\begin{eqnarray}
& & \Gamma^r_{rr} = \Gamma^r_{r\phi} = \Gamma^{\phi}_{rr} = \Gamma^r_{\phi r} = \Gamma^{\phi}_{\phi \phi}=  0 \nolabel \\
& & \Gamma^{\phi}_{r\phi} = \Gamma^{\phi}_{\phi r} = {1 \over r} \nolabel \\
& & \Gamma^r_{\phi \phi} = -r \label{eq:christoffelsymbolsforpolarcoordinates}
\end{eqnarray}
and in Cartesian coordinates the connection was trivial:
\begin{eqnarray}
\Gamma^i_{jk} = 0 \; \forall i,j,k \label{eq:christoffelsymbolsforcartesiancoordinates}
\end{eqnarray}

\subsection{Transformation of the Connection}
\label{sec:transformationoftheconnection}

In our primary example in section \ref{sec:paralleltransport} the connection term for $\nabla_{\phi} w^r$ was $-w^{\phi} r$.  We saw in (\ref{eq:covderivtransformedwithconnectionsubtractingoffbadterm}) that under the transformation from polar to Cartesian coordinates this term created a ``counter-term" that exactly cancelled the non-linear/non-tensorial term created when ${\partial w^r \over \partial \phi}$ transformed.  We claimed in section \ref{sec:paralleltransport} that this is indicative of a general pattern - that the transformation law of the connection cancels out the non-tensorial transformation of the partial derivative.  Now we prove this.  

We saw in (\ref{eq:transformationofderivativeofvector}) that, in general
\begin{eqnarray}
{\partial w^i \over \partial y^j} = {\partial x^k \over \partial y^j} {\partial y^i \over \partial x^l} {\partial v^l \over \partial x^k} +  {\partial x^k \over \partial y^j} {\partial^2 y^i \over \partial x^k \partial x^l} v^l \label{eq:howconnectiontransformsfirst}
\end{eqnarray}
where $\bf w\it$ is the vector in coordinates $y^i$ and $\bf v\it$ is the vector in coordinates $x^i$.  Let's denote the connection for the $x$ coordinate system $\Gamma^i_{jk}$, and the connection for the $y$ coordinate system $\tilde \Gamma^i_{jk}$.  If we then denote the basis vectors
\begin{eqnarray}
\bf e\it_i = {\partial \over \partial x^i} \qquad \rm and \it \qquad \bf a\it_i = {\partial \over \partial y^i}
\end{eqnarray}
we can write (cf. (\ref{eq:covderivofbasisvecs})) 
\begin{eqnarray}
\nabla_{\bf e\it_i} \bf e\it_j = \Gamma^k_{ij} \bf e\it_k \label{eq:transformconn1}
\end{eqnarray}
However, using (\ref{eq:changeofbasisvectorusingchainrulethingy}) we can write 
\begin{eqnarray}
\bf e\it_j = {\partial \over \partial x^j} = {\partial y^i \over \partial x^j} {\partial \over \partial y^i} = {\partial y^i \over \partial x^j} \bf a\it_i
\end{eqnarray}
Plugging this into both sides of (\ref{eq:transformconn1}) we get
\begin{eqnarray}
\nabla_{\bf e\it_i} \bf e\it_j = \Gamma^k_{ij} \bf e\it_k &\Rightarrow & \nabla_{\bf e\it_i}  \bigg({\partial y^l \over \partial x^j}\bf a\it_l\bigg)  = \Gamma^k_{ij} {\partial y^l \over \partial x^k} \bf a\it_l \nolabel \\
&\Rightarrow& \nabla_{\bf e\it_i}\bigg({\partial y^l \over \partial x^j}\bigg) \bf a\it_l + \bigg({\partial y^p \over \partial x^j} \bigg) \nabla_{\bf e\it_i} \bf a\it_p = \Gamma^k_{ij} {\partial y^l \over \partial x^k} \bf a\it_l \nolabel \\
&\Rightarrow& {\partial^2 y^l \over \partial x^i \partial x^j} \bf a\it_l + {\partial y^p \over \partial x^j} {\partial y^k \over \partial x^i} \nabla_{\bf a\it_k} \bf a\it_p = \Gamma^k_{ij} {\partial y^l \over \partial x^k} \bf a\it_l \nolabel \\
&\Rightarrow & {\partial^2 y^l \over \partial x^i \partial x^j} \bf a\it_l + {\partial y^p \over \partial x^j}{\partial y^k \over \partial x^i}\tilde \Gamma^l_{kp} \bf a\it_l = \Gamma^k_{ij} {\partial y^l \over \partial x^k} \bf a\it_l \nolabel \\
&\Rightarrow & {\partial^2 y^l \over \partial x^i \partial x^j} + {\partial y^p \over \partial x^j}{\partial y^k \over \partial x^i}\tilde \Gamma^l_{kp}  = \Gamma^k_{ij} {\partial y^l \over \partial x^k} \nolabel \\
&\Rightarrow & {\partial^2 y^l \over \partial x^i \partial x^j} {\partial x^q \over \partial y^l} + {\partial y^p \over \partial x^j} {\partial y^k \over \partial x^i} {\partial x^q \over \partial y^l}\tilde \Gamma^l_{kp} = \Gamma^k_{ij} {\partial y^l \over \partial x^k} {\partial x^q \over \partial y^l} = \Gamma^k_{ij} \delta^q_k
\end{eqnarray}
Or
\begin{eqnarray}
\Gamma^q_{ij} = {\partial y^p \over \partial x^j} {\partial y^k \over \partial x^i} {\partial x^q \over \partial y^l} \tilde \Gamma^l_{kp} + {\partial^2 y^l \over \partial x^i \partial x^j} {\partial x^q \over \partial y^l} \label{eq:howconnectiongammatransforms}
\end{eqnarray}
This is the transformation law for a connection coefficient - notice that the first term on the right hand side is what we would expect if $\Gamma^i_{jk}$ did transform like a tensor.  The second term on the right is the non-linear term that prevents it from being tensorial.  

So, starting with a general covariant derivative (\ref{eq:covariantderivativeintrosectionderivofvecalongvec}), we can transform it using (\ref{eq:howconnectiongammatransforms}) and (\ref{eq:howconnectiontransformsfirst}),\footnote{For notational simplicity we are taking everything with a tilde to be in $y$ coordinates and everything without a tilde to be in the $x$ coordinates.},\footnote{We will be making extensive use of the identity 
\begin{eqnarray}
\delta^i_j = {\partial x^i \over \partial y^k} {\partial y^k \over \partial x^j} \label{eq:identityfordeltafunctionintermsofpartialderivativescontractedtogether}
\end{eqnarray}
etc.}
\begin{eqnarray}
\tilde \nabla_{\bf \tilde u\it} \bf \tilde w\it &=& \tilde u^j \bigg({\partial \tilde w^i \over \partial y^j} + \tilde \Gamma^i_{jk} \tilde w^k\bigg){\partial \over \partial y^i} \nolabel \\
&=& {\partial y^j \over \partial x^{\phi}} u^{\phi} \bigg({\partial y^i \over \partial x^{\alpha}} {\partial x^{\beta} \over \partial y^j} {\partial w^{\alpha} \over \partial x^{\beta}} + {\partial x^{\alpha} \over \partial y^j} {\partial^2 y^i \over \partial x^{\alpha} \partial x^{\beta}} w^{\beta} \nolabel \\
& & + \bigg[{\partial x^{\alpha} \over \partial y^k} {\partial x^{\beta} \over \partial y^j} {\partial y^i \over \partial x^{\alpha}} \Gamma^{\gamma}_{\beta \alpha} + {\partial y^i \over \partial x^{\gamma}} {\partial^2 x^{\gamma} \over \partial y^j \partial y^k} \bigg] {\partial y^k \over \partial x^{\delta}} w^{\delta}\bigg) {\partial x^{\epsilon} \over \partial y^i} {\partial \over \partial x^{\epsilon}} \nolabel \\
&=& u^{\phi} {\partial y^j \over \partial x^{\phi}} {\partial y^i \over \partial x^{\alpha}} {\partial x^{\beta} \over \partial y^j} {\partial x^{\epsilon} \over \partial y^i} {\partial w^{\alpha} \over \partial x^{\beta}} {\partial \over \partial x^{\epsilon}} \nolabel \\
& & + u^{\phi} {\partial y^j \over \partial x^{\phi}} {\partial x^{\alpha} \over \partial y^j}{\partial x^{\epsilon} \over \partial y^i}{\partial^2 y^i \over \partial x^{\alpha} \partial x^{\beta}}w^{\beta} {\partial \over \partial x^{\epsilon}} \nolabel \\
& & + u^{\phi} {\partial y^j \over \partial x^{\phi}}{\partial x^{\alpha} \over \partial y^k}{\partial x^{\beta} \over \partial y^j}{\partial y^i \over \partial x^{\gamma}}{\partial y^k \over \partial x^{\delta}}{\partial x^{\epsilon} \over \partial y^i}\Gamma^{\gamma}_{\beta \alpha} w^{\delta} {\partial \over \partial x^{\epsilon}} \nolabel \\
& & + u^{\phi} {\partial y^j \over \partial x^{\phi}}{\partial y^i \over \partial x^{\gamma}}{\partial y^k \over \partial x^{\delta}}{\partial x^{\epsilon} \over \partial y^i}{\partial^2 x^{\gamma} \over \partial y^j \partial y^k}w^{\delta} {\partial \over \partial x^{\epsilon}} \nolabel \\
&=& u^{\phi}\delta^{\beta}_{\phi} \delta^{\epsilon}_{\alpha} {\partial w^{\alpha} \over \partial x^{\beta}} {\partial \over \partial x^{\epsilon}} + u^{\phi} \delta^{\alpha}_{\phi} {\partial x^{\epsilon} \over \partial y^i} {\partial^2 y^i \over \partial x^{\alpha} \partial x^{\beta}} w^{\beta} {\partial \over \partial x^{\epsilon}} \nolabel \\
& & + u^{\phi} \delta^{\beta}_{\phi} \delta^{\alpha}_{\delta} \delta^{\epsilon}_{\gamma}\Gamma^{\gamma}_{\beta \alpha} w^{\delta} {\partial \over \partial x^{\epsilon}} + u^{\phi} {\partial y^j \over \partial x^{\phi}} \delta^{\epsilon}_{\gamma} {\partial y^k \over \partial x^{\delta}}{\partial^2 x^{\gamma} \over \partial y^j \partial y^k} w^{\delta} {\partial \over \partial x^{\epsilon}} \nolabel \\
&=& u^{\beta} {\partial w^{\alpha} \over \partial x^{\beta}} {\partial \over \partial x^{\alpha}} + u^{\alpha} {\partial x^{\epsilon} \over \partial y^i} {\partial^2 y^i \over \partial x^{\alpha} \partial x^{\beta}} w^{\beta} {\partial \over \partial x^{\epsilon}} + u^{\beta} \Gamma^{\epsilon}_{\beta \alpha} w^{\alpha} {\partial \over \partial x^{\epsilon}} + u^{\phi} {\partial y^j \over \partial x^{\phi}} {\partial y^k \over \partial x^{\delta}} {\partial^2 x^{\gamma} \over \partial y^j \partial y^k} w^{\delta} {\partial \over \partial x^{\gamma}} \nolabel \\
&=& u^{\beta}\bigg({\partial w^{\alpha} \over \partial x^{\beta}} + \Gamma^{\alpha}_{\beta \delta} v^{\delta}\bigg) {\partial \over \partial x^{\alpha}} + u^{\alpha}\bigg[ {\partial x^{\delta} \over \partial y^{\gamma}}{\partial^2 y^{\gamma} \over \partial x^{\alpha} \partial x^{\beta}} + {\partial y^{\gamma} \over \partial x^{\alpha}}{\partial y^{\epsilon} \over \partial x^{\beta}}{\partial^2 x^{\delta} \over \partial y^{\gamma} \partial y^{\epsilon}}\bigg] w^{\beta} {\partial \over \partial x^{\delta}} \nolabel \\
&=& \nabla_{\bf u\it} \bf w\it + u^{\alpha}\bigg[ {\partial x^{\delta} \over \partial y^{\gamma}}{\partial^2 y^{\gamma} \over \partial x^{\alpha} \partial x^{\beta}} + {\partial y^{\epsilon} \over \partial x^{\beta}}{\partial y^{\gamma} \over \partial x^{\alpha}}{\partial^2 x^{\delta} \over \partial y^{\gamma} \partial y^{\epsilon}}\bigg] w^{\beta} {\partial \over \partial x^{\delta}} \label{eq:covderiviscovwithextratermthatwillvanish}
\end{eqnarray}
So, if the second term in square brackets is equal to $0$, then the covariant derivative does in fact transform properly.  With that in mind, let's look at the second term in the square brackets:
\begin{eqnarray}
{\partial y^{\epsilon} \over \partial x^{\beta}} {\partial y^{\gamma} \over \partial x^{\alpha}} {\partial^2 x^{\delta} \over \partial y^{\gamma} \partial y^{\epsilon}} &=& {\partial y^{\epsilon} \over \partial x^{\beta}} {\partial y^{\gamma} \over \partial x^{\alpha}} {\partial \over \partial x^{\gamma}} {\partial x^{\delta} \over \partial y^{\epsilon}} \nolabel \\
&=& {\partial y^{\epsilon} \over \partial x^{\beta}} {\partial^2 x^{\delta} \over \partial x^{\alpha} \partial y^{\epsilon}} \label{eq:secondterminvarianceofcovderivderiv}
\end{eqnarray}
Now, starting with (\ref{eq:identityfordeltafunctionintermsofpartialderivativescontractedtogether}), we can find the identity
\begin{eqnarray}
\delta^i_j &=& {\partial x^i \over \partial y^k} {\partial y^k \over \partial x^j} \nolabel \\
\Rightarrow  0 &=& {\partial \over \partial x^l}\bigg[ {\partial x^i \over \partial y^k} {\partial y^k \over \partial x^j}\bigg] \nolabel \\
&=& {\partial x^i \over \partial y^k} {\partial^2 y^k \over \partial x^l \partial x^j} + {\partial y^k \over \partial x^j} {\partial^2 x^i \over \partial x^l \partial y^k} \nolabel \\
\Rightarrow {\partial x^i \over \partial y^k} {\partial^2 y^k \over \partial x^l \partial x^j} &=& -{\partial y^k \over \partial x^j} {\partial^2 x^i \over \partial x^l \partial y^k} \label{eq:secondterminvarianceofcovderivderiv2}
\end{eqnarray}
Plugging this into (\ref{eq:secondterminvarianceofcovderivderiv}) we have
\begin{eqnarray}
{\partial y^{\epsilon} \over \partial x^{\beta}} {\partial^2 x^{\delta} \over \partial x^{\alpha} \partial y^{\epsilon}} = -{\partial x^{\delta} \over \partial y^{\gamma}} {\partial^2 y^{\gamma} \over \partial x^{\alpha} \partial x^{\beta}}
\end{eqnarray}
which exactly cancels the first term in the square brackets in (\ref{eq:covderiviscovwithextratermthatwillvanish}).  So, we finally have
\begin{eqnarray}
\tilde \nabla_{\bf \tilde u\it} \bf \tilde w\it = \nabla_{\bf u \it} \bf w\it
\end{eqnarray}
The covariant derivative is in fact covariant.  Neither the partial derivative nor the connection transform in a covariant way, but when they are added together, the non-linear parts exactly cancel making the sum covariant.  

\subsection{Other Connections}
\label{sec:otherconnections}

Admittedly there is something a bit arbitrary about how we defined ``parallel transport" in section \ref{sec:paralleltransport}.  We said there that a vector is parallel transported along a line in such a way that it maintains its direction in $\mathbb{R}^2$.  While this may seem like the only (or at least the most) sensible way of defining it, keep in mind that we are ultimately interested in much more general manifolds than $\mathbb{R}^2$.  Before generalizing again to an arbitrary manifold (as we did in sections \ref{sec:connections} and \ref{sec:transformationoftheconnection}), let's consider an alternative connection in $\mathbb{R}^2$.  

Rather than demand that a transported vector keep its direction constant in $\mathbb{R}^2$, let's say (for example) that the transported vector retains the angle it makes with the radial vector:
\begin{center}
\includegraphics[scale=.6]{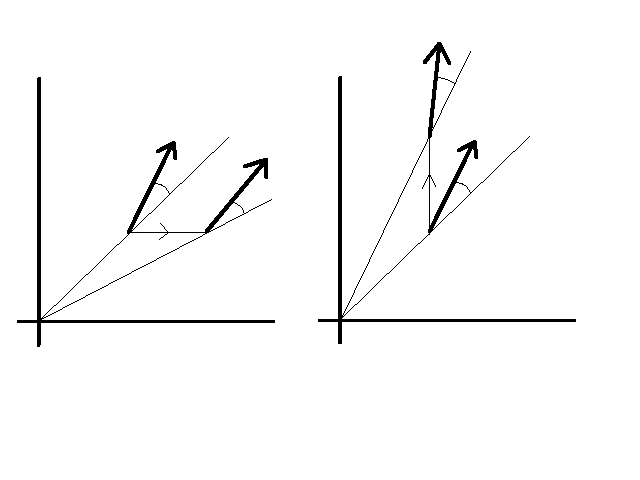}
\end{center}
We can work out the details of this in a similar way as we did to get (\ref{eq:transformationofuincartcords}), (\ref{eq:usubrdoesntchange}), (\ref{eq:howusubthetachangeswithr}), (\ref{eq:uranduphiwhenqisalongtheta}).  Rather than work through the details, we merely cite the results (you should work this out to verify on your own).  Starting with Cartesian coordinates, moving a small distance in the $x$ direction ($x\rightarrow x+ \delta x$) gives
\begin{eqnarray}
u^x \rightarrow u^x + \delta x {y \over x^2 + y^2} u^y \nolabel \\
u^y \rightarrow u^y - \delta x {y \over x^2 + y^2} u^x
\end{eqnarray}
And moving a small distance in the $y$ direction ($y\rightarrow y+ \delta y$) gives
\begin{eqnarray}
u^x \rightarrow u^x - \delta y {x \over x^2 + y^2} u^y \nolabel \\
u^y \rightarrow u^y + \delta y {x \over x^2 + y^2} u^x
\end{eqnarray}
So, in the same way that we got (\ref{eq:christoffelsymbolsforpolarcoordinates}), we now have
\begin{eqnarray}
\Gamma^1_{12} = \Gamma^x_{xy} &=& - {y \over x^2 + y^2}\nolabel \\
\Gamma^2_{11} = \Gamma^y_{xx} &=& {y \over x^2 + y^2} \nolabel \\
\Gamma^1_{22} = \Gamma^x_{yy} &=& {x \over x^2 + y^2} \nolabel \\
\Gamma^2_{21} = \Gamma^y_{y1} &=& -{x \over x^2 + y^2} 
\end{eqnarray}
and all the rest are $0$.  

Now we want to find the connection coefficients in polar coordinates.\footnote{One might be tempted to merely transform them directly to polar coordinates, for example
\begin{eqnarray}
{x \over x^2+y^2} = {r\cos\phi \over r^2 \cos^2\phi + r^2 \sin^2\phi} = {\cos\phi \over r}
\end{eqnarray}
However this would merely be the polar coordinate representation of the change of an $x$ direction basis vector \it in the $y$ direction\rm.  To find the change of, say, an $r$ direction basis vector in the $\theta$ direction, we must use the standard transformation law for the connection coefficients, equation (\ref{eq:howconnectiongammatransforms}).}  To do this we use (\ref{eq:howconnectiongammatransforms}).  Again, we encourage you to work this out on your own, but we merely quote the results.  Under $r \rightarrow r+\delta r$,
\begin{eqnarray}
u^r &\rightarrow& u^r \nolabel \\
u^{\phi} &\rightarrow& u^{\phi} 
\end{eqnarray}
And under $\phi \rightarrow \phi+ \delta \phi$, 
\begin{eqnarray}
u^r &\rightarrow& u^r \nolabel \\
u^{\phi} &\rightarrow& u^{\phi} + {1 \over r} \delta \phi u^r
\end{eqnarray}
So the only non-zero connection coefficient is
\begin{eqnarray}
\Gamma^{\phi}_{\phi r} = -{1 \over r}
\end{eqnarray}

The important thing to realize is that this connection is just as good as any other.  The connection doesn't explicitly say anything about the space - rather the connection defines our arbitrary convention for moving vectors around to take derivatives.  We get to make these rules up however we want - "parallel" can mean anything we want it to mean.  

We will however see soon that on manifolds with metrics there is one connection that has a special meaning.  We will discuss this later.  

\subsection{Geodesics}
\label{sec:geodesics}

As a brief review, recall that when we started our exposition of calculus on manifolds in section \ref{sec:paralleltransport}, part of our fundamental problem was that there is no natural way of comparing vectors in $T_p\mathcal{M}$ to vectors in $T_q\mathcal{M}$ for $p\neq q$.  This ambiguity in comparing vectors in different tangent spaces led to the non-tensorial transformation law (\ref{eq:transformationofderivativeofvector}).  We amended this problem through parallel transportation (cf. page  \pageref{pagewherewetalkaboutparalleltransporting} ff) - a way of defining how a vector at one point is to be moved to a vector at another point.  Then the transported vector can be compared to the actual vector at the new point and a derivative can be defined.  We called the ``instructions" for how to transport a vector the "connection" (cf page \pageref{pagewithfirstuseofthewordconnection} and section \ref{sec:connections}).  We considered two examples of connections in $\mathbb{R}^2$: the trivial connection (\ref{eq:christoffelsymbolsforcartesiancoordinates}) (which was (\ref{eq:christoffelsymbolsforpolarcoordinates})) in polar coordinates) and the non-trivial connection in section \ref{sec:otherconnections}.  The first of these examples lines up with our intuitive notion of ``parallel" transportation, where a vector doesn't change direction as you move it around.  However, as we pointed out, the way we parallel transport is ultimately arbitrary.  Keep in mind that for the more general manifolds we will be interested in later, we don't have the luxury of being able to trust our intuition about things.  There is nothing natural which says that (\ref{eq:christoffelsymbolsforpolarcoordinates}) (keeping their direction the same) provides a better way of moving vectors around $\mathbb{R}^2$ than the connection in section \ref{sec:otherconnections} (keeping the angle with the radial vector the same).  

So what does a connection, or a specific set of instructions about how to parallel transport, tell us?  Recall that the point of parallel transportation is that it provides a way of moving a vector around a manifold ``without changing it" (page \pageref{pagewherewetalkaboutparalleltransporting}).  Or, in the language we have learned to this point, a vector $\bf u\it$ is parallel transported in the direction of a vector $\bf v\it$ if the covariant derivative of $\bf u\it$ in the direction of $\bf v\it$ vanishes (cf equation (\ref{eq:firstcovderivativeexplanation})).  This (obviously) means that if we start with $\bf u\it$ and move in the direction of $\bf v\it$, then $\bf u\it$ won't change - the actual vector at any new point will be equal to the parallel transported vector at the new point.  Of course, by ``won't change" we mean according to our arbitrary definition of what it means for something to not change (notice that (\ref{eq:firstcovderivativeexplanation}) depends on the connection - we can make it anything we want by making our connection anything we want).  

To be more explicit, we may start with a vector $\bf u\it(p)$ (that is part of a vector field) at point $p\in \mathcal{M}$.  Then move to some point $q \in \mathcal{M}$ that is in the $\bf v\it$ direction from $p$, parallel transporting $\bf u\it(p)$ with you using whatever connection you have chosen.  Then, when you get to $q$, find the actual value of $\bf u\it(q)$, and compare it to the $\bf u\it(p)$ you have parallel transported with you.  If they are the same then the covariant derivative of the vector field $\bf u\it$ vanishes in the direction $\bf v\it$ from $p$ to $q$.\footnote{We are, of course, assuming that the path from $p$ to $q$ is infinitesimal - our language is relaxed for simplicity.}  

This discussion allows us to define an important idea.  Imagine that the vector field $\bf u\it(p)$ is defined along a curve $q(\tau):[a,b]\rightarrow \mathcal{M}$ in $\mathcal{M}$ (it can be defined elsewhere on $\mathcal{M}$ as well - we are only interested in the definition along the curve for now).  Let's take $\bf u\it(q(\tau))$ to represent the motion of some particle moving on $\mathcal{M}$, where $\bf u\it(q(\tau))$ is the particle's velocity at the point $q(\tau) \in \mathcal{M}$, and therefore (cf section \ref{sec:flowandlie}).
\begin{eqnarray}
\bf u\it(q(\tau)) = {d \bf q\it(\tau) \over d\tau}
\end{eqnarray}
(where $\bf q\it(\tau)$ is simply the coordinate representation of $q(\tau)$).  

By both physical reasoning and mathematical reasoning, we know that if left alone a particle will simply move in the straightest line it can.  In other words it's not going to change its velocity for no reason and it's going to follow some extremum path in going from one point to another.  In other words, its velocity vector $\bf u\it(q(\tau))$ will be parallel transported at each point, \it in the direction it is moving\rm.  We can capture this more precisely by saying that the particle will travel along the path represented by $\bf u\it(q(\tau))$ if
\begin{eqnarray}
u^i \nabla_i \bf u\it = \nabla_{\bf u\it} \bf u\it = 0 \label{eq:originalequationforgeodesics}
\end{eqnarray}
Using the definition of the covariant derivative (\ref{eq:covariantderivativeintrosectionderivofvecalongvec}) to get
\begin{eqnarray}
\nabla_{\bf u\it} \bf u\it = 0 &\Longrightarrow& u^j\bigg({\partial u^i \over \partial x^j} + \Gamma^i_{jk} u^k\bigg) {\partial \over \partial x^i} = 0 \nolabel \\
&\Longrightarrow& \bigg(u^j {\partial \over \partial x^j} u^i + \Gamma^i_{jk}u^j u^k\bigg){\partial \over \partial x^i} = 0 \nolabel \\
&\Longrightarrow& \bigg( {d q^j \over d\tau} {\partial \over \partial x^j}{ d q^i \over d\tau} + \Gamma^i_{jk} {dq^j \over d\tau} {dq^k \over d\tau} \bigg) {\partial \over \partial x^i} = 0 \nolabel \\
\end{eqnarray}
The values $q^j$ are simply coordinates, so the first term above can be simplified resulting in
\begin{eqnarray}
&\Longrightarrow & \bigg( {d \over d\tau} {d q^i \over d\tau} + \Gamma^i_{jk} {dq^j \over d\tau} {dq^k \over d\tau} \bigg){\partial \over \partial x^i} = 0
\end{eqnarray}
which will vanish in general for
\begin{eqnarray}
{d^2 q^i \over d\tau^2} + \Gamma^i_{jk} {d q^j \over d \tau} {d q^k \over d\tau} = 0 \label{eq:geodesicdifferentialequation}
\end{eqnarray}
for every $i$.  So, (\ref{eq:geodesicdifferentialequation}) is a differential equation for $\bf q\it(\tau)$, and the solutions will be a curve in $\mathcal{M}$ which gives rise to a vector field that is parallel transported along the path $q(\tau)$.  Furthermore, this path will be the equivalent to an extremum path for the particle to follow - or put less formally, the path that corresponds to the ``straightest line" along $\mathcal{M}$.  We call such paths \bf geodesics\rm.  They are the natural path a particle will follow when moving through $\mathcal{M}$, given some particular connection.

Of course, as we have indicated several times, this means that the ``straightest line" the particle will follow is  completely dependent on the connection.  To see a few examples of this, let's once again consider the familiar $\mathbb{R}^2$, starting with the trivial connection (\ref{eq:christoffelsymbolsforcartesiancoordinates}).  Because all of the connection coefficients vanish this will simply give\footnote{We are switching from $q^i$, the coordinates of the points of the curve, to simple coordinates - nothing is lost in this switch because the $q^i$ are nothing more than coordinates to begin with.}
\begin{eqnarray}
{d^2 x \over d\tau^2} = 0 \nolabel \\
{d^2 y \over d\tau^2} = 0  
\end{eqnarray}
Which has the straightforward solutions
\begin{eqnarray}
x(\tau) = A\tau+B \nolabel \\
y(\tau) = C\tau+D
\end{eqnarray}
where $A,B,C,D$ are simply constants of integration.  Obviously the geodesics in this case will be straight lines.  

We can set up the equation with the same connection in polar coordinates (\ref{eq:christoffelsymbolsforpolarcoordinates}), getting
\begin{eqnarray}
& &{d^2 r \over d \tau^2} - r \dot \phi^2 = 0 \nolabel \\
& &{d^2 {\phi} \over d \tau^2} + {2 \over r} \dot r \dot \phi = 0 
\end{eqnarray}
This is harder to solve, but you can show that the solutions are of the form
\begin{eqnarray}
r(t) &=& \sqrt{(A\tau + B)^2 + (C\tau+D)^2} = \sqrt{(q^x)^2+(q^y)^2} \nolabel \\
{\phi}(r) &=& \tan^{-1}\bigg({C\tau+D \over A\tau+B}\bigg) = \tan^{-1}\bigg({q^y \over q^x}\bigg)
\end{eqnarray}
Graphing these will once again give straight lines in $\mathbb{R}^2$.  So, we have shown that the straightest lines in $\mathbb{R}^2$ given $\Gamma^i_{jk} = 0$ (in Cartesian coordinates) will be straight lines.  

But what about the connection in section \ref{sec:otherconnections}?  In this case the Cartesian equations are very difficult to solve, but the polar equations are easy:
\begin{eqnarray}
& &{d^2 {\phi} \over d\tau^2} - {1 \over r} \dot \phi \dot r = 0 \nolabel \\
& &{d^2 r \over d\tau^2} = 0
\end{eqnarray}
We can solve the second easily:
\begin{eqnarray}
r(\tau) = At+B
\end{eqnarray}
Then, plugging this into the first we get
\begin{eqnarray}
\phi(\tau) = {1 \over 2} CA\tau^2 + CB\tau + D
\end{eqnarray}
There are several classes of solutions to this, depending on the values of $A,B,C$ and $D$.  We provide a few examples.  For $A=B=D=1$ and $C=0$ we get straight radial lines from the center:
\begin{center}
\includegraphics[scale=.6]{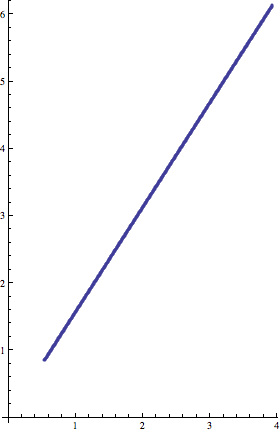} \label{pagewithfirstweirdconnectiongeodescigraphs}
\end{center}
For $A=0$ and $B=C=D=1$ we have circles
\begin{center}
\includegraphics[scale=.6]{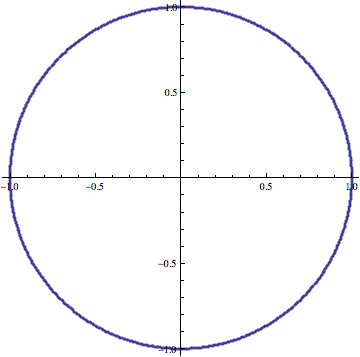}
\end{center}
And for $A=B=C=D=1$ we have spirals
\begin{center}
\includegraphics[scale=.6]{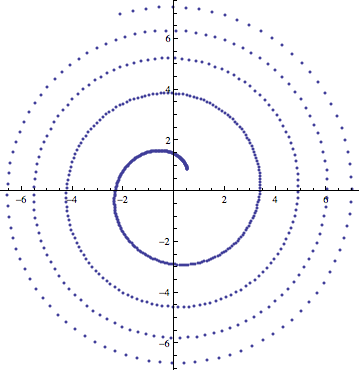}
\end{center}

So by changing the connection we have radically altered the geodesics, or ``straightest lines" in this space.  The physical meaning of the connection is likely not clear at this point.  As we have defined it, the connection is nothing more than a precise statement of our (arbitrary) convention for moving vectors around in order to take a covariant derivative.  It is therefore not obvious why such a convention would have such radical geometric effects.  

While we will explore this in much greater detail later, we briefly address it now.  The point is in the fact that a parallel transport moves the vector ``without changing it".  Considering the connection in section \ref{sec:connections}, on one hand we can think of this as making things move in a strange way around the normal flat $\mathbb{R}^2$ by changing the geodesics.  A physical analogy for this is in electromagnetism.  Consider some section of flat spacetime with a non-zero value for the electromagnetic field.  There is no curvature, and therefore the spacetime geodesics are straight lines.  However a particle that carries electric charge will not move in a straight line through this space - the geodesics it travels on are altered by the presence of the field.  In other words the metric is unchanged as in the example here (the metric in our copy of $\mathbb{R}^2$ in this section is the flat metric), but there is an additional connection added to the space that has nothing to do with the metric, and therefore particles that "see" this connection follow non-straight line geodesics.

As one final comment, consider the geodesic equation (\ref{eq:geodesicdifferentialequation}) (rewritten in terms of the coordinates $x^i$ rather than a path $q^i$)
\begin{eqnarray}
\nabla_{\bf u\it} \bf u\it = {d^2 x^i \over d \tau^2} + \Gamma^i_{jk} {dx^i \over d\tau} {dx^k \over d\tau} = 0
\end{eqnarray}
Let's consider an arbitrary reparameterization
\begin{eqnarray}
\tau \longrightarrow \lambda = \lambda(\tau)
\end{eqnarray}
Under this,
\begin{eqnarray}
{d \over d\tau} \longrightarrow {d \lambda \over d\tau} {d \over d\lambda} = \lambda' {d \over d \lambda}
\end{eqnarray}
And
\begin{eqnarray}
{d^2 \over d\tau^2} \longrightarrow \bigg(\lambda'{d \over d\lambda}\bigg)\bigg(\lambda'{d \over d\lambda}\bigg) = (\lambda')^2 {d^2 \over d\lambda^2} + \lambda' \lambda'' {d \over d\lambda}
\end{eqnarray}
So under this transformation the geodesic equation becomes
\begin{eqnarray}
{d^2 x^i \over d \tau^2} + \Gamma^i_{jk} {dx^j \over d\tau} {dx^k \over d\tau}=0  &\longrightarrow& (\lambda')^2 {d^2 x^i \over d \lambda^2} + \lambda' \lambda'' {d x^i \over d \lambda} + (\lambda')^2\Gamma^i_{jk}  {dx^j \over d\tau} {dx^k \over d\tau} = 0 \nolabel \\
&\longrightarrow& (\lambda')^2 \bigg({d^2 x^i \over d\lambda^2} + \Gamma^i_{jk} {dx^j \over d\lambda} {dx^k \over d\lambda}\bigg) = -\lambda'\lambda'' {dx^i \over d\lambda} \nolabel \\
&\longrightarrow& {d^2 x^i \over d\lambda^2} + \Gamma^i_{jk} {dx^j \over d\lambda} {dx^k \over d\lambda} = -{\lambda'' \over \lambda'} {dx^i \over d\lambda}
\end{eqnarray}
If we set $f\equiv -{\lambda'' \over \lambda'}$ then this is equivalent to
\begin{eqnarray}
\nabla_{\bf u\it} \bf u\it = f \bf u\it \label{eq:generalizedsortofgeodesicconstraint}
\end{eqnarray}
rather than our original constraint (\ref{eq:originalequationforgeodesics}).  The difference is that (\ref{eq:generalizedsortofgeodesicconstraint}) demands that the vector be transported in such a way that it is always parallel to itself.  Equation (\ref{eq:originalequationforgeodesics}) on the other hand demands that it be transported along itself without changing speed.  The solutions to (\ref{eq:originalequationforgeodesics}) and (\ref{eq:generalizedsortofgeodesicconstraint}) will be the same paths, but with different parameterizations.  If we eventually want to take the parameter to be time it will be helpful to choose a parameterization which leads to (\ref{eq:originalequationforgeodesics}) instead of (\ref{eq:generalizedsortofgeodesicconstraint})

\subsection{Torsion and the Relationship Between $\nabla_{\bf u\it}\bf v\it$ and $\mathcal{L}_{\bf u\it}\bf v\it$}
\label{sec:torsionandliederiv}

Before moving on to consider the relationship between the connection and the geometry of a manifold in greater details, we consider a few aspects of connections that we have only considered implicitly so far.  We now make them explicit. 

First, let's briefly review what we said about Lie derivatives in sections \ref{sec:flowandlie} and \ref{sec:anotherperspectiveliederivatives}.\footnote{You are encouraged to reread those sections before moving on.}  The essential idea of a Lie derivatives is that, given two vector fields, we can take the derivative of one (at a point) in the direction of the other.  Because we were talking about about infinitesimal displacements, there was not a problem in talking about moving from a point $p\in \mathcal{M}$ to a point $q(\epsilon,p)$ via the infinitesimal displacement due to a vector:
\begin{eqnarray}
\bf q\it(\epsilon, \bf x\it(p)) = \bf x\it(p) + \epsilon \bf v\it(\bf x\it(p))
\end{eqnarray}
(cf equation (\ref{eq:expandtofirstorderliederiv})).  

Recall that the idea behind a Lie derivative $\mathcal{L}_{\bf u\it}\bf v\it$ is to use the flow induced by $\bf u\it$ to provide a tangent mapping for $\bf v\it$ - it was through this tangent mapping that we were able to compare vectors in different tangent spaces.  The geometrical interpretation of the Lie derivative was then a statement of the failure of a rectangle made from two different paths to close (cf picture on page \pageref{pagewithpictureofliederivativenonclosure} and surrounding discussion, including discussion in section \ref{sec:anotherperspectiveliederivatives}).

Now let's consider how the Lie derivative relates to the covariant derivative.  First we write out both forms:
\begin{eqnarray}
\mathcal{L}_{\bf u\it}\bf v\it &=& [\bf u\it,\bf v\it] =\bigg( u^j {\partial v^i \over \partial x^j} - v^j {\partial u^i\over \partial x^j}\bigg){\partial \over \partial x^i} \nolabel \\
\nabla_{\bf u\it} \bf v\it &=& u^j \bigg({\partial v^i \over \partial x^j} + \Gamma^i_{jk}v^k\bigg) {\partial \over \partial x^i}
\end{eqnarray}
Surely there must be some way of relating them to each other.  The similarity is the presence of partial derivatives.  In $\mathcal{L}_{\bf u\it}\bf v\it$ there is an antisymmetrized pair of partial derivatives, so a natural guess may be to try an antisymmetrized covariant derivative:
\begin{eqnarray}
\nabla_{\bf u\it}\bf v\it - \nabla_{\bf v\it}\bf u\it &=& u^j \bigg({\partial v^i \over \partial x^j} + \Gamma^i_{jk}v^k\bigg) {\partial \over \partial x^i} - v^j \bigg({\partial u^i \over \partial x^j} + \Gamma^i_{jk}u^k\bigg) {\partial \over \partial x^i} \nolabel \\
&=& \bigg(u^j {\partial v^i \over \partial x^j} - v^j {\partial u^i \over \partial x^j} +\Gamma^i_{jk} u^j v^k - \Gamma^i_{jk} v^j u^k\bigg){\partial \over \partial x^i} \nolabel \\
&=& \mathcal{L}_{\bf u\it} \bf v\it + \big(\Gamma^i_{jk} - \Gamma^i_{kj}\big) u^j v^k {\partial \over \partial x^i} \label{eq:firsttimeweseeantisymmsumofconnections}
\end{eqnarray}
So, if the connection is symmetric ($\Gamma^i_{jk} = \Gamma^i_{kj}$), then
\begin{eqnarray}
\nabla_{\bf u\it} \bf v\it - \nabla_{\bf v\it}\bf u\it = \mathcal{L}_{\bf u\it}\bf v\it
\end{eqnarray}

But when the connection is not symmetric we have the relation (in components)
\begin{eqnarray}
\nabla_{\bf u\it} v^i - \nabla_{\bf v\it} u^i -\mathcal{L}_{\bf u\it} v^i = (\Gamma^i_{jk} - \Gamma^i_{kj}) u^jv^k \label{eq:covderliedertorsion}
\end{eqnarray}

So what is the meaning of this antisymmetric part of the connection?  To see it, consider a point $p\in \mathcal{M}$ with a geodesic passing through it in the 
\begin{eqnarray}
\boldsymbol{\epsilon} = \epsilon^i{\partial \over \partial x^i}
\end{eqnarray}
direction ($\boldsymbol{\epsilon}$ is an infinitesimal vector).\footnote{The picture is misleading because we have drawn $\boldsymbol{\epsilon}$ with a finite length, rather than an infinitesimal length.  Think of the $\boldsymbol{\epsilon}$ vector at $p$ that is drawn as merely pointing in the correct direction - the actual vector does not have the finite length that the picture indicates.}
\begin{center}
\includegraphics[scale=.5]{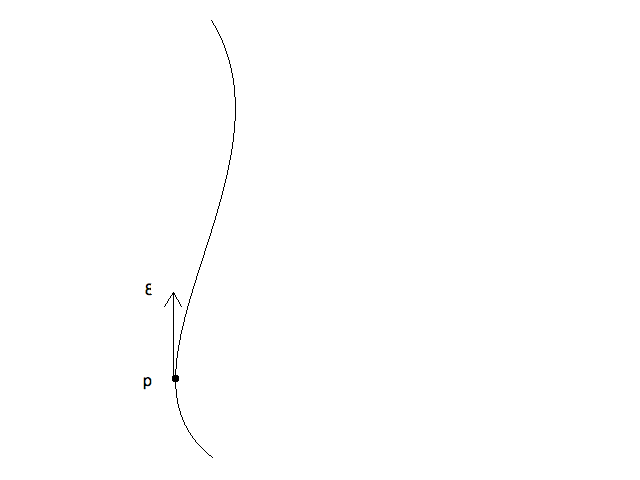}
\end{center}
Now consider another point $q$ separated from $p$ by a small distance $\delta$, or in other words separated from $\bf x\it(p)$ by a small displacement vector $\boldsymbol{\delta}$.\footnote{The vector $\boldsymbol{\delta}$ will act as a displacement vectors in the exact same way as $\bf v\it^{(i)}$ did in equations (\ref{eq:twoinfinitesimalpaths1}) and (\ref{eq:twoinfinitesimalpaths2}).  Again you should think of $\boldsymbol{\delta}$ as being infinitesimal.  We cannot draw an infinitesimal vector, so keep in mind that the points $p$ and $q$ are only separated by a very small distance.}
\begin{center}
\includegraphics[scale=.5]{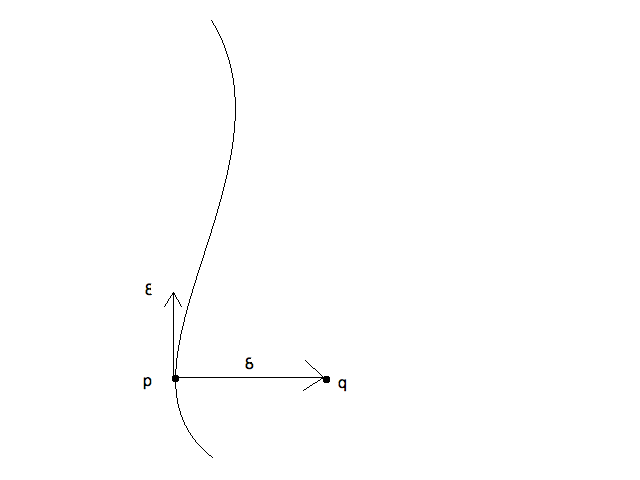}
\end{center}
Now parallel transport $\boldsymbol{\epsilon}$ to $q$ along $\boldsymbol{\delta}$, getting
\begin{eqnarray}
\epsilon^i \rightarrow \epsilon'^i = \epsilon^i - \Gamma^i_{jk}\delta^j \epsilon^k
\end{eqnarray}
\begin{center}
\includegraphics[scale=.5]{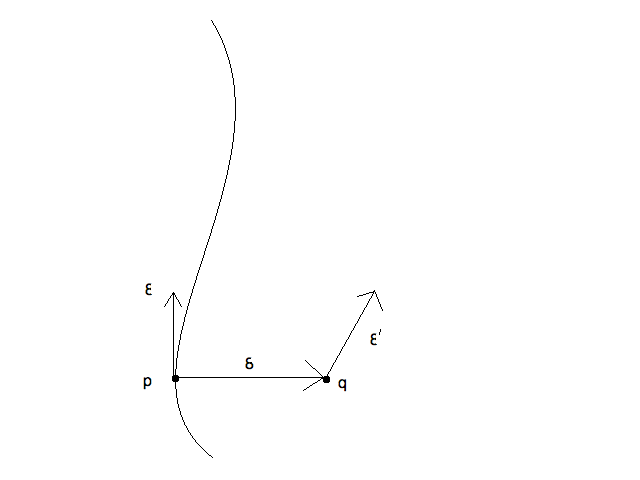}
\end{center}
Of course the vector $\epsilon^i - \Gamma^i_{jk}\delta^j \epsilon^k$ at $q$ is exactly parallel to $\epsilon^i$ at $p$ by definition \it because we parallel transported it\rm.  We drew it in a way that indicates that it may have twisted in some sense when it was transported, but that doesn't matter - the notion of parallel is defined by the connection.  The vectors $\epsilon^i$ and $\epsilon^i - \Gamma^i_{jk}\delta^j\epsilon^k$ should be considered parallel. 

Now consider a geodesic through $q$ in the direction of $\boldsymbol{\epsilon}'$.  
\begin{center}
\includegraphics[scale=.5]{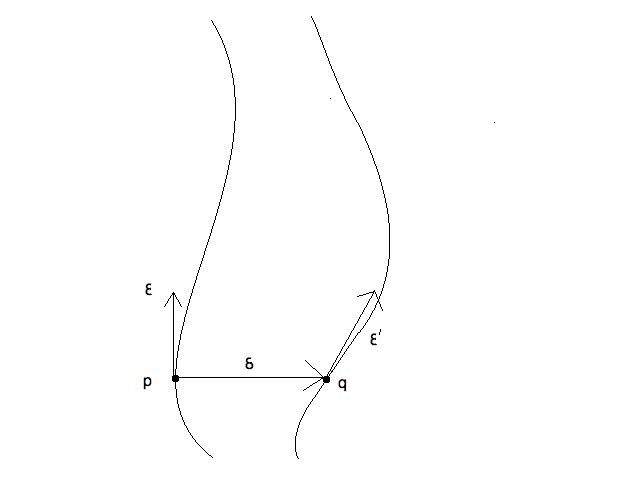}
\end{center}
We can parallel transport $\boldsymbol{\delta}$ along the geodesic through $p$.  If we parallel transport it only the infinitesimal displacement $\boldsymbol{\epsilon}$, we get
\begin{eqnarray}
\delta^i \rightarrow \delta'^i = \delta^i - \Gamma^i_{jk} \epsilon^j \delta^k
\end{eqnarray}
\begin{center}
\includegraphics[scale=.5]{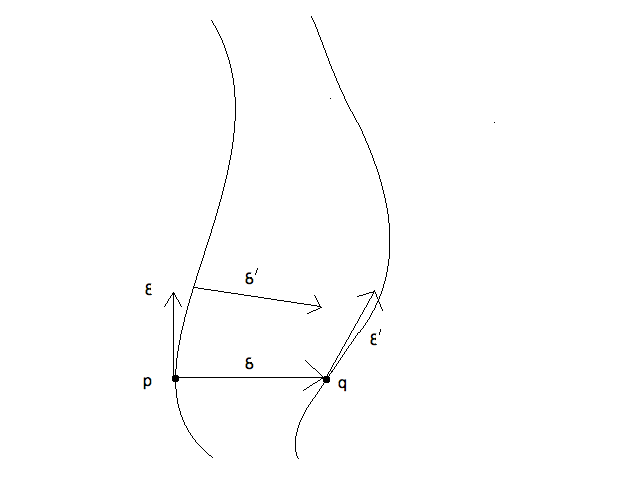}
\end{center}
As mentioned above, these vectors are infinitesimal, and therefore their finite appearance in the above pictures is misleading.  Because they are infinitesimal, we can redraw this more accurately as
\begin{center}
\includegraphics[scale=.5]{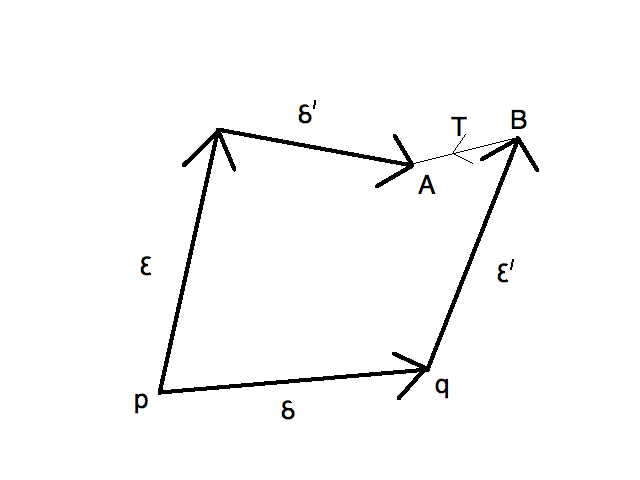}
\end{center}
So, as the vector $\boldsymbol{\delta}$ starts off attached to the geodesic through $q$.  However, as it moves along the geodesic through $q$, the two geodesics (through $q$ and $p$) begin to \it twist \rm away from each other, resulting in the parallel transported $\boldsymbol{\delta}$ becoming detached from the geodesic through $q$.  The degree of separation is given by the vector $\bf T\it$ in the diagram.  To find the exact form of $\bf T\it$, we subtract the vector from $p$ to $B$ from the vector from $p$ to $A$.  From $p$ to $A$ is given by
\begin{eqnarray}
x^i(p) + \epsilon^i + \delta'^i =x^i(p)+ \epsilon^i + \delta^i - \Gamma^i_{jk}\epsilon^j\delta^k
\end{eqnarray}
and from $p$ to $B$ is
\begin{eqnarray}
x^i(p) + \delta^i + \epsilon'^i = x^i(p) + \delta^i + \epsilon'^i - \Gamma^i_{jk}\delta^j\epsilon^k
\end{eqnarray}
And the difference between them is
\begin{eqnarray}
\big( x^i(p) + \epsilon^i + \delta^i - \Gamma^i_{jk} \epsilon^j\delta^k\big) &-& \big(x^i(p) + \delta^i + \epsilon^i - \Gamma^i_{jk}\delta^j\epsilon^k\big) \nolabel \\
&=& (\Gamma^i_{jk} - \Gamma^i_{kj}) \delta^j\epsilon^k \nolabel \\
&\equiv & T^i_{jk} \delta^j \epsilon^k \label{eq:definitionoftorsiontensor774}
\end{eqnarray}
So, $T^i_{jk} = \Gamma^i_{jk} - \Gamma^i_{kj}$ represents how much one geodesic twists away from a nearby geodesic.  For this reason we call $T^i_{jk}$ the \bf torsion tensor\rm.  If two nearby geodesics stay near each other along the entire geodesic, then there is no twisting towards or away from each other, and the torsion vanishes.  A connection in which the geodesics have this property is said to be torsion free.  Such a connection is symmetric.  A non-symmetric connection will have torsion.  Also, obviously the torsion tensor is totally antisymmetric:
\begin{eqnarray}
T^i_{jk} = -T^i_{kj}
\end{eqnarray}

Also, notice that $T^i_{jk}$ is the exact expression we found on the right hand side of (\ref{eq:covderliedertorsion}) above.  So finally, we have the relationship (in components)
\begin{eqnarray}
\nabla_{\bf u\it} v^i - \nabla_{\bf v\it} u^i - \mathcal{L}_{\bf u\it }v^i = T^i_{jk} u^jv^k \label{eq:antisymcovlieandtorrel}
\end{eqnarray}
Or 
\begin{eqnarray}
\nabla_{\boldsymbol{\delta}} \epsilon^i - \nabla_{\boldsymbol{\epsilon}} \delta^i - \mathcal{L}_{\boldsymbol{\delta}} \epsilon^i = T^i_{jk}\epsilon^j\delta^k \label{eq:antisymcovlieandtorrel2}
\end{eqnarray}
We can take (\ref{eq:antisymcovlieandtorrel}) (and/or (\ref{eq:antisymcovlieandtorrel2})) as the \it definition \rm of torsion.  

Recall from section \ref{sec:flowandlie} that the Lie derivative defined the non-closure of a rectangle (cf page \pageref{pagewithpictureofliederivativenonclosure}).  The idea is that if you start at $p \in \mathcal{M}$ and take two infinitesimal displacement vectors $\boldsymbol{\epsilon}$ and $\boldsymbol{\delta}$, you can transport each along the other in two ways: the tangent mapping as with the Lie derivative, or with parallel transport as with the covariant derivative.  Denoting a parallel transported vector with a $||$ subscript and a vector that has been tangent mapped with a $m$ subscript, we have
\begin{center}
\includegraphics[scale=.6]{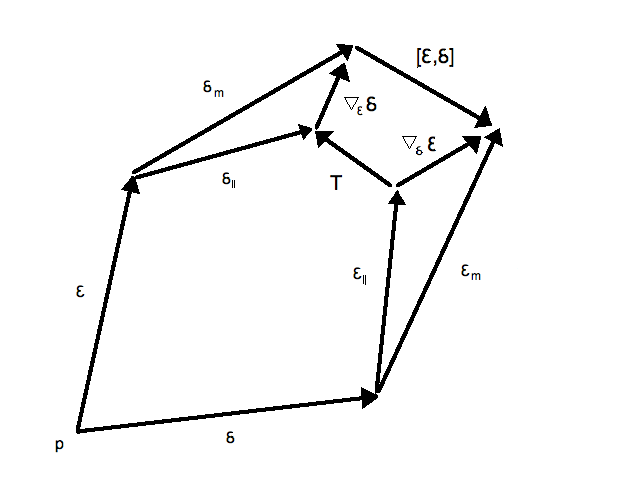}
\end{center}
Equation (\ref{eq:antisymcovlieandtorrel}) can simply be read off of this as a simple vector diagram.  

We can look at $T^i_{jk}$ in yet another way.  Looking again at the equation for geodesics (\ref{eq:geodesicdifferentialequation})
\begin{eqnarray}
{d^2 q^i \over d\tau^2} + \Gamma^i_{jk} {dq^j \over d\tau} {dq^k \over d\tau} = 0
\end{eqnarray}
As with any tensor, we can break $\Gamma^i_{jk}$ up into a symmetric and and antisymmetric part:
\begin{eqnarray}
\Gamma^i_{(jk)} &=& {1 \over 2} (\Gamma^i_{jk} + \Gamma^i_{kj}) \nolabel \\
\Gamma^i_{[jk]} &=& {1 \over 2} (\Gamma^i_{jk} - \Gamma^i_{kj}) = {1 \over 2}T^i_{jk} \nolabel \\
\Gamma^i_{jk} &=& \Gamma^i_{(jk)} + \Gamma^i_{[jk]}
\end{eqnarray}
So, (\ref{eq:geodesicdifferentialequation}) becomes
\begin{eqnarray}
{d^2q^i \over d\tau} + \Gamma^i_{jk} {dq^j \over d\tau} {dq^k \over d\tau} &=& {d^2q^i \over d\tau} + \big(\Gamma^i_{(jk)} +\Gamma^i_{[jk]}\big) {dq^j \over d\tau} {dq^k \over d\tau} \nolabel \\
&=& {d^2q^i \over d\tau} + \Gamma^i_{(jk)}{dq^j \over d\tau} {dq^k \over d\tau} + \Gamma^i_{[jk]}{dq^j \over d\tau} {dq^k \over d\tau} \nolabel \\
&=& {d^2q^i \over d\tau} + \Gamma^i_{(jk)}{dq^j \over d\tau} {dq^k \over d\tau} + {1 \over 2}T^i_{jk}{dq^j \over d\tau} {dq^k \over d\tau}
\end{eqnarray}
The last term in the third line will vanish because it is a sum over all antisymmetric indices.  This tells us that it is only the \it symmetric \rm part of the connection that contributes to the geodesics.  

In other words, for a given connection, the geodesics, or ``straightest possible lines" are defined entirely by the symmetric part.  The antisymmetric part plays a different role - namely that of torsion.  The antisymmetric part defines how much the geodesics twist relative to each other.  We can picture this as
\begin{center}
\includegraphics[scale=.7]{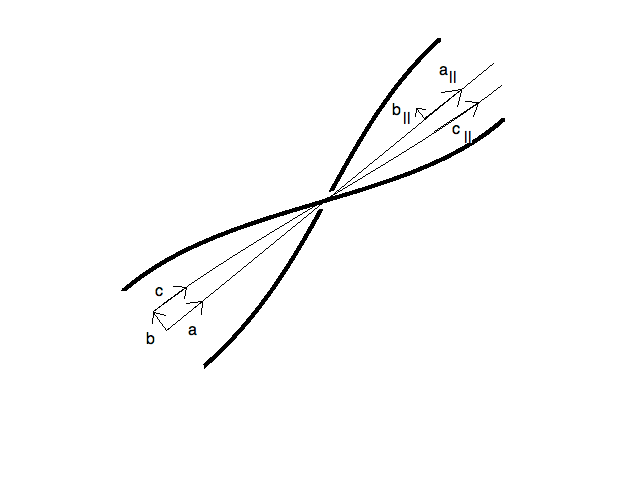}
\end{center}

To see this, consider the two connections from sections (\ref{sec:connections}) and (\ref{sec:otherconnections}), given (respectively) by
\begin{eqnarray}
& &\Gamma^{\phi}_{r \phi} = \Gamma^{\phi}_{\phi r} = {1 \over r} \nolabel \\
& &\Gamma^r_{\phi \phi} = -r
\end{eqnarray}
and 
\begin{eqnarray}
\Gamma^{\phi}_{\phi r} = -{1 \over r}
\end{eqnarray}
(all other connection coefficients in both vanish).  The first is obviously symmetric, so there is no antisymmetric part at all.  On the other hand, for the second we have
\begin{eqnarray}
\Gamma^{\phi}_{(\phi r)} &=& {1 \over 2}(\Gamma^{\phi}_{\phi r} + \Gamma^{\phi}_{r \phi}) = {1 \over 2} \bigg(-{1 \over r} + 0 \bigg) = -{1 \over 2r} = \Gamma^{\phi}_{(r \phi)} \nolabel \\
\Gamma^{\phi}_{[\phi r]} &=& {1 \over 2} (\Gamma^{\phi}_{\phi r} - \Gamma^{\phi}_{r \phi}) = {1 \over 2} \bigg(-{1 \over r} - 0\bigg) = -{1 \over 2r} = -\Gamma^{\phi}_{[r\phi]} 
\end{eqnarray}

So for the first one there is no torsion.  This makes sense because the geodesics are simply straight lines, which we can plainly see don't twist relative to each other.  

On the other hand, the second one does have torsion given by
\begin{eqnarray}
T^{\phi}_{\phi r} &=& -{1 \over r} \nolabel \\
T^{\phi}_{r \phi} &=& {1 \over r}
\end{eqnarray}
By studying the geodesics given in section \ref{sec:otherconnections}, you can see that they do indeed twist away from each other as you spiral out from the center.  

\subsection{The Metric Connection}
\label{sec:metricconnection}

We started chapter \ref{sec:chapwithmet}, ``Manifolds with Metrics", with a discussion of metrics, which is not surprising.  But once we got to section \ref{sec:paralleltransport}, you may have noticed that we have hardly mentioned the word ``metric" since.  We mentioned at the beginning of section \ref{sec:paralleltransport} that metrics are still our ultimate focus, and that has remained true.  Now that we have spent an adequate amount of time discussing connections, we are finally in a position to tie metrics back in.  

We pointed out when we introduced connections that a connection can be anything - it is entirely up to us to define parallel transport however we want.  We exploited this freedom to define a nice and ``expected" connection which led to straight line geodesics in section \ref{sec:connections}, and then arbitrarily define a completely different connection in \ref{sec:otherconnections} which led to much more exotic geodesics (straight lines, circles, and spirals).  However at the end of section \ref{sec:otherconnections}, we mentioned that if a manifold has a metric, then there is a particular connection (or rather class of connections) that is special.  Such a connection is said to be a ``metric compatible", or a ``metric compatible connection".  

Our approach will be to assume a manifold $\mathcal{M}$ comes with a metric $g_{ij}$, and then use the metric to put certain conditions on $\Gamma^i_{jk}$.  Recall from section \ref{sec:metrics} that the point of a metric is to provide a map from two vectors, say $\bf u\it$ and $\bf v\it$, to the reals:
\begin{eqnarray}
g : T_p\mathcal{M} \otimes T_p\mathcal{M} \longrightarrow \mathbb{R}
\end{eqnarray}
of the (component) form 
\begin{eqnarray}
g(\bf v\it,\bf u\it) = g_{ij} v^iu^j
\end{eqnarray}
(cf equation (\ref{eq:generalformofmetricinnerproduct})).  Let's say we start with $\bf u\it$ and $\bf v\it$ in $T_p\mathcal{M}$ with inner product $g(\bf u\it,\bf v\it) = g_{ij} v^iu^j$.  We can then choose some other arbitrary vector $\bf w\it$ in $T_p\mathcal{M}$.  There will be some geodesic in the direction of $\bf w\it$ through $p$.  
\begin{center}
\includegraphics[scale=.6]{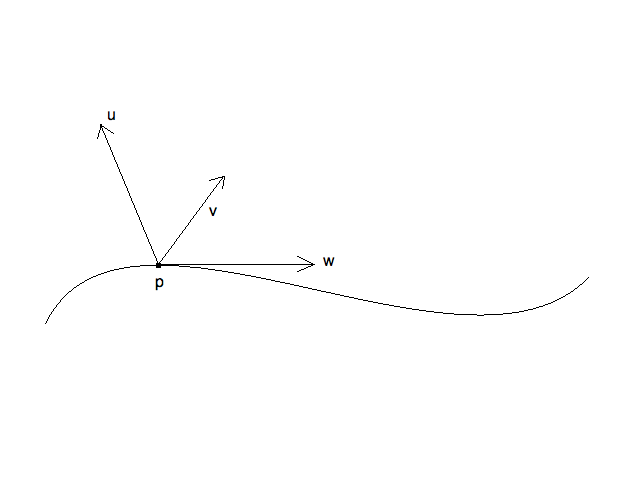}
\end{center}
Let's then parallel transport $\bf u\it$ and $\bf v\it$ along this geodesic:
\begin{center}
\includegraphics[scale=.6]{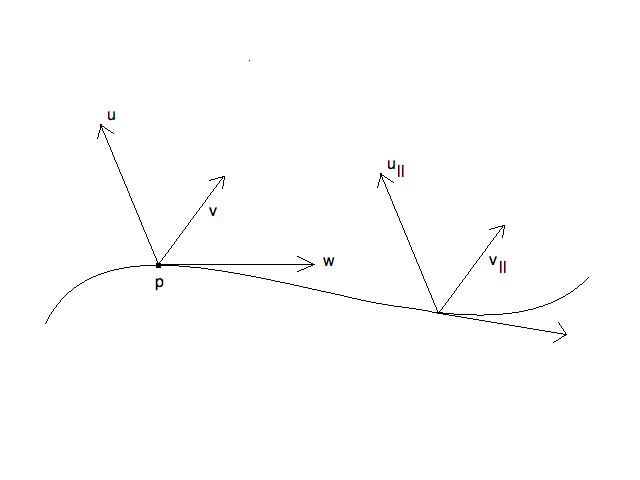}
\end{center}
The condition we will impose will be that the inner product does not change as you parallel transport along the geodesic.  

So, because we are parallel transporting, by definition we have
\begin{eqnarray}
\nabla_{\bf w\it} \bf u \it = \nabla_{\bf w\it} \bf v\it = 0
\end{eqnarray}
To preserve the inner product along the geodesic, we must have
\begin{eqnarray}
0 &=& \nabla_{\bf w\it} \big( g(\bf v\it, \bf u\it,)\big) \nolabel \\
&=& \nabla_{\bf w\it} \big( g_{ij} v^i u^j\big) \nolabel \\
&=& w^k \nabla_k \big( g_{ij} v^i u^j\big) \nolabel \\
&=& w^k \nabla_k (g_{ij}) v^i u^j + w^k g_{ij} \nabla_k(v^i) u^j + w^k g_{ij} v^i \nabla_k (u^j) \nolabel \\
&=& w^k v^i u^j \nabla_k g_{ij}
\end{eqnarray}
Because we want this to be true for any vectors $\bf u\it$ and $\bf v\it$, we must therefore demand
\begin{eqnarray}
\nabla_k g_{ij} = 0 \label{eq:covderivofmetriciszeroparforgravasgaugesecpar}
\end{eqnarray}
Using equation (\ref{eq:mostgeneralcovderoftensorpossible}), we can write this as
\begin{eqnarray}
\nabla_k g_{ij} = 0 \qquad \Rightarrow \qquad {\partial g_{ij} \over \partial x^k} - \Gamma^m_{ki} g_{mj} - \Gamma^m_{kj} g_{im} = 0 \label{eq:metriccompcond1}
\end{eqnarray}
If (\ref{eq:metriccompcond1}) is satisfied by $\Gamma^i_{jk}$, then the connection is said to be a \bf metric-compatible \rm connection.  

Next we can exploit the fact that $g_{ij}$ is symmetric and write all three ($k,i,j$) cyclic permutations of (\ref{eq:metriccompcond1}):
\begin{eqnarray}
& &{\partial g_{ij} \over \partial x^k} - \Gamma^m_{ki} g_{mj} - \Gamma^m_{kj} g_{mi} = 0 \nolabel \\
& &{\partial g_{jk} \over \partial x^i} - \Gamma^m_{ij} g_{mk} - \Gamma^m_{ik} g_{mj} = 0 \nolabel \\
& &{\partial g_{ki} \over \partial x^j} - \Gamma^m_{jk} g_{mi} - \Gamma^m_{ji} g_{mk} = 0
\end{eqnarray}
Now take minus the first one plus the second plus the third (trust us):
\begin{eqnarray}
0&=& -\bigg({\partial g_{ij} \over \partial x^k} - \Gamma^m_{ki} g_{mj} - \Gamma^m_{kj} g_{mi}\bigg) \nolabel \\
& & + \bigg({\partial g_{jk} \over \partial x^i} - \Gamma^m_{ij} g_{mk} - \Gamma^m_{ik} g_{mj}\bigg) \nolabel \\
& & +\bigg( {\partial g_{ki} \over \partial x^j} - \Gamma^m_{jk} g_{mi} - \Gamma^m_{ji} g_{mk}\bigg) \nolabel \\
&=& -{ \partial g_{ij} \over \partial x^k} + {\partial g_{jk} \over \partial x^i} + {\partial g_{ki} \over \partial x^j} \nolabel \\
& & + \big(\Gamma^m_{ki} - \Gamma^m_{ik}\big) g_{mj} + \big(\Gamma^m_{kj} - \Gamma^m_{jk} \big)g_{mi} \nolabel \\
& & - \big( \Gamma^m_{ij} + \Gamma^m_{jk}\big) g_{mk} \nolabel \\
&=& -{ \partial g_{ij} \over \partial x^k} + {\partial g_{jk} \over \partial x^i} + {\partial g_{ki} \over \partial x^j} \nolabel \\
& & + T^m_{ki} g_{mj} + T^m_{kj} g_{mi} \nolabel \\
& & - 2\Gamma^m_{(ij)} g_{mk} \label{eq:stepinfindingmetricconnection}
\end{eqnarray} 
We can solve (\ref{eq:stepinfindingmetricconnection}) with
\begin{eqnarray}
& & 2\Gamma^m_{(ij)} g_{mk} = {\partial g_{jk} \over \partial x^i} + {\partial g_{ki} \over \partial x^j} - {\partial g_{ij} \over \partial x^k} + T^m_{ki} g_{mj} + T^m_{kj} g_{mi} \nolabel \\
& \Rightarrow & \Gamma^m_{(ij)} g_{mk} g^{kn} = {1 \over 2}g^{kn} \bigg({\partial g_{jk} \over \partial x^i} + {\partial g_{ki} \over \partial x^j} - {\partial g_{ij} \over \partial x^k} + {1 \over 2}\big(T^m_{ki} g_{mj} + T^m_{kj} g_{mi}\big) \bigg) \nolabel \\
& \Rightarrow & \Gamma^n_{(ij)} = {1 \over 2} g^{kn}\bigg({\partial g_{jk} \over \partial x^i} + {\partial g_{ki} \over \partial x^j} - {\partial g_{ij} \over \partial x^k}\bigg) +{1 \over 2}(T^n_{ki} + T^n_{kj}) \label{eq:metriccompatibilityconstraintonsymmetricpart}
\end{eqnarray}

But in general, we have
\begin{eqnarray}
\Gamma^n_{ij} &=& \Gamma^n_{(ij)} + \Gamma^n_{[ij]} \nolabel \\
&=& {1 \over 2} g^{kn}\bigg({\partial g_{jk} \over \partial x^i} + {\partial g_{ki} \over \partial x^j} - {\partial g_{ij} \over \partial x^k}\bigg) +{1 \over 2}(T^n_{ki} + T^n_{kj}) + \Gamma^n_{[ij]} \nolabel \\
&=& {1 \over 2} g^{kn}\bigg({\partial g_{jk} \over \partial x^i} + {\partial g_{ki} \over \partial x^j} - {\partial g_{ij} \over \partial x^k}\bigg) +{1 \over 2}(T^n_{ki} + T^n_{kj}) + {1 \over 2} (\Gamma^n_{ij} - \Gamma^n_{jk}) \nolabel \\
&=& {1 \over 2} g^{kn}\bigg({\partial g_{jk} \over \partial x^i} + {\partial g_{ki} \over \partial x^j} - {\partial g_{ij} \over \partial x^k}\bigg) +{1 \over 2}(T^n_{ki} + T^n_{kj} + T^n_{ij})
\end{eqnarray}
The second term with the sum of three torsion tensors is called the \bf contorsion \rm tensor.  

If we choose a connection that is symmetric, however, the contorsion tensor vanishes.  In this case we are left with
\begin{eqnarray}
\Gamma^n_{ij} = {1 \over 2} g^{kn}\bigg({\partial g_{jk} \over \partial x^i} + {\partial g_{ki} \over \partial x^j} - {\partial g_{ij} \over \partial x^k}\bigg) \label{eq:levicivitaconstraint}
\end{eqnarray}
If a connection satisfies (\ref{eq:levicivitaconstraint}) (implying that there is no torsion), then it is called the \bf Levi-Civita Connection\rm.  We will see that Levi-Civita connections play a central role in a tremendous amount of physics.  In fact, general relativity is specifically a theory of Levi-Civita connections.  

As a brief comment, the Levi-Civita connection coefficients are also called the \bf Christoffel Symbols \rm in a great deal of general relativity and differential geometry literature.  

Another way of defining a Levi-Civita connection is that it is a metric compatible connection in which the torsion vanishes.  Notice that if a connection satisfies the metric compatibility constraint (\ref{eq:metriccompcond1}), there is only a constraint put on its \it symmetric \rm part ((\ref{eq:metriccompcond1}) lead to (\ref{eq:metriccompatibilityconstraintonsymmetricpart})).  Therefore a connection can be metric compatible (its symmetric part satisfies (\ref{eq:metriccompatibilityconstraintonsymmetricpart})), but it has an antisymmetric part, making it not a Levi-Civita connection.  Also, a symmetric connection may not necessarily satisfy (\ref{eq:metriccompatibilityconstraintonsymmetricpart}), making it symmetric but not metric-compatible.  Therefore a connection may be symmetric but not Levi-Civita.  However, a Levi-Civita connection is by definition symmetric and metric-compatible.  So, we have the following Venn diagram of connections:
\begin{center}
\includegraphics[scale=.6]{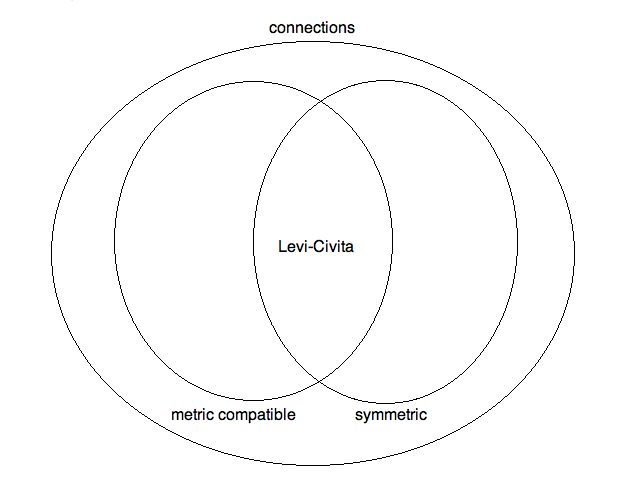}
\end{center}

For the next large part of these notes whenever we talk about a connection we will mean a Levi-Civita connection.  Eventually we will generalize to talk about non Levi-Civita connections, and we will announce that change at the appropriate time.  But until further notice, any connection can be assumed to be Levi-Civita, and therefore given by (\ref{eq:levicivitaconstraint}).  

As a few examples, consider the metrics we found in section \ref{sec:metrics}.  We had $\mathbb{R}^2$ in Cartesian coordinates:
\begin{eqnarray}
g_{ij} = \delta_{ij} = 
\begin{pmatrix}
1 & 0 \\ 0 & 1
\end{pmatrix}
\end{eqnarray}
You can plug this into (\ref{eq:levicivitaconstraint}) and get
\begin{eqnarray}
\Gamma^i_{jk} = 0 \; \forall \; i,j,k \label{eq:allconnvanforcart}
\end{eqnarray}
which agrees with what we found in (\ref{eq:christoffelsymbolsforcartesiancoordinates}).  

Leading up to equation (\ref{eq:metrictransfromcarttopolar}) we found that the Cartesian metric on $\mathbb{R}^2$ transforms to
\begin{eqnarray}
g_{ij} = 
\begin{pmatrix}
1 & 0 \\ 0 & r^2
\end{pmatrix}
\end{eqnarray}
in polar coordinates.  You are encouraged to work out (\ref{eq:levicivitaconstraint}) for this on your own, and you will see that indeed the non-vanishing components are
\begin{eqnarray}
& &\Gamma^{\phi}_{r \phi} = \Gamma^{\phi}_{\phi r} = {1\over r} \nolabel \\
& &\Gamma^r_{\phi \phi} = -r
\end{eqnarray}
as in (\ref{eq:christoffelsymbolsforpolarcoordinates}).\footnote{Notice the consistency here.  We found that the metric (\ref{eq:euclidianmetricintwodimensionsforexmaple}) transformed as a tensor to give (\ref{eq:metrictransfromcarttopolar}) in polar coordinates.  Also we know from (\ref{eq:allconnvanforcart}) that Cartesian coordinates give you a vanishing connection, which is also what we found in (\ref{eq:christoffelsymbolsforcartesiancoordinates}).  And, we can use the transformation law for a connection (\ref{eq:howconnectiongammatransforms}) to transform (\ref{eq:allconnvanforcart}) into (\ref{eq:metrictransfromcarttopolar}).  }

You are encouraged to work out on your own the fact that the Minkowski metric (in 3+1 dimensions)
\begin{eqnarray}
g_{ij} = \eta_{ij} = 
\begin{pmatrix}
-1 & 0 & 0 & 0 \\
0 & 1 & 0 & 0 \\
0 & 0 & 1 & 0 \\
0 & 0 & 0 & 1
\end{pmatrix}
\end{eqnarray}
gives 
\begin{eqnarray}
\Gamma^i_{jk} = 0 \; \forall \; i,j,k \label{eq:theconnectioncoefficientsassociatedwiththeminkowskimetricvanish}
\end{eqnarray}

Next is the metric for a circle (\ref{eq:metricforcircle})
\begin{eqnarray}
g_{ij} = r^2
\end{eqnarray}
(it is only $1\times 1$ because a circle is $1$ dimensional).  This will give
\begin{eqnarray}
\Gamma^i_{jk} = 0 \; \forall \; i,j,k 
\end{eqnarray}

Less trivial is the ellipse metric (\ref{eq:metricforoval})
\begin{eqnarray}
g_{ij} = r^2(4\cos^2\theta + \sin^2\theta)
\end{eqnarray}
which gives
\begin{eqnarray}
\Gamma^{\theta}_{\theta \theta} = -{3\sin(2\theta) \over 5+4\cos(2\theta)}
\end{eqnarray}
(again, the ellipse is only $1$ dimensional so there is only one value each index can take - the $\theta$ value).  

Next is the metric for $S^2$ give in (\ref{eq:metricfors2inmetricsection}),
\begin{eqnarray}
g_{ij} = 
\begin{pmatrix}
1 & 0 \\ 0 & \sin^2\theta
\end{pmatrix}
\end{eqnarray}
This will give non vanishing components
\begin{eqnarray}
& &\Gamma^{\theta}_{\phi \phi} = -\sin\theta \cos\theta \nolabel \\
& &\Gamma^{\phi}_{\theta \phi} = \Gamma^{\phi}_{\phi \theta} = \cot\theta \label{eq:connectionfornormals2}
\end{eqnarray}

Or there is the deformed $S^2$ in (\ref{eq:metricfordeformeds2}),
\begin{eqnarray}
g_{ij} = 
\begin{pmatrix}
{1 \over 2}\big((\lambda^2+1) - (\lambda^2 - 1)\cos(2\theta)\big) & 0 \\
0 & \sin^2\theta \\
\end{pmatrix}
\end{eqnarray}
Here the non vanishing components will be
\begin{eqnarray}
\Gamma^{\theta}_{\theta \theta} &=& {(\lambda^2 - 1)\sin(2\theta) \over (\lambda^2+1) - (\lambda^2 - 1)\cos(2\theta)} \nolabel \\
\Gamma^{\theta}_{\phi \phi} &=& -{2 \cos\theta \sin \theta \over (\lambda^2+1) - (\lambda^2 - 1)\cos(2\theta)} \nolabel \\
\Gamma^{\phi}_{\theta \phi} &=& \Gamma^{\phi}_{\phi \theta} = \cot\theta \label{eq:connectionforeggs2}
\end{eqnarray}
Notice that (\ref{eq:connectionforeggs2}) reduces to (\ref{eq:connectionfornormals2}) for $\lambda = 1$, as expected.  

Finally we have the torus metric (\ref{eq:metricfortorusinmetricsection}),
\begin{eqnarray}
g_{ij} = 
\begin{pmatrix}
r^2 & 0 \\ 0 & (R+r\cos\theta)^2
\end{pmatrix}
\end{eqnarray}
This will give non-vanishing components
\begin{eqnarray}
& &\Gamma^{\phi}_{\phi \phi} = {(R+r\cos\theta)\sin\theta \over r} \nolabel \\
& &\Gamma^{\phi}_{\theta \phi} = \Gamma^{\phi}_{\phi \theta} = -{r \sin \theta \over (R+r\cos\theta)}
\end{eqnarray}

\subsection{Metric Geodesics}

Before concluding this section we consider one final idea.  In section \ref{sec:geodesics} we considered geodesics, or the ``straightest possible path" in the manifold.  We said there that a geodesic is the path a particle will follow if not acted on by any other forces.  We now revisit this idea.  

Consider two points on a manifold $\mathcal{M}$, $p$ and $q$.  We want to find the path between them that extremizes the distance.  If we assume that $\mathcal{M}$ has a metric then the idea of distance is well defined.  Specifically, the infinitesimal distance between two points is given by
\begin{eqnarray}
ds = \sqrt{g_{ij} dx^i dx^j}
\end{eqnarray}

Generalizing this, the distance $\mathcal{I}$ from $p$ to $q$ will then be the integral from $p$ to $q$ of some path, where we can parameterize the path $x^i(\tau)$ by the parameter $\tau$:
\begin{eqnarray}
\mathcal{I} = \int_p^q \sqrt{g_{ij}{dx^i(\tau)\over d\tau}{dx^j(\tau)\over d\tau}} d\tau \label{eq:firstequationforgeodesicdistancetobeextremized}
\end{eqnarray}
Notice that $\mathcal{I}$ doesn't depend on the parameterization (i.e., if we change the parameterization , $\tau \rightarrow \tau'(\tau)$, the integral remains identical).  

This is simply a calculus of variations problem, and we therefore find the extremized path by demanding that the first order term of the variation of the path vanish\footnote{We assume the reader is familiar with variational calculations such as the one here.  If you are not we strongly encourage you to spend some time reviewing such problems.},\footnote{Keep in mind that the metric is position dependent ($g_{ij} = g_{ij}(x^i(\tau))$) and therefore the variation of the path includes a variation of the metric.  Furthermore, because the metric is a function of $x^i$, we will use the chain rule to take the variation of the metric:
$$\delta g_{ij}(x^i) = {\partial g_{ij} \over \partial x^k} \delta x^k $$}:
\begin{eqnarray}
0 &=& \delta \mathcal{I} \nolabel \\
&=& \delta \int_p^q \bigg(g_{ij}{dx^i \over d\tau}{dx^j\over d\tau}\bigg)^{{1 \over 2}} d\tau \nolabel \\
&=& \int_p^q \bigg(g_{ij}{dx^i\over d\tau} {dx^j\over d\tau}\bigg)^{-{1\over 2}}\bigg(\delta g_{ij} {dx^i\over d\tau} {dx^j\over d\tau} + g_{ij}{d(\delta x^i)\over d\tau} {dx^j\over d\tau}+g_{ij}{dx^i\over d\tau} {d(\delta x^j)\over d\tau}\bigg) d\tau \nolabel \\
&=& \int_p^q \bigg(g_{ij}{dx^i\over d\tau} {dx^j\over d\tau}\bigg)^{-{1\over 2}}\bigg( {\partial g_{ij} \over \partial x^k} \delta x^k {dx^i\over d\tau} {dx^j\over d\tau} + 2g_{ij} {dx^i\over d\tau} {d(\delta x^j)\over d\tau}\bigg)d\tau
\end{eqnarray}
Because (\ref{eq:firstequationforgeodesicdistancetobeextremized}) is independent of the parameterization, we can now choose our parameterization so as to fix
\begin{eqnarray}
g_{ij} {dx^i \over d\tau} {dx^j\over d\tau} = 1
\end{eqnarray}
without loss of generality.  So we are left with the requirement
\begin{eqnarray}
0 &=& \int_p^q \bigg( {\partial g_{ij} \over \partial x^k} \delta x^k {dx^i\over d\tau} {dx^j\over d\tau} + 2g_{ij} {dx^i\over d\tau} {d(\delta x^j)\over d\tau}\bigg)d\tau \nolabel
\end{eqnarray}
We can integrate the second term by parts and with appropriate boundary conditions ignore the surface term.  So, 
\begin{eqnarray}
0 &=& \int_p^q \bigg( {\partial g_{ij} \over \partial x^k} \delta x^k {dx^i\over d\tau} {dx^j\over d\tau} - 2{d \over d\tau} \bigg[g_{ij} {dx^i \over d\tau}\bigg] \delta x^j\bigg)d \tau \nolabel \\
&=& \int_p^q \bigg({\partial g_{ij} \over \partial x^k} \delta x^k {dx^i\over d\tau} {dx^j\over d\tau} - 2g_{ij} {d^2 x^i \over d\tau^2} \delta x^j - 2 {\partial g_{ij} \over \partial \tau} {dx^i \over d\tau} \delta x^j\bigg)d\tau \nolabel \\
&=& \int_p^q \bigg({\partial g_{ij} \over \partial x^k} \delta x^k {dx^i\over d\tau} {dx^j\over d\tau} - 2g_{ij} {d^2 x^i \over d\tau^2} \delta x^j - 2 {\partial g_{ij} \over \partial x^k}{dx^k \over d\tau} {dx^i \over d\tau} \delta x^j\bigg)d\tau \nolabel \\
&=& \int_p^q \delta x^k\bigg( {\partial g_{ij} \over \partial x^k} {dx^i \over d\tau}{dx^j\over d\tau} - 2g_{ik} {d^2 x^i \over d\tau^2} - 2 {\partial g_{ik} \over \partial x^j} {d x^j \over d \tau}{dx^i\over d\tau}\bigg) d\tau \nolabel
\end{eqnarray}
So $\delta \mathcal{I} = 0$ as long as (dividing by $2$ to make the result more transparent)
\begin{eqnarray}
g_{ik} {d^2 x^i \over d\tau^2} + \bigg({\partial g_{ik} \over \partial x^j}-{1 \over 2}{\partial g_{ij} \over \partial x^k}\bigg){dx^j\over d\tau}{dx^i \over d\tau}&=& 0  \nolabel \\
\Rightarrow  g^{nk} g_{ik}{d^2 x^i \over d\tau^2} + g^{nk}\bigg({\partial g_{ik} \over \partial x^j}-{1 \over 2}{\partial g_{ij} \over \partial x^k}\bigg){dx^j\over d\tau}{dx^i \over d\tau}&=& 0  \nolabel \\
\Rightarrow {d^2 x^n \over d\tau^2} + g^{nk}\bigg({\partial g_{ik} \over \partial x^j}-{1 \over 2}{\partial g_{ij} \over \partial x^k}\bigg){dx^j\over d\tau}{dx^i \over d\tau}&=& 0  \nolabel \\
\Rightarrow {d^2 x^n \over d\tau^2}+ {1 \over 2} g^{nk}\bigg( {\partial g_{ik} \over \partial x^j} + {\partial g_{jk} \over \partial x^i} - {\partial g_{ij} \over \partial x^k}\bigg) {dx^i \over d\tau} {dx^j \over d\tau} &=& 0 \nolabel \\
\Rightarrow {d^2 x^n \over d\tau^2} + \Gamma^n_{ij} {dx^i \over d\tau} {dx^j \over d\tau} &=& 0 
\end{eqnarray}
which is exactly what we had before in equation (\ref{eq:geodesicdifferentialequation}) in section \ref{sec:geodesics}.  

\subsection{Normal Coordinates}
\label{sec:normalcoordinates}

For a given metric, when working with the corresponding Levi-Civita connection we can choose a particular coordinate system which simplifies things greatly.   

Consider a point $p \in \mathcal{M}$.  If we take a very small neighborhood of $p$, there will be an infinite number of geodesics through $p$:
\begin{center}
\includegraphics[scale=.6]{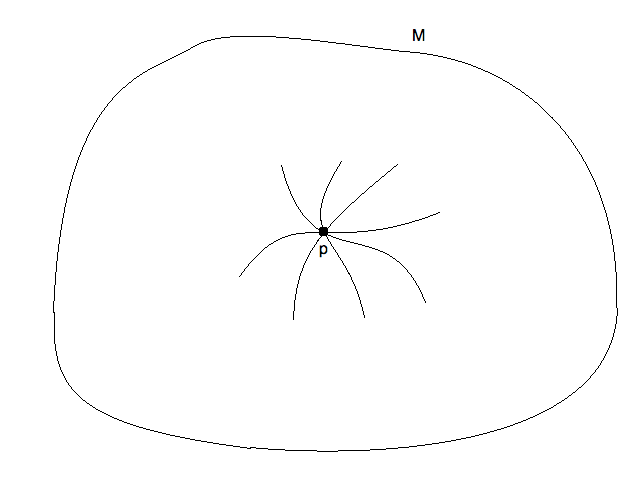}
\end{center}
Recall from the discussion at the end of section \ref{sec:geodesics} that the geodesic equation
\begin{eqnarray}
\nabla_{\bf u\it} \bf u\it = 0
\end{eqnarray}
not only specifies the path, but because the right hand side is $0$, also specifies the parameterization.  For this reason, every geodesic at $p$ specifies (and is specified by) a vector $\bf v\it \in T_p\mathcal{M}$.  In fact, the geodesics emanating from $p$ are one to one with elements of $T_p\mathcal{M}$.  

Consider a point $q \in \mathcal{M}$ that is very near to $p$.  Because we have fixed the right hand side of the geodesic equation to be $0$ (thus specifying the parameterization of a given geodesic), the point $q$ will specify \it one single \rm geodesic starting at $p$, denoted $q_{p,q}(\tau)$ that satisfies (by definition)
\begin{eqnarray}
q_{p,q}(\tau)\big|_{\tau = 0} &=& p \nolabel \\
q_{p,q}(\tau)\big|_{\tau = 1} &=& q \label{eq:tauequalszeroandtauequalsoneconstraintforpandq}
\end{eqnarray}
\begin{center}
\includegraphics[scale=.6]{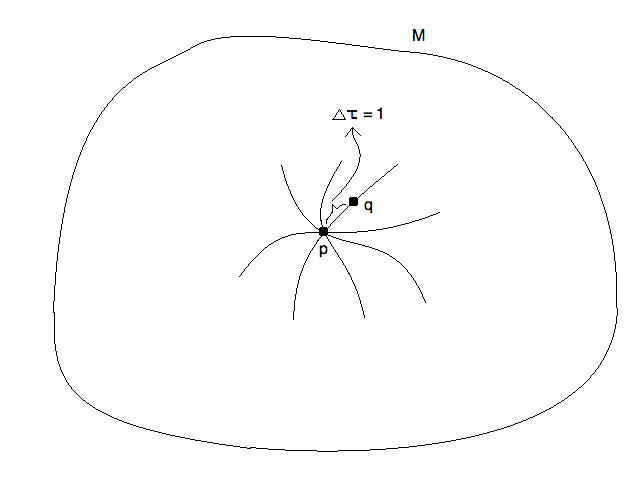}
\end{center}

Now, using the one to one relationship between the geodesics emanating from $p$ and tangent vectors in $T_p\mathcal{M}$, we can use this geodesic $q_{p,q}(\tau)$ to specify a particular vector $\bf v\it_p$ which corresponds (in coordinates) to $q_{p,q}(\tau)$:
\begin{eqnarray}
\bf v\it_{p,q} = {d \bf q\it_{p,q}(\tau) \over d\tau} \bigg|_{\tau = 0}
\end{eqnarray}
\begin{center}
\includegraphics[scale=.6]{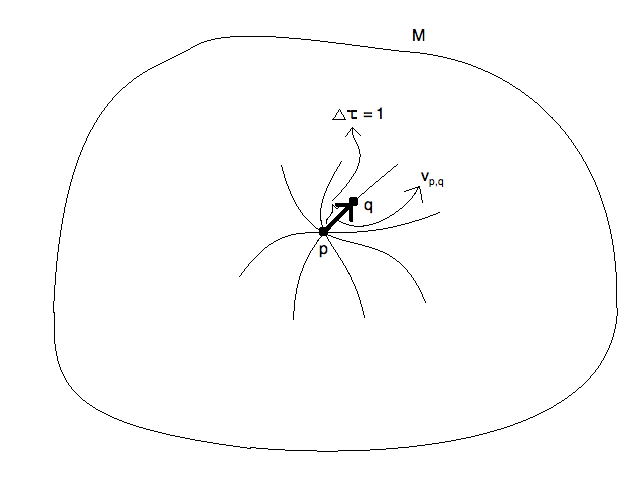}
\end{center}
We can then take the vector $\bf v\it$, or more properly the \it components \rm of the vector $\bf v\it$, to be the components of the point $q$.  This type of coordinate system based on geodesics in the neighborhood of $p$ is called the \bf Normal Coordinate \rm system of $p$, and is only valid in a neighborhood of $p$.  

If we denote the normal coordinates as $\bf x\it$, then obviously $\bf x\it(p) = 0$.  From the discussion following equation (\ref{eq:vectorsaregeneratorsofcurvesthroughapoint}), we can then define the coordinates of $q$ (where $q$ is near $p$) to be
\begin{eqnarray}
\bf x\it(q) = e^{\bf v\it_{p,q}}\bf x\it(p) = \bf v\it_{p,q}
\end{eqnarray}
where $\bf v\it_{p,q}$ is defined above.  

Then, an arbitrary point ``between" $p$ and $q$ on the path $q_{p,q}(\tau)$ will have coordinates
\begin{eqnarray}
\bf q\it_{p,q}(\tau) = \bf x\it(q_{p,q}(\tau)) = \tau \bf v\it_{p,q} \label{eq:normalcoord}
\end{eqnarray}
(notice this easily satisfies both equations in (\ref{eq:tauequalszeroandtauequalsoneconstraintforpandq})).  

So, within a sufficiently small neighborhood of $p$ we can use normal coordinates to specify any point.  Now consider the geodesic equation in normal coordinates $\bf x\it$.  It reads
\begin{eqnarray}
& & {d^2 x^i \over d\tau^2} + \Gamma^i_{jk} (x^i) {d x^j \over d\tau} {dx^k \over d\tau} = 0 \nolabel \\
&\Rightarrow& {d^2 (\tau v^i_{p,q}) \over d\tau^2} + \Gamma^i_{jk} (\tau v^i_{p,q}) {d (\tau v^j_{p,q}) \over d\tau} {d (\tau v^k_{p,q}) \over d\tau} = 0 \nolabel \\
&\Rightarrow& \Gamma^i_{jk} (\tau v^i_{p,q}) v^j_{p,q} v^k_{p,q} = 0
\end{eqnarray}
(where we are stating the position dependence of the $\Gamma^i_{jk}$ terms explicitly).  The only way for this to be satisfied in general is for
\begin{eqnarray}
\Gamma^i_{jk} = 0 \label{eq:christoffelvanishesinnormalcoordinates}
\end{eqnarray}
for every $i,j,k$.  In other words, we have shown that if we use normal coordinates the connection coefficients all vanish at $p$.  Consequently the covariant derivatives all become partial derivatives.  

Also, by comparing (\ref{eq:christoffelvanishesinnormalcoordinates}) to (\ref{eq:levicivitaconstraint}), we can see that $\Gamma^i_{jk} = 0$ implies 
\begin{eqnarray}
\partial_i g_{jk} = 0 \label{eq:firstderivofmetricvanishesinnormalcoords}
\end{eqnarray}
for every $i,j,k$.  

As a warning, these relationships are \it only \rm true for $p$.  Therefore, in these coordinates, there is no guarantee that the connection will vanish at any other point, and there is no guarantee that any derivatives of the connection will vanish even at $p$.  

We won't always be using normal coordinates, and you shouldn't assume we are in normal coordinates unless we explicitly say so.  We will find them useful for several calculations, but because they only hold for a specific point (at a time), they are only useful to get information about that point.  

\section{The Hodge Star and Integration}

Before moving on to discuss curvature, which is the primary point of this chapter, it will prove useful to first consider another important concept that will aid us greatly as we proceed.  The ideas we will be discussing here are the first in an enormously rich and deep brach of mathematics.  We will revisit these ideas many, many times throughout the rest of these notes and the rest of this series.  For now we will content ourselves with a very superficial discussion.  

\subsection{The Hodge Star}
\label{sec:hodgestar}

Recall\footnote{We strongly suggest you reread section \ref{sec:forms} at this point.} from (\ref{eq:combin}) that the dimension of a given set of $p$-forms is
\begin{eqnarray}
\dim(\Lambda^p \mathcal{M}) = 
\begin{pmatrix}
n \\ p
\end{pmatrix} = {n! \over p!(n-p)!}
\end{eqnarray}
where $n$ is the dimension of $\mathcal{M}$.  Using the standard identity from combinators for the binomial coefficient,
\begin{eqnarray}
\begin{pmatrix}
n \\p
\end{pmatrix} = 
\begin{pmatrix}
n \\ n-p
\end{pmatrix}
\end{eqnarray}
We have that
\begin{eqnarray}
\dim(\Lambda^p\mathcal{M}) = \dim(\Lambda^{n-p}\mathcal{M})
\end{eqnarray}
Specifically, the fact that $\dim(\Lambda^p\mathcal{M}) = \dim(\Lambda^{n-p}\mathcal{M})$ indicates that the number of basis covectors for $\Lambda^p\mathcal{M}$ is equal to the number of basis covectors for $\Lambda^{n-p}\mathcal{M}$.  We can therefore say that the basis for $\Lambda^p\mathcal{M}$ is isomorphic to the basis for $\Lambda^{n-p}\mathcal{M}$.  When we first saw this in section \ref{sec:forms} we didn't have the machinery to exploit this symmetry.  Now that we understand metrics we are able to understand it more fully. 

The idea is that we want to define a map that takes a basis element in $\Lambda^p\mathcal{M}$ to a basis element in $\Lambda^{n-p}\mathcal{M}$ in a well-defined way.  We do this by using the \bf Hodge Star Oparator\rm
\begin{eqnarray}
\star  : \Lambda^p\mathcal{M} \rightarrow \Lambda^{n-p}\mathcal{M}
\end{eqnarray}
Given a $p$ form 
\begin{eqnarray}
\omega = {1 \over p!} \omega_{i_1 i_2 \cdots i_p} dx^{i_1} \wedge dx^{i_2} \wedge \cdots \wedge dx^{i_p}
\end{eqnarray}
we operate on $\omega$ with $\star$ according to
\begin{eqnarray}
\star \omega = {\sqrt{|g|} \over p!(n-p)!} \omega_{i_1 i_2 \cdots i_p} \epsilon^{i_1 i_2 \cdots i_p}_{j_{p+1} j_{p+2} \cdots j_{n}} dx^{j_{p+1}} \wedge dx^{j_{p+2}} \wedge \cdots \wedge dx^{j_n} \label{eq:definitionofthehodgestar}
\end{eqnarray}
where we have used the metric to lower $n-p$ of the indices on the purely antisymmetric tensor:
\begin{eqnarray}
\epsilon^{i_1 i_2 \cdots i_p}_{j_{p+1} j_{p+2} \cdots j_{n}} = g_{j_{p+1}k_{p+1}}g_{j_{p+2}k_{p+2}} \cdots g_{j_n k_n} \epsilon^{i_1i_2\cdots i_p k_{p+1} k_{p+2} \cdots k_n}
\end{eqnarray}

But care must be taken here.  We must be careful about the ordering between the upper and lower indices.  We will take as our convention that the lower indices come in order after the upper indices.  For example, we know that $\epsilon^{123} = 1$ and $\epsilon^{132} = -1$.  If we assume that we are in three dimensional Euclidian space, the metric $g_{ij} = \delta_{ij}$, so 
\begin{eqnarray}
\epsilon^{ij}_k = \delta_{kn}\epsilon^{ijn}
\end{eqnarray}
So 
\begin{eqnarray}
\epsilon^{12}_3 = \delta_{3n}\epsilon^{12n} = \epsilon^{123} = 1
\end{eqnarray}
and 
\begin{eqnarray}
\epsilon^{13}_2 = \delta_{2n}\epsilon^{13n} = \epsilon^{132} = -1
\end{eqnarray}
Similarly,
\begin{eqnarray}
\epsilon^1_{23} = \delta_{2m} \delta_{3n} \epsilon^{1mn} = \epsilon^{123} = 1
\end{eqnarray}
and
\begin{eqnarray}
\epsilon^1_{32} = \delta_{3m}\delta_{2n} \epsilon^{1mn} = \epsilon^{132} = -1
\end{eqnarray}
and so on.  

As a few examples of this, consider again working in $\mathbb{R}^3$ with Euclidian metric.  Let's start with the $0$-form ${1 \over 0!}\omega$.  The determinant of the metric $g_{ij} = \delta_{ij}$ is $\delta = 1$.  In this case, $n=3$ and $p=0$.  Starting with $\epsilon^{ijk}$ we need to lower all three indices, which simply gives $\epsilon_{ijk}$.  So, equation (\ref{eq:definitionofthehodgestar}) gives
\begin{eqnarray}
\star \omega &=& {1 \over 0! (3-0)!} \omega \epsilon_{ijk} dx^i \wedge dx^j \wedge dx^k \nolabel \\
&=& {1 \over 6} \omega (dx \wedge dy \wedge dz + dy \wedge dz \wedge dx + dz \wedge dx \wedge dy \nolabel \\
& & \qquad - dx \wedge dz \wedge dy - dz \wedge dy \wedge dx - dy \wedge dx \wedge dz) \nolabel \\
&=& \omega \; dx \wedge dy \wedge dz \in \Lambda^3\mathbb{R}^3
\end{eqnarray}
which is the natural guess for the $3$ form in $\mathbb{R}^3$ that a $0$ form corresponds to - there is only one of each so of course they map to each other.  

Next consider an arbitrary $1$-form $\boldsymbol{\omega} = \omega_i dx^i = \omega_1 dx + \omega_2 dy + \omega_3 dz$:
\begin{eqnarray}
\star \boldsymbol{\omega} &=& {1 \over 1!(3-1)!} \omega_i \epsilon^i_{jk} dx^j \wedge dx^k \nolabel \\
&=& {1 \over 2}\big( \omega_1( dx^2 \wedge dx^3 - dx^3 \wedge dx^2) + \omega_2( dx^3 \wedge dx^1 - dx^1 \wedge dx^3) + \omega_3 (dx^1 \wedge dx^2 - dx^2 \wedge dx^1)\big) \nolabel \\
&=& \omega_1 dx^2 \wedge dx^3 + \omega_2 dx^3 \wedge dx^1 + \omega_3 dx^1 \wedge dx^2 \in \Lambda^2\mathbb{R}^3
\end{eqnarray}

Next we look at the arbitrary $2$-form $\boldsymbol{\omega} = {1 \over 2} \omega_{ij} dx^i \wedge dx^j$:
\begin{eqnarray}
\star \boldsymbol{\omega} &=& {1 \over 2!(3-2)!}  \omega_{ij} \epsilon^{ij}_k dx^k \nolabel \\
&=& {1 \over 2} \big(\omega_{12}dx^3 - \omega_{21} dx^3 + \omega_{31}dx^2 - \omega_{13}dx^2 + \omega_{23} dx^1 - \omega_{32}dx^1\big) \nolabel \\
&=&   \omega_{12} dx^3 + \omega_{31}dx^2 + \omega_{23}dx^1 \in \Lambda^1\mathbb{R}^3
\end{eqnarray}

And finally, the arbitrary $3$-form $\boldsymbol{\omega} = {1 \over 3!} \omega_{ijk} dx^i \wedge dx^j \wedge dx^k$:
\begin{eqnarray}
\star \boldsymbol{\omega} &=& {1 \over 3!(3-3)!} \omega_{ijk} \epsilon^{ijk} \nolabel \\
&=& {1 \over 6} (\omega_{123} + \omega_{231} + \omega_{312} - \omega_{132} - \omega_{321} - \omega_{213}) \nolabel \\
&=& \omega_{123} \in \Lambda^0 \mathbb{R}^3
\end{eqnarray}

On the other hand, consider $\mathbb{R}^4$ with a Minkowski metric $g_{ij} = \eta_{ij} = diag(-1,1,1,1)$.  Now a $0$-form will become a $4$-form - you can work out the details yourself (the indices run from $0$ to $3$, where $x^0 = t$):
\begin{eqnarray}
\star \omega &=& {1 \over 0!(4-0)!} \omega \eta_{ai} \eta_{bj} \eta_{ck} \eta_{dl} \epsilon^{abcd} dx^i \wedge dx^j \wedge dx^k \wedge dx^l \nolabel \\
&=& - \omega dt \wedge dx \wedge dy \wedge dz
\end{eqnarray}
You are encouraged to work out the rest in Minkowski space on your own.  

As a few other examples in three dimensional Euclidian space, consider two covectors $\omega_i dx^i$ and $\alpha_i dx^i$.  In equation (\ref{eq:earlierequationwheretwoomegaswedgedtogetherlookslikecrossproduct}) we commented that the wedge product between them looks like the cross product.  Specifically,
\begin{eqnarray}
\boldsymbol{\omega} \wedge \boldsymbol{\alpha} = \det
\begin{pmatrix}
dy \wedge dz & dx \wedge dz & dx \wedge dy \\
\omega_1 & \omega_2 & \omega_3 \\
\alpha_1 & \alpha_2 & \alpha_3
\end{pmatrix}
\end{eqnarray}
Consider now the quantity $\star (\boldsymbol{\omega} \wedge \boldsymbol{\alpha})$.  You can write out the details yourself (expand out the determinant and then apply $\star$ on each term):
\begin{eqnarray}
\star (\boldsymbol{\omega} \wedge \boldsymbol{\alpha}) = \det
\begin{pmatrix}
dx & dy & dz \\
\omega_1 & \omega_2 & \omega_3 \\
\alpha_1 & \alpha_2 & \alpha_3
\end{pmatrix}
\end{eqnarray}
which is the exact form of the cross product in three dimensions.  

You are encouraged to convince yourself of the following generalized relations in three dimensions, but we merely quote the results (we are denoting a $p$-form as $\omega_{(p)}$).  
\begin{eqnarray}
\rm gradient\it :\qquad \boldsymbol{\nabla} \omega_{(0)} &=& d \omega_{(0)} \nolabel \\
\rm curl\it:\qquad \boldsymbol{\nabla} \times \boldsymbol{\omega}_{(1)} &=& \star d \boldsymbol{\omega}_{(1)} \nolabel \\
\rm divergence\it:\qquad \boldsymbol{\nabla} \cdot \boldsymbol{\omega}_{(1)} &=& \star d \star \boldsymbol{\omega}_{(1)} \label{eq:gradcurlanddivincoordandformnotation}
\end{eqnarray}
The difference between the left hand sides and the right hand sides of (\ref{eq:gradcurlanddivincoordandformnotation}) is that the right hand sides are unique to $3$ dimensional Euclidian space, whereas the left hand sides are totally general for any arbitrary manifold.  We will see how they allow us to generalize things like Maxwell's equations later.  

\subsection{Invariant Volume Elements}
\label{sec:invariantvolumeelement}

We make one final brief comment before moving on to discuss curvature\footnote{It would be helpful to reread section \ref{sec:intdiffforms} at this point.}.  Consider an $n$ dimensional manifold $\mathcal{M}$ with coordinate functions $x^i$ for $i=1,\ldots,n$.  At any given point $p$ a basis for $\Lambda^1T_p\mathcal{M}$, the cotangent space at $p$, will be the set
\begin{eqnarray}
dx^i \qquad i=1,\cdots ,n
\end{eqnarray}
We can then put these together to form a volume element as in equation (\ref{eq:volumeformsincartesianandspherical}).  In Cartesian coordinates for $\mathbb{R}^3$ this was
\begin{eqnarray}
dx \wedge dy \wedge dz \label{eq:volumeforminr3usingcartcordasdf}
\end{eqnarray}
and in spherical coordinates for $\mathbb{R}^3$ this was
\begin{eqnarray}
r^2 \sin \phi dr \wedge d\theta \wedge d\phi \label{eq:volumeforminr3usingcartcordasdf2}
\end{eqnarray}

More generally we can write the volume form as
\begin{eqnarray}
h\; dx^1 \wedge dx^2 \wedge \cdots \wedge dx^n
\end{eqnarray}
where $h$ is some (scalar) function that makes the integral a legitimate volume form.  It was $h=1$ for Cartesian coordinates and $h = r^2 \sin \phi$ in spherical coordinates.  

But, consider the transformation law for this - under the transformation from coordinate functions $x^i$ to coordinate functions $y^i$, we have (cf equation (\ref{eq:definitionofdeterminant}))
\begin{eqnarray}
h(x^i) dx^1 \wedge \cdots \wedge dx^n = h'(y^i) \bigg| {\partial x \over \partial y}\bigg| dy^1 \wedge \cdots \wedge dy^n
\end{eqnarray}
where the lines $|\; |$ indicate a determinant.  We left off the indices on the determinant term both for  brevity and to emphasize that it is a determinant and therefore a scalar - it has no indices.  You should note that the $x$ and $y$ do have multiple components and a determinant is being taken over those components.  

Let us make the permanent definition
\begin{eqnarray}
h(x^i) = \sqrt{|g|}
\end{eqnarray}
where $|g|$ is the determinant of $g_{ij}$ as expressed in the $x^i$ coordinate functions.  The transformation law for $g_{ij}$ is (denoting the metric in $y^i$ coordinates as $\tilde g_{ij}$)
\begin{eqnarray}
g_{ij} = {\partial y^a \over \partial x^i} {\partial y^b \over \partial x^j} \tilde g_{ab}
\end{eqnarray}
So the determinant
\begin{eqnarray}
|g| = \bigg| {\partial y^a \over \partial x^i} {\partial y^b \over \partial x^j} \tilde g_{ab}\bigg| = \bigg| {\partial y^a \over \partial x^i} \bigg| \bigg| {\partial y^b \over \partial x^j} \bigg| |\tilde g| = \bigg|{\partial y \over \partial x} \bigg|^2 |\tilde g|
\end{eqnarray}
And therefore
\begin{eqnarray}
\sqrt{|g|} = \bigg|{\partial y \over \partial x}\bigg| \sqrt{|\tilde g|}
\end{eqnarray}
And therefore the entire volume form we have defined transforms as
\begin{eqnarray}
\sqrt{|g|} dx^1 \wedge \cdots \wedge dx^n &=& \bigg|{\partial y \over \partial x}\bigg| \bigg|{\partial x \over \partial y}\bigg| \sqrt{|\tilde g|} dy^1 \wedge \cdots \wedge dy^n \nolabel \\
&=& \sqrt{|\tilde g|} dy^1 \wedge \cdots \wedge dy^n
\end{eqnarray}
So it is indeed invariant.  

This is consistent with what we did above with (\ref{eq:volumeforminr3usingcartcordasdf}) and (\ref{eq:volumeforminr3usingcartcordasdf2}).  The determinant of the Euclidian metric in Cartesian coordinates, $g_{ij} = \delta_{ij}$, is obviously $\sqrt{|g|} = 1$, which is what we had.  Then, from (\ref{eq:sphericalcoordinatemetricinr3}), for spherical coordinates we have
\begin{eqnarray}
\sqrt{|g|} = \sqrt{\det 
\begin{pmatrix}
1 & 0 & 0 \\ 0 & r^2 \sin^2\phi & 0 \\ 0 & 0 & r^2
\end{pmatrix} } = \sqrt{r^4 \sin^2 \phi} = r^2 \sin \phi
\end{eqnarray}
which is also what we had above for spherical coordinates.  

The purpose of this is that we now have a way of defining an integral over any (orientable) manifold $\mathcal{M}$ that is invariant under coordinate transformations.  We will denote this invariant volume element the volume form $\Omega_{\mathcal{M}}$, and if we want to integrate an arbitrary function $f$ over $\mathcal{M}$, we define the integral to be
\begin{eqnarray}
\int_{\mathcal{M}} f \Omega_{\mathcal{M}} = \int_{\mathcal{M}} f \sqrt{|g|} dx^1 dx^2 \cdots dx^n
\end{eqnarray}

Finally, consider the Hodge dual of the $0$-form $1$:
\begin{eqnarray}
\star 1 &=& {\sqrt{|g|} \over n!} \epsilon_{j_1j_2\cdots j_n} dx^{j_1} \wedge dx^{j_2} \wedge \cdots \wedge dx^{j_n}\nolabel \\
&=& \sqrt{|g|} dx^1 \wedge dx^2 \wedge \cdots \wedge dx^n
\end{eqnarray}
So we can rewrite the integral of an arbitrary function $f$ over $\mathcal{M}$ as
\begin{eqnarray}
\int_{\mathcal{M}} f \star 1
\end{eqnarray}

\subsection{Hodge Star Properties}

Consider a $p$-form $\boldsymbol{\omega} = {1 \over p!} \omega_{i_1i_2\cdots i_p}dx^{i_1}\wedge dx^{i_2} \wedge \cdots \wedge dx^{i_p}$ on an $n$ dimensional manifold $\mathcal{M}$ with Riemannian metric.  For simplicity we will take $\sqrt{|g|} = 1$.  Taking the dual:
\begin{eqnarray}
\star \boldsymbol{\omega} &=& {1 \over p!(n-p)!} \omega_{i_1i_2 \cdots i_p} \epsilon^{i_1i_2\cdots i_p}_{j_{p+1}j_{p+1}\cdots j_n} dx^{j_{p+1}}\wedge dx^{j_{p+2}} \wedge \cdots \wedge dx^{j_n}
\end{eqnarray}
Let's then take the Hodge dual of this:
\begin{eqnarray}
\star \star \boldsymbol{\omega} &=& {1 \over (n-p)!(n-(n-p))!}\omega_{i_1 \cdots i_p} \epsilon^{i_1\cdots i_p}_{j_{p+1}j_{p+1}\cdots j_n} \epsilon^{j_{p+1}j_{p+1}\cdots j_n}_{k_1 \cdots k_p} dx^{k_1} \wedge \cdots \wedge dx^{k_p} \nolabel\\
&=& {\omega_{i_1\cdots i_p} \over p!(n-p)!}  \epsilon^{i_1\cdots i_p m_{p+1}\cdots m_n}\epsilon^{j_{p+1}\cdots j_n n_1\cdots n_p} g_{m_{p+1}j_{p+1}} \cdots g_{m_nj_n} g_{n_1k_1}\cdots g_{n_pk_p}dx^{k_1} \wedge \cdots \wedge dx^{k_p} \nolabel \\
&=& {(-1)^{p(n-p)} \over p!(n-p)!} \omega_{i_1\cdots i_p} \epsilon^{i_1\cdots i_p m_{p+1}\cdots m_n} \epsilon^{ n_1\cdots n_p j_{p+1}\cdots j_n} g_{m_{p+1}j_{p+1}} \cdots g_{m_nj_n} g_{n_1k_1}\cdots g_{n_pk_p}dx^{k_1} \wedge \cdots \wedge dx^{k_p} \nolabel \\
&=& {(-1)^{p(n-p)} \over p!(n-p)!} \omega_{i_1\cdots i_p} \epsilon^{i_1\cdots i_p m_{p+1}\cdots m_n} \epsilon_{k_1\cdots k_pm_{p+1}\cdots m_n} dx^{k_1} \wedge \cdots \wedge dx^{k_p} \nolabel \\
&=& {(-1)^{p(n-p)} \over p!(n-p)!} \omega_{i_1\cdots i_p} (n-p)! dx^{i_1} \wedge \cdots \wedge dx^{i_p} \nolabel \\
&=& {(-1)^{p(n-p)} \over p!} \omega_{i_1 \cdots i_p} dx^{i_1} \wedge \cdots \wedge dx^{i_p} \nolabel \\
&=& (-1)^{p(n-p)} \boldsymbol{\omega}
\end{eqnarray}
So $\star \star$ is the identity operator up to a sign\footnote{To see the $5^{th}$ equality write out a few simple examples.}.  This is expected since $\star$ takes a form of rank $p$ to $n-p$, and then $n-p$ back to $n-(n-p) = p$.  

It is instructive to work out that on a pseudo-Riemannian manifold with a Lorentzian metric (one negative and the rest positive so that $\det(g) = -1$) we have
\begin{eqnarray}
\star \star = (-1)^{1+p(n-p)}
\end{eqnarray}

This makes finding the following identity maps easy:
\begin{eqnarray}
1 &=& (-1)^{p(n-p)} \star \star \qquad \qquad Riemannian \nolabel \\
1 &=& (-1)^{1+ p(n-p)} \star \star \qquad \qquad Lorentzian \label{eq:staroperatoridentityoperators}
\end{eqnarray}
And therefore
\begin{eqnarray}
\star^{-1} &=& (-1)^{p(n-p)} \star \qquad \qquad Riemannian \nolabel \\
\star^{-1} &=& (-1)^{1+p(n-p)} \star \qquad \qquad Lorentzian
\end{eqnarray}

Now consider two $p$-forms (on an $n$-dimensional manifold) $\boldsymbol{\omega}$ and $\boldsymbol{\alpha}$.  We discussed the wedge product of $\boldsymbol{\omega}$ and $\boldsymbol{\alpha}$ in section \ref{sec:forms}.  Obviously if $p$ is greater than ${n\over 2}$ then $\boldsymbol{\omega} \wedge \boldsymbol{\alpha}$ will vanish identically.  However, consider the quantity $\boldsymbol{\omega} \wedge \star \boldsymbol{\alpha}$:
\begin{eqnarray}
\boldsymbol{\omega} \wedge \star \boldsymbol{\alpha} &=& { 1\over p!}\omega_{i_1\cdots i_p} dx^{i_1} \wedge \cdots \wedge dx^{i_p} \wedge {\sqrt{|g|} \over p!(n-p)!} \alpha_{j_1\cdots j_p}\epsilon^{j_1\cdots j_p}_{k_{p+1}\cdots k_n} dx^{k_{p+1}} \wedge \cdots\wedge dx^{k_n} \nolabel \\
&=& {1 \over (p!)^2 (n-p)!} \omega_{i_1 \cdots i_p} \alpha_{j_1 \cdots j_p} \epsilon^{j_1 \cdots j_p}_{k_{p+1}\cdots k_n} \sqrt{|g|} dx^{i_1} \wedge \cdots \wedge dx^{i_p} \wedge dx^{k_{p+1}} \wedge \cdots \wedge dx^{k_n} \nolabel \\
&=& {1 \over (p!)^2 (n-p)!} \omega_{i_1 \cdots i_p} \alpha^{j_1 \cdots j_p} \epsilon_{j_1 \cdots j_p k_{p+1} \cdots k_n}  \epsilon^{i_1 \cdots i_p k_{p+1}\cdots  k_n} \sqrt{|g|} dx^1 \wedge \cdots \wedge dx^n \nolabel \\
&=& { 1\over p!} \omega_{i_1\cdots i_p} \alpha^{i_1 \cdots i_p} \Omega_{\mathcal{M}} \label{eq:toreferaesdkfhasdfto1}
\end{eqnarray}
where, of course, $\Omega_{\mathcal{M}}$ is the invariant volume form on the manifold.  Notice that had we swapped $\boldsymbol{\omega}$ and $\boldsymbol{\alpha}$ in the above expression we would have gotten
\begin{eqnarray}
\boldsymbol{\alpha} \wedge \star \boldsymbol{\omega} &=& {1 \over p!} \alpha_{i_1 \cdots i_p} \omega^{i_1 \cdots i_p} \Omega_{\mathcal{M}}
\end{eqnarray}
And therefore
\begin{eqnarray}
\boldsymbol{\omega} \wedge \star \boldsymbol{\alpha} = \boldsymbol{\alpha}\wedge \star \boldsymbol{\omega} \label{eq:omegawedgestaralplhaequalsalphawedgestaromega}
\end{eqnarray}
And because this is a well-defined $n$ form, its integral 
\begin{eqnarray}
\int_{\mathcal{M}} \boldsymbol{\omega}\wedge \star \boldsymbol{\alpha} \label{eq:toreferaesdkfhasdfto2}
\end{eqnarray}
is well defined.  It will prove useful to define an inner product between two forms according to (\ref{eq:toreferaesdkfhasdfto2}).  In other words, the inner product between an $n$-form $\boldsymbol{\alpha}$ and an $n$-form $\boldsymbol{\beta}$ is 
\begin{eqnarray}
(\boldsymbol{\alpha} , \boldsymbol{\beta}) \equiv \int \boldsymbol{\alpha} \wedge \star \boldsymbol{\beta} \label{eq:innerproductbetweentwonforms}
\end{eqnarray}

Now consider the exterior derivative of $\boldsymbol{\omega}\wedge \star \boldsymbol{\alpha}$ where $\boldsymbol{\alpha}$ is a $p$ form and $\boldsymbol{\omega}$ is a $p-1$ form:
\begin{eqnarray}
d (\boldsymbol{\omega}\wedge \star \boldsymbol{\alpha}) &=& (d\boldsymbol{\omega}) \wedge \boldsymbol{\star \alpha} + (-1)^{p-1}\boldsymbol{\omega} \wedge (d\star \boldsymbol{\alpha})
\end{eqnarray}
(cf rules for exterior derivative on page \pageref{pagewithrulesforexteriorderivative}).  Then (assuming this is a Riemannian manifold) we can insert the identity operator (first equation in (\ref{eq:staroperatoridentityoperators})).  Noting that $d \star \boldsymbol{\alpha}$ is an $n-p+1$ form, this gives
\begin{eqnarray}
d(\boldsymbol{\omega} \wedge \star \boldsymbol{\alpha}) &=& (d\boldsymbol{\omega}) \wedge \star \boldsymbol{\alpha} + (-1)^{p-1}(-1)^{ (n-p+1)(n-(n-p+1))}\boldsymbol{\omega} \wedge (\star \star d \star \boldsymbol{\alpha}) \nolabel \\
&=& (d\boldsymbol{\omega}) \wedge \star \boldsymbol{\alpha} - (-1)^{ np+n+1 }\boldsymbol{\omega} \wedge (\star \star d \star \boldsymbol{\alpha}) \label{eq:extderivofomegawedgestaralpha}
\end{eqnarray}
Now define the operator
\begin{eqnarray}
(-1)^{np+n+1} \star d \star \boldsymbol{\alpha} \equiv d^{\dagger} \boldsymbol{\alpha}
\end{eqnarray}
where
\begin{eqnarray}
d^{\dagger} = (-1)^{np+n+1}\star d\star \label{eq:adjointexteriorderiv}
\end{eqnarray}
Now we can write (\ref{eq:extderivofomegawedgestaralpha}) as
\begin{eqnarray}
d (\boldsymbol{\omega} \wedge \star \boldsymbol{\alpha}) &=& (d\boldsymbol{\omega}) \wedge \star \boldsymbol{\alpha} - \boldsymbol{\omega} \wedge( \star d^{\dagger} \boldsymbol{\alpha})
\end{eqnarray}
Finally, integrate $d(\boldsymbol{\omega} \wedge \star \boldsymbol{\alpha})$ over a manifold without a boundary $\partial\mathcal{M} \equiv 0$.  By Stokes theorem this will vanish.  So,
\begin{eqnarray}
\int_{\mathcal{M}} d(\boldsymbol{\omega}\wedge \star \boldsymbol{\alpha}) &=& \int_{\mathcal{M}} \big((d\boldsymbol{\omega} )\wedge \star \boldsymbol{\alpha} - \boldsymbol{\omega} \wedge (\star d^{\dagger} \boldsymbol{\alpha})\big) \nolabel\\
&=& \int_{\partial\mathcal{M}} \boldsymbol{\omega} \wedge \star \boldsymbol{\alpha} \nolabel \\
&=& 0
\end{eqnarray}
And therefore
\begin{eqnarray}
\int_{\mathcal{M}} (d\boldsymbol{\omega}) \wedge \star \boldsymbol{\alpha} = \int_{\mathcal{M}} \boldsymbol{\omega} \wedge (\star d^{\dagger} \boldsymbol{\alpha}) \label{eq:resultthatisusfullater}
\end{eqnarray}
This result will be useful later.  

Looking briefly at the operator (\ref{eq:adjointexteriorderiv}), which is called the \bf Adjoint Exterior Derivative\rm, consider what it does to a $p$-form.  The $\star$ takes the $p$-form to an $n-p$ form.  The exterior derivative then takes the $n-p$ form to an $n-p+1$ form.  The next $\star$ then takes the $n-p+1$ form to an $n-(n-p+1) = p+1$ form.  So,
\begin{eqnarray}
d^{\dagger}: \Lambda^p\mathcal{M} \longrightarrow \Lambda^{p+1} \mathcal{M}
\end{eqnarray}
Finally consider the square of $d^{\dagger}$:
\begin{eqnarray}
d^{\dagger} d^{\dagger} &=& \star d \star \star d \star \propto \star d^2 \star \equiv 0 \label{eq:ddaggerisnilpotent}
\end{eqnarray}
So $d^{\dagger}$ is, like $d$, nilpotent.  

\subsection{The Hodge Decomposition and Harmonic Forms}

Before moving on, we will plant a few seeds for ideas that will be significant (and remarkably powerful) later in this series.  We discussed cohomology in some detail previously (section \ref{sec:cohomology}).  Recall that the "point" (so to speak) of cohomology stems from the nilpotency of the exterior derivative:
\begin{eqnarray}
d^2 = 0
\end{eqnarray}
We used this to draw a distinction between forms that are exact\footnote{A form $\boldsymbol{\omega}$ is exact if it can be written globally as the exterior derivative of another form: $\boldsymbol{\omega} = d \alpha$.} and consequently closed\footnote{A form $\boldsymbol{\omega}$ is closed if its exterior derivative vanishes: $d\boldsymbol{\omega} = 0$.} trivially, and forms that are closed but not exact.  We saw that the existence of forms that are closed but not exact indicates a "hole" of some sort, or in other words that the manifold is topologically non-trivial.  

In the previous section we defined a new differential operator, $d^{\dagger}$, and in (\ref{eq:ddaggerisnilpotent}) showed that $d^{\dagger}$ is nilpotent.  This leads us to consider the cohomology of $d^{\dagger}$, which we from now on call the \bf co-exterior derivative\rm.  

Following what we did with the cohomology of $d$, we define a form as \bf coexact \rm if it can be \it globally \rm written as the co-exterior derivative of another form.  In other words, $\boldsymbol{\omega}$ is coexact if there exists a form $\boldsymbol{\alpha}$ such that
\begin{eqnarray}
\boldsymbol{\omega} = d^{\dagger} \boldsymbol{\alpha}
\end{eqnarray}
holds globally.\footnote{Again, the requirement that this be global is extremely important.}  We then define a form as \bf coclosed \rm if 
\begin{eqnarray}
d^{\dagger} \boldsymbol{\omega} = 0
\end{eqnarray}

Then, just as in section \ref{sec:cohomology}, we call the set of all coexact $n$-forms $B^n_{d^{\dagger}}(\mathcal{M})$ and the set of all coclosed $n$-forms $Z^n_{d^{\dagger}}(\mathcal{M})$.\footnote{From now on we will use a $d$ or $d^{\dagger}$ subscript to specify whether we are doing $d$ cohomology or $d^{\dagger}$ cohomology.}  This allows us to define an alternative cohomology group
\begin{eqnarray}
H^n_{d^{\dagger}}(\mathcal{M}) \equiv Z^n_{d^{\dagger}}(\mathcal{M}) / B^n_{d^{\dagger}}(\mathcal{M}) \label{eq:alternativecohomologygroup}
\end{eqnarray}
The "meaning" of this construction of cohomology is similar to the meaning of the $d$ cohomology of section \ref{sec:cohomology}, and we won't delve into the details here,\footnote{We will later in a much more general setting.  As you can imagine we are skimming the peaks of enormous mountains here.} other than to point out that $d$ mapped $n$ forms to $n+1$ forms, and $d^{\dagger}$ maps $n$ forms to $n-1$ forms, which changes the direction of the $d^{\dagger}$ cohomology.  

However, there is an important relationship between exact forms and coexact forms.  We said in equation (\ref{eq:innerproductbetweentwonforms}) that the inner product between two $n$-forms is given by
\begin{eqnarray}
(\boldsymbol{\alpha} , \boldsymbol{\beta}) = \int \boldsymbol{\alpha} \wedge\star \boldsymbol{\beta}
\end{eqnarray}
So consider the inner product between a closed form an a co-closed form:
\begin{eqnarray}
(d \boldsymbol{\alpha}, d^{\dagger} \boldsymbol{\beta}) = \int d\boldsymbol{\alpha} \wedge \star d^{\dagger} \boldsymbol{\beta}
\end{eqnarray}
where clearly $\boldsymbol{\alpha}$ is an $n-1$ form and $\boldsymbol{\beta}$ is an $n+1$ form.  Now according to (\ref{eq:resultthatisusfullater}) we can rewrite this as
\begin{eqnarray}
(d \boldsymbol{\alpha}, d^{\dagger} \boldsymbol{\beta}) &=& \int d\boldsymbol{\alpha} \wedge \star d^{\dagger} d^{\dagger} \boldsymbol{\beta} \nolabel \\
&=& \int d\boldsymbol{\alpha} \wedge 0  \nolabel\\
&\equiv & 0
\end{eqnarray} 
So, exact forms and coexact forms are orthogonal to each other.  This tells us that we can decompose an arbitrary form into a part that is exact, a part that is coexact, and something else that is neither exact nor coexact but thus far undefined.  In other words, if we define the set of all exact forms as $B^n_d(\mathcal{M})$ and the set of all coexact forms as $B^n_{d^{\dagger}}(\mathcal{M})$, 
\begin{eqnarray}
\Lambda^n\mathcal{M} = B^n_d(\mathcal{M}) \oplus B^n_{d^{\dagger}}(\mathcal{M}) \oplus (?)
\end{eqnarray}
where we have left the final part a question mark because we assume that there must exist forms that are neither exact nor coexact, but we don't know what kind of forms they are.  So, we want to find the set of forms that are orthogonal (according to the inner product (\ref{eq:innerproductbetweentwonforms})) to both $B^n_d(\mathcal{M})$ and $B^n_{d^{\dagger}}(\mathcal{M})$.  

Consider a form $\boldsymbol{\gamma}$.  We want to find the most general $\boldsymbol{\gamma}$ that satisfies both of the following:\footnote{We write them in opposite order with foresight - to make the result follow in a more obvious way.  This is allowed because it is an inner product - $(\boldsymbol{\alpha}, \boldsymbol{\beta}) = (\boldsymbol{\beta}, \boldsymbol{\alpha})$ in general.}
\begin{eqnarray}
0 &=& (d\boldsymbol{\alpha}, \boldsymbol{\gamma}) = \int d\boldsymbol{\alpha} \wedge \star \boldsymbol{\gamma} \nolabel \\
0 &=& (\boldsymbol{\gamma},d^{\dagger} \boldsymbol{\beta}) = \int  \boldsymbol{\gamma}\wedge \star d^{\dagger} \boldsymbol{\beta} 
\end{eqnarray}
Then, using (\ref{eq:resultthatisusfullater}) on both, we have
\begin{eqnarray}
\int \boldsymbol{\alpha} \wedge \star d^{\dagger} \boldsymbol{\gamma} &=& 0 \nolabel \\
\int d \boldsymbol{\gamma} \wedge \star \boldsymbol{\beta} &=& 0
\end{eqnarray}
So, in order for this to be satisfied for arbitrary $\boldsymbol{\alpha}$ and $\boldsymbol{\beta}$, it must be the case that
\begin{eqnarray}
d\boldsymbol{\gamma} = d^{\dagger} \boldsymbol{\gamma} = 0 \label{eq:dgammaequalsddaggergammaequalszero}
\end{eqnarray}
So, $\boldsymbol{\gamma}$ must be both closed and coclosed.  But we can make an important observation - if $\boldsymbol{\gamma}$ is an $p$-form, then $d\boldsymbol{\gamma}$ is a $p+1$ form and $d^{\dagger}\boldsymbol{\gamma}$ is a $p-1$ form.  Furthermore, both $dd^{\dagger} \boldsymbol{\gamma}$ and $d^{\dagger}d\boldsymbol{\gamma}$ will be $p$-forms.  If $\boldsymbol{\gamma}$ is closed and coclosed, then we can take
\begin{eqnarray}
(d^{\dagger}d + dd^{\dagger}) \boldsymbol{\gamma} = (d+d^{\dagger})^2 \boldsymbol{\gamma} \equiv \Delta \boldsymbol{\gamma} =  0
\end{eqnarray}
where 
\begin{eqnarray}
\Delta \equiv (d+d^{\dagger})^2 = d^{\dagger}d + dd^{\dagger}
\end{eqnarray}
is called the \bf Laplacian \rm operator.\footnote{It is called the Laplacian because if you write it out in three dimensional Euclidian space with metric $g_{ij} = \delta_{ij}$ it is simply 
$$\vec \nabla^2 f(x,y,z) = \bigg({\partial^2 \over \partial x^2} + {\partial^2 \over \partial y^2}+{\partial^2\over\partial z^2}\bigg) f(x,y,z)$$So, $\Delta$ is simply the generalization of $\vec \nabla^2$ to an arbitrary $n$-dimensional manifold with an arbitrary metric.}  

So, if $\boldsymbol{\gamma}$ is both closed and coclosed, then $\Delta\boldsymbol{\gamma} = 0$.  On the other hand, if we assume that 
\begin{eqnarray}
\Delta\boldsymbol{\gamma} = 0 \label{eq:capitaldeltagammaequalszero}
\end{eqnarray}
then it must be the case that $\boldsymbol{\gamma}$ is both closed and coclosed.  To see this assume the contrary, that it is closed but not coclosed.  This means that $d^{\dagger} \boldsymbol{\gamma}$ is not zero, but is equal to some other form, say $d^{\dagger}\boldsymbol{\gamma} = \boldsymbol{\omega}$.  So, 
\begin{eqnarray}
0 &=& \Delta \boldsymbol{\gamma} \nolabel \\
&=& (d^{\dagger}d + dd^{\dagger})\boldsymbol{\gamma} \nolabel \\
&=& dd^{\dagger} \boldsymbol{\gamma} \nolabel \\
&=& d \boldsymbol{\omega}
\end{eqnarray}
Because this is not true in general we have a contradiction, and after using the same argument where we assume $\boldsymbol{\omega}$ is coclosed but not closed, we can conclude 
\begin{eqnarray}
d\boldsymbol{\gamma}=d^{\dagger}\boldsymbol{\gamma} = 0 \iff \Delta \boldsymbol{\gamma} = 0
\end{eqnarray}
So we have traded the two constraints (\ref{eq:dgammaequalsddaggergammaequalszero}) for the single constraint (\ref{eq:capitaldeltagammaequalszero}).  

We denote any form that satisfies 
\begin{eqnarray}
\Delta \boldsymbol{\gamma} = 0
\end{eqnarray}
a \bf Harmonic Form\rm.  We denote the set of all harmonic $n$-forms on $\mathcal{M}$ as $H^n_{\Delta}(\mathcal{M})$

So finally, because we know that $B^n_d(\mathcal{M})$ is orthogonal to $B^n_{d^{\dagger}}(\mathcal{M})$, and both are orthogonal to 
$H^n_{\Delta}(\mathcal{M})$, we have the decomposition
\begin{eqnarray}
\Lambda^n(\mathcal{M}) = B^n_d(\mathcal{M}) \oplus B^n_{d^{\dagger}}(\mathcal{M}) \oplus  H^n_{\Delta}(\mathcal{M}) \label{eq:arbitraryformdecomposesintoexactcoexactandharmonic}
\end{eqnarray}
This decomposition is called the \bf Hodge Decomposition\rm.  It essentially says that just as any matrix can be decomposed into an antisymmetric part, a traceless symmetric part, and a pure trace, any form can be decomposed into an exact form, a coexact form, and a harmonic form.  This is true of all forms on an arbitrary differentiable manifold.  

\subsection{A Bit More Topology}

A natural question to ask why 'why have we gone to the trouble of decomposing an arbitrary form into these three orthogonal parts?' (\ref{eq:arbitraryformdecomposesintoexactcoexactandharmonic}).  

Consider a $n$-form $\boldsymbol{\omega}$ that is in the $d$-Homology group $H^n_d(\mathcal{M})$.  From section \ref{sec:cohomology} we know that this means that $\boldsymbol{\omega}$ is closed:
\begin{eqnarray}
d \boldsymbol{\omega} = 0
\end{eqnarray}
but there does not exist any $n-1$ form $\boldsymbol{\alpha}$ such that
\begin{eqnarray}
\boldsymbol{\omega} = d \boldsymbol{\alpha}
\end{eqnarray}
or in other words, $\boldsymbol{\omega}$ is not exact.  Knowing from the previous section that we can decompose an arbitrary form $\boldsymbol{\omega}$ into the exterior derivative of an $n-1$ form $\boldsymbol{\alpha}$, the coexterior derivative of an $n+1$ form $\boldsymbol{\beta}$, and a harmonic form $\boldsymbol{\gamma}$, we write
\begin{eqnarray}
\boldsymbol{\omega} = d\boldsymbol{\alpha} + d^{\dagger} \boldsymbol{\beta} + \boldsymbol{\gamma}
\end{eqnarray}
The fact that $\boldsymbol{\omega}$ is closed means that
\begin{eqnarray}
d \boldsymbol{\omega} = d^2 \boldsymbol{\alpha} + dd^{\dagger} \boldsymbol{\beta} + d \boldsymbol{\gamma} = dd^{\dagger} \boldsymbol{\beta} = 0
\end{eqnarray}
But then, using the inner product (\ref{eq:innerproductbetweentwonforms}) and the relationship (\ref{eq:resultthatisusfullater}), this implies
\begin{eqnarray}
0&=& (0,\boldsymbol{\beta}) \nolabel \\
&=& (dd^{\dagger} \boldsymbol{\beta},  \boldsymbol{\beta}) \nolabel \\
&=& (d^{\dagger} \boldsymbol{\beta},d^{\dagger} \boldsymbol{\beta}) 
\end{eqnarray}
which implies
\begin{eqnarray}
d^{\dagger} \boldsymbol{\beta} = 0
\end{eqnarray}
and therefore $\boldsymbol{\omega} \in H^n_d(\mathcal{M})$ is generally written as
\begin{eqnarray}
\boldsymbol{\omega} = d \boldsymbol{\alpha} + \boldsymbol{\gamma}
\end{eqnarray}

Now consider some other element $\boldsymbol{\omega}' \in H^n_d(\mathcal{M})$.  By definition this means that there exists some form $\boldsymbol{\theta}$ such that
\begin{eqnarray}
\boldsymbol{\omega}' = \boldsymbol{\omega} + d\boldsymbol{\theta} = d\boldsymbol{\alpha} + \boldsymbol{\gamma} + d \boldsymbol{\theta} = d(\boldsymbol{\alpha} + \boldsymbol{\theta}) + \boldsymbol{\gamma} = d \boldsymbol{\psi} + \boldsymbol{\gamma}
\end{eqnarray}
where 
\begin{eqnarray}
\boldsymbol{\psi} \equiv  \boldsymbol{\alpha}+\boldsymbol{\theta}
\end{eqnarray}
Notice that the only difference between
\begin{eqnarray}
\boldsymbol{\omega} = d\boldsymbol{\alpha}+\boldsymbol{\gamma}
\end{eqnarray}
and
\begin{eqnarray}
\boldsymbol{\omega}' = d\boldsymbol{\psi}+\boldsymbol{\gamma}
\end{eqnarray}
is in the exact form.  The harmonic form is the same in both cases.  In other words, every representative of a given equivalence class in $H^n_d(\mathcal{M})$ has the same harmonic form, which means that there is only one harmonic form for a given equivalence class in $H^n_d(\mathcal{M})$.  

What all of this allows us to say is that given any element $\boldsymbol{\omega} \in H^n_d(\mathcal{M})$ such that $\boldsymbol{\omega}$ decomposes as written above, we can choose $\boldsymbol{\theta}$ above to be $\boldsymbol{\theta} = - \boldsymbol{\alpha}$, so 
\begin{eqnarray}
\boldsymbol{\omega}' = \boldsymbol{\gamma}
\end{eqnarray}
This means that every equivalence class, or element, of $H^n_d(\mathcal{M})$, has a single harmonic form which can represent it.  So,
\begin{eqnarray}
H^n_d(\mathcal{M}) \subset H^n_{\Delta}(\mathcal{M}) \label{eq:dhnmsubsetdeltahnm}
\end{eqnarray}

Looking at the definition of the $n^{th}$ cohomology group,
\begin{eqnarray}
H^n_d(\mathcal{M}) = Z^n_d(\mathcal{M}) / B^n_d(\mathcal{M})
\end{eqnarray}
Because $\boldsymbol{\gamma}$ is harmonic as we have assumed, it is therefore also closed (and coclosed).  Therefore
\begin{eqnarray}
\boldsymbol{\gamma} \in Z^n_d(\mathcal{M})
\end{eqnarray}
However, by definition $\boldsymbol{\gamma}$ is not exact, and therefore
\begin{eqnarray}
\boldsymbol{\gamma} \not \in B^n_d(\mathcal{M})
\end{eqnarray}
And so, by the definition of $H^n_d(\mathcal{M})$, this means that $\boldsymbol{\gamma}$ is a non-trivial element of $H^n_d(\mathcal{M})$, so
\begin{eqnarray}
H^n_{\Delta}(\mathcal{M}) \subset H^n_d(\mathcal{M}) \label{eq:deltahnmsubsetdhnm}
\end{eqnarray}

Finally, comparing (\ref{eq:dhnmsubsetdeltahnm}) to (\ref{eq:deltahnmsubsetdhnm}), we have the remarkable result
\begin{eqnarray}
H^n_{\Delta}(\mathcal{M}) = H^n_d(\mathcal{M})
\end{eqnarray}
where the $=$ should be understood as stating an isomorphism.  What this result tells us is that there is exactly one harmonic $n$-form for every equivalence class of $H^n_d(\mathcal{M})$.  

Recall from section \ref{sec:theeulercharacteristicandbettinumbers} that we had the string of equalities
\begin{eqnarray}
b^n(\mathcal{M}) = dim(H^n_d(\mathcal{M})) = dim(H_n(\mathcal{M})) = b_n(\mathcal{M})
\end{eqnarray}
where $b^n(\mathcal{M}) = b_n(\mathcal{M})$ where the Betti numbers and $H_n(\mathcal{M})$ is the $n^{\th}$ homology group of $\mathcal{M}$.  We then showed that the Euler number could be written as
\begin{eqnarray}
\chi(\mathcal{M}) = \sum_{i=0}^{\infty} (-1)^ib^i(\mathcal{M})
\end{eqnarray}
We can now add to this string of equalities
\begin{eqnarray}
dim(H^n_{\Delta} (\mathcal{M})) = dim(H^n_d(\mathcal{M})) = b^n(\mathcal{M})
\end{eqnarray}
And therefore,
\begin{eqnarray}
\chi(\mathcal{M}) = \sum_{i=0}^{\infty} (-1)^i dim(H^n_{\Delta}(\mathcal{M})) \label{eq:relationshipbetweeneulernumberandharmonicforms}
\end{eqnarray}
Once again we see a remarkable equation - the left hand side is a purely topological quantity, while the right hand side is a purely analytic property.  The left hand side has absolutely nothing to do with anything geometrical, whereas the right hand side can't even be defined without a specific metric!  

We have derived (\ref{eq:relationshipbetweeneulernumberandharmonicforms}) for a variety of reasons, and as you might imagine it is a very simple and very specific case of a much more general idea.  It will be useful in some of the physical applications we will consider later, but it is also to provide a simple illustration of much, much more profound and far reaching ideas.  Later in this series we will devote a tremendous amount of time considering these types of relationships between global topological properties and local differential/geometric properties.  We will find that such analytic relationships provide a profound gateway between topology and geometry, and will be a primary tool in more advanced topics, especially string theory.

% Discuss where we are going with this - tie it into the betti number stuff from the topology chapter and foreshadow the global analysis stuff.  

\section{Curvature}

\subsection{First Intuitive Idea of Curvature}

We are finally able to talk about curvature, one of the most fundamental ideas in both geometry and, as we will see, physics.  

As a less technical and more intuitive introduction to this topic, we begin with a more qualitative discussion of how curvature is measured.  To understand the meaning of both the Lie derivative and the torsion tensor, we started at a point $p \in \mathcal{M}$ and moved to another point on $\mathcal{M}$ through two different paths.  In general, the result of going around the two paths is different, and this difference is the quantity we are interested in.  With the Lie derivative, the lack of closure of the relevant rectangle indicated the degree to which one vector field changes as you move along another.  With the torsion tensor, the lack of closure of the relevant rectangle indicates the degree to which two geodesics twist relative to each other.  

Learning by example, let us consider a similar situation.  However, instead of merely moving along two different paths, let us carry something along the two paths.  Consider, as usual, $\mathbb{R}^2$.  Let's take a random vector in $\mathbb{R}$ and, using the trivial connection ($\Gamma^i_{jk} = 0$), let's parallel transport it along two different paths: first in the $\bf e\it_x$ direction some distance $\delta x$, then the $\bf y\it$ direction some distance $\delta y$,
\begin{center}
\includegraphics[scale=.7]{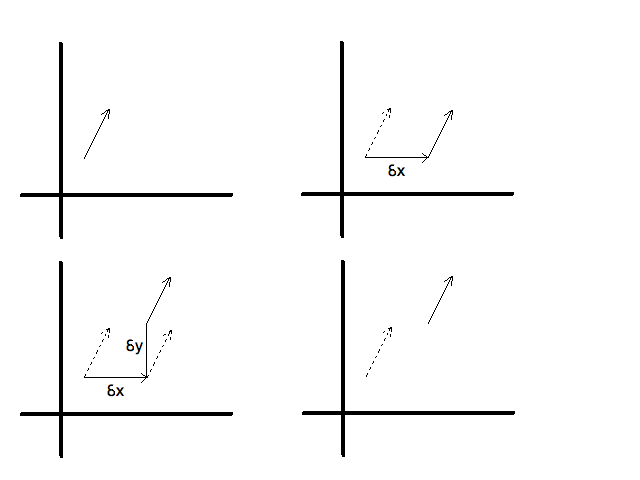}
\end{center}
Then we can do this in the opposite order,
\begin{center}
\includegraphics[scale=.7]{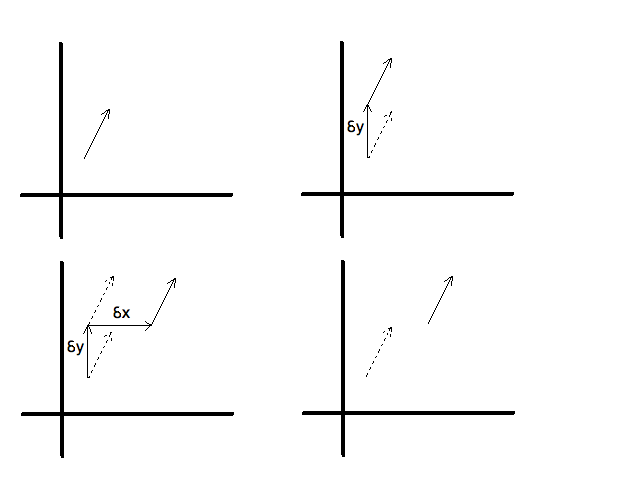}
\end{center}
Notice that after both paths the vector is the same.  In other words, the difference in the vector after being parallel transported through each path is $0$.  

On the other hand, consider $S^2$ with an arbitrary tangent vector at the north pole.  
\begin{center}
\includegraphics[scale=.5]{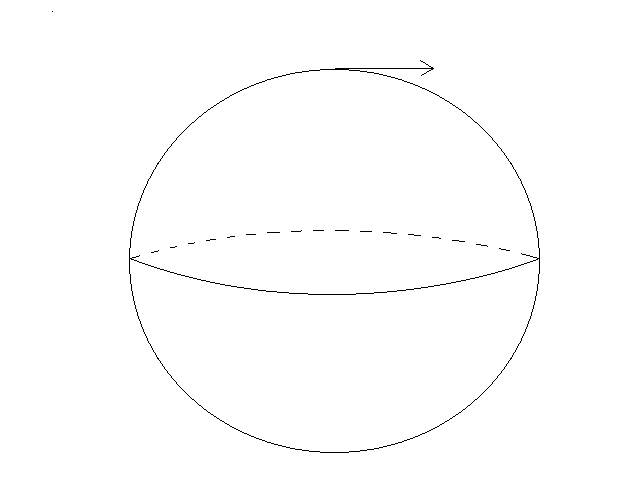}
\end{center}
We will define parallel transport such that the angle between the vector we are transporting and the tangent vector along the curve we are transporting along maintain the same angle.  With that said, let's parallel transport along the left side of the diagram down to the equator:
\begin{center}
\includegraphics[scale=.5]{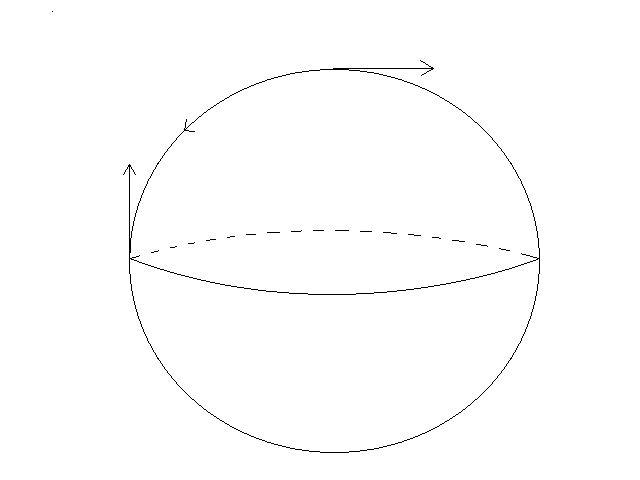}
\end{center}
then transport this around to the front of the sphere:
\begin{center}
\includegraphics[scale=.5]{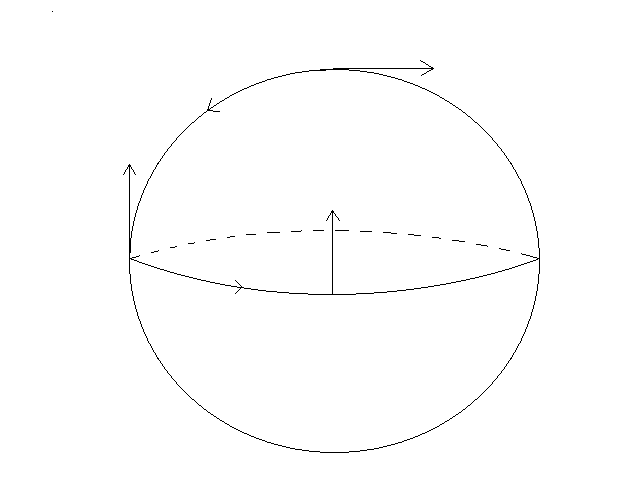}
\end{center}
We can then transport down the right side, then to the front:
\begin{center}
\includegraphics[scale=.6]{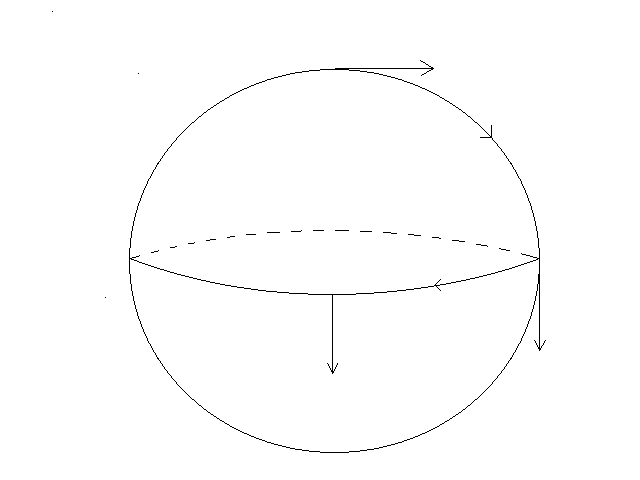}
\end{center}
So, under these two paths the result is different.  A moment's thought will make it clear that the reason for this difference between $\mathbb{R}^2$ and $S^2$ is due to the fact that $S^2$ is curved, whereas $\mathbb{R}^2$ is not.  

This is the approach we will take to study curvature.  We will parallel transport an arbitrary vector at an arbitrary point on $\mathcal{M}$ around two different paths.  We will take the difference in the resulting vector at the new point to be a measure of the curvature of $\mathcal{M}$.  

\subsection{The Riemann Tensor}
\label{sec:riemanntensor}

We now do mathematically exactly what we did in the previous section pictorially.  As we noted above, we have found tremendous success in moving around rectangles.  Previously, in studying the meaning of the Lie derivative and torsion tensor we simply used displacement vectors to move a \it point \rm around the rectangle.  Now we will move a vector around it.  

Recall from section \ref{sec:torsionandliederiv} that as long as we assume torsion vanishes ($\Gamma^i_{jk} = \Gamma^i_{kj}$)\footnote{The calculation that follows, as well as our entire discussion of curvature, will depend on the assumption that there is no torsion.  We will eventually generalize this, but for now torsionless manifolds are sufficient for our purposes.}, parallel transporting a point along the infinitesimal vectors $\boldsymbol{\epsilon}$ then $\boldsymbol{\delta}$ will send you to the same point as parallel transporting along $\boldsymbol{\delta}$ and then $\boldsymbol{\epsilon}$.  

So consider a vector $\bf v\it$ at $p \in \mathcal{M}$, along with vertices at $p$, $p+\epsilon$, $p+\delta$, and $p+\delta+\epsilon$,
\begin{center}
\includegraphics[scale=.65]{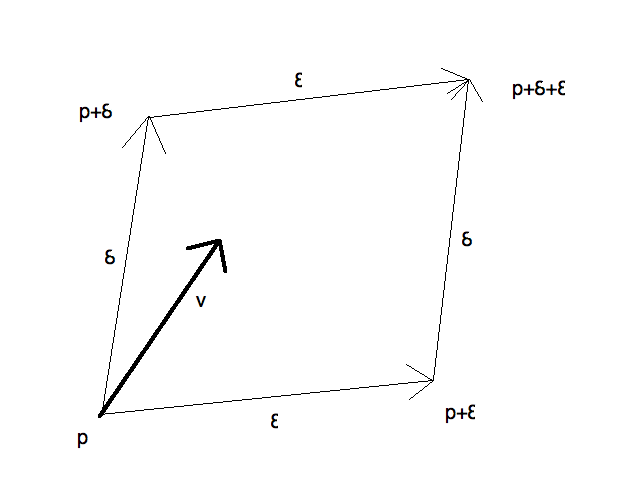}
\end{center}
We can parallel transport $\bf v\it$ along $\boldsymbol{\epsilon}$ first, getting (from (\ref{eq:introtogamma}))
\begin{eqnarray}
v^i(p) \longrightarrow v'^i(p+\epsilon) = v^i(p) - \Gamma^i_{jk}(p)\epsilon^jv^k(p)
\end{eqnarray}
\begin{center}
\includegraphics[scale=.65]{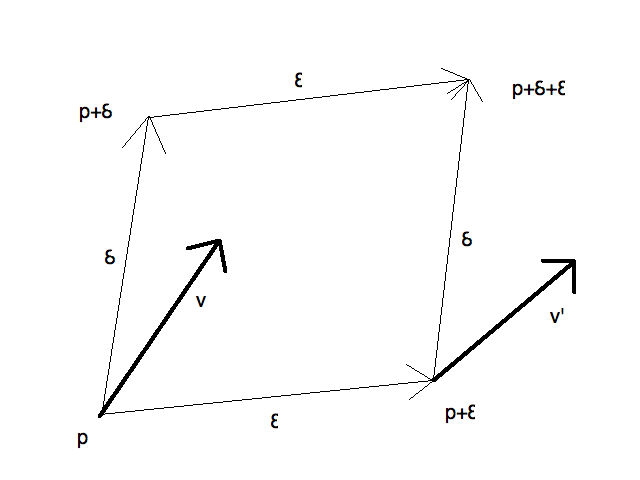}
\end{center}
Then we can parallel transport $\bf v\it'$ to $p+\delta \epsilon$, getting\footnote{We don't have to both with parallel transporting $\boldsymbol{\delta}$ to the point $p+\epsilon$ because we have assumed that there is no torsion on our manifold.  We could redo all of this, carefully transporting $\boldsymbol{\delta}$ and $\boldsymbol{\epsilon}$ as well as $\bf v\it$, but because there is no torsion it wouldn't change our answer.}
\begin{eqnarray}
v'^i(p+\epsilon) \longrightarrow v''^i(p+ \epsilon + \delta) &=& v'^i(p+\epsilon) - \Gamma^i_{jk}(p+\epsilon) \delta^j v'^k(p+\epsilon) \nolabel \\
&=& v^i(p) - \Gamma^i_{jk}(p)\epsilon^jv^k(p) \nolabel \\
& & - \big( \Gamma^i_{jk}(p) + \epsilon^l\partial_l \Gamma^i_{jk}(p)\big)\delta^j \big( v^k(p) - \Gamma^k_{mn} (p) \epsilon^m v^n(p)\big) \nolabel \\
&=& v^i - \Gamma^i_{jk}\epsilon^jv^k - \Gamma^i_{jk}\delta^jv^k + \Gamma^i_{jk}\Gamma^k_{mn}\delta^j\epsilon^m v^n - \partial_l\Gamma^i_{jk}\epsilon^l\delta^j v^k   \nolabel \\ \label{eq:riemannfirstpath}
\end{eqnarray}
(where we have kept terms only to first order in $\boldsymbol{\epsilon}$ and $\boldsymbol{\delta}$, and we dropped the argument $(p)$ in the last line):
\begin{center}
\includegraphics[scale=.65]{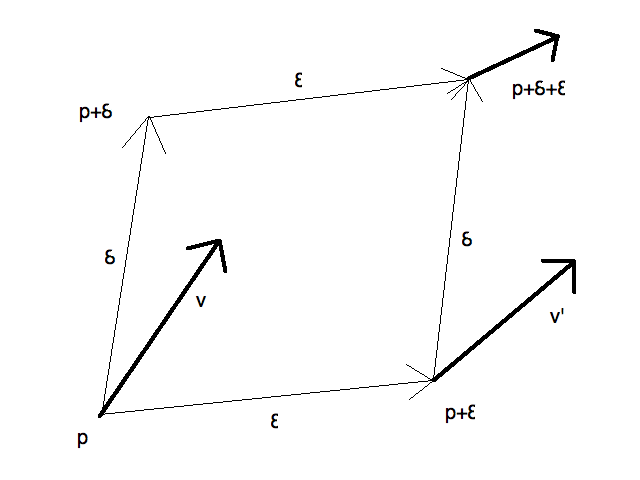}
\end{center}

We can then go in the opposite order, first getting
\begin{eqnarray}
v^i(p) \longrightarrow v'''^i(p+\delta) = v^i(p) - \Gamma^i_{jk}(p)\delta^jv^k(p)
\end{eqnarray}
\begin{center}
\includegraphics[scale=.65]{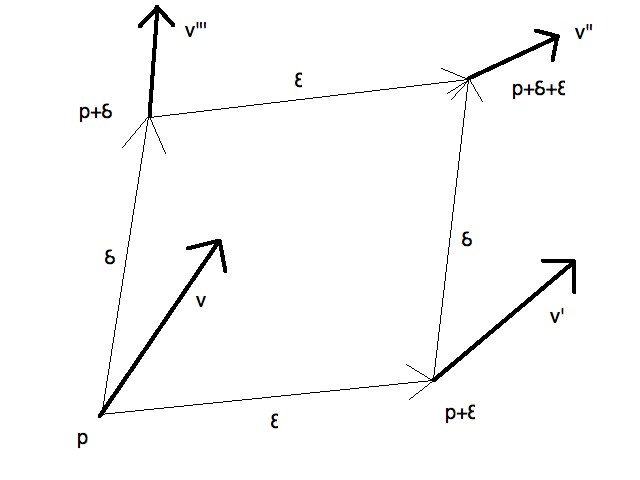}
\end{center}
and finally getting
\begin{eqnarray}
v'''^i(p+\delta) \longrightarrow v''''^i =  v^i - \Gamma^i_{jk}\delta^jv^k - \Gamma^i_{jk}\epsilon^jv^k + \Gamma^i_{jk} \Gamma^k_{mn} \epsilon^j \delta^m v^n- \partial_l \Gamma^i_{jk} \delta^l\epsilon^j v^k \label{eq:riemannsecondpath}
\end{eqnarray}
\begin{center}
\includegraphics[scale=.65]{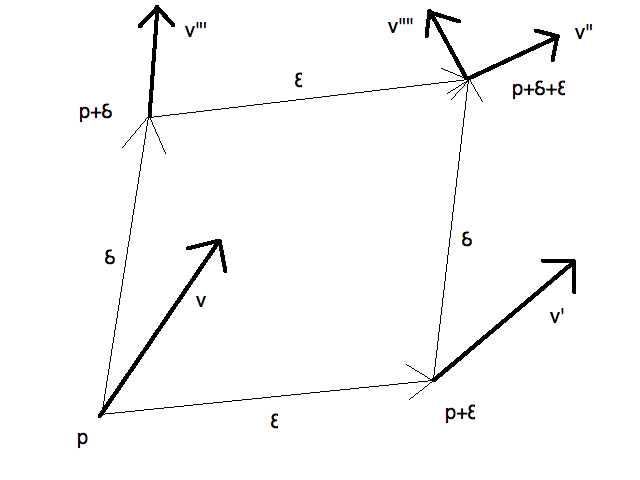}
\end{center}

We want to know how much the vectors at $p+\delta+\epsilon$ differ, and we therefore take the difference between (\ref{eq:riemannfirstpath}) and (\ref{eq:riemannsecondpath}):
\begin{eqnarray}
v''''^i - v''^i &=& \big( v^i - \Gamma^i_{jk}\delta^jv^k - \Gamma^i_{jk}\epsilon^jv^k + \Gamma^i_{jk} \Gamma^k_{mn} \epsilon^j \delta^m v^n- \partial_l \Gamma^i_{jk} \delta^l\epsilon^j v^k \big) \nolabel \\
& & - \big(v^i - \Gamma^i_{jk}\epsilon^jv^k - \Gamma^i_{jk}\delta^jv^k + \Gamma^i_{jk}\Gamma^k_{mn}\delta^j\epsilon^m v^n - \partial_l\Gamma^i_{jk}\epsilon^l\delta^j v^k \big) \nolabel \\
&=& \Gamma^i_{jk}\Gamma^k_{mn}\epsilon^j\delta^m v^n - \partial_l \Gamma^i_{jk} \delta^l \epsilon^j v^k - \Gamma^i_{jk} \Gamma^k_{mn} \delta^j \epsilon^m v^n + \partial_l \Gamma^i_{jk} \epsilon^l \delta^j v^k \nolabel \\
&=& \big( \partial_j \Gamma^i_{kl} - \partial_k \Gamma^i_{jl}+\Gamma^i_{jm} \Gamma^m_{kl} - \Gamma^i_{km}\Gamma^m_{jl} \big) \epsilon^j \delta^k v^l \nolabel \\
&\equiv & R^i_{ljk} \epsilon^j\delta^k v^l \label{eq:howriemannshowsupinvectorthing}
\end{eqnarray}
where we switched around the summed dummy indices to get the second to last line.  
The tensor
\begin{eqnarray}
R^i_{jkl} = \partial_k \Gamma^i_{lj} - \partial_l\Gamma^i_{kj} + \Gamma^i_{km}\Gamma^m_{lj} - \Gamma^i_{lm}\Gamma^m_{kj} \label{eq:equationofriemanncurvaturetensor}
\end{eqnarray}
is called the \bf Riemann Tensor\rm, and it is a measure of how different a vector is after it has been moved around two different paths.  Or, based on our considerations in the previous section, it is a measure of the curvature of the manifold:
\begin{center}
\includegraphics[scale=.65]{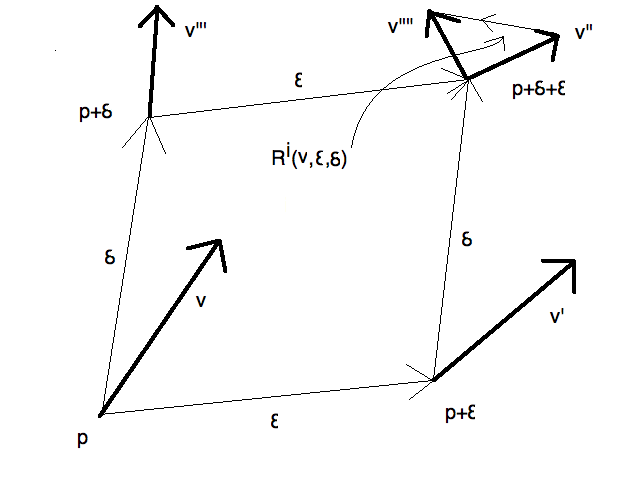} \label{eq:lastpictureinriemanntensor1derivation}
\end{center}

Before moving on, we briefly consider the symmetries of the Riemann tensor.  Notice that in the picture above we denoted the Riemann tensor as $R^i (\bf v\it, \boldsymbol{\epsilon}, \boldsymbol{\delta})$ instead of $R^i_{jkl}v^i \epsilon^k \delta^l$.  In doing this, we are merely emphasizing the fact that the Riemann tensor has three covariant indices and one contracovariant index, which we can take to mean that it is an object that maps three vectors to one vector:
\begin{eqnarray}
R^i_{jkl} :  T_p(\mathcal{M}) \otimes T_p(\mathcal{M}) \otimes T_p(\mathcal{M}) \longrightarrow T_p(\mathcal{M})
\end{eqnarray}
Specifically, $R^i(\bf v\it, \boldsymbol{\epsilon}, \boldsymbol{\delta}) = R^i_{jkl} v^j \epsilon^k \delta^l$.  

From this we can plainly see several symmetries and antisymmetries.  The most obvious is between the $\boldsymbol{\epsilon}$ and $\boldsymbol{\delta}$ vectors - if we had switched the order the $R^i(\bf v\it, \boldsymbol{\epsilon}, \boldsymbol{\delta})$ vector would point in the opposite direction.  We should therefore suspect that $R^i_{jkl}$ is antisymmetric in the last two indices.  Looking at (\ref{eq:equationofriemanncurvaturetensor}), we see that this is indeed the case.  So, 
\begin{eqnarray}
R^i_{jkl} = -R^i_{jlk} \label{eq:antisymmetryoflasttwoindicesofRiemanntensor}
\end{eqnarray}

Another useful identity is to use the metric to lower the first index of the Riemann tensor:
\begin{eqnarray}
R_{ijkl} = g_{in}R^n_{jkl}
\end{eqnarray}
The symmetries of this new tensor\footnote{Don't worry about the physical meaning of this tensor for now.  We are merely quoting these results for completeness.} are
\begin{eqnarray}
R_{iklm} &=& -R_{ikml} = -R_{kilm} \nolabel \\
R_{iklm} &=& R_{lmik} \nolabel \\
R_{iklm} + R_{ilmk} + R_{imkl} &=& 0 \label{eq:symmetryandantisymmetryinloweredR}
\end{eqnarray} 

Finally, if we work in normal coordinates (cf section \ref{sec:normalcoordinates}) we can show that
\begin{eqnarray}
R^i_{jkl} &=& \partial_k \Gamma^i_{lj} - \partial_l\Gamma^i_{kj} \nolabel \\
R_{ijkl} &=& {1 \over 2}\big[\partial_j \partial_k g_{il} - \partial_i\partial_k g_{lj} - \partial_j\partial_lg_{ik} + \partial_i\partial_l g_{kj}\big] \label{eq:loweredriemanntensorinnormalcoords}
\end{eqnarray}

Before moving on, we illustrate the meaning of the Riemann tensor one other way.  If we start with a vector $\bf v\it$ at $p$ and parallel transport to point $p+\epsilon$ we can think of this as a Taylor expansion, but instead of using the normal partial derivative we use the covariant derivative:
\begin{eqnarray}
\bf v\it(p) \longrightarrow \bf v\it(p+\epsilon) = \bf v\it(p) + \nabla_{\boldsymbol{\epsilon}} \bf v\it(p) + \cdots
\end{eqnarray}
Then dragging this along to $p+\epsilon+\delta$, 
\begin{eqnarray}
\bf v\it(p+\epsilon) &\longrightarrow& \bf v\it(p+\epsilon+\delta) \nolabel \\
&=& \bf v\it(p) + \nabla_{\boldsymbol{\delta}}\bf v\it(p) + \nabla_{\boldsymbol{\epsilon}} \bf v\it(p) + \nabla_{\boldsymbol{\delta}} \nabla_{\boldsymbol{\epsilon}} \bf v \it(p) + \cdots
\end{eqnarray}
On the other hand, had we done $\delta$ then $\epsilon$, we would have
\begin{eqnarray}
\bf v\it(p) \longrightarrow \bf v\it(p+\delta+\epsilon) = \bf v\it(p) + \nabla_{\boldsymbol{\epsilon}} \bf v\it(p) + \nabla_{\boldsymbol{\delta}} \bf v\it(p) + \nabla_{\boldsymbol{\epsilon}} \nabla_{\boldsymbol{\delta}} \bf v\it(p) + \cdots
\end{eqnarray}
So the difference in these two vectors (to first order in $\delta$ and $\epsilon$)
\begin{eqnarray}
\big(\bf v\it(p) + \nabla_{\boldsymbol{\delta}}\bf v\it(p) + \nabla_{\boldsymbol{\epsilon}} \bf v\it(p) + \nabla_{\boldsymbol{\delta}} \nabla_{\boldsymbol{\epsilon}} \bf v \it(p)\big) &-& \big( \bf v\it(p) + \nabla_{\boldsymbol{\epsilon}} \bf v\it(p) + \nabla_{\boldsymbol{\delta}} \bf v\it(p) + \nabla_{\boldsymbol{\epsilon}} \nabla_{\boldsymbol{\delta}} \bf v\it(p) \big) \nolabel \\
&=& \nabla_{\boldsymbol{\delta}} \nabla_{\boldsymbol{\epsilon}} \bf v\it(p) - \nabla_{\boldsymbol{\epsilon}} \nabla_{\boldsymbol{\delta}} \bf v\it(p) \nolabel \\
&=& [\nabla_{\boldsymbol{\delta}} , \nabla_{\boldsymbol{\epsilon}}] \bf v\it(p)
\end{eqnarray}
If we write this expression out (in components)
\begin{eqnarray}
[\nabla_{\boldsymbol{\delta}} , \nabla_{\boldsymbol{\epsilon}}]v^k &=& \nabla_{\boldsymbol{\delta}}\nabla_{\boldsymbol{\epsilon}} v^k - \nabla_{\boldsymbol{\epsilon}}\nabla_{\boldsymbol{\delta}}v^k \nolabel \\
&=& \delta^i\epsilon^j \nabla_i\nabla_j v^k - \epsilon^i\delta^j \nabla_i\nabla_j v^k \nolabel \\
&=& \delta^i\epsilon^j\nabla_i\bigg( {\partial v^k \over \partial x^j} + \Gamma^k_{jl}v^l\bigg) - \epsilon^i\delta^j \nabla_i \bigg({\partial v^k \over \partial x^j} + \Gamma^k_{jl}v^l\bigg) \nolabel \\
&=& \delta^i\epsilon^j\bigg( {\partial \over \partial x^j}( \nabla_i v^k) + \Gamma^k_{jl} \nabla_i v^l - {\partial \over \partial x^i} (\nabla_j v^k) - \Gamma^k_{il} \nabla_j v^l\bigg) \nolabel \\
&=& \delta^i\epsilon^j \bigg[{\partial \over \partial x^j}\bigg({\partial v^k \over \partial x^i} + \Gamma^k_{il}v^l\bigg) + \Gamma^k_{jl} \bigg({\partial v^l \over \partial x^i} + \Gamma^l_{im}v^m\bigg) \nolabel \\
& & \qquad - {\partial \over \partial x^i}\bigg({\partial v^k \over \partial x^j} + \Gamma^k_{jl}v^l\bigg) - \Gamma^k_{il} \bigg({\partial v^l \over \partial x^j} + \Gamma^l_{jm}v^m\bigg)\bigg] \nolabel \\
&=& \delta^i\epsilon^j \big[ \partial_j\partial_i v^k+ v^l\partial_j\Gamma^k_{il} + \Gamma^k_{il}\partial_j v^l + \Gamma^k_{jl}\partial_i v^l + \Gamma^k_{jl}\Gamma^l_{im}v^m \nolabel \\
& & \qquad - \partial_i\partial_j v^k - v^l\partial_i\Gamma^k_{jl} - \Gamma^k_{jl}\partial_i v^l - \Gamma^k_{il}\partial_j v^l - \Gamma^k_{il}\Gamma^l_{jm} v^m\big] \nolabel \\
&=& \delta^i\epsilon^j \big[ v^l\partial_j\Gamma^k_{il} - v^l\partial_i\Gamma^k_{jl} + \Gamma^k_{jm}\Gamma^m_{il} v^l - \Gamma^k_{im}\Gamma^m_{jl}v^l\big] \nolabel \\
&=& \delta^i\epsilon^j v^l \big[\partial_j\Gamma^k_{il} - \partial_i \Gamma^k_{jl} + \Gamma^k_{jm}\Gamma^m_{il} - \Gamma^k_{im}\Gamma^m_{jl}\big] \nolabel\\
&=& \delta^i \epsilon^j v^l R^k_{lji} \label{eq:seconderivationofriemanntensor}
\end{eqnarray}
which is exactly what we had above in (\ref{eq:equationofriemanncurvaturetensor}).  So we can tentatively say that
\begin{eqnarray}
R^k_{lji} \delta^i \epsilon^j v^l = [\nabla_{\boldsymbol{\delta}}, \nabla_{\boldsymbol{\epsilon}}] v^k
\end{eqnarray}
or in the other notation mentioned above, 
\begin{eqnarray}
R^i(\bf v\it, \boldsymbol{\delta}, \boldsymbol{\epsilon}) =  [\nabla_{\boldsymbol{\delta}}, \nabla_{\boldsymbol{\epsilon}}] v^k
\end{eqnarray}
We included the word "tentatively" above because there is one major assumption we made in the derivation of (\ref{eq:seconderivationofriemanntensor}) - namely that the relevant rectangle is actually closed.  The point is that we are comparing the vector $\bf v\it$ after going to the same point around two different loops.  But if you don't arrive at the same point after going around those two loops then the comparison of the resulting $\bf v\it$ is meaningless.  If, for example, 
\begin{eqnarray}
\mathcal{L}_{\boldsymbol{\delta}} \boldsymbol{\epsilon} \neq 0
\end{eqnarray}
then this won't work because the resulting $\bf v\it$'s will be at two different points (cf pictures on pages \pageref{eq:lastpictureinriemanntensor1derivation} and \pageref{pagewithpictureofliederivativenonclosure}).  Fixing this problem is very easy - if the Lie derivative is non-zero, then the locations of the parallel transported $\bf v\it$'s after each loop will be separated by a distance $\mathcal{L}_{\boldsymbol{\delta}} \boldsymbol{\epsilon}$, and therefore we can parallel transport one of the resulting $\bf v\it$'s to line up with the other by simply using the same Taylor expansion we used to "move" the $\bf v\it$'s to get (\ref{eq:seconderivationofriemanntensor}).  In other words, the corrected expression for the Riemann tensor is
\begin{eqnarray}
R^k_{lji} \delta^i \epsilon^j v^l = R^k(\bf v\it, \boldsymbol{\delta}, \boldsymbol{\epsilon}) &=& [\nabla_{\boldsymbol{\delta}}, \nabla_{\boldsymbol{\epsilon}}] v^k - \nabla_{\mathcal{L}_{\boldsymbol{\delta}} \boldsymbol{\epsilon}} v^k \nolabel \\
&=& \big([\nabla_{\boldsymbol{\delta}}, \nabla_{\boldsymbol{\epsilon}}]  - \nabla_{[\boldsymbol{\delta}, \boldsymbol{\epsilon}]}\big) v^k \label{eq:mostgeneraldefinitionofriemanntensorwithliederivativecorrection}
\end{eqnarray}
We can take this to be the general definition of the Riemann tensor.  In fact, this is a general expression for any curvature tensor, whether it comes from the Levi-Civita connection or not.  

\subsection{Second Intuitive Idea of Curvature}
\label{sec:secondintuitiveideaofcurvature}

The Riemann tensor is the first, and in many ways the most fundamental, notion of curvature we will work with.  However there is another way to quantify curvature that, while related to the Riemann tensor, has a unique interpretation.  

Consider two particles in $\mathbb{R}^2$ at points $p$ and $q$ with Euclidian metric $g_{ij} = \delta_{ij}$, initially separated by a distance $a$, each moving parallel to each other along the $x$ axis towards the origin:
\begin{center}
\includegraphics[scale=.6]{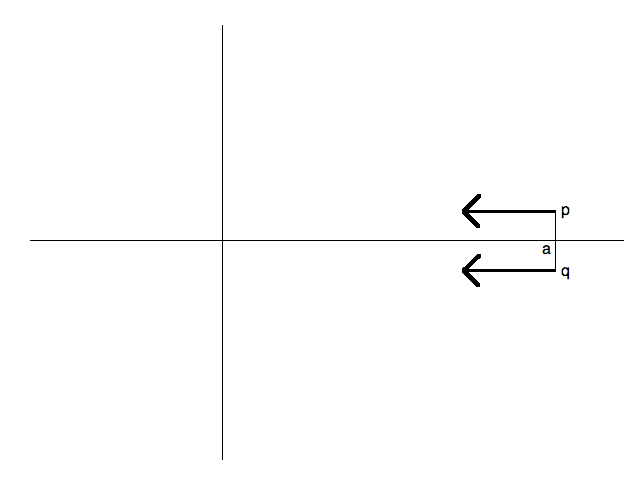}
\end{center}
If we take the geodesics in $\mathbb{R}^2$ with $g_{ij} = \delta_{ij}$ through $p$ and $q$ in the direction of these vectors, parallel transporting the vectors along those geodesics (respectively) will give, at a later time, 
\begin{center}
\includegraphics[scale=.6]{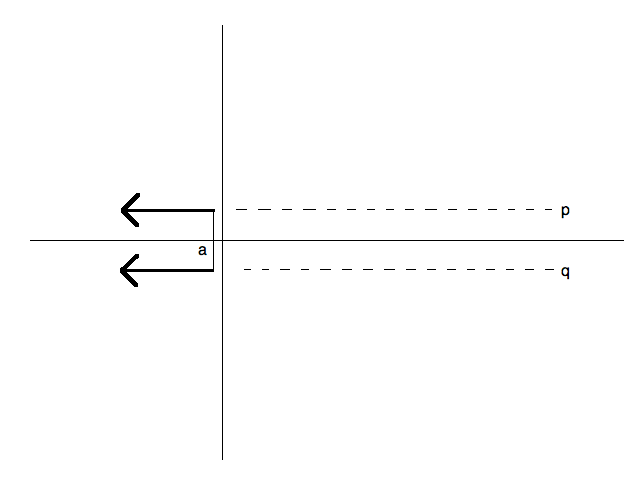}
\end{center}
They are still separated by the distance $a$.  

Now consider a similar situation but instead of $\mathbb{R}^2$ we work with $S^2$ with metric (\ref{eq:metricfors2inmetricsection}).  Start with $p$ and $q$ at the ``equator" and point both vectors towards the ``north pole".  
\begin{center}
\includegraphics[scale=.6]{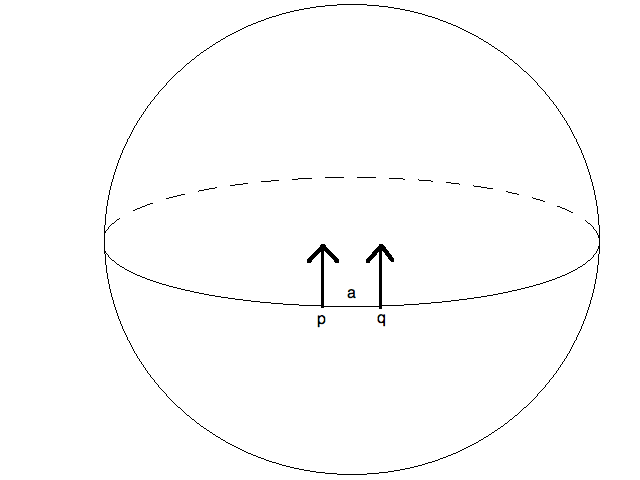}
\end{center}
If we parallel transport these in a similar way, we end up with\footnote{Please forgive the sloppy artwork.}
\begin{center}
\includegraphics[scale=.6]{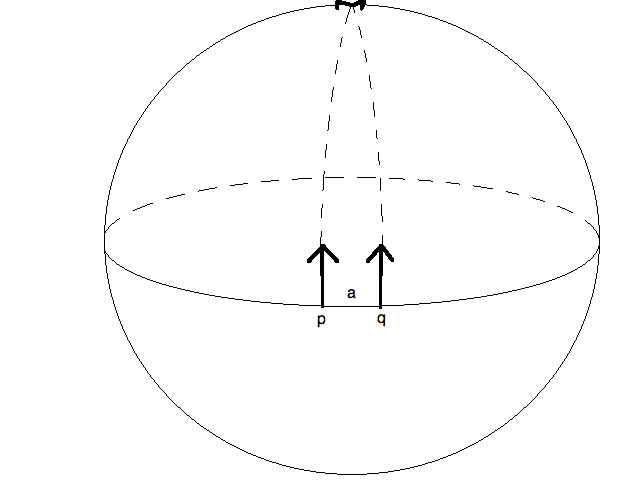}
\end{center}
Notice that in this case the vectors, which started out parallel and separated by a distance $a$, are now at the same point.  

This property, of vectors starting out parallel to each other, each moving along parallel lines, but still intersecting, is a direct consequence of the Riemann tensor.  The vector initially at $p$ can be parallel transported along as drawn in the above pictures.  On the other hand, it can be parallel transported along $a$ to the point $q$, resulting in vector $q$, which is then parallel transported to the new point.  This is equivalent to what we did to derive the Riemann tensor, only instead of parallel transporting the vector along two paths of a rectangle, here we have parallel transported it along two paths of a triangle.  The result is the same - the degree to which they are no longer parallel is a measurement of curvature, and is quantified by the Riemann tensor.  In this sense, we have done nothing new.  

However, we can approach this a different way.  Looking again at $\mathbb{R}^2$, consider a volume $\mathcal{V}$ located between $p$ and $q$\footnote{Of course, because we are in $2$ dimensions, a ``volume" is an area.  We are saying ``volume" because we will eventually generalize this to an arbitrary manifold.}:
\begin{center}
\includegraphics[scale=.6]{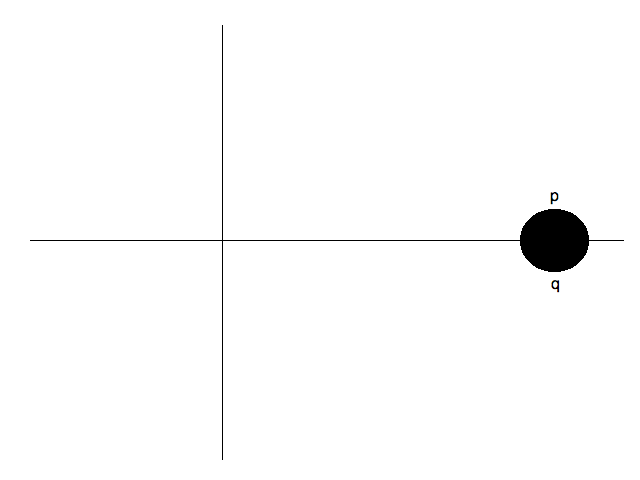}
\end{center}
With a metric this becomes straightforward to write the volume of this - assuming $p$ and $q$ are still separated by a distance $a$, the volume (area) will be (assuming it is a circle, which is a reasonable assumption at this point) ${\pi a^2 \over 4}$.  Now choose any point in the volume and choose a vector at that point.  We can parallel transport this vector to every other point in the volume to define a class of vectors - one for each point in $\mathcal{V}$.  
\begin{center}
\includegraphics[scale=.6]{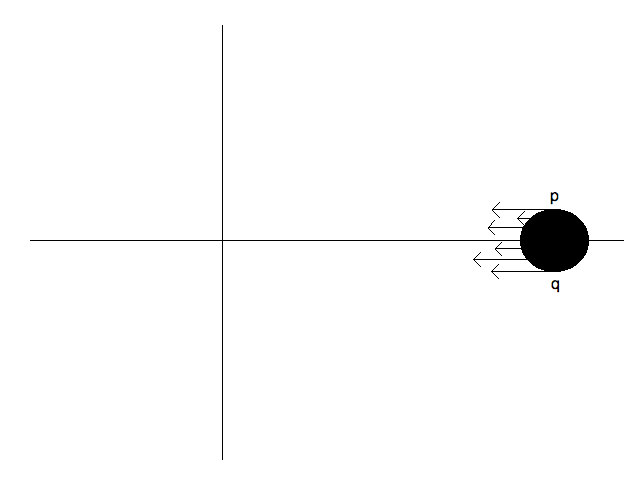}
\end{center}
If we parallel transport each of these points along their respective vector, we end up with, at a later time,
\begin{center}
\includegraphics[scale=.6]{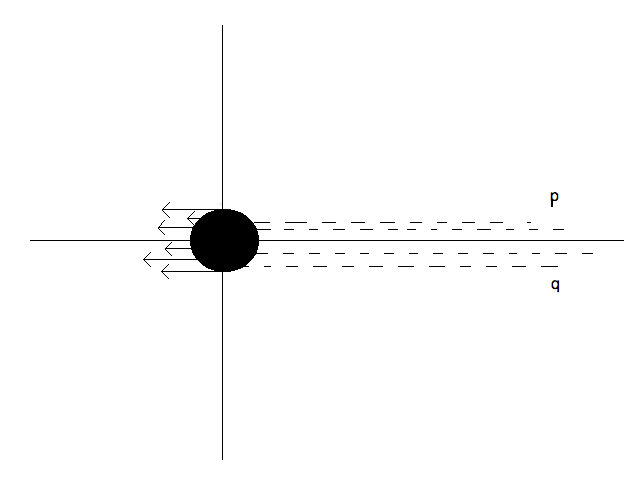}
\end{center}
the volume will be the same.  

On the other hand, if we had done the same thing on $S^2$ as before,
\begin{center}
\includegraphics[scale=.6]{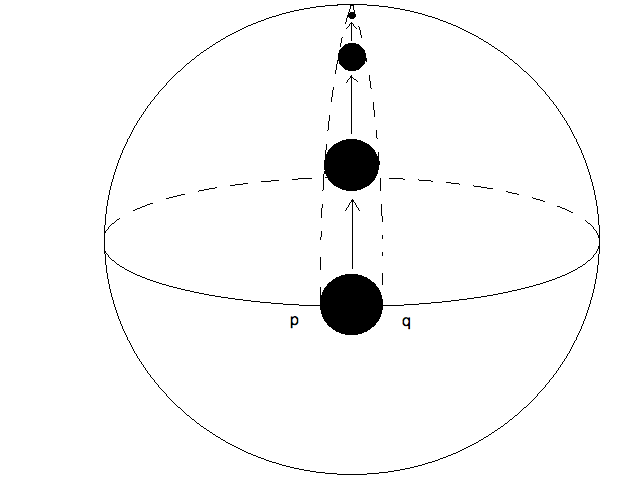}
\end{center}
we find (from this admittedly crude graphical approach) that the volume changes as each point of $\mathcal{V}$ moves along its geodesic.  

Of course, we could have started with a volume near the north pole and parallel transported each point in $\mathcal{V}$ down towards the equator, and the result would have been $\mathcal{V}$ increasing in size.  

So on the flat $\mathbb{R}^2$ the volume did not change, whereas on the curved $S^2$ the volume did change.  It turns out that this property, volume deviation, is yet another quantitative measurement of curvature.  The specific quantity by which the volume changes is measured by what is called the \bf Ricci Tensor\rm.  As you might imagine, it is intimately related to the Riemann tensor.  However, in the next section we will derive this volume deviation term independently and then show that/how it is related to the Riemann tensor.  

\subsection{Ricci Tensor}
\label{sec:riccitensorsection11278943}

We want to make what we said in the previous section concrete.  We start with a collection of particles clumped together near each other in some part of a manifold $\mathcal{M}$.  At the beginning, their relative positions are fixed.  Furthermore they are all moving in the same direction initially (each velocity vector is initially parallel to every other velocity vector), and therefore the initial velocity between any two particles is initially zero.  However, because we know from the previous section that as this collection of particles moves along a collection of geodesics the relative positions of any two of them may change, the \it second \rm derivative of position, or the relative acceleration, between any two points may be non-zero initially.  We therefore want to find the acceleration between two arbitrary points.  To do this we will follow a calculation similar to what we did in section \ref{sec:riemanntensor}.  

Let's repeat the general argument that led to (\ref{eq:howriemannshowsupinvectorthing}) with a few slight changes.  We begin with a vector $\bf v\it$ at $p \in \mathcal{M}$.  We will interpret this as a velocity vector, and we can then take the geodesic through $p$ in the direction of $\bf v\it$:
\begin{center}
\includegraphics[scale=.6]{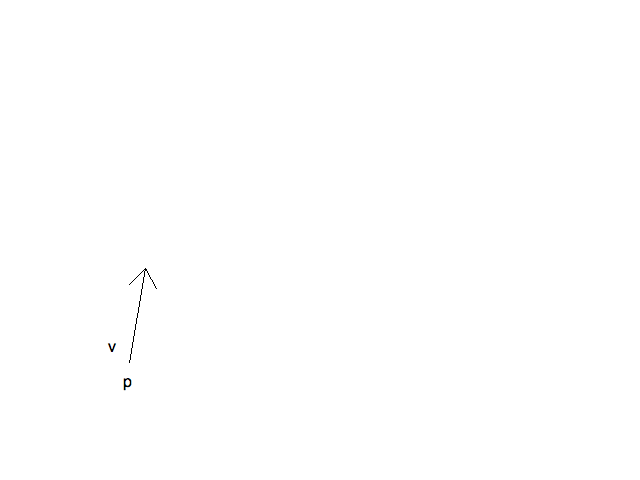}
\end{center}
We can parallel transport $\bf v\it$ along the geodesic defined by $\bf v\it$ for a small time $\Delta \epsilon$ (where $\Delta \epsilon$ is now an infinitesimal scalar, not an infinitesimal vector).  In other words, we are parallel transporting $\bf v\it$ a distance $\Delta \epsilon \bf v\it$:
\begin{center}
\includegraphics[scale=.6]{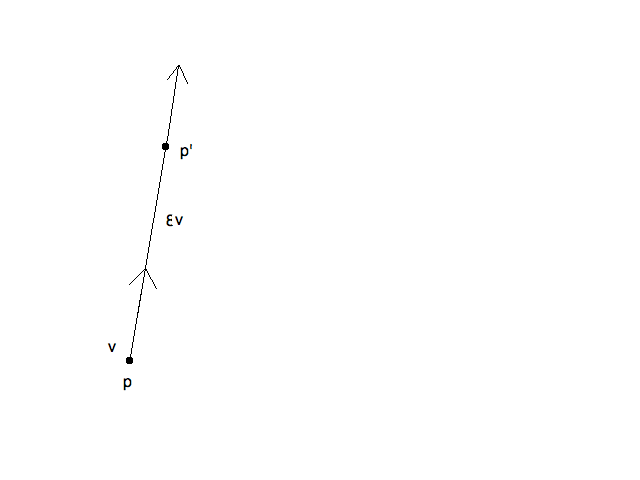}
\end{center}
(Notice that we are doing essentially the same calculation that led to (\ref{eq:howriemannshowsupinvectorthing}) except the vector we are parallel transporting is in the same direction as our geodesic).  

We can also take another point $q$ near $p$, separated by a very small vector $\Delta \epsilon \bf u\it$ (again $\Delta \epsilon$ is an infinitesimal scalar and $\bf u\it$ is a finite vector), parallel transport $\bf v\it$ to $q$ so that the vector at $p$ and the vector at $q$ are parallel, and then let the vector at $q$ follow its geodesic for a time $\Delta \epsilon$,
\begin{center}
\includegraphics[scale=.6]{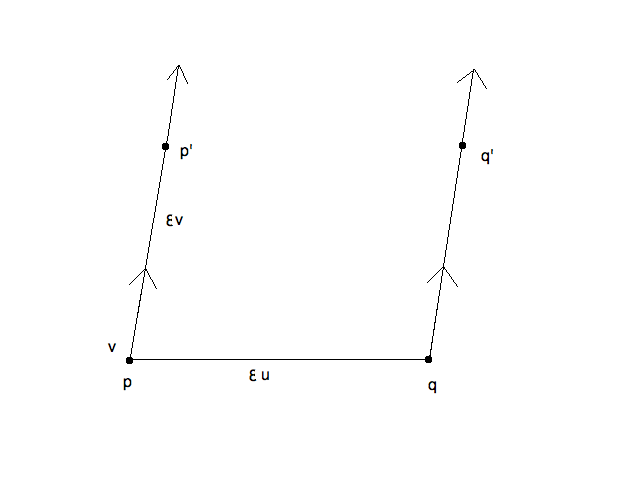}
\end{center}
Our interest (at this point) is the relative velocities of the particles after the time $\Delta \epsilon$.  We can find this by taking the \it difference \rm in velocity at $p'$ and $q'$, which we find by parallel transporting the velocity at $p'$ to $q'$:
\begin{center}
\includegraphics[scale=.6]{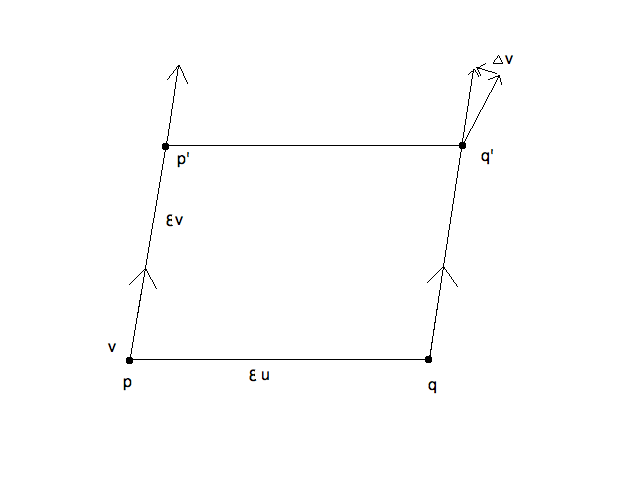}
\end{center}

Comparing this situation to (\ref{eq:howriemannshowsupinvectorthing}) we have that $\Delta \bf v\it$ is given by
\begin{eqnarray}
\Delta v^i = (\Delta \epsilon)^2 R^i_{ljk} u^j v^k v^l \label{eq:ricci21}
\end{eqnarray}

As stated above, we are interested in the relative acceleration of these particles.  Acceleration is simply the change in velocity over change in time, so we can define the acceleration vector $\bf a\it$ as
\begin{eqnarray}
a^i =\lim_{\Delta \epsilon \rightarrow 0} { \Delta v^i \over \Delta \epsilon} \label{eq:defofa}
\end{eqnarray}
and therefore (\ref{eq:ricci21}) is 
\begin{eqnarray}
\lim_{\Delta \epsilon \rightarrow 0} {a^i \over \Delta \epsilon} = R^i_{ljk} u^j v^k v^l
\end{eqnarray}
This is the equation of geodesic deviation.  What it says is nothing really new after section \ref{sec:riemanntensor}.  If the Riemann tensor vanishes, two objects traveling along geodesics don't accelerate relative to each other.\footnote{This is a physical interpretation of the Euclidian postulate that parallel lines don't intersect.  We are doing non-Euclidian geometry where parallel lines can intersect.}  If the Riemann tensor is non-zero then they will accelerate relative to each other.  This relative acceleration is a measurement of how curved a space is.  

Now we attempt to recreate what we did in the second half of section \ref{sec:secondintuitiveideaofcurvature} with the small volumes.  Imagine a small ball of initial radius $R_0$ along every direction.  If we wait a time $\Delta \epsilon$ (as above), the new radius along the $i^{th}$ dimension will be\footnote{This is nothing but the standard ``physics I" expression $$ \Delta x = x_0 + v_0 t + {1 \over 2} a t^2$$ where $v_0$ is zero because the particles start off along parallel geodesics and $a^i$ is given in equation (\ref{eq:defofa})}
\begin{eqnarray}
r^i(\Delta \epsilon) = R_0 + {1 \over 2} a^i (\Delta \epsilon)^2
\end{eqnarray}
So, from this we can easily calculate
\begin{eqnarray}
\dot r^i(\Delta \epsilon) &=& a^i \Delta \epsilon \nolabel \\
\ddot r^i(\Delta \epsilon) &=& a^i
\end{eqnarray}
where the dot represents a derivative with respect to time $\Delta \epsilon$.  This gives us the relationships
\begin{eqnarray}
r^i(\Delta \epsilon)\bigg|_{\Delta \epsilon =0} &=& R_0 \nolabel \\
\dot r^i (\Delta \epsilon)\bigg|_{\Delta \epsilon = 0} &=& 0 \nolabel \\
\ddot r^i(\Delta \epsilon)\bigg|_{\Delta \epsilon =0} &=& a^i \label{eq:rrdotandrdoubledotatepsilontimegoestozero}
\end{eqnarray}
So
\begin{eqnarray}
\lim_{\Delta \epsilon \rightarrow 0} {\ddot r^i(\Delta \epsilon) \over r^i(\Delta \epsilon)} &=& {a^i \over R_0} \nolabel \\
&=&\lim_{\Delta \epsilon \rightarrow 0}\bigg( {\Delta \epsilon \over R_0}\bigg) {a^i \over \Delta \epsilon} \nolabel \\
&=&\lim_{\Delta \epsilon \rightarrow 0} \bigg({\Delta \epsilon \over R_0}\bigg) R^i_{ljk} u^j v^k v^l
\end{eqnarray}
Furthermore, because $r^i$ is the radius between the starting points, we can without loss of generality take $\bf u\it$ to be a unit vector in the $i^{th}$ direction.  

We can get rid of the limit on the right hand side by taking $R_0 = \Delta\epsilon$, giving
\begin{eqnarray}
\lim_{\Delta \epsilon \rightarrow 0} {\ddot r^i(\Delta \epsilon) \over r^i(\Delta \epsilon)} = R^i_{lik} v^k v^l \label{eq:rdoubledotoverrequalsriemanntensorequation}
\end{eqnarray}
Where we have taken $\bf u\it$ to be a unit vector in the $i^{th}$ direction, and no summation is implied by the repeated $i$ on either side.  

So this volume started off an $n$-dimensional sphere of radius $R_0$.  This will have volume\footnote{Notice that this reduces to the familiar $\pi (R_0)^2$ for a circle and ${4 \over 3} \pi (R_0)^3$ for the volume of $S^2$.}
\begin{eqnarray}
V_n = {\pi^{{n\over 2}} \over \Gamma({n\over 2}+1)} (R_0)^n
\end{eqnarray}
(you can check this formula by simply doing the integrals, or by checking any introductory text on basic geometry), where $\Gamma (n)$ is the Euler gamma function.  

In our case, however, the radius along each dimension is changing, and therefore instead of an $n$-dimensional sphere with radius $R_0$, we have an $n$-dimensional ellipsoid with radii $r^i(\Delta \epsilon)$ for every $i$.  The volume of this ellipsoid is
\begin{eqnarray}
V_n =  {\pi^{{n\over 2}} \over \Gamma({n\over 2}+1)} \prod_{i=1}^d r^i(\Delta \epsilon)
\end{eqnarray}
where $\Gamma(n)$ is the Euler gamma function (you can again check this expression in any intro geometry text).  Then we can take a derivative with respect to $\Delta \epsilon$, getting (leaving out the $\Delta \epsilon$ dependence for notational simplicity) 
\begin{eqnarray}
\dot V_n =  {\pi^{{n\over 2}} \over \Gamma({n\over 2}+1)} \sum_{i=1}^d \prod_{j=1}^d {r^j\over r^i} \dot r^i
\end{eqnarray}
Then
\begin{eqnarray}
\ddot V_n =  {\pi^{{n\over 2}} \over \Gamma({n\over 2}+1)} \sum_{i=1}^d \bigg( {\ddot r^i \over r^i} + 2 \sum_{k=1}^{i-1} {\dot r^i \dot r^k \over r^i r^k}\bigg) \prod_{j=1}^d r^j
\end{eqnarray}
So,
\begin{eqnarray}
{\ddot V_n \over V_n} =  \sum_{i=1}^d \bigg( {\ddot r^i \over r^i} + 2 \sum_{k=1}^{i-1} {\dot r^i \dot r^k \over r^i r^k}\bigg) 
\end{eqnarray}
Taking the $\Delta\epsilon\rightarrow 0$ limit and plugging in (\ref{eq:rrdotandrdoubledotatepsilontimegoestozero}),
\begin{eqnarray}
\lim_{\Delta \epsilon\rightarrow 0} {\ddot V_n \over V_n} =\lim_{\Delta \epsilon\rightarrow 0} \sum_{i=1}^d {\ddot r^i \over r^i}
\end{eqnarray}
and then plugging in (\ref{eq:rdoubledotoverrequalsriemanntensorequation}),
\begin{eqnarray}
\lim_{\Delta \epsilon\rightarrow 0} {\ddot V_n \over V_n} = \lim_{\Delta \epsilon\rightarrow 0}\sum_{i=1}^d {\ddot r^i \over r^i} = \sum_{i=1}^d R^i_{lik} v^k v^l = R^i_{lik} v^k v^l \equiv R_{lk} v^l v^k
\end{eqnarray}
where we have invoked the normal summation convention in the second to last equality, and in the final line we have defined the rank 2 tensor
\begin{eqnarray}
R_{lk} \equiv R^i_{lik}
\end{eqnarray}
This tensor is called the \bf Ricci Tensor\rm, and as promised in the last section it tells us the quantity by which the volume is changing.  And, as we indicated in the last section, it is clearly directly related to the Riemann tensor.  Specifically, the meaning is that given some volume of particles forming a volume
\begin{eqnarray}
V_{n,0} = V_n\bigg|_{\Delta \epsilon \rightarrow 0}
\end{eqnarray}
if every point of the volume is moved along an initially parallel set of geodesics (all parallel to $\bf v\it$), the volume will remain unchanged in a flat space where $R^i_{jlk} = 0$:
\begin{eqnarray}
\ddot V_{n,0} = 0
\end{eqnarray}
The first derivative of $V_n$ will always be zero because every point starts off along parallel geodesics and therefore no two points are initially moving relative to each other.  However, they are initially \it accelerating \rm relative to each other, and therefore in a space where $R_{ij}$ is not zero,
\begin{eqnarray}
\ddot V_{n,0} = (R_{ij}v^iv^j) V_{n,0} \label{eq:vdoubledotequalsriccivivjtimesvsubnzero}
\end{eqnarray}
Again, we will postpone examples until later.  

\subsection{Third Intuitive Idea of Curvature}

Alice and Bob are standing on a two dimensional manifold, but they don't know anything about the manifold.  Alice suggests an experiment - she happens to have a rope of length $L$.  She holds one end and tells Bob to take the other and walk away from her until the rope is pulled tight.  Then, she says, her plan is for her to stand still while Bob walks in a big circle, keeping the rope pulled tight, therefore staying the same distance from her on the manifold.  She also tells Bob to keep track of exactly how far he walks before he comes back to the same point.  Because the rope has length $L$, they predict that he will walk $2\pi L$ units before returning to where he started.  

However, after doing this experiment Bob finds that surprisingly he has actually travelled $4 L$ units!  Returning to talk things over with Alice, they realize what is going on.  They conclude that they must be standing on a sphere with radius ${2L \over \pi}$.  The circumference (at the widest) point of such a sphere would be $2 \pi {2L \over \pi} = 4L$, the exact distance he actually walked.  Because $L$ is equal to one quarter of the circumference of the sphere, when Bob pulled the rope tight, he had actually walked a quarter of the way around.  If Alice had been standing on the ``top" of the sphere, Bob was standing on the equator.  
\begin{center}
\includegraphics[scale=.5]{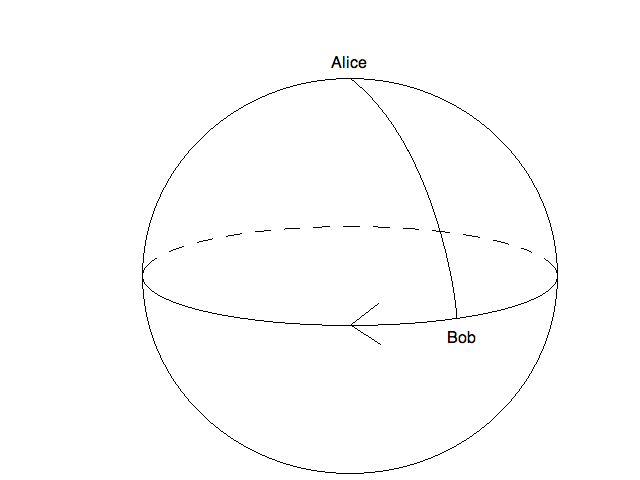}
\end{center}
The reason the circumference of the circle Bob walked around was not what they expected from their ``flat space" intuition is because the surface was actually curved.  In other words, another measure of curvature is the deviation of the volume of the boundary of a space from what you would expect in flat space.  

More generally, if we take a point $p$ on an $n$ dimensional manifold $\mathcal{M}$ and then take the collection of all points a particular distance away, that subspace will have an $n-1$ dimensional volume.  One could calculate what they would guess that volume to be \it if \rm they were in flat space.  But if the space is curved there will be some deviation.  The specific measure of curvature we will be interested in is the lowest order correction to the $n-1$ volume.  We will make this more precise in the next section.  

\subsection{The Ricci Scalar}

To make the idea of the previous section precise, we begin with some point $p \in \mathcal{M}$ (we will be assuming $\mathcal{M}$ has dimension $n$).  We will define an $n-1$ volume by taking every point some distance $\lambda$ away from $p$:
\begin{center}
\includegraphics[scale=.6]{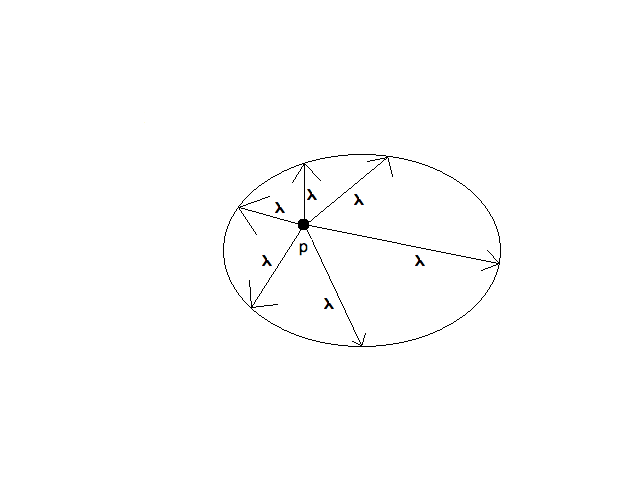}
\end{center}
This will define an $n-1$ subspace homeomorphic to $S^{n-1}\subset \mathcal{M}$ (we will denote this subspace $S^{n-1}$ for the rest of this section).  Because we will approach this problem by defining everything in terms of an enclosure of $p$, maintaining generality will demand that we only use terms evaluated at $p$.  This will make our calculation quite difficult and we will therefore proceed slowly.  

To begin with, we will define coordinates on $S^{n-1}$, which we denote $\theta^i$ (for $i=1,\ldots , n-1$).  We map a given point $\boldsymbol{\theta}$ to $\mathcal{M}$ with the coordinate functions on $\mathcal{M}$:
\begin{eqnarray}
S^{n-1} &\longrightarrow& \mathcal{M} \nolabel \\
\boldsymbol{\theta} &\longrightarrow& x^a(\boldsymbol{\theta})
\end{eqnarray}
The collection of all points $\bf x\it(\boldsymbol{\theta})$ for all $\boldsymbol{\theta}$ is then the set of all points of $S^{n-1}$.  

Our ultimate interest is in the $n-1$ volume of $S^{n-1}$, which we find by integrating over the invariant $n-1$ volume form (cf section \ref{sec:invariantvolumeelement})\footnote{With the appropriate handling of multiple coordinate patches with partitions of unity, etc. - cf section \ref{sec:intdiffforms}.}
\begin{eqnarray}
\Omega_{S^{n-1}} = \sqrt{|\tilde g|} \; d\theta^1 \wedge d\theta^2 \wedge \cdots \wedge d\theta^{n-1}
\end{eqnarray}
where $\tilde g_{ij}$ is the metric on $S^{n-1}$.\footnote{For this section we will primarily use $a,b,c,\ldots$ for coordinates in $\mathcal{M}$ and $i,j,k,\ldots$ for coordinates on $S^{n-1}$.}  

So, working from $p \in \mathcal{M}$, we want to find $\tilde g_{ij}$ at all points $p+\lambda$ (along all possible geodesics through $p$) in all directions.  We assume that we have the metric $g_{ab}$ on $\mathcal{M}$, and we can therefore use the pullback from $x^a(\theta^i)$ to find the induced metric $\tilde g_{ij}$ on $S^{n-1}$.  However, as we said above, we must be careful because we want to define everything in terms of what we know at $p$, which will demand some work because clearly $p \not \in S^{n-1}$.  

For an arbitrary point on $S^{n-1}$ (at $p+\lambda$), we have the general pullback given by (cf section \ref{sec:pullbacks})
\begin{eqnarray}
\tilde g_{ij}(p+\lambda) = {\partial x^a(p+\lambda) \over \partial \theta^i} {\partial x^b(p+\lambda) \over \partial \theta^j} g_{ab}(p+\lambda) \label{eq:firsttermformetricpullbackricciscalar}
\end{eqnarray}
We must therefore expand all three term on the right hand side around $p$.  

Starting with the derivative terms on the right, the exact coordinates of $x^a(p+\lambda)$ will be given by
\begin{eqnarray}
x^a(p+\lambda) = x^a(p) + \lambda {\partial x^a(p) \over \partial \lambda} + {1 \over 2} \lambda^2 {\partial^2 x^a (p) \over \partial \lambda^2} + {1 \over 3!} \lambda^3 {\partial^3 x^a(p) \over \partial \lambda^3} + \cdots 
\end{eqnarray}
This will be valid to arbitrary $\lambda$, but we can simplify it by assuming that $\lambda$ parameterizes a geodesic.  In this case we have (cf equation (\ref{eq:geodesicdifferentialequation}))
\begin{eqnarray}
{\partial^2 x^a \over \partial \lambda^2} &=&- \Gamma^a_{bc} {\partial x^b \over \partial \lambda} {\partial x^c \over \partial \lambda} \nolabel \\
{\partial^3 x^a \over \partial \lambda^3} &=& -2 \Gamma^a_{bc} {\partial x^b \over \partial \lambda} {\partial^2 x^c \over \partial \lambda^2} - {\partial \Gamma^a_{bc} \over \partial \lambda} {\partial x^b \over \partial \lambda} {\partial x^c \over \partial \lambda} \nolabel \\
&=& -2\Gamma^a_{bc} {\partial x^b \over \partial \lambda}{\partial^2 x^c \over \partial \lambda^2} - \partial_d\Gamma^a_{bc} {\partial x^d \over \partial \lambda} {\partial x^b \over \partial \lambda} {\partial x^c \over \partial \lambda}
\end{eqnarray}
where $\partial_a = {\partial \over \partial x^a}$.  So (all terms are evaluated at $p$ so we drop the arguments)
\begin{eqnarray}
x^a(p+\lambda) = x^a + \lambda {\partial x^a \over \partial \lambda} -  {\lambda^2 \over 2}\Gamma^a_{bc} {\partial x^b \over \partial \lambda} {\partial x^c \over \partial \lambda} - {\lambda^3 \over 6} \partial_d \Gamma^a_{bc} {\partial x^d \over \partial \lambda} {\partial x^b \over \partial \lambda} {\partial x^c \over \partial \lambda} - {\lambda^3 \over 3} \Gamma^a_{bc}{\partial x^b \over \partial \lambda} {\partial x^c \over \partial \lambda} + \cdots  \nolabel \\
\end{eqnarray}
Then if we use normal coordinates at $p$ we can take (at $p$) $\bf x\it = \lambda \bf v\it(\theta^i)$, where $\bf v\it$ is some vector at $p$ in the direction of a particular $\boldsymbol{\theta} \in S^{n-1}$.  So, from section \ref{sec:normalcoordinates} we have
\begin{eqnarray}
{\partial^2 x^a \over \partial \lambda^2} &=& 0 \nolabel \\
\Gamma^a_{bc} &=& 0 \nolabel \\
x^a(p) &=& 0 \nolabel \\
{\partial x^a \over \partial \lambda} &=& v^a
\end{eqnarray}
So finally,
\begin{eqnarray}
x^a(p+\lambda) &=& \lambda v^a - {\lambda^3 \over 6} \partial_d\Gamma^a_{bc} v^d v^b v^c \nolabel \\
&=&\lambda v^a - {\lambda^3 \over 18} \partial_{(d}\Gamma^a_{bc)} v^dv^bv^c
\end{eqnarray}
where have symmetrized the sum in the last term.  

So, finally, the partial derivative terms in (\ref{eq:firsttermformetricpullbackricciscalar}) are given by
\begin{eqnarray}
{\partial x^a \over \partial \theta^i} = \lambda {\partial v^a \over \partial \theta^i} - {\lambda^3 \over 6} \partial_{(d} \Gamma^a_{bc)} v^d v^b{\partial v^c \over \partial \theta^i}
\end{eqnarray}
where the product rule with the partial derivative on the three $v$'s has resulted in the factor of $3$.  

The metric term in (\ref{eq:firsttermformetricpullbackricciscalar}) is a bit easier to expand:
\begin{eqnarray}
g_{ab}(p+\lambda) &=& g_{ab}(p) + \lambda {\partial g_{ab}(p) \over \partial \lambda} + {1 \over 2} \lambda^2 {\partial^2 g_{ab}(p) \over \partial \lambda^2} + \cdots \nolabel \\
&=& g_{ab} + \lambda {\partial x^e \over \partial \lambda} \partial_e g_{ab} + {1 \over 2}\lambda^2 \bigg({\partial x^c \over \partial \lambda} {\partial x^d \over \partial \lambda} \partial_c \partial_d g_{ab} + {\partial x^c \over \partial \lambda} \partial_c \bigg({\partial x^d \over \partial \lambda}\bigg)\partial_d g_{ab}\bigg) \nolabel \\
&=& g_{ab} + {1 \over 2} \lambda^2 v^c v^d \partial_c \partial_d g_{ab}
\end{eqnarray}
where we have used (\ref{eq:firstderivofmetricvanishesinnormalcoords}).  

So we can finally piece together (\ref{eq:firsttermformetricpullbackricciscalar}) to form $\tilde g_{ij}$.  Multiplying this out and keeping terms to only fourth order in $\lambda$ (we will switch dummy indices around quite a bit in what follows), 
\begin{eqnarray}
\tilde g_{ij} &=& {\partial x^a(p+\lambda) \over \partial \theta^i} {\partial x^b(p+\lambda) \over \partial \theta^j} g_{ab}(p+\lambda) \nolabel \\
&=& \bigg[\lambda {\partial v^a \over \partial \theta^i} - {\lambda^3 \over 6}\partial_{(d} \Gamma^a_{ec)} v^d v^e {\partial v^c \over \partial \theta^i}\bigg]\bigg[\lambda {\partial v^b \over \partial \theta^j} - {\lambda^3 \over 6}\partial_{(f} \Gamma^b_{gh)} v^f v^g {\partial v^h \over \partial \theta^j}\bigg]\bigg[g_{ab} + {\lambda^2 \over 2}v^k v^l\partial_k\partial_l g_{ab}\bigg]\nolabel \\
&=& \lambda^2 {\partial v^a \over \partial \theta^i}{\partial v^b \over \partial \theta^j} g_{ab}- {\lambda^4 \over 6} {\partial v^a \over \partial \theta^i} \partial_{(f} \Gamma^b_{gh)} v^f v^g{\partial v^h \over \partial \theta^j} g_{ab} - {\lambda^4 \over 6} \partial_{(d}\Gamma^a_{ec)} v^d v^e {\partial v^c \over \partial \theta^i} {\partial v^b \over \partial \theta^j} g_{ab} + {\lambda^4 \over 2} {\partial v^a \over \partial \theta^i} {\partial v^b \over \partial \theta^j} v^k v^l\partial_k\partial_l g_{ab} \nolabel \\
&=& \lambda^2 {\partial v^a \over \partial \theta^i} {\partial v^b \over \partial \theta^j} g_{ab} + {\lambda^4 \over 2} \bigg[{\partial v^e \over \partial \theta^i}{\partial v^b \over \partial \theta^j} v^c v^d \partial_c \partial_d g_{eb} -{1 \over 3} \partial_{(c}\Gamma^a_{de)} v^c v^d g_{ab} \bigg({\partial v^e \over \partial \theta^i} {\partial v^b \over \partial \theta^j} + {\partial v^b \over \partial \theta^i}{\partial v^e \over \partial \theta^j}\bigg)\bigg] \nolabel \\
&=& \lambda^2 {\partial v^a \over \partial \theta^i} {\partial v^b \over \partial \theta^j} g_{ab} + {\lambda^4 \over 2} \bigg[{\partial v^e \over \partial \theta^i}{\partial v^b \over \partial \theta^j} v^c v^d \partial_c \partial_d g_{eb} - {1 \over 3}\partial_{(c}\Gamma^a_{de)} v^c v^d g_{ab} \bigg\{{\partial v^e \over \partial \theta^i} ,{\partial v^b \over \partial \theta^j}\bigg\}\bigg] \nolabel \\
&=& \lambda^2 {\partial v^a \over \partial \theta^i} {\partial v^b \over \partial \theta^j} g_{ab} + {\lambda^4 \over 2} \bigg[{1 \over 2} \bigg\{{\partial v^e \over \partial \theta^i},{\partial v^b \over \partial \theta^j}\bigg\} v^c v^d \partial_c \partial_d g_{eb} - {1 \over 3}\partial_{(c}\Gamma^a_{de)} v^c v^d g_{ab} \bigg\{{\partial v^e \over \partial \theta^i} ,{\partial v^b \over \partial \theta^j}\bigg\}\bigg] \nolabel \\
&=& \lambda^2 {\partial v^a \over \partial \theta^i} {\partial v^b \over \partial \theta^j} g_{ab} + {\lambda^4 \over 4}v^c v^d \bigg\{{\partial v^e \over \partial \theta^i},{\partial v^b \over \partial \theta^j}\bigg\} \bigg[\partial_c\partial_dg_{eb} -{ 2\over 3} \partial_{(c} \Gamma^a_{de)} g_{ab}\bigg] 
\end{eqnarray}
where the curly brackets are, as usual, the anticommutator\footnote{So $\{A,B\} = AB + BA$.}

Looking at the term in brackets in the last line (and remembering that in normal coordinates $\partial_ag_{bc} = 0$),
\begin{eqnarray}
\partial_c\partial_dg_{eb} - {2\over 3}\partial_{(c}\Gamma^a_{de)} g_{ab} &=& \partial_c\partial_dg_{eb} - {1 \over 3} (\partial_c\partial_eg_{bd} - \partial_c\partial_dg_{be} + \partial_c\partial_b g_{ed}) \nolabel \\
&=& {2\over 3}R_{ecdb}
\end{eqnarray}
where the last line is the Riemann tensor in normal coordinates (cf equation (\ref{eq:loweredriemanntensorinnormalcoords})).  

So now we can write
\begin{eqnarray}
\tilde g_{ij} = \lambda^2 {\partial v^a \over \partial \theta^i} {\partial v^b \over \partial \theta^j} g_{ab} + {\lambda^4 \over 6} v^c v^d \bigg\{{\partial v^e \over \partial \theta^i} , {\partial v^b \over \partial \theta^j}\bigg\} R_{ecdb}
\end{eqnarray}

With foresight, let's look more closely at the original metric $g_{ab}$ on $\mathcal{M}$.  Consider using, instead of the coordinates $x^a$ on $\mathcal{M}$, generalized spherical coordinates with radial direction $\lambda$ and angular coordinates $\theta^i$.  Denote this metric on $\mathcal{M}$ $G_{ab}$ ($G_{ab}$ will be an $n\times n$ matrix).  We can then do a normal coordinate transformation to find (if $i=1$ is the radial direction)
\begin{eqnarray}
G_{11} &=& {\partial x^a \over \partial \lambda}{\partial x^b \over \partial \lambda} g_{ab} = v^a v^b g_{ab} \equiv 1 \nolabel \\
G_{1c} &=& {\partial x^a \over \partial \lambda} {\partial x^b \over \partial \theta^c} g_{ab} = \lambda v^a {\partial v^b \over \partial \theta^c} g_{ab} = {\lambda \over 2} {\partial \over \partial \theta^c}\bigg( v^a v^b g_{ab}\bigg) = {\lambda \over 2} {\partial \over \partial \theta^c}(1) = 0 \nolabel \\
G_{cd} &=& {\partial x^a \over \partial \theta^c} {\partial v^b \over \partial \theta^d} g_{ab} = \lambda^2 {\partial v^a \over \partial \theta^c}{\partial v^b \over \partial \theta^d} g_{ab} \equiv \lambda^2 h_{cd} \label{eq:ricciscalar5}
\end{eqnarray}
where $c\neq 1$ in the second line and $c,d \neq 1$ in the third.  So $G_{ab}$ is the block diagonal matrix
\begin{eqnarray}
G_{ab} \dot = 
\begin{pmatrix}
1 & 0 & \cdots & &  \\
0 & &  & \\
\vdots &  & \lambda^2 h_{ab} & \\
 & & & 
\end{pmatrix}
\end{eqnarray}
and therefore
\begin{eqnarray}
G^{ab} \dot = 
\begin{pmatrix}
1 & 0 & \cdots & &  \\
0 & &  & \\
\vdots &  & {1 \over \lambda^2}h^{ab} & \\
 & & & 
\end{pmatrix}
\end{eqnarray}
We can then transform back to the $x^a$ coordinates:
\begin{eqnarray}
g^{ab} &=& {\partial x^a \over \partial \lambda} {\partial x^b \over \partial \lambda} G^{11} + {\partial x^a \over \partial \theta^i} {\partial x^b \over \partial \theta^j} G^{ij} \nolabel \\
&=& v^a v^b + \lambda^2 {\partial v^a \over \partial \theta^i} {\partial v^b \over \partial \theta^j} {1 \over \lambda^2} h^{ij} \nolabel \\
&=& v^a v^b + {\partial v^a \over \partial \theta^i} {\partial v^b \over \partial \theta^j} h^{ij} \nolabel \\
\end{eqnarray}
This then implies
\begin{eqnarray}
h^{ij} {\partial v^a \over \partial \theta^i} {\partial v^b \over \partial \theta^j} = g^{ab} - v^a v^b \label{eq:randomrelationshipforricciscalarderivation}
\end{eqnarray}

So now,
\begin{eqnarray}
\tilde g_{ij} = (\lambda^2) h_{ij} + (\lambda^2)^2\bigg[ {1 \over 6} v^c v^d \bigg\{{\partial v^e \over \partial \theta^i} , {\partial v^b \over \partial \theta^j}\bigg\} R_{ecdb}\bigg] \label{eq:whatweneedtotakethedeterminantofricciscalar}
\end{eqnarray}

To find the determinant $\tilde g$, we start with the general relationship that can be found in any introductory linear algebra text.  For matrices $X$ and $A$, 
\begin{eqnarray}
\det ( A + \epsilon X) = \det (A) (1 + \epsilon Tr(A^{-1}X))
\end{eqnarray}
Using this we can easily compute the determinant of (\ref{eq:whatweneedtotakethedeterminantofricciscalar}), getting (the determinant of $h_{ij}$ is $h$)
\begin{eqnarray}
\tilde g = h \bigg( 1 + {\lambda^2 \over 6} v^c v^d \bigg\{{\partial v^e \over \partial \theta^i} , {\partial v^b \over \partial \theta^j}\bigg\} h^{ij}R_{ecdb}\bigg)
\end{eqnarray}
We can simplify this using (\ref{eq:randomrelationshipforricciscalarderivation}):
\begin{eqnarray}
\tilde g &=&  h \bigg( 1 + {\lambda^2 \over 6} v^c v^d \bigg\{ {\partial v^e \over \partial \theta^i} , {\partial v^b \over \partial \theta^j}\bigg\} h^{ij} R_{ecdb}\bigg) \nolabel \\
&=&  h \bigg( 1 + {\lambda^2 \over 3} v^c v^d (g^{eb} - v^ev^b) R_{ecdb}\bigg) \nolabel \\
&=&  h \bigg(1 + {\lambda^2 \over 3} v^c v^d g^{eb}R_{ecdb} - {\lambda^2 \over 3} v^c v^d v^e v^b R_{ecdb}\bigg)
\end{eqnarray}
The last term involves a symmetric sum over $R_{ecdb}$, which is antisymmetric in the first two and last two indices (cf (\ref{eq:symmetryandantisymmetryinloweredR})) and it therefore vanishes.  So
\begin{eqnarray}
\tilde g &=& h \bigg( 1 + {\lambda^2 \over 3} v^c v^d g^{eb} R_{ecdb}\bigg) \nolabel \\
&=& h \bigg( 1 - {\lambda^2 \over 3} v^c v^d R^b_{cbd}\bigg) \nolabel \\
&=&  h \bigg( 1 - {\lambda^2 \over 3} v^c v^d R_{cd}\bigg) 
\end{eqnarray}
So then, finally, the $n-1$ dimensional volume form is
\begin{eqnarray}
\Omega_{S^{n-1}} &=& \sqrt{|\tilde g|} d\theta^1 \wedge \cdots \wedge d\theta^{n-1} \nolabel \\
&=& \sqrt{h} \sqrt{1 - {\lambda^2 \over 3} v^c v^dR_{cd}}\; d\theta^1 \wedge \cdots \wedge d\theta^{n-1} \nolabel \\
&\approx& \sqrt{h}\bigg( 1 - {\lambda^2 \over 6} v^c v^d R_{cd}\bigg) \; d\theta^1 \wedge \cdots \wedge d\theta^{n-1} \nolabel \\
&=& \sqrt{h}\; d\theta^1 \wedge \cdots \wedge d\theta^{n-1} - {\lambda^2 \over 6} \sqrt{h} v^c v^d R_{cd} \; d\theta^1 \wedge \cdots \wedge d\theta^{n-1}
\end{eqnarray}

So finally, we can find the $n-1$ volume in the curved space, $V_c$:
\begin{eqnarray}
V_c = \int \Omega_{S^{n-1}} &=& \int \sqrt{h} d\theta^1 \cdots d\theta^{n-1} - {\lambda^2 \over 6} R_{cd} \int \sqrt{h} v^c v^d d\theta^1 \cdots d\theta^{n-1} \label{eq:firstexpressionofcurvevolumericciscalar}
\end{eqnarray}
The first term has no dependence on the Riemann or Ricci tensors - in other words it represents the volume contribution in flat space, $V_f$.  In the second term we are integrating the product of $\bf v\it$ over all angles.  But from the first equation in (\ref{eq:ricciscalar5}) we know that $\bf v\it$ is invariant under rotations.  Therefore the integral over them must be simply proportional to the metric:\footnote{This is a standard trick in differential geometry - the metric plays the role of the identity in tensor equations.}
\begin{eqnarray}
\int \sqrt{h} v^c v^d d\theta^1 \cdots d\theta^{n-1}  \propto g^{cd}
\end{eqnarray}
Inserting a proportionality constant and contracting both sides with the metric:
\begin{eqnarray}
& & g_{cd} \int \sqrt{h} v^c v^d d\theta^1 \cdots d\theta^{n-1}  = ag_{cd}g^{cd} \nolabel \\
&\Rightarrow& \int \sqrt{h} (1) d\theta^1 \cdots d\theta^{n-1} = an \nolabel \\
&\Rightarrow& V_f = an \nolabel \\
&\Rightarrow& a = {V_f \over n}
\end{eqnarray}

With all of this, (\ref{eq:firstexpressionofcurvevolumericciscalar}) can be simplified as
\begin{eqnarray}
V_c = V_f - {\lambda^2 \over 6} R_{cd}g^{cd} a = V_f - {\lambda^2 \over 6} {V_f \over n} R = V_f \bigg( 1 - {\lambda^2 \over 6n}R\bigg)
\end{eqnarray}
where 
\begin{eqnarray}
R \equiv R_{ab}g^{ab} \label{eq:firstandonlydefinitionofricciscalar11}
\end{eqnarray}
is called the \bf Ricci Scalar\rm.  Its meaning is that it is the lowest order correction to a volume compared to what would be expected in flat space.  

\subsection{Fourth Intuitive Idea of Curvature}
\label{sec:fourthint}

The technical derivation of the final curvature tensor we are interested in is much more self-explanatory than the previous three.  We therefore keep our comments here brief.  Imagine that you are interested in the curvature of a manifold, but not the entire manifold.  Rather, you are interested in "slices" of it.  For example, imagine a tube:
\begin{center}
\includegraphics[scale=.6]{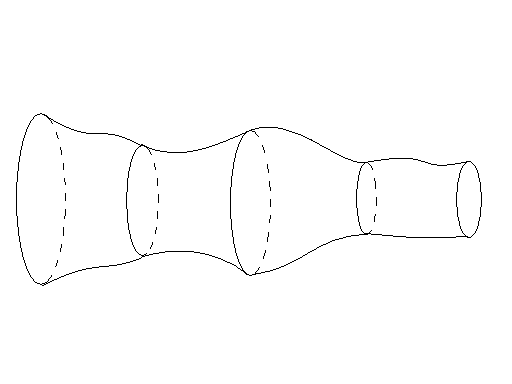}
\end{center}
(imagine it going off in both directions forever).  You may have information about the curvature across the whole thing, but let's say that you don't want that much information.  Instead you only want information about the curvature at a single slice:
\begin{center}
\includegraphics[scale=.6]{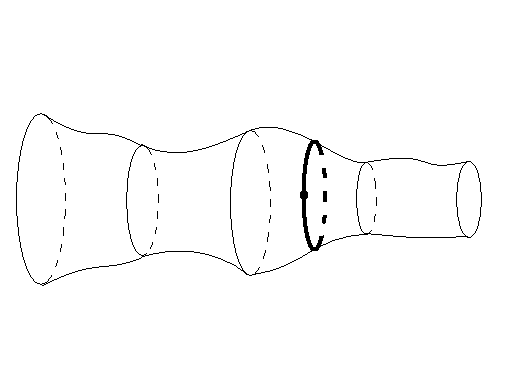}
\end{center}
You can specify this slice by choosing a unit vector at a point and taking the slice orthogonal to it:
\begin{center}
\includegraphics[scale=.6]{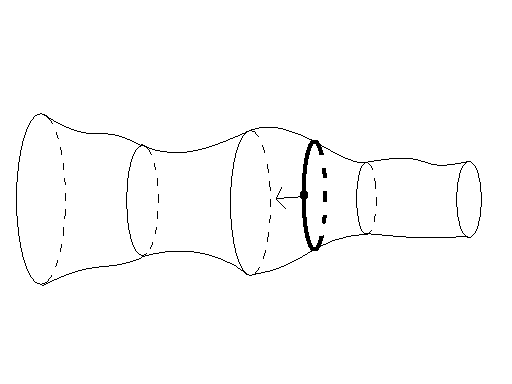}
\end{center}

So, this final notion of curvature doesn't provide a new way of measuring curvature, but rather introduces the notion of "sectional curvature".  

\subsection{The Einstein Tensor}
\label{sec:fourthint2}

Consider a point $p \in \mathcal{M}$.  We will start with the Riemann tensor $R_{abcd}$ and contract to get the Ricci scalar.  However we aren't interested in the Ricci scalar on all of $\mathcal{M}$.  Instead we are only interested in the curvature of the subspace of $\mathcal{M}$ that is orthogonal to some $\bf v\it \in T_p\mathcal{M}$ (of unit length, $g_{ab}v^av^b = 1$).  So whereas normally we would get the Ricci scalar from $R_{abcd}$ by contracting it as
\begin{eqnarray}
g^{ac}g^{bd}R_{abcd} = R
\end{eqnarray}
that won't work here.  Rather, we must project off from each of these contractions the component that is parallel to $\bf v\it$.  Therefore we replace $g^{ab}$ with $g^{ab} - v^av^b$ in the above contraction.  We will denote this ``directional" curvature value $-2G_{ab} v^av^b$:\footnote{Don't worry about the factor of $-2$ - it's not important at this point and we merely include it for agreement with what will come later.  You can think of it as an irrelevant scale factor for now.}
\begin{eqnarray}
-2 G_{bd}v^bv^d &=&(g^{ac} - v^av^c)(g^{bd} - v^bv^d)R_{abcd} \nolabel \\
&=& g^{ac}g^{bd}R_{abcd} - g^{ac}v^bv^dR_{abcd} - g^{bd}v^av^cR_{abcd} + v^av^cv^bv^dR_{abcd} \nolabel \\
&=& R - 2g^{ac} v^b v^d R_{abcd} \nolabel \\
&=& R - 2 v^b v^d R_{bd} \nolabel \\
&=& R v^bv^dg_{bd} - 2 v^b v^d R_{bd} \nolabel \\
&=& (Rg_{bd} - 2R_{bd}) v^bv^d \nolabel \\
\Longrightarrow G_{bd} &=& R_{bd} - {1 \over 2}g_{bd}R \label{eq:introductionofeinsteintensorwithminus2factor}
\end{eqnarray}

This new tensor,
\begin{eqnarray}
G_{ab} = R_{ab} - {1 \over 2} g_{ab}R \label{eq:firstandonlydefinitionofEinsteintensor}
\end{eqnarray}
is called the \bf Einstein Tensor\rm.  So, given an $n$-dimensional manifold with Ricci scalar $R_{(n)}$ and a vector $\bf v\it \in T_p\mathcal{M}$ for some $p \in \mathcal{M}$, the $n-1$ dimensional subspace of $\mathcal{M}$ orthogonal to $\bf v\it$ will have curvature
\begin{eqnarray}
R_{(n-1)} = G_{ab} v^av^b
\end{eqnarray}

This notion of "sectional curvature" will play a very important role in general relativity.  Specifically the vectors $\bf v\it$ that we will use to define an orthogonal subspace will be timelike vectors.  Given a particular unit timelike vector, the subspace orthogonal to it will be the purely spatial region for a given instant in time.  General relativity will say that the curvature of a spatial "slice" of spacetime, given by the Einstein tensor, is proportional to the energy density in the spacetime.  We will discuss this in more detail later.  

\section{Examples of Curvature Tensors}

We will now go through the manifolds we have been considering, that we know the metrics for, and compute each curvature tensor for them.  We summarize our results so far, starting with the metric, in the following table:

\begin{center}
\begin{tabular}{|c|c|c|}\hline Name & Tensor & Definition \\\hline  & &  \\Metric & $g_{ij}$ & $g_{ij}(p)dx^i \otimes dx^j$\\  & & \\\hline  & & \\ Connection & $\Gamma^i_{ij}$ & ${1\over 2} g^{il}(\partial_j g_{kl} + \partial_k g_{lj} - \partial_lg_{jk})$\\  & &  \\\hline  & &  \\ Riemann Tensor & $R^i_{jkl}$ & $\partial_k\Gamma^i_{lj} - \partial_l\Gamma^i_{kj} + \Gamma^i_{km}\Gamma^m_{lj} - \Gamma^i_{lm}\Gamma^m_{kl}$\\  & &  \\\hline & & \\ Ricci Tensor & $R_{ij}$ & $R^k_{ikj}$\\ & &  \\\hline & & \\ Ricci Scalar & $R$ & $g^{ij}R_{ij}$\\ & &  \\\hline \end{tabular}
\end{center}

We have found the metrics and connections for several manifolds already (cf sections \ref{sec:inducedmetrics} and \ref{sec:metricconnection}).  We will now go through each of those manifolds and compute the curvature tensor for each.  

\subsection{$S^1$}

We found the metric on the circle by embedding it in $\mathbb{R}^2$ 
\begin{eqnarray}
x &=& r\cos\theta \nolabel \\
y &=& r\sin\theta
\end{eqnarray}
and using the pullback.  We previously found that the metric for the circle is 
\begin{eqnarray}
ds^2 = r^2 d\theta^2
\end{eqnarray}
or
\begin{eqnarray}
g_{ij} = g_{\theta \theta} = r^2
\end{eqnarray}
As shown above this lead to 
\begin{eqnarray}
\Gamma^i_{ij} = 0 \qquad \forall \; i,j,k
\end{eqnarray}
And therefore all of the curvature tensor vanish.  Therefore, the circle has no curvature.  

This result may seem surprising.  We will therefore discuss what it means for a manifold to be "flat" in the next section.  To get an initial intuitive feeling for why it is flat, recall how we derived the Riemann tensor.  We parallel transported a vector around two paths to the same point and took the difference.  If you imagine doing this on a circle, it is clear that the only two paths from point $p$ to point $q$ are the two directions around the circle, and obviously in both cases the vector will be the same in the end.  We will discuss this notion of flatness (and others) in section \ref{sec:meaningofflat}.  

\subsection{$S^1$ Stretched Out}

The next manifold we considered was the "stretched out" circle or the ellipse, which we mapped using
\begin{eqnarray}
x&=&2r\cos\theta \nolabel \\
y&=&r\sin\theta 
\end{eqnarray}
Even though we pointed out that this is actually identical to the circle (up to coordinate transformation) we will still consider it separately to show this equivalence explicitly.  
The metric was
\begin{eqnarray}
ds^2 = r^2 (4\cos^2\theta + \sin^2\theta) d\theta^2
\end{eqnarray}
or 
\begin{eqnarray}
g_{ij} = g_{\theta\theta} = r^2(4\cos^2\theta+\sin^2\theta)
\end{eqnarray}

We found for this that the connection coefficient is
\begin{eqnarray}
\Gamma^{\theta}_{\theta\theta} = -{3\sin(2\theta) \over 5+4\cos(2\theta)}
\end{eqnarray}
It is then straightforward to show that
\begin{eqnarray}
R^i_{jkl} = 0 \qquad \forall\; i,j,k,l
\end{eqnarray}
So all of the curvature tensors vanish and we see that the ellipse also has no curvature.  

\subsection{$S^2$}

The next manifold we considered was the sphere, which we mapped using
\begin{eqnarray}
x&=& \sin\theta\cos\phi \nolabel \\
y&=& \sin\theta\sin\phi \nolabel \\
z&=& \cos\theta \label{eq:exampleofcurvaturetensorss2parameterization}
\end{eqnarray}
The metric was
\begin{eqnarray}
ds^2 = d\theta^2 + \sin^2\theta d\phi^2
\end{eqnarray}
or
\begin{eqnarray}
g_{ij} \dot = 
\begin{pmatrix}
1 & 0 \\
0 & \sin^2\theta
\end{pmatrix}
\end{eqnarray}
We then found that the non-zero connection coefficients were
\begin{eqnarray}
\Gamma^{\theta}_{\phi \phi} &=& -\sin\theta\cos\theta \nolabel \\
\Gamma^{\phi}_{\theta \phi} &=& \Gamma^{\phi}_{\phi \theta} = \cot\theta
\end{eqnarray}

From these we can compute the Riemann tensor, which has non-vanishing components 
\begin{eqnarray}
R^{\phi}_{\theta\phi \theta} &=& -R^{\phi}_{\theta \theta \phi} = 1 \nolabel \\
R^{\theta}_{\phi \theta \phi} &=& -R^{\theta}_{\phi \phi \theta} = \sin^2\theta
\end{eqnarray}
It is good that we found that $R^i_{jkl}$ is non-vanishing for the sphere - it was the example we used to define $R^i_{jkl}$, so if it was zero we'd be in trouble!  

Next, the Ricci tensor has non-vanishing components
\begin{eqnarray}
R_{\theta \theta} &=& 1 \nolabel \\
R_{\phi \phi} &=& \sin^2\theta
\end{eqnarray}
And finally the Ricci scalar is
\begin{eqnarray}
R &=& g^{\theta\theta}R_{\theta\theta} + g^{\theta\phi}R_{\theta\phi} + g^{\phi\theta}R_{\phi\theta}+g^{\phi\phi}R_{\phi\phi} \nolabel \\
&=& 1 + {1 \over \sin^2\theta} \sin^2\theta \nolabel \\
&=& 2
\end{eqnarray}
So the sphere has constant curvature.  This isn't surprising - every point on $S^2$ is "sloped" the same as any other point.  There is no variation in how curved it is from point to point.  

Before moving on, notice that 
\begin{eqnarray}
R_{ij} \dot = 
\begin{pmatrix}
1 & 0 \\
0 & \sin^2\theta
\end{pmatrix} = g_{ij}
\end{eqnarray}
The fact that it is proportional to (in fact, equal) the metric is not surprising.  On any manifold of constant curvature $R$, 
\begin{eqnarray}
R = g^{ij}R_{ij} = const &\Longrightarrow& R_{ij} \propto g_{ij}  \nolabel\\
&\Longrightarrow& R = g^{ij}R_{ij} = g^{ij}\alpha g_{ij} = \alpha n
\end{eqnarray}
where $\alpha$ is the proportionality constant and $n$ is the dimension of the manifold.  Specifically, 
\begin{eqnarray}
\alpha = {R\over n}
\end{eqnarray}
So, we know that here $R=2$ and $n=2$, so $\alpha = {2 \over 2} = 1$, and therefore
\begin{eqnarray}
R_{ij} = \alpha g_{ij} = g_{ij}
\end{eqnarray}
we we found above.  

As is clear from (\ref{eq:exampleofcurvaturetensorss2parameterization}), we choose this sphere to have radius $1$.  You can go back and convince yourself that had we taken the radius to be $r$, we would have had
\begin{eqnarray}
R = {2 \over r^2} \label{eq:firsttimeweseethatthecurvscalisequaltotwooverrsquared}
\end{eqnarray}
So the larger the radius the smaller the curvature.  This makes sense.  On a small sphere the curvature would be easy to notice.  On a larger sphere (like the Earth), the surface looks flat (i.e. less curved).  

Finally, we can compute the Einstein tensor:
\begin{eqnarray}
G_{ij} &=& R_{ij} - {1 \over 2} g_{ij}R \nolabel \\
&=& 
\begin{pmatrix}
1 & 0 \\
0 & \sin^2\theta
\end{pmatrix} - {1 \over 2} 
\begin{pmatrix}
1 & 0 \\
0 & \sin^2\theta
\end{pmatrix} 2 \nolabel \\
&=& 0 \label{eq:firstexampleofvanishingeinsteintensorparfors2closepar}
\end{eqnarray}
This makes sense - for a given vector somewhere on $S^2$, the subspace orthogonal will be a circle $S^1$, which we know from above is flat.  

\subsection{$S^2$ Stretched Out}

The stretched out $S^2$ given by
\begin{eqnarray}
x &=& \sin\theta\cos\phi \nolabel \\
y &=& \sin\theta\sin\phi \nolabel \\
z &=& \lambda \cos\theta
\end{eqnarray}
is more interesting.  We found above that the metric is
\begin{eqnarray}
ds^2 = {1 \over 2}\big((\lambda^2+1)-(\lambda^2-1)\cos(2\theta)\big)d\theta^2 + \sin^2\theta d\phi^2
\end{eqnarray}
or
\begin{eqnarray}
g_{ij} \dot = 
\begin{pmatrix}
{1 \over 2}\big((\lambda^2+1)-(\lambda^2-1)\cos(2\theta)\big) & 0 \\
0 & \sin^2\theta
\end{pmatrix}
\end{eqnarray}
We also found the non-zero connection coefficients to be
\begin{eqnarray}
\Gamma^{\theta}_{\theta\theta} &=& {(\lambda^2-1)\sin(2\theta) \over (\lambda^2+1)-(\lambda^2-1)\cos(2\theta)} \nolabel \\
\Gamma^{\theta}_{\phi \phi} &=& -{2\cos\theta\sin\theta \over (\lambda^2+1) - (\lambda^2-1)\cos(2\theta)}\nolabel \\
\Gamma^{\phi}_{\theta\phi} &=& \Gamma^{\phi}_{\phi\theta} = \cot\theta
\end{eqnarray}

Then (with either a lot of tedious work or a computer), it is straightforward to calculate the non-vanishing components of Riemann:
\begin{eqnarray}
R^{\phi}_{\theta\phi\theta} &=& -R^{\phi}_{\theta\theta\phi} = {2\lambda^2 \over (\lambda^2+1)-(\lambda^2-1)\cos(2\theta)} \nolabel \\
R^{\theta}_{\phi\phi\theta} &=& -R^{\theta}_{\phi\theta\phi} = {-4\lambda^2\sin^2\theta \over \big((\lambda^2+1)-(\lambda^2-1)\cos(2\theta)\big)^2}
\end{eqnarray}
The Ricci tensor has non-vanishing components
\begin{eqnarray}
R_{\theta\theta} &=& {2\lambda^2 \over (\lambda^2 +1) - (\lambda^2 - 1)\cos(2\theta)} \nolabel \\
R_{\phi\phi}&=& {4\lambda^2\sin^2\theta \over \big( (\lambda^2+1) - (\lambda^2-1)\cos(2\theta)\big)^2}
\end{eqnarray}
And finally, the Ricci curvature scalar is
\begin{eqnarray}
{8\lambda^2 \over \big((\lambda^2+1)-(\lambda^2-1)\cos(2\theta)\big)^2}
\end{eqnarray}

To try to give a bit more intuition about what the curvature scalar represents, let's look at it more closely for various values of $\lambda$.\footnote{It will be very helpful to read over the section on this manifold in section \ref{sec:inducedmetrics}, especially the pictures for various $\lambda$.}  First, notice that all three tensors (Riemann, Ricci, and the scalar) all reduce to their values for the un-stretched $S^2$ in the previous section for $\lambda=1$.  Next, notice that the scalar has no $\phi$ dependence.  This is because the distortion due to $\lambda$ is entirely in the $z$ direction, and because $\phi$ is the azimuthal angle it is invariant under distortions along the $z$ axis.  From the pictures in section \ref{sec:inducedmetrics} it is clear that for any fixed $\theta$ all points are symmetric around the $z$ axis.  

We can graph the curvature for various values of $\lambda$ and $\theta$:
\begin{center}
\includegraphics[scale=1.2]{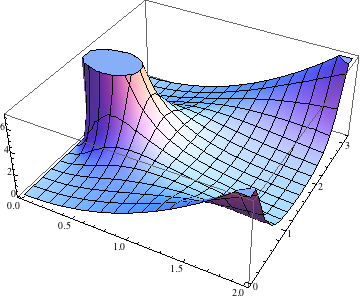}
\end{center}
We have taken $\lambda$ (the axis on the bottom left) to run from $0$ to $2$, and $\theta$ (the bottom right axis) to run from $0$ to $\pi$.  Notice that at $\lambda=1$ the curvature is constant (at $2$) as expected (this is the standard sphere).  For $\lambda =2$, the sphere is elongated along the $z$ axis (cf picture of egg on page \pageref{pictureofeggelongatedalongzaxis}).  As expected the curvature takes the greatest values at $\theta = 0$ and $\theta = \pi$, and the least value at $\theta = {\pi \over 2}$.  This is exactly what we would expect - the curvature is greatest in the $\pm z$ directions and least at $z=0$.  

For $\lambda \rightarrow0$ (cf picture of pancake on page \pageref{eqpictureofpancakeshortatzaxiswithalmbda}), the curvature is very small (approaching $0$) for $\theta = 0$ through close to $\theta = {\pi \over 2}$, which is what we would expect.  For $\lambda = 0$, the curvature will become infinite at $\theta = {\pi \over 2}$ and zero elsewhere.  

Finally we can form the Einstein tensor.  But before doing so, let's think about what we'd expect it to be.  With the sphere we pointed out that a cross section is $S^1$.  And, as we saw above, $S^1$ is always flat.  So even on this deformed sphere we would expect that the Einstein tensor vanishes.  So, testing our intuition:
\begin{eqnarray}
G_{ij} &=& R_{ij} - {1 \over 2} g_{ij} R \nolabel \\
&=& 
\begin{pmatrix}
{2\lambda^2 \over (\lambda^2+1)-(\lambda^2-1)\cos2\theta} & 0\\
0 & {4 \lambda^2\sin^2\theta \over \big( (\lambda^2 + 1)-(\lambda^2-1)\cos(2\theta)\big)^2}
\end{pmatrix} \nolabel \\
& & - {1 \over 2} 
\begin{pmatrix}
{1 \over 2}\big((\lambda^2 + 1) - (\lambda^2 - 1)\cos(2\theta) & 0 \\
0 & \sin^2\theta
\end{pmatrix}{8 \lambda^2 \over \big((\lambda^2+1) - (\lambda^2-1)\cos(2\theta)\big)^2} \nolabel \\
&=& 
\begin{pmatrix}
{2\lambda^2 \over (\lambda^2+1)-(\lambda^2-1)\cos2\theta} & 0\\
0 & {4 \lambda^2\sin^2\theta \over \big( (\lambda^2 + 1)-(\lambda^2-1)\cos(2\theta)\big)^2}
\end{pmatrix} \nolabel \\
& & - 
\begin{pmatrix}
{2\lambda^2 \over (\lambda^2+1)-(\lambda^2-1)\cos2\theta} & 0\\
0 & {4 \lambda^2\sin^2\theta \over \big( (\lambda^2 + 1)-(\lambda^2-1)\cos(2\theta)\big)^2}
\end{pmatrix} \nolabel \\
&=& 0
\end{eqnarray}

\subsection{Torus}

Finally, we considered the torus $T^2$ with map
\begin{eqnarray}
x &=& (R+r\cos\theta) \cos\phi \nolabel \\
x &=& (R+r\cos\theta)\sin\phi \nolabel \\
z &=& r\sin\theta
\end{eqnarray}
This gave metric
\begin{eqnarray}
ds^2 = r^2d\theta^2 + (R+r\cos\theta)^2 d\phi^2
\end{eqnarray}
or
\begin{eqnarray}
g_{ij} \dot = 
\begin{pmatrix}
r^2 & 0 \\
0 & (R+r\cos\theta)^2
\end{pmatrix}
\end{eqnarray}
This metric gave us the non-vanishing components of the connection:
\begin{eqnarray}
\Gamma^{\phi}_{\phi \phi} &=& {(R+r\cos\theta)\sin\theta \over r} \nolabel \\
\Gamma^{\phi}_{\theta \phi} &=& \Gamma^{\phi}_{\phi \theta} = -{r\sin\theta \over R+r\cos\theta}
\end{eqnarray}

This connection will then give
\begin{eqnarray}
R^i_{jkl} = 0 \qquad \forall \; i,j,k,l
\end{eqnarray}
and therefore all curvature tensors vanish.  Thus $T^2$ is a flat manifold.  Again, you can begin to see why by considering a vector being parallel transported around different paths to the same point.  You should convince yourself that the final vector will not depend on the path taken.  

\subsection{$S^2 \otimes \mathbb{R}$}
\label{sec:s2otimesbbr}

All of our previous examples have been in either one or two dimensions.  While this is helpful to get some initial intuition it is not nearly general enough.  We therefore conclude by adding to the manifolds we have been considering one final three dimensional example.  

Consider the three dimensional space generated by "dragging" a sphere along the real line.  In other words, at any point you could move in the two dimensional space of $S^2$ \it as well as \rm the additional degree of freedom of $\mathbb{R}$.  We can think of this as the subspace in $\mathbb{R}^4$ defined by the map:
\begin{eqnarray}
x &=& \sin\theta\cos\phi \nolabel \\
y &=& \sin\theta\sin\phi \nolabel \\
z &=& \cos\theta \nolabel \\
w &=& \psi
\end{eqnarray}
where 
\begin{eqnarray}
\theta &\in& [0,\pi) \nolabel \\
\phi &\in& [0,2\pi) \nolabel \\
\psi &\in& [0,\pi)
\end{eqnarray}
This is essentially attaching an $S^2$ to every point on $\mathbb{R}$ between $0$ and $\pi$.  We can then compute the metric:
\begin{eqnarray}
ds^2 &=& d\theta^2 + \sin^2\theta d\phi^2 + d\psi^2 \nolabel \\
&=& ds^2_{S^2} + d\psi^2  \label{eq:flattensorwithS2crossRinexamplesofcurvtenssec}
\end{eqnarray}

Before moving on to compute the remaining values, this metric shouldn't be a surprise - all we have done is add the single \it flat \rm one dimensional degree of freedom.  We would therefore expect the metric to have this form.  Furthermore, because this has added only a flat degree of freedom, we don't expect the curvature to differ drastically from the sphere by itself.  

With that said, we can begin computing to see if we're right.  The non-vanishing connection terms are:
\begin{eqnarray}
\Gamma^{\phi}_{\phi\theta} &=& \Gamma^{\phi}_{\theta\phi} = \cot\theta \nolabel \\
\Gamma^{\theta}_{\phi \phi} &=& -\cos\theta\sin\theta
\end{eqnarray}
The non-vanishing Riemann tensor values are:
\begin{eqnarray}
R^{\theta}_{\phi\theta\phi} &=& -R^{\theta}_{\phi\phi\theta} = \sin^2\theta \nolabel \\
R^{\phi}_{\theta\phi\theta} &=& -R^{\phi}_{\theta\theta\phi} = 1
\end{eqnarray}
The Ricci tensor is then
\begin{eqnarray}
R_{ij} \dot = 
\begin{pmatrix}
1 & 0 & 0 \\
0 & \sin^2\theta & 0 \\
0 & 0 & 0
\end{pmatrix}
\end{eqnarray}
And the curvature scalar is
\begin{eqnarray}
R = 2
\end{eqnarray}
So, just as we suspected, the curvature is exactly the same as with $S^2$ by itself.  However we can compute
\begin{eqnarray}
G_{ij} \dot = 
\begin{pmatrix}
0 & 0 & 0 \\
0 & 0 & 0 \\
0 & 0 & -1
\end{pmatrix} \label{eq:finalexampleofcurvaturetensorseinsteintensorinwhatweareconsideringflat}
\end{eqnarray}

To see what this means, consider a unit vector $\bf v\it$ in the $\theta$ or $\phi$ directions (tangent to the sphere).  This will give
\begin{eqnarray}
G_{ij} v^iv^j = 0
\end{eqnarray}
This is exactly what we found above in (\ref{eq:firstexampleofvanishingeinsteintensorparfors2closepar}).  A cross section of a vector in the $\theta$ or $\phi$ direction will be a cross section of $S^2$, which is $S^1$, which we know is flat.  

However, if $\bf v\it$ is a unit vector in the $\psi$ direction, we have
\begin{eqnarray}
G_{ij} v^iv^j = -1
\end{eqnarray}
And, from our definition of $G_{ij}$ (cf top line in equation (\ref{eq:introductionofeinsteintensorwithminus2factor})) we know that the "sectional curvature" we are interested in is
\begin{eqnarray}
-2G_{ij} = -1 \Longrightarrow R_{(2)} = 2
\end{eqnarray}
This is what we should expect.  A cross section orthogonal to given vector in the $\psi$ direction will simply be a copy of $S^2$, which has curvature $2$.  

We could conclude here, but we will investigate manifolds of this type a bit more.  One common type of problem in physics (namely general relativity) is that we have some information about Einstein's tensor and we want to know the metric.  You are welcome to write out $G_{ij}$ entirely in terms of the metric, but you will find it to be an extraordinarily complicated expression.  Furthermore, you will find that, given $G_{ij}$, solving for $g_{ij}$ involves solving a non-linear partial differential equation that, in general, can't be solved.  

But there is still much we can do.  Let's consider (\ref{eq:finalexampleofcurvaturetensorseinsteintensorinwhatweareconsideringflat}) to be the flat, or homogeneous case.  In other words, if we define the matrix
\begin{eqnarray}
G_0 = 
\begin{pmatrix}
0 & 0 & 0 \\
0 & 0 & 0 \\
0 & 0 & -1
\end{pmatrix}
\end{eqnarray}
Then, taking $G_{ij}$ to be a collection of first and second derivatives of $g_{ij}$, we interpret
\begin{eqnarray}
G_{ij} \dot = G_0
\end{eqnarray}
to be a homogeneous differential equation for $g_{ij}$.  The solutions will then obviously be (\ref{eq:flattensorwithS2crossRinexamplesofcurvtenssec}).  

Now let's say that $G_{ij} \neq G_0$, but rather there is some source term on the right hand side that is "driving" the geometry\footnote{recall that the \it topology \rm $S^2\otimes \mathbb{R}$ can, via homeomorphism, have a huge array of metrics and therefore a huge array of geometries.  In this sense the source term will change the geometry but not the topology, and the manifold will always be $S^2 \otimes \mathbb{R}$.} of $S^2\otimes \mathbb{R}$ in the same way that a source term in Maxwell's equation "drives" the electric and magnetic fields.  We will call this source term $T_{ij}$.  So, our inhomogeneous differential equation for $g_{ij}$ is now
\begin{eqnarray}
G_{ij} = T_{ij} \label{eq:firstexampleofeinstenequationforsemirealisticcaseinexamplesofcurvaturetensorssection}
\end{eqnarray}

As we said above, this cannot be easily solved for arbitrary $T_{ij}$.  However, it is often possible to make certain physical assumptions that can simplify the solution.  For example, we might assume based on a symmetry argument that the sphere must not be deformed into an egg or a pancake, etc.  Therefore any term on the $ds^2_{S^2}$ part of the metric must act out front as a radial term.  Furthermore, we may assume that the source acts homogeneously throughout $S^2$, and therefore the source can only depend on $\psi$.  

With these assumptions, we guess that the metric will have the form
\begin{eqnarray}
a(\psi) ds^2_{S^2} + d\psi^2 \label{eq:s2crossrexamplemetricwithacoeff}
\end{eqnarray}
where $a(\psi)$ is some unknown function.  

We can redo the calculations of the curvature tensors, and we will find that now
\begin{eqnarray}
G_{ij} \dot = 
\begin{pmatrix}
{a\ddot a \over (1+\dot a^2)^2} & 0 & 0 \\
0 & {a\ddot a \sin^2 \theta \over (1+\dot a^2)^2} & 0 \\
0 & 0 & -{1\over a^2}
\end{pmatrix}
\end{eqnarray}
where the dot represents a derivative with respect to $\psi$.  Also, noticing that $a(\psi)$ plays the role of the radius in (\ref{eq:s2crossrexamplemetricwithacoeff}), and in $G_{ij}$ here we see that for constant radius the sectional curvature ($G_{33}$) is $-{1 \over a^2}$, so the curvature is ${2\over a^2}$ (because of the factor of $-2$ in the definition of $G_{ij}$).  This agrees with what we had above in (\ref{eq:firsttimeweseethatthecurvscalisequaltotwooverrsquared}).  

Let's assume that $a(\psi)$ is a physically meaningful quantity, and we can use it to define the source term that is driving our geometry.  In other words, we can write $T_{ij}$ in terms of $a(\psi)$.  Let's then say that some sort of physical considerations tell us that the source term has the form\footnote{It may seem strange for us to introduce this mysterious "geometrical source term" without any explanation about where it comes from.  For now just take for granted that some sort of physical reasoning can give it to you (at least in terms of a set of physical parameters like $a(\psi)$, and the equation (\ref{eq:firstexampleofeinstenequationforsemirealisticcaseinexamplesofcurvaturetensorssection}) can be formed in a relatively straightforward way.}
\begin{eqnarray}
T_{ij} \dot = 
\begin{pmatrix}
{-\omega^2 a^2 \over (1+\dot a^2)^2} & 0 & 0 \\
0 & {-\omega^2 a^2 \sin^2\theta \over(1+\dot a^2)^2}   & 0 \\
0 & 0 & -{1\over a^2}
\end{pmatrix} \label{eq:firstexampleansatzins2crossr1examplesection1}
\end{eqnarray}

So, by making the "physical" assumptions about the symmetry and homogeneity on $S^2$, we have traded the unsolvable differential equation for $g_{ij}$ for a much easier set of equations for a physical parameter $a(\psi)$.  Reading from the above equations for $G_{ij}$ and $T_{ij}$, we actually only have one unique equation:
\begin{eqnarray}
a\ddot a &=& -\omega^2 a^2 \qquad \Longrightarrow \qquad \ddot a = -\omega^2 a
\end{eqnarray}
which has solutions
\begin{eqnarray}
a(\psi) = A\sin(\omega\psi) + B\cos(\omega\psi)
\end{eqnarray}
The values of $A$ and $B$ can then be found via boundary conditions.  For example, $B=0$ and $\omega=1$ gives
\begin{eqnarray}
ds^2 = \sin(\psi) ds^2_{S^2} + d\psi^2
\end{eqnarray}
This corresponds to starting with a singularity (an $S^2$ with zero radius) at $\psi = 0$, the sphere growing to have unit radius at $\psi = {\pi \over 2}$, and then crunching back down to a singularity at $\psi = \pi$:
\begin{center}
\includegraphics[scale=.7]{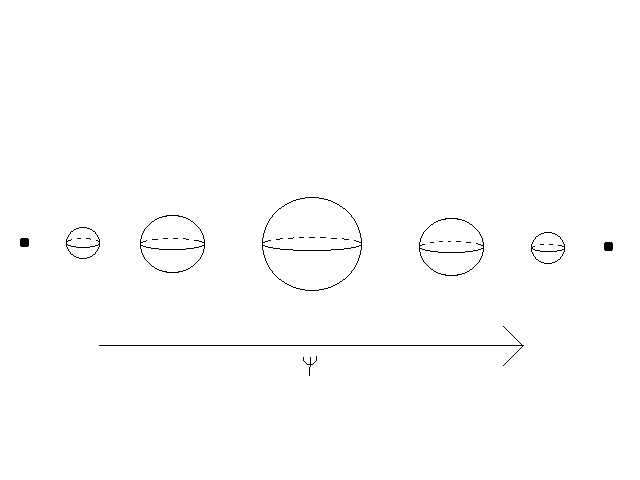}
\end{center}

\section{Concluding Thoughts on Differential Geometry}

Before concluding this chapter we briefly discuss what we have done and what we still have to do.  In this chapter we essentially added a single structure, the metric, to the general topological manifolds $\mathcal{M}$ from the prior chapters.  This structure did two things - it fixed the structure of the manifold so that we don't have the enormous homeomorphism redundancies discussed previously, and it greatly enhanced our ability to do calculus on $\mathcal{M}$.  

One of the most important consequences of what the metric does is that it gives us a precise way of discussing curvature - the Riemann and Ricci tensors and the curvature scalar provide a powerful tool for probing a manifold, and the "sectional curvature" that the Einstein tensor provides will also be very useful.  

We have now spent nearly 300 pages doing math.  You have likely noticed that there has hardly been a single word of physics.  In the next few chapters, however, we will begin to use what we have learned so far to (finally) do some physics.  As the title of this paper (and this series) indicates, our interest is ultimately particle physics.  However, we will spend some time appearing to be distracted from that goal by studying electrodynamics, general relativity and cosmology.  

The reason for this is two-fold.  The first reason is that the current status of high energy particle physics and cosmology leaves them, at times, intimately linked.  String theory (our long term goal with this series), which is a theory of quantum gravity, is perhaps the most glaring example of this.  Many of the most important and exciting aspects of string theory are ultimately cosmological, and vice versa.  Therefore it will be necessary when studying string theory later to have some grasp of general relativity.  

The second reason is that our goal with this series is to provide not only an overview of the technical aspects of particle physics, but also provide some intuition.  We have developed a great deal of topology and geometry, but not nearly enough for string theory or even a respectable stab at gauge theories.  And while the geometric and topological tools we will develop in order to study more advanced topics are extraordinarily powerful and elegant, is is very, very easy to completely lose sight of their meaning.  No physicist should feel comfortable "plugging and chugging" with equations he or she doesn't understand.  

Furthermore, providing intuition with the geometrical and topological ideas we will be discussing later becomes increasingly difficult, especially when we make the jump to complex and algebraic geometry and look at the related topological concepts.  Therefore, in order to ease the process of understanding the more advanced topics, we will invest time now to understanding the relatively simple ideas we have discussed thus far.  General relativity in particular provides a very nice and (somewhat) straightforward application of the math thus far outlined.  Understanding it should give a deeper intuition about how these ideas come into play in physics.  

As an indication of how this will help, recall that we said above that general relativity is fundamentally a theory of the Levi-Civita connection.  Certain physical considerations demand that the connections be torsion free and metric compatible when studying gravitation, and this greatly simplifies the situation.  When we begin studying general gauge theories later, however, we won't have the luxury of only considering the Levi-Civita case - the connections there are much more general and can be much more complicated.  If you don't have a good grasp of the Levi-Cevitia connection, curvature, and how geometry affects physics, you will likely be lost when trying to make sense of the generalizations.  

So with that said, we will spend several chapters now doing physics before transitioning back to math in the next paper in this series.  We hope this is somewhat refreshing after several hundred pages of math, and that it helps make the ideas developed so far more concrete.  

\section{References and Further Reading}

The primary reference for this chapter was \cite{nakahara}, though we also made extensive use of \cite{carroll}, \cite{dinverno}, \cite{lee}, and most notably \cite{schutzDG}.  The derivation of the Ricci tensor and Ricci scalar was found in \cite{loveridge}.  For further reading we recommend \cite{abraham}, \cite{besse}, \cite{frankel}, \cite{gockeler}, \cite{helgason}, \cite{hicks}, \cite{isham}, \cite{jost}, \cite{nash}, and \cite{spivak}.  

\part{Physics}

\chapter{Electrodynamics}
\label{sec:electrodynamics}

The first physical topic we will discuss is electrodynamics.  We do this for a variety of reasons.  First of all, it provides a nice physical application of some of the mathematical ideas we have discussed, and ties them to something you should already be familiar with.  Second, the ideas developed here will he helpful when we discuss general relativity and cosmology.  Third, the underlying ideas of electrodynamics (believe it or not) play a huge role in string theory, and here is as good a time as any to introduce them.  Fourth, the long term goal of this paper is to build a more thorough understanding of gauge theories, and as we saw in \cite{Firstpaper}, electrodynamics is the simplest and most straightforward gauge theory.  We will consider it in a fuller geometric context later, but introducing it now at an intermediate step will help the transition to the more complete picture.  

\section{Formalisms of Electrodynamics}

The primary point of this section is not to necessarily provide any real mathematical or physical insight, but merely to illustrate how the mathematics we have developed previously (primarily differential forms and cohomology) can help clarify and generalize a physical theory.  

\subsection{Electrodynamics and Vectors}
\label{sec:electrodynamicsandvectors}

\subsubsection{Maxwell's Equations}

Electrodynamics is based almost entirely on Maxwell's equations:\footnote{As usual we work in units where $c=1$.}
\begin{eqnarray}
\boldsymbol{\nabla} \cdot \vec B &=& 0 \nolabel \\
\boldsymbol{\nabla} \times \vec E +{\partial \vec B \over \partial t} &=& 0 \nolabel \\
\boldsymbol{\nabla} \cdot \vec E &=& \rho  \nolabel \\
\boldsymbol{\nabla} \times \vec B -  {\partial \vec E \over \partial t}&=&  \vec J  \label{eq:maxwell}
\end{eqnarray}
where $\vec E$ are $\vec B$ are the electric and magnetic fields, respectively, $\vec J$ (current per area) is the source current, and $\rho$ (charge per volume) is the electric charge density.  The first two are the source free, or homogeneous, equations, and the last two are the inhomogeneous source equations.  

\subsubsection{The Continuity Equation}

An important consequence of Maxwell's equations is the charge conservation law.  Consider the last of (\ref{eq:maxwell}), and take the divergence of both sides:
\begin{eqnarray}
\boldsymbol{\nabla} \times \vec B - {\partial \vec E \over \partial t} = \vec J &\Longrightarrow & \boldsymbol{\nabla} \cdot \boldsymbol{\nabla}\times \vec B - {\partial \over \partial t} \boldsymbol{\nabla} \cdot \vec E = \boldsymbol{\nabla} \cdot \vec J \nolabel \\
&\Longrightarrow & \boldsymbol{\nabla} \cdot \vec J +{\partial \rho \over \partial t}  = 0 \label{eq:chargeconservation1}
\end{eqnarray}
where we used the third of (\ref{eq:maxwell}) and the fact that the divergence of a curl always vanishes to get the last line.  This equation is the conservation equation which says that charge must be conserved.  

\subsubsection{The Gauge Potentials}

In any introductory or graduate course on electromagnetism one usually introduces the scalar and vector potentials $\phi$ and $\vec A$, and then the electric and magnetic fields are defined by
\begin{eqnarray}
\vec E &=& - {\partial \vec A \over \partial t} - \boldsymbol{\nabla} \phi \nolabel \\
\vec B &=& \boldsymbol{\nabla} \times \vec A \label{eq:eandbfieldsdefinedintermsofscalarandvectorpotentials}
\end{eqnarray}
In the introductory courses this is motivated for two reasons: it often makes solving Maxwell's equations easier, and it essentially removes half of them.  Consider the right hand side of the first homogeneous equation, written in terms of the scalar and vector potential:
\begin{eqnarray}
\boldsymbol{\nabla} \cdot \vec B &=& \boldsymbol{\nabla}\cdot (\boldsymbol{\nabla} \times \vec A) \nolabel \\
&\equiv & 0
\end{eqnarray}
The final equality is an identity - the divergence of a curl vanishes automatically.  So, with $\vec B$ defined as in (\ref{eq:eandbfieldsdefinedintermsofscalarandvectorpotentials}) the first of Maxwell's theories are automatically satisfies - no more work need be done.  

Similarly for the second homogeneous equation:
\begin{eqnarray}
\boldsymbol{\nabla} \times \vec E +{\partial \vec B \over \partial t} &=& \boldsymbol{\nabla} \times \bigg(- {\partial \vec A \over \partial t} - \boldsymbol{\nabla} \phi\bigg) +  {\partial \over \partial t} (\boldsymbol{\nabla} \times \vec A) \nolabel \\
&=& -{\partial \over \partial t} (\boldsymbol{\nabla}\times \vec A) - \boldsymbol{\nabla}\times \boldsymbol{\nabla}\phi + {\partial \over \partial t}(\boldsymbol{\nabla}\times \vec A) \nolabel \\
&\equiv& 0
\end{eqnarray}
where the first and last term cancel and the second vanishes because the curl of a gradient automatically vanishes.  

So when $\vec E$ and $\vec B$ are written in terms of $\phi$ and $\vec A$, the homogeneous Maxwell equations are automatically satisfied.  A course in electrodynamics then typically consists of learning about a billion tricks to solve the inhomogeneous equations for $\phi$ and $\vec A$.  

\subsubsection{Gauge Transformations}

Another important idea is that of a gauge transformation.  For a given $\phi$ and $\vec A$, we can make the transformations
\begin{eqnarray}
\phi \longrightarrow \phi' &=& \phi - {\partial \chi \over \partial t} \nolabel \\
\vec A \longrightarrow \vec A' &=& \vec A + \boldsymbol{\nabla} \chi \label{eq:firsttimeIdefineagaugetransformation}
\end{eqnarray}
where $\chi$ is an arbitrary (scalar) function.  Plugging these new values into (\ref{eq:eandbfieldsdefinedintermsofscalarandvectorpotentials}) gives
\begin{eqnarray}
\vec E' &=& -{\partial \vec A' \over \partial t} - \boldsymbol{\nabla} \phi' \nolabel \\
&=& -{\partial \over \partial t} (\vec A + \boldsymbol{\nabla} \chi) - \boldsymbol{\nabla}(\phi - {\partial \chi \over \partial t}) \nolabel \\
&=& -{\partial \vec A \over \partial t} - {\partial \over \partial t} \boldsymbol{\nabla} \chi - \boldsymbol{\nabla} \phi + {\partial \over \partial t} \boldsymbol{\nabla} \chi \nolabel \\
&=& -{\partial \vec A \over \partial t} - \boldsymbol{\nabla} \phi \nolabel \\
&=& \vec E \label{eq:invarianceofEundergauge}
\end{eqnarray}
and
\begin{eqnarray}
\vec B' &=& \boldsymbol{\nabla} \times \vec A' \nolabel \\
&=& \boldsymbol{\nabla} \times (\vec A + \boldsymbol{\nabla}\chi) \nolabel \\
&=& \boldsymbol{\nabla} \times \vec A + \boldsymbol{\nabla} \times \boldsymbol{\nabla} \chi \nolabel \\
&=& \boldsymbol{\nabla}\times \vec A \nolabel \\
&=& \vec B \label{eq:invarianceofBundergauge}
\end{eqnarray}

So in trying to solve Maxwell's equations for $\phi$ and $\vec A$, we are free to make any gauge transformation we want to simplify finding solutions.  

\subsubsection{Application of Stokes Theorem}

Finally, consider the first inhomogeneous Maxwell equation:
\begin{eqnarray}
\boldsymbol{\nabla} \cdot \vec E = \rho \label{eq:gausslawdifferentialform}
\end{eqnarray}
We can use this to find the exact expression for the electric field $\rho$.  Consider a very small charged particle with charge density $\rho$ and total charge $Q$ at the center of a three dimensional ball $B^3$.  We can integrate both sides of (\ref{eq:gausslawdifferentialform}) through the ball to get the total charge:
\begin{eqnarray}
\int_{B^3}dV \boldsymbol{\nabla} \cdot \vec E &=& \int_{B^3} dV \rho
\end{eqnarray}
where $dV$ is the three dimensional volume element.  The right hand side will clearly be the total charge $Q$.  On the left hand side we can use the well-known divergence theorem:\footnote{As learned in any introductory calculus series.}
\begin{eqnarray}
\int_{B^3} dV \boldsymbol{\nabla} \cdot \vec E &=& \int_{\partial_3 B^3} d\vec A\cdot  \vec E 
\end{eqnarray}
where $\partial_3 B^3$ is the boundary operator (taking $B^3$ to the boundary, or $S^2$), (cf section \ref{sec:boundaryoperator}) and $d\vec A$ is the area element on the surface of the $S^2$.  We then assume that $\vec E$ depends only on the radial direction from the charged particle, and therefore it is constant across $S^2$.  So,
\begin{eqnarray}
\int_{\partial_3B^3} d\vec A \cdot \vec E &=& \vec E \cdot \int_{\partial_3B^3} d\vec A \nolabel \\
&=& E(r)(4\pi r^2)
\end{eqnarray}
where we have assumed that $\vec E$ has only a radial direction (and hence depends only on $r$), and therefore the angle between $\vec E$ and $d\vec A$ is always $0$ (and $\cos(0) = 1$), so we get simply the magnitude of $\vec E$ (the direction is understood to be in the $r$ direction).  So, finally,
\begin{eqnarray}
& & E(r) (4\pi r^2) = \int_{B^3} dV \rho = Q \nolabel \\
&\Longrightarrow & E(r) = { Q \over 4\pi r^2}
\end{eqnarray}
which is the standard expression for an electric field $E(r)$ a distance $r$ from a point charge $Q$.  

A similar calculation using the first of (\ref{eq:maxwell}) indicates that there are no such things as magnetic charges\footnote{in classical electrodynamics.}  How they might be incorporated into electrodynamics is not clear in the formalism we are currently presenting.  We will see later how such a thing may be built in naturally.  

Furthermore, how might electromagnetism be generalized to an arbitrary manifold?  We have been assuming in this section that these fields exist on $\mathbb{R}^3$ with time as a parameter.  This is an extremely limiting constraint, but finding the correct generalization is not straightforward at this point.  

\subsubsection{Electrodynamic Lagrangian}

As with any physical theory, we would ultimately like to put it in the form of an action.\footnote{Admittedly this approach isn't typically useful in classical electrodynamics.  However doing so at this stage will make generalizing the action easier later.}  Starting with the energy of an electromagnetic field\footnote{This expression can be found in any introductory text on E\&M.}
\begin{eqnarray}
V = {1 \over 2} (|\vec E|^2 - |\vec B|^2)
\end{eqnarray}
We can write this out in terms of $\phi$ and $\vec A$ getting
\begin{eqnarray}
{1 \over 2} \bigg[ \bigg({\partial \vec A\over \partial t}\bigg)^2 + (\boldsymbol{\nabla}\phi)^2 + |\boldsymbol{\nabla}\cdot \vec A|^2 + 2 {\partial \vec A \over \partial t} \cdot \boldsymbol{\nabla} \phi - \boldsymbol{\nabla}^2 (\vec A)^2\bigg]
\end{eqnarray}
Then, including a source term $-\rho \phi + \vec J \cdot \vec A$, you can show (with a great deal of tedium) that variation of the Lagrangian
\begin{eqnarray}
\mathcal{L} = {1 \over 2} \bigg[ \bigg({\partial \vec A\over \partial t}\bigg)^2 + (\boldsymbol{\nabla}\phi)^2 + |\boldsymbol{\nabla}\cdot \vec A|^2 + 2 {\partial \vec A \over \partial t} \cdot \boldsymbol{\nabla} \phi - \boldsymbol{\nabla}^2 (\vec A)^2\bigg] + \rho \phi - \vec J \cdot \vec A \label{eq:firstEandMLagrangian}
\end{eqnarray}
will give Maxwell's equations.  

However, this approach is not particularly interesting (or useful), and we therefore won't pursue it.  We mention it now merely to provide something to compare our later (more useful) results to.

\subsection{Electrodynamics and Tensors}

\subsubsection{Maxwell's Equations}

One major difference between classical mechanics and classical electrodynamics is that classical mechanics is not an automatically (special) relativistic theory - electrodynamics is.  Special relativity works not with $3$ dimensional space with time as a parameter, but with $4$ dimensional \it spacetime\rm.  It begins with the Minkowski metric\footnote{While the Minkowski metric generally introduced in an introductory course on special relativity, it should have a much deeper and richer connotation after the previous chapter of these notes!},\footnote{We will work with the convention that Greek indices run over all four spacetime components whereas Latin indices run only over spatial components.}
\begin{eqnarray}
\eta_{\mu \nu} \dot = 
\begin{pmatrix}
-1 & 0 & 0 & 0 \\
0 & 1 & 0 & 0 \\
0 & 0 & 1 & 0 \\
0 & 0 & 0 & 1
\end{pmatrix}
\end{eqnarray}
We then replace the spatial vector potential $\vec A$ and scalar potential $\phi$ with the single \it spacetime \rm $4$-vector \rm Potential\rm
\begin{eqnarray}
A^{\mu} \dot = 
\begin{pmatrix}
\phi \\ \vec A
\end{pmatrix}= 
\begin{pmatrix}
\phi \\ A^1 \\ A^2 \\A^3
\end{pmatrix}
\end{eqnarray}
(note that $A_{\mu}$ will be $A_{\mu} \dot = (-\phi, \vec A)^T$, where the minus sign in the first component of $\eta_{\mu \nu}$).  It is then straightforward, though very tedious, to show that
\begin{eqnarray}
F_{\mu \nu} = \partial_{\mu} A_{\nu} - \partial_{\nu} A_{\mu} \label{eq:definitionofeandmfieldstrength}
\end{eqnarray}
where 
\begin{eqnarray}
F_{\mu \nu} \dot = 
\begin{pmatrix}
0 & -E_x & -E_y & -E_z \\
E_x & 0 & B_z & -B_y \\
E_y & -B_z & 0 & B_x \\
E_z & B_y & -B_x & 0
\end{pmatrix} \label{eq:firstfieldstrengthtensor}
\end{eqnarray}
is the \bf Field Strength Tensor\rm.  We can recover the original fields easily:
\begin{eqnarray}
E_i &=& F_{0i} \nolabel \\
&=& \partial_0A_i - \partial_iA_0 \nolabel \\
&=& - {\partial \over \partial t} A_i - \partial_i \phi \nolabel \\
\Longrightarrow \vec E &=& -\boldsymbol{\nabla} \phi - {\partial \vec A \over \partial t} \label{eq:efromphiandvecA}
\end{eqnarray}
which is what we had above\footnote{The reason for the minus sign on the time derivative in the third line is that $\partial^{\mu} \dot = (\partial^0 , \partial^1, \partial^2, \partial^3)^T$, and lowering the index puts a minus sign on the time derivative.} in (\ref{eq:eandbfieldsdefinedintermsofscalarandvectorpotentials}).  You can convince yourself that
\begin{eqnarray}
B_k = F_{ij}\epsilon^{ijk} \label{eq:bfromfepsi}
\end{eqnarray}

Now, consider the quantity $\partial_{\mu} F_{\nu \lambda}$:
\begin{eqnarray}
\partial_{\mu}F_{\nu \lambda} &=& \partial_{\mu}(\partial_{\nu}A_{\lambda} - \partial_{\lambda}A_{\nu}) \nolabel \\
&=& \partial_{\mu} \partial_{\nu}A_{\lambda} - \partial_{\mu}\partial_{\lambda}A_{\nu}
\end{eqnarray}
If we symmetrize this, we find
\begin{eqnarray}
\partial_{(\mu} F_{\nu \lambda)} &=& \partial_{\mu}F_{\nu \lambda} + \partial_{\nu}F_{\lambda \mu} + \partial_{\lambda}F_{\mu \nu} \nolabel \\
&=& \partial_{\mu} \partial_{\nu}A_{\lambda} - \partial_{\mu}\partial_{\lambda}A_{\nu}  + \partial_{\nu} \partial_{\lambda}A_{\mu} - \partial_{\nu} \partial_{\mu} A_{\lambda}  + \partial_{\lambda}\partial_{\mu}A_{\nu} - \partial_{\lambda}\partial_{\nu}A_{\mu} \nolabel \\
&\equiv& 0 \label{eq:tensorformulationdemonstrationofhomogeneousmaxequ}
\end{eqnarray}
So what does this equation mean?  You can try this for various components yourself to see that this gives
\begin{eqnarray}
& & {\partial B_z \over \partial t} + {\partial E_y \over \partial x} - {\partial E_x \over \partial y} = 0 \nolabel \\
& & {\partial B_y \over \partial t} + {\partial E_x \over \partial z} - {\partial E_z \over \partial x} = 0 \nolabel \\
& & {\partial B_z \over \partial t} + {\partial E_z \over \partial y} - {\partial E_y \over \partial z} = 0 \nolabel \\
& & {\partial B_x \over \partial x} + {\partial B_y \over \partial y} + {\partial B_z \over \partial z} = 0
\end{eqnarray}
which are exactly the homogeneous Maxwell equations from (\ref{eq:maxwell}).  So once again, half of Maxwell's equations are satisfied automatically if we define things in terms of the potential $A_{\mu}$.  

Writing the inhomogeneous Maxwell equations requires the introduction of the source $4$-vector 
\begin{eqnarray}
J^{\mu} \dot = (\rho, \vec J)^T
\end{eqnarray}
Then, considering the quantity $\partial_{\nu} F^{\mu \nu}$, we have\footnote{The field strength $F^{\mu \nu}$ with upper indices is the field strength $F_{\mu \nu}$ with lowered indices raised by the metric $\eta_{\mu \nu}$, and therefore there will be minus signs to keep track of.}
\begin{eqnarray}
\partial_{\nu} F^{\mu \nu} &=& \partial_{\nu}\partial^{\mu} A^{\nu} - \partial_{\nu}\partial^{\nu} A^{\mu} \nolabel \\
&=& \partial_{\nu} \partial^{\mu} A^{\nu} - \partial^2 A^{\mu}
\end{eqnarray}
Then, looking at the quantity $\partial_{\nu} F^{\mu \nu} = J^{\mu}$, plugging in different index values gives
\begin{eqnarray}
{\partial E_x \over \partial x} + {\partial E_y \over \partial y} + {\partial E_z \over \partial x} &=& \rho \nolabel \\
{\partial B_y \over \partial x} - {\partial B_x \over \partial y} - {\partial E_z \over \partial t} &=& J_z \nolabel \\
{\partial B_x \over \partial z} - {\partial B_z \over \partial x} - {\partial E_y \over \partial t} &=& J_y \nolabel \\
{\partial B_z \over \partial y} - {\partial B_y \over \partial z} - {\partial E_x \over \partial t} &=& J_x \label{eq:tensorformulationofMaxwellsourceequations}
\end{eqnarray}
which are exactly the inhomogeneous Maxwell equations from (\ref{eq:maxwell}).  

So, in terms of the field strength $F_{\mu\nu}$, Maxwell's equations can be rewritten as
\begin{eqnarray}
\partial_{(\mu} F_{\nu \lambda)} &=& 0 \nolabel \\
\partial_{\nu} F^{\mu \nu} &=& J^{\mu} \label{eq:maxwell2}
\end{eqnarray}
However, because of the definition of the field strength (\ref{eq:definitionofeandmfieldstrength}), the first of these is merely an identity and contributes nothing to the physics.  In this sense, electrodynamics in terms of the $\vec E$ and $\vec B$ is a theory with four fundamental \it vector \rm equations, whereas electrodynamics in terms of $F_{\mu \nu}$ (which is written in terms of $A_{\mu}$) is a theory with one fundamental \it tensor \rm equation.

\subsubsection{The Continuity Equation}

Once again, we can start with the inhomogeneous Maxwell equation and take the derivative of both sides:
\begin{eqnarray}
\partial_{\mu} \partial_{\nu} F^{\mu \nu} = \partial_{\mu} J^{\mu}
\end{eqnarray}
The right hand side is a symmetric sum over an antisymmetric tensor and therefore vanishes, leaving
\begin{eqnarray}
\partial_{\mu} J^{\mu} = 0
\end{eqnarray}
Writing this out:
\begin{eqnarray}
\partial_{\mu}J^{\mu} &=& -\partial_0 \rho + \partial_1J_1 + \partial_2 J_2 +\partial_3J_3 \nolabel \\
&=& \boldsymbol{\nabla} \cdot \vec J - {\partial \rho \over \partial t} \nolabel \\
&=& 0 \label{eq:chargeconservationrelatvisticversion}
\end{eqnarray}
You can see that it is simply the charge conservation equation (\ref{eq:chargeconservation1}).\footnote{Don't worry about the sign difference - it is simply part of how the values are defined.}  

\subsubsection{Gauge Transformations}

Next we discuss gauge transformations.  Again, $A_{\mu}$ doesn't uniquely define a given $F_{\mu \nu}$.  Any transformation of the form
\begin{eqnarray}
A_{\mu} \longrightarrow A'_{\mu} = A_{\mu} + \partial_{\mu} \chi
\end{eqnarray}
leaves $F_{\mu\nu}$ unchanged:
\begin{eqnarray}
F'_{\mu \nu} &=& \partial_{\mu} A'_{\nu} - \partial_{\nu} A'_{\mu} \nolabel \\
&=& \partial_{\mu} (A_{\nu} + \partial_{\nu} \chi) - \partial_{\nu}(A_{\mu} + \partial_{\mu} \chi) \nolabel \\
&=& \partial_{\mu} A_{\nu} - \partial_{\nu} A_{\mu} + \partial_{\mu}\partial_{\nu} \chi - \partial_{\mu} \partial_{\nu} \chi \nolabel \\
&=& \partial_{\mu} A_{\nu} - \partial_{\nu} A_{\mu} \nolabel \\
&=& F_{\mu \nu}
\end{eqnarray}

Notice that the form of Maxwell's equations in (\ref{eq:maxwell}) are specific to three plus one dimensional space.  However, the form in (\ref{eq:maxwell2}) are not.  For this reason we take (\ref{eq:maxwell2}) to \it define \rm Maxwell theory in general (regardless of the manifold).  

With that said we are able to glean some insights into Maxwell theory in dimensions not equal to $3+1$.  For example, consider $2+1$ dimensions.  $F_{\mu \nu}$ will now be a $3\times 3$ matrix.  We can find it by simply truncating (\ref{eq:firstfieldstrengthtensor}) to
\begin{eqnarray}
F_{\mu \nu} \dot = 
\begin{pmatrix}
0 & -E_x & -E_y \\
E_x & 0 & B \\
E_y & -B & 0
\end{pmatrix}
\end{eqnarray}
where we removed the subscripts on $B$ because it is now a scalar.  So, interestingly, the magnetic field is not a vector in $2+1$ dimensions - it is a scalar.  The electric field, however, is still a vector.  

In $4+1$ dimensions, it is not clear how to generalize the magnetic part.  But what is clear is that the electric field will \it still \rm be a vector, and the magnetic field will be a $4\times 4$ antisymmetric tensor - not a vector (or a scalar).  

Continuing with this, in $n+1$ spacetime dimensions, $\vec E$ will \it always \rm be an $n$-dimensional vector, but the magnetic field will always be an $n\times n$ antisymmetric tensor.\footnote{An antisymmetric $2\times 2$ tensor (as in the $2+1$ dimensional case) has only one degree of freedom which is why it was a scalar.  An antisymmetric $3\times 3$ tensor (as in our familiar $3+1$ dimensions) has three degrees of freedom, which is why we see it as a vector.  In general an antisymmetric $n\times n$ tensor has ${1 \over 2} n(n-1)$ independent components.}

The formalism of this section is much neater than the formalism of the previous section.  The most important aspect is that this formalism is manifestly relativistic.  Another nice aspect is that it it valid in arbitrary dimensions.  

However we still have the shortcoming that it isn't always obvious how to do electrodynamics on an arbitrary manifold.  The next section will provide a powerful formalism that allows for electrodynamics to be done on any manifold.  

\subsubsection{Electrodynamic Lagrangian}

Writing out a Lagrangian in this manifestly relativistic formulation is much easier.  As discussed in \cite{Firstpaper}, the Lagrangian that produces Maxwell's equations (and is equivalent to (\ref{eq:firstEandMLagrangian}))
\begin{eqnarray}
\mathcal{L} = -{1 \over 4} F_{\mu \nu} F^{\mu \nu} - J^{\mu} A_{\mu} \label{eq:maxwelllagrangianintensorformulationadsfasdfaeqgr545}
\end{eqnarray}
will give Maxwell's equations (\ref{eq:maxwell2})\footnote{Again, see the first chapter \cite{Firstpaper} for how this is done.}

We don't need to pursue this any more - we merely wanted to illustrate the vast improvement this notation offers over the vector notation (as in (\ref{eq:firstEandMLagrangian})).  

\subsection{Electrodynamics and Forms}

\subsubsection{Maxwell's Equations}

The frequent reference to antisymmetric tensors in the previous section should imply that we can use differential forms for what we are doing.  

To begin with, consider the $1$-form 
\begin{eqnarray}
\bf A\it_{(1)} = A_{\mu} dx^{\mu}
\end{eqnarray}
Consider taking the exterior derivative (cf section \ref{sec:exteriorderivatives}) of this:
\begin{eqnarray}
d \bf A\it_{(1)} &=& \partial_{\mu} A_{\nu} dx^{\mu} \wedge dx^{\nu} \nolabel \\
&=& {1 \over 2} (\partial_{\mu}A_{\nu} - \partial_{\nu}A_{\mu}) dx^{\mu} \wedge dx^{\nu} \nolabel \\
&\equiv& {1 \over 2} F_{\mu \nu} dx^{\mu} \wedge dx^{\nu} \label{eq:firststatementofexteriorderivofAtogetFmunu}
\end{eqnarray}
where we have defined the two form
\begin{eqnarray}
\bf F\it_{(2)} = F_{\mu \nu} dx^{\mu} \wedge dx^{\nu}
\end{eqnarray}
which is, of course, an antisymmetric tensor.  Clearly this identical to the previous two sections, where $\bf A\it_{(1)}$ is the potential and $\bf F\it_{(2)}$ is the field strength, with 
\begin{eqnarray}
F_{\mu\nu} = \partial_{\mu} A_{\nu} - \partial_{\nu} A_{\mu}
\end{eqnarray}
What is so nice about this is that the form of the exterior derivative exactly captures the form of the field strength from before (compare (\ref{eq:firststatementofexteriorderivofAtogetFmunu}) to (\ref{eq:eandbfieldsdefinedintermsofscalarandvectorpotentials}) and (\ref{eq:definitionofeandmfieldstrength}))

Identifying $\bf F\it_{(2)}$ as the field strength, we can make the identification\footnote{We are switching from $t,x,y,z$ to $x^0,x^1,x^2,x^3$ for generality.}
\begin{eqnarray}
\bf F\it_{(2)} &=& E_1dx^1 \wedge dx^0 + E_2 dx^2 \wedge dx^0 + E_3 dx^3 \wedge dx^0 \nolabel \\
& &  -B_1 dx^2 \wedge dx^3 - B_2 dx^3 \wedge dx^1 - B_3 dx^1 \wedge dx^2 
\end{eqnarray}
You are encouraged to show that this reproduces the results in (\ref{eq:efromphiandvecA}) and (\ref{eq:bfromfepsi}).  

Next, because of the nilpotency of the exterior derivative (\ref{eq:nilpotentencyofd}) we have
\begin{eqnarray}
d \bf F\it_{(2)} = d (d\bf A\it_{(1)}) = d^2 \bf A\it_{(1)} \equiv 0
\end{eqnarray}
Writing this out in components gives
\begin{eqnarray}
d \bf F\it_{(2)} &=& \partial_{\mu} F_{\nu \lambda} dx^{\mu}\wedge dx^{\nu} \wedge dx^{\lambda}  \nolabel \\
&=& {1 \over 3} (\partial_{\mu} F_{\nu \lambda} + \partial_{\nu} F_{\lambda \mu} + \partial_{\lambda} F_{\mu \nu}) dx^{\mu} \wedge dx^{\nu} \wedge dx^{\lambda} \nolabel \\
&=& 0
\end{eqnarray}
Comparing this to (\ref{eq:tensorformulationdemonstrationofhomogeneousmaxequ}) we see that they are identical.  So the nilpotency of $d$ has given us the homogeneous Maxwell equations!  You can write this out in component form to show that they reproduce (\ref{eq:maxwell}).  

We now want to write the inhomogeneous Maxwell equations.  There is, however, a difficulty - they will involve derivatives of $\bf F\it_{(2)}$, but the only derivative we have to work with with differential forms is the exterior derivative, and because of its nilpotency we can't use it on $\bf F\it_{(2)}$.  The solution to this problem lies in the Hodge star (cf section \ref{sec:hodgestar}).  

Let's begin by taking the Hodge star of $\bf F\it_{(2)}$:
\begin{eqnarray}
\star \bf F\it_{(2)} &=& \star F_{\mu \nu} dx^{\mu} \wedge dx^{\nu} \nolabel \\
&=& {1 \over 2!(4-2)!} F_{\mu \nu} \epsilon^{\mu \nu}_{\lambda \sigma} dx^{\lambda} \wedge dx^{\sigma} \nolabel \\
&=& {1 \over 4} F_{\mu \nu} \eta_{\lambda\tau} \eta_{\sigma\upsilon} \epsilon^{\mu \nu \tau \upsilon} dx^{\lambda} \wedge dx^{\sigma} \nolabel \\
&=& E_1 dx^2 \wedge dx^3 + E_2 dx^3 \wedge dx^1 + E_3 dx^1 \wedge dx^2 \nolabel \\
& & -B_1 dx^1 \wedge dx^0 - B_2 dx^2 \wedge dx^0 - B_3 dx^3 \wedge dx^0 \label{eq:starFinformnotation}
\end{eqnarray}

Then we can take the exterior derivative of this:
\begin{eqnarray}
d\star \bf F\it_{(2)} &=& \partial_1 E_1 dx^1 \wedge dx^2 \wedge dx^3 + \partial_0 E_1 dx^0 \wedge dx^2 \wedge dx^3 \nolabel \\
& & + \partial_2 E_2 dx^2 \wedge dx^3 \wedge dx^1 + \partial_0 E_2 dx^0 \wedge dx^3 \wedge dx^1 \nolabel \\
& & + \partial_3E_3 dx^3 \wedge dx^1 \wedge dx^2 + \partial_0 E_3 dx^0 \wedge dx^1 \wedge dx^2 \nolabel \\
& & -\partial_2 B_1 dx^2 \wedge dx^1 \wedge dx^0 - \partial_3 B_1 dx^3 \wedge dx^1 \wedge dx^0 \nolabel \\
& & -\partial_1B_2 dx^1 \wedge dx^2 \wedge dx^0 - \partial_3 B_2 dx^3 \wedge dx^2 \wedge dx^0 \nolabel \\
& & - \partial_1 B_3 dx^1 \wedge dx^3 \wedge dx^0 - \partial_2 B_3 dx^2 \wedge dx^3 \wedge dx^0 \nolabel \\
&=& (\partial_1E_1 + \partial_2 E_2 + \partial_3E_3) dx^1 \wedge dx^2 \wedge dx^3 \nolabel \\
& & + (\partial_0 E_3 + \partial_2 B_1 - \partial_1B_2) dx^0 \wedge dx^1 \wedge dx^2 \nolabel \\
& & +(\partial_0E_2+ \partial_1B_3 - \partial_3B_1)dx^0 \wedge dx^3 \wedge dx^1 \nolabel \\
& & +(\partial_0 E_1 + \partial_3B_2 - \partial_2B_3)dx^0 \wedge dx^2 \wedge dx^3
\end{eqnarray}
Comparing each of these lines to (\ref{eq:tensorformulationofMaxwellsourceequations}) we see that if we define
\begin{eqnarray}
\bf J\it_{(1)} = \rho dx^0 + J_1 dx^1 + J_2dx^2+J_3dx^3
\end{eqnarray}
then
\begin{eqnarray}
\star \bf J\it_{(1)} &=& \rho dx^1 \wedge dx^2 \wedge dx^3 + J_1 dx^0 \wedge dx^2 \wedge dx^3 \nolabel \\
& & J_2 dx^0 \wedge dx^3 \wedge dx^1 + J_3 dx^0 \wedge dx^1 \wedge dx^2
\end{eqnarray}
Then, finally, the equation
\begin{eqnarray}
d\star \bf F\it_{(2)} = \star \bf J\it_{(1)} \label{eq:formversionofinhomogmaxwell}
\end{eqnarray}
represents the inhomogeneous Maxwell equations.  

\subsubsection{The Continuity Equation}

Taking the exterior derivative of both sides of (\ref{eq:formversionofinhomogmaxwell}) we have
\begin{eqnarray}
d^2 \star \bf F\it_{(2)} = d \star \bf J\it_{(1)}
\end{eqnarray}
The left hand side vanishes due to nilpotency, and we can write out the right hand side:
\begin{eqnarray}
d\star \bf J\it_{(1)} &=& (\partial_0\rho - \partial_1J_1 - \partial_2J_2- \partial_3J_3) dx^0 \wedge dx^1 \wedge dx^2 \wedge dx^3 \nolabel \\
&=& 0 \label{eq:conservationofchargeinformformalism}
\end{eqnarray}
Comparing this to (\ref{eq:chargeconservationrelatvisticversion}) we see that, as expected, charge is conserved.  

\subsubsection{Gauge Transformations}

Next we define a gauge transformation.  This is made particularly easy by the nilpotency of the exterior derivative.  We leave $\bf F\it_{(2)}$ unchanged by the addition of the exterior derivative of any arbitrary exact $0$-form $\chi$:
\begin{eqnarray}
\bf A\it_{(1)}\longrightarrow \bf A\it'_{(1)} = \bf A\it_{(1)} + d\chi_{(0)} \label{eq:gaugetransformationinformsforaharonov}
\end{eqnarray}
Then, almost trivially,
\begin{eqnarray}
\bf F\it'_{(2)} &=& d\bf A\it'_{(1)} \nolabel \\
&=& d ( \bf A\it_{(1)} + d\chi_{(0)}) \nolabel \\
&=& d\bf A\it_{(1)} + d^2 \chi_{(0)} \nolabel \\
&=& d\bf A\it_{(1)} \nolabel \\
&=& \bf F\it_{(2)} \label{eq:aharonovgaugeleavesFunchangedinformlanguage}
\end{eqnarray}

\subsubsection{Application of Stokes Theorem}

Finally, consider again the derivation of the strength of the electric field a distance $r$ from a point charge $Q$ from section \ref{sec:electrodynamicsandvectors}.  We commented there that the version of Gauss' Law we were using isn't generalized to arbitrary dimension in an obvious way.  Now that we are working in terms of forms this generalization becomes extremely easy.  Specifically we can define the charge in a region $B^n$ (an $n$-dimensional Ball) as the integral over the source term $\bf J\it_{(3)}$:
\begin{eqnarray}
Q &=& \int_{B^n}\star  \bf J\it_{(1)}  \nolabel \\
&=& \int_{B^n} d \star \bf F\it_{(2)}  \nolabel \\
&=& \int_{\partial B^n} \star \bf F\it_{(2)}
\end{eqnarray}
where we have used Stokes theorem (\ref{eq:stokestheoremcompactform}).  

Then, noting from (\ref{eq:starFinformnotation}) that the electric part of the two form $ \star\rm F\it_{(2)}$ is the purely spatial $2$-form (across $dx^i \wedge dx^j$ where $i,j \neq 0$), which we interpret to mean the spatial $n-1$ sphere $S^{n-1}$ which forms the boundary of $B^n$, or $\partial B^n$.  Then, by the analogous argument as in section \ref{sec:electrodynamicsandvectors} we can assume that at a constant radius $r$ around the center of $B^n$ the electric field is equal, and therefore we can bring it out, leaving only an integral across the boundary $\partial B^n$.  So, finally,
\begin{eqnarray}
Q = E(r) \int_{\partial B^n} dV
\end{eqnarray}
where the integral simply gives the volume of the sphere $S^{n-1}$ of radius $r$, which is\footnote{Again, you can find this in any text on basic geometry.}
\begin{eqnarray}
V(S^{n-1}) = {2 \pi^{{n\over2}} \over \Gamma({n\over2})}r^{n-1}
\end{eqnarray}
So,
\begin{eqnarray}
E(r) = {\Gamma({n \over 2}) \over 2\pi^{{n\over 2}}} {Q \over r^{n-1}}
\end{eqnarray}

\subsubsection{Electrodynamic Lagrangian}

Finally we can write out the Lagrangian in terms of forms.  The Lagrangian is defined as the integrand of the action, which is (in a relativistic formalism) an integral over all spacetime.  Therefore we should seek a well-defined volume form.  In light of the discussion surrounding (\ref{eq:toreferaesdkfhasdfto1}) and (\ref{eq:toreferaesdkfhasdfto2}), we guess that the kinetic term
\begin{eqnarray}
\mathcal{L}_{Kin} \propto \bf F\it_{(2)} \wedge \star \bf F\it_{(2)}
\end{eqnarray}
which we know will be a well defined volume form, is the best guess.  Writing this out in Minkowski space,
\begin{eqnarray}
\bf F\it_{(2)} \wedge \star \bf F\it_{(2)} &=&{1 \over 4} (F_{\mu \nu} dx^{\mu} \wedge dx^{\nu}) \wedge (F_{\alpha\beta} \epsilon^{\alpha\beta}_{\gamma\delta} dx^{\gamma}\wedge dx^{\delta}) \nolabel \\
&=& {1 \over 4} F_{\mu \nu} F_{\alpha \beta} \epsilon^{\alpha \beta}_{\gamma\delta} ( dx^{\mu} \wedge dx^{\nu} \wedge dx^{\gamma} \wedge dx^{\delta}) \nolabel \\
&=& {1 \over 4} F_{\mu \nu} F_{\alpha\beta} \epsilon_{\zeta \omega \gamma\delta} g^{\zeta \alpha} g^{\omega \beta} \epsilon^{\mu \nu \gamma \delta} (dx^0 \wedge dx^1 \wedge dx^2 \wedge dx^3) \nolabel \\
&=& {1 \over 4} F_{\mu \nu} F^{\zeta \omega} \epsilon_{\zeta \omega \gamma \delta} \epsilon^{\mu \nu \gamma \delta} (dx^0 \wedge dx^1 \wedge dx^2 \wedge dx^3) \nolabel \\
&=& {1 \over 2} F_{\mu \nu} F^{\mu \nu}  (dx^0 \wedge dx^1 \wedge dx^2 \wedge dx^3)
\end{eqnarray}
Comparing this with (\ref{eq:maxwelllagrangianintensorformulationadsfasdfaeqgr545}) we see that the appropriate kinetic term for the electromagnetic field is
\begin{eqnarray}
\mathcal{L}_{Kin} = -{1 \over 2} \bf F\it_{(2)} \wedge \star \bf F\it_{(2)}
\end{eqnarray}

For the source term, we know from (\ref{eq:maxwelllagrangianintensorformulationadsfasdfaeqgr545}) that the source $\bf J\it_{(1)}$ must somehow couple to the potential form $\bf A\it_{(1)}$.  These are both $1$-forms, so we again guess that the source term should be of the form
\begin{eqnarray}
\mathcal{L}_{Source} \propto \bf A\it_{(1)} \wedge \star \bf J\it_{(1)}
\end{eqnarray} 
(recall from (\ref{eq:omegawedgestaralplhaequalsalphawedgestaromega}) that $\bf A\it_{(1)} \wedge \star \bf J\it_{(1)} = \bf J\it_{(1)} \wedge \star \bf A\it_{(1)}$).  Writing this out gives
\begin{eqnarray}
\bf A\it_{(1)} \wedge \star \bf J\it_{(1)} &=& {1 \over 6} (A_{\mu} dx^{\mu})\wedge (J_{\nu} \epsilon^{\nu}_{\alpha\beta\gamma} dx^{\alpha} \wedge dx^{\beta} \wedge dx^{\gamma}) \nolabel \\
&=& {1 \over 6} A_{\mu} J_{\nu} \epsilon^{\nu}_{\alpha\beta\gamma} (dx^{\mu} \wedge dx^{\alpha} \wedge dx^{\beta} \wedge dx^{\gamma}) \nolabel \\
&=& {1 \over 6} A_{\mu} J_{\nu} \epsilon_{\delta \alpha\beta\gamma} g^{\delta\nu} \epsilon^{\mu \alpha \beta\gamma} dx^0 \wedge dx^1 \wedge dx^2 \wedge dx^3) \nolabel \\
&=& A_{\mu}J^{\mu} dx^0 \wedge dx^1 \wedge dx^2 \wedge dx^3
\end{eqnarray}

So the appropriate Lagrangian to agree with (\ref{eq:maxwelllagrangianintensorformulationadsfasdfaeqgr545}) will be
\begin{eqnarray}
\mathcal{L} = - {1 \over 2} \bf F\it_{(2)} \wedge \star \bf F\it_{(2)} - \bf A\it_{(1)} \wedge \star \bf J\it_{(2)}
\end{eqnarray}

\subsection{Summary of Electrodynamic Formalisms}

We can summarize everything we have done in the past three sections as follow:

\begin{center}
\begin{tabular}{|c||c|c|c|}
\hline
 & Classical & Tensor & Forms \\
\hline 
Maxwell & \begin{tabular}{c}$\boldsymbol{\nabla}\cdot \vec B = 0$ \\ $\boldsymbol{\nabla} \cdot \vec E = \rho$ \\$\boldsymbol{\nabla} \times \vec E + {\partial \vec B \over \partial t} = 0$ \\ $\boldsymbol{\nabla} \times \vec B - {\partial \vec E \over \partial t} = \vec J$\end{tabular} & \begin{tabular}{c}$\partial_{(\mu}F_{\nu\lambda)} = 0$ \\ \\$\partial_{\nu}F^{\mu \nu} = J^{\mu}$\end{tabular} & $d\star \bf F\it = \bf J\it$\\
\hline 
Potential & \begin{tabular}{c}$\vec E = -{\partial \vec A \over \partial t} - \boldsymbol{\nabla} \phi$ \\$\vec B = \boldsymbol{\nabla} \times \vec A$\end{tabular} & $F_{\mu\nu} = \partial_{\mu}A_{\nu} - \partial_{\nu}A_{\mu}$ & $\bf F\it = d\bf A\it$ \\
\hline 
Gauge Trans. & \begin{tabular}{c}$\phi \rightarrow \phi - {\partial \chi \over \partial t}$ \\$\vec A \rightarrow \vec A + \boldsymbol{\nabla} \chi$\end{tabular} & $A_{\mu} \rightarrow A_{\mu} + \partial_{\mu} \chi$ & $\bf A\it \rightarrow \bf A\it + d \bf \chi\rm$\\
\hline 
Stokes & $\int_{B^3} dV \boldsymbol{\nabla} \cdot \vec E = \int_{\partial_3 B^3} d\vec A \cdot \vec E$ & Same & $\langle \partial c|\omega\rangle = \langle c|d \omega\rangle$ \\
\hline 
Lagrangian &very ugly (\ref{eq:firstEandMLagrangian}) & $-{1 \over 4} F_{\mu\nu}F^{\mu\nu} - J_{\mu}A^{\mu}$ & $-{1 \over 2} \bf F\it \wedge \star \bf F\it - \bf A\it \wedge \star \bf J\it$ \\
\hline
\end{tabular}
\end{center}

The advantages in the use of forms is very obvious here, not only in its efficiency, but also in the fact that they provide an index free formulation that is completely independent of the manifold these fields exist on.  This will give us the ability to consider, in the next few sections, how electromagnetic fields might behave on a topologically non-trivial manifold.  

\section{The Aharonov-Bohm Effect}

The Aharonov-Bohm affect provides a very nice illustration of marriage of mathematics and physics.  Our presentation of it will be  brief, but it will provide a nice illustration of how the formalism developed thus far can be used.  

The classical Lorentz force law on a particle of electric charge $q$ is
\begin{eqnarray}
\vec F = q (\vec E + \vec v \times \vec B) \label{eq:lorentzforcelawforEandB}
\end{eqnarray}
Therefore, if both the $\vec E$ and $\vec B$ fields are both zero, the particle experiences no electromagnetic force.  

Furthermore, as mentioned above, we can write the electric and magnetic fields in terms of the vector and scalar potentials $\vec A$ and $\phi$ as (cf equation (\ref{eq:eandbfieldsdefinedintermsofscalarandvectorpotentials}))
\begin{eqnarray}
\vec E &=& -{\partial \vec A \over \partial t} - \boldsymbol{\nabla}\phi \nolabel \\
\vec B &=& \boldsymbol{\nabla} \times \vec A
\end{eqnarray}
And, as we mentioned above, the vector and scalar potentials are not unique.  Under a gauge transformation (cf (\ref{eq:firsttimeIdefineagaugetransformation}))
\begin{eqnarray}
\phi &\longrightarrow& \phi' = \phi - {\partial \chi \over \partial t} \nolabel \\
\vec A &\longrightarrow& \vec A' = \vec A + \boldsymbol{\nabla}\chi
\end{eqnarray}
the electric and magnetic fields remain unchanged (cf (\ref{eq:invarianceofEundergauge}) and (\ref{eq:invarianceofBundergauge})):
\begin{eqnarray}
\vec E &\longrightarrow& \vec E \nolabel \\
\vec B &\longrightarrow& \vec B
\end{eqnarray}
So, classically speaking, because it is only the $\vec E$ and $\vec B$ fields that show up in (\ref{eq:lorentzforcelawforEandB}), it is only $\vec E$ and $\vec B$ that are physically measurable.  The potentials, which don't uniquely define $\vec E$ and $\vec B$, are a nice mathematical tool to help simplify solving Maxwell's equations, but they aren't considered to be truly physical fields.  

Furthermore, we know from \cite{Firstpaper} that the gauge symmetry in classical electromagnetism is a remnant of the deeper $U(1)$ symmetry in the field Lagrangians.  The gauge transformation for fields $\psi$ and $\bar \psi$ is
\begin{eqnarray}
\psi \longrightarrow e^{i\alpha}\psi \nolabel \\
\bar \psi \longrightarrow \bar \psi e^{-i\alpha}
\end{eqnarray}
where $e^{-i\alpha}$ is an arbitrary element of $U(1)$.  And because in quantum mechanics the only physically measurable quantity is the amplitude,
\begin{eqnarray}
\bar \psi \psi \longrightarrow \bar \psi e^{-i\alpha}e^{i\alpha} \psi = \bar \psi \psi \label{eq:aharonovpsisquaredisequalwhentransformed}
\end{eqnarray}
it appears that the gauge potentials are indeed not physically measurable quantities - they are nothing more than phases.  

However, this turns out not to be the case.  Consider some region of $n$ dimensional space $\mathcal{U}$.  We first take $\mathcal{U}$ to be topologically trivial (say, $\mathcal{U} = \mathbb{R}^n$).  Because $\mathcal{U}$ is topologically trivial, all cohomology groups (greater than $0$) will be trivial:
\begin{eqnarray}
H^m(\mathcal{U}) = 0 \qquad \forall \;\; m>0 \label{eq:arahonovHmzllzero}
\end{eqnarray}
Recall from sections \ref{sec:cohomologygroups}-\ref{sec:meaningofnthcohomologygroup} that the definition of the $m^{th}$ cohomology group $H^m(\mathcal{U})$ is
\begin{eqnarray}
H^m(\mathcal{U}) = Z^m(\mathcal{U})/B^m(\mathcal{U}) \label{eq:defofcohomologyinbohmsection}
\end{eqnarray}
where $Z^m(\mathcal{U})$ is the set of all closed $m$ forms on $\mathcal{U}$ ($m$ forms $\omega_m$ satisfying $d\omega_m = 0$) and $B^m(\mathcal{U})$ is the set of all exact $m$ forms on $\mathcal{U}$ ($m$ forms $\omega_m$ that can be written as the exterior derivative of an $m-1$ form $\alpha_{m-1}$.  So $\omega_m = d \alpha_{m-1}$).  

Recall our discussion from section \ref{sec:meaningofnthcohomologygroup}.  There we discussed on a factor group $G/H$ creates an equivalence class of elements of $G$ where two elements of $G$ are said to be equivalent if their difference is in $H$.  So, in the definition of a cohomology group (\ref{eq:defofcohomologyinbohmsection}), two closed forms are considered "equivalent" if their difference is an exact form.  In other words, two closed $m$ forms $\beta_m$ and $\gamma_m$ are equivalent, or "cohomologous", if their difference is exact:
\begin{eqnarray}
\beta_m - \gamma_m = d\alpha_{m-1}
\end{eqnarray}
for some $m-1$ form $\alpha{m-1}$.  
We can restate this as saying that $\beta_m$ and $\gamma_m$ are cohomologous if one can be expressed in terms of the other plus an exact form:
\begin{eqnarray}
\beta_m = \gamma_m + d\alpha_{m-1} \label{eq:aharonovbetamequalsgammamplusdbetamminus1}
\end{eqnarray}

But, as we've said, the cohomology groups for a topologically trivial space like $\mathcal{U} = \mathbb{R}^n$ are trivial (equal to zero, the identity element, cf equation (\ref{eq:arahonovHmzllzero})), we know that \it all \rm closed forms must be exact:
\begin{eqnarray}
Z^m(\mathcal{U}) = B^m(\mathcal{U})
\end{eqnarray}
and so for \it any \rm two $m$ forms $\beta_m$ and $\gamma_m$ we can find a $\beta_{m-1}$ to satisfy equation (\ref{eq:aharonovbetamequalsgammamplusdbetamminus1}).  In other words, all $m$ forms are cohomologous to each other in that any two $m$ forms can be related by the addition of an exact form.  

Now let's translate this into electromagnetism in $\mathcal{U}$.  Consider some potential $1$ form $\bf A\it$ and field strength $2$ form
\begin{eqnarray}
\bf F\it = d \bf A\it
\end{eqnarray}
defined in $\mathcal{U}$.  As discussed above, we can perform a gauge transformation to $\bf A\it$ (equation (\ref{eq:gaugetransformationinformsforaharonov})) that leaves the field strength unchanged (equation (\ref{eq:aharonovgaugeleavesFunchangedinformlanguage})):
\begin{eqnarray}
\bf A\it \longrightarrow \bf A\it' = \bf A\it + d \chi \qquad \longrightarrow \qquad \bf F\it \longrightarrow \bf F\it' = \bf F\it
\end{eqnarray}
And because the space these fields live in ($\mathcal{U}$) is topologically trivial and therefore has trivial non-zero cohomology groups, we can consider any two potentials we want ($\bf A\it$ and $\bf A\it'$) and they will be cohomologous - there will be a gauge transformation that relates them to each other.  

So, consider letting $\bf F\it = 0$ in all of $\mathcal{U}$.  This means that
\begin{eqnarray}
\bf F\it = d\bf A\it = 0
\end{eqnarray}
We know that it must be a constant (there is no solution to $d \bf A\it = 0$ on \it all \rm of $\mathbb{R}^n$ other than $\bf A\it =$ a constant.  And if $\bf A\it $ is a constant on all of $\mathcal{U}$, the field at every point will have the same phase, and therefore equation (\ref{eq:aharonovpsisquaredisequalwhentransformed}) indicates that there is nothing physically measurable about the potential.  

Now, however, let's let $\mathcal{U}$ be topologically non-trivial.  This may mean that, for example,
\begin{eqnarray}
H^1(\mathcal{U}) \neq 0
\end{eqnarray}
This means that when the field strength is zero ($\bf F\it = 0$), there may be a solution to $d\bf A\it = 0$ for $\bf A\it$ that is not a constant (cf example starting on page \pageref{pagewherewediscusshowcohomologyworks}).  And for a non-zero gauge field that is not a constant, the phase may be different at different points.  

An experiment has been set up involving sending particles through a double slit as usual, except behind the double slit, between the slit and the screen, there is an extremely thin solenoid (a wire parallel to the slits that runs through the area the electrons are moving through) with a non-zero field \it inside \rm the solenoid, but a zero field outside it.  The electrons are shielded from the interior of the solenoid.  
\begin{center}
\includegraphics[scale=.7]{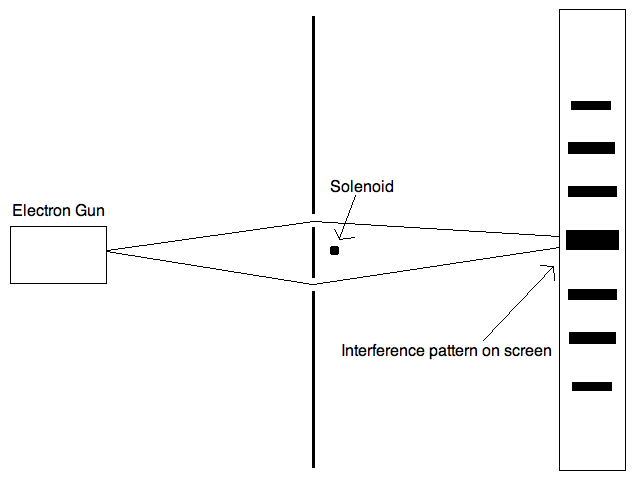}
\end{center}
When the field inside the solenoid is zero, the field \it and \rm the potential outside the solenoid is zero.  However, when the field inside the solenoid is non-zero, the field outside the solenoid is still zero but the potential outside the solenoid is non-zero.  The effect of this is that, with the solenoid field non-zero, the space becomes topologically non-trivial.  It is $\mathbb{R}^3 - \mathbb{R}$ where the $\mathbb{R}$ is the one dimensional space the solenoid is running through.\footnote{The necessity that the solenoid be extremely thin is the major challenge in carrying out this experiment.}  So, because the topology is non-trivial, we can have non-constant solutions to $d\bf A\it = 0$, and therefore the phase of the field on each side of the solenoid need not be the same.  So (\ref{eq:aharonovpsisquaredisequalwhentransformed}) would now be
\begin{eqnarray}
\bar \psi\psi  \longrightarrow \bar \psi e^{-i\alpha} e^{i\alpha'} \psi = e^{i(\alpha' - \alpha)} \bar \psi \psi
\end{eqnarray}

This experiment has been carried out, and it has been observed that the interference pattern (which depends on the phases of the electrons) does in fact shift when the field inside the solenoid is turned on.  This affect is called the \bf Aharonov-Bohm Effect\rm.  The point of the effect is that it demonstrates that at the quantum level it is the \it potentials\rm, not the \it field strenghs\rm, that are truly fundamental.  We will see this point made in much greater force in the next paper in this series, but for now what we have said will suffice.  

There is vastly more we could say about the Aharonov-Bohm effect and related ideas, and anyone reading who is more familiar with the topic will likely be annoyed that we haven't said more.  However, once we have the necessary machinery (in the next paper in this series), we will return to this topic in much greater detail.  

\section{References and Further Reading}

The formalisms for electrodynamics can be found in almost any of the differential geometry texts listed in chapter \ref{sec:chapwithmet}.  We primarily used \cite{griffiths} and \cite{appel}, though \cite{griffiths} and \cite{schwinger} are also very good.  

For further reading on the various applications of geometry and topology to electromagnetism, there are very interesting ideas in \cite{alvarez}, \cite{henneaux}, \cite{ranada}.  

\chapter{Gravity}
\label{sec:gravity}

It may seem strange to include a chapter on general relativity, Einstein's (classical) theory of gravitation, in a series on (quantum) particle physics.  But as we mentioned earlier there are several reasons it is an appropriate topic at this point.  First of all we want to understand particle physics as comprehensively as possible, and a quantum theory of gravity is a necessary part of that.  When we eventually get to the leading candidate for quantum gravity, string theory, it will be necessary to understand the theory we are trying to quantize.  The second major reason to discuss general relativity now is that it is a very good way to illustrate the geometrical ideas we have discussed so far in this paper.  Providing some intuition with what we have done will be helpful when we discuss much more complicated geometrical and topological ideas later.  

With that said, we will approach general relativity as follows.  We will begin by discussing Newtonian mechanics in the more geometrical language we have been using.  Then we will discuss special relativity\footnote{We assume you have some familiarity with special relativity.} in geometrical language.  Then we will discuss what lead Einstein to make the leap from special relativity to general relativity, and what the consequences his theory are.  Therefore, we begin with Newton.  

\section{Newtonian Geometry}

Simply put, classical, or Newtonian mechanics is geometrically very boring - everything takes place on $\mathbb{R}^3$ with Euclidian metric $g_{ij} = \delta_{ij}$.  Because we have a metric, we can define the "distance" between two arbitrary points (cf section \ref{sec:TheMetricandthePythagoreanTheorem}):
\begin{eqnarray}
ds^2 = dx^2+dy^2+dz^2 \label{eq:distancebetweentwopointsnewtoniangeometrysection}
\end{eqnarray}
We take this distance to be an intrinsic part of the space, and we know that it should not change if different coordinates are chosen.  For example an observer using Cartesian coordinates and an observer using spherical coordinates should not report different distances between two points.  Nor should two observers standing in different locations.  

Furthermore, notice that the Newtonian geometry looks only at spatial dimensions.  Time is treated as a separate parameter, not on the same geometrical footing as space.  This is why Newtonian physics takes place in \it three \rm dimensions, not four.  

Also, because 
\begin{eqnarray}
{\partial \delta_{ij} \over \partial x^k} = 0
\end{eqnarray}
all of the connection coefficients vanish:
\begin{eqnarray}
\Gamma^i_{jk} = 0
\end{eqnarray}
and therefore (obviously) this space is flat.  This also means that we can write out the geodesic equations very easily (equation (\ref{eq:geodesicdifferentialequation})):
\begin{eqnarray}
{d^2 x^i \over d \tau^2} = 0
\end{eqnarray}
which is solved to give straight lines:
\begin{eqnarray}
x^i (\tau) = C^i_1 \tau + C^i_2 \label{eq:straightline}
\end{eqnarray}
where $C^i_j$ are constants of integration depending on boundary conditions.  Equation (\ref{eq:straightline}) is clearly the equation of a line and therefore we have proven the well known fact that "the shortest distance between two points is a straight line" (in $\mathbb{R}^3$ with $g_{ij} = \delta_{ij}$).  

So at any point we can define a set of geodesics through that point (any solution to (\ref{eq:straightline}) that passes through the point), and because we know that there is no curvature, we can take the vector corresponding to any geodesic and parallel transport it to any other point in the space.  Because the curvature vanishes the vector at the new point will be independent of the path taken and therefore there is a \it unique \rm vector at each point.  So imagine starting with a set of $3$ basis vectors  at some point.  We can parallel transport each of these three vectors to every other point in $\mathbb{R}^3$, thus defining a rigid coordinate system at every point.  

Furthermore, the fact that all particles move in geodesics through $\mathbb{R}^3$, which are straight lines, is exactly equivalent to Newton's first law - objects in motion stay in motion unless acted on by an external force.  This is entirely contained by the statement "in the absence of external forces objects follow geodesics".  

This notion of a "rigid coordinate system at every point" is at the heart of Newton's view of space.  This is also what motivates the classical Galilean transformations between two observers moving relative to each other:
\begin{eqnarray}
t' &=& t \nolabel \\
x' &=& x - vt \nolabel \\
y' &=& y \nolabel \\
z' &=& z  \label{eq:galileantransformations}
\end{eqnarray}
where the observer in the primed frame is moving in the $x$ direction with velocity $v$ and their origins coincide at time $t=0$.  Both observers have identical notions of time and clearly the distance between two points $(x_1,y_1,z_1)$ and $(x_2,y_2,z_2)$ will be the same:
\begin{eqnarray}
\sqrt{(x'_1-x'_2)^2+(y'_1-y'_2)^2+(z'_1-z'_2)^2} &=& \sqrt{((x_1-vt)-(x_2-vt))^2+(y_1-y_2)^2+(z_1-z_2)^2} \nolabel \\
&=& \sqrt{(x_1-x_2)^2+(y_1-y_2)^2+(z_1-z_2)^2}
\end{eqnarray}

In summary, the geometry of Newton ($\mathbb{R}^3$ with metric $\delta_{ij}$) has profound implications - the fact that it is flat allows for a unique set of "parallel" vectors across the entire space, creating a fixed, rigid, unmoving, un-dynamical background, as well as define the straight line geodesics all particles travel along.  As a result, all observers will measure the same time no matter what they are doing, and the distance they measure between two points will always be the same regardless of what they are doing.  

We now look to see what geometric changes Einstein made in jumping from Newtonian mechanics to special relativity.  

\section{Minkowski Geometry}

In many ways, special relativity is the clearest demonstration of the true brilliance of Einstein.  General relativity may demand more mathematical acrobatics, but it was the geometrical leap made in moving from Newton's view of space and time to the single continuous notion of \it spacetime \rm that was truly demonstrated Einstein's genius.  The fundamental idea of this paradigm shift is that, rather than all physics being carried out on $\mathbb{R}^3$ with metric $\delta_{ij}$, physics occurs on $\mathbb{R}^4$, but rather than with the Euclidian metric $\delta_{ij}$, instead with the \it Lorentz \rm metric
\begin{eqnarray}
\eta_{ij} =
\begin{pmatrix}
-1 & 0 & 0 & 0 \\
0 & 1 & 0 & 0 \\
0 & 0 & 1 & 0 \\
0 & 0 & 0 & 1
\end{pmatrix}
\end{eqnarray}
The most obvious result of this is that time is now treated as a dimension in the geometry of the universe, rather than merely a parameter.  

Another equally important consequence is that the simple Galilean transformations (\ref{eq:galileantransformations}) must be modified.  The reason for this is that it is no longer merely the spatial distance between two locations (\ref{eq:distancebetweentwopointsnewtoniangeometrysection}) that is preserved, but rather the \it spacetime \rm distance between two \it events\rm:\footnote{We will be taking $c=1$ in everything that follows.}
\begin{eqnarray}
ds^2 = \eta_{ij}dx^i dx^j = -dt^2+dx^2+dy^2+dz^2
\end{eqnarray}
This space is called Minkowsi space.  The set of transformations which leave this interval unchanged is no longer (\ref{eq:galileantransformations}) but the Lorentz transformaions:
\begin{eqnarray}
t' &=& {t-vx \over \sqrt{1-v^2}} \nolabel \\
x' &=& {x-vt \over \sqrt{1-v^2}} \nolabel \\
y' &=& y \nolabel \\
z' &=& z
\end{eqnarray}
where again the observer in the primed frame is moving in the $x$-direction with velocity $v$ and their origins coincide at $t=0$.  

Notice that with the metric defined as above, it is possible for a vector to have positive, negative, or even zero norm.  For example, a unit vector in the time direction but no component in a spatial component: $\bf v\it = (1,0,0,0)^T$, will have norm
\begin{eqnarray}
\eta_{ij} v^iv^j = -1
\end{eqnarray}
On the other hand, a vector with no time component but a spatial component, for example $\bf v\it = (0,1,0,0)^T$, will have norm
\begin{eqnarray}
\eta_{ij} v^i v^j = 1
\end{eqnarray}
And finally, consider the vector $\bf v\it = (1,1,0,0)^T$.  It will have norm
\begin{eqnarray}
\eta_{ij} v^iv^j = -1+1 = 0
\end{eqnarray}
We call vectors with negative norm \bf timelike\rm, vectors with positive norm \bf spacelike\rm, and vectors with zero norm \bf lightlike \rm or \bf null\rm.  

The geodesic equations are again very simple:
\begin{eqnarray}
{d^2 x^i \over d \tau^2} = 0
\end{eqnarray}
which has solutions
\begin{eqnarray}
x^i(\tau) = C^i_1\tau+C^i_2
\end{eqnarray}
So the geodesics are again straight lines in $\mathbb{R}^4$.  

At any point $p \in \mathbb{R}^4$ we can choose a family of geodesics passing through $p$, each of which will correspond to a vector that is either timelike, spacelike, or lightlike.  Notice that if an observer is on a timelike geodesic, there does not exist a continuous Lorentz transformation that will transform him to a spacelike or timelike geodesic.  in fact, it is not possible to use a continuous Lorentz transformation to transform between \it any \rm two of the three types of vectors.  We therefore consider the manifold $\mathbb{R}^4$ at any given point and vector to be divided into three distinct sections - the future light cone, the past light cone, and "elsewhere".\footnote{Any introductory text on special relativity will contain numerous helpful illustrations of all of this.  We omit them both for brevity and because we are assuming some knowledge of special relativity.}  This leads to the well known result that all material particles travel along geodesics that are everywhere timelike (negative norm) and all massless particles travel on null, or lightlike (zero norm) geodesics.  

For an arbitrary timelike geodesic $x^i(\tau) = C^i_1\tau+C^i_2$ we can find the velocity vector
\begin{eqnarray}
v^i = {d x^i \over d\tau} = C^i_1
\end{eqnarray}
with norm
\begin{eqnarray}
\eta_{ij} v^iv^j = \eta_{ij} {dx^i\over d \tau}{dx^j\over d \tau} = k
\end{eqnarray}
where by assumption $k<0$ (because we have assumed that the geodesic is timelike).  If we introduce a new parameter $\lambda = \lambda(\tau)$ such that
\begin{eqnarray}
\bigg( {d\lambda \over d\tau}\bigg)^2 = k
\end{eqnarray}
then 
\begin{eqnarray}
\eta_{ij} {dx^i \over d\tau}{dx^j \over d\tau} = \eta_{ij}\bigg({d\lambda \over d\tau}\bigg)^2 {dx^i\over d\lambda}{dx^j\over d\lambda} = k \Longrightarrow \eta_{ij}  {dx^i\over d\lambda}{dx^j\over d\lambda} = 1
\end{eqnarray}
So, when parameterized with $\lambda$ this timelike vector has unit length.  We call this parameter the \bf proper time \rm parameter.  Physically it is the time as measured by an observer in the same inertial frame as the observer on the geodesic $x^i$.  

Furthermore, we have
\begin{eqnarray}
{d \eta_{ij} \over dx^k} = 0
\end{eqnarray}
and therefore the connection vanishes, as does the curvature.  This means that for a given vector at a given point, we can parallel transport this vector to any other point in $\mathbb{R}^4$ resulting in a unique vector at the new point.  For a given observer in some location with some velocity, the set of all observers in any location with parallel velocity are considered to be in the same inertial frame.  These inertial frames are, as you are likely familiar, a central theme of special relativity.  The consequence of these ideas is that there is no such thing as an absolute notion of simultaneity, no absolute notion of "motionless", etc.  Two things can be said to be simultaneous, or motionless, relative to a given inertial frame, but nothing more.  

Of course, the construction of special relativity has the shortcoming of being unable to discuss non-inertial, or accelerating frames.  This is the domain of general relativity.  Before discussing general relativity, however, we discuss one final aspect of special relativity - the energy momentum tensor.  Because almost no book adequately explains the meaning of the energy momentum tensor, we will build it up in steps, starting with the non-relativistic "stress tensor" and generalizing from there.  

\section{The Energy-Momentum Tensor}

\subsection{The Stress Tensor}

Consider some infinitesimal volume of material in three dimensional space which for simplicity we will take to be a rectangle.  We take the volume of this rectangle to be $dV$.  Because the boundary of the rectangle (a rectangular box) is homeomorphic to $S^2$, we can take each face of the rectangle to have an oriented area element:
\begin{center}
\includegraphics[scale=.7]{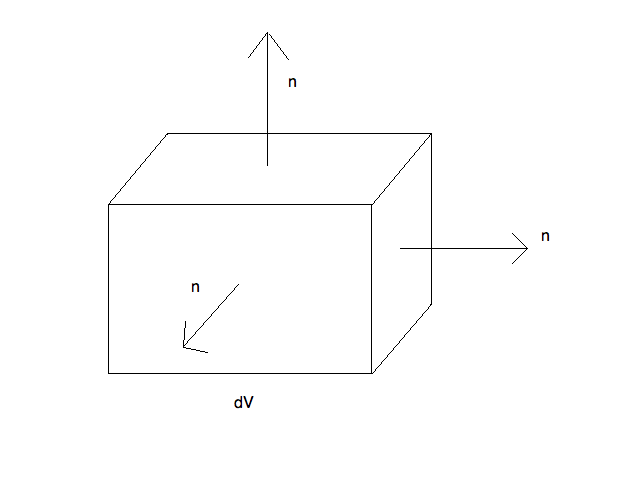}
\end{center}
We will choose the convention illustrated where the positive oriented area unit vector $\bf n\it $ is outward from the volume.  

There are two types of forces that can act on $dV$ - volume forces and surface forces.  Volume forces are forces that act on the entire body at once and are therefore proportional to $dV$.  Gravity and electromagnetism are examples of volume forces.  The gravitational force on this infinitesimal volume will be
\begin{eqnarray}
d\bf F\it_{grav} = \rho_m \bf g\it dV
\end{eqnarray}
where $\rho_m$ is the volume mass density, and the electric force is
\begin{eqnarray}
d\bf F\it_{elec} = \rho_e \bf E\it dV
\end{eqnarray}
where $\rho_e$ is the volume charge density.  Volume forces generally very familiar and we will therefore not focus on them. 

Surface forces are forces that act on a particular surface of the volume.  The most common type of surface force is pressure, or the force per unit area.  The force due to pressure is then proportional to the area element on which the pressure is applied:
\begin{eqnarray}
dF_{pressure} \propto dA \label{eq:fproptodA}
\end{eqnarray}

Furthermore, there are two types of surface forces - pressure and shear.  Pressure is a force with components perpendicular to the surface:
\begin{center}
\includegraphics[scale=.7]{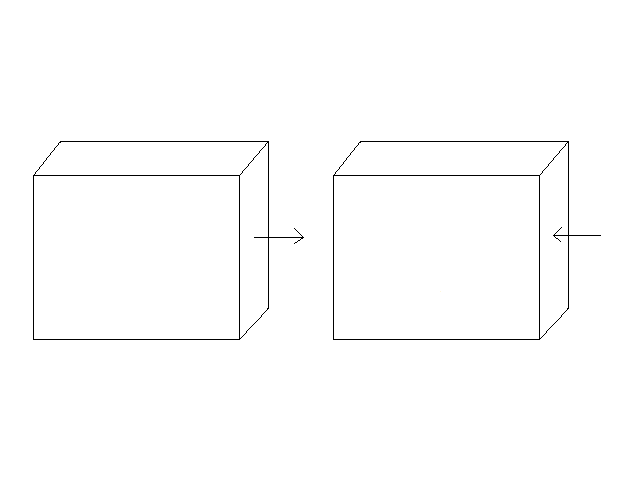}
\end{center}
(we include both positive and negative pressure for clarity).  Shear force is a force with components perpendicular to the surface:
\begin{center}
\includegraphics[scale=.7]{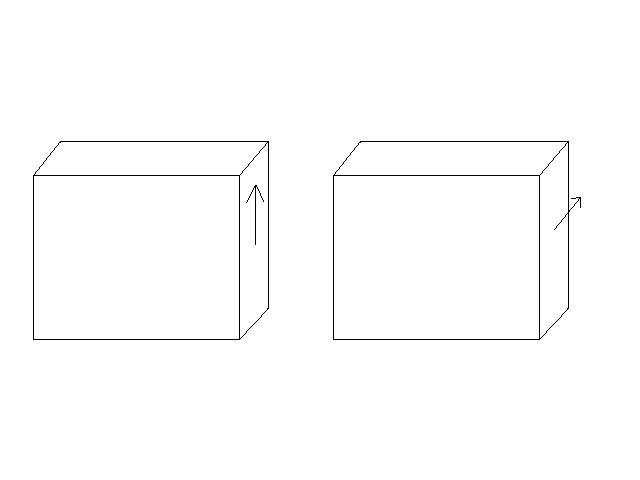}
\end{center}
Obviously in three dimensions the pressure force spans only a single dimension (parallel to $\bf n\it$) whereas the shear force spans a two dimensional space.  

Any surface force is proportional to the area on which the force acts.  We therefore consider the ratio of the force to the area, which we call the \bf Stress\rm.  There are two types of stresses - the pressure stress and the shearing stress.  The pressure stress is
\begin{eqnarray}
stress = {F_P \over A} = P \label{eq:stressisfpovera}
\end{eqnarray}
where $P$ is the pressure and $F_P$ is the pressure force, and the shearing stress is
\begin{eqnarray}
stress = {F_S \over A} \label{eq:stressisfsovera}
\end{eqnarray}
where $F_S$ is the shearing force.  We can use these relationships to discuss the force per area, or in other words the force on a small area $dA$ with unit normal $\bf n\it$.  We will use the notation 
\begin{eqnarray}
d\bf A\it = \bf n\it dA
\end{eqnarray}
Our goal is to write the force on the infinitesimal piece $d\bf F\it$ as a function of a given volume element, so we are looking for a function $d\bf F\it = d\bf F\it (d \bf A\it)$.  

To discover the form of this function, consider three area elements forming a triangle at any arbitrary point in $dV$:
\begin{center}
\includegraphics[scale=.8]{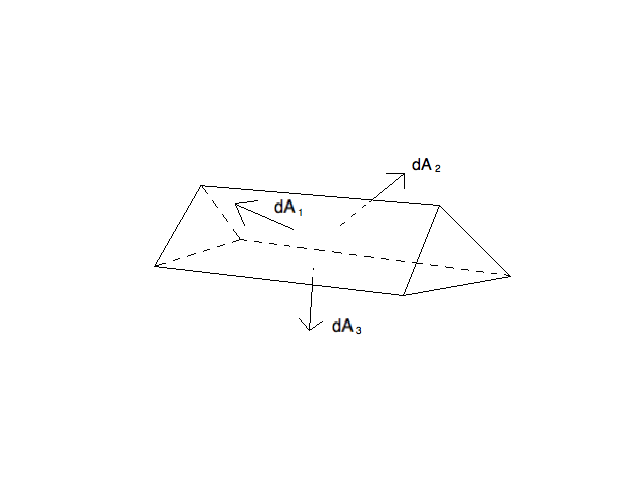}
\end{center}
From the side view, this is
\begin{center}
\includegraphics[scale=.8]{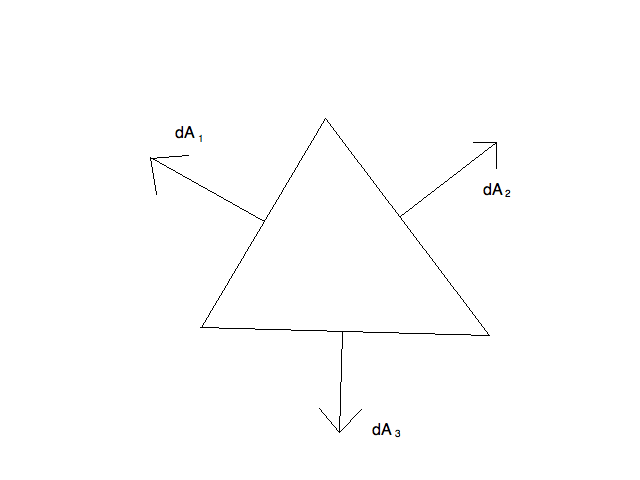}
\end{center}
Because this is a closed triangle, we can immediately write down the relation
\begin{eqnarray}
d\bf A\it_1 + d\bf A\it_2 + d\bf A\it_3 = 0 \label{eq:dAsumequals0}
\end{eqnarray}

Also, from both the general form of (\ref{eq:stressisfpovera}) and (\ref{eq:stressisfsovera}), we can deduce that $d\bf F\it(d\bf A\it)$ is proportional to the area element $dA$ (cf (\ref{eq:fproptodA})) and therefore
\begin{eqnarray}
d\bf F\it (\alpha d\bf A\it) = \alpha d\bf F\it(d\bf A\it) \label{eq:falphadaeqalphafda}
\end{eqnarray}
where $\alpha$ is some arbitrary constant.  

Finally, consider the net force on this triangle from Newton's second law:
\begin{eqnarray}
d\bf F\it(d\bf A\it_1) + d\bf F\it(d\bf A\it_2) + d\bf F\it(d\bf A\it_3) = d(m \bf a\it)
\end{eqnarray}
Now consider an identical triangle but scale it by a factor of $\lambda$ in all three dimensions.  The three surface terms on the left are proportional to area and therefore they will scale by a factor of $\lambda^2$.  However the mass term is proportional to $dV$ and will therefore scale by $\lambda^3$.  So, the net force on this scaled triangle will be
\begin{eqnarray}
\lambda^2(d\bf F\it(d\bf A\it_1) + d\bf F\it(d\bf A\it_2) + d\bf F\it(d\bf A\it_3)) = \lambda^3d(m \bf a\it)
\end{eqnarray}
which simplifies to
\begin{eqnarray}
d\bf F\it(d\bf A\it_1) +d \bf F\it(d\bf A\it_2) +d \bf F\it(d\bf A\it_3) = \lambda d(m \bf a\it)
\end{eqnarray}
But because this must be valid for \it any \rm value of $\lambda$, we must have $\bf a\it$ = 0.  So, we have the remarkable property that
\begin{eqnarray}
d\bf F\it(d\bf A\it_1) + d\bf F\it(d\bf A\it_2) + d\bf F\it(d\bf A\it_3)  = 0 \label{eq:fdAsumequals0}
\end{eqnarray}

By Newton's third law we know that
\begin{eqnarray}
d\bf F\it (-d\bf A\it) = -d \bf F\it (d\bf A\it) \label{eq:fmdaeqmdfa}
\end{eqnarray}
And combining (\ref{eq:dAsumequals0}), (\ref{eq:fdAsumequals0}), and (\ref{eq:fmdaeqmdfa}), we have
\begin{eqnarray}
d\bf F\it( d\bf A\it_1 + d \bf A\it_2) &=&d \bf F\it(-d\bf A\it_3) \nolabel \\
&=& -d\bf F\it( d\bf A\it_3) \nolabel \\
&=& d\bf F\it(d\bf A\it_1) + d\bf F\it(d\bf A\it_2)
\end{eqnarray}
Combining this with (\ref{eq:falphadaeqalphafda}) we have that $d\bf F\it(d\bf A\it)$ is a linear function in $d\bf A\it$.  We therefore write the components of $d\bf F\it$ in terms of the components of $d\bf A\it$ as follows:
\begin{eqnarray}
dF^i(d\bf A\it) = \sigma^i_j dA^j = \sigma^i_j n^j dA \label{eq:stresstensor1}
\end{eqnarray}
(where $j$ is obviously summed) where $\sigma^i_j$ is some $3\times 3$ matrix.  The matrix $\sigma^i_j$ is called the \bf Stress Tensor \rm of the volume.  

The meaning of the stress tensor is straightforward from the definition (\ref{eq:stresstensor1}).  Namely, at any point in $dV$, the force applied to an area element $\bf n\it dA$ is given by the stress tensor via (\ref{eq:stresstensor1}).  So, if you know the stress tensor for a medium, you can choose any arbitrary area element anywhere in the medium, and the product of the stress tensor and the area element unit vector will be the force acting on that area element.  

To get a better feel for this tensor, consider a unit area element pointing in the positive $x$ direction:
\begin{eqnarray}
d\bf A\it = \bf n\it^1 dA
\end{eqnarray}
This makes (\ref{eq:stresstensor1}) very simple - the $x$ component of the force on this area that is normal to $\bf n\it^1$ is:
\begin{eqnarray}
dF^1 (d\bf A\it) = \sigma^1_1 dA
\end{eqnarray}
Or inverting this, 
\begin{eqnarray}
\sigma^1_1 = {dF^1 \over dA}
\end{eqnarray}
which is the pressure stress in the $x$ direction.

Similarly, we can see that $\sigma^i_i$ for any $i$ is the pressure stress in the $i^{th}$ direction.  

Consider again $d\bf A\it = \bf n\it^1 dA$, but now consider the $y$ component of the force on this area that is normal to $\bf n\it^1$ is:
\begin{eqnarray}
dF^2 (d\bf A\it) = \sigma^2_1 dA
\end{eqnarray}
Inverting this,
\begin{eqnarray}
\sigma^2_1 = {dF^2 \over dA}
\end{eqnarray}
which is the shear stress in the $y$ direction on the area element.  

More generally, $\sigma^i_j$ is the shear stress in the $i^{th}$ direction on an area element normal to the $j^{th}$ direction.  

As a final comment, consider the net torque of some volume of radius $r$ around, say, the $z$ axis.  The net torque will be
\begin{eqnarray}
\tau^3 = (\sigma^2_1 - \sigma^1_2) r dA = {d L^3 \over dt}
\end{eqnarray}
where $L^3$ is the $z^{th}$ component of angular momentum.  If we once again re-scale the volume, $r$ and $dA$ together get a factor of $\lambda^3$, whereas $L^3$ gets a factor of $\lambda^4$.  So,
\begin{eqnarray}
(\sigma^2_1 - \sigma^1_2) rdA = \lambda {dL^3 \over dt}
\end{eqnarray}
and therefore $\sigma^2_1 - \sigma^1_2 = 0$, or
\begin{eqnarray}
\sigma^2_1 = \sigma^1_2
\end{eqnarray}
Continuing this with the other components, we see that $\sigma^i_j$ must be a symmetric tensor, giving it a total of 6 independent components.  

\subsection{The Energy-Momentum Tensor}

We now want to generalize from the purely spatial stress tensor, which was a $3\times 3$ tensor, to a more general form that incorporates time as well as space.  To see how to do this, consider the meaning of the $\sigma^i_j$ component of the stress tensor:
\begin{eqnarray}
\sigma^i_j = {dF^i \over dA}
\end{eqnarray}
or
\begin{eqnarray}
dF^i = \sigma^i_j n^j dA \label{eq:dfequalssigmanda}
\end{eqnarray}
where $dA$ is the area element normal to the the $\bf n\it$, the unit vector in the $j^{th}$ direction.  We want to generalize from the purely spatial stress tensor to a \it spacetime \rm tensor, which means that we want to incorporate time.  

This means that if we start with a \it spatial \rm unit vector $\bf n\it$, the infinitesimal element normal to it will no longer merely be an area element, but will also have a time component.  In other words, it will no longer be $d\bf A\it = \bf n\it dA$, but now will instead be
\begin{eqnarray}
d\bf S\it = \bf n\it dS = \bf n\it dAdt
\end{eqnarray}
where $d\bf S\it$ is a spacetime interval.  

Newton's laws make this generalization very easy.  Recall that the definition of force is
\begin{eqnarray}
\bf F\it = {d \bf p\it \over dt}
\end{eqnarray}
This suggests that we should generalize (\ref{eq:dfequalssigmanda}) as
\begin{eqnarray}
dF^i = \sigma^i_j n^j dA \qquad \longrightarrow \qquad dp^i = \sigma^i_j n^j dAdt = \sigma^i_j n^j dS \label{eq:generalizingdfsigmandatosigmandadt}
\end{eqnarray}
Because we assumed that $\bf n\it$ is a purely spatial vector, each of the indices here are spatial.  We can interpret the component $\sigma^i_j$ very similarly to how we interpreted the components of the stress tensor.  Rather than $\sigma^i_j$ being $i^{th}$ component of the force on an area element normal to the $j^{th}$ direction, we instead take $\sigma^i_j$ as the $i$ direction \it flux \rm of the momentum through an area element normal to the $j^{th}$ direction.\footnote{Where "flux" relates to the momentum flow \it per time\rm.}  As far as $\sigma^i_j$ with spatial indices, this definition is identical to the definition of the stress tensor in the previous section.  We are merely emphasizing "momentum per time" rather than "force".  

But because we are working with spacetime, we must allow for the possibility that $\bf n\it$ be a time-like vector as well.  However, the infinitesimal element normal to a time-like $\bf n\it$ will have no time-component - it is purely spatial.  Therefore we have in this case
\begin{eqnarray}
d\bf S\it = \bf n\it dS = \bf n\it dV
\end{eqnarray}
where $dV$ is a standard spatial volume element.  The form of (\ref{eq:generalizingdfsigmandatosigmandadt}) suggests that we generalize this as
\begin{eqnarray}
dp^0 = \sigma^0_j n^j dV = \sigma^0_0 n^0 dV
\end{eqnarray}
where the $0^{th}$ component represents the time component as usual.  But we know from relativity that  time component of the momentum four vector is energy (hence "energy-momentum four vector").  We therefore identify $p^0$ with energy, making $\sigma^0_0$ the energy density per unit volume.  

So, we know what $\sigma^0_0$ represents, and we know what $\sigma^i_j$ for $i,j=1,2,3$ represents.  What about $\sigma^i_0$?  We know that $\sigma^i_0$ will correspond to a time-like normal vector, and therefore the relevant spacetime element will be purely spatial ($dV$).  So, this corresponds to
\begin{eqnarray}
dp^i = \sigma^i_0 dV
\end{eqnarray}
So, 
\begin{eqnarray}
\sigma^i_0 = {dp^i \over dV}
\end{eqnarray}
is the $i^{th}$ component of the momentum density.  

On the other hand, the component $\sigma^0_i$ can be interpreted from
\begin{eqnarray}
dp^0 = \sigma^0_i n^i dS
\end{eqnarray}
where $i$ is a spatial index.  The element normal to a spatial $\bf n\it$ is $dAdt$, so this is
\begin{eqnarray}
dp^0 = \sigma^0_i dAdt
\end{eqnarray}
from which we can see that
\begin{eqnarray}
\sigma^0_i = {dp^0 \over dAdt}
\end{eqnarray}
is the energy flux per unit time through the area element $dA$.  

This $4\times 4$ tensor is called the \bf Energy-Momentum Tensor\rm, and it will play a very large role in general relativity.  We have used the notation $\sigma^i_j$, but the more common notation is $T^{\mu}_{\nu}$ where we are using the greek indices because they run over all \it spacetime \rm values.  

Because we are talking about energy and momentum, you may guess that there must be some sort of conservation law involved.  To see how this arises, consider the expression $\partial_{\mu} T^{\mu}_{0}$:
\begin{eqnarray}
\partial_{\mu} T^{\mu}_{0} &=& -\partial_0 T^0_0 + \partial_iT^i_0 \nolabel \\
&=& -\partial_0\bigg({dp^0 \over dV}\bigg) + \partial_i\bigg({dp^i \over dV}\bigg) \nolabel \\
&=& {d \over dV} (\partial_{\mu}p^{\mu}) \nolabel \\
&=& 0
\end{eqnarray}
where the last equality holds by conservation of energy-momentum $\partial_{\mu}p^{\mu} = 0$.  Next we consider $\partial_{\mu}T^{\mu}_i$:
\begin{eqnarray}
\partial_{\mu} T^{\mu}_i &=& -\partial_0 T^0_i + \partial_jT^j_i \nolabel \\
&=& -\partial_0\bigg({dp^0 \over dAdt}\bigg) + \partial_j \bigg({dp^j \over dAdt}\bigg) \nolabel \\
&=& {d \over dAdt}(\partial_{\mu}p^{\mu}) \nolabel \\
&=& 0
\end{eqnarray}

In other words, the energy-momentum tensor provides four independently conserved currents, with conservation laws given by
\begin{eqnarray}
\partial_{\mu} T^{\mu}_{\nu} = 0
\end{eqnarray}

We have already discussed the meaning of each component of the energy-momentum tensor.  We now comment on the quantity $T^{\mu}_{\nu} n^{\nu}$ for unit four-vector $n^{\nu}$.  This can easily be seen by the general expression (\ref{eq:generalizingdfsigmandatosigmandadt}):
\begin{eqnarray}
dp^{\mu} = T^{\mu}_{\nu} n^{\nu} dS
\end{eqnarray}
The expression $T^{\mu}_{\nu} n^{\nu}$ is simply the energy-momentum flux through the spacetime interval $dS$ normal to $n^{\mu}$.  But now consider using the Minkowski metric $\eta_{\mu\nu}$ to lower the first index of $T^{\mu}_{\nu}$ giving
\begin{eqnarray}
T_{\mu \nu} = \eta_{\mu\alpha} T^{\alpha}_{\nu}
\end{eqnarray}

Now consider the expression $T_{\mu\nu} n^{\nu}$:
\begin{eqnarray}
T_{\mu\nu} n^{\nu} &=& \eta_{\mu\alpha}T^{\alpha}_{\nu}n^{\nu} \nolabel \\
&=& \eta_{\mu\alpha} { dp^{\alpha} \over dS} \nolabel \\
&=& {dp_{\mu} \over dS} \label{eq:tmununnuequalsdpmuds}
\end{eqnarray}
This is, as we have already discussed, simply the energy-momentum flux through the infinitesimal spacetime volume $dS$ (with the spacetime index lowered by the metric), or in other words the spacetime density of the energy-momentum.  Now consider taking the dot product of (\ref{eq:tmununnuequalsdpmuds}) with $n^{\mu}$:
\begin{eqnarray}
{dp_{\mu} \over dS}n^{\mu} &=& {d \over dS} ( p_{\mu} n^{\mu})
\end{eqnarray}
This is the spacetime density of the component of the energy-momentum four vector in the direction of $n^{\mu}$.  So what does this mean?  

As we said above, $T_{\mu\nu} n^{\nu}$ is the spacetime density of the energy-momentum four vector of the material described by $T^{\mu}_{\nu}$ in the spacetime interval normal to $n^{\mu}$.  If we take $n^{\mu}$ to be time-like, we can interpret it as representing the motion of a particular inertial observer relative to the material described by $T^{\mu}_{\nu}$.  Then, the spacetime interval normal to $n^{\mu}$ will be an infinitesimal box of the material as observed by the $n^{\mu}$ observer.  The energy-momentum density of this infinitesimal box will be given by the four vector 
\begin{eqnarray}
T_{\mu\nu} n^{\nu} = {dp_{\mu} \over dS}
\end{eqnarray}
The component of the energy momentum tensor in the direction of $n^{\mu}$ can then naturally be interpreted as the \it energy density \rm of the material as observed by the $n^{\mu}$ observer.  In other words,
\begin{eqnarray}
{dp_{\mu} \over dS} n^{\mu} 
\end{eqnarray}
is the energy density of the material as measured by the $n^{\mu}$ observer.  And therefore, we can take
\begin{eqnarray}
T_{\mu\nu}n^{\mu}n^{\nu} 
\end{eqnarray}
to be the energy density of the material as observed by the $n^{\mu}$ observer.  

To illustrate this, consider letting $n^{\mu}$ be a unit vector in the time dimension, 
\begin{eqnarray}
n^{\mu} \dot = 
\begin{pmatrix}
1 & 0 & 0 & 0
\end{pmatrix}
\end{eqnarray}
This represents an observer with no spatial translation relative to the material described by $T^{\mu}_{\nu}$ - in other words an observer in the rest frame of the material.  This will give
\begin{eqnarray}
T_{\mu\nu} n^{\mu}n^{\nu} = T_{00} = -T^0_0
\end{eqnarray}
which is (minus) the energy density - exactly what we would expect.  

This generalizes nicely to an arbitrary inertial observer whose motion is described by the four-vector $v^{\mu}$.  The expression
\begin{eqnarray}
T_{\mu\nu} v^{\mu}v^{\nu} = energy \label{eq:energymomentumcontractedwithavectorgivesenergydensity1127}
\end{eqnarray}
the energy density of the material as measured by an inertial observer with motion $v^{\mu}$ relative to the material.  

As a brief preview of where this is going, recall from sections \ref{sec:fourthint} and \ref{sec:fourthint2} that given a manifold $\mathcal{M}$ and some vector $v^{\mu}$ on that manifold, we can consider the subspace of $\mathcal{M}$ that is normal to $v^{\mu}$.  If we have a metric on $\mathcal{M}$, we can form the Einstein tensor $G_{\mu \nu}$.  The scalar curvature of the submanifold normal to $v^{\mu}$ is then given by
\begin{eqnarray}
R_{submanifold} = G_{\mu\nu} v^{\mu}v^{\nu} = curvature
\end{eqnarray}

Einstein's general theory of relativity essentially says that energy = curvature.  In other words, for some observer with motion described by the time-like four vector $v^{\mu}$, the subspace normal to that observer has curvature equal to the energy density.  Einstein therefore set\footnote{cf section \ref{sec:s2otimesbbr}.}
\begin{eqnarray}
G_{\mu\nu} \propto T_{\mu\nu}
\end{eqnarray}
So, inserting a proportionality constant $\kappa$ and contracting both sides with some vector $v^{\mu}$, we have
\begin{eqnarray}
G_{\mu\nu} v^{\mu}v^{\nu} = \kappa T_{\mu\nu} v^{\mu}v^{\nu} \qquad \Longrightarrow \qquad (curvature)= \kappa(energy) \label{eq:initialstatementofmeaningofEinsteinfieldequationsbeforerelativisticformluationssection}
\end{eqnarray}
We will discuss this in much more depth below.  We mention it now merely to indicate how this will show up in physics.  

\subsection{The Relativistic Formulation}
\label{sec:therelativisticforumlation}

In the previous two sections we derived the energy-momentum tensor using a fairly intuitive approach so that the meaning of each component could be seen and understood.  However, recall from any introductory lecture covering Noether's theorem that both momentum and energy are conserved currents arising from symmetries in a Lagrangian.  Specifically, energy is a conserved quantity resulting from invariance under time translations, while momentum is a conserved quantity resulting from invariance under spatial translations.  So if we consider a relativistic Lagrangian that is symmetric under \it spacetime \rm translations, we should arrive at a set of conserved quantities that match the energy-momentum tensor from the previous sections.  We now set out to do this.  

Consider some arbitrary Lagrangian density that is a functional of the field $\phi$.  The action is given by
\begin{eqnarray}
S = \int d^nx\mathcal{L}(\phi,\partial_{\mu} \phi)
\end{eqnarray}
The equations of motion are then given by the relativistic field version of the Euler-Lagrange equation:
\begin{eqnarray}
\partial_{\mu}\bigg({\partial \mathcal{L} \over \partial(\partial_{\mu} \phi)}\bigg) - {\partial \mathcal{L} \over \partial \phi} = 0 \label{eq:relverofeleq}
\end{eqnarray}

Now consider a spacetime translation
\begin{eqnarray}
x^{\mu} \longrightarrow x^{\mu} + a^{\mu} \label{eq:spacetimetranslation1}
\end{eqnarray}
This will induce the transformation
\begin{eqnarray}
\mathcal{L}(x^{\mu}) &\longrightarrow& \mathcal{L}(x^{\mu}+a^{\mu}) \nolabel \\
&=& \mathcal{L}(x^{\mu}) + a^{\nu}\partial_{\nu} \mathcal{L} \label{eq:1}
\end{eqnarray}
But at the same time, (\ref{eq:spacetimetranslation1}) induces
\begin{eqnarray}
\phi(x^{\mu}) &\longrightarrow& \phi(x^{\mu}+a^{\mu}) \nolabel \\
&=& \phi(x^{\mu}) + a^{\nu}\partial_{\nu} \phi(x^{\mu})
\end{eqnarray}
So, under (\ref{eq:spacetimetranslation1}) we have
\begin{eqnarray}
\delta \phi &=& a^{\nu}\partial_{\nu} \phi 
\end{eqnarray}
These transformations will then induce 
\begin{eqnarray}
\mathcal{L}(\phi, \partial_{\mu} \phi) &\longrightarrow& \mathcal{L}(\phi+\delta \phi, \partial_{\mu}\phi+\delta(\partial_{\mu}\phi)) \nolabel \\
&=& \mathcal{L}(\phi, \partial_{\mu}\phi) + \delta \phi {\partial \mathcal{L} \over \partial \phi} + (\partial_{\mu}\delta \phi) {\partial \mathcal{L} \over \partial(\partial_{\mu} \phi)} \label{eq:expansionoflagrantianspacetimetranslation5}
\end{eqnarray}
and if the Euler-Lagrange equation (\ref{eq:relverofeleq}) is satisfied, this is
\begin{eqnarray}
\mathcal{L}(\phi, \partial_{\mu}\phi) &=& \mathcal{L}(\phi, \partial_{\mu}\phi) + \delta \phi \partial_{\mu}\bigg({\partial \mathcal{L} \over \partial (\partial_{\mu} \phi)}\bigg) +  (\partial_{\mu}\delta \phi) {\partial \mathcal{L} \over \partial(\partial_{\mu} \phi)} \nolabel \\
&=& \mathcal{L}(\phi, \partial_{\mu} \phi) + \partial_{\mu} \bigg({\partial \mathcal{L} \over \partial(\partial_{\mu}\phi)}\delta \phi \bigg) \label{eq:2}
\end{eqnarray}

So, according to (\ref{eq:1}), the transformation (\ref{eq:spacetimetranslation1}) induces
\begin{eqnarray}
\delta \mathcal{L} = a^{\mu} \partial_{\mu} \mathcal{L}
\end{eqnarray}
But according to (\ref{eq:2}) it also induces
\begin{eqnarray}
\delta \mathcal{L} = \partial_{\mu} \bigg({\partial \mathcal{L} \over \partial(\partial_{\mu}\phi)}\delta \phi \bigg) = \partial_{\mu} \bigg({\partial \mathcal{L} \over \partial(\partial_{\mu}\phi)}a^{\nu}\partial_{\nu}\phi \bigg)
\end{eqnarray}
Equating these we get
\begin{eqnarray}
& & a^{\mu}\partial_{\mu}\mathcal{L} = \partial_{\mu}\bigg({\partial \mathcal{L} \over \partial(\partial_{\mu}\phi)} a^{\nu}\partial_{\nu}\phi\bigg) \nolabel \\
&\Longrightarrow& \partial_{\mu}\bigg( {\partial \mathcal{L} \over \partial(\partial_{\mu}\phi)} a^{\nu}\partial_{\nu}\phi - a^{\mu}\mathcal{L} \bigg) = 0 \nolabel \\
&\Longrightarrow& \partial_{\mu}\bigg( {\partial \mathcal{L} \over \partial(\partial_{\mu}\phi)} a^{\nu}\partial_{\nu}\phi - \eta^{\mu}_{\nu} \mathcal{L}\bigg) a^{\nu} = 0 \label{eq:partialmudeltamunucallminusetcequals0}
\end{eqnarray}
(where $\eta^{\mu}_{\nu}$ is the metric with one index raised).  We identify the quantity in the large parentheses the \bf Energy-Momentum Tensor\rm, 
\begin{eqnarray}
T^{\mu}_{\nu} =   {\partial \mathcal{L} \over \partial(\partial_{\mu}\phi)} \partial_{\nu}\phi - \eta^{\mu}_{\nu} \mathcal{L} \label{eq:firstfullstatementoftheenergymomentumtensorinflatspace}
\end{eqnarray}
which clearly satisfies (from (\ref{eq:partialmudeltamunucallminusetcequals0}))
\begin{eqnarray}
\partial_{\mu} T^{\mu}_{\nu} = 0 \label{eq:partialmuTmunuequalszero000}
\end{eqnarray}
meaning that it is a conserved quantity.  The interpretation of each component is identical to the interpretation in the previous section.  

To see this consider the Lagrangian with two of the spatial dimensions suppressed (for simplicity),
\begin{eqnarray}
\mathcal{L} = -{1 \over 2} \partial_{\mu} \phi \partial^{\mu} \phi - {1 \over 2} m^2 \phi^2 = -{1 \over 2} (-\dot \phi^2 + \phi'^2) - {1 \over 2}m^2 \phi^2 \label{eq:lagrangianforscalarwithonlyonespatialdim}
\end{eqnarray}
So,
\begin{eqnarray}
T^0_0 &=& {\partial \mathcal{L} \over \partial\dot \phi} \dot \phi - \bigg[ -{1 \over 2} (-\dot \phi^2 + \phi'^2) - {1 \over 2}m^2 \phi^2\bigg] \nolabel \\
&=& \dot \phi^2 - {1 \over 2} \dot \phi^2 + {1 \over 2} \phi'^2 + {1 \over 2}m^2 \phi^2 \nolabel \\
&=& {1 \over 2} (\dot \phi^2 + \phi'^2) + {1 \over 2} m^2\phi^2
\end{eqnarray}
which you can recognize as the Hamiltonian density, or total energy density, of the system - exactly what we would expect from $T^0_0$.\footnote{If you don't understand why this is the Hamiltonian, reread the first chapter of \cite{Firstpaper} - you can work out the Hamiltonian from (\ref{eq:lagrangianforscalarwithonlyonespatialdim}) using the Legendre transformation discussed there.}  You can furthermore work out the other components of $T^{\mu}_{\nu}$ using (\ref{eq:lagrangianforscalarwithonlyonespatialdim}) to see that they have the same meaning in terms of the energy and momentum.  

\subsection{The Modern Formulation}
\label{sec:themodernformulation}

As a final comment, we want to take a closer look at the energy-momentum tensor and the action.  To begin with, consider a Lagrangian 
\begin{eqnarray}
\mathcal{L} = \mathcal{L}(\phi, \partial_{\mu}\phi)
\end{eqnarray}
The general form such a Lagrangian will take will be
\begin{eqnarray}
\mathcal{L} = -{1 \over 2}g^{\mu\nu}\partial_{\mu}\phi\partial_{\nu}\phi - V(\phi) \label{eq:generalformsuchalagrangianwilltakewillbe}
\end{eqnarray}
So to build the energy-momentum tensor from this involves first calculating (from (\ref{eq:firstfullstatementoftheenergymomentumtensorinflatspace}))
\begin{eqnarray}
{\partial \mathcal{L} \over \partial (\partial_{\mu}\phi)} = -g^{\mu\nu} \partial_{\nu}\phi = -\partial^{\mu}\phi
\end{eqnarray}
Then
\begin{eqnarray}
{\partial \mathcal{L} \over \partial (\partial_{\mu}\phi)} \partial_{\nu}\phi = -\partial^{\mu}\phi \partial_{\nu}\phi
\end{eqnarray}
So we have the general form
\begin{eqnarray}
T^{\mu}_{\nu} = -\partial^{\mu}\phi\partial_{\nu}\phi - g^{\mu}_{\nu}\mathcal{L} \label{eq:dfshjkasdfjkhsadfkhjdsfiuuyfdiuyiuy719}
\end{eqnarray}

Now, recall that the action involves an integral over all spacetime.  The primary idea of general relativity is that the spacetime metric becomes a dynamical field (along with all physical fields like $\phi$, etc.).  Therefore the geometry of the spacetime manifold is is unspecified \it a priori\rm.  For this reason we don't merely integrate $\mathcal{L}$ over spacetime as
\begin{eqnarray}
S = \int dxdydzdt\; \mathcal{L}
\end{eqnarray}
Instead we integrate over the invariant volume element (cf section \ref{sec:invariantvolumeelement}),
\begin{eqnarray}
S = \int dxdydzdt\; \sqrt{|g|}\mathcal{L}
\end{eqnarray}
We therefore define a "new" Lagrangian density $\mathscr{L}$ defined by
\begin{eqnarray}
\mathscr{L} \equiv \sqrt{|g|} \mathcal{L}
\end{eqnarray}

Now consider the expression
\begin{eqnarray}
{\partial \mathscr{L} \over \partial g^{\mu\nu}} &=& {\partial (\sqrt{|g|}\mathcal{L}) \over \partial g^{\mu\nu}} \nolabel \\
&=& \sqrt{|g|}{\partial \mathcal{L} \over \partial g^{\mu\nu}} +  {\partial \sqrt{|g|} \over \partial g^{\mu\nu}} \mathcal{L} \label{eq:newderivativewithtensordensitywithtwotermsweneedtoderivative}
\end{eqnarray}
The derivative in the first term can be calculated easily from (\ref{eq:generalformsuchalagrangianwilltakewillbe}):
\begin{eqnarray}
{\partial \mathcal{L}\over \partial g^{\mu\nu}} &=& {1\over 2}{\partial \over \partial g^{\mu\nu}} (g^{\alpha\beta} \partial_{\alpha}\phi\partial_{\beta}\phi + 2V(\phi)) \nolabel \\
&=& {1\over 2} \partial_{\mu}\phi\partial_{\nu}\phi
\end{eqnarray}

Now we need to calculate the derivative in the second term in (\ref{eq:newderivativewithtensordensitywithtwotermsweneedtoderivative}).  This is
\begin{eqnarray}
{\partial \sqrt{|g|} \over \partial g^{\mu\nu}} &=& {\partial \over \partial g^{\mu\nu}} (|g|)^{1/2} \nolabel \\
&=& {1 \over 2}(|g|)^{-1/2} {\partial |g| \over \partial g^{\mu\nu}} 
\end{eqnarray}
Now we make use of a general relation from linear algebra\footnote{This can be found in any introductory text on linear algebra in the chapter on determinants.},
\begin{eqnarray}
{\partial |g| \over \partial g^{\mu\nu}} = |g|g_{\mu\nu} \label{eq:derivativeofmatrixdetwithrespectoelements}
\end{eqnarray}
So
\begin{eqnarray}
{\partial \sqrt{|g|} \over \partial g^{\mu\nu}} = {|g| g_{\mu\nu} \over 2\sqrt{|g|}} = {1 \over 2}\sqrt{|g|} g_{\mu\nu} \label{eq:variationofsquarerootofabsolutevalueofmetricwithrespecttometric}
\end{eqnarray}

Now we can plug these derivatives into (\ref{eq:newderivativewithtensordensitywithtwotermsweneedtoderivative}), getting
\begin{eqnarray}
{\partial \mathscr{L} \over \partial g^{\mu\nu}} &=& \sqrt{|g|}\bigg( {1 \over 2} \partial_{\mu}\phi\partial_{\nu}\phi + {1 \over 2}g_{\mu\nu}\mathcal{L}\bigg) \nolabel \\
&=& -{1 \over 2} \sqrt{|g|}( -\partial_{\mu}\phi\partial_{\nu}\phi - g^{\mu\nu}\mathcal{L}) \nolabel \\
&=& -{1 \over 2} \sqrt{|g|} T_{\mu\nu}
\end{eqnarray}
where we used equation (\ref{eq:dfshjkasdfjkhsadfkhjdsfiuuyfdiuyiuy719}) in the last line.  This allows us to write 
\begin{eqnarray}
T_{\mu\nu} = -{2 \over \sqrt{|g|}} {\partial \mathscr{L} \over \partial g^{\mu\nu}} \label{eq:fundamentalexpressionofenergymomentumtensor}
\end{eqnarray}
It turns out that the energy-momentum tensor as defined in (\ref{eq:fundamentalexpressionofenergymomentumtensor}) holds in any arbitrary geometry and for arbitrary fields.\footnote{Though we only did this calculation using scalar fields $\phi$, it is valid with any set of physical fields and with any potential.}  We will therefore take it to be the most fundamental \it defining \rm statement of the energy-momentum tensor.  We will find this expression extremely helpful throughout this notes as well as the others in the series.  

\subsection{The Meaning of the Modern Formulation}
\label{sec:themeaningofthemodernformulation}

The derivation of (\ref{eq:fundamentalexpressionofenergymomentumtensor}) above is admittedly lacking in rigor.  We "derived" it only by showing that it leads to the same result \it in flat space \rm as the Noether's theorem approach in the preceding section \it for scalar fields only\rm.  We will therefore spend a bit of time discussing it in more detail now.  We will first derive it in a different way, then discuss what the implications of each approach is.  

The fundamental idea behind general relativity will be that we take the spacetime metric to be a physical and dynamic field, \it in addition \rm to the physical fields.  So, instead of the Lagrangian\footnote{We are using all covariant indices for simplicity and for later convenience.}
\begin{eqnarray}
\mathcal{L} = \mathcal{L}(\phi_{\mu}, \partial_{\nu}\phi_{\mu})
\end{eqnarray}
(where $\phi_{\mu}$ is a vector field for generality), we have the Lagrangian
\begin{eqnarray}
\mathcal{L} = \mathcal{L}(\phi_{\mu}, \partial_{\nu}\phi_{\mu}, g_{\mu\nu}) \label{eq:modappwithg}
\end{eqnarray}
where $g_{\mu\nu}$ is the spacetime metric.  Consider a change of coordinates from, say, $x^{\mu}$ to $y^{\mu}$.  This transformation matrix for this will be
\begin{eqnarray}
R^{\mu}_{\nu} \equiv {\partial x^{\mu} \over \partial y^{\nu}} \label{eq:modapp1}
\end{eqnarray}
where the determinant of $R^{\mu}_{\nu}$ is denoted
\begin{eqnarray}
\det(R^{\mu}_{\nu}) = R
\end{eqnarray}

Now consider again the Lagrangian with the addition of the invariant volume term,
\begin{eqnarray}
\mathscr{L} = \sqrt{|g|} \mathcal{L}
\end{eqnarray}
Under the transformation (\ref{eq:modapp1}) the invariant volume term will transform according to
\begin{eqnarray}
\sqrt{|g|} &=& \sqrt{|\det(g_{\mu\nu})|} \nolabel \\
&=& \sqrt{\bigg|\det\bigg({ \partial x^{\alpha} \over \partial y^{\mu}}{\partial x^{\beta} \over \partial y^{\nu}} g_{\alpha\beta}\bigg)\bigg|} \nolabel \\
&=& \sqrt{|\det(R^{\alpha}_{\mu}R^{\beta}_{\nu}g_{\alpha\beta})|} \nolabel \\
&=& \sqrt{R^2|\det(g_{\alpha\beta})|} \nolabel \\
&=& R\sqrt{|g|}
\end{eqnarray}
So, under (\ref{eq:modapp1}) we have
\begin{eqnarray}
\mathscr{L} \longrightarrow R \mathscr{L} \label{eq:modapphowLtransformsJac}
\end{eqnarray}

On the other hand we can look at this transformation from another perspective.  The functional dependence of $\mathscr{L}$ is indicated in (\ref{eq:modappwithg}):
\begin{eqnarray}
\mathscr{L} = \mathscr{L} (\phi_{\mu}, \partial_{\mu}\phi_{\nu}, g_{\mu\nu})
\end{eqnarray}
We can apply the same transformation (\ref{eq:modapp1}) to each of these.  For $\phi_{\mu}$ and $g_{\mu\nu}$ this is easy (using the bar to indicate the $y$ coordinates):
\begin{eqnarray}
\phi_{\mu} &\longrightarrow& \bar \phi_{\mu} = {\partial x^{\alpha} \over \partial y^{\mu}} \phi_{\alpha} = R^{\alpha}_{\mu} \phi_{\alpha} \nolabel \\
g_{\mu\nu} &\longrightarrow& \bar g_{\mu\nu} = {\partial x^{\alpha} \over \partial y^{\mu}} {\partial x^{\beta} \over \partial y^{\nu}} g_{\alpha\beta} = R^{\alpha}_{\mu} R^{\beta}_{\nu} g_{\alpha\beta}
\end{eqnarray}
With $\partial_{\mu}\phi_{\nu}$ we must be more careful - we must do what we did in (\ref{eq:transformationofderivativeofvector})):
\begin{eqnarray}
\partial_{\mu}\phi_{\nu} &\longrightarrow& \bar \partial_{\mu}\bar \phi_{\nu} \nolabel \\
&=& {\partial \over \partial y^{\mu}} \phi_{\nu} \nolabel \\
&=& \bigg({\partial x^{\alpha} \over \partial y^{\mu}} {\partial \over \partial x^{\alpha}} \bigg) {\partial x^{\gamma} \over \partial y^{\nu}} \phi_{\gamma} \nolabel \\
&=& {\partial x^{\alpha} \over \partial y^{\mu}} {\partial x^{\gamma} \over \partial y^{\nu}} \partial_{\alpha} \phi_{\gamma} + {\partial x^{\alpha} \over \partial y^{\mu}} {\partial^2 x^{\gamma} \over \partial x^{\alpha} \partial y^{\nu}} \phi_{\gamma} \nolabel \\
&=& {\partial x^{\alpha} \over \partial y^{\mu}} {\partial x^{\gamma} \over \partial y^{\nu}} \partial_{\alpha} \phi_{\gamma} + {\partial x^{\alpha} \over \partial y^{\mu}} {\partial y^{\eta} \over \partial x^{\alpha}} {\partial^2 x^{\gamma} \over \partial y^{\eta} \partial y^{\nu}} \phi_{\gamma} \nolabel \\
&=& {\partial x^{\alpha} \over \partial y^{\mu}} {\partial x^{\gamma} \over \partial y^{\nu}} \partial_{\alpha} \phi_{\gamma} + {\partial^2 x^{\gamma} \over \partial y^{\mu} \partial y^{\nu}} \phi_{\gamma} \nolabel \\
&=& R^{\alpha}_{\mu} R^{\gamma}_{\nu} \partial_{\alpha}\phi_{\gamma} + {\partial^2 x^{\gamma} \over \partial y^{\mu} \partial y^{\nu}} \phi_{\gamma} \nolabel \\
\end{eqnarray}
Writing the second expression in terms of $R^{\mu}_{\nu}$ is a bit trickier because it can be written as
\begin{eqnarray}
{\partial^2 x^{\gamma} \over \partial y^{\mu} \partial y^{\nu}} = \partial_{\mu} R^{\gamma}_{\nu} = {\partial^2x^{\gamma} \over \partial y^{\nu} \partial y^{\mu}} = \partial_{\nu} R^{\gamma}_{\mu}
\end{eqnarray}
So, to make sure we are as general as possible, we write it as
\begin{eqnarray}
\bar \partial_{\mu} \bar \phi_{\nu} = R^{\alpha}_{\mu} R^{\gamma}_{\nu} \partial_{\alpha} \phi_{\gamma} + {1 \over 2} (\partial_{\mu} R^{\gamma}_{\nu} + \partial_{\nu}R^{\gamma}_{\mu}) \phi_{\gamma}
\end{eqnarray}

So, combining these results we have
\begin{eqnarray}
\bar \phi_{\mu} &=& R^{\alpha}_{\mu} \phi_{\alpha} \nolabel \\
\bar \partial_{\mu} \bar \phi_{\nu} &=& R^{\alpha}_{\mu} R^{\gamma}_{\nu} \partial_{\alpha} \phi_{\gamma} + {1 \over 2} (\partial_{\mu} R^{\gamma}_{\nu} + \partial_{\nu}R^{\gamma}_{\mu})\phi_{\gamma} \nolabel \\
\bar g_{\mu\nu} &=& R^{\alpha}_{\mu}R^{\beta}_{\nu} g_{\alpha\beta} \label{eq:modappfieldstrans}
\end{eqnarray}

Now consider making the transformation $R^{\mu}_{\nu}$ to $\mathscr{L} = \mathscr{L}(\phi_{\mu}, \partial_{\mu}\phi_{\nu}, g_{\mu\nu})$.  Combining (\ref{eq:modappfieldstrans}) and (\ref{eq:modapphowLtransformsJac}) we have
\begin{eqnarray}
\mathscr{\bar L} (\bar \phi_{\mu}, \bar \partial_{\mu} \bar \phi_{\nu}, \bar g_{\mu\nu}) = R\mathscr{L}(\phi_{\mu}, \partial_{\mu}\phi_{\nu}, g_{\mu\nu}) \label{eq:modappbarLequalsDL}
\end{eqnarray}
We now do something that will seem very strange, but will be very helpful.  We take the derivative of both sides of (\ref{eq:modappbarLequalsDL}) with respect to the transformation $R^{\mu}_{\nu}$.  Doing this will require the result from linear algebra (\ref{eq:derivativeofmatrixdetwithrespectoelements}), so the right side will be
\begin{eqnarray}
{\partial \over \partial R^{\rho}_{\sigma}} (R\mathscr{L}) = (R^{-1})^{\sigma}_{\rho} R\mathscr{L} \label{eq:derivativeofbarLequalsRLequation}
\end{eqnarray}
(the $\mathscr{L}$ part has no dependence on $R^{\mu}_{\nu}$).  

The derivative of the left hand side of (\ref{eq:modappbarLequalsDL}) can then be expanded using the chain rule:
\begin{eqnarray}
{\partial \over \partial R^{\rho}_{\sigma}} (\mathscr{\bar L}) &=& {\partial \bar \phi_{\mu} \over \partial R^{\rho}_{\sigma}} {\partial \mathscr{\bar L} \over \partial \bar \phi_{\mu}} + {\partial (\bar \partial_{\mu} \bar \phi_{\nu}) \over \partial R^{\rho}_{\sigma}}{\partial \mathscr{\bar L} \over \partial (\bar \partial_{\mu} \bar \phi_{\nu})} + {\partial (g_{\mu\nu}) \over \partial R^{\rho}_{\sigma}} {\partial \mathscr{\bar L} \over \partial (g_{\mu\nu})} \label{eq:modappchainrule1}
\end{eqnarray}
But from (\ref{eq:modappfieldstrans}) we can find each of the derivatives of the fields in this expression:
\begin{eqnarray}
{\partial \bar \phi_{\mu} \over \partial R^{\rho}_{\sigma}} &=& {\partial \over \partial R^{\rho}_{\sigma}} (R^{\alpha}_{\mu} \phi_{\alpha}) \nolabel \\
&=& \delta^{\alpha}_{\rho} \delta_{\mu}^{\sigma} \phi_{\alpha}\nolabel \\
&=& \delta^{\sigma}_{\mu} \phi_{\rho} \nolabel \\
{\partial (\bar \partial_{\mu} \bar \phi_{\nu}) \over \partial R^{\rho}_{\sigma}} &=& {\partial \over \partial R^{\rho}_{\sigma}} (R^{\alpha}_{\mu} R^{\gamma}_{\nu} \partial_{\alpha} \phi_{\gamma} + {1 \over 2} (\partial_{\mu} R^{\gamma}_{\nu} + \partial_{\nu}R^{\gamma}_{\mu})\phi_{\gamma}) \nolabel \\
&=& R^{\alpha}_{\mu}\delta^{\gamma}_{\rho}\delta_{\nu}^{\sigma}\partial_{\alpha} \phi_{\gamma} + \delta^{\alpha}_{\rho} \delta_{\mu}^{\sigma} R^{\gamma}_{\nu} \partial_{\alpha}\phi_{\gamma} \nolabel \\
&=& \delta^{\sigma}_{\nu} R^{\alpha}_{\mu} \partial_{\alpha} \phi_{\rho} + \delta^{\sigma}_{\mu} R^{\alpha}_{\nu} \partial_{\rho} \phi_{\alpha} \nolabel \\
{\partial (\bar g_{\mu\nu}) \over \partial R^{\rho}_{\sigma}} &=& {\partial \over \partial R^{\rho}_{\sigma}} (R^{\alpha}_{\mu}R^{\beta}_{\nu} g_{\alpha\beta}) \nolabel \\
&=& R^{\alpha}_{\mu} \delta^{\beta}_{\rho} \delta_{\nu}^{\sigma} g_{\alpha\beta} + \delta^{\alpha}_{\rho} \delta^{\sigma}_{\mu} R^{\beta}_{\nu} g_{\alpha\beta} \nolabel \\
&=& \delta^{\sigma}_{\nu} R^{\alpha}_{\mu} g_{\alpha\rho} + \delta^{\sigma}_{\mu} R^{\beta}_{\nu} g_{\rho\beta} \nolabel \\
&=& (\delta^{\sigma}_{\nu} R^{\alpha}_{\mu} + \delta^{\sigma}_{\mu} R^{\alpha}_{\nu}) g_{\alpha\rho}
\end{eqnarray}
Now we can plug these into (\ref{eq:modappchainrule1}) to get
\begin{eqnarray}
{\partial \over \partial R^{\rho}_{\sigma}}(\mathscr{\bar L}) &=& \delta^{\sigma}_{\mu} \phi_{\rho} {\partial \mathscr{\bar L} \over \partial \bar \phi_{\mu}} + (\delta^{\sigma}_{\nu} R^{\alpha}_{\mu} \partial_{\alpha} \phi_{\rho} + \delta^{\sigma}_{\mu} R^{\alpha}_{\nu} \partial_{\rho} \phi_{\alpha}){\partial \mathscr{\bar L} \over \partial (\bar \partial_{\mu} \bar \phi_{\nu})} + (\delta^{\sigma}_{\nu} R^{\alpha}_{\mu} + \delta^{\sigma}_{\mu} R^{\alpha}_{\nu}) g_{\alpha\rho} {\partial \mathscr{\bar L} \over \partial (g_{\mu\nu})} \nolabel \\
&=& (R^{-1})^{\sigma}_{\rho} R \mathscr{L}
\end{eqnarray}
where we used (\ref{eq:derivativeofbarLequalsRLequation}) in the last line.  

Now let's take the particular case where $y^{\mu} = x^{\mu}$, or the identity.  This makes
\begin{eqnarray}
R^{\mu}_{\nu} = (R^{-1})^{\mu}_{\nu} = \delta^{\mu}_{\nu}
\end{eqnarray}
and
\begin{eqnarray}
R = 1
\end{eqnarray}
So, we now have
\begin{eqnarray}
\delta^{\sigma}_{\rho}\mathscr{L} &=& \delta^{\sigma}_{\mu} \phi_{\rho} {\partial \mathscr{ L} \over \partial  \phi_{\mu}} + (\delta^{\sigma}_{\nu}\delta^{\alpha}_{\mu} \partial_{\alpha}\phi_{\rho} + \delta^{\sigma}_{\mu}\delta^{\alpha}_{\nu}\partial_{\rho}\phi_{\alpha}){\partial \mathscr{ L} \over \partial (\partial_{\mu} \phi_{\nu})} + (\delta^{\sigma}_{\nu}\delta^{\alpha}_{\mu} + \delta^{\sigma}_{\mu}\delta^{\alpha}_{\nu})g_{\alpha\rho} {\partial \mathscr{L} \over \partial (g_{\mu\nu})} \nolabel \\
&=& \phi_{\rho}{\partial \mathscr{L} \over \partial \phi_{\sigma}} + \partial_{\mu}\phi_{\rho}{\partial \mathscr{L} \over \partial (\partial_{\mu} \phi_{\sigma})} + \partial_{\rho}\phi_{\mu}{\partial \mathscr{L} \over \partial (\partial_{\sigma} \phi_{\mu})} + g_{\mu\rho} {\partial \mathscr{L} \over \partial g_{\mu\sigma}} + g_{\mu\rho}{\partial \mathscr{L} \over \partial g_{\sigma\mu}}  \nolabel \\
&=& \phi_{\rho}{\partial \mathscr{L} \over \partial \phi_{\sigma}} + \partial_{\mu}\phi_{\rho}{\partial \mathscr{L} \over \partial (\partial_{\mu} \phi_{\sigma})} + \partial_{\rho}\phi_{\mu}{\partial \mathscr{L} \over \partial (\partial_{\sigma} \phi_{\mu})} + 2g_{\mu\rho} {\partial \mathscr{L} \over \partial g_{\mu\sigma}}  \nolabel \\
\end{eqnarray}
We can then use the Euler-Lagrange equation on the first term, followed by an integration by parts:
\begin{eqnarray}
\phi_{\rho} {\partial \mathscr{L} \over \partial \phi_{\sigma}} &=& \phi_{\rho} \partial_{\mu}\bigg({\partial \mathscr{L} \over \partial (\partial_{\mu} \phi_{\sigma})}\bigg) \nolabel \\
&=& -(\partial_{\mu} \phi_{\rho}) {\partial \mathscr{L} \over \partial(\partial_{\mu}\phi_{\sigma})}
\end{eqnarray}
But this will cancel with the second term above, leaving
\begin{eqnarray}
\delta^{\sigma}_{\rho}\mathscr{L} = \partial_{\rho}\phi_{\mu} {\partial \mathscr{L} \over \partial (\partial_{\sigma} \phi_{\mu})} + 2 {\partial \mathscr{L} \over \partial g^{\rho}_{\sigma}}
\end{eqnarray}
Rearranging this and writing it in terms of $\mathcal{L}$ rather than $\mathscr{L}$, this is
\begin{eqnarray}
2 {\partial \mathscr{L} \over \partial g^{\rho\sigma}} = -\sqrt{|g|} {\partial \mathcal{L} \over \partial (\partial^{\sigma} \phi_{\mu})} (\partial_{\rho} \phi_{\mu}) + \sqrt{|g|}\delta_{\sigma\rho}\mathcal{L}
\end{eqnarray}
Or, finally,
\begin{eqnarray}
-{2 \over \sqrt{|g|}} {\partial \mathscr{L} \over \partial g_{\rho\sigma}} = {\partial \mathcal{L} \over \partial (\partial^{\sigma}\phi_{\mu})} (\partial_{\rho}\phi_{\mu})- \delta_{\sigma\rho}\mathcal{L}
\end{eqnarray}
Comparing this with (\ref{eq:firstfullstatementoftheenergymomentumtensorinflatspace}), we can identify\footnote{There is the small difference of the delta function here and the metric there, but had we been more rigorous they would have been the same.  We are merely trying to communicate the essential idea, not the details.} 
\begin{eqnarray}
-{2 \over \sqrt{|g|}} {\partial \mathscr{L} \over \partial g^{\rho\sigma}} = T_{\rho\sigma}
\end{eqnarray}
just as in (\ref{eq:fundamentalexpressionofenergymomentumtensor}).  

So what is going on here?  The energy-momentum tensor is essentially a measurement of the flow of energy and momentum (which viewed relativistically are part of the same 4-vector) through spacetime, and we can therefore think of it as a conserved Noether current.  The generators of this conserved current are the invariance of $\mathcal{L}$ (or $\mathscr{L}$) under spacetime translations.  However, a spacetime translation is nothing more than a change of coordinates.  So, by imposing (\ref{eq:modappbarLequalsDL}) we are essentially imposing the invariance of the Lagrangian under spacetime translations.  

However, we can also think of a spacetime translation as a change in the metric.  Consider such a change, $g_{\mu\nu} \longrightarrow g_{\mu\nu}+\delta g_{\mu\nu}$.  This will result in
\begin{eqnarray}
\mathscr{L}(g_{\mu\nu}) &\longrightarrow& \mathscr{L}(g_{\mu\nu}+\delta g_{\mu\nu}) \nolabel \\
&=& \mathscr{L}(g_{\mu\nu}) + \delta g_{\mu\nu} {\partial \mathscr{L} \over \partial g_{\mu\nu}} + \cdots \nolabel \\
&=& \mathscr{L}(g_{\mu\nu}) - {1 \over 2}  \sqrt{|g|} \delta g_{\mu\nu} T^{\mu\nu} + \cdots
\end{eqnarray}
So, if you change the spacetime, you change the physics (obviously).  And to first order, the change in the physics is given by the energy-momentum tensor.  This is perhaps the most fundamental meaning of $T_{\mu\nu}$, and stems from the derivation of both the relativistic derivation of section \ref{sec:therelativisticforumlation} and the derivation in this section.  It is essentially a statement of the "reaction" of the matter fields to a change in the coordinates (cf the Noether current derivation above), which is essentially a local change in the metric.  

\section{General Relativity}

Introductions to general relativity typically approach the subject by one (or both) of two roads.  The first is the axiomatic approach, which is the approach Einstein originally followed.  This involves taking the physical assumptions upon which pre-general relativity physics is built and showing how they led Einstein to postulate the theory.  The second road, while less intuitive, is more fundamental.  It begins, like most any other physical theory, with an action.  Both roads are very useful for different reasons.  We will therefore consider both, beginning with the axiomatic approach.  This will allow us to write out the complete field equations for general relativity, thereby specifying the complete theory.  Then, before looking at the action principle, we will look at the meaning of Einstein's fields equations, as well as some properties, consequences, and various solutions.  The we will conclude with the action/Lagrangian of general relativity.  

\subsection{The Equivalence Principle}
\label{sec:theequivalenceprinciple}

To begin our exposition of this idea, we will begin with an idea from introductory physics.  Imagine an observer, $S$, "in space" (the absence of any force fields like gravitational or electromagnetic), in an inertial frame.  Then, imagine a test object in the same frame as the observer.  Obviously the test object will be stationary relative to this observer, and therefore Newton's second law for this system is
\begin{eqnarray}
\bf F\it_{net} = 0
\end{eqnarray}
If the test object is experiencing an applied force $\bf f\it_{applied}$, this would be modified as
\begin{eqnarray}
\bf F\it_{net} = \bf f\it_{applied} = m\bf a\it \label{eq:equalenceprincipleequation3}
\end{eqnarray}

Now, consider another observer, $S'$, accelerating relative to the first observer.\footnote{Of course there is a fundamental difference in these two observers in that $S$ is in an inertial frame whereas $S'$ is not.  Therefore special relativity doesn't apply to $S'$.  Observer $S$ will not be able to tell the difference between his frame and any other inertial frame, but $S'$ will be able to "feel" the acceleration of his frame and will therefore know that he is accelerating relative to $S$.}  We will take their axes to line up at $t=0$, and the acceleration to be in the $z$ direction with acceleration $a_z$.  Now, because $S'$ is accelerating relative to the frame of $S$, $S'$ is also accelerating relative to the test object.  However, relative to $S'$, the test object is accelerating in the \it negative \rm $z$-direction.  
\begin{center}
\includegraphics[scale=.75]{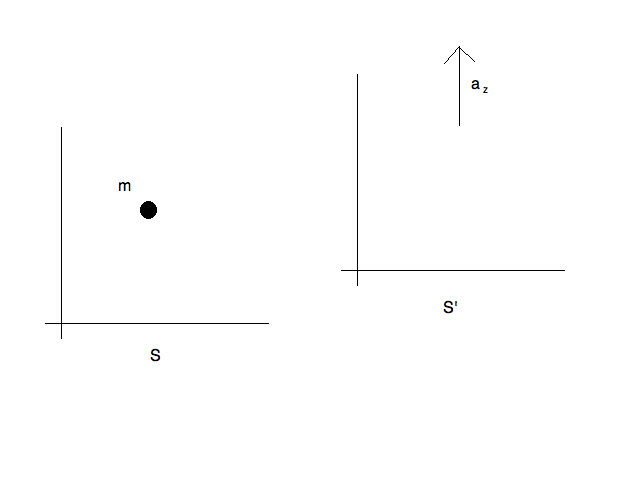}
\end{center}
Relative to $S'$, the the test object is accelerating.  Therefore, according to Newton's second law, the test object has a non-zero net force.  Because in classical mechanics these aren't actual forces but rather are a result of the frame of $S'$, they are called \bf Fictional Forces\rm, or sometimes \bf Pseudo-Forces\rm.  However, let's take them seriously and write out Newton's law for the object in this frame.  It is accelerating \it downward \rm (in the $z$ direction) with acceleration $a_z$, it has mass $m$, and so
\begin{eqnarray}
\bf F\it_{net} = -ma_z \bf e\it_z = m\bf a\it
\end{eqnarray}
where $\bf e\it_z$ is a unit vector in the positive $z$ direction.  If the object is experiencing an applied force $\bf f\it_{applied}$, this would be modified as
\begin{eqnarray}
\bf F\it_{net} = -ma_z \bf e\it_z + \bf f\it_{applied} = m\bf a\it \label{eq:equivalenceprincipleequation1}
\end{eqnarray}
We could rewrite (\ref{eq:equivalenceprincipleequation1}) as
\begin{eqnarray}
\bf F\it'_{net}  = \bf f\it_{applied} \label{eq:equivalenceprincipleequation2}
\end{eqnarray}
where
\begin{eqnarray}
\bf F\it'_{net} = \bf F\it_{net}+ma_z \bf e\it_z \label{eq:equivalenceprinciple5}
\end{eqnarray}
Note that (\ref{eq:equivalenceprincipleequation2}) has exactly the same form as (\ref{eq:equalenceprincipleequation3}).  In other words, if we are working with the "net force equals mass times acceleration" form of Newton's second law, we can generalize to non-inertial frames by replacing "net force" term on the left hand side with the fictional force terms.  The right hand side of the equation is then the same as in the inertial frame case.  

Now consider what will appear to be a very different situation.  Imagine an observer sitting on a planet, and therefore in the presence of a gravitational field.  We choose the observer's axes so that gravity is pointing in the negative $z$ direction with gravitational acceleration to be $g$.\footnote{On the surface of the Earth this would be $g=9.8\; m/s^2$.}  If there are no other forces acting on the object, Newton's second law gives
\begin{eqnarray}
\bf F\it_{net} = -mg\bf e\it_z = m\bf a\it
\end{eqnarray}
If there are any other forces acting on the object ($\bf f\it_{applied}$), this will be
\begin{eqnarray}
\bf F\it_{net} = -mg \bf e\it_z + \bf f\it_{applied} = m\bf a\it \label{eq:equivalenceprinciple8}
\end{eqnarray}
We can take the gravitational force term to the left hand side, getting
\begin{eqnarray}
\bf F\it'_{net} = \bf f\it_{applied}  \label{eq:equivalenceprinciple7}
\end{eqnarray}
where 
\begin{eqnarray}
\bf F\it'_{net} = \bf F\it_{net} + mg\bf e\it_z \label{eq:equivalenceprinciple4}
\end{eqnarray}

Now compare $\bf F\it'_{net}$ in (\ref{eq:equivalenceprinciple4}) to $\bf F\it'_{net}$ in (\ref{eq:equivalenceprinciple5}):
\begin{eqnarray}
(\ref{eq:equivalenceprinciple5})  \qquad & & \bf F\it'_{net} = \bf F\it_{net} + m a_z \bf e\it_z \nolabel \\
(\ref{eq:equivalenceprinciple4})  \qquad & & \bf F\it'_{net} = \bf F\it_{net} + mg \bf e\it_z
\end{eqnarray}
If we take the acceleration due to gravity to be $g=a_z$, we see a remarkable coincidence - \textit{the inertial mass in (\ref{eq:equivalenceprinciple5}) is equal to the gravitational mass in (\ref{eq:equivalenceprinciple4})}!  Of course this coincidence may not seem particularly profound.  But we will see that, in fact, it is this coincidence that forms almost the entire foundation of general relativity.  In fact, it is so important that it has its own name - \bf The Equivalence Principle\rm.  The equivalence principle can actually be stated several ways.  This is simply the first.   

Now consider an observer in a box with no windows and no means of observing anything outside of the box.  The observers can feel that he is accelerating towards the "floor" of the box\footnote{Of course he calls that particular wall of the box the "floor" merely because that is where he is accelerating.}.  He happens to have a test object of known mass $m$ with him (he carries it around with him in case he ends up inside a box like this one).  He holds it in front of him, lets go and of course it accelerates towards the ground.  He measures the acceleration of the object to be $a$.  He concludes that Newton's law for this object is
\begin{eqnarray}
\bf F\it_{net} = -ma \bf e\it_z
\end{eqnarray}
where once again $\bf e\it_z$ is simply a unit vector pointing towards the "ceiling".  Then, being somewhat familiar with Newton's laws in non-inertial frames, he realizes that there is an ambiguity in his situation.  Specifically, he wonders of he is in a box out "in space" that is accelerating at $a$ (with no other force fields around), or if he is in a box sitting still in the presence of a gravitational field with gravitational acceleration $a$.  After a few minutes of pondering he realizes a somewhat frustrating consequence of the equivalence principle.  If there was a slight difference in the inertial mass and the gravitational mass of the particle, perhaps he could devise some experiment to see which mass was coming into play in his box (assuming he knew the mass(s) of the object before getting into the box).  However, because gravitational mass equals inertial mass, he realizes despairingly, it is impossible for him to tell the difference.  Until he can get out of the box and look, \textit{there is no difference from his point of view between sitting still in a gravitational field and accelerating in the absence of a gravitational field}.  

Now consider another observer in the same type of box.  He doesn't feel any acceleration.  He also carries a test mass with him, so he holds it in front of him, let's go, and of course it remains floating in front of him - no motion at all (relative to him).  He initially concludes that he must be out in space (away from any force fields), and in an inertial frame.  Being familiar with special relativity, he despairs that because he can't see out of the box, he doesn't know anything about his motion other than the fact that it is inertial.  After a moments thought, however, he realizes that the situation is much worse.  If there is an observer in a gravitational field with gravitational acceleration $g$, but \textit{the observer is in "free fall" in this field}, then the observer's acceleration will obviously be $g$.  This means that, in the frame of the planet causing the gravitational field, the net force on the test mass is 
\begin{eqnarray}
\bf F\it_{net} = -mg\bf e\it_z
\end{eqnarray}
But, because the frame of this observer is accelerating at $g$ as well, the fictional force term will be $mg\bf e\it_z$.  So, equation (\ref{eq:equivalenceprinciple7}) becomes
\begin{eqnarray}
\bf F\it'_{net} = \bf F\it_{net} + mg\bf e\it_z = -mg\bf e\it_z + mg\bf e\it_z = 0
\end{eqnarray}
And therefore equation (\ref{eq:equivalenceprinciple8}) is
\begin{eqnarray}
\bf f\it_{applied} = 0
\end{eqnarray}
So, this observer realizes, not only does he not know what inertial frame he is in, but also he doesn't even know if he is in an inertial frame.  He could be in free fall the gravitational field of a planet!\footnote{This is not only frustrating because of a fundamental scientific and epistemological limitation, but also because he might be about to slam into a planet.}

These considerations bring us to another statement of the equivalence principle.  Specifically, \textit{It is impossible to distinguish between acceleration in the absence of gravitation from remaining stationary in the presence of gravitation}.  

Of course, the observer in the second box realizes that if the box were very large, and he had two test masses, he may be able to determine which situation he is in.  The following diagram illustrates how:
\begin{center}
\includegraphics[scale=.5]{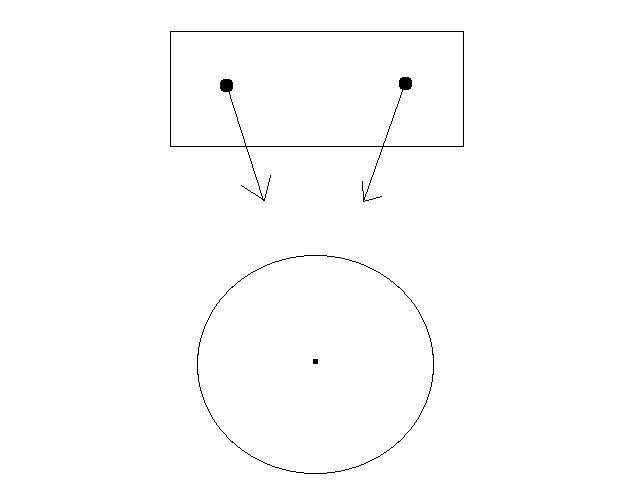} \label{twotestmassesgetclosetoeachotherfornonlocalliftexperiment}
\end{center}
The planet (the circle) pulls both the box and the test masses towards its center.  If the box is big enough, the observer in the box would be able to drop the two test masses and see if they move towards each other as they fall.  If the box is not sufficiently large, however, this is not possible.  

So, in light of this realization, we modify our statement of the equivalence principle above: \textit{It is impossible to distinguish between acceleration in the absence of gravitation from remaining stationary in the presence of gravitation, in a sufficiently small region of spacetime.}

The equivalence principle we have discussed so far is really nothing more than a coincidence of Newtonian theory.  It was the similarity of equations (\ref{eq:equivalenceprinciple5}) and (\ref{eq:equivalenceprinciple4}) that lead us (and the observers in the boxes) to realize the equivalence of various scenarios.  Of course, if it turned out that gravitational mass, $m_g$, is not equal to inertial mass, $m_i$, Newtonian theory would have no problem.  We would simply have two different mass terms in (\ref{eq:equivalenceprinciple5}) and (\ref{eq:equivalenceprinciple4}).  However, countless tests have been done to test this fact, and to date there is no evidence that there is a difference.  In other words, as far as we can tell experimentally,
\begin{eqnarray}
m_g = m_i
\end{eqnarray}
It was this that lead Einstein to make the underlying assumption that led to his general theory.  Rather than take the equality of $m_g$ and $m_i$ as an interesting coincidence of Newtonian theory, he decided to make it a fundamental postulate.  

The reason this is so important is that we can reason from what we know of the equivalence principle so far to yet another statement of it.  Recall from introductory physics that the force of a planet of mass $M$ on a test mass of mass $m$ is 
\begin{eqnarray}
F_{grav} = G{Mm \over r^2} 
\end{eqnarray}
Setting up Newton's second law,
\begin{eqnarray}
G{Mm \over r^2} = ma
\end{eqnarray}
The $m$ on the right side of this equation is the inertial mass - whereas the mass on the left side is the gravitational mass.  But because of the equivalence principle we know that they are equal, and therefore $m$ divides out of the equation.  
\begin{eqnarray}
a = G{M \over r^2}
\end{eqnarray}
So, the acceleration is independent of the mass of the object.  This leads to the alternative statement of the equivalence principle: \textit{The motion of an object in a gravitational field is independent of its mass or composition}.  

This statement, however, has extraordinarily profound implications.  This is because, if the motion of an object in a gravitational field is independent of its mass or composition, \textit{It is impossible to shield any object from the effects of gravitation - in other words, gravity couples to everything}.  

To see why this is so important, consider electromagnetism\label{pagewherewefirststartcmparinggravitytoeandminequivalenceprinciplesection}.  How can the strength of the electromagnetic field be measured?  The simplest way is simply to take two particles of known mass and composition - one that is uncharged and one that has a known charge.  The uncharged particle will not interact with the electromagnetic field and we can therefore take its motion to be "inertial".  It is then sensical and well defined to measure the acceleration of the charged particle \it relative to \rm the uncharged particle, and in doing so we know the force on the charged particle and through the equations of electromagnetism the strength of the electromagnetic field can be determined.  

However we don't have this luxury with gravity.  There is no equivalence principle for electromagnetism.  We cannot say that the motion of an object in an electromagnetic field is independent of its composition - it depends very much on the charge.  We cannot say that the electromagnetic field couples to everything - it doesn't couple to things that don't carry electric charge.  An observer trapped in a box can tell the difference between inertial motion outside of an electromagnetic field and "free fall" inside an electromagnetic field by observing the motion of objects with different composition and charge.  

But with gravity, not only is the motion of the test object completely independent of the composition, but also it is completely impossible to shield anything from gravity to act as an "inertial" object.  In other words, whereas with electromagnetism an uncharged object can act as the inertial object from which to compare the acceleration of the charged object, with gravity we have no such inertial objects because everything is affected by gravity.  

So, if everything is affected by gravity, and if the motion of everything is the same regardless of its composition, it becomes possible to set up a set of "preferred" world lines, or curves in spacetime, along which any arbitrary object will travel (and are therefore \it geodesic \rm lines).  And because these world lines don't depend at all on the objects traveling through spacetime, a natural conclusion is that the world lines don't depend on the objects, but instead depend on the spacetime itself.  And finally, this leads to the radical conclusion that the influence of gravity is not due to an field \it in spacetime \rm as in electromagnetism, but rather is due entirely to properties of the spacetime.  So, we have elevated spacetime from merely the background in which physics happens to a dynamical player in physics.  
Specifically, \textit{The spacetime metric is not flat as in special relativity ($\eta_{\mu\nu}$).  Rather, the motion of particles in a gravitational field are the geodesics on a curved spacetime}.  

This leap, from the assumption that gravity is a field like the electromagnetic field, to the assumption that in fact there is no gravitational field, but rather gravity is the effect of the curvature of spacetime, is a direct consequence of the equivalence principle, the fundamental basis of general relativity, and along with quantum mechanics one of the most radical paradigm shifts in the whole of science.  

Recall that in special relativity a key idea is that of an inertial frame.  It is not possible (in special relativity) to uniquely choose an inertial frame that is "at rest" - in fact there is no notion of absolute rest or absolute motion.  However (in special relativity), it is possible to single out frames that are accelerating - there is a notion of absolute acceleration.  This is what allowed us to find the strength of the electromagnetic field - we can tell the difference between non-accelerating and accelerating.  But, with our new understanding of gravity, nothing can be shielded from gravity, and therefore there is no way of giving a meaningful definition of "the acceleration due to gravity".  

However, this doesn't mean that we must give up a notion of a preferred collection of reference frames.  Recall above that the observer in the box that is in free fall in the gravitational field will experience no acceleration, and there will be zero net force on an object in this frame.  This leads to the notion that we should \it define \rm "non-accelerating" as "freely falling in a gravitational field".  

Of course, this definition appears at first to be backwards.  We are saying that an object sitting still on the surface of the Earth is actually accelerating, whereas an object that is accelerating down towards the Earth at $9.8\; m/s^2$ is not actually accelerating.  However it is actually the most sensible definition.  Recall that in our new understanding, gravity is not a "force" but rather represents a deviation from a flat metric.  So, consider the object sitting still on the surface of the Earth.  Free fall would dictate that the geodesic it "wants" to follow is the one that moves it through spacetime towards the Earth's center.  However, because of the normal force from the surface of the earth, the object doesn't follow this geodesic.  Because in our new paradigm, gravity is not a force, the only true force acting on the object is the normal force of the ground.  Therefore, with this single force, the object is actually accelerating away from the center of the Earth!  The fact that, when an observer is standing still on the surface of the earth, he \it feels \rm the force of the earth on his feet is further evidence of this.  On the other hand, the observer freely falling in the gravitational field doesn't "feel" any acceleration.\footnote{Consider the Vomit Comet.}  So, our designation of free call as non-accelerating and stationary as accelerating is actually the most sensible way to think!  

To emphasize this, recall from section \ref{sec:geodesics} ff that a geodesic is the path a particle will follow in the absence of any other forces.  Any deviation from a geodesic is caused by an external force.  So, we can take an object with no force to be an object following a geodesic.  As we said, an object following a geodesic will freely fall and "accelerate" (in the Newtonian sense) towards the center of the Earth.  Because it is following a geodesic, it isn't experiencing an external force, and therefore is "non-accelerating".  However, the object sitting still on the surface of the earth \it is \rm deviating from a geodesic, and therefore is experiencing an external force, and therefore is accelerating.  

As a brief comment, the previous considerations allow us to make a final observation.  Consider an observer in free fall in a gravitational field.  This observer is by definition not accelerating, but rather is merely following a geodesic without any external forces.   But according to the equivalence principle, this is absolutely identical to being in an inertial frame in the absence of any gravitational field.  So, we have the conclusion that, for any gravitational field, there is a reference frame (namely, the free fall frame) in which \textit{the effects of gravity are completely non-existant}.  For an observer in a free fall frame, there essentially is no gravity whatsoever, and there is no local\footnote{Of course, he could look around to see if there is a planet nearby, or if he had enough space he could set up an experiment like in the picture on page \pageref{twotestmassesgetclosetoeachotherfornonlocalliftexperiment}.  By "local" we mean that he cannot do these types of experiments that require looking beyond his immediate surroundings - he's trapped in a small box like the observers discussed above.} experiment he can do to detect any gravity.  So, in such a frame gravity can be completely ignored (locally).  

However, other forces like electromagnetism don't have an equivalence principle, and therefore there does not exist a privileged frame frame where the electromagnetic field can be ignored.  It is always possible to choose an object that is uncharged, or with a different charge, to compare one's motion to.  This difference between gravity and all other forces is a particularly important quirk of the equivalence principle.  We will see that it has radical physical consequences when we attempt to create a geometrical theory of non gravitational forces.  The primary difference, as we will see, between gravity and the other forces is that gravity is a theory of the geometry "of" spacetime, whereas the other forces are theories of the geometry "in" spacetime.  Another way of putting this is that gravity involves altering the spacetime metric and consequently the metric connection is altered.  In this sense general relativity is a theory of the \bf Levi-Civita Connection\rm, (cf section \ref{sec:metricconnection}).  The other forces, on the other hand, are theories where the spacetime metric is unchanged, but the connection takes on a form \it other than \rm the Levi-Civita connection - the fields create a non-metric connection on top of the metric connection.  To make better sense of this, consider the rather strange example of the connection in section \ref{sec:otherconnections}, with geodesics graphed starting on page \pageref{pagewithfirstweirdconnectiongeodescigraphs}.  Defining the connection required us to use the standard Euclidian metric on $\mathbb{R}^2$, but we were choosing a connection unrelated to the Euclidian metric.  And as a result, despite the space being flat (according to the metric), the geodesics were not the usual straight lines we'd expect in flat $\mathbb{R}^2$.  The meaning of this (which we will spend the next paper discussing in much, much greater detail) is that the connection we chose, which was not the metric connection, acted as a "field" which changed the geodesics the particle was traveling under.  In other words, there are some particles that may "see" that connection and others that wouldn't.  Any particle that could see the connection would follow those geodesics.  Any particle that couldn't see that connection would follow the usual flat space straight line geodesics.  However, if the space wasn't flat - i.e. if we changed the metric of the space from Euclidian to non-flat, it wouldn't be possible for any particle to be oblivious, and all particles would follow the non-straight line geodesics.  Adding a field of some sort to the non-flat space would simply cause even more exotic geodesics.  It is \it this \rm that outlines the difference between gravity and the other forces.  All other forces create an additional connection in addition to the spacetime metric connection, and only those particles that can see that additional connection will alter their geodesics accordingly\footnote{We will call "seeing" these connections "charge".}.  Those particles that cannot see the additional connection will ignore it and follow the geodesic any particle would follow were the additional connection not there.  On the other hand, because gravity represents a deviation in the connection \it of the spacetime metric\rm, any particle that is in spacetime (which is, of course, all particles) will "see" it and therefore the geodesics of \it all \rm particles will be affected.  This is the essential content of the equivalence principle - nothing can be shielded from gravity because gravity affects spacetime itself.  

All of this brings us to our first mathematical result of this section.  We are claiming that gravity is the result of deviations from a flat spacetime metric.  And therefore, the absence of "gravity" is actually the absence of curvature.  So, in the absence of curvature the spacetime metric is flat.  And therefore the geodesics are "straight lines" through spacetime - corresponding to solutions to 
\begin{eqnarray}
{d^2 x^{\mu} \over d\tau^2} = 0
\end{eqnarray}
(cf section \ref{sec:geodesics}, where the connection coefficients vanish because the metric is flat, cf equation (\ref{eq:levicivitaconstraint})).  However, in the presence of spacetime curvature, the metric is not flat, and according to (\ref{eq:geodesicdifferentialequation}), the geodesic equations will be
\begin{eqnarray}
{d^2 x^{\alpha} \over d\tau^2} + \Gamma^{\alpha}_{\mu\nu} {dx^{\mu} \over d\tau}{dx^{\nu} \over d\tau} = 0 \label{eq:geodesicequationinequivalenceprinciplesection}
\end{eqnarray}
The solutions to this will obviously not be straight lines for $\Gamma^{\alpha}_{\mu\nu} \neq 0$.  

As a brief comment before moving on, consider what we have done from a more mathematical point of view.  In special relativity the underlying theme was \it all inertial frames are equivalent\rm.\footnote{This was called the \bf Principle of Special Covariance\rm, or as the \bf Principle of Lorentz Covariance\rm.}  Mathematically this was stated as the idea that all observers related by Lorentz transformations are equivalent.  A Lorentz transformation is defined as a transformation leaving the metric $\eta_{\mu\nu}$ unchanged.  In other words, the physical statement "all inertial observers are equivalent" is the same as the more mathematical statement "observers related by transformations preserving the metric".  However, we are now allowing for transformations that \it don't \rm preserver the metric - the whole point is that we are considering the metric to be dynamical, and the deviations from flat are interpreted as gravity!  So, we are now talking about a more general set of observers - not merely the ones who see the same metric as in special relativity.  This leads to another perspective on general relativity - the \textbf{Principle of General Covariance}, which says that \textit{all observers are equivalent}.  

\subsection{Einstein's Field Equations}

This now raises the question - what causes the metric to curve.  We have essentially said
\begin{eqnarray}
gravity \; = \; curvature
\end{eqnarray}
but what causes curvature?  Finding the geodesic equations requires knowing $\Gamma^{\alpha}_{\mu\nu}$, and finding $\Gamma^{\alpha}_{\mu\nu}$ requires knowing the metric.  So, how can we find the metric, and what makes it curved?  When will it be flat? In other words,
\begin{eqnarray}
gravity \; = \; curvature\; = \; ?
\end{eqnarray}
Most introductory texts or notes on general relativity will, at this point, dive into a discussion of a idea called \bf Mach's Principle\rm.  This is somewhat misleading, however, because while Mach's thinking initially inspired Einstein, Einstein ultimately rejected the central notions of Machian thought, and general relativity is not a truly Machian theory.  Rather than discuss Mach's principle, we mention some of the features that survived in general relativity.  

In special relativity, the principle of special covariance stated that all inertial frames are equivalent.  The meaning of this is that there is no such thing as absolute motion or absolute rest.  No observer can be said to be truly and absolutely "at rest" or truly "in motion".  Only relative velocities can be measured.  However, in the context of special relativity, it is very easy to differentiate between inertial and non-inertial observers.  This is precisely the limitation of special covariance - it is limited to inertial frames (frames related by transformations that preserve the metric $\eta_{\mu\nu}$).  Einstein's leap was to reject this distinction between inertial and non-inertial frames - he said that not only is it true that motion is relative, but also it is true that acceleration is relative.  In other words, one cannot talk about absolute acceleration - only acceleration \it relative to something else\rm.  If one were to be in a universe with no other matter in it, acceleration wouldn't be possible (the idea of acceleration couldn't even be defined).  However, because we do live in a universe with matter it it, we can talk about acceleration in a meaningful way \it relative \rm to that matter.  And because it is matter that allows us to talk about acceleration, Einstein guessed that it is specifically this matter that is the \it cause \rm of acceleration!  And from special relativity we know that matter is nothing more than energy ($E=mc^2$), he made the guess that the underlying equation of general relativity should be
\begin{center}
gravity \; = \; curvature \; = \; energy
\end{center}
Or in mathematical terms (cf equation (\ref{eq:initialstatementofmeaningofEinsteinfieldequationsbeforerelativisticformluationssection})),
\begin{eqnarray}
G_{\mu\nu} = \kappa T_{\mu\nu} \label{eq:einsteinfieldequationwithkappaundefined}
\end{eqnarray}
(where $\kappa$ is a proportionality constant to be determined later, $G_{\mu\nu}$ is the Einstein tensor, and $T_{\mu\nu}$ is the energy-momentum tensor).  Equations (\ref{eq:einsteinfieldequationwithkappaundefined}) are called the \rm Einstein Field Equations\rm.  They are a set of differential equations for the metric (cf the end of section \ref{sec:s2otimesbbr}) which explain how curvature is governed by energy/mass.  

Notice that (\ref{eq:einsteinfieldequationwithkappaundefined}) has the exact form we stated it would have in (\ref{eq:initialstatementofmeaningofEinsteinfieldequationsbeforerelativisticformluationssection}).  

The implications of equation (\ref{eq:einsteinfieldequationwithkappaundefined}) are profound.  They state that, on the one hand, the motion of matter in the universe is determined by the geometry of the universe.  On the other hand the geometry of the universe is determined by the distribution of matter in the universe.  We will see the results of this in several examples of solutions to (\ref{eq:einsteinfieldequationwithkappaundefined}) later in this chapter.   

\subsection{The Correspondence Principle and Newtonian Gravity}

Einstein's general relativity, as encapsulated in equation (\ref{eq:einsteinfieldequationwithkappaundefined}), leads us to ask two questions.  The first is simply what the value of $\kappa$ should be.  The second is the slightly more complicated question of whether or not general relativity is consistent to Newtonian gravity in the appropriate limit.  We treat these two questions together because answering one will answer the other.  Namely, by showing that (\ref{eq:einsteinfieldequationwithkappaundefined}) is consistent with Newton's "one over $r$ squared" law, we will be able to see what the value of $\kappa$ should be.  

We can rewrite (\ref{eq:einsteinfieldequationwithkappaundefined}) in terms of the definition of the Einstein tensor (cf equation (\ref{eq:firstandonlydefinitionofEinsteintensor})), getting
\begin{eqnarray}
R_{\mu\nu} - {1 \over 2} Rg_{\mu\nu} = \kappa T_{\mu\nu}
\end{eqnarray}
We can take a trace of both sides (contracting the indices with the metric) and, denoting the trace of the energy momentum tensor $g^{\mu\nu}T_{\mu\nu} \equiv T$,
\begin{eqnarray}
g^{\mu\nu}\bigg(R_{\mu\nu} - {1 \over 2}Rg_{\mu\nu}\bigg) = \kappa T
\end{eqnarray}
Then, using (\ref{eq:thetraceofthemetricisequaltothedimensionofthemanifold}) (along with the assumption that we are working with $4$ dimensional spacetime) and (\ref{eq:firstandonlydefinitionofricciscalar11}), we have
\begin{eqnarray}
R - {1 \over 2} R 4 = \kappa T
\end{eqnarray}
or
\begin{eqnarray}
R = -\kappa T
\end{eqnarray}
We can plug this back into (\ref{eq:einsteinfieldequationwithkappaundefined}), which effectively swaps the role of $R_{\mu\nu}$ and $T_{\mu\nu}$:
\begin{eqnarray}
R_{\mu\nu} + {1 \over 2} \kappa T g_{\mu\nu} = \kappa T_{\mu\nu}
\end{eqnarray}
or
\begin{eqnarray}
R_{\mu\nu} = \kappa T_{\mu\nu} - {1 \over 2} \kappa T g_{\mu\nu} \label{eq:onourwaytoNewtonianfromGR1}
\end{eqnarray}

Now consider a time-like unit vector $t^{\mu}$.  We can contract both sides of (\ref{eq:onourwaytoNewtonianfromGR1}) with this vector,
\begin{eqnarray}
R_{\mu\nu} t^{\mu}t^{\nu} = \kappa T_{\mu\nu} t^{\mu}t^{\nu} - {1 \over 2} \kappa T g_{\mu\nu} t^{\mu}t^{\nu} \label{eq:onourwaytoNewtonianfromGR2}
\end{eqnarray}
On the right hand side, the first term is simply the energy density observed by someone in the $t^{\mu}$ frame (cf equation (\ref{eq:energymomentumcontractedwithavectorgivesenergydensity1127})).  We will take it to simply designate the mass density, which we will denote $\rho$.  The second term on the right hand side is the dot product of $t^{\mu}$ with itself, which is $1$ because we assumed $t^{\mu}$ is a unit vector ($g_{\mu\nu} t^{\mu}t^{\nu} = t_{\mu}t^{\mu} = t^2 =1$).  So, (\ref{eq:onourwaytoNewtonianfromGR2}) is
\begin{eqnarray}
R_{\mu\nu} t^{\mu}t^{\nu} = \kappa \rho - {1 \over 2} \kappa T
\end{eqnarray}
Then, we can choose our vector $t^{\mu}$ to correspond to the rest frame of the matter $T_{\mu\nu}$ describes, meaning that the diagonal elements other than $T_{00}$ vanish.  Then, because $T_{00}$ is the energy density in the rest frame of the matter, $T_{00} = \rho$, and therefore $T = \rho$.  
\begin{eqnarray}
R_{\mu\nu}t^{\mu}t^{\nu} = \kappa \rho - {1 \over 2} \kappa \rho = {1 \over 2} \kappa \rho
\end{eqnarray}
or
\begin{eqnarray}
R_{\mu\nu}t^{\mu}t^{\nu} = {1 \over 2} \kappa \rho \label{eq:onourwaytoNewtoinafromGR3}
\end{eqnarray}

Now consider that the matter density is a sphere of radius $r$ and mass $M$ with uniform mass density (so $\rho V = M$).  The volume of this sphere will be
\begin{eqnarray}
V = {4 \over 3} \pi r^3
\end{eqnarray}
We can take a first and second time derivative of this, getting
\begin{eqnarray}
\dot V &=& 4 \pi r^2 \dot r \nolabel \\
\ddot V &=& 4\pi r^2 \ddot r + 8 \pi r \dot r^2
\end{eqnarray}
Now, considering this in the context of what we did in section \ref{sec:riccitensorsection11278943}, equation (\ref{eq:rrdotandrdoubledotatepsilontimegoestozero}) tells us that the second term here vanishes, leaving
\begin{eqnarray}
\ddot V = 4 \pi r^2 \ddot r
\end{eqnarray}
Then, plugging (\ref{eq:vdoubledotequalsriccivivjtimesvsubnzero}) into (\ref{eq:onourwaytoNewtoinafromGR3}) we have 
\begin{eqnarray}
{\ddot V \over V} = {1 \over 2} \kappa \rho = {1 \over 2} \kappa {M \over V}
\end{eqnarray}
or
\begin{eqnarray}
4\pi r^2 \ddot r = {1 \over 2} \kappa M
\end{eqnarray}
Or, the acceleration $a = \ddot r$ is
\begin{eqnarray}
a = { \kappa \over 8 \pi} {M \over r^2}
\end{eqnarray}
This is exactly the Newtonian expression for acceleration in a gravitational field if we set
\begin{eqnarray}
\kappa = 8 \pi G \label{eq:definitionofkappaineinsteinfieldequation}
\end{eqnarray}
where $G$ is Newton's gravitational constant.  

So, we have succeeded in not only showing that general relativity does indeed reduce to Newtonian gravity in the low curvature limit, but also finding the proportionality constant.  So, the full Einstein field equations can finally be written as
\begin{eqnarray}
G_{\mu\nu} = 8 \pi G T_{\mu\nu}
\end{eqnarray}
or if we work in "natural" units where $G=1$, the more familiar form,
\begin{eqnarray}
G_{\mu\nu} = 8\pi T_{\mu\nu} \label{eq:EinsteinsFullFieldEquations}
\end{eqnarray}

\subsection{The Function of Einstein's Field Equations}

There are several ways we can view (\ref{eq:EinsteinsFullFieldEquations}).  The first is "from right to left" - they are a set of equations to give you the geometry once the matter distribution ($T_{\mu\nu}$) is known.  In other words, in this view, we use whatever physical information we have to piece together the energy momentum tensor.  Then, by setting $G_{\mu\nu}$ equal to whatever this turns out to be, we have a set of differential equations for $g_{\mu\nu}$ which we can solve to completely determine the geometry.  This approach, while nice in principle, is rarely practical.  The Einstein tensor ends up being a highly non-linear expression of the derivatives of $g_{\mu\nu}$, and it is almost never possible to solve it for $g_{\mu\nu}$ directly.  

The second way to look at (\ref{eq:EinsteinsFullFieldEquations}) is "from left to right" - starting with some metric we compute $G_{\mu\nu}$ and consequently know everything about the matter distribution by simply reading off $T_{\mu\nu}$.  However, this is also rarely useful because it is rare that an arbitrarily chosen metric will yield a physically meaningful energy momentum tensor.  It is possible that this approach results in something useful or interesting, but that is not the norm.  

Finally, there is the way that has proven to be the most useful.  Typically, it is not the case that we know either side of (\ref{eq:EinsteinsFullFieldEquations}) completely but nothing at all about the other side.  We usually know a little about the matter and a little about the geometry.  This allows us to make a simultaneous choice of the general form of $g_{\mu\nu}$ and $T_{\mu\nu}$, which is called an \bf ansatz\rm.  Einstein's equation then provides a set of constraints between the two.  The example we did in section \ref{sec:s2otimesbbr} is a good example of this.  We were able to guess the general form of the metric in (\ref{eq:s2crossrexamplemetricwithacoeff}), and the form of the energy-momentum tensor in equation in (\ref{eq:firstexampleansatzins2crossr1examplesection1}).  We didn't have complete information about the metric or about the matter distribution, but by the ansatz we assumed we ended up with a much nicer differential equation which acted as a constraint on both sides.  How this works physically should be more clear when we do examples shortly. 

So, the way we will approach problems in general relativity is to assume a form of both the metric and the matter distribution, and then plug everything in to get a differential equation.  If our ansatz was well chosen, the solutions to the differential equation will give a good physically meaningful answer.  

\subsection{The Cosmological Constant and Additional Comments}

Before looking at actual solutions to (\ref{eq:EinsteinsFullFieldEquations}) we make a few more comments about the general structure of Einstein's field equations.  

It turns out that there is a way that (\ref{eq:EinsteinsFullFieldEquations}) could be modified.  This is through the addition of a \bf Cosmological Constant \rm term, denoted $\Lambda$.  This term is simply added to the field equation:
\begin{eqnarray}
& & G_{\mu\nu} = R_{\mu\nu} - {1 \over 2} R g_{\mu\nu} = 8\pi T_{\mu\nu} \nolabel \\
&\Longrightarrow& G_{\mu\nu} + \Lambda g_{\mu\nu} = R_{\mu\nu} - {1 \over 2} Rg_{\mu\nu} + \Lambda g_{\mu\nu} = T_{\mu\nu}
\end{eqnarray}
Initially Einstein was looking for solutions that would represent a static universe because he was unaware of the evidence that the universe had a beginning.  No such static universe solution exists for (\ref{eq:EinsteinsFullFieldEquations}) in a universe with matter, so he added the cosmological constant term.\footnote{However it has been said that Einstein later called the addition of the cosmological constant the biggest mistake of his life.}  If the cosmological constant term is chosen in a particular way, static solutions can exist but they are unstable.  With Hubble's discovery of an expanding universe and the resulting big bang theory, static solutions became unimportant and Einstein rejected the need for $\Lambda$.  

However, despite the initial apparent lack of necessity, $\Lambda$ has managed to stick around for a variety of reasons.  One of the most interesting reasons is if we consider the situation with no matter:
\begin{eqnarray}
G_{\mu\nu} + \Lambda g_{\mu\nu} = R_{\mu\nu} - {1 \over 2} Rg_{\mu\nu} + \Lambda g_{\mu\nu}  = 0
\end{eqnarray}
We can then bring the cosmological constant term to the other side,
\begin{eqnarray}
G_{\mu\nu} = R_{\mu\nu} - {1 \over 2} Rg_{\mu\nu} = -\Lambda g_{\mu\nu}
\end{eqnarray}
In this case we can interpret the cosmological as an alternative energy momentum tensor, $T_{\mu\nu} = -\Lambda g_{\mu\nu}$.  So, even in the absence of matter there is a non-zero energy momentum tensor, and we therefore interpret $\Lambda$ as the energy density of the vacuum.  

This interpretation is actually very convenient.  Recall from \cite{Firstpaper} that the probability for virtual particles to appear in the vacuum is non-zero.  This implies that there must be something in the vacuum which can cause this, so a non-zero energy density in the vacuum is an appropriate concept.  Quantum field theoretic considerations and renormalization imply that the value of $\Lambda$ is approximately equal to the fourth power of the Planck mass $m_p$ 
\begin{eqnarray}
\Lambda \approx m_p^4
\end{eqnarray}
where $m_p \approx 10^{19}$ GeV.  However, observations of the universe indicate that this is incorrect, and rather the actual value of $\Lambda$ is smaller by a factor of at least $10^{120}$ - this is called the \bf Cosmological Constant Problem\rm.  This is, to date, the largest discrepancy between theory and experiment in the entirety of science and is considered to be one of the most fundamental and important unanswered questions in physics.  

There is quite a bit more we could say about Einstein's equation, especially about how it can be generalized.  There are all kinds of terms we could add to it, but with the exception of the cosmological constant, they are almost always omitted for a variety of reasons, and we therefore won't bother discussing them.  

\section{The Schwarzschild Solution}

We are now in a position to talk about actual solutions to (\ref{eq:EinsteinsFullFieldEquations}).  We will begin with a discussion of the simplest case - the absence of matter, or the vacuum ($T_{\mu\nu} = 0$).  

\subsection{General Vacuum Solutions}

If there is no matter, we can take $T_{\mu\nu} = 0$, and therefore (\ref{eq:EinsteinsFullFieldEquations}) is (including the cosmological constant term for generality)
\begin{eqnarray}
R_{\mu\nu} - {1 \over 2} R g_{\mu\nu} + \Lambda g_{\mu\nu} = 0
\end{eqnarray}
Now contract both sides with the metric,
\begin{eqnarray}
& & g^{\mu\nu} R_{\mu\nu} - {1 \over 2} R g^{\mu\nu}g_{\mu\nu} + \Lambda g^{\mu\nu}g_{\mu\nu} = 0 \nolabel \\
&\Longrightarrow& R - 2 R + 4 \Lambda = 0 \nolabel \\
&\Longrightarrow& R = 4\Lambda
\end{eqnarray}
Or if the cosmological constant is zero,
\begin{eqnarray}
R=0
\end{eqnarray}
But, in the case of $\Lambda =0$, we also have from Einstein's equation with $T_{\mu\nu}=0$
\begin{eqnarray}
Rg_{\mu\nu} = 2R_{\mu\nu}
\end{eqnarray}
and therefore $R=0$ implies
\begin{eqnarray}
R_{\mu\nu} = 0 \label{eq:noenergymomentumtensorimpliesricciflat}
\end{eqnarray}
The most obvious solution to (\ref{eq:noenergymomentumtensorimpliesricciflat}) is the Minkowski metric
\begin{eqnarray}
\eta_{\mu\nu} \dot = 
\begin{pmatrix}
-1 & 0 & 0 & 0 \\
0 & 1 & 0 & 0 \\
0 & 0 & 1 & 0 \\
0 & 0 & 0 & 1
\end{pmatrix}
\end{eqnarray}
(recall that we showed that the connection coefficients all vanish for the metric $\eta_{\mu\nu}$ in equation (\ref{eq:theconnectioncoefficientsassociatedwiththeminkowskimetricvanish}), and therefore the Riemann tensor trivially vanishes, and therefore the Ricci tensor also trivially vanishes).  So flat space is indeed a solution to Einstein's equations in the absence of matter.  

We call the class of all solutions to (\ref{eq:noenergymomentumtensorimpliesricciflat}) to \bf Vacuum Solutions \rm to Einstein's equations.  And while flat Minkowski space is the simplest example, there are other non-trivial examples.  Consider, for example, the metric near a planet.  If the planet has radius $R$, then a point a distance $r > R$ will contain no matter and therefore $T_{\mu\nu} = 0$, but obviously the curvature will be non-flat.  So there must be a vacuum solution that is not flat.  We now find such a solution for the simplest case.  

\subsection{The Schwarzschild Solution}

We want to find the vacuum solution for Einstein's equations in the vicinity of a planet.  To make the problem as simple as possible, we will assume that the planet is a perfect sphere (which is certainly not an unreasonable assumption) with radius $R$ and total mass $M$.  This means that we can assume that the metric is spherically symmetric.  

Furthermore we will assume that the solution is static - this means two things.  First, it means that the metric doesn't depend explicitly on time:
\begin{eqnarray}
\partial_0 g_{\mu\nu} = 0
\end{eqnarray}
Second, it means that the $g_{0i}$ components must vanish:
\begin{eqnarray}
g_{0i} = g_{i0} = 0
\end{eqnarray}
(if these components were non-zero then spatial displacement would involve temporal displacement, which would contradict the assumption that our solution is static.  

Because we are working with a spherically symmetric metric we will work in spherical coordinates.  The general form of the metric in spherical coordinates is
\begin{eqnarray}
ds^2 = -dt^2 + dr^2 + r^2 (d\theta^2 + \sin^2\theta d\phi^2)
\end{eqnarray}
In order to preserve the spherical symmetry, the most general ansatz we can choose is
\begin{eqnarray}
ds^2 = -a(r) dt^2 + b(r) dr^2 + c(r) r^2 (d\theta^2 + \sin^2\theta d\phi^2)
\end{eqnarray}
We can replace $r$ by any function of $r$ without disturbing the spherical symmetry, so we have the freedom to set $c(r) = 1$ without losing any generality.  So, our ansatz is
\begin{eqnarray}
ds^2 = -a(r) dt^2 + b(r) dr^2 + r^2 (d\theta^2 + \sin^2\theta d\phi^2) 
\end{eqnarray}
However, it will be convenient for reasons that will be clear later if we make the replacements $a(r) = e^{2\nu (r)}$ and $b(r) = e^{2\lambda (r)}$, making our ansatz
\begin{eqnarray}
ds^2 = -e^{2\nu} dt^2 + e^{2\lambda} dr^2 + r^2 (d\theta^2 + \sin^2\theta d\phi^2)\label{eq:schwarzchildansatz1}
\end{eqnarray}

We can plug (\ref{eq:schwarzchildansatz1}) into the equations for the connection, Riemann tensor, and then Ricci tensor (a tedious exercise we leave to you, but encourage you to either work out by hand or at least write a computer program to do - it is more instructive to work this out than you might think), results in the Ricci tensor having the following non-zero components, which when combined with (\ref{eq:noenergymomentumtensorimpliesricciflat}) gives the following differential equations (the prime represents a derivative with respect to $r$):
\begin{eqnarray}
R_{00} &=& \bigg( -\nu'' + \lambda' \nu' - \nu'^2 - {2\nu' \over r} \bigg) e^{2\nu - 2\lambda} = 0 \nolabel \\
R_{11} &=& \nu'' - \lambda' \nu' + \nu'^2 - {2 \lambda' \over r} = 0 \nolabel \\
R_{22} &=& (1+r\nu' - r\lambda')e^{-2\lambda} - 1 =0 \nolabel \\
R_{33} &=& R_{22} \sin^2 \theta = 0 \label{eq:firstricciequationstogetschwarzschild}
\end{eqnarray}
The first of (\ref{eq:firstricciequationstogetschwarzschild}) implies
\begin{eqnarray}
-\nu'' + \lambda' \nu' - \nu'^2 - {2\nu' \over r} = 0
\end{eqnarray}
so adding this to the second of (\ref{eq:firstricciequationstogetschwarzschild}) then gives
\begin{eqnarray}
& & \nu'' - \nu'' + \lambda'\nu' - \lambda'\nu' - \nu'^2+\nu'^2 - {2 \over r}(\nu' + \lambda') = 0 \nolabel \\
&\Longrightarrow& \nu' + \lambda' = 0 \label{eq:nuprimepluslambdaprimeequals0}
\end{eqnarray}
Plugging this into the third of (\ref{eq:firstricciequationstogetschwarzschild}) gives
\begin{eqnarray}
& & (1+r\nu' - r\lambda')e^{-2\lambda} - 1 = 0 \nolabel \\
&\Longrightarrow& (1+r\nu' + r\nu')e^{-2\lambda} - 1 = 0 \nolabel \\
&\Longrightarrow& (1+2r\nu')e^{-2\lambda} = 1  \label{eq:onourwaytoschwarz}
\end{eqnarray}

Next, looking at (\ref{eq:nuprimepluslambdaprimeequals0}) we have
\begin{eqnarray}
& & \nu' + \lambda' = 0 \nolabel \\
&\Longrightarrow & {\partial \over \partial r} (\nu + \lambda) = 0 \nolabel \\
&\Longrightarrow & \nu + \lambda = constant = C
\end{eqnarray}
This means that the coefficient of, say, $g_{00}$ in the metric can be written
\begin{eqnarray}
-e^{2\nu} = -e^{2(C - \lambda)} = -e^{2C}e^{-2\lambda}
\end{eqnarray}
However, we assume that as $r\rightarrow \infty$ the metric should reduce to the Minkowski metric.  If we take $\lambda \rightarrow 0$ as $r \rightarrow \infty$, we get that $C$ must be $0$.  So,
\begin{eqnarray}
\nu + \lambda = C = 0 \qquad \Longrightarrow \qquad \nu = -\lambda \label{eq:nuequalsminuslambdaschwarz}
\end{eqnarray}

Plugging this into (\ref{eq:onourwaytoschwarz}) gives
\begin{eqnarray}
& & (1+ 2r\nu') e^{2\nu} = 1 \nolabel \\
&\Longrightarrow& {\partial \over \partial r} (re^{2\nu}) = 1 \nolabel \\
&\Longrightarrow& re^{2\nu} = r + A \nolabel \\
&\Longrightarrow& e^{2\nu} = 1 + {A \over r} \label{eq:schwarzwherewechoose2masintegrationconstant}
\end{eqnarray}
where $A$ is a constant of integration.  Then, using (\ref{eq:nuequalsminuslambdaschwarz}),
\begin{eqnarray}
& & e^{-2\lambda} = 1+{A\over r} \nolabel \\
&\Longrightarrow & e^{2\lambda} = \bigg(1+{A \over r}\bigg)^{-1}
\end{eqnarray}

So, finally, the metric (\ref{eq:schwarzchildansatz1}) is
\begin{eqnarray}
ds^2 = -\bigg(1 + {A \over r}\bigg) dt^2 + \bigg(1 + {A \over r}\bigg)^{-1} dr^2 +  r^2 (d\theta^2 + \sin^2\theta d\phi^2) \label{eq:schwarzschildmetric}
\end{eqnarray}
This solution is called the \bf Schwarzschild Metric\rm.  The first thing to notice about it is that in the limit where $r\rightarrow \infty$ (as we get infinitely far away from the planet), we have ${A \over r} \rightarrow 0$, and therefore (\ref{eq:schwarzschildmetric}) reduces to the Minkowski metric sufficiently far from the planet - exactly as we would expect.  

\subsection{Correspondence of Schwarzschild and Newton}

So what is the meaning of (\ref{eq:schwarzschildmetric}), and what is the value of $A$?  To see this we need to review some introductory physics first.  Recall that Newton's law of gravitation says that the gravitational force between a body of mass $M$ and a body of mass $m$ is given by
\begin{eqnarray}
\bf F\it = G{Mm \over r^2} \bf \hat r\it
\end{eqnarray}
where $G$ is Newton's gravitational constant and $\bf \hat r\it$ is a unit vector in the direction of the vector between the objects.  Or, the gravitational field due to a body of mass $M$ is
\begin{eqnarray}
\bf g\it = \bf a\it = G{M \over r^2}\bf \hat r\it \label{eq:schwarznewtonian1}
\end{eqnarray}
where $\bf a\it$ is the acceleration an object in the field of $M$ will experience.  We can also express this in terms of the gravitational potential $\Phi$,
\begin{eqnarray}
\bf g\it = \bf a\it = \boldsymbol{\nabla} \Phi \qquad where \qquad \Phi = -G {M \over r} \label{eq:schwarznewtonian2}
\end{eqnarray}

Now consider a particle in the Newtonian limit -  moving very slowly in a gravitational field weak enough to be approximated by a perturbation from $\eta_{\mu\nu}$, and where the metric is static (unchanging in time).  The assumption that the particles are moving slowly means that if we are parameterizing the spacetime coordinates $x^{\mu}$ of the particles with $\tau$, 
\begin{eqnarray}
{dx^i \over d\tau} << {dx^0 \over d\tau}
\end{eqnarray}
We therefore take ${dx^i \over d\tau} = 0$.  Now the geodesic equation will be
\begin{eqnarray}
{d^2 x^{\mu} \over d\tau^2} + \Gamma^{\mu}_{00} \bigg({dx^0 \over d\tau}\bigg)^2 = 0 \label{eq:geodesicequation1onthewaytonewtonianschwarz}
\end{eqnarray}
Now, the assumption that the metric is static (and therefore $\partial_0 g_{\mu\nu} = 0$) allows us to simplify the connection coefficient $\Gamma^{\mu}_{00}$:
\begin{eqnarray}
\Gamma^{\mu}_{00} &=& {1 \over 2} g^{\mu\lambda}(\partial_0g_{\lambda 0} + \partial_0g_{0\lambda} - \partial_{\lambda} g_{00}) \nolabel \\
&=&- {1 \over 2} g^{\mu\lambda} \partial_{\lambda}g_{00} \label{eq:newtonainlimitschwarz1}
\end{eqnarray}
Next, we use the assumption that the gravitational field is weak enough to be approximated by a perturbation from the flat metric - in other words we write
\begin{eqnarray}
g_{\mu\nu} = \eta_{\mu\nu} + h_{\mu\nu} \label{eq:schwartzhperturbonmetric}
\end{eqnarray}
where $h_{\mu\nu}$ is small ($|h_{\mu\nu}|<<1$).  

Now plugging all of this into (\ref{eq:newtonainlimitschwarz1}) we have
\begin{eqnarray}
\Gamma^{\mu}_{00} = -{1 \over 2} \eta^{\mu\lambda} \partial_{\lambda} h_{00}
\end{eqnarray}
where we neglect the $h^{\mu\nu}$ term because, again, $|h_{\mu\nu}|<<1$.  

The geodesic equation (\ref{eq:geodesicequation1onthewaytonewtonianschwarz}) is
\begin{eqnarray}
{d^2 x^{\mu} \over d\tau^2} = {1 \over 2} \eta^{\mu\lambda} \partial_{\lambda} h_{00} \bigg({dx^0 \over d\tau}\bigg)^2 \label{eq:anotherrequationonthewaytonewtonianpotentialasdfasdf7}
\end{eqnarray}
We have assumed that the metric is static and therefore its time derivative vanishes.  This means that the  $\mu=0$ component of this equation is
\begin{eqnarray}
{d^2 x^0 \over d\tau^2} = {1 \over 2} \eta^{00}\partial_0 h_{00}\bigg({dx^0 \over d\tau}\bigg)^2 = 0
\end{eqnarray}
which implies
\begin{eqnarray}
{dx^0 \over d\tau} = constant = C
\end{eqnarray}

Now consider the spatial part of (\ref{eq:anotherrequationonthewaytonewtonianpotentialasdfasdf7}).  The spatial part of the Minkowski metric $\eta_{\mu\nu}$ is just the identity matrix, so (\ref{eq:anotherrequationonthewaytonewtonianpotentialasdfasdf7}) is
\begin{eqnarray}
{d^2 x^i \over \partial \tau^2} = {1 \over 2} C^2\delta^{ij}\partial_j h_{00}  = {1 \over 2} C^2 \partial_i h_{00}
\end{eqnarray}
If we choose our $\tau$ parameterization so that $x^0 = \tau$, we get $C=1$ and this is
\begin{eqnarray}
{d^2 x^i \over d (x^0)^2} = {1 \over 2} \partial_i h_{00}
\end{eqnarray}
We can recognize the term on the left hand side of this as the acceleration the particle will undergo and therefore we set
\begin{eqnarray}
{d^2 x^i \over d(x^0)^2} = a^i
\end{eqnarray}
and so
\begin{eqnarray}
a^i = {1 \over 2} \partial_i h_{00}
\end{eqnarray}
Comparing this to (\ref{eq:schwarznewtonian2}), we see that we should identify
\begin{eqnarray}
h_{00} = -2\Phi = 2G{M \over r}
\end{eqnarray}
And therefore (finally), by (\ref{eq:schwartzhperturbonmetric}), 
\begin{eqnarray}
g_{00} &=& \eta_{00} + h_{00} \nolabel \\
&=&  \eta_{00} - 2\Phi \nolabel \\
&=& -1 + G{2M \over r} \nolabel \\
&=& -\bigg(1-G{2M \over r}\bigg)
\end{eqnarray}
where $M$ is the mass of the planet.  Obviously comparing this to our result for the Schwarzschild metric in equation (\ref{eq:schwarzschildmetric}) shows that it is identical (we could have done similar calculations to get the radial metric coefficient as well).  So, we can see that the value of the integration constant in (\ref{eq:schwarzwherewechoose2masintegrationconstant}) is
\begin{eqnarray}
A = -2GM
\end{eqnarray}

So, the true Schwarzschild metric for the vacuum around a planet of mass $M$ is
\begin{eqnarray}
ds^2 = -\bigg(1 - {2M \over r}\bigg) dt^2 + \bigg(1 - {2M \over r}\bigg)^{-1} dr^2 +  r^2 (d\theta^2 + \sin^2\theta d\phi^2)  \label{eq:finalschwarzschildmetric}
\end{eqnarray}
where we have gone back to units where $G=1$.  

\section{Geodesics of the Schwarzschild Metric}

Now that Einstein's equations have given us the vacuum metric solution for a spherically symmetric static massive body, we can get down the more interesting work of seeing what this metric implies by plugging the metric into the geodesic equations.  We will look at two examples of how an object will move under the gravitational influence of such a massive body, including motion around black holes and the perihelion of Mercury.  

The fact that "spherically symmetric" and "static" is a reasonable approximation for a great deal of what we see in space lends itself to the enormous explanatory power of the Schwarzschild metric, and the examples we will look at provide illustrations of some of the early experimental successes of Einstein's theory.  

\subsection{Black Holes}

There is one quirk that can be immediately seen the form of the Schwarzschild metric - namely there is a radius at which there is a singularity.  Note that when $r=2M$
\begin{eqnarray}
g_{00} &=& 0 \nolabel \\
g_{11} &=& \infty
\end{eqnarray}
Furthermore, note that for $r>2M$ the metric has the normal Minkowski $(-,+,+,+)$ signature.  In that the sign is the relativistic distinction between a "time" dimension and a "spatial" dimension, this is a necessary feature.  However, on the other side of the singularity ($r<2M$), the time dimension becomes positive and the radial dimension becomes negative - in other words the time dimension "becomes spatial" and the radial dimension "becomes time-like".  

What would this value be for, say, the earth?  Restoring all of the constants to standard units, the $g_{00}$ coefficient is
\begin{eqnarray}
g_{00} = -\bigg(1 - {2GM \over c^2 r}\bigg)
\end{eqnarray}
So, the actual "singularity" radius is
\begin{eqnarray}
r = {2GM \over c^2} = {2(6.67\times 10^{-11}{m^3 \over kg s^2})(5.97\times 10^{24} kg) \over (2.99 \times 10^{8}{m\over s})^2} = 8.91 \times 10^{-3} m
\end{eqnarray}
which obviously isn't a substantially large radius, despite the size the earth.  We don't have to worry about running into this radius accidentally.  For the sun, the value is about $2.98\times 10^3m$, or a little less than 2 miles.  But given that the radius of the sun is more than $432,000$ miles, we don't need to worry about running into this radius, either.  

But you can imagine a body with much, much greater mass than the Earth, in which case the radius would be much greater.  And if the mass density of the object is great enough, it is possible that this "singularity radius" is greater than the radius of the body.  Such an object is called a \bf Black Hole\rm, and the radius we have been calling the "singularity radius" is called the \bf Event Horizon\rm.  

So what are the properties of black holes, and how do things move near them?  We can find this by considering the geodesics of the Schwarzschild solution both inside and outside the $r=2M$ (with natural units again) event horizon.  

Let's start outside the event horizon with a particle falling into the black hole radially (so that $d\phi = d\theta = 0$).  We'll consider an observer watching this happen from a large distance away.  The observer will parameterize what he sees with the parameter $\tau$.  To find the geodesics we use the geodesic equation (\ref{eq:geodesicdifferentialequation}).  Starting with the time (or zero) component, this is
\begin{eqnarray}
& & {d^2 x^0 \over d\tau^2} + \Gamma^0_{\mu\nu} {dx^{\mu} \over d\tau}{dx^{\nu} \over d\tau} = 0 \nolabel \\
&\Longrightarrow& {d^2 x^0 \over d\tau^2} + {1 \over 2} g^{k0}\bigg({\partial g_{\nu k} \over \partial x^{\mu}} + {\partial g_{k\mu} \over \partial x^{\nu}} - {\partial g_{\mu\nu} \over \partial x^k}\bigg){dx^{\mu} \over d\tau}{dx^{\nu} \over d\tau} = 0 \nolabel \\
&\Longrightarrow& {d^2 x^0 \over d\tau^2} + {1 \over 2} g^{00}\bigg({\partial g_{\nu 0} \over \partial x^{\mu}} + {\partial g_{0\mu} \over \partial x^{\nu}} - {\partial g_{\mu\nu} \over \partial x^0} \bigg){dx^{\mu} \over d\tau}{dx^{\nu} \over d\tau} = 0 
\end{eqnarray}
The third term always vanishes because there is no $x^0=t$ dependence in any component of the metric, and the first two terms are only non-zero when $g_{00}$ is involved and when the derivative is with respect to $x^1=r$.  So, the geodesic equation is
\begin{eqnarray}
& & {d^2x^0 \over d\tau^2} + {1 \over 2} g^{00} \bigg({\partial g_{00} \over \partial x^1} {dx^1 \over d\tau}{dx^0 \over d\tau} + {\partial g_{00} \over \partial x^1} {dx^0 \over d\tau}{dx^1 \over d\tau}\bigg) = 0 \nolabel \\
&\Longrightarrow& {d^2x^0 \over d\tau^2} + g^{00} {\partial g_{00} \over \partial x^1} {dx^1 \over d\tau} {dx^0 \over d\tau} = 0 \nolabel \\
&\Longrightarrow& {d^2x^0 \over d\tau^2} + g^{00} {\partial g_{00} \over \partial \tau} {dx^0 \over \partial \tau} = 0 \nolabel \\
&\Longrightarrow& {d^2x^0 \over d\tau^2} + {1 \over g_{00}} {\partial g_{00} \over \partial \tau} {dx^0 \over \partial \tau} = 0 \nolabel \\
&\Longrightarrow& g_{00}{d^2x^0 \over d\tau^2} + {\partial g_{00} \over \partial \tau} {dx^0 \over \partial \tau} = 0 \nolabel \\
&\Longrightarrow& {d \over d\tau} \bigg( g_{00} {dx^0 \over d\tau}\bigg) = 0 \nolabel \\
&\Longrightarrow& g_{00} \dot t = k \label{eq:wherekisdefinedinblackholegeodesicsection}
\end{eqnarray}
where $k$ is some integration constant and the dot represents a derivative with respect to $\tau$.  Or writing this out,
\begin{eqnarray}
-\bigg( 1-{2M\over r}\bigg) \dot t = k \label{eq:blackholegeodesic1}
\end{eqnarray}

We can then assume that the parameter $\tau$ is such that the norm of the four-velocity $v^{\mu} = {dx^{\mu} \over d\tau}$ has unit length:
\begin{eqnarray}
g_{\mu\nu}v^{\mu}v^{\nu} = 1 \qquad &\Longrightarrow& \qquad  g_{00} (v^0)^2 + g_{11}(v^1)^2 = 1 \nolabel \\
&\Longrightarrow& \qquad (g_{00} v^0)^2 + g_{00}g_{11} (v^1)^2 = g_{00} \nolabel \\
&\Longrightarrow& \qquad k^2 - (v^1)^2 = 1-{2M \over r}
\end{eqnarray}
where we have recognized $\dot t = v^0$ and used the exact form of the metric to see that $g_{00}g_{11} = -1$ (cf (\ref{eq:finalschwarzschildmetric})).  Rearranging this,
\begin{eqnarray}
v^1 = {dx^1 \over d\tau} = \sqrt{k^2-1+{2M\over r}} \label{eq:blackholegeodesic2}
\end{eqnarray}

Now, the quantity that is of interest to us is the "$r$ velocity", or the derivative of the $r = x^1$ component with respect to the $t=x^0$ component, ${dr \over dt} = {dx^1 \over dx^0}$.  But, we can write this as
\begin{eqnarray}
{dr \over dt} = {(dr/d\tau) \over (dt/d\tau)} = {\dot r \over \dot t} = {v^1 \over \dot t} \label{eq:blackholegeodesic5}
\end{eqnarray}
Plugging in (\ref{eq:blackholegeodesic1}) and (\ref{eq:blackholegeodesic2}) this is
\begin{eqnarray}
{v^1 \over \dot t} = { \sqrt{k^2-1+{2M\over r}} \over { k \over \big(1-{2M\over r}\big)}} = {dr \over dt}
\end{eqnarray}
Rearranging this,
\begin{eqnarray}
{dt \over dr} = -k\bigg(1-{2M \over r}\bigg)^{-1}\bigg(k^2 - 1 + {2M \over r}\bigg)^{-1/2} \label{eq:blackholegeodesic3}
\end{eqnarray}
Consider the behavior the observer will see when the particle gets close to the event horizon, so that $r=2M+\epsilon$ (it should already be clear that the first term in big parentheses above is singular in the limit $\epsilon \rightarrow 0$).  Expanding (\ref{eq:blackholegeodesic3}) to first order in $\epsilon$ gives (sparing the tedious Taylor expansions),
\begin{eqnarray}
{dt \over dr} = -k\bigg(1-{2M \over 2M+\epsilon}\bigg)^{-1}\bigg(k^2 - 1 + {2M \over 2M+\epsilon}\bigg)^{-1/2} \approx -{2M \over \epsilon} = -{2M \over r-2M}
\end{eqnarray}
Integrating this,
\begin{eqnarray}
& & \int dt = -2M \int dr {1 \over r-2M} \nolabel \\
&\Longrightarrow& t = -2M \ln (r-2M) + Const.
\end{eqnarray}
Graphing this gives (for $r>2M$)
\begin{center}
\includegraphics[scale=.7]{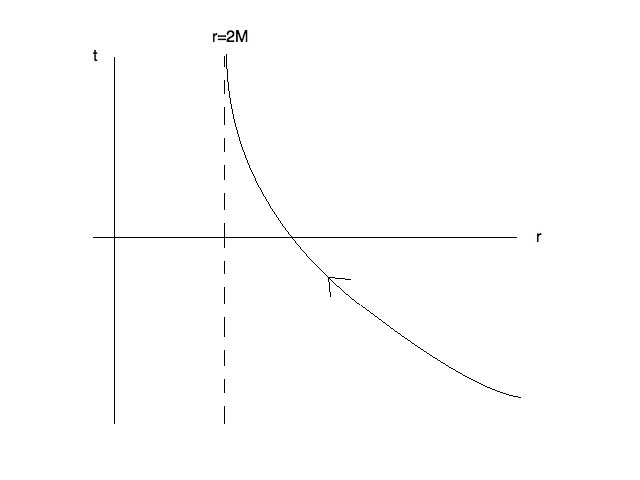}
\end{center}
So, as $r\rightarrow 2M$, we have $t\rightarrow \infty$.  In other words, to the observer watching the particle fall into the black hole, he sees it take an infinite amount of time for the particle to reach the event horizon.  He will watch the object fall forever.  

But certainly this must be incorrect!  What could cause the object to \it slow down \rm as it gets closer and closer to something that becomes more and more attractive?  Intuition (and agreement with Newton) would demand that it move faster and faster (relative to the black hole, and therefore relative to an observer stationary relative to the black hole) as it gets closer and closer to the event horizon.  

The solution to this apparent paradox is in the fact that we are speaking relative to what the observer \it sees\rm.  Keep in mind that what the observer sees is the light reflecting from the object back to him.  However, as the object gets closer and closer to the event horizon, the gravitational pull (i.e. spacetime curvature) gets stronger and stronger, and therefore the light rays have a harder and harder time getting back to the observer.  As the observer gets closer to $r=2M$, the light is less and less able to get from the falling particle to the observer, and therefore the particle appears to take an infinite amount of time to fall in.  

But what about an observer falling into the black hole?  What will he see?\footnote{Assuming he is concerned with mathematical physics while preparing to be crushed to death by a black hole.}  We can calculate this by simply considering the geodesic equations not with an arbitrary parameter $\tau$, but with the proper time observed by the falling particle.  In this case $x^0 = t = \tau$, and therefore we have (from (\ref{eq:blackholegeodesic5})),
\begin{eqnarray}
{dr \over dt} = {(dr/d\tau) \over (dt/d\tau)} = {(dr/d\tau) \over (d\tau /d\tau)} = {\dot r \over 1} = \dot r = v^1
\end{eqnarray}
So, 
\begin{eqnarray}
v^1 = -\sqrt{k^2 - 1 + {2M \over r}} = {dr \over dt}
\end{eqnarray}
or
\begin{eqnarray}
{dt \over dr} = -\bigg(k^2 - 1 + {2M \over r}\bigg)^{-1/2}
\end{eqnarray}
If we once again take $r=2M+\epsilon$, notice that we won't have the singular behavior like we did above.  Rather than actually using the $\epsilon$ approximation we used above, let's try to integrate this directly:
\begin{eqnarray}
\int dt &=& -\int dr {1 \over \sqrt{k^2 - 1 + {2M \over r}}} \nolabel \\
&=& -\int dr \sqrt{  {r \over r(k^2-1) + 2M} } \nolabel \\
&\equiv& -\int dr \sqrt{r \over rA + B} \label{eq:blackholegeodesic6}
\end{eqnarray}
where $A$ and $B$ are defined in the obvious way indicated.  The exact form of this integral is fairly ugly.  However, notice that the integral becomes very, very simple if we can take $A=0$.  While this may initially seem like a bit of a cop out, let's consider the physical meaning of such an imposition.  

If we are to take $A = k^2 - 1 = 0$, this is the same as setting $k=\pm 1$.  Looking back at the definition of $k$ in equation (\ref{eq:wherekisdefinedinblackholegeodesicsection}), it is defined by
\begin{eqnarray}
g_{00} \dot t = -\bigg(1 - {2M \over r}\bigg) \dot t = k
\end{eqnarray}
where the dot represents a derivative with respect to the parameter $\tau$.  If we assume that the particle starts from rest a long way from the black hole (so that $r$ is large), the value $k$ becomes the initial value of the $0,0$ component of the metric.  And if we assume that the particle starts a long way from the black hole, we would expect it to be the value of the flat Minkowski metric - it should be $1$.  So, not only is $k=1$ a reasonable constraint - it is in fact the \it most \rm reasonable value for $k$ to take.  

With that imposed, our integral (\ref{eq:blackholegeodesic6}) is then
\begin{eqnarray}
t &=& -\int dr \sqrt{{r \over rA + B}} \nolabel \\
&=& -{1 \over \sqrt{B}}\int dr r^{1/2} \nolabel \\
&=& -{3 \over 2\sqrt{B}} r^{3/2} + Const.
\end{eqnarray}
When graphed, this gives 
\begin{center}
\includegraphics[scale=.7]{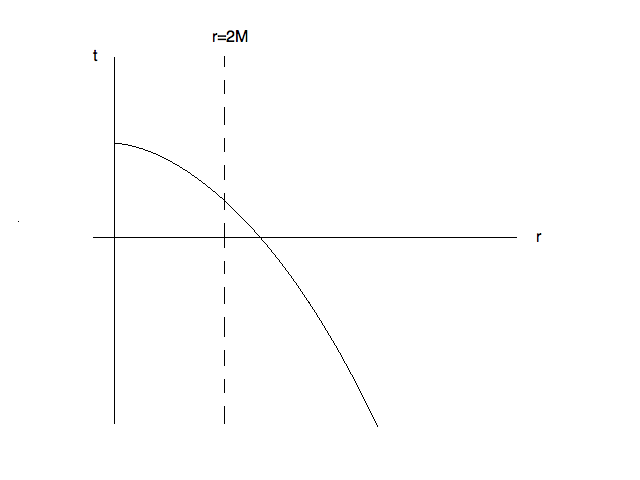}
\end{center}
So, an observer falling with the particle will go right past the event horizon to be promptly crushed inside the black hole.  

There is, as you would imagine, volumes more we could say about black holes.  But because this is merely meant to be an introduction to the basic ideas of general relativity (which we are including largely as an illustration of the geometry we considered in the first part of these notes), we won't continue our discussion.  

\subsection{Perihelion of Mercury}

Next we consider another illustration of the applications of general relativity.  But in order to appreciate this particular application, we first review (what should be) a familiar calculation from classical mechanics.  The following several pages will be a return to entirely Newtonian theory - forget about general relativity until we bring it up again.  The purpose of this is to (later) compare the Newtonian result to its relativistic generalization.  

We could derive the following (Newtonian) result by using Newton's laws directly, but instead we will take the Lagrangian approach.  The end result is the same.  Newton proposed the "one over $r$ squared" law for gravitation, where the force on an object of mass $m$ by an object of mass $M$ separated by a distance $\bf r\it = r \bf \hat r \it$ (where $\bf \hat r\it$ is the unit vector in the direction from $M$ to $m$ in a coordinate system where the center of $M$ is taken to be the origin - note however that $\bf r\it$ should be thought of as a position vector, not merely the radial component of a position vector) is 
\begin{eqnarray}
\bf F\it = -{Mm \over r^2} \bf \hat r\it
\end{eqnarray}
(where we are still taking $G=1$ and the minus sign is because the force on $m$ is towards the origin, which is the opposite direction as $\bf \hat r\it$).  If we take an arbitrary position vector in this coordinate system to be $\bf r\it$, then Newton's Law for this system is
\begin{eqnarray}
-{Mm \over r^2} \bf \hat r\it = m \bf \ddot r\it
\end{eqnarray}

Now consider the angular momentum $\bf L\it$ for this system.  This will be generally defined by
\begin{eqnarray}
\bf L\it = \bf r\it \times m \bf \dot r\it
\end{eqnarray}
where the $\times$ means the usual vector cross product in three spatial dimensions.  Consider the time derivative of $\bf L\it$:
\begin{eqnarray}
{d \bf L\it \over dt} = {d \over dt}(\bf r\it \times m \bf \dot r\it) = \bf r\it \times m \bf \ddot r\it + \bf \dot r \it \times m \bf \dot r\it
\end{eqnarray}
The second term vanishes because any vector crossed with itself is zero.  Then, plugging in Newton's law for the first term,
\begin{eqnarray}
\bf r\it \times m \bf \ddot r\it = \bf r\it \times \bigg(-{Mm \over r^2} \bf \hat r\it\bigg) \equiv 0
\end{eqnarray}
because $\bf r\it$ and $\bf \hat r\it$ are in the same direction and therefore their cross product vanishes as well.  So, we have
\begin{eqnarray}
{d \bf L\it \over dt} = 0
\end{eqnarray}
so the angular momentum is conserved.  This means that, while $m$ may rotate around $M$, it will not leave the plane it starts in.  We therefore proceed with the assumption that $m$ will move in a plane.  

We will write out the Lagrangian for $m$ in polar coordinates ($R$, $\phi$) to derive its equations of motion.  We can write the potential term for Newton's gravitational force law as
\begin{eqnarray}
V = -{Mm \over R}
\end{eqnarray}
So that 
\begin{eqnarray}
\bf F\it =- \boldsymbol{\nabla} V =-\bf \hat r\it { \partial \over \partial R} \bigg(-{Mm \over R}\bigg) = -{Mm \over R^2} \bf \hat r\it
\end{eqnarray}
Then the kinetic term will be
\begin{eqnarray}
T &=& {1 \over 2} m (\dot x^2 + \dot y^2) \nolabel \\
&=& {1 \over 2} m \bigg[\bigg({\partial \over \partial t}(R\cos\phi)\bigg)^2 + \bigg({\partial \over \partial t}(R\sin\phi)\bigg)^2\bigg] \nolabel \\
&=& \cdots \nolabel \\
&=& {1 \over 2} m (\dot R^2 + R^2\dot \phi^2)
\end{eqnarray}
So, the full Lagrangian is
\begin{eqnarray}
L = T-V = {1 \over 2} m(\dot R^2 + R^2 \dot \phi^2) + {Mm \over R}
\end{eqnarray}
We can write out the equations of motion for this, starting with the $\phi$ equations:
\begin{eqnarray}
{\partial L \over \partial \phi} &=& 0 \nolabel \\
{d \over dt} {\partial L \over \partial \dot \phi} &=& {d \over dt}(mR^2 \dot \phi) = 0
\end{eqnarray}
Integrating the second equation gives
\begin{eqnarray}
R^2\dot \phi = l \label{eq:periangmomcons1}
\end{eqnarray}
where $l$ is some integration constant.  Equation (\ref{eq:periangmomcons1}) is a statement of the conservation of angular momentum.  

The $R$ equations of motion then come from
\begin{eqnarray}
{\partial L \over \partial R} &=& mR\dot \phi^2 - {Mm \over R^2} \nolabel \\
{d \over dt}{\partial L \over \partial \dot R} &=& m\ddot R \nolabel \\
{d \over dt} {\partial L \over \partial \dot R} - {\partial L \over \partial R} = 0 \qquad &\Longrightarrow& \qquad \ddot R - R\dot \phi^2 + {M \over R^2} = 0 \label{eq:pariangmomcons2}
\end{eqnarray}

So our two equations of motion are (\ref{eq:periangmomcons1}) and (\ref{eq:pariangmomcons2}).  We can plug the first into the second:
\begin{eqnarray}
\ddot R - {l^2 \over R^3} + {M \over R^2} = 0 \label{eq:pariangmomcons3}
\end{eqnarray}

Now, introduce the variable 
\begin{eqnarray}
u \equiv {1 \over R} \label{eq:pariangmomcons4}
\end{eqnarray}
Now,
\begin{eqnarray}
\dot R = {\partial \over \partial t} R = {\partial \over \partial t}\bigg({1 \over u}\bigg) = -{\dot u \over u^2} 
\end{eqnarray}
However, we can re-express this as
\begin{eqnarray}
-{1 \over u^2} \dot u &=& -{1 \over u^2} {\partial u \over \partial t} \nolabel \\
&=& -{1 \over u^2} {\partial u \over \partial \phi} {\partial \phi \over \partial t} \nolabel \\
&=& -{1 \over u^2} \dot \phi {\partial u \over \partial \phi} \nolabel \\
&=& -{1 \over u^2} {l \over R^2}{ \partial u \over \partial \phi} \nolabel \\
&=& -{1 \over u^2} l u^2 {\partial u \over \partial \phi} \nolabel \\
&=& - l {\partial u \over \partial \phi} \label{eq:pariangmomcons9}
\end{eqnarray}
where we used (\ref{eq:periangmomcons1}) and (\ref{eq:pariangmomcons4}).  Then, 
\begin{eqnarray}
\ddot R &=& {\partial \over \partial t} \bigg( -{\dot u \over u^2}\bigg) \nolabel \\
&=& -{ u^2 \ddot u - 2u\dot u^2 \over u^4} \nolabel \\
&=& 2{\dot u^2 \over u^3} - {\ddot u \over u^2} \nolabel \\
&=& {2 \over u^3} \bigg( {\partial u \over \partial t}\bigg)^2 - {1 \over u^2} {\partial \over \partial t}{\partial u\over \partial t} \nolabel \\
&=& {2 \over u^3} \bigg( lu^2 {\partial u \over \partial \phi}\bigg)^2 - {1 \over u^2} lu^2 {\partial \over \partial \phi} \bigg(lu^2{\partial u \over \partial \phi}\bigg) \nolabel \\
&=& 2l^2 u \bigg({\partial u \over \partial \phi}\bigg)^2 - l^2 \bigg(u^2 {\partial^2 u \over \partial \phi^2} + 2u {\partial u \over \partial \phi}{\partial u \over \partial \phi}\bigg) \nolabel \\
&=& 2l^2 u \bigg({\partial u \over \partial \phi}\bigg)^2 - l^2 u^2{\partial^2 u \over \partial \phi^2} - 2l^2u \bigg({\partial u \over \partial \phi}\bigg)^2 \nolabel \\
&=& -l^2 u^2 {\partial^2 u \over \partial \phi^2} \label{eq:pariangmomcons5}
\end{eqnarray}

Now, plugging (\ref{eq:pariangmomcons5}) and (\ref{eq:pariangmomcons4}) into (\ref{eq:pariangmomcons3}), we have
\begin{eqnarray}
-l^2 u^2 {\partial^2 u \over \partial \phi^2} - l^2 u^3 + M u^2 = 0
\end{eqnarray}
or
\begin{eqnarray}
{\partial^2 u \over \partial \phi^2} + u - {M \over l^2} = 0 \label{eq:Binet}
\end{eqnarray}
Equation (\ref{eq:Binet}) is called the \bf Binet equation\rm, and it is a differential equation for $u={1 \over R}$ whose solutions give the equation for the orbital motion $R = R(\phi)$.  The solutions are\footnote{It is relatively straightforward to solve this equation so we leave the details to you.} 
\begin{eqnarray}
u(\phi) = {1 \over R(\phi)} = {M \over l^2} + A\cos(\phi - B)
\end{eqnarray}
where $A$ and $B$ are constants of integration.  Rewriting this for $R$,
\begin{eqnarray}
R(\phi) = {l^2 \over M+Al^2 \cos(\phi - B)} = { {l^2 \over M} \over 1+{Al^2 \over M} \cos(\phi - B)}
\end{eqnarray}
Or, defining $L \equiv {l^2 \over M}$ and $E \equiv {Al^2 \over M}$, 
\begin{eqnarray}
R(\phi) = {L \over 1+ E \cos(\phi-B)}
\end{eqnarray}
You should recognize this as the equation for an ellipse, so long as $0<E<1$.  If $E=0$ this is the equation for a circle:
\begin{eqnarray}
R(\phi) = L
\end{eqnarray}
If $E$ is between $0$ and $1$, however, the graph is an ellipse along the axis defined by the angle $B$.  
\begin{center}
\includegraphics[scale=.8]{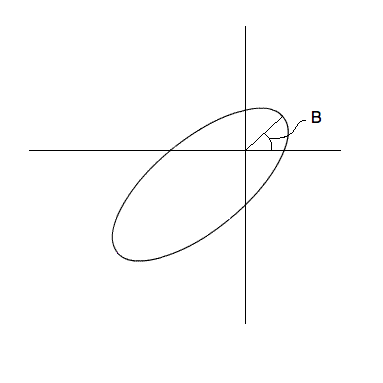}
\end{center}
The closer $E$ gets to $1$ the "longer" the ellipse gets.  The angle $B$ specifies the angle where the graph comes the closest to the origin - or the point where the orbiting object $m$ comes the closest to $M$.  This location, the point of closest approach, is called the \bf perihelion \rm of the orbit.  

So, this shows us that according to Newton's gravitational law, objects orbit planets in ellipses.  

For $E$ greater than or equal to $1$, the graph is no longer an ellipse but rather is a parabola:
\begin{center}
\includegraphics[scale=.7]{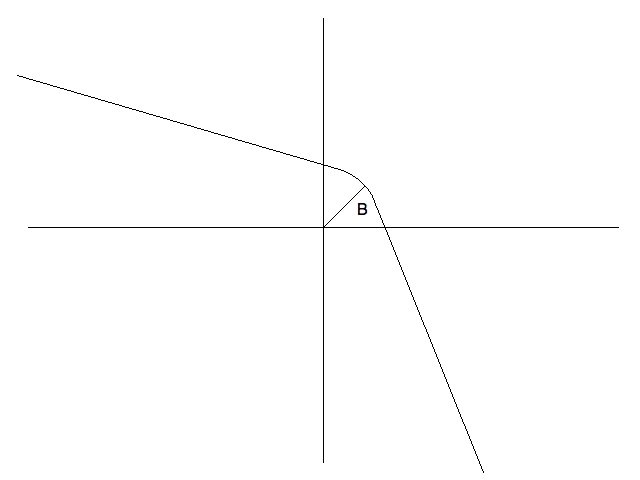}
\end{center}
This is the case where $m$ comes close enough to $M$ for its trajectory to be affected, but not close enough for orbit or to fall into $M$.  

In either case, both $A$ and $B$ are constants, the mass of the planet $M$ is assumed to be constant, and the value $l$ is constant (because angular momentum is conserved).  So, according to Newtonian gravitation, $m$ will move through the same elliptical orbit forever - it will never leave the plane it starts in and the perihelion will never change.  

But this lead to an observational discrepancy.  It was observed that the perihelion of the planet Mercury was not constant - its $B$ value was precessing, or rotating around the sun.  This type of affect could be loosely explained using Newtonian gravitation once the other planets were taken into affect - it is possible for the remaining mass in the solar system to cause the perihelion of one planet to precess.  However, the rate of the precession was nowhere near what Newtonian theory could justify.  A number of ad hoc approaches were considered, but none were successful.  

However, Einstein showed that general relativity predicts exactly the observed precession of Mercury's perihelion.  We will now reproduce this calculation.  The gravity of the sun, which is a spherical body, can be modeled by the Schwarzschild Metric:\footnote{Of course, because the radius of the sun is much, much larger than the radius of its event horizon, we won't have to worry about Mercury's orbit approaching it.}
\begin{eqnarray}
ds^2 = -\bigg(1-{2M \over r}\bigg) dt^2 + \bigg(1-{2M \over r}\bigg)^{-1} dr^2 + r^2 (d\theta^2 + \sin^2\theta d\phi^2)
\end{eqnarray}
In (\ref{eq:wherekisdefinedinblackholegeodesicsection}) we showed that one of the equations of motion for a particle moving near a Schwarzschild body is
\begin{eqnarray}
{d \over d\tau} \bigg[\bigg(1-{2M \over r}\bigg) \dot t\bigg] = 0
\end{eqnarray}
We can use the geodesic equation to find the remaining equations.  It turns out that while the $\mu=1$ equation is difficult to compute, it is not necessary.  Therefore, we'll start with the $\mu=2$ equation:
\begin{eqnarray}
& & {d^2 x^2 \over d\tau^2} + \Gamma^2_{\mu\nu} {dx^{\mu} \over d\tau}{dx^{\nu} \over d\tau} = 0 \nolabel \\
& & {d^2 x^2 \over d\tau^2} + {1 \over 2} g^{k2}\bigg({\partial g_{\nu k } \over \partial x^{\mu}} + {\partial g_{k\mu} \over \partial x^{\nu}} - {\partial g_{\mu\nu} \over \partial x^k}\bigg){dx^{\mu} \over d\tau}{dx^{\nu} \over d\tau} = 0  \nolabel \\
& & {d^2 x^2 \over d\tau^2} + {1 \over 2} g^{22}\bigg({\partial g_{\nu 2} \over \partial x^{\mu}} + {\partial g_{2\mu} \over \partial x^{\nu}} - {\partial g_{\mu\nu} \over \partial x^2} \bigg) {dx^{\mu} \over d\tau}{dx^{\nu} \over d\tau} = 0 \nolabel \\
& & 2g_{22}{d^2 x^2 \over d\tau^2} + 2{\partial g_{22} \over \partial x^1} \dot x^2 \dot x^1 - {\partial g_{33} \over \partial x^2} (\dot x^3)^2 = 0 \nolabel \\
& & 2r^2 \ddot \theta + 2(2r) \dot r\dot \theta - 2r^2\sin\theta\cos\theta \dot\phi^2 = 0 \nolabel \\
& & {d \over d\tau} (r^2 \dot \theta) - r^2\sin\theta\cos\theta\dot\phi^2 = 0
\end{eqnarray}

Then, finally, the $\mu=3$ equation:
\begin{eqnarray}
& & {d^2 x^3 \over d\tau^2} + \Gamma^3_{\mu\nu} {dx^{\mu} \over d\tau}{dx^{\nu} \over d\tau} = 0 \nolabel \\
& & {d^2 x^3 \over d\tau^2} + {1 \over 2} g^{k3}\bigg({\partial g_{\nu k } \over \partial x^{\mu}} + {\partial g_{k\mu} \over \partial x^{\nu}} - {\partial g_{\mu\nu} \over \partial x^k}\bigg){dx^{\mu} \over d\tau}{dx^{\nu} \over d\tau} = 0  \nolabel \\
& & \cdots \nolabel \\
& & {d \over d\tau}(r^2\sin^2\theta\dot\phi) = 0
\end{eqnarray}
So, our equations of motion from the metric are
\begin{eqnarray}
{d \over d\tau}\bigg[\bigg(1-{2M \over r}\bigg)\dot t\bigg] &=& 0 \nolabel \\
{d \over d\tau} (r^2\dot \theta) - r^2\sin\theta\cos\theta\dot\phi^2 &=& 0 \nolabel \\
{d \over d\tau} (r^2\sin^2\theta\dot\phi) &=& 0
\end{eqnarray}

We found that angular momentum was conserved in Newtonian motion, allowing us to restrict motion to a plane.  Let's see if this is a solution here - consider letting $\theta$ be fixed at $\theta = {\pi \over 2}$.  So, $\dot \theta = 0$,\footnote{However, $\dot \theta = 0$ does not necessarily mean that all higher derivatives of $\theta$ are zero as well - we will have to determine whether or not this is the case.} and our second equation of motion becomes
\begin{eqnarray}
{d \over d\tau} (r^2\dot \theta) = r^2 \ddot \theta + 2r\dot r \dot \theta = r^2\ddot \theta = 0
\end{eqnarray}
We can take further derivatives of this equation to show that indeed all higher derivatives of $\theta$ are in fact zero.  So, planar motion is in fact possible, and we can take $\theta = {\pi \over 2}$ exactly in all that follows.  

This will make the third of our equations of motion
\begin{eqnarray}
{d \over d\tau}(r^2 \dot \phi) = 0 \qquad \Longrightarrow \qquad r^2 \dot \phi = l
\end{eqnarray}
where $l$ is the constant of integration.  This is again the conservation of angular momentum.  

We can also integrate the first of the equations of motion to get
\begin{eqnarray}
\bigg(1-{2M \over r}\bigg)\dot t = k \label{eq:perihelion7}
\end{eqnarray}
where $k$ is a constant of integration.  

Now let's consider the motion of Mercury using the proper time, so $\tau = s$.  This value for $\tau$ will make the ${ds \over d\tau} = 1$, and therefore we can rewrite the metric as (leaving $d\theta = 0$ because we have fixed $\theta$)
\begin{eqnarray}
& & ds^2 = -\bigg(1-{2M \over r}\bigg) dt^2 + \bigg(1-{2M \over r}\bigg)^{-1} dr^2 + r^2 (d\theta^2 + \sin^2\theta d\phi^2) \nolabel \\
&\Longrightarrow& 1 = -\bigg(1-{2M \over r}\bigg) \dot t^2 + \bigg(1-{2M \over r}\bigg)^{-1} \dot r^2 + r^2 \dot \phi^2
\end{eqnarray}
Plugging in (\ref{eq:perihelion7}) this is
\begin{eqnarray}
1 = -\bigg(1-{2M \over r}\bigg)^{-1} k^2 + \bigg(1-{2M \over r}\bigg)^{-1} \dot r^2 + r^2 \dot \phi^2 \label{eq:perihelion9}
\end{eqnarray}

Now we make the same $u \equiv {1 \over r}$ substitution we did in the previous section (cf (\ref{eq:pariangmomcons4})).  Then, using the definition of $u$ along with (\ref{eq:pariangmomcons9}) and (\ref{eq:pariangmomcons5}), we can rearrange (\ref{eq:perihelion9}) as
\begin{eqnarray}
\bigg({\partial u \over \partial \phi}\bigg)^2 + u^2 = {k^2-1 \over l} + {2M \over l^2} + 2Mu^3
\end{eqnarray}
This can't be easily solved, but if we take the derivative of this with respect to $\phi$ we get
\begin{eqnarray}
{\partial^2 u \over \partial \phi^2} + u - {M \over l^2} = 3Mu^2 \label{eq:Binet2}
\end{eqnarray}
Comparing this with (\ref{eq:Binet}) we can see that the left hand side is the exact same, while the right and side has the $3Mu^2$ term.  

Equation (\ref{eq:Binet2}) is the relativistic version of Binet's equation.  Consider the ratio of the additional $3Mu^2$ term to the constant expression from the original Binet equation:
\begin{eqnarray}
{3Mu^2 \over M/l^2} = 3u^2 l^2 = {3l^2 \over r^2}
\end{eqnarray}
For Mercury this ratio is on the order of $10^{-7}$ - thus the additional term is very small.  

Because the general relativistic correction is very small we can solve the relativistic Binet equation perturbatively.  Again sparing the details of solving it,\footnote{Finding the solution is straightforward using standard approximation techniques.} we merely give the solution:
\begin{eqnarray}
R(\phi) = {L \over 1+E \cos(\phi(1-\epsilon)-B)}
\end{eqnarray}
where $\epsilon$ is the small value $\epsilon \equiv {2M^2 \over l^2}$.  Graphing this gives
\begin{center}
\includegraphics[scale=1]{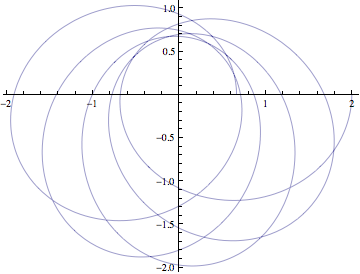}
\end{center}
As you can see, the motion is still elliptical.  However, the $(1-\epsilon)$ term has the effect of shifting the perihelion by a small amount on each revolution.  The value of this precessional shift predicted by general relativity lined up almost exactly with what was observed by Mercury's motion.    

\subsection{Concluding Thoughts on Schwarzschild Geodesics}

As we have illustrated, the true physics of gravity is tied up in the geodesic equations.  We start with the metric, which defines the geometry of the spacetime, and then use the metric to calculate the Levi-Civita connection, which in turn gives us the differential equations for the geodesics.  These differential equations take the place of Newton's gravitational law and Newton's second law (which are also differential equations).  We have seen that in the limit where gravity is weak, the geodesic differential equations of general relativity reduce to Newton's laws.  However, for strong gravitational fields there are relativistic corrections to the Newtonian equations.  

So, whereas in Newtonian physics, Newton's laws provide the differential equations of motion whose solutions define the behavior of a physical system, in general relativity the Levi-Civita connection (as defined by metric) takes over to define the geodesics.  And whereas in Newton's laws the form of the differential equation came about from observation (i.e. $F = G{Mm \over r^2}$), general relativity provides a way of calculating it a priori - using Einstein's field equations $G_{\mu\nu} = \kappa T_{\mu\nu}$.  While Einstein's field equations can rarely be solved directly, the use of an ansatz (an educated guess) can make solving them possible, resulting in the metric in terms of the energy distribution in a region of spacetime.  And once we have the metric, we are able to calculate anything else we want.

\section{Cosmology}

d

\subsection{Energy Momentum Tensor for the Universe}

d

\subsection{General History of the Universe Through Present}

d

\subsection{Observable Matter and Dark Matter}

d

\subsection{Cosmological Constant and Dark Energy}

d

\subsection{The Future of the Universe}

d

\section{General Relativity as a Gauge Theory}
\label{sec:generalrelativityasagaugetheory}

Before concluding this chapter we would like to tie in everything we have done with general relativity into our overall agenda with this series of papers.  One reason for this is that our larger goal is to understand gauge theories.  The idea of a gauge theory\footnote{Review \cite{Firstpaper} if you're not familiar with this.} is that given some Lagrangian $\mathcal{L}$ with some global symmetry, gauging the transformation destroys the symmetry.  Restoring the symmetry requires the introduction of gauge fields $A_{\mu}$ which don't transform tensorially.  Through minimal coupling we introduce the gauge fields into our Lagrangian by replacing partial derivatives with covariant derivatives containing the gauge fields.\footnote{All of this should have a radically deeper meaning to you after having read the contents of this paper so far.}  Then, so that the gauge fields don't become merely background fields that decouple from all interactions we introduce a field strength term $F_{\mu\nu}$.  

A second reason to look at gravity as a gauge theory is that eventually we will want to talk about quantum theories of gravity.  And, as discussed in \cite{Firstpaper}, writing out a Lagrangian for a theory is the first step in quantizing it.  So, having a Lagrangian for general relativity will be a necessary tool when we get to theories of quantum gravity.  

\subsection{Relationship Between Gauge Theory and Geometry - A First Look}
\label{sec:relationshipbetweengaugetheoryandgeometryafirstlook}

First of all, we're calling this section merely "A First Look" because the true relationship between gauge theory and geometry is one of the most profoundly deep and rich fields in all of mathematical physics.  In fact, this entire paper is merely a primer for one to \it begin \rm learning how gauge theory and geometry converge.  We will have more to say about this in the last chapter of this paper, and the next paper in this series will be a much, much more detailed treatment of geometrical gauge theory.  

Before moving on with gravity as a gauge theory, we'll spend some time considering how gauge theories are generally set up for non-gravitational forces (the standard model forces).  This will guide our intuition for gravitation as a gauge theory.  We'll do this by first reviewing the geometry of this section, and then comparing this to how gauge theories were set up in \cite{Firstpaper}.  

\subsubsection{Review of Differential Geometry} 

In chapter \ref{sec:chapwithmet}, after introducing metrics, we discussed the notion of parallel transport.\footnote{You are encouraged to go back and re-read section \ref{sec:connectionsandcovariantderivatives} at this point.}  In order to define the derivative of a tensor field we must have some notion of what "parallel" means.  In other words, we must be able to compare tensors in two different tangent spaces in order to define a derivative in a meaningful way.  This problem was manifested by the fact that partial derivatives don't transform in a tensorial way (cf equation (\ref{eq:transformationofderivativeofvector})):
\begin{eqnarray}
\bigg({\partial v^i \over \partial x^j}\bigg) \longrightarrow {\partial x^k \over \partial x'^j} {\partial x'^i \over \partial x^l} \bigg({\partial v^l \over \partial x^k}\bigg) +  {\partial x^k \over \partial x'^j} {\partial^2 x'^i \over \partial x^k \partial x^l} v^l \label{eq:gaugetheoryandgeometry1}
\end{eqnarray}
Notice that the first term is in fact tensorial while the presence of the second non-linear term makes this transformation non-linear/non-tensorial.  

We can rewrite this in terms of the transformation we are actually invoking:
\begin{eqnarray}
U^i_j \equiv {\partial x^i \over \partial x'^j}
\end{eqnarray}
where the inverse is
\begin{eqnarray}
(U^{-1})^i_j \equiv {\partial x'^i \over \partial x^j}
\end{eqnarray}
In terms of this transformation, (\ref{eq:gaugetheoryandgeometry1}) is
\begin{eqnarray}
\bigg({\partial v^i \over \partial x^j}\bigg) \longrightarrow U^k_j \bigg({\partial v^l \over \partial x^k}\bigg) (U^{-1})^i_l + v^l U^k_j \partial_k (U^{-1})^i_l \label{eq:gaugetheoryandgeometry3}
\end{eqnarray}

In order to correct this nonlinear term, we introduced a \it connection\rm, $\Gamma^i_{jk}$ which we defined to have a certain transformation law:
\begin{eqnarray}
\Gamma^q_{ij} \longrightarrow {\partial x'^p \over \partial x^j} {\partial x'^k \over \partial x^i}{\partial x^q \over \partial x'^l} \Gamma^l_{kp} + {\partial^2 x'^l \over \partial x^i \partial x^j} {\partial x^q \over \partial x'^l}
\end{eqnarray}
Again, the first term in the transformation law is tensorial.  The second, however, is a non-linear term which makes the connection non-tensorial.  We can of course re-write this as:
\begin{eqnarray}
\Gamma^q_{ij} \longrightarrow (U^{-1})^p_j (U^{-1})^k_i U^q_l \Gamma^l_{kp} + U^q_l \partial_i (U^{-1})^l_j \label{eq:gaugethoeryandgeometry4}
\end{eqnarray}

Comparing (\ref{eq:gaugetheoryandgeometry3}) and (\ref{eq:gaugethoeryandgeometry4}) we see the non-linear terms are identical (except for the $v$ in (\ref{eq:gaugetheoryandgeometry3})).  

Next, in order to make a derivative that does transform in a tensorial way (and therefore has a well defined way of comparing tensors in different tangent spaces), we replaced the partial derivative with the covariant derivative (cf equation (\ref{eq:firstexampleofacovariantderivativeofavectorfield}) ff)
\begin{eqnarray}
{\partial v^i \over \partial x^j} \longrightarrow D_j v^i = {\partial v^i \over \partial x^j} + \Gamma^i_{jk} v^k
\end{eqnarray}
Then, using the transformation laws (\ref{eq:gaugetheoryandgeometry3}) and (\ref{eq:gaugethoeryandgeometry4}), it is straightforward (though tedious, cf equations (\ref{eq:covderiviscovwithextratermthatwillvanish}) and (\ref{eq:secondterminvarianceofcovderivderiv})) to show that
\begin{eqnarray}
{\partial v^i \over \partial x^j} + \Gamma^i_{jk}v^k \longrightarrow U^a_j\bigg({\partial v^b \over \partial x^a} + \Gamma^b_{ak}v^k \bigg) (U^{-1})^i_b = {\partial x^a \over \partial x'^j} {\partial x'^i \over \partial x^b}\bigg({\partial v^b \over \partial x^a} + \Gamma^b_{ak}v^k \bigg)
\end{eqnarray}
which has exactly the tensorial form we would expect.  So, by introducing the connection $\Gamma^i_{jk}$ we have preserved the tensorial nature of the derivative, and in doing so allowed a derivative that compares tensors in different tangent spaces in a meaningful way.  

So, to summarize, the problem was that partial derivatives don't transform tensorially in a geometrically non-trivial space:
\begin{eqnarray}
{\partial v^i \over \partial x^j} \longrightarrow {\partial x^k \over \partial x'^j} {\partial x'^j \over \partial x^l} {\partial v^l \over \partial x^k} + {\partial x^k \over \partial x'^j}{\partial^2 x'^i \over \partial x^k \partial x^l} v^l \label{eq:difgeorev1}
\end{eqnarray}
To fix this we define the connection $\Gamma^i_{jk}$ which also transforms in a non-tensorial way:
\begin{eqnarray}
\Gamma^q_{ij} \longrightarrow {\partial x'^p \over \partial x^j} {\partial x'^k \over \partial x^i}{\partial x^q \over \partial x'^l} \Gamma^l_{kp} + {\partial^2 x'^l \over \partial x^i \partial x^j} {\partial x^q \over \partial x'^l} \label{eq:difgeorev1.5}
\end{eqnarray}
And then, forming the covariant derivative 
\begin{eqnarray}
D_j v^i = {\partial v^i \over \partial x^j} + \Gamma^i_{jk} v^k \label{eq:difgeorev2}
\end{eqnarray}
we find that $D_jv^i$ transforms in a tensorial way.  

\subsubsection{Review of Gauge Theory}

We now revisit what we did what gauge theories in \cite{Firstpaper}.  You no doubt suspect that there is a strong correlation between the geometry of this paper and the physics of the previous - if for no other reason than the fact that we have things called "covariant derivatives" in both places.  

In a gauge theory we begin with a Lagrangian $\mathcal{L}$ (or $\mathscr{L}$) that possesses some global symmetry, $\phi \rightarrow U \phi$ where $U$ is some transformation matrix.  The transformation may be a spacetime transformation, like a translation, rotation or a Lorentz boost, or it may be a gauge transformation, like $U(1)$, $SU(2)$, $SU(3)$, etc.  In either case, the field is free to "move" through some space, whether spacetime or gauge space.  The only difference is that spacetime is considered an \it external \rm degree of freedom where the field can move, and the gauge spaces are considered \it internal \rm degrees of freedom where the field can move.  Keep in mind that gauge groups are Lie groups, which are groups that correspond to a manifold - a geometrical space.  So, on one hand the field can move around spacetime, but on the other hand it can "move" around the gauge space in the exact same way.  

It is in this sense that we must realize that gauge theory demands the introduction (or, the assumption) of some very elaborate geometries.  For example, consider a field whose Lagrangian is symmetric under a $U(1)$ transformation, like the Lagrangian for a Dirac spinor (i.e. an electron).  The field is certainly able to move around spacetime (translations, rotations, boosts), but it can also change its $U(1)$ phase.  And because the gauge/Lie group $U(1)$ is topologically/geometrically a unit circle $S^1$ in the complex plane, we can envision the true space the physical field is allowed to move in as not merely the familiar $1+3$ dimensional Minkowski spacetime $\mathcal{M}^4$, but actually the more elaborate space $\mathcal{M}^3 \otimes S^1$.  

So, the degrees of freedom of the field are spacetime translations, spacetime rotations, spacetime Lorentz boosts, and $U(1)$ rotations.  And in demanding the theory to be relativistically invariant, we are demanding that the Lagrangian be invariant under \it global \rm translations, rotations, and Lorentz boosts.  Consider the case of making these local - this means that the translation, rotation, or boost may change from point to point in spacetime.  It should therefore be plain that the difference between a global translations, rotations, and boosts and a local translations, rotations, and boosts is exactly the difference between inertial frames and non-inertial frames!  If the way the field translates, rotates, or boosts depends on the spacetime location, that means that it may be changing speed, direction, etc.  So, gauging the spacetime symmetries of the field is identically what we need to make the jump from special relativity to general relativity (more on this later).  

And the same is true with making a gauge symmetry local.  The field can be transformed through its $U(1)$ space globally, which means that the field is altered the same at every point in space and time.  By gauging the $U(1)$, we are essentially allowing the field to be "non-inertial" in the $S^1$ part of the space it is free to move in.  \it This \rm is precisely the geometrical meaning of gauging a symmetry.  

And, whereas making the spacetime degrees of freedom local allowed for non-inertial frames which made the spacetime geometry non-flat (cf section \ref{sec:theequivalenceprinciple}), thus demanding the introduction of a connection (cf the previous section as summarized in equations (\ref{eq:difgeorev1})-(\ref{eq:difgeorev2})) and curved geodesics (cf equation (\ref{eq:geodesicequationinequivalenceprinciplesection}) at the end of section (\ref{sec:theequivalenceprinciple})), gauging the gauge symmetry allows for "non-inertial" frames in the Lie space, making the gauge geometry "non-flat", thus demanding the introduction of a gauge field (which is identical to a connection).  And because the gauge fields play the role of force carrying particles, they do indeed alter the geodesics of the field.  

So, with gauge theory, we started with a Lagrangian $\mathcal{L}$ that was invariant under a global Lie group ($G$) symmetry.  This is essentially demanding that all "inertial" frames in $\mathcal{M}^4 \otimes G$ are equivalent.  By gauging the symmetry we are allowing "non-inertial frames" in the internal degrees of freedom $G$.  This, as expected (by analogy with (\ref{eq:difgeorev1})), results in a non-linear transformation of the Lagrangian, which demands that we introduce a connection.  Mathematically, it is identical to the connection we discussed in chapter \ref{sec:chapwithmet}, physically it is a gauge field.  

To illustrate this with a simple example, consider the process of making the $U(1)$ local as in \cite{Firstpaper}.  We begin with the Dirac Lagrangian 
\begin{eqnarray}
\mathcal{L} = \bar \psi (i \gamma^{\mu} \partial_{\mu} - m) \psi
\end{eqnarray}
which has a global $U(1)$ symmetry:
\begin{eqnarray}
\psi \longrightarrow e^{i\alpha} \psi \qquad and \qquad \bar \psi \longrightarrow  \bar \psi e^{-i\alpha}
\end{eqnarray}
so under this transformation,
\begin{eqnarray}
\mathcal{L} = \bar \psi (i \gamma^{\mu} \partial_{\mu} - m) \psi \longrightarrow  \bar \psi (i \gamma^{\mu} \partial_{\mu} - m) \psi
\end{eqnarray}
Making the symmetry local had the result of adding a non-linear term when the transformation is carried out:
\begin{eqnarray}
\mathcal{L} =\bar \psi (i \gamma^{\mu} \partial_{\mu} - m) \psi \longrightarrow \bar \psi (i \gamma^{\mu} \partial_{\mu} - m - \gamma^{\mu} \partial_{\mu} \alpha(x)) \psi \label{eq:reviewofgaugetheories1}
\end{eqnarray}
To correct this, we introduced a gauge field $A_{\mu}$ which we defined to transform under $U(x) = e^{i\alpha(x)}$ according to
\begin{eqnarray}
A_{\mu} \longrightarrow U(x) A_{\mu} U^{-1}(x) + iU(x) \partial_{\mu} U^{-1}(x) = A_{\mu} - \partial_{\mu} \alpha(x) \label{eq:reviewofgaugetheories2}
\end{eqnarray}
We introduce the gauge field through the covariant derivative:
\begin{eqnarray}
D_{\mu} \equiv \partial_{\mu} + iA_{\mu}
\end{eqnarray}
and by replacing the partial derivatives in the Lagrangian with covariant derivatives, we restore the symmetry:
\begin{eqnarray}
\mathcal{L} = \bar \psi (i \gamma^{\mu} D_{\mu} - m )\psi \longrightarrow \bar \psi(i\gamma^{\mu} D_{\mu} - m)\psi
\end{eqnarray}

So, as outlined above, the role of the connection in non-flat geometries is absolutely and in all ways identical to the role of the gauge field in particle physics.  Comparing the relevant equations, with geometry we had partial derivatives transform non-tensorially (\ref{eq:gaugetheoryandgeometry3}):
\begin{eqnarray}
{\partial v^i \over \partial x^j} \longrightarrow U^k_j \bigg({\partial v^l \over \partial x^k}\bigg) (U^{-1})^i_l + v^l U^k_j \partial_k (U^{-1})^i_l
\end{eqnarray}
whereas with gauge theory we had a non-tensorial term in our Lagrangian (\ref{eq:reviewofgaugetheories1}):
\begin{eqnarray}
\bar \psi (i \gamma^{\mu} \partial_{\mu} - m) \psi \longrightarrow \bar \psi (i \gamma^{\mu} \partial_{\mu} - m - \gamma^{\mu} \partial_{\mu} \alpha(x)) \psi
\end{eqnarray}
Notice that the non-tensorial term in both these expressions is essentially the same (other than some constants).\footnote{This is more clear if we note that in the first equation, $U$ (like all transformations) can be written as the exponentiation of some generators $T$, $U(x) = e^{i\alpha(x)T}$, so $U\partial U^{-1}$ will be $$ U\partial U^{-1} = e^{i\alpha(x)T}\partial e^{-i\alpha(x)T} = e^{i\alpha(x)T}e^{-i\alpha(x)T}(-i\partial \alpha(x) T) = -iT\partial\alpha(x)$$
where $T =1$ for the $U(1)$ example we are working with here, because the generator for the Abelian $U(1)$ is simply the identity.}

Then, with geometry we introduce a connection that also transforms non-tensorially (\ref{eq:gaugethoeryandgeometry4}):
\begin{eqnarray}
\Gamma^q_{ij} \longrightarrow (U^{-1})^p_j (U^{-1})^k_i U^q_l \Gamma^l_{kp} + U^q_l \partial_i (U^{-1})^l_j 
\end{eqnarray}
whereas with gauge theory we had the non-tensorial gauge field which transforms according to (\ref{eq:reviewofgaugetheories2})
\begin{eqnarray}
A_{\mu} \longrightarrow U(x) A_{\mu} U^{-1}(x) + iU(x) \partial_{\mu} U^{-1}(x) 
\end{eqnarray}
Again, notice that the non-tensorial term in both cases is the same (other than some constants) - $U\partial U^{-1}$.  

Finally, with geometry we build the covariant derivative as
\begin{eqnarray}
D_j v^i = {\partial v^i \over \partial x^j} + \Gamma^i_{jk} v^k
\end{eqnarray}
whereas with gauge theory we had
\begin{eqnarray}
D_{\mu} \phi = \partial_{\mu} \phi + i A_{\mu} \phi
\end{eqnarray}
where $A_{\mu}$ and $\Gamma^i_{jk}$ play exactly the same roles.  

All of this was to provide the beginnings of a link between differential geometry and gauge theory.  The purpose of spending several pages on this, however, was not really to enlighten you to the geometrical aspects of gauge theory (although we hope we have at least begun to do that), but rather to provide some intuition on how we may proceed with making general relativity a gauge theory.  As we said above, we will spend considerably more time talking about the geometry of gauge theory.  

But we have gone as far as we need to go down this road (for now), and we return to general relativity as a gauge theory.  

\subsection{Gravity Coupled to Scalar Fields}
\label{sec:gravitycoupledtoscalarfields}

We're going to begin this topic by coupling gravity to scalar fields.  While not particularly interesting physically, this will provide a nice introduction to the topic of gravity as a gauge theory, and it will be a nice stepping stone for when we begin to study gauge theories from a geometrical context later.  

We'll begin with a real scalar field with Lagrangian
\begin{eqnarray}
\mathcal{L} = -{1 \over 2} \partial^{\mu} \phi \partial_{\mu}\phi - {1 \over 2} m^2\phi^2
\end{eqnarray}
As we mentioned in section (\ref{sec:themodernformulation}), the idea underlying general relativity is that the spacetime metric becomes a dynamic field along with $\phi$.  This means that we don't have our spacetime geometry build in \it a priori\rm.  We therefore define the action not as the integral merely over $dtdxdydz$, but over the invariant volume element $dtdxdydz\sqrt{|g|}$.  We do this by modifying our Lagrangian from $\mathcal{L}$ to $\mathscr{L}$ where
\begin{eqnarray}
\mathscr{L} = \sqrt{|g|} \mathcal{L} = -{1 \over 2} \sqrt{|g|} (\partial^{\mu}\phi\partial_{\mu}\phi + m^2\phi^2) \label{eq:scalarlagrangianwithinvariantvolumeleementingravasgaugetheorysec}
\end{eqnarray}
So, our action is
\begin{eqnarray}
S = \int dtdxdydz \mathscr{L} = \int dtdxdydz \sqrt{|g|} \mathcal{L}
\end{eqnarray}
We will, as in section (\ref{sec:themodernformulation}), work with $\mathscr{L}$ as our Lagrangian. 

We indicated above that the role of the gauge group in general relativity is the collection spacetime translations, rotations, and Lorentz Boosts.  Obviously our scalar Lagrangian here is indeed invariant under such (global) transformations.  Following what we usually do for gauge theories, we would gauge the symmetry in (\ref{eq:scalarlagrangianwithinvariantvolumeleementingravasgaugetheorysec}), introduce a gauge field through a covariant derivative, and then include a kinetic term for the gauge field.  But, it is in trying this that we run up against a small barrier - the scalar field $\phi$ transforms trivially under translations, rotations, and boosts - it has no spacetime indices.  So, the covariant derivative is (boringly) equal to the partial derivative:
\begin{eqnarray}
D_{\mu} = \partial_{\mu}
\end{eqnarray}

However, as we have said, general relativity consists of making the spacetime metric a dynamical field.  And, because we are working with $\mathscr{L}$ instead of $\mathcal{L}$, our matter is certainly coupled to this new dynamical field ($g_{\mu\nu}$).\footnote{Of course, the inner product in $\partial^{\mu}\phi\partial_{\mu} = g_{\mu\nu} \partial^{\mu} \phi \partial^{\nu} \phi$ also couples $\phi$ to the metric.}  So while we're off to an uninteresting start, we can still proceed because it is clear that our action has no term containing a derivative of $g_{\mu\nu}$, and therefore we have the same problem we had with the gauge fields $A_{\mu}$ - it has no dynamics.  The equations of motion prevent the metric from being a dynamical field, and because the whole point of general relativity is that the metric \it is \rm a dynamic field, we should follow what we did in \cite{Firstpaper} and add a kinetic term for the metric.  

But we must be careful\label{pagewherewealsotalkaboutthedifferencebetweenyangmillsfieldsandspacetimefieldsforGRasgaugetheorysection}.  Keep in mind that there is a fundamental difference between the fields $A_{\mu}$ of the standard model and the field $g_{\mu\nu}$ - namely that $A_{\mu}$ is a field \it existing in \rm spacetime, whereas $g_{\mu\nu}$ \it is spacetime itself\rm.  This is our first indication that mirroring what we did for the standard model forces may not be our best bet.  

What we need is a good guess for the kinetic term for $g_{\mu\nu}$.  First of all, it must contain derivatives of the metric in order to be a kinetic term.  Second, because we are forming a Lagrangian, it must be a scalar.  So, what we need is a scalar that contains geometric information and also consists of derivatives of the metric.  

Having gone through the contents of this paper, coming up with a guess should be easy!  The most obvious guess for a scalar that has derivatives of $g_{\mu\nu}$ and contains geometric information is the \it Ricci scalar\rm, $R = g^{\mu\nu}R_{\mu\nu}$, a quantity naturally provided by differential geometry.  So, our guess for the term we add to the Lagrangian to give the metric dynamics is
\begin{eqnarray}
\mathscr{L}_H = \sqrt{|g|} R
\end{eqnarray}
where the $H$ in the subscript stands for Hilbert, who was the first to propose this action.  Now, if we denote (\ref{eq:scalarlagrangianwithinvariantvolumeleementingravasgaugetheorysec}) the "matter Lagrangian", $\mathscr{L}_M$, our total action is
\begin{eqnarray}
S &=& \int dtdxdydz (\mathscr{L}_M + \kappa \mathscr{L}_H)  \nolabel \\
&=& \int dtdxdydz \sqrt{|g|} \bigg( -{1 \over 2} \partial^{\mu} \phi \partial_{\mu}\phi - {1 \over 2} m^2\phi^2 + \kappa R\bigg) \label{eq:fullGRactionforscalarfieldincludinghilbertterm}
\end{eqnarray}
where $\kappa$ is some constant of proportionality.  

From sections \ref{sec:themodernformulation} and \ref{sec:themeaningofthemodernformulation} (specifically equation (\ref{eq:fundamentalexpressionofenergymomentumtensor})) we know that the variation of the matter part of this action with respect to the metric will result in the energy momentum tensor.  So what will the variation of $R$ with respect to $g_{\mu\nu}$ give?  

We ultimately want to find $\delta S_H$, where 
\begin{eqnarray}
S_H = \int dtdxdydz \mathscr{L}_H = \int d^4x \mathscr{L}_H
\end{eqnarray}
and 
\begin{eqnarray}
S_H[g_{\mu\nu}] \longrightarrow S_H[g_{\mu\nu}+\delta g_{\mu\nu}] = S_H[g_{\mu\nu}] + \delta S_H[g_{\mu\nu}]
\end{eqnarray}
so that
\begin{eqnarray}
\delta S_H[g_{\mu\nu}] = S_H[g_{\mu\nu}+\delta g_{\mu\nu}] - S_H[g_{\mu\nu}]
\end{eqnarray}
So, writing this out,
\begin{eqnarray}
\delta S_H &=& \int d^4x \delta (\sqrt{|g|} R) \nolabel \\
&=& \int d^4x \delta (\sqrt{|g|} g^{\mu\nu} R_{\mu\nu}) \nolabel \\
&=& \int d^4x ( g^{\mu\nu}R_{\mu\nu} \delta \sqrt{|g|} + \sqrt{|g|} R_{\mu\nu} \delta g^{\mu\nu} + \sqrt{|g|} g^{\mu\nu} \delta R_{\mu\nu}) \label{eq:deltaSHilbertequals0}
\end{eqnarray}
We'll start with the last term, $\delta R_{\mu\nu}$.  Recall that the Ricci tensor $R_{\mu\nu}$ is a contraction of the Riemann tensor, so
\begin{eqnarray}
\delta R_{\mu\nu} = \delta R^{\rho}_{\mu\rho\nu}
\end{eqnarray}
Then, using the definition of the Riemann tensor (\ref{eq:equationofriemanncurvaturetensor}), this is
\begin{eqnarray}
\delta R^{\rho}_{\mu\rho\nu} &=& \delta( \partial_{\rho} \Gamma^{\rho}_{\nu\mu} - \partial_{\nu}\Gamma^{\rho}_{\rho\mu} + \Gamma^{\rho}_{\rho \sigma}\Gamma^{\sigma}_{\nu\mu} - \Gamma^{\rho}_{\nu\sigma}\Gamma^{\sigma}_{\rho\mu}) \nolabel \\
&=& \partial_{\rho} (\delta \Gamma^{\rho}_{\nu\mu} )- \partial_{\nu} (\delta \Gamma^{\rho}_{\rho\mu}) +\Gamma^{\rho}_{\rho\sigma} (\delta \Gamma^{\sigma}_{\nu\mu}) + (\delta \Gamma^{\rho}_{\rho\sigma}) \Gamma^{\sigma}_{\nu\mu} - \Gamma^{\rho}_{\nu\sigma} (\delta \Gamma^{\sigma}_{\rho\mu}) - (\delta \Gamma^{\rho}_{\nu\sigma})\Gamma^{\sigma}_{\rho\mu} \nolabel \\ \label{eq:deltaRprhomurhonu}
\end{eqnarray}
The obvious step at this point would be to write out the connection coefficients in terms of the metric $g_{\mu\nu}$ to get the variation.  This approach works just fine, but instead we will take a simpler (and more clever) approach.  

Consider the variation of the connection $\Gamma^{\rho}_{\mu\nu}$.  By definition this comes from
\begin{eqnarray}
\Gamma^{\rho}_{\mu\nu} [g_{\alpha\beta}] \longrightarrow \Gamma^{\rho}_{\mu\nu}[g_{\alpha\beta} + \delta g_{\alpha\beta}] = \Gamma^{\rho}_{\mu\nu} [g_{\alpha\beta}] + \delta \Gamma^{\rho}_{\mu\nu} [g_{\alpha\beta}]
\end{eqnarray}
And so
\begin{eqnarray}
\delta \Gamma^{\rho}_{\mu\nu} [g_{\alpha\beta}] = \Gamma^{\rho}_{\mu\nu} [g_{\alpha\beta}+\delta g_{\alpha\beta}] - \Gamma^{\rho}_{\mu\nu} [g_{\alpha\beta}]
\end{eqnarray}
Now, while it is true that $\Gamma^{\rho}_{\mu\nu}$ is not a tensor, we pointed out in section \ref{sec:torsionandliederiv} that the \it difference \rm between two connections is a tensor.  We can therefore take $\delta \Gamma^{\rho}_{\mu\nu}$ to be a type $(1,2)$ tensor.  And because it is a tensor, we can take a covariant derivative.  Using the generalized form of the covariant derivative, equation (\ref{eq:mostgeneralcovderoftensorpossible}).  This is
\begin{eqnarray}
\nabla_{\gamma} (\delta \Gamma^{\rho}_{\mu\nu}) = \partial_{\gamma} (\delta \Gamma^{\rho}_{\mu\nu}) + \Gamma^{\rho}_{\gamma \sigma} (\delta \Gamma^{\sigma}_{\mu\nu}) - \Gamma^{\sigma}_{\gamma\mu}(\delta \Gamma^{\rho}_{\sigma\nu}) - \Gamma^{\sigma}_{\gamma\nu}(\delta \Gamma^{\rho}_{\mu\sigma})
\end{eqnarray}

Now consider the difference in two such terms:
\begin{eqnarray}
\nabla_{\gamma}(\delta \Gamma^{\rho}_{\mu\nu}) - \nabla_{\mu}(\delta \Gamma^{\rho}_{\gamma\nu}) &=& \partial_{\gamma} (\delta \Gamma^{\rho}_{\mu\nu}) - \partial_{\mu} (\delta \Gamma^{\rho}_{\gamma\nu}) + \Gamma^{\rho}_{\gamma \sigma} (\delta \Gamma^{\sigma}_{\mu\nu}) - \Gamma^{\rho}_{\mu \sigma} (\delta \Gamma^{\sigma}_{\gamma\nu}) \nolabel \\
& & -\Gamma^{\sigma}_{\gamma\mu}(\delta \Gamma^{\rho}_{\sigma\nu}) + \Gamma^{\sigma}_{\mu\gamma}(\delta \Gamma^{\rho}_{\sigma\nu}) -  \Gamma^{\sigma}_{\gamma\nu}(\delta \Gamma^{\rho}_{\mu\sigma}) +  \Gamma^{\sigma}_{\mu\nu}(\delta \Gamma^{\rho}_{\gamma\sigma}) \nolabel \\
&=& \partial_{\gamma} (\delta \Gamma^{\rho}_{\mu\nu}) - \partial_{\mu} (\delta \Gamma^{\rho}_{\gamma\nu}) + \Gamma^{\rho}_{\gamma \sigma} (\delta \Gamma^{\sigma}_{\mu\nu}) \nolabel \\
& & - \Gamma^{\rho}_{\mu \sigma} (\delta \Gamma^{\sigma}_{\gamma\nu}) - \Gamma^{\sigma}_{\gamma\nu}(\delta \Gamma^{\rho}_{\mu\sigma}) +  \Gamma^{\sigma}_{\mu\nu}(\delta \Gamma^{\rho}_{\gamma\sigma}) \nolabel \\
\end{eqnarray}
Comparing this to (\ref{eq:deltaRprhomurhonu}) we see that they are the same!  So,
\begin{eqnarray}
\delta R_{\mu\nu} = \delta R^{\rho}_{\mu\rho\nu} = \nabla_{\rho}(\delta \Gamma^{\rho}_{\mu\nu}) - \nabla_{\nu}(\delta \Gamma^{\rho}_{\rho\mu})
\end{eqnarray}

So, the last term in (\ref{eq:deltaSHilbertequals0}) is now
\begin{eqnarray}
\int d^4x \sqrt{|g|} g^{\mu\nu} \delta R_{\mu\nu} &=& \int d^4x \sqrt{|g|} g^{\mu\nu} \big( \nabla_{\rho}(\delta \Gamma^{\rho}_{\mu\nu}) - \nabla_{\nu} (\delta \Gamma^{\rho}_{\rho\mu})\big)
\end{eqnarray}
Then, using the fact that the covariant derivative of the metric vanishes by definition (cf equation (\ref{eq:covderivofmetriciszeroparforgravasgaugesecpar})), this is 
\begin{eqnarray}
\int d^4x \sqrt{|g|} \nabla_{\sigma} \big( g^{\mu\nu} (\delta \Gamma^{\sigma}_{\mu\nu}) - g^{\mu\sigma}(\delta \Gamma^{\rho}_{\rho\mu})\big)
\end{eqnarray}
This is an integral over a total (covariant) derivative and therefore is equal to a boundary term which we can take to be zero (as we always do in physics).  

So, the expression we are trying to find, (\ref{eq:deltaSHilbertequals0}), is now
\begin{eqnarray}
\delta S_H = \int d^4x (g^{\mu\nu} R_{\mu\nu} \delta \sqrt{|g|} + \sqrt{|g|} R_{\mu\nu}\delta g^{\mu\nu}) \label{eq:gravasgaugetheorydeltaSsubHequationwearetryingtosolveequaltozero}
\end{eqnarray}
We now work with the first term.  Evaluating this requires the use of (\ref{eq:variationofsquarerootofabsolutevalueofmetricwithrespecttometric}):
\begin{eqnarray}
{\partial \sqrt{|g|} \over \partial g_{\alpha\beta}} = {1 \over 2} \sqrt{|g|} g^{\alpha\beta}
\end{eqnarray}
from which it is clear that
\begin{eqnarray}
\delta \sqrt{|g|} = {1 \over 2} \sqrt{|g|} g^{\alpha\beta} \delta g_{\alpha\beta}
\end{eqnarray}
So, we can now write (\ref{eq:gravasgaugetheorydeltaSsubHequationwearetryingtosolveequaltozero}) as
\begin{eqnarray}
\delta S_H &=&\int d^4x\bigg( g^{\mu\nu} R_{\mu\nu} {1 \over 2} \sqrt{|g|} g^{\alpha\beta} \delta g_{\alpha\beta} + \sqrt{|g|} R_{\mu\nu} \delta g^{\mu\nu}\bigg)
\end{eqnarray}
Notice that the metric variation part of the first term has lowered indices whereas the metric variation in the second term has raised indices.  We correct this using (\ref{eq:thetraceofthemetricisequaltothedimensionofthemanifold}):
\begin{eqnarray}
g^{\mu\nu}g_{\mu\nu} = n \qquad &\Longrightarrow& \qquad \delta (g^{\mu\nu}g_{\mu\nu}) = 0 \nolabel \\
&\Longrightarrow& \qquad g^{\mu\nu}\delta g_{\mu\nu} + g_{\mu\nu}\delta g^{\mu\nu} = 0 \nolabel \\
&\Longrightarrow& \qquad g^{\mu\nu} \delta g_{\mu\nu} = -g_{\mu\nu} \delta g^{\mu\nu}
\end{eqnarray}
and therefore, finally,
\begin{eqnarray}
\delta S_H &=& \int d^4x\bigg( g^{\mu\nu} R_{\mu\nu} {1 \over 2} \sqrt{|g|} g^{\alpha\beta} \delta g_{\alpha\beta} + \sqrt{|g|} R_{\mu\nu} \delta g^{\mu\nu}\bigg)  \nolabel \\
&=& \int d^4x \bigg( -{1 \over 2}R \sqrt{|g|} g_{\alpha\beta} \delta g^{\alpha\beta} + \sqrt{|g|} R_{\mu\nu} \delta g^{\mu\nu} \bigg) \nolabel \\
&=& \int d^4x \sqrt{|g|} \delta g^{\mu\nu} \bigg( R_{\mu\nu} - { 1\over 2}R g_{\mu\nu}\bigg) \nolabel \\
&=& \int d^4x \sqrt{|g|} \delta g^{\mu\nu} G_{\mu\nu}
\end{eqnarray}
where $G_{\mu\nu}$ is the Einstein tensor (cf equation (\ref{eq:firstandonlydefinitionofEinsteintensor})).  So, amazingly, the the equation of motion for the Hilbert action without a matter Lagrangian is (for arbitrary $\delta g_{\mu\nu}$) is
\begin{eqnarray}
G_{\mu\nu} = 0
\end{eqnarray}
which is Einstein's equation in the absence of matter, as expected (cf equation (\ref{eq:EinsteinsFullFieldEquations})).  

Or, if we use the full action (\ref{eq:fullGRactionforscalarfieldincludinghilbertterm}),
\begin{eqnarray}
\delta S = \delta \int d^4x (\mathscr{L}_M + \kappa \mathscr{L}_H) = \int d^4x (T_{\mu\nu} + \kappa G_{\mu\nu}) = 0
\end{eqnarray}
So, if we use (\ref{eq:definitionofkappaineinsteinfieldequation}) to get the appropriate value for $\kappa$, we have Einstein's field equation exactly (cf equation (\ref{eq:EinsteinsFullFieldEquations})):
\begin{eqnarray}
G_{\mu\nu} = 8\pi T_{\mu\nu}
\end{eqnarray}
Thus, we have confirmed that the Hilbert action is the correct term for the dynamics of the metric, as well as that 
\begin{eqnarray}
S = \int d^4 x(\mathscr{L}_M + \kappa \mathscr{L}_H) 
\end{eqnarray}
is the correct action for scalar fields and gravitation.  

\subsection{Non-Coordinate Bases}
\label{sec:noncoordinatebases}

We would now like to move on to writing a Lagrangian coupling fermions to spacetime via general relativity so as to give fermions gravitational interactions.  Doing this, however, requires a bit of formalism first.  

Thinking back to chapter \ref{sec:differentialmanifolds}, recall that part of our definition of an $n$-dimensional differentiable manifold $\mathcal{M}$ was that, on some patch of the manifold that is (locally) homeomorphic to and open subset of $\mathbb{R}^n$, we can define coordinate functions $\bf x\it$ which map points $p \in \mathcal{M}$ into $\mathbb{R}^n$:
\begin{eqnarray}
p \in \mathcal{M} \longrightarrow \bf x\it(p) \in \mathbb{R}^n
\end{eqnarray}
Of course it may not always be possible to cover the entirety of $\mathcal{M}$ with a single coordinate neighborhood, and therefore there may be neighborhoods on $\mathcal{M}$ that are covered by two different coordinate functions - say $\bf x\it$ and $\bf y\it$.  We then demanded that the function
\begin{eqnarray}
\bf y\it ( x^{-1}(p)): \bf x\it(p) \longrightarrow \bf y\it(p)
\end{eqnarray}
be infinitely differentiable and have infinitely differentiable inverse.  This was the primary content of section \ref{sec:formaldefofmanifolds}.  

Then in section \ref{sec:tangentspacesandframes} and following we discussed the tangent space and cotangent space, which were copies of $\mathbb{R}^n$, which were attached at every point of $\mathcal{M}$.  We then showed that in some neighborhood of $\mathcal{M}$ (that is homeomorphic to an open subset of $\mathbb{R}^n$), we can use the coordinate functions to define frames, or bases, for these spaces.  Namely,
\begin{eqnarray}
{\partial \over \partial x^i}
\end{eqnarray}
spans the tangent space, whereas
\begin{eqnarray}
dx^i
\end{eqnarray}
spans the cotangent space.  Of course these bases satisfy
\begin{eqnarray}
dx^i\bigg({\partial \over \partial x^j}\bigg) = \delta^i_j \label{eq:noncordbasis4}
\end{eqnarray}

We have been able to make considerable progress considering such frames, or bases, of tangent spaces.  However, there is no \it a priori \rm reason from choosing a tangent space basis that coincides with the coordinates in this way.  As you should be well aware from linear algebra, \it any \rm basis of $\mathbb{R}^n$ is just as good as any other, and we are therefore free to choose whatever basis we want.  

With that said, we will therefore consider tangent and cotangent space bases that are not based on the coordinate functions - such frames are aptly called \bf Non-Coordinate Bases\rm.  Returning to the notation where a greek index indicates a spacetime index, let's make the switch from a basis ${\partial \over \partial x^{\mu}}$ to some other basis.  We'll call the new basis $\bf  e\it_a$, where the latin index (instead of a greek index) is because this basis doesn't have any relation to the spacetime coordinates.  Of course, because all bases are equally good as any other, we can write the new basis at any point as a linear combination of the coordinate basis.  In other words we can write
\begin{eqnarray}
\bf  e\it_a =  e_a^{\mu} {\partial \over \partial x^{\mu}} \label{eq:noncoordbasis1}
\end{eqnarray}
When written this way, the non-coordinate basis has both a spacetime index and a non-coordinate basis index (more on this later).  The lower latin index labels the non-coordinate basis vector, while the upper greek index labels the spacetime (manifold) component of that basis vector.  

We can then demand that the non-coordinate basis be orthonormal at each point according to whatever orthonormal may mean on that manifold.  For example, if we are working on a space with an Euclidian metric signature\footnote{This doesn't necessarily mean it has a Euclidian metric - it merely means that each of the diagonal components of the metric have the same sign, positive or negative.}, the inner product between any two of the non-coordinate basis vectors will be (cf equation (\ref{eq:generalformofmetricinnerproduct}))
\begin{eqnarray}
g(\bf  e\it_a, \bf  e\it_b) = g_{\mu\nu}  e^{\mu}_a  e^{\nu}_b =  e^{\mu}_a  e_{b\mu} = \delta_{ab} \label{eq:noncordbasis2}
\end{eqnarray}
whereas if the metric has a Lorentz signature\footnote{Simply meaning that the diagonal elements don't have the same sign.}, this is
\begin{eqnarray}
g(\bf  e\it_a, \bf  e\it_b) = g_{\mu\nu}  e^{\mu}_a  e^{\nu}_b =  e^{\mu}_a  e_{b\mu} = \eta_{ab} \label{eq:noncordbasis3}
\end{eqnarray}
The nomenclature typically used for such an orthonormal non-coordinate basis $\bf  e\it_a$ is a \bf Vielbein\rm.  And, just as it isn't always possible to use a single coordinate system to cover a manifold and therefore isn't always possible to use a single coordinate basis for every point on the manifold, it may not be possible to use a single vielbein to cover an entire manifold.  We will therefore need to talk about transformation laws between vielbein's on coordinate neighborhood overlaps.  

Note that a vielbein is a very simple idea - we haven't done anything profound here.  All we're doing is renaming things, nothing more.  Make sure you understand the simple nature of what we've done in this section so far before moving on - if it seems complicated at all then you're missing something.  

Moving on, we can of course invert (\ref{eq:noncoordbasis1}):\footnote{This of course assumes that $ e\it^{\mu}_a$, which forms an $n\times n$ matrix, is invertible - we will assume this from now on.}
\begin{eqnarray} 
\bf  e\it_a =  e^{\mu}_a{\partial \over \partial x^{\mu}} &\Longrightarrow& ( e^{-1})^a_{\nu} \bf  e\it_a = ( e^{-1})^a_{\nu}  e^{\mu}_a {\partial \over \partial x^{\mu}} \nolabel \\
&\Longrightarrow& ( e^{-1})^a_{\nu} \bf  e\it_a = {\partial \over \partial x^{\nu}} \label{eq:noncordbasis5}
\end{eqnarray}
where we have introduced the inverse vielbein, which has the spacetime lowered and the non-coordinate index raised (the opposite of the vielbein).  The inverse obviously satisfies
\begin{eqnarray}
 e\it^{\mu}_a( e^{-1})^a_{\nu} = \delta_{\nu}^{\mu} \qquad and \qquad  e^{\mu}_a( e^{-1})^b_{\mu} = \delta^b_a \label{eq:noncordbasis7}
\end{eqnarray}

We can also use the inverse vielbein to invert (\ref{eq:noncordbasis2}) and (\ref{eq:noncordbasis3}):
\begin{eqnarray}
g_{\mu\nu}  e^{\mu}_a  e^{\nu}_b = \delta_{ab} &\Longrightarrow& g_{\mu\nu}  e^{\mu}_a  e^{\nu}_b ( e^{-1})^a_{\alpha} ( e^{-1})^b_{\beta} = \delta_{ab}( e^{-1})^a_{\alpha} ( e^{-1})^b_{\beta} \nolabel \\
&\Longrightarrow& g_{\alpha\beta} = \delta_{ab}( e^{-1})^a_{\alpha} ( e^{-1})^b_{\beta} 
\end{eqnarray}
or
\begin{eqnarray}
g_{\alpha\beta} = \eta_{ab}( e^{-1})^a_{\alpha}( e^{-1})^b_{\beta}
\end{eqnarray}
These are extremely powerful relationships between the vielbein's and the spacetime metric - namely they make obvious that the metric can be written in terms of the vielbein and the flat space metric exactly.  

Furthermore, the inverse vielbein allows us to write a non-coordinate basis for the cotangent space:
\begin{eqnarray}
(\bf  e\it^{-1})^a = ( e^{-1})^a_{\mu} dx^{\mu}
\end{eqnarray}
Or inverting this,
\begin{eqnarray}
dx^{\mu} = e^{\mu}_a(\bf  e\it^{-1})^a 
\end{eqnarray}
Obviously this preserves the relationship (\ref{eq:noncordbasis4}):
\begin{eqnarray}
\delta^{\mu}_{\nu} = dx^{\mu}{\partial \over \partial x^{\nu}} &=& e^{\mu}_a(\bf  e\it^{-1})^a ( e^{-1})^b_{\nu} \bf  e\it_b \nolabel \\
&=& e^{\mu}_a ( e^{-1})^b_{\nu} (\bf  e\it^{-1})^a \bf  e\it_b \nolabel \\
&=& e^{\mu}_a ( e^{-1})^b_{\nu} \delta^a_b \nolabel \\
&=& e^{\mu}_a ( e^{-1})^a_{\nu} \nolabel \\
&=& \delta^{\mu}_{\nu}
\end{eqnarray}

We will from now on drop the $-1$ notation from the inverse vielbein.  It should be understood that when the greek index is lowered and the latin index is raised it is inverse, whereas when the greek is raised and the latin is lowered it is not the inverse.  

Next, note that we can express any arbitrary vector in terms of the vielbein.  Consider the vector (in terms of the coordinate basis) $\bf v\it = v^{\mu} {\partial \over \partial x^{\mu}}$.  In addition to switching the coordinate basis as in (\ref{eq:noncordbasis5}), we can write the components of $\bf v\it$ as
\begin{eqnarray}
v^{\mu} \longrightarrow v^a =  e^a_{\mu} v^{\mu}
\end{eqnarray}
The vector is then 
\begin{eqnarray}
\bf v\it = v^a \bf  e\it_a
\end{eqnarray}
Of course plugging in the known values for each of these terms recovers the original vector:
\begin{eqnarray}
\bf v\it = v^a\bf  e\it_a = v^{\mu} e^a_{\mu}  e^{\nu}_a{\partial \over \partial x^{\nu}} = v^{\mu} {\partial \over \partial x^{\mu}} \label{eq:noncordbasis8}
\end{eqnarray}

We can do this same thing for an arbitrary tensor.  For example, we can write
\begin{eqnarray}
T^{\mu\nu\cdots}_{\alpha\beta\cdots} = e^{\mu}_ae^{\nu}_b \cdots e^{\alpha}_ce^{\beta}_d \cdots T^{ab\cdots}_{cd\cdots}
\end{eqnarray}
We want to reiterate at this point that there is nothing profound or even particularly interesting about what we are doing.  We are merely relabeling things - nothing more.  

However, we are now able to begin to get into the real meat of using a non-coordinate basis, or a vielbein - transformation laws.  What is particularly interesting is that, because the vielbein is completely independent of the spacetime manifold coordinates, it can be transformed independently of the spacetime coordinates.  In other words, we can transform the latin indices without worrying about the greek indices.  The meaning of this is that we're changing the vielbein without changing the coordinate functions.  Keep in mind that a vielbein makes no reference whatsoever to the spacetime manifold coordinates.  And therefore at a given point on the spacetime manifold, in choosing a vielbein, we are choosing a completely arbitrary set of basis vectors that don't have anything whatsoever to do with the coordinates.  Therefore, changing, or transforming, to some other completely arbitrary set of basis vectors that also have nothing whatsoever to do with the coordinates doesn't require that we think about the coordinates.  In a sense, by working with the vielbein we have "lifted" ourselves from the manifold into another space.  

To see what this space is, keep in mind that the only rule we are imposing for the vielbein is equation (\ref{eq:noncordbasis3})\footnote{From now on we will not make reference to the Euclidian signature version, but rather talk about the Lorentz signature version only.  The Euclidian analogue is obvious.} - the vielbein must be orthonormal at each point.  So, the only constraint is that the vielbein we transform to must be orthonormal.  Therefore, we can make absolutely any transformation on the vielbein that preserves the relationship (\ref{eq:noncordbasis3}).  In other words, we are allowed any transformation that preserves the Minkowski metric - and we know exactly what types of transformations these are - Lorentz transformations!  So, given any vielbein at any point (which, of course, has nothing to do with the spacetime coordinates at that point), we can transform that vielbein to any other vielbein as long as the transformation is a Lorentz transformation \it on the latin vielbein indices\rm.  So, denoting Lorentz transformations on the vielbein as $\Lambda^a_b$, this means
\begin{eqnarray}
\bf e\it_a \longrightarrow \bf e\it'_a = \Lambda^b_a\bf e\it_b
\end{eqnarray}
and in order to preserve (\ref{eq:noncordbasis7}) we have
\begin{eqnarray}
\bf e\it^a \longrightarrow \bf e\it'^a = (\Lambda^{-1})^a_b \bf e\it^b
\end{eqnarray}
And, as we have said repeatedly, no such transformation has any affect on the actual spacetime coordinates, and therefore no spacetime indices need to be transformed.  In other words, these Lorentz transformations are completely \it internal\rm.  For example in (\ref{eq:noncordbasis5}), this internal Lorentz transformation will take
\begin{eqnarray}
{\partial \over \partial x^{\mu}} = e^a_{\mu} \bf e\it_a \longrightarrow e'^a_{\mu} \bf e\it'_a &=& e^b_{\mu}(\Lambda^{-1})^a_b \Lambda_a^c \bf e\it_c \nolabel \\
&=& e^b_{\mu} \delta^c_b \bf e\it_c \nolabel \\
&=& e^b_{\mu} \bf e\it_b \nolabel \\
&=& {\partial \over \partial x^{\mu}}
\end{eqnarray}
So, indeed it is the case that transformation on this \it internal \rm space have no affect on anything in spacetime.  Furthermore, it should be clear that the internal space we are working with is specifically the space of all Lorentz transformations on the vielbein space!  In other words, it is as if we have attached a copy of the Lorentz group $SO(1,3)$ to every point on the spacetime manifold, and so by performing a Lorentz transformation on the vielbein at every point, we are assigning an element of the Lorentz group to every element of spacetime.  In other words, what we have is a local, or gauged, Lorentz group.  

Furthermore, just as the internal vielbein transformations don't have any affect on coordinate transformations, transformations that take place on the spacetime indices don't have any affect on the vielbein coordinates.  Speaking more physically, as we have said the vielbein at a particular point has absolutely no relation to the coordinates.  Therefore changing the coordinates should obviously have no affect on the vielbein.  Such a spacetime coordinate transformation will only act on the greek spacetime indices.  So, looking at (\ref{eq:noncoordbasis1}), we have for arbitrary coordinate transformation ${\partial x^{\mu} \over \partial x'^{\nu}}$:
\begin{eqnarray}
\bf e\it_a = e^{\mu}_a {\partial \over \partial x^{\mu}} \longrightarrow e'^{\mu}_a {\partial \over \partial x'^{\mu}} &=& e^{\nu}_a {\partial x'^{\mu}\over \partial x^{\nu}} {\partial x^{\alpha} \over \partial x'^{\mu}} {\partial \over \partial x^{\alpha}} \nolabel \\
&=& e^{\nu}_a \delta^{\alpha}_{\nu} {\partial \over \partial x^{\alpha}} \nolabel \\
&=& e^{\nu}_a {\partial \over \partial x^{\nu}} \nolabel \\
&=& \bf e\it_a
\end{eqnarray}

As interesting as all this may be, it is still the case that we've done little more than complicated our notation.  We have demonstrated via transformation laws that there is a rich geometrical structure in using the vielbein's (namely attaching a copy of $SO(1,3)$ to every point on the manifold), but anything we can write with spacetime indices we can write with vielbein indices and vice versa.  However, the real divergence in the two notations comes when we try to take derivatives.  

As usual, when working with arbitrary manifolds there is no automatic way to compare tensors in two different tangent spaces, and we therefore must include a connection term to form the covariant derivative.  But consider the vector in vielbein coordinates rather than spacetime coordinates.  This takes on values in vielbein space $SO(1,3)$, not in spacetime.  Therefore, while the covariant derivative will still need a connection, it clearly can't simply be the spacetime metric's Levi-Civita connection $\Gamma^{\rho}_{\mu\nu}$.  Instead, we introduce a connection that "lives" on the vielbein space, $\omega^a_{\mu b}$ (note that it has one spacetime index and two internal vielbein indices).  Because this is a connection with reference to the \it internal \rm space of Lorentz transformations, which are actually just rotations, we call this connection the \bf Spin Connection\rm.  Then, we take the covariant derivative of $v^a$ using the spin connection:
\begin{eqnarray}
D_{\mu} v^a = \partial_{\mu} v^a + \omega^a_{\mu b} v^b
\end{eqnarray}
Or, more generally (cf (\ref{eq:mostgeneralcovderoftensorpossible}))
\begin{eqnarray}
D_{\mu}T^{ab\cdots}_{cd\cdots} = \partial_{\mu}T^{ab\cdots}_{cd\cdots} + \omega^a_{\mu n} T^{nb\cdots}_{cd\cdots} + \omega^b_{\mu n} T^{an\cdots}_{cd \cdots} + \cdots - \omega^n_{\mu c} T^{ab\cdots}_{nd\cdots} - \omega^n_{\mu d} T^{ab\cdots}_{cn\cdots} - \cdots
\end{eqnarray}

Or, if we had an expression with both a greek spacetime index and a latin internal vielbein index, like $V^{\alpha a}$, the covariant derivative would be
\begin{eqnarray}
D_{\mu} V^{\alpha a} = \partial_{\mu} V^{\alpha a} + \Gamma^{\alpha}_{\mu \nu} V^{\nu a} + \omega^a_{\mu b} V^{\alpha b}
\end{eqnarray}
where $\Gamma^{\alpha}_{\mu\nu}$ is the Levi-Civita connection.  

Now, the whole point of a tensor quantity is that it doesn't depend on how it is written - this ties into the fundamental law of physics that physics shouldn't depend on how we choose to describe the universe.  Therefore there should be some relationship between the spin connection and the Levi-Civita connection.  In other words, there should be a way to relate the following covariant derivatives
\begin{eqnarray}
D_{\mu} v^{\nu} &=& \partial_{\mu} v^{\nu} + \Gamma^{\nu}_{\mu\alpha} v^{\alpha} \nolabel \\
D_{\mu} v^a &=& \partial_{\mu} v^a + \omega^a_{\mu b}v^b
\end{eqnarray}
because they are saying the exact same things.  

To find this relationship, first consider a vector field written in out in an index free fashion:
\begin{eqnarray}
\nabla v &=& (D_{\mu} v^{\nu} )\; dx^{\mu} \otimes {\partial \over \partial x^{\nu}} \nolabel \\
&=& (\partial_{\mu} v^{\nu} + \Gamma^{\nu}_{\mu\alpha} v^{\alpha})\; dx^{\mu} \otimes {\partial \over \partial x^{\nu}} 
\end{eqnarray}
Now do the same thing but use a vielbein index on the vector:
\begin{eqnarray}
\nabla v &=& (D_{\mu} v^a) \; dx^{\mu} \otimes \bf e\it_a \nolabel \\
&=& (\partial_{\mu} v^a + \omega^a_{\mu b}v^b) \; dx^{\mu} \otimes \bf e\it_a 
\end{eqnarray}
Now transform the vielbein index back into a spacetime index:
\begin{eqnarray}
\nabla v &=& (\partial_{\mu} v^a + \omega^a_{\mu b}v^b) \; dx^{\mu} \otimes \bf e\it_a \nolabel \\
&=& \big(\partial_{\mu} (v^{\alpha}e^a_{\alpha}) + \omega^a_{\mu b} (v^{\alpha} e^b_{\alpha})\big) dx^{\mu} \otimes e^{\beta}_a {\partial \over \partial x^{\beta}} \nolabel \\
&=& e^{\beta}_a\big(e^a_{\alpha} \partial_{\mu} v^{\alpha} +v^{\alpha} \partial_{\mu} e^a_{\alpha}+ e^b_{\alpha} \omega^a_{\mu b} v^{\alpha} \big) dx^{\mu} \otimes {\partial \over \partial x^{\beta}} \nolabel \\
&=& (\partial_{\mu} v^{\alpha} + v^{\sigma} e^{\alpha}_a \partial_{\mu} e^a_{\sigma} + e^{\alpha}_a e^b_{\sigma} \omega^a_{\mu b} v^{\sigma}) dx^{\mu} \otimes {\partial \over \partial x^{\alpha}}
\end{eqnarray}
Then, because the tensorial nature of physics demands that $\nabla v$ be the same no matter how we describe things, we must have
\begin{eqnarray}
(\partial_{\mu} v^{\alpha} + v^{\sigma} e^{\alpha}_a \partial_{\mu} e^a_{\sigma} + e^{\alpha}_a e^b_{\sigma} \omega^a_{\mu b} v^{\sigma}) dx^{\mu} \otimes {\partial \over \partial x^{\alpha}}  = (\partial_{\mu} v^{\alpha} + \Gamma^{\alpha}_{\mu\sigma} v^{\sigma})\; dx^{\mu} \otimes {\partial \over \partial x^{\alpha}} 
\end{eqnarray}
From this we get
\begin{eqnarray}
\Gamma^{\alpha}_{\mu\sigma}v^{\sigma} = v^{\sigma} e^{\alpha}_a \partial_{\mu} e^a_{\sigma} + e^{\alpha}_a e^b_{\sigma} \omega^a_{\mu b} v^{\sigma}
\end{eqnarray}
or dropping the $v^{\sigma}$ common to each term,
\begin{eqnarray}
\Gamma^{\alpha}_{\mu\sigma} =  e^{\alpha}_a \partial_{\mu} e^a_{\sigma} + e^{\alpha}_a e^b_{\sigma} \omega^a_{\mu b} 
\end{eqnarray}
Or, inverting this,
\begin{eqnarray}
\omega^a_{\mu b} = e^a_{\nu} e^{\sigma}_b \Gamma^{\nu}_{\mu \sigma} - e^{\sigma}_b \partial_{\mu} e^a_{\sigma}
\end{eqnarray}

So, the complete covariant derivative of a vector field $v^{\mu}$ written in the vielbein basis is
\begin{eqnarray}
D_{\mu} v^a = \partial_{\mu} v^a + \omega^a_{\mu b} v^b = \partial_{\mu} v^a + (e^a_{\nu} e^{\sigma}_b \Gamma^{\nu}_{\mu \sigma} - e^{\sigma}_b \partial_{\mu} e^a_{\sigma}) v^b
\end{eqnarray}

Recall that the point of a covariant derivative is to ensure that the derivative term transforms covariantly, or in a nice tensorial way.  This allows us to see what the transformation law for the connection must be.  Under a spacetime coordinate transformation it should be clear that there will be no problems - the derivative isn't acting on anything with a spacetime index and therefore we won't get any additional terms.  This tells us that the spacetime index on the spin connection $\omega^a_{\mu b}$ transforms tensorially.  

However, under an internal vielbein Lorentz transformation, there will be an additional term because the partial derivative acts on the $v^a$, which has an internal latin index.  So, assuming that the transformation law for the spin connection is a tensorial term plus a non-linear term (as usual), it will be
\begin{eqnarray}
\omega^a_{\mu b} \longrightarrow \omega'^a_{\mu b} = \Lambda^a_c \Lambda^d_b \omega^c_{\mu d} + \Omega^a_{\mu b}
\end{eqnarray}
(where $\Omega^a_{\mu b}$ is the term we want to find), we can write the transformation for the covariant derivative as
\begin{eqnarray}
D_{\mu} v^a \partial_{\mu} v^a + \omega^a_{\mu b} v^b &\longrightarrow& \partial_{\mu} (\Lambda^a_b v^b) + (\Lambda^a_c\Lambda^d_b \omega^c_{\mu d} + \Omega^a_{\mu b}) \Lambda^b_e v^e \nolabel \\
&=& \Lambda^a_b \partial_{\mu} v^b + \Lambda^a_c \Lambda^d_b \Lambda^b_e \omega^c_{\mu d} v^e + v^b \partial_{\mu} \Lambda^a_b + \Lambda^b_e \Omega^a_{\mu b} v^e  \nolabel \\
&=& \Lambda^a_b \partial_{\mu} v^b + \Lambda^a_b \omega^b_{\mu d} v^d + v^b(\partial_{\mu} \Lambda^a_b + \Lambda^c_b \Omega^a_{\mu c}) \nolabel \\
&=& \Lambda^a_b (D_{\mu} v^b) + v^b(\partial_{\mu} \Lambda^a_b + \Lambda^c_b \Omega^a_{\mu c})
\end{eqnarray}
So, we get a covariant transformation as long as the last term vanishes:
\begin{eqnarray}
\partial_{\mu} \Lambda^a_b + \Lambda^c_b \Omega^a_{\mu c} = 0
\end{eqnarray}
or
\begin{eqnarray}
\Omega^a_{\mu c} =- (\Lambda^{-1})^b_c \partial_{\mu} \Lambda^a_b
\end{eqnarray}
And so the spin connection transforms according to
\begin{eqnarray}
\omega^a_{\mu b} \longrightarrow \omega'^a_{\mu b} = \Lambda^a_c \Lambda^d_b \omega^c_{\mu d} - (\Lambda^{-1})^b_c \partial_{\mu} \Lambda^a_b  \label{eq:GRferm6}
\end{eqnarray}
under an internal Lorentz transformation.  

We now have the tools to discuss how fermions can be coupled to gravity.  

\subsection{Gravity Coupled to Fermions}
\label{sec:gravitycoupledtofermions}

We have done the bulk of the work necessary to couple fermions to spacetime in the previous sections, and we will therefore treat this topic briefly.  We know that after gauging the symmetry and introducing a covariant derivative we will need a field strength.  We worked out what this will be in section \ref{sec:gravitycoupledtoscalarfields} - the Hilbert Lagrangian
\begin{eqnarray}
\mathscr{L}_H = \sqrt{|g|} R
\end{eqnarray}
where $R$ is the Ricci curvature scalar.  

Next, we look at the Lagrangian for a Dirac fermion (including the invariant volume form part):
\begin{eqnarray}
\mathscr{L}_D = \sqrt{|g|} \bar \psi (i \gamma^{\mu}\partial_{\mu} - m)\psi \label{eq:GRferm1}
\end{eqnarray}

As discussed above in the Review of Gauge Theory part of section \ref{sec:relationshipbetweengaugetheoryandgeometryafirstlook}, the gauge group for general relativity will be the Lorentz group.  We'll walk through the usual steps in gauging this symmetry with hopes that doing so makes our exposition easier to follow.  

The fermions will transform under the spinor representation of the Lorentz group (cf \cite{Firstpaper}, section 3.1.5), called the \bf spinor representation\rm, and we denote a general Lorentz transformation in the spinor representation as $S(\Lambda)$: so
\begin{eqnarray}
\psi \longrightarrow S(\Lambda) \psi \qquad and \qquad \bar \psi \longrightarrow \bar \psi S^{-1}(\Lambda)
\end{eqnarray}  
So, under a \it global \rm Lorentz transformation the Dirac Lagrangian (\ref{eq:GRferm1}) will transform as
\begin{eqnarray}
\mathscr{L}_D = \sqrt{|g|} \bar \psi (i \gamma^{\mu}\partial_{\mu} - m)\psi &\longrightarrow& \sqrt{|g|} \bar \psi  S^{-1}(\Lambda) (i \gamma^{\mu}\partial_{\mu} - m)S(\Lambda) \psi \nolabel \\
&=&\sqrt{|g|} \bar \psi S^{-1}(\Lambda) S(\Lambda) (i \gamma^{\mu}\partial_{\mu} - m)\psi  \nolabel \\
&=& \sqrt{|g|} \bar \psi (i \gamma^{\mu}\partial_{\mu} - m)\psi \nolabel \\
&=& \mathscr{L}_D
\end{eqnarray}
So, $\mathscr{L}_D$ is indeed invariant under a global Lorentz transformation.  

However, if we gauge the symmetry and let $S(\Lambda)$ depend on spacetime (a \it local \rm Lorentz transformation), we have (suppressing the $(\Lambda)$ for notational simplicity)
\begin{eqnarray}
\mathscr{L}_D = \sqrt{|g|} \bar \psi (i \gamma^{\mu}\partial_{\mu} - m)\psi &\longrightarrow& \sqrt{|g|} \bar \psi  S^{-1} (i \gamma^{\mu}\partial_{\mu} - m)S \psi \nolabel \\
&=& \sqrt{|g|} \bar \psi S^{-1} (i \gamma^{\mu}S\partial_{\mu} + i \gamma^{\mu} (\partial_{\mu}S) - m) \psi \label{eq:GRferm3}
\end{eqnarray}
which as expected has a non-linear term that we need to cancel.  We achieve this cancellation by introducing a connection, or a gauge field, via a covariant derivative.  We assume the form of this covariant derivative to be
\begin{eqnarray}
D_{\mu} = \partial_{\mu} + \Omega_{\mu} \label{eq:GRferm2}
\end{eqnarray}
where $\Omega_{\mu}$ is the connection we have yet to determine.  

However it is at this point that we need to be careful.  The close analogy we have been maintaining between the differential geometry developed in this paper the gauge theory formalism we developed in the previous paper would indicate that because we are coupling the fermions to spacetime, we should use the connection of spacetime, the Levi-Civita connection $\Gamma^{\rho}_{\mu\nu}$ as our gauge field.  However this turns out to be incorrect.  

The reason this is incorrect is that the Levi-Civita connection acts on vectors with a spacetime index.  For example in section \ref{sec:connections} (i.e. equation (\ref{eq:firstexampleofacovariantderivativeofavectorfield})) the Levi-Civita connection was being used to form the "gauge field", or covariant derivative, for a vector with indices relating to the coordinates of the manifold.  

But our fermion fields $\psi$ and $\bar \psi$ don't have spacetime manifold indices.  As we discussed in \cite{Firstpaper}, the degree of freedom of a fermion, called spin, is not a rotation through spacetime.  Rather, it is a rotation through "spinor space".  This is an \it internal \rm degree of freedom.  Therefore the Levi-Civita connection is not the tool we need (at least, not exactly).  

However, as you no doubt have realized, this idea of an internal degree of freedom can be handled extremely well with the vielbein formalism developed in the previous section (\ref{sec:noncoordinatebases}).  The essential idea of that section was to introduce a vielbein, or basis at each point that was unrelated to the coordinates of the spacetime manifold, and by writing quantities in terms of the vielbein, we were working with an internal space.  This lead to the notion of a spin connection which allowed us to form a covariant derivative for objects that had no spacetime index but rather had an internal index.  And that is \it exactly \rm what we have with fermions!  They have no spacetime index\label{theyhavenospacetimeindexbuttheyhaveaninternalindex}, but they have an internal \it spin \rm index.  Therefore, by simply replacing $\Lambda^a_b$ from section \ref{sec:noncoordinatebases} with the spinor representation of the Lorentz group, $S^a_b$, we can form a covariant derivative for $\psi$.  

First, we note that like any transformation, we can write a Lorentz transformation $S$ as the exponentiation of the Lorentz algebra generators (where the generators are the spinor representation generators), which we denote $\Sigma^{ab}$.  So
\begin{eqnarray}
S = e^{{ i \over 2} \alpha_{ab}\Sigma^{ab}}
\end{eqnarray}
(the factor of $1/2$ is conventional) where $\alpha_{ab}$ are the parameters for the Lorentz transformation.\footnote{It is straightforward to show that both $\Sigma^{ab}$ will be an antisymmetric $4 \times 4$ matrix, which has 6 real independent components corresponding to three rotations and three boosts.}

Then, working with the spin connection for our internal spinor space, we expand the connection in terms of the generators (from (\ref{eq:GRferm2})):
\begin{eqnarray}
\Omega_{\mu} = {i \over 2} \omega^a_{\mu b} \Sigma_a^b
\end{eqnarray}
(where we have lowered one of the indices on the generator matrix to keep our notation on the spin connection consistent with the previous section - this is no problem because we can of course raise and lower indices at will with the metric.  

So, our covariant derivative is 
\begin{eqnarray}
D_{\mu} \psi = \partial_{\mu} \psi + {i \over 2} \omega^a_{\mu b} \Sigma^b_a \psi \label{eq:GRferm5}
\end{eqnarray}
We leave it to you to show that replacing the partial derivative in $\mathscr{L}_D$ with (\ref{eq:GRferm5}) along with the transformation for the spin connection (\ref{eq:GRferm6}) exactly cancels the non-linear term, making $\mathscr{L}_D$ with the spin connection covariant derivative invariant under the local Lorentz transformation.  

So, finally, we can write out the action for fermions coupled to gravity:
\begin{eqnarray}
S &=& \int d^4x \sqrt{|g|} \big( \bar \psi (i \gamma^{\mu} D_{\mu} - m) \psi + \kappa R\big) \nolabel \\
&=& \int d^4x \sqrt{|g|} \big( \bar \psi (i \gamma^{\mu} (\partial_{\mu} + {i \over 2} \omega^a_{\mu b} \Sigma^b_a) - m) \psi + \kappa R\big) 
\end{eqnarray}
Or for a theory containing both scalar fields and bosons,
\begin{eqnarray}
S = \int d^4x \sqrt{|g|} \bigg( - {1 \over 2} \partial^{\mu} \phi \partial_{\mu} \phi - {1 \over 2} m^2 \phi^2 + \bar \psi (i \gamma^{\mu} (\partial_{\mu} + {i \over 2} \omega^a_{\mu b} \Sigma^b_a) - m) \psi + \kappa R\bigg) \nolabel \\ \label{eq:fullGRasgaugetheoryLagrangianscalarsandfermions}
\end{eqnarray}

\section{References and Further Reading}

The primary general sources for this chapter were \cite{carroll} and \cite{dinverno}.  Most of the discussions of the meaning of the various concepts came from \cite{baez}.  The discussions of the equivalence principle and Newtonian correspondences came from \cite{carroll}.  The discussions of solutions to the geodesic equations came from \cite{dinverno}.  We following \cite{dirac} to find the Schwarzschild solution.  We followed \cite{goldstein} and \cite{taylor} in discussing the stress tensor and energy momentum tensor, and \cite{lovelock} in deriving the modern formulation of the energy momentum tensor.  For general relativity as a gauge theory we used \cite{carroll} and \cite{nakahara}.  

For further reading in special relativity we recommend \cite{chamseddine}, \cite{french}, \cite{naberSR}, and \cite{woodhouseSR}.  For further reading in general relativity we recommend \cite{besse}, \cite{hartle}, \cite{misner}, \cite{shutzGR}, \cite{wald}, and \cite{woodhouseGR}

\chapter{Concluding Thoughts on Geometry, Topology, Gauge Theories, and Gravity}
\label{sec:finalchap}

We'll conclude with a few comments relating to the gauge theory of general relativity that, while not necessary to understanding Einstein's class theory or the derivation or meaning of (\ref{eq:fullGRasgaugetheoryLagrangianscalarsandfermions}), will help tremendously in preparing us for where we are going with this series.  

You may have noticed that we did a slight slide of hand in our derivation of the spin connection in the previous section.  Specifically, on page \pageref{theyhavenospacetimeindexbuttheyhaveaninternalindex}, we pointed out that the similarity between the "internal" spinor space and the "internal" vielbein space of section \ref{sec:noncoordinatebases} (in the paragraph beginning with "However, as you no doubt ...").  But, recall that the idea of a vielbein was that $\bf e\it_a$ forms a basis for the \it tangent space \rm of the manifold.  They are nothing more than spacetime vectors defined at a point - and obviously the spinor space is not the same as a tangent space.  So we made a fundamental change from section \ref{sec:noncoordinatebases} to section \ref{sec:gravitycoupledtofermions}.  

Recall that in section \ref{sec:noncoordinatebases} we commented that by introducing the vielbein we have, in essence, attached a copy of the Lorentz group $SO(1,3)$ to every point on the spacetime manifold $\mathcal{M}$.  However, this additional space arose naturally, even necessarily, from the fact that the basis for the tangent space at a point is completely arbitrary and any basis will work as well as any other, and therefore the $SO(1,3)$ in this framework acts on the basis \it vectors\rm, and is therefore the vector representation of $SO(1,3)$, not the spinor representation.  However with the vielbein that lead to the spin connection, the internal indices are not simply elements in the tangent space - they are in spinor space.  And, unlike the tangent space which arises necessarily from the manifold, the addition of a spinor space is somewhat artificial - there is no pre-existing reason to add a spinor space to a manifold.  

So, let's take stock of what we have.  We have a manifold\footnote{In physical language this is spacetime.}, and naturally associated with the manifold is a tangent space and a cotangent space.  Then there are tensors of various ranks that can live in the tangent and cotangent space\footnote{These are our physical fields.}.  In order to take the derivatives of tensor fields on the manifold we must be able to compare tensors in different tangent spaces.  We do this by introducing the Levi-Civita connection through the covariant derivative.  The covariant derivative then replaces the partial derivative for differentiating tensor fields that live in the vector and covector space of the manifold.  
Also, as a direct result of the structure of the tangent space at every point of the manifold, we have a symmetry associated with our ability to choose any basis of the tangent space at each point.  All that is required is that the basis be orthonormal - this leads to the symmetry being $SO(1,3)$ (or $SO(n,m)$ for arbitrary manifold with Lorentz signature and $SO(n)$ for arbitrary manifold with Euclidian signature).  We can treat this $SO(1,3)$ etc. as an \it internal \rm degree of freedom, but again this degree of freedom has arisen very naturally from the structure of the manifold - no new ideas are introduced in arriving here.  Then, as we saw, we can express any spacetime tensor, rather than in terms of the manifold coordinates, in terms of this internal "vielbein" space.  Again, doing this doesn't require any new ideas - this has all risen naturally from the structure of the manifold.  

On the other hand, as we have shown via the spin connection, we can go through all of this again but in a way that doesn't arise naturally from the structure of the manifold.  We can start with some manifold with all the same things as above (tangent and cotangent spaces, tensors, a tangent space induced vielbein space, etc.).  But, we may then introduce something other than merely the transformation group on the tangent space basis vectors.  We may introduce, say, the spinor representation of $SO(1,3)$ as we did in the previous section, or we may introduce absolutely any other space we want.  We may write a basis at each point for our "artificial" space we've attached.  We can then express any element of this space in terms of the basis we have chosen at each point.  

So we have two different types of "things" attached to our manifold.  One is the tangent/cotangent space structure which arises naturally and is automatically built in to our geometry.  The other is a somewhat "artificial" space we attach to the manifold.  In both cases we can talk about fields "living" in a vector space that is defined at every point of the manifold.  Introducing some new notation, we will refer to the former the \bf Tangent Bundle \rm (which consists of the tangent space, the cotangent space, and all the higher dimensional tensor spaces resulting from this in the natural way), and the latter "artificial" spaces will be referred to as \bf Fibre Bundles\rm.  

Looking back to page \pageref{firstmentiontangentbundlesthisisforthesectionongravasgaugetheory} in section \ref{sec:tangentspacesandframes}, we briefly mentioned this idea.  Specifically, if our manifold is $\mathcal{M}$, the tangent bundle $T\mathcal{M}$ is the product $\mathcal{M}\otimes \mathbb{R}^n$ (cf equation (\ref{eq:firsttimewementionthetangentbundlethisisforGRasgaugetheoyrsection})).  For these more general fibre bundles we are introducing now, if the space we are artificially attaching is denoted $\mathcal{G}$, the total space is (sort of\footnote{This "sort of" is enormously important and we will discuss it in much, much, \it much \rm greater depth in the next paper in this series.  In reality the total space is only \it locally \rm $\mathcal{M}\otimes \mathcal{G}$ - but more on that (including what that means) later.}) $\mathcal{M}\otimes \mathcal{G}$.  And, just as an individual element of the tangent bundle is a single tangent space at a point, a single element of the fibre bundle is a single \bf fibre \rm at a point.  So, the copy of $\mathcal{G}$ at $p \in \mathcal{M}$ is the \it fibre \rm at $p$.  The underlying manifold $\mathcal{M}$ is called the \bf Base Space \rm of the total structure, which we call the \bf Total Space\rm.  

Another important point is that the total spaces, consisting of the base space, the tangent bundle, and the fibre bundles, can genuinely be viewed as single larger spaces.  As we will see, the details on how everything works together can be very complicated, but it is still the case that these total spaces are well defined geometries that we can view as a single geometrical object.  

We also mentioned on page \pageref{firstmentiontangentbundlesthisisforthesectionongravasgaugetheory} the idea of a \bf projection map\rm.  The idea was that, at a given point $p \in \mathcal{M}$ there is a copy of the tangent space $\mathbb{R}^n$, and therefore any tangent vector at $p$, denoted $(p, \bf v\it)$, can be "projected" to the point of $\mathcal{M}$ the vector is attached to - or in other words $\bf v\it$ is projected to the point of $\mathcal{M}$ such that $\bf v\it \in T_p\mathcal{M}$.  We call this projection map $\pi$:
\begin{eqnarray}
\pi: T_p\mathcal{M} &\longrightarrow& \mathcal{M} \nolabel \\
(p,\bf v\it) &\longmapsto& p
\end{eqnarray}
Then, we can use the inverse of $\pi$ at every point to recover the entire tangent space:
\begin{eqnarray}
\pi^{-1}(p) = T_p\mathcal{M}
\end{eqnarray}

We will make extensive use of the same idea for general fibre bundles.  At every point $p \in \mathcal{M}$ we can attach the vector space $\mathcal{G}$.  Then, some element $\bf g\it \in \mathcal{G}$ cane be projected down to the point to which it is attached with $\pi$, and from $p \in \mathcal{M}$ we can recover the entire fibre:
\begin{eqnarray}
\pi^{-1}(p) = \mathcal{G}
\end{eqnarray}

Again, we want to emphasize that the \it difference \rm between what we are calling the tangent bundle and fibre bundle is that the tangent bundle is intimately associated with the base space manifold, whereas the fibre bundle is not - the fibre bundle is defined entirely independent of the base space.  So, a total space $\mathcal{S}$ with base manifold $\mathcal{M}$, tangent bundle $T\mathcal{M}$, and fibre bundle $\mathcal{G}$ may be written\footnote{Again, the $\mathcal{G}$ part of this may only \it locally \rm have a $\mathcal{G}$ attached to it in this way - for now don't worry about it.  We just want to mention this for readers who already have some familiarity with these concepts.}
\begin{eqnarray}
\mathcal{S} = \mathcal{M} \otimes T\mathcal{M} \otimes \mathcal{G}
\end{eqnarray}
Or, if we want, we can attach multiple fibre bundles:
\begin{eqnarray}
\mathcal{S} = \mathcal{M} \otimes T\mathcal{M} \otimes \mathcal{G}_1 \otimes \mathcal{G}_2 \otimes \cdots
\end{eqnarray}
You can likely imagine that the total space
\begin{eqnarray}
\mathcal{S} = \mathcal{M}^4 \otimes T\mathcal{M}^4 \otimes SU(3) \otimes SU(2) \otimes U(1)
\end{eqnarray}
which is of course viewed (in this paradigm) as a single geometrical space, is particularly interesting.  

As we noted in section \ref{sec:noncoordinatebases}, an arbitrary vector with spacetime indices $v^{\mu}$ can be written in terms of the vielbein indices, $v^a = v^{\mu} \bf e\it_{\mu}$.  This is a direct consequence of the fact (that we have repeated several times in this section) that the vielbein space is a direct consequence of the manifold itself and is intimately tied to it.  For an arbitrary fibre bundle, however (as we have also repeated several times in this section), there is absolutely no necessary correlation whatsoever between the geometry of the base manifold and the geometry of the fibre.  Therefore it isn't in general possible to write a spacetime vector in terms of the basis for the fibre.  This brings us to the notion of two fundamentally different types of fields (cf the comparison between gravity and electromagnetism starting on page \pageref{pagewherewefirststartcmparinggravitytoeandminequivalenceprinciplesection} in section \ref{sec:theequivalenceprinciple})

The first type of field are the fields we have been discussing all along in these notes - tensor fields that have spacetime indices.  These are fields that live in the $T\mathcal{M}$ and the higher dimensional tensor spaces that we form naturally from $T\mathcal{M}$.  Vectors, forms, matrices, and so on - anything with a spacetime index is such a field.  Again, these are fields that "live" in the tangent and cotangent spaces, which (again) are spaces that are deeply and intimately linked to the manifold itself.  

The second type of field are fields that "live" in fibre bundles.  These are fields that don't necessarily carry any spacetime indices and therefore can't be expressed as tensors with spacetime indices.  These are fields that don't arise as a result of any natural structure on the base manifold - they must be put "on top of" the manifold.  

We already discussed this distinction a bit in sections \ref{sec:theequivalenceprinciple} and page \pageref{pagewherewealsotalkaboutthedifferencebetweenyangmillsfieldsandspacetimefieldsforGRasgaugetheorysection} of section \ref{sec:gravitycoupledtoscalarfields}.  The content of the equivalence principle can be stated more mathematically by simply saying that gravity is entirely a result of things relating to the first type of field.  On the other hand, as we mentioned in section \ref{sec:theequivalenceprinciple}, there is no equivalence principle for the other forces, like electromagnetism.  This can be stated more mathematically by simply saying that the other forces (like electromagnetism) are the result of things relating to the second type of fields.  

As usual, we are skimming the peaks of mountains here.  There is much, much more we can say about all of this, and indeed much more that we will say.  The notion of fibre bundles over manifolds and the types of structures that can be built out of and on the resulting total spaces provides one of the richest and most profound areas in all of mathematics and physics, as well as providing a way of revealing deep relationships between gauge theories and general relativity and between geometry and topology. 

And so it is here that we conclude these notes.  We have outlined the geometrical and topological ideas necessary to make sense (at least at an introductory level) of general relativity and several aspects of electromagnetic theory.  And, by applying the mathematical ideas to the physical situations we have, we have gained a shadowy glimpse into much fuller and richer structure of fundamental physical theories.  The next paper in this series will begin to sharpen this glimpse, and by its end we should have a much more solid grasp of the profound relationship between physics and mathematics.

\end{document}